\documentclass{sig-alternate-05-2015}

\usepackage{tikz}
\usetikzlibrary{decorations.pathreplacing}
\usetikzlibrary{patterns}
\usepackage{subcaption}

\usepackage[utf8]{inputenc} 
\usepackage[T1]{fontenc}

\usepackage[vlined, ruled, commentsnumbered, linesnumbered]{algorithm2e}
\usepackage{url}
\usepackage{color}
\usepackage{listings}
\usepackage{enumitem}

\usepackage{amsmath}
\usepackage{amssymb}
\usepackage{xspace}

\usepackage[binary-units]{siunitx}
\usepackage{booktabs}
\usepackage{tabularx}
\usepackage{placeins}
\usepackage{balance}

\usepackage{needspace}
\usepackage{fixltx2e}

\DeclareSIUnit{\byte}{\text{\ensuremath{Bytes}}}

\newcommand{\fig}{Figure\xspace}

\newcommand{\tab}{Table\xspace}

\newcommand{\Sec}{Section\xspace}

\newcommand{\Gl}{Guideline\xspace}
\newcommand{\Gls}{Guidelines\xspace}
\newcommand{\etest}{Expectation Test\xspace}
\newcommand{\eg}{e.g.\@\xspace}
\newcommand{\ie}{i.e.\@\xspace}
\newcommand{\etal}{\textit{et~al.\@}\xspace}

\newcommand{\cf}{cf.\@\xspace}
\newcommand{\vs}{vs.\@\xspace}
\newcommand{\versus}{versus\@\xspace}

\newcommand{\mpibarrier}{\texttt{MPI\_Barrier}\xspace}
\newcommand{\mpisend}{\texttt{MPI\_\-Send}\xspace}
\newcommand{\mpirecv}{\texttt{MPI\_\-Recv}\xspace}
\newcommand{\mpibcast}{\texttt{MPI\_\-Bcast}\xspace}
\newcommand{\mpiallgather}{\texttt{MPI\_\-Allgather}\xspace}

\newcommand{\mpipack}{\texttt{MPI\_\-Pack}\xspace}
\newcommand{\mpiunpack}{\texttt{MPI\_\-Unpack}\xspace}

\newcommand{\mpicontig}{\texttt{MPI\_\-Type\_\-contiguous}\xspace}
\newcommand{\mpivector}{\texttt{MPI\_\-Type\_\-vector}\xspace}
\newcommand{\mpihvector}{\texttt{MPI\_\-Type\_\-hvector}\xspace}
\newcommand{\mpiindex}{\texttt{MPI\_\-Type\_\-indexed}\xspace}
\newcommand{\mpiblock}{\texttt{MPI\_\-Type\_\-create\_\-indexed\_\-block}\xspace}
\newcommand{\mpistruct}{\texttt{MPI\_\-Type\_\-create\_\-struct}\xspace}
\newcommand{\mpiresized}{\texttt{MPI\_\-Type\_\-cre\-ate\_\-resized}\xspace}
\newcommand{\mpicommit}{\texttt{MPI\_\-Type\_\-commit}\xspace}

\newcommand{\mpiint}{\texttt{MPI\_\-INT}\xspace}
\newcommand{\mpichar}{\texttt{MPI\_\-CHAR}\xspace}
\newcommand{\mpibyte}{\texttt{MPI\_\-BYTE}\xspace}
\newcommand{\mpidouble}{\texttt{MPI\_\-DOUBLE}\xspace}
\newcommand{\mpishort}{\texttt{MPI\_\-SHORT}\xspace}

\newcommand{\dtfont}[1]{\texttt{#1}}

\newcommand{\dtcontig}{\dtfont{Contiguous}\xspace}
\newcommand{\dttiled}{\dtfont{Tiled}\xspace}
\newcommand{\dtblock}{\dtfont{Block}\xspace}
\newcommand{\dtbucket}{\dtfont{Bucket}\xspace}
\newcommand{\dtalternating}{\dtfont{Alternating}\xspace}
\newcommand{\dttiledhet}{\dtfont{Tiled-heterogeneous}\xspace}
\newcommand{\dttiledstruct}{\dtfont{Tiled-struct}\xspace}

\newcommand{\dtstiled}{\ensuremath{AB}\xspace}
\newcommand{\dtsblock}{\ensuremath{A\bar{B}}\xspace}
\newcommand{\dtsbucket}{\ensuremath{\bar{A}B}\xspace}
\newcommand{\dtsalternating}{\ensuremath{\bar{A}\bar{B}}\xspace}
\newcommand{\dtstiledhet}{\ensuremath{AB\bar{T}}\xspace}

\newcommand{\dtdtiled}{\dtfont{Tiled\,(\dtstiled)}\xspace}
\newcommand{\dtdblock}{\dtfont{Block\,(\dtsblock)}\xspace}
\newcommand{\dtdbucket}{\dtfont{Bucket\,(\dtsbucket)}\xspace}
\newcommand{\dtdalternating}{\dtfont{Alternating\,(\dtsalternating)}\xspace}
\newcommand{\dtdtiledhet}{\dtfont{Tiled-hetero\-gene\-ous\,(\dtstiledhet)}\xspace}

\newcommand{\ddtcontig}{\dtfont{Contiguous-subtype}\xspace}

\newcommand{\ddttiledvector}{\dtfont{Tiled-vector}\xspace}
\newcommand{\ddtvectortiled}{\dtfont{Vector-tiled}\xspace}
\newcommand{\ddtblockindexed}{\dtfont{Block-indexed}\xspace}
\newcommand{\ddtalternatingindexed}{\dtfont{Alternating-indexed}\xspace}
\newcommand{\ddtalternatingrepeated}{\dtfont{Alternating-repeated}\xspace}
\newcommand{\ddtalternatingstruct}{\dtfont{Alternating-struct}\xspace}
\newcommand{\ddtrowcolfullindexed}{\dtfont{RowCol-fully-indexed}\xspace}
\newcommand{\ddtrowcolcontiguousandindexed}{\dtfont{RowCol-contiguous-and-in\-dexed}\xspace}
\newcommand{\ddtrowcolstruct}{\dtfont{RowCol-struct}\xspace}

\newcommand{\variantone}{\textbf{Variant~1}\xspace}
\newcommand{\varianttwo}{\textbf{Variant~2}\xspace}

\newcommand{\contig}{\mathsf{contig}\xspace}
\newcommand{\packed}{\mathsf{packed}\xspace}
\newcommand{\normal}{\mathsf{normal}\xspace}
\newcommand{\mpi}{\mathsf{MPI}\xspace}

\newcommand{\guidelt}{\ensuremath{\preceq}\xspace}
\newcommand{\guidesim}{\ensuremath{\simeq}\xspace}

\newcommand{\mvapich}{MVAPICH\xspace}
\newcommand{\infiniband}{InfiniBand\xspace}
\newcommand{\gcc}{gcc\xspace}
\newcommand{\mpirun}{\texttt{mpirun}\xspace}
\newcommand{\mpiruns}{\texttt{mpirun}s\xspace}
\newcommand{\machone}{\emph{Jupiter}\xspace}
\newcommand{\machtwo}{\emph{VSC-3}\xspace}
\newcommand{\jupiternecmpi}{NEC\,MPI-1.3.1\xspace}
\newcommand{\jupitermvapich}{MVAPICH2-2.1\xspace}
\newcommand{\jupiteropenmpi}{OpenMPI-1.10.1\xspace}
\newcommand{\jupitergccold}{\gcc~4.4.7\xspace}
\newcommand{\jupitergccnew}{\gcc~4.9.2\xspace}
\newcommand{\vscintelmpi}{Intel\,MPI-5.1.3\xspace}
\newcommand{\vscintelcomp}{Intel~16.0.1\xspace}

\newcommand{\runtime}{run-time\xspace}

\newcommand{\runtimes}{run-times\xspace}

\newcommand{\pingpong}{Ping-pong\xspace}

\newcommand{\VARdatasize}{\ensuremath{m}\xspace}

\newcommand{\VARblocksize}{\ensuremath{A}\xspace}
\newcommand{\VARstride}{\ensuremath{B}\xspace}
\newcommand{\extent}{extent\xspace}
\newcommand{\datasize}{datasize\xspace}
\newcommand{\stride}{stride\xspace}

\newcommand{\blocksize}{blocksize\xspace}
\newcommand{\blocksizes}{blocksizes\xspace}

\newcommand{\Bytes}{Bytes\xspace}

\newcommand{\VARnrep}{\textit{nrep}\xspace} 
\newcommand{\VARnmpirun}{\textit{r}\xspace}

\newcommand{\myhead}[1]{\paragraph{#1}}
\newcommand{\DEdescr}{\Needspace{7\baselineskip}\myhead{\textbf{Experiment}}\noindent}
\newcommand{\DEhypo}{\Needspace{2\baselineskip}\myhead{\textbf{Expectation}}}
\newcommand{\DEhypos}{\Needspace{7\baselineskip}\myhead{\textbf{Expectations}}}
\newcommand{\DEresults}{\myhead{\textbf{Results}}}
\newcommand{\DEtypes}{\myhead{\textbf{Type description}}}
\newenvironment{myitemize}
  {\begin{enumerate}[topsep=3pt,itemsep=3pt,partopsep=0ex,parsep=0ex,leftmargin=15pt]}
  {\end{enumerate}}

\newenvironment{expitemize}
  {\begin{itemize}[topsep=0pt,itemsep=0ex,partopsep=0ex,parsep=0ex,leftmargin=*,label=\textperiodcentered]}
  {\end{itemize}}

\newenvironment{dtabular}[1]
  {\begin{small}\begin{tabular}{#1}}
  {\end{tabular}\end{small}}

\usepackage{etoolbox}
\makeatletter
\patchcmd{\maketitle}{\@copyrightspace}{}{}{}
\makeatother

\definecolor{lightgray}{RGB}{183,183,183}

\newcounter{exptestno}
\DeclareRobustCommand{\exptest}[1]{%
   \refstepcounter{exptestno}%
   \theexptestno\label{#1}}

\newcommand{\subexptest}[1]{\subsubsection{\etest \exptest{#1}}\label{sec:#1}}
\newcommand{\appexp}[1]{\subsection{Expectation Test \ref{#1} (Sect. \ref{sec:#1})}}

\newcommand{\expparam}[2]{%
#2: \hspace{.3em} #1
}

\newlength{\mydescpadding}
\newcommand{\appexpdesc}[2]{%
\vspace*{\mydescpadding}
\hrule
\vspace*{\mydescpadding}
\noindent
#1 
\vspace*{\mydescpadding}
\hrule
\vspace*{\mydescpadding}
#2 
\vspace*{\mydescpadding}
\hrule
}

\def\sharedaffiliation{%
\end{tabular}
\begin{tabular}{c}}

\hyphenchar\font=\string"7F
\begin{document}

\title{MPI Derived Datatypes: Performance Expectations and Status
  Quo\thanks{This work was supported by the Austrian FWF project
    ``Verifying self-consistent MPI performance guidelines'' (P25530),
    and co-funded by the European Commission through the EPiGRAM
    project (grant agreement no.\ 610598).}}

\numberofauthors{3}
\author{\alignauthor Alexandra Carpen-Amarie
  \email{carpenamarie@par.tuwien.ac.at}
  \alignauthor Sascha Hunold 
  \email{hunold@par.tuwien.ac.at} 
  \alignauthor Jesper Larsson Tr\"aff
  \email{traff@par.tuwien.ac.at}
  \sharedaffiliation
  \affaddr{TU Wien}\\ \affaddr{Faculty of Informatics, Institute of
    Information Systems}\\ \affaddr{Research Group Parallel
    Computing}\\ \affaddr{Favoritenstrasse 16/184-5, 1040 Vienna,
    Austria}\\ }

\maketitle

\begin{abstract}
We examine natural expectations on communication performance using MPI
derived datatypes in comparison to the baseline, ``raw'' performance of
communicating simple, non-contiguous data layouts. We show that
common MPI libraries sometimes violate these datatype performance
expectations, and discuss reasons why this happens, but also show cases where MPI libraries perform
well.  Our findings are in many ways surprising and disappointing.
First, the performance of derived datatypes is sometimes worse than
the semantically equivalent packing and unpacking using the corresponding
MPI functionality. Second, the communication performance equivalence
stated in the MPI standard between a single contiguous datatype and
the repetition of its constituent datatype does not hold
universally. Third, the heuristics that are typically employed by MPI
libraries at type-commit time are insufficient to enforce natural
performance guidelines, and better type normalization heuristics may have a 
significant performance impact. We show cases where all the 
MPI type constructors are necessary to achieve
the expected performance for certain data layouts. We describe our
benchmarking approach to verify the datatype performance
guidelines, and present extensive verification results for different MPI libraries.
\end{abstract}

\section{Introduction}
\label{sec:intro}

The derived or user-defined datatype mechanism is a powerful, integral
feature of MPI that enables communication of possibly structured,
non-contiguous, and non-homogeneous (with different constituent basic
types) application data with any of the MPI communication operations,
without the need for tedious, explicit, possibly time- and
space-consuming manual packing between intermediate communication
buf\-fers~\cite[Chapter 4]{MPI-3.1}.

Characterizing the expected and actual performance of MPI
communication with structured, non-contiguous data is a difficult
problem that has been addressed in many studies~\cite{Traff11:typeguide,Traff00:skampi,SchneiderGerstenbergerHoefler14}.
We extend and complement this research using a different
approach. MPI derived datatypes can be viewed as a mechanism for
serializing the access to non-contiguous data layouts. Data elements
stored non-contiguously in memory have to be sent or received in a
certain order. Serialization can be, and in applications often
is~\cite{HoeflerGottlieb10}, handled manually by \emph{packing} and
\emph{unpacking} the data via contiguous, intermediate buffers of
elements of basic datatypes in the desired order, upon which MPI
communication operations are then performed. Alternatively, the given
non-contiguous data layout and access order can be described by a
derived datatype, and the serialization is handled transparently by the
MPI library implementation. There are three interrelated issues
determining the performance of the data serialization and the derived
datatype mechanism:
\begin{description}
\item[Issue 1]
\label{iss:raw}
How expensive is it \emph{per se} to access and serialize data stored
in certain (non-)regular patterns in memory?
\item[Issue 2]
\label{iss:datatype}
How well do specific MPI libraries handle the
serialization using derived datatypes? Does the performance depend on
the type of communication operation?
\item[Issue 3]
\label{iss:normalization}
How do different derived datatype descriptions of the \emph{same layout}
affect serialization cost?
\end{description}

The first issue has to do with the data layout itself, and access
performance is dependent on both the specific data layout, as well as
on the memory system and other factors of the underlying system, and
on how well the serialization can be implemented to exploit such
capabilities (cache, vectorization, prefetching). Because of this
essential dependence on both system capabilities and on what is
possible for each particular access pattern, it does not seem possible
to state system-independent expectations or guidelines \emph{a priori}
on the costs of processing and communicating structured
data. Nevertheless, it is enlightening for users to have means to
measure the difference in communication performance with differently
structured data layouts.

The second issue focuses on the quality of the MPI implementation for
accessing structured layouts. The MPI standard itself does not
prescribe how the datatype mechanism has to be implemented.  It does,
however, interrelate communication and datatype constructors in a way
that makes it possible to formulate and check concrete expectations on
the performance of the derived datatype mechanism. We explain and
examine such expectations in the paper.

The third issue is solely related to the quality of the MPI library. 
With the given MPI datatype constructors~\cite[Chapter
4]{MPI-3.1}, it is easy to see that the same layout can be described
in an infinite number of ways (almost all of which are trivial and
irrelevant). However, for a given application layout there are often
competing, non-contrived ways of describing it. We can
compare the communication performance with such different
descriptions. It might be sensible to expect that an MPI library
ensures that performance is more or less the same, no matter how the
user chooses to describe the given layout. We will
argue why this is a reasonable expectation, and discuss why it cannot be
(easily) fulfilled.

We discuss benchmarking of the MPI derived datatype mechanism in an
attempt to characterize both the ``raw performance'' of communication
with structured data (Issue~\ref{iss:raw}), as well as to develop
means for verifying the expected performance of certain uses of the
derived datatype mechanism. We focus on three different (meta)
performance guidelines, previously discussed by
Gropp~\etal~\cite{Traff11:typeguide}, but give more precise
formulations and implementations here. We then use our benchmarks to
evaluate concrete MPI libraries and systems.  Our benchmarks are
synthetic, but parameterized to make it possible to investigate
patterns that are relevant for applications. Most of the patterns and
derived datatype descriptions that are considered here are natural and
deliberately quite simple. Other synthetic patterns, in part derived
from applications, have been used in other
studies~\cite{Traff00:skampi,SchneiderGerstenbergerHoefler14}.  Schulz
\etal used derived datatypes for piggybacking small headers on larger
messages, and contrast the performance achieved with derived datatypes
against uses of the MPI pack/unpack functionality~\cite{SchulzBS08}.

We believe that the capability of transparently communicating structured data
 is a strong (and rather unique) feature of MPI. It is
therefore important to ensure good and consistent performance of
communication with derived datatypes. The larger purpose of this study
is to prevent unrealistic performance
expectations, but also to make developers and application programmers
aware of concrete performance problems in given MPI libraries and
systems.  Much work has been done over the past decades in improving
the communication performance with derived
datatypes~\cite{BynaGroppSunThakur03,PrabhuGropp15,RossMillerGropp03,RossLathamGroppLuskThakur09,SchneiderKjolstadHoefler13,Traff99:flattening}.
For instance, it has been shown, in many different variations, that
piece-wise packing of structured layouts described by derived
datatypes can be performed
efficiently~\cite{PrabhuGropp15,SchneiderKjolstadHoefler13,Traff99:flattening}.
This is important for efficient pipelining and overlapping of data
accesses and communication. Likewise, some developments focused on
exploiting the memory hierarchy~\cite{BynaGroppSunThakur03} and
 the communication capabilities for strided,
non-contiguous data communication~\cite{WuWyckoffPanda04}.

Many of the experiments in this paper are concerned with
Issue~\ref{iss:normalization}. The expectation is that MPI libraries
(at the \mpicommit operation) compute a good, internal representation
of the user-specified datatype, which was termed \emph{type
  normalization}~\cite{Traff11:typeguide}. In this paper, we put
more emphasis on showing that a strong datatype normalization can be
advantageous performance-wise. It has recently been shown that optimal type
normalization of derived datatypes into tree-structured
representations is possible in polynomial time, but
costly~\cite{Traff16:alltypetree,Traff15:mpilinear,Traff14:normalization}.
The latter two papers show that normalization costs are moderate, if
the \mpistruct constructor is left out. However, as our examples will
show, the normalization that is required in order to get the expected
performance for certain layouts requires this constructor, even if the
layout is homogeneous, \ie, it only consists of data items of
the same basic datatype.

The paper is structured in two main parts. In Section~\ref{sec:raw},
the focus is mostly on Issue~\ref{iss:raw}, where communication
performance for simple layouts described in simple ways is contrasted
with performance with unstructured data. In
Section~\ref{sec:guide}, we formalize relative performance
expectations as MPI performance guidelines~\cite{Traff10:selfcons},
and use them to structure the experiments. The focus here is on
different descriptions of the same simple layouts as used in the first
part, and on the performance of derived datatypes \versus packing and
unpacking with the \mpipack and \mpiunpack operations.

\section{Characterizing Datatype\\ Performance}
\label{sec:raw}

We first attempt to estimate the additional overhead (if any) in
communicating non-contiguous data by comparing to the time for
communicating the same amount of data from a contiguous memory
buffer. In other words, the focus is on the differences in
communication performance caused by different types of (non-)regular
layouts, and not on the way that MPI handles such layouts. However,
these concerns cannot be completely separated.  We use MPI derived
datatypes to describe our non-contiguous layouts, and thus do not
attempt to estimate derived datatype overheads in any absolute way by
comparing to any ``best possible'' way of copying non-contiguous data
layouts between contiguous buffers (``packing and unpacking by
hand'').  The reasons for this are twofold. First, it is not at all
obvious what the best possible way to manually pack some complex
non-contiguous layout into a contiguous buffer for some specific system
is. Second, such a comparison is not necessarily fair, since the
derived datatype mechanism makes it possible to interact with the
communication system, for instance by pipelining large non-contiguous
buffers by partial
packing~\cite{PrabhuGropp15,SchneiderKjolstadHoefler13,Traff99:flattening}
and/or by exploiting hardware capabilities for non-contiguous data
communication. Such optimizations that are possible for the MPI
implementation via the MPI derived datatype mechanism are
difficult to perform at the application level.

To establish a baseline performance, we consider non-contiguous,
blockwise, strided data layouts with a given serialization order of a
given number of $n$ elements (of some predefined, basic datatype
corresponding to a programming language type), and measure the
communication performance for different $n$. We consider what we think
are the simplest such layouts, and use what we think are the MPI
implementation friendliest (non-nested) derived datatypes to describe
fixed blocks of $k$ elements, such that the complete layouts are
described by $n/k$ successive, contiguous repetitions (count argument
in the MPI operations) of these blocks. The baseline performance
delivered by an MPI library is the time for communicating $n$
contiguous elements of the same basic type. We describe the basic
layouts in \Sec~\ref{sec:basiclayouts}.

\subsection{Communication Patterns}

Derived datatypes can be used with all types of MPI communication
operations, but may behave differently in different contexts. We
therefore benchmark with three different types of communication
operations in order to get an idea of whether this is the case. The
$n$ elements are communicated either from a contiguous buffer, or as a
non-contiguous layout described by a derived 
datatype as
outlined above. We use the following communication operations and
patterns:
\begin{myitemize}
\item
\emph{Point-to-point} communication with blocking \mpisend and
\mpirecv operations.
\item
The \emph{asymmetric (rooted) collective} \mpibcast on $p$ processes.
\item
The \emph{symmetric (non-rooted) collective} \mpiallgather on $p$ processes.
\end{myitemize}
We do not benchmark one-sided communication performance
with structured data. One reason is that with the one-sided
communication model, descriptions of derived datatypes may have to be
transferred between processes, and MPI libraries may differ too much in
the way this is handled.

\subsection{Basic, Static Datatype Layouts}
\label{sec:basiclayouts}

\begin{figure}[t]
\centering

\begin{tikzpicture}[yscale=0.47]
	\def\vertStride{3.0}

        \def\startH{0.0}

	\def\Hzero{0.5}

	\node[left] at (0,\Hzero+0.5) {\normalsize{\dtcontig}};

	\foreach \x in {0,1,2,3,4,5,6,7,8,9}
	\draw[fill = lightgray] (\x*.5,\Hzero) rectangle (\x*.5+.5,\Hzero+1);

	\node at (5.5,\Hzero+0.5) {...};

	\def\Hone{\startH-\vertStride+1}

	\node[left] at (0,\Hone+0.5) {\normalsize{\dtdtiled}};

	\foreach \x in {0} 
	\draw[fill = lightgray] (\x,\Hone) rectangle (\x+1.5,\Hone+1);

	\foreach \x in {0, 0.5, ..., 2.0} 
	\draw (\x,\Hone) rectangle (\x+0.5,\Hone+1);

	\node at (5.5,\Hone+0.5) {...};

	\node (A) at (-0.1,\Hone-0.2) {};
	\node (B) at (2.6,\Hone-0.2) {};
	\path[<->, font=\small] (A) edge node[below]{\normalsize{$B$}} (B);

	\node (C) at (-0.1,\Hone+1.2) {};
	\node (D) at (1.6,\Hone+1.2) {};
	\path[<->, font=\small] (C) edge node[above]{\normalsize{$A$}} (D);

	\def\Hthree{\Hone-\vertStride}

	\node[left] at (0,\Hthree+0.5) {\normalsize{\dtdblock}};

	\foreach \x in {0, 2.0}
	\draw[fill = lightgray] (\x,\Hthree) rectangle (\x+1.5,\Hthree+1);

	\foreach \x in {0, 0.5, ..., 4.5} 
	\draw (\x,\Hthree) rectangle (\x+0.5,\Hthree+1);

	\node at (5.5,\Hthree+0.5) {...};

	\node (A) at (-0.1,\Hthree-0.2) {};
	\node (B) at (2.1,\Hthree-0.2) {};
	\path[<->, font=\small] (A) edge node[below]{\normalsize{$B_1$}} (B);

	\node (C) at (-0.1,\Hthree+1.2) {};
	\node (D) at (1.6,\Hthree+1.2) {};
	\path[<->, font=\small] (C) edge node[above]{\normalsize{$A$}} (D);

	\node (A) at (1.9,\Hthree-0.2) {};
	\node (B) at (5.1,\Hthree-0.2) {};
	\path[<->, font=\small] (A) edge node[below]{\normalsize{$B_2$}} (B);

	\node (C) at (1.9,\Hthree+1.2) {};
	\node (D) at (3.6,\Hthree+1.2) {};
	\path[<->, font=\small] (C) edge node[above]{\normalsize{$A$}} (D);

	\def\Htwo{\Hthree-\vertStride}

	\node[left] at (0,\Htwo+0.5) {\normalsize{\dtdbucket}};

	\draw[fill = lightgray] (0,\Htwo) rectangle (1.0,\Htwo+1);
	\draw[fill = lightgray] (2.5,\Htwo) rectangle (4.5,\Htwo+1);

	\foreach \x in {0, 0.5, ..., 4.5} 
	\draw (\x,\Htwo) rectangle (\x+0.5,\Htwo+1);

	\node at (5.5,\Htwo+0.5) {...};

	\node (A) at (-0.1,\Htwo-0.2) {};
	\node (B) at (2.6,\Htwo-0.2) {};
	\path[<->, font=\small] (A) edge node[below]{\normalsize{$B$}} (B);

	\node (C) at (-0.1,\Htwo+1.2) {};
	\node (D) at (1.1,\Htwo+1.2) {};
	\path[<->, font=\small] (C) edge node[above]{\normalsize{$A_1$}} (D);

	\node (A) at (2.4,\Htwo-0.2) {};
	\node (B) at (5.1,\Htwo-0.2) {};
	\path[<->, font=\small] (A) edge node[below]{\normalsize{$B$}} (B);

	\node (C) at (2.4,\Htwo+1.2) {};
	\node (D) at (4.6,\Htwo+1.2) {};
	\path[<->, font=\small] (C) edge node[above]{\normalsize{$A_2$}} (D);

	\def\Hfour{\Htwo-\vertStride}

	\node[left] at (0,\Hfour+0.5) {\normalsize{\dtdalternating}};

	\draw[fill = lightgray] (0,\Hfour) rectangle (1.0,\Hfour+1);
	\draw[fill = lightgray] (2.0,\Hfour) rectangle (4,\Hfour+1);

	\foreach \x in {0, 0.5, ..., 4.5} 
	\draw (\x,\Hfour) rectangle (\x+0.5,\Hfour+1);

	\node at (5.5,\Hfour+0.5) {...};

	\node (A) at (-0.1,\Hfour-0.2) {};
	\node (B) at (2.1,\Hfour-0.2) {};
	\path[<->, font=\small] (A) edge node[below]{\normalsize{$B_1$}} (B);

	\node (C) at (-0.1,\Hfour+1.2) {};
	\node (D) at (1.1,\Hfour+1.2) {};
	\path[<->, font=\small] (C) edge node[above]{\normalsize{$A_1$}} (D);

	\node (A) at (1.9,\Hfour-0.2) {};
	\node (B) at (5.1,\Hfour-0.2) {};
	\path[<->, font=\small] (A) edge node[below]{\normalsize{$B_2$}} (B);

	\node (C) at (1.9,\Hfour+1.2) {};
	\node (D) at (4.1,\Hfour+1.2) {};
	\path[<->, font=\small] (C) edge node[above]{\normalsize{$A_2$}} (D);

\end{tikzpicture}
\caption{Basic, static layouts with parameters chosen according to
  \variantone and with $A=\num{3}$. Serialization of the basetype elements
  is from left to right.}
\label{fig:k-layouts}
\end{figure}
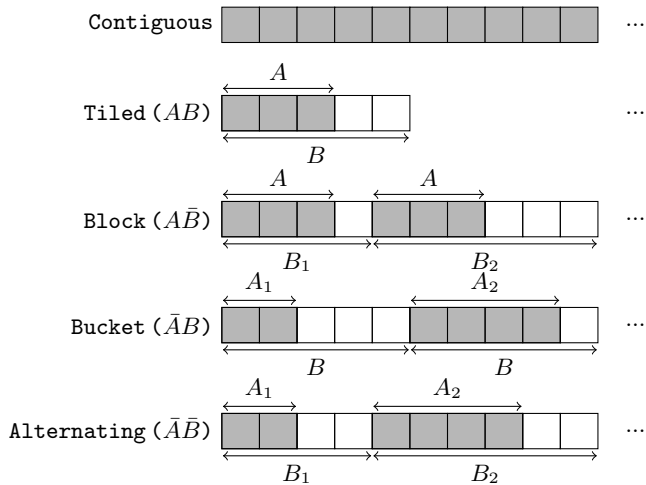

We first experiment with the parameterized, blockwise layouts
described below. These layouts are \emph{static}, by which we mean that
derived datatypes of $k$ basetype elements are set up in advance and
used for the whole sequence of experiments. When a total of $n$ elements, stored
according to either of these layouts, is to be communicated, the
count argument in the communication operations is adjusted down to
$n/k$. The layouts thus consist of regularly strided, but structured,
non-consecutive \emph{blocks}. We illustrate these types of blocks in
Figure~\ref{fig:k-layouts}, and state below the MPI datatype constructors
used to describe them. All layouts consist of contiguous
smaller units of elements; we use $A$ for the number of elements in a
unit. Units are strided with some stride~$B$, and mostly we require
$B>A$. In either of the layouts the number of elements in a unit may
vary (different $A$ values), or the strides of the units may vary
(different $B$ values), or both.  We use an ($AB$) notation for
this. We use $A$ (respectively $B$) when the \blocksize (respectively
\stride) is fixed over the units, and $\bar{A}$ (respectively
$\bar{B}$) when the \blocksize (respectively \stride) varies between
the units. Our basic layouts are as follows.

\begin{description}[itemsep=0ex,labelindent=0ex,labelwidth=0ex]
\item[\dtcontig:] is a contiguous buffer of elements described by a
  predefined, basic MPI datatype (no derived datatype).
\item[\dtdtiled:] is a contiguous unit of $A$ elements repeated with a
  stride of $B$ elements, requirement $B>A$ (the case $B=A$ would be a
  contiguous layout). In MPI, the datatype is constructed using
  \mpicontig with a count of $A$ and a call to \mpiresized to obtain
  the extent $B$. A block in this case has $k=A$ elements and an
  extent of $B$ elements.
\item[\dtdblock:] consists of two contiguous units of $A$ elements
  with alternating strides $B_1$ and $B_2$, requirement $B_1\neq B_2$,
  and $B_1,B_2>A$ (otherwise the layout would be as above).  The
  description in MPI is done using \mpiblock and \mpiresized. This
  block has $k=2A$ elements and an extent of $B_1+B_2$ elements.
\item[\dtdbucket:] consists of two alternating, contiguous units of
  $A_1$ and $A_2$ elements, with a regular stride $B$, requirement
  $B>A_1,A_2$. The MPI description is formulated with \mpiindex. This
  block has $k=A_1+A_2$ elements and an extent of $2B$ elements.
\item[\dtdalternating:] consists of two alternating, contiguous units
  of $A_1$ and $A_2$ elements, with strides $B_1$ and $B_2$,
  respectively. In MPI, the datatype is described with \mpiindex.
  This block has $k=A_1+A_2$ elements and an extent of $B_1+B_2$
  elements.
\end{description}

All blocks can be defined over arbitrary predefined, basic MPI
datatypes. It seems natural to assume that communication performance,
regardless of the type of communication, should not depend on which
basic type is used, but only on the amount of data
communicated. However, with MPI this assumption is problematic, since
different basic types have different semantics (doubles, integers,
characters), and MPI may have to handle different basic types
differently. In most systems and situations, this will probably not be
the case, but measurements with different basetypes have to be
performed.

\subsection{Benchmarking Setup}

In the following, we give an overview of the hardware and software
setup used for our experiments.

\subsubsection{System and MPI Libraries}

\begin{table}[t]
  \centering
  \caption{Hardware and software used in the experiments.}
  \label{tab:machines}
  \begin{scriptsize}
  \begin{tabular}{l@{\hskip .1in}l}
    \toprule
    machine  & 36 $\times$ Dual Opteron 6134 @ \SI{2.3}{\giga\hertz}  \\
             &  \infiniband QDR MT26428   \\ 
    machine name & \machone \\
    \midrule 
    MPI libraries &  \jupiternecmpi, \jupitermvapich, \jupiteropenmpi \\
    Compiler & \jupitergccold, \jupitergccnew (flags \texttt{-O3})\\
    \bottomrule
  \end{tabular}
  \end{scriptsize}
\end{table}
The experiments have been conducted on a 36~node Linux cluster called
\machone, where each node is equipped with two Opteron~6134 processors (see
\tab~\ref{tab:machines}). The nodes are interconnected using an
\infiniband QDR network. 
We have benchmarked the datatype performance for three MPI libraries,
namely \jupiternecmpi, \jupitermvapich and \jupiteropenmpi; in the
paper we show results for the former two (see the appendix for
\jupiteropenmpi results). The benchmarks, for which the results are
shown in the paper, have been compiled using \jupitergccold. We have
examined the datatype performance after compiling with \jupitergccnew,
to check whether the compiler version is an experimental
factor. However, we have not seen any effects by using \jupitergccnew.

\subsubsection{Benchmarking Communication Patterns}

We now explain how the benchmarking of the different communication
patterns was done, in particular, which times have been measured.

In each benchmark (\pingpong or collective), we compare different
datatypes for the same total communication volume. In addition, the
measured \runtimes do not include the datatype setup times, and thus,
they represent the communication times (latencies) only.

For the \pingpong experiments, we first synchronize the two involved
MPI processes with an \mpibarrier. Then, a message is sent from one
process, received by the other and returned using \mpisend and
\mpirecv operations, and each process measures the time taken for the
two operations. The time to perform a \pingpong is then computed as
the maximum over the local \runtimes of both processes. The \pingpong
measurement is repeated \VARnrep times within one \mpirun call. Then,
we repeat this \pingpong test over \VARnmpirun calls to
\mpirun~\cite{Traff14:benchmarking}. In our \pingpong experiments, we
used $\VARnrep=\num{100}$ and $\VARnmpirun=5$.

When benchmarking the collective communication operations (\mpibcast,
\mpiallgather), we also synchronize the processes before each
collective call with \mpibarrier. All processes call the collective
operation and measure the \runtime (latency) locally. This measurement
scheme, consisting of an \mpibarrier and the timing of a collective call,
is repeated \VARnrep times. The \runtimes (latencies) of the
collective calls from each process are sent (reduced) to (on) the root
process, and the \runtime for a each collective call is computed as
the maximum \runtime over all processes. As \mpirun can be an
experimental factor, we repeat this experiment \VARnmpirun times. For
details, we refer the reader to Algorithm~1 from
Hunold~\etal~\cite{Traff14:benchmarking}.
Since the \runtime of collective calls becomes relatively long for the
larger message sizes in our experiments (\eg, around \SI{1}{\second}
for \mpiallgather), we cannot afford to execute \num{100} repetitions
for every experiment. Moreover, the variance of the \runtime for such
larger message sizes is relatively small. We therefore reduce the
number of repetitions (of collective calls) per test case depending on
the \datasize \VARdatasize:
\[
    \VARnrep= 
\begin{cases}
    100 & \text{if } \VARdatasize \le \SI{32}{\kilo\byte},\\
    50   & \text{if } \SI{32}{\kilo\byte}< \VARdatasize \le \SI{320}{\kilo\byte}, \\
    20   & \text{if } \VARdatasize > \SI{320}{\kilo\byte}.\\
\end{cases}
\]
Each datatype experiment with a collective call has been measured for
$\VARnmpirun=\num{5}$~\mpiruns.

\subsubsection{Data Processing}

When conducting a single datatype experiment, we obtain \VARnmpirun
datasets, each containing \VARnrep measurements. For each \mpirun, we
compute the median of the \VARnrep \runtimes. Then, we calculate the
mean, minimum, and maximum values over these \VARnmpirun median
\runtimes. These values will be used in the plots, \ie, the error
bars in the bar graphs denote the minimum and maximum of the
\VARnmpirun median \runtimes.

\subsection{Experimental Results}

\begin{table}[t]
  \centering
  \caption{\label{tab:test_cases_basic}Basic layout variants (\cf~\fig~\ref{fig:k-layouts}).}
  \begin{small}
  \begin{tabular}{lll}
    \toprule
    Layout & \variantone & \varianttwo \\
    \midrule
    \dtdtiled & $B_{\,\,\,}=A+2$ & $B_{\,\,\,}=3A$\\
    \midrule
    \dtdblock & $B_1=A+1$ & $B_1=2A$ \\
              & $B_2=A+3$ & $B_2=4A$ \\
    \midrule
    \dtdbucket & $A_1= A-1$ & $A_1=A/2$ \\
           & $A_2=A+1$ & $A_2=3/2 A$ \\
           & $B_{\,\,\,}=A+2$ & $B_{\,\,\,}=3A$ \\
    \midrule
    \dtdalternating & $A_1=A-1$  &  $A_1=A/2$ \\
           & $ A_2=A+1$ & $A_2=3/2A$ \\
           & $B_1=A+1$ & $B_1=2A$ \\
           & $B_2=A+3$ & $B_2=4A$\\
    \bottomrule
  \end{tabular}
  \end{small}  
\end{table}

\begin{figure*}[!t]
\centering
\begin{subfigure}{.33\linewidth}
\centering
\includegraphics[width=\linewidth]{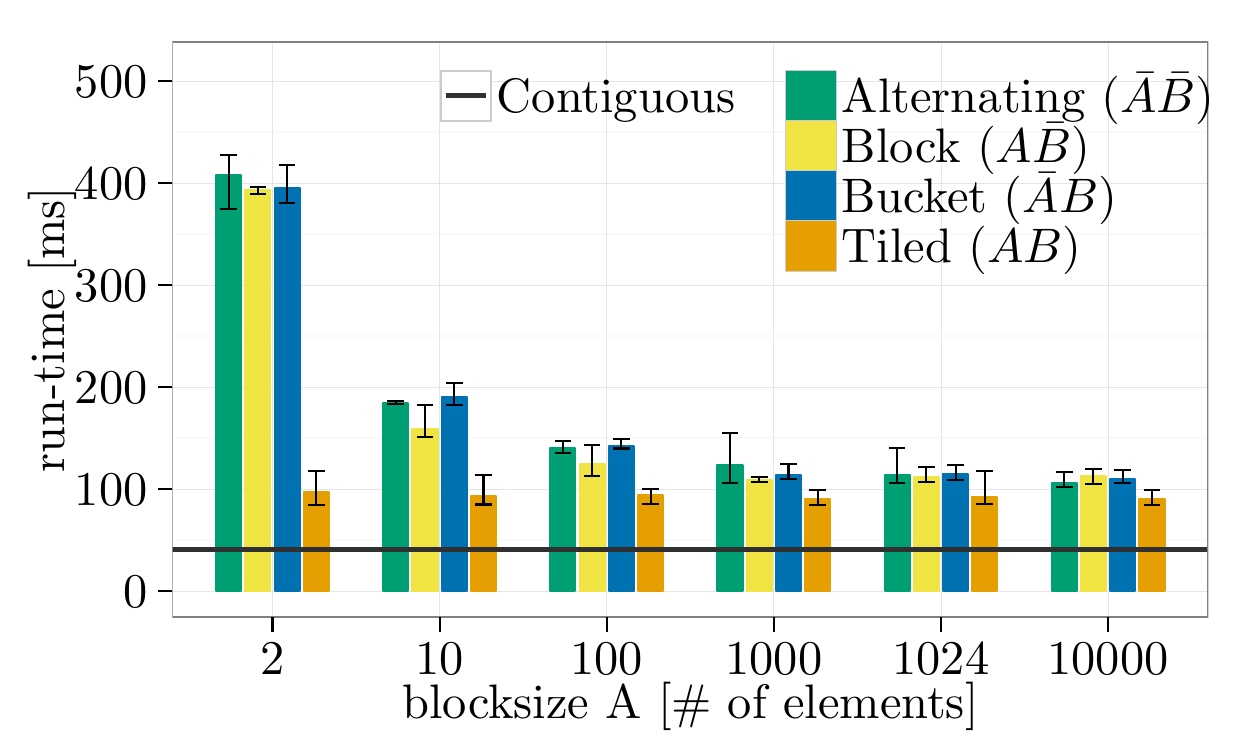}
\caption{%
\label{fig:exp:allgather-nlarge-32p-nec}%
\mpiallgather, \jupiternecmpi%
}%
\end{subfigure}%
\hfill%
\begin{subfigure}{.33\linewidth}
\centering
\includegraphics[width=\linewidth]{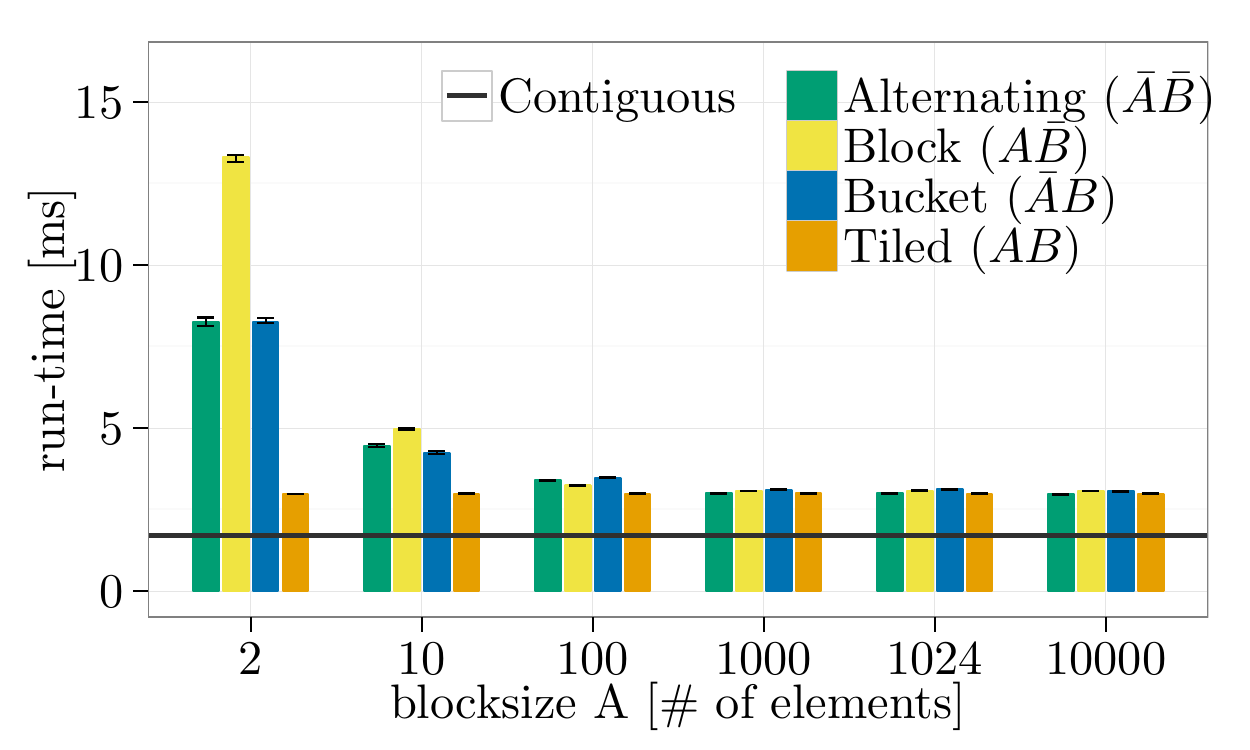}
\caption{%
\label{fig:exp:bcast-nlarge-32p-nec}%
\mpibcast, \jupiternecmpi%
}%
\end{subfigure}%
\hfill%
\begin{subfigure}{.33\linewidth}
\centering
\includegraphics[width=\linewidth]{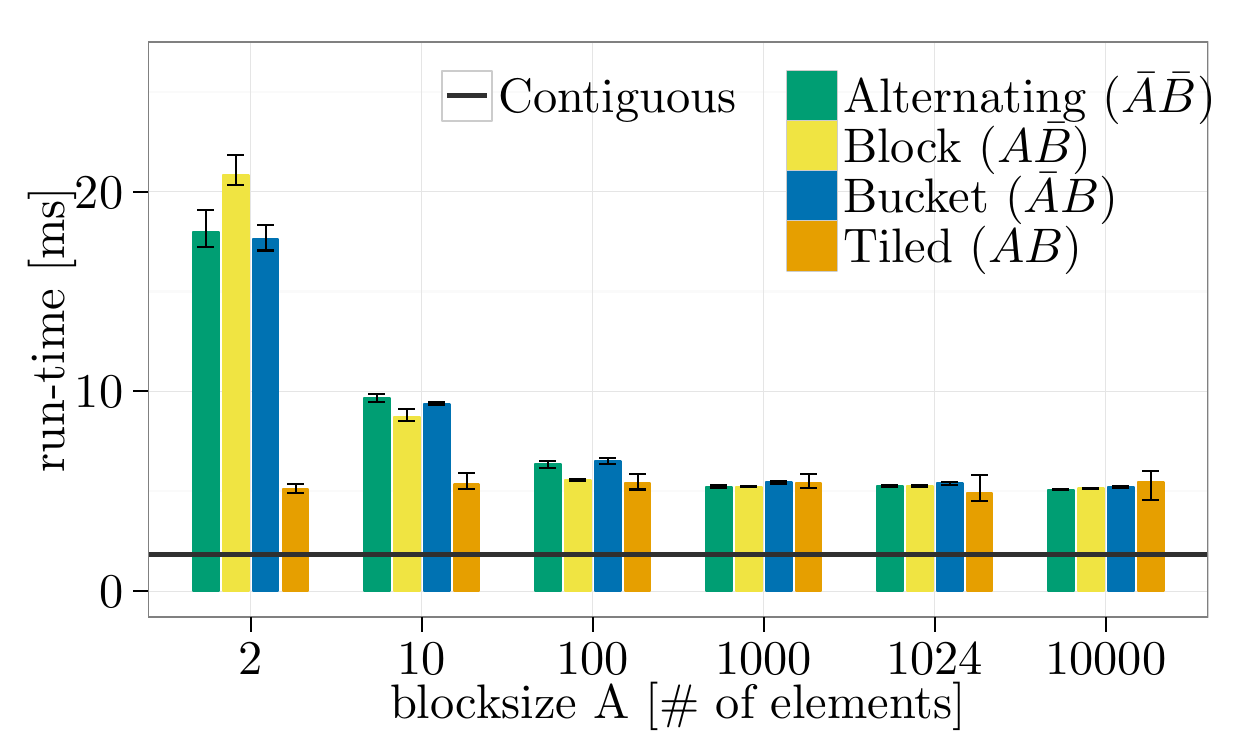}
\caption{%
\label{fig:exp:pingpong-nlarge-32p-nec}%
\pingpong, \jupiternecmpi%
}%
\end{subfigure}%
\\%
\begin{subfigure}{.33\linewidth}
\centering
\includegraphics[width=\linewidth]{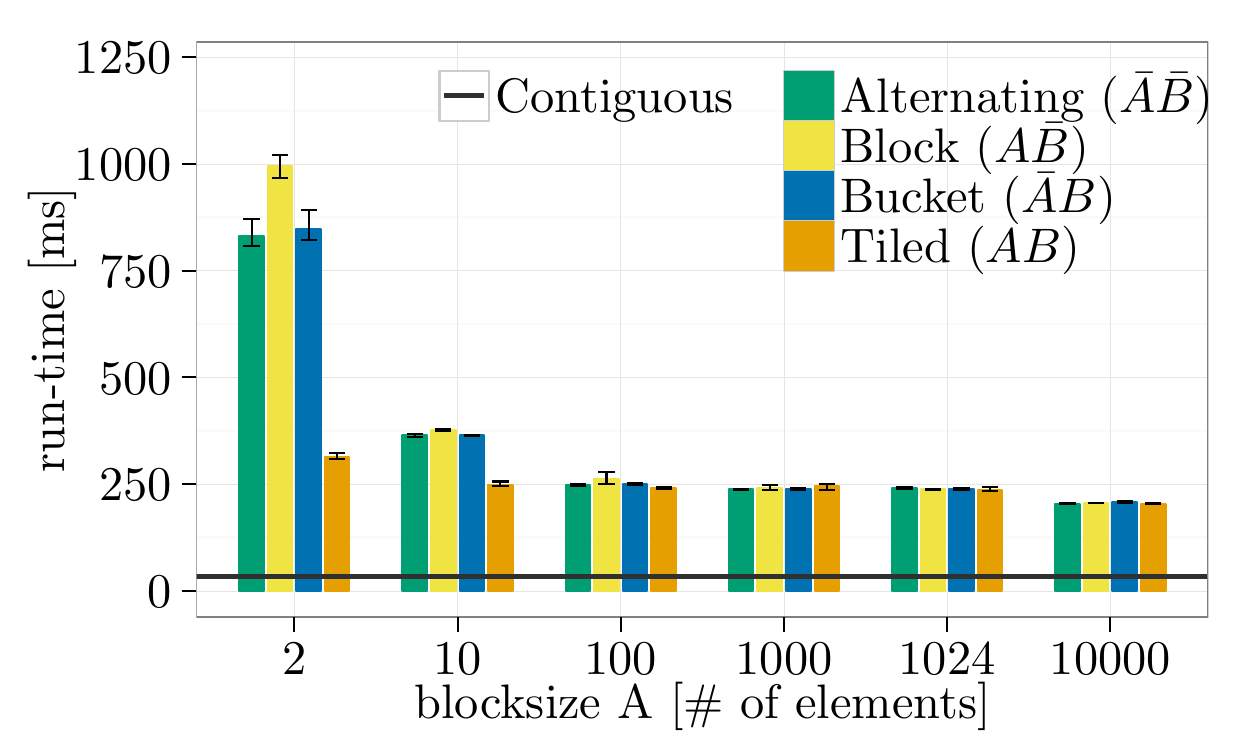}
\caption{%
\label{fig:exp:allgather-nlarge-32p-mvapich}%
\mpiallgather, \jupitermvapich%
}%
\end{subfigure}%
\hfill%
\begin{subfigure}{.33\linewidth}
\centering
\includegraphics[width=\linewidth]{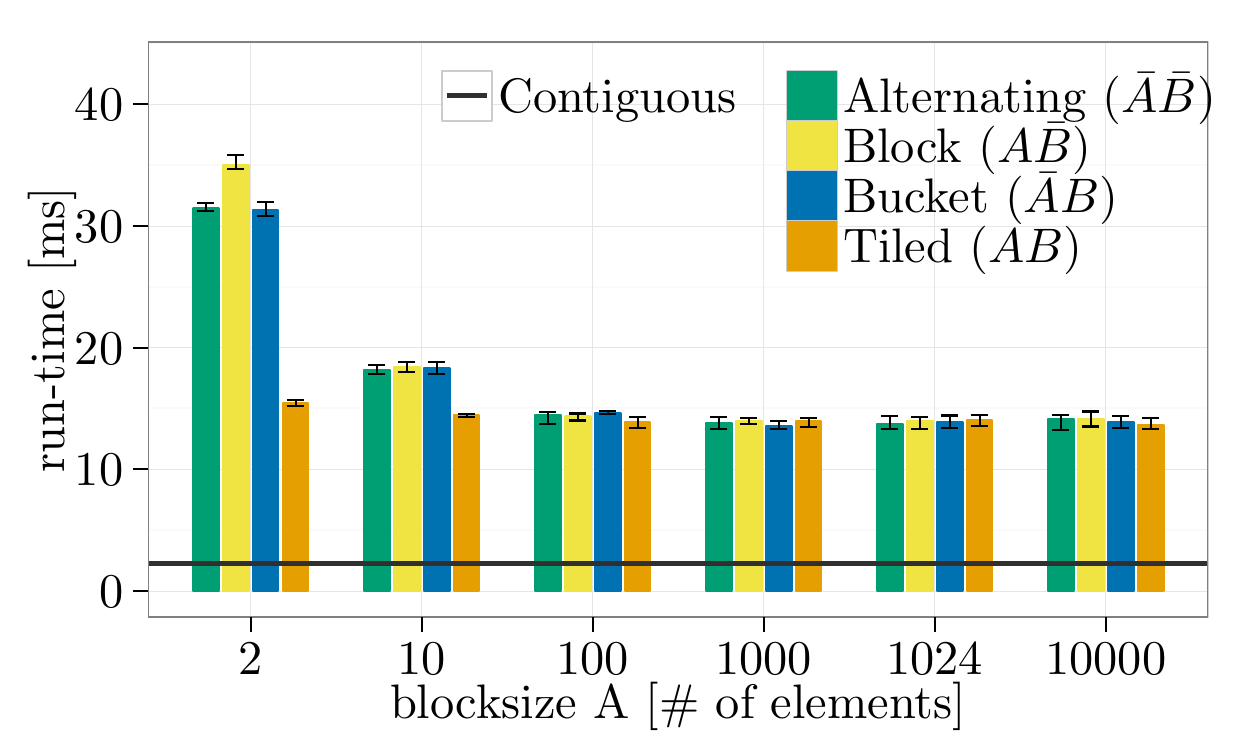}
\caption{%
\label{fig:exp:bcast-nlarge-32p-mvapich}%
\mpibcast, \jupitermvapich%
}%
\end{subfigure}%
\hfill%
\begin{subfigure}{.33\linewidth}
\centering
\includegraphics[width=\linewidth]{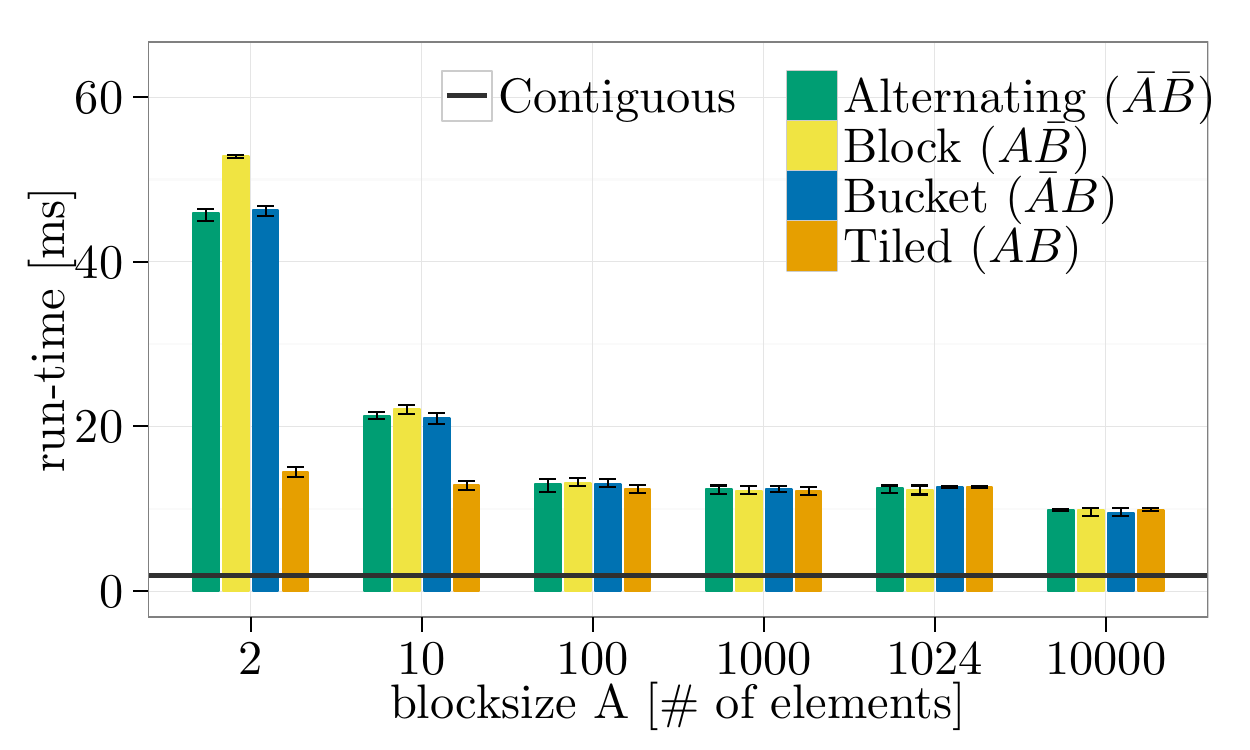}
\caption{%
\label{fig:exp:pingpong-nlarge-32p-mvapich}%
\pingpong, \jupitermvapich%
}%
\end{subfigure}%
\caption{\label{fig:exp:layouts-nlarge-32p}  Contiguous \vs typed,  $\VARdatasize=\SI{2.56}{\mega\byte}$, element datatype: \mpiint, \num{32x1}~processes (\num{2x1} for \pingpong), \variantone.}
\end{figure*}

We now summarize our findings that characterize the costs of
communicating simple, structured data in comparison to communicating
the same amount of contiguous data.  We have used the basic datatype
\mpiint as the element basetype. We have experimented with two
variants for each of the basic layouts, which are summarized in
\tab~\ref{tab:test_cases_basic}. All layouts in both variants are
defined using the unit size parameter $A$, and the values of the $A$'s
and $B$'s are chosen such that all \variantone layouts have the same
total extent of $n+2n/A$, and all \varianttwo layouts have the same
total extent of $3n$.

We describe all experiments by stating the (derived) datatypes used,
the reference (baseline) layout against which we evaluate, our
expectations (hypotheses) on the performance, and then show and
comment on the results.

This gives rise to a very large amount of experimental data, and we
cannot show all our results here. A few exemplary results are included
in the paper; most other results can be found in the appendix.

\subexptest{exptest:basic_layouts}

This is our basic experiment to measure the ``raw'' performance of the
simple, non-contiguous layouts of \fig~\ref{fig:k-layouts}. We experiment
with different message sizes (fixed number of elements in the
layouts) and vary the \blocksize parameter $A$. We use both variants 
\variantone and \varianttwo for determining the remaining parameters in
the layouts.

\DEdescr
\begin{dtabular}{ll}
  \toprule
  Reference Layout & \dtcontig \\
  Compared Layouts & \dtdtiled, \dtdblock \\
                   & \dtdbucket, \dtdalternating \\
  \midrule
  \blocksize~\VARblocksize  & $\num{2},\num{10},\num{100},\num{1000},\num{1024}, \num{10000}$ \\
  \datasize~\VARdatasize & \SI{3200}{\Bytes}, \SI{2560000}{\Bytes}\\
  comm. patterns   & \pingpong, \mpibcast, \mpiallgather \\
  layout variant   & 1 \& 2 \\
  \# of processes   & \num{32x1}, \num{1x16}, \\
  (\#nodes $\times$ \#cores) & (\num{2x1}, \num{1x2} for \pingpong) \\
  \bottomrule
\end{dtabular}

\DEtypes %
We compare the datatype layouts depicted in \fig~\ref{fig:k-layouts}.

\DEhypos 
We expect all communication operations with the non-contiguous layouts
to be slower than using the \dtcontig layout.  We expect this
difference to become smaller when increasing the \blocksize $A$. It is
interesting to find out how large the difference to the contiguous
baseline performance is, how the performance is changing between the
layouts, how it depends on the type of communication, and whether
there are differences between the MPI libraries.

\newpage

\DEresults %
The tiled layout indeed gives the best performance among the four
non-contiguous layouts for all three communication patterns. The
differences between the layouts are the largest for small values of
the \blocksize parameter $A$. For all libraries, there is a large
difference between process configurations when all MPI processes are
on the same node and when they are on different nodes (not shown here,
see appendix). For the \jupiternecmpi library, the performance with
non-contiguous data, especially \dtdtiled, for processes on the same
node is close to the raw performance with contiguous data. The
libraries show a large difference in the way they handle
non-contiguous data. While the raw performance with contiguous data is
comparable among libraries, there is about a factor of two (and more)
difference for the non-contiguous layouts, see
\fig~\ref{fig:exp:layouts-nlarge-32p}. The \jupiternecmpi library
performs best, as it handles non-contiguous layouts with a tolerable
overhead.

\subexptest{exptest:tiled_het}

MPI provides predefined, basic datatypes corresponding to the basic C
and Fortran programing language types. These basic types have
different semantic content, and communication performance may differ
for data of different basic types. Knowing when this is the case is
a valuable information to the application programmer. In particular, we
investigate how consecutive buffers consisting of different semantic units
perform in comparison to raw, uninterpreted bytes 
(described by \mpibyte).

\pagebreak

\DEdescr
\begin{tabular}{ll}
  \toprule
  Reference Layout & \dtcontig buffer of \mpibyte\\
  Compared Layout & \dtdtiledhet \\
  \midrule
  \blocksize~\VARblocksize  & $\num{2}, \num{6}, \num{8},\num{10}, \num{16}, \num{100}, \num{128}, \num{200}$ \\
  \stride~\VARstride & \VARblocksize \\
  \datasize~\VARdatasize & \SI{48000}{\Bytes}, \SI{1500000}{\Bytes} \\
  comm. patterns   & \pingpong \\
  layout variant   & 1 \\
  \# of processes   & \num{2x1}, \num{1x2} \\
  \bottomrule
\end{tabular}

\DEtypes %
The unit of the \dtdtiledhet layout consists of different basetypes
$T_1, T_2, \ldots$ with \blocksize $A$ and \stride $B$, where the
stride of each block is given in units of the corresponding basetype.
The layout is shown in \fig~\ref{fig:test_tiled_het}. 
The unit can be described in MPI
using \mpistruct. It is required that $B\geq A$, where equality is
allowed. In our experiments, we use a contiguous layout consisting of \mpichar,
\mpiint, \mpidouble, \mpishort, with $B=A$ and with
$A$ varying between \num{2} and \num{200}.

\begin{figure}[h!]
\centering
\begin{tikzpicture}[scale=0.45]
  \def\toff{-1.5}
  \draw (0,\toff+0) rectangle (11.5,\toff+1);
  \foreach \xoff in {0,1.5,...,6} {
  \foreach \x in {0}
  \draw[fill = lightgray] (\xoff+\x,\toff+0) rectangle (\xoff+\x+.5,\toff+1);
  \foreach \x in {.5}
  \draw[pattern=north west lines, pattern color=black] (\xoff+\x,\toff+0) rectangle (\xoff+\x+.5,\toff+1);
  \foreach \x in {1.0}
  \draw[pattern=dots, pattern color=black] (\xoff+\x,\toff+0) rectangle (\xoff+\x+.5,\toff+1);
  }
  \node at (9.5,\toff+0.5) {...};

  \draw [<->]
  (0,\toff-.5) -- node[anchor = north, yshift=-1.0]{$A=B$}
  (1.5,\toff-.5);

  \draw [decorate,decoration={brace,amplitude=5pt,mirror}]
  (0,\toff-1.5) -- node[anchor = north west, xshift=-5, yshift=-3.0]{\mpistruct}
  (1.5,\toff-1.5);
  
  \def\toff{0}
  \draw (0,\toff) rectangle (11.5,\toff+1);
  \foreach \hoff in {0,2,4,6}
  \foreach \x in {0,.25,.5,.75,1,1.25,1.5,1.75}
  \draw[fill=gray] (\x+\hoff,\toff) rectangle (\x+\hoff+.25,\toff+1);
  \node at (9.5,\toff+0.5) {...};

\end{tikzpicture}
\caption{\dtcontig buffer of bytes (\mpibyte) (top) \vs \dtdtiledhet
  with $A$=$B$ (bottom).}
\label{fig:test_tiled_het}
\end{figure}

\DEhypo 

Unless the underlying system indeed requires a different handling of
different basic datatypes, we would expect that a contiguous, dense
layout of different basic types performs as well as the corresponding
amount of unstructured bytes in the MPI communication
operations. Since the \mpistruct constructor respects alignment
constraints for the basic MPI datatypes (in a sense, these are
semantic constraints), there might be differences when calling
\mpistruct to set up the \dttiledhet layout leads to a non-contiguous
layout. This may degrade communication performance.

\DEresults
Our results confirm the hypothesis. The cases with a large difference
between the reference and the compared layouts can be explained by
alignment constraints that cause the \dttiledhet layout to become
larger than expected, and non-contiguous. This is the same for all
three libraries. As the results are not surprising, the corresponding
figures are omitted.

\subsection{Summary}

Our basic layouts can be used to gain insight about the performance
when communicating non-contiguous data. We used only two \blocksize
and stride variants (see Table~\ref{tab:test_cases_basic}), and
observed that the qualitative performance differences are similar. It
may thus not be required to check a very large number of other
variants. For this reason, we only use the \variantone layouts in the
next section. It is noteworthy that there are surprisingly large
performance differences in handling structured data between the
MPI libraries.

\section{Performance Expectations}
\label{sec:guide}

We now investigate relative performance gains (or the opposite) by
using MPI derived datatypes, that is, the second and third issue
raised in Section~\ref{sec:intro}. We first formulate more precisely
what it is reasonable to expect, and benchmark with the aim of
verifying or falsifying these expectations. Our expectations take the
form of self-consistent performance
guidelines~\cite{Traff10:selfcons}.

An MPI performance guideline states that a certain MPI operation for
some given problem size $n$ should not be slower than some equivalent
MPI way of performing the same operation on the same problem size
(with all other things being equal). If the MPI operation is slower
than the composition of other MPI constructs implementing the same
functionality, the operation could be replaced. This is clearly
something that an application programmer should not have to
do. Verifying such guidelines that interrelate different operations
and features of the MPI standard provides a strong means of verifying
that a given MPI implementation is ``sane''.  
The verification of a set of guidelines can give valuable hints to the
application programmer on how to use features of the MPI standard
best.

For MPI datatypes, it seems impossible to say anything absolute about
the communication performance for different layouts. But it may well
be possible to formulate expectations about how the \emph{same} layout
is handled when it is described with different datatype constructors
and different MPI operations.

A first guideline, which is directly derived from the MPI
standard~\cite[Section 4.1.11]{MPI-3.1}, states that
\begin{eqnarray}
\label{eqn:contig}
\mpi\_X(c,t) & \guidesim & \mpi\_X(1,\contig(c,t))
\end{eqnarray}
excluding the time for setting up and committing the contiguous type
on the right-hand side, \ie, an MPI communication operation $X$ should
have a similar latency when transferring either $c$ elements of type
$t$ or one derived datatype of a contiguous block of $c$ elements.  If
either side of the equation would be faster than the other, the
application programmer could easily switch between them.
Thus, it can be expected that
an MPI implementation delivers similar performance for both equation
sides. As we will see, the argument is not correct, since the
\mpicommit operation can perform global optimizations on the
contiguous representation of the layout that may lead to a better
performance than possible with $c$ repetitions of the block described
by $t$. Since this cannot be controlled (or queried) by the
application, the two sides of \Gl~(\ref{eqn:contig}) may actually be
doing different things.

The next guidelines state that whatever implicit packing and unpacking
of non-contiguous data (that may be necessary inside an MPI
communication operation) is performed at least as efficiently as
explicitly packing and unpacking the whole communication buffer before
and after the communication operation using 
\mpipack and \mpiunpack~\cite[Section 4.2]{MPI-3.1}.  In a good MPI
library, we would expect many cases when the left-hand side performs
significantly better than the right-hand side. Thus:
\begin{eqnarray}
\label{eqn:pack}
\mpi\_X(c,t) & \guidelt & \mpipack(c,t) + \nonumber \\
& & \mpi\_X(1,\packed(c,t))
\end{eqnarray}
for an MPI sending operation~$X$, meaning that the performance of the
left-hand side is expected to be at least as good as the performance
of the right-hand side (all other things being equal). Similarly for
an MPI receiving operation~$Y$:
\begin{eqnarray}
\label{eqn:unpack}
\mpi\_Y(c,t) & \guidelt & \mpi\_Y(1,\packed(c,t)) + \nonumber \\
& & \mpiunpack(c,t)
\end{eqnarray}

The right-hand sides have the disadvantages (1) of requiring an extra
buffer for the intermediate, contiguous packing unit, (2) of
preventing direct communication of large contiguous parts of the
datatype and (3) of preventing pipelining of packing and unpacking in
the communication operations (as well as all other dynamic
optimizations, and optimizations that exploit communication hardware
support). Therefore it should not be recommended. We would expect that
MPI libraries trivially fulfill these guidelines with equality, and
would hope to find relevant cases where the left-hand sides are much
faster than the right-hand sides.

Any data layout can be described in an infinite number of ways with
the available MPI datatype constructors. This is easy to see, for
instance $\contig(1,t)$ describes the same layout as $t$ itself for
any datatype $t$. For any given data layout, each MPI library will
have layout descriptions that lead to the best communication performance.
The \mpicommit operation provides a handle for the MPI library to
transform the datatype given by the user into a better (if possible),
internal description. This process is called \emph{datatype
  normalization}~\cite{Traff11:typeguide}, and we call this best,
alternative representation of a layout described by datatype $t$ its
\emph{normalized form} $\normal(t)$. The expectation is that an MPI
library will indeed attempt to find a good normalized form at \mpicommit
time (if not, the user could do better by deriving the normalized form by
himself and setting up the datatype in that way), which is formalized as the
following \emph{datatype normalization} performance guideline:
\begin{eqnarray}
\label{eqn:normal}
\mpi\_X(c,t) & \guidelt & \mpi\_X(c,\normal(t))
\end{eqnarray}
That is, we expect the performance of a communication operation $X$ with
datatype $t$ to be no worse than what can be achieved with the best,
normalized description of the layout. The guideline is tricky,
since the user may not readily be able to see what is the best way to
describe a layout in a given situation. But in many cases he can give
a good guess, and the guideline states that we would expect the
\mpicommit operation to do as well.

The normalization heuristics typically applied by MPI libraries
replace more general type constructors (struct) with more specific
ones (index or index block), collapse nested constructors, and
identify large contiguous segments, where such replacements are
applicable. Explicit
descriptions of common type normalization
heuristics can be found in~\cite{KjolstadHoeflerSnir11,KjolstadHoeflerSnir12,PrabhuGropp15,RossMillerGropp03,SchneiderKjolstadHoefler13}. As
we will see in the following, there are natural layout descriptions
that are \emph{not} normalized by these heuristics, leading to severe
violations of the guideline.

\subsection{Communication Patterns}

In our experiments, we will use the same three types of communication
operations as in \Sec~\ref{sec:raw}. In order to verify
\Gls~(\ref{eqn:pack}) and~(\ref{eqn:unpack}), we extend the benchmarks
with \mpipack and \mpiunpack operations to achieve the same semantics
as when datatype arguments were used directly in the communication
calls:
\begin{myitemize}
\item \pingpong (\cf
  Schneider~\etal~\cite{SchneiderGerstenbergerHoefler14}): Ping side:
  \mpipack followed by \mpisend followed by \mpirecv followed by
  \mpiunpack.  Pong side: \mpirecv followed by \mpiunpack followed by
  \mpipack followed by \mpisend.
\item Asymmetric (rooted) collective, e.g., \mpibcast on $p$
  processes. Root: \mpipack followed by \mpibcast. Non-roots:
  \mpibcast followed by \mpiunpack.
\item Symmetric (non-rooted) collective, e.g., \mpiallgather on $p$
  processes. All processes call \mpipack, followed by \mpiallgather,
  then all processes perform $p$ successive \mpiunpack operations on
  the received, packed blocks.
\end{myitemize}

In the \mpiallgather pattern, the successive unpacking of the
received blocks is necessary, since the catenation of packing units is
not a packing unit~\cite[Section 4.2]{MPI-3.1}, so even if the
received $p$ packed blocks do form a contiguous piece of memory, it is
not correct to unpack it with only one \mpiunpack operation.

\subsection{Experimental Results}

The structure of experiments is guided by the guidelines, and we state
for each experiment what our expectations (hypotheses) are, and
comment on whether the results support or falsify them.  As baseline
we use in most cases the simple layouts of Section~\ref{sec:raw}.  We
report results only for the \mpiint basetype and the \variantone basic
layouts.

\subexptest{exptest:pack_unpack}
For \Gls~(\ref{eqn:pack}) and~(\ref{eqn:unpack}), we first use the
layouts of \Sec~\ref{sec:raw} with the same values for $A$ and $B$ and
compare the benchmark performance with datatype communication against
the performance with explicit pack and unpack operations.

\DEdescr%
\begin{dtabular}{ll}
  \toprule
  Reference Layouts & \dtdtiled, \dtdblock\\
                   & \dtdbucket, \dtdalternating  \\
  Compared Layouts  & same layouts, but using \\
                   &  \mpipack and \mpiunpack \\
  \midrule
  \blocksize~\VARblocksize  & $\num{2}, \num{10}, \num{10000}$ \\
  \datasize~\VARdatasize & \SIrange{64000}{2560000}{\Bytes}\\
  comm. patterns   & \pingpong, \mpiallgather, \mpibcast \\
    \# of processes   & \num{32x1}, \num{1x16} \\
  			& (\num{2x1}, \num{1x2} for \pingpong) \\
  \bottomrule
\end{dtabular}

\DEhypo

We do not expect any MPI library to significantly violate the
guidelines \Gls~(\ref{eqn:pack}) and~(\ref{eqn:unpack}) , but with
these simple layouts we hope to see cases where an MPI library
performs significantly better with datatypes than with explicit
packing and unpacking.

\DEresults

\begin{figure*}[!t]
\centering
\begin{subfigure}{.24\linewidth}
\centering
\includegraphics[width=\linewidth]{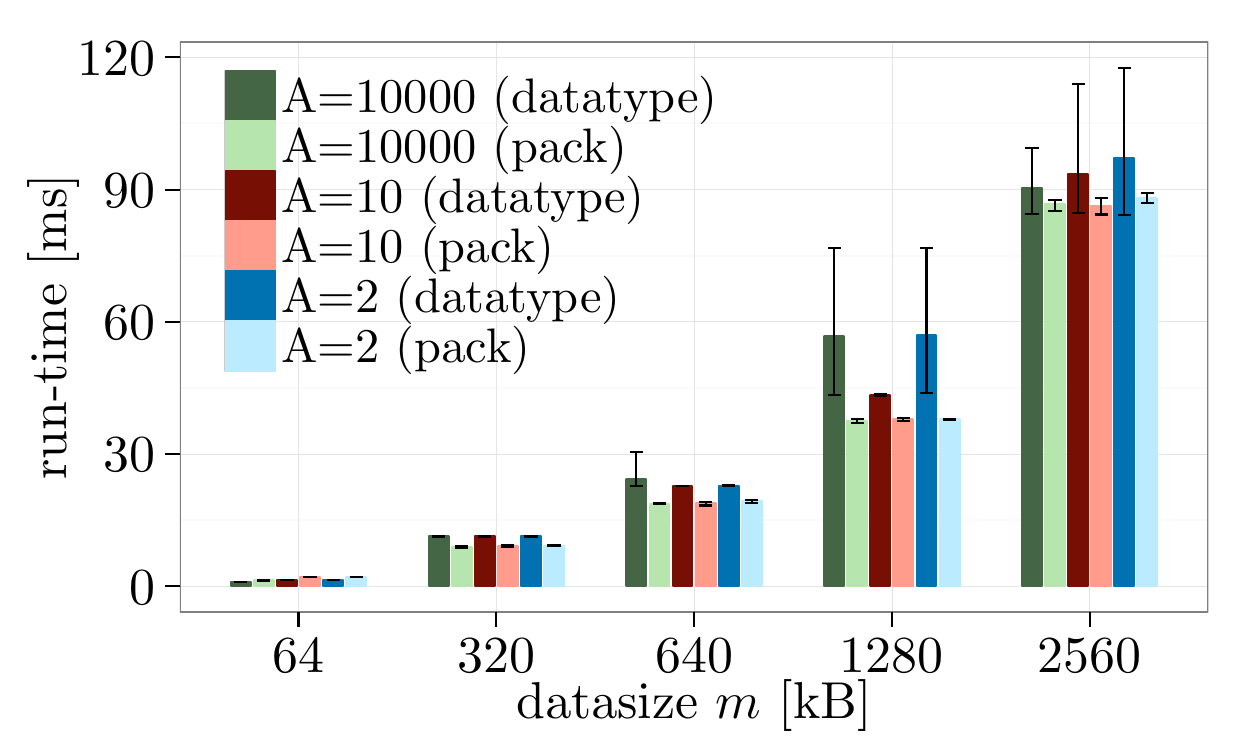}
\caption{%
\label{fig:exp:allgather-pack-tiled-32x1-nec}%
\dttiled, \jupiternecmpi%
}%
\end{subfigure}%
\hfill%
\begin{subfigure}{.24\linewidth}
\centering
\includegraphics[width=\linewidth]{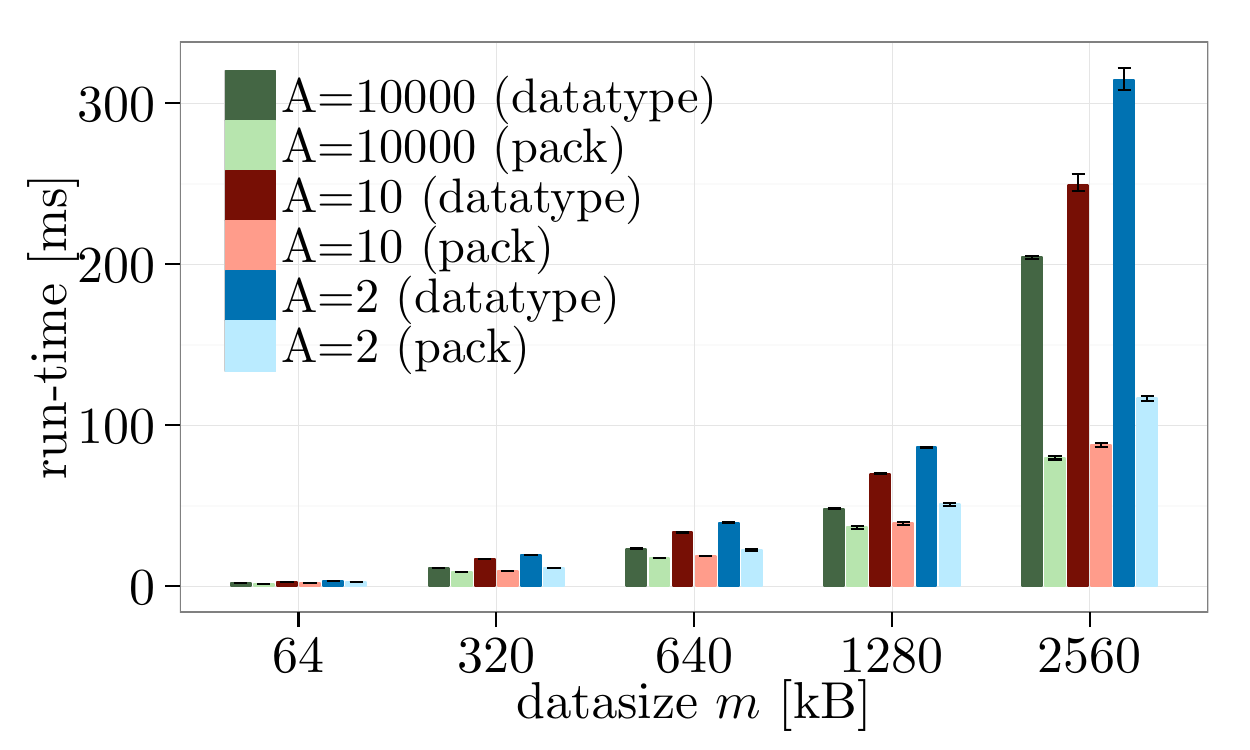}
\caption{%
\label{fig:exp:allgather-pack-tiled-32x1-mvapich}%
\dttiled, \jupitermvapich%
}%
\end{subfigure}%
\hfill%
\begin{subfigure}{.24\linewidth}
\centering
\includegraphics[width=\linewidth]{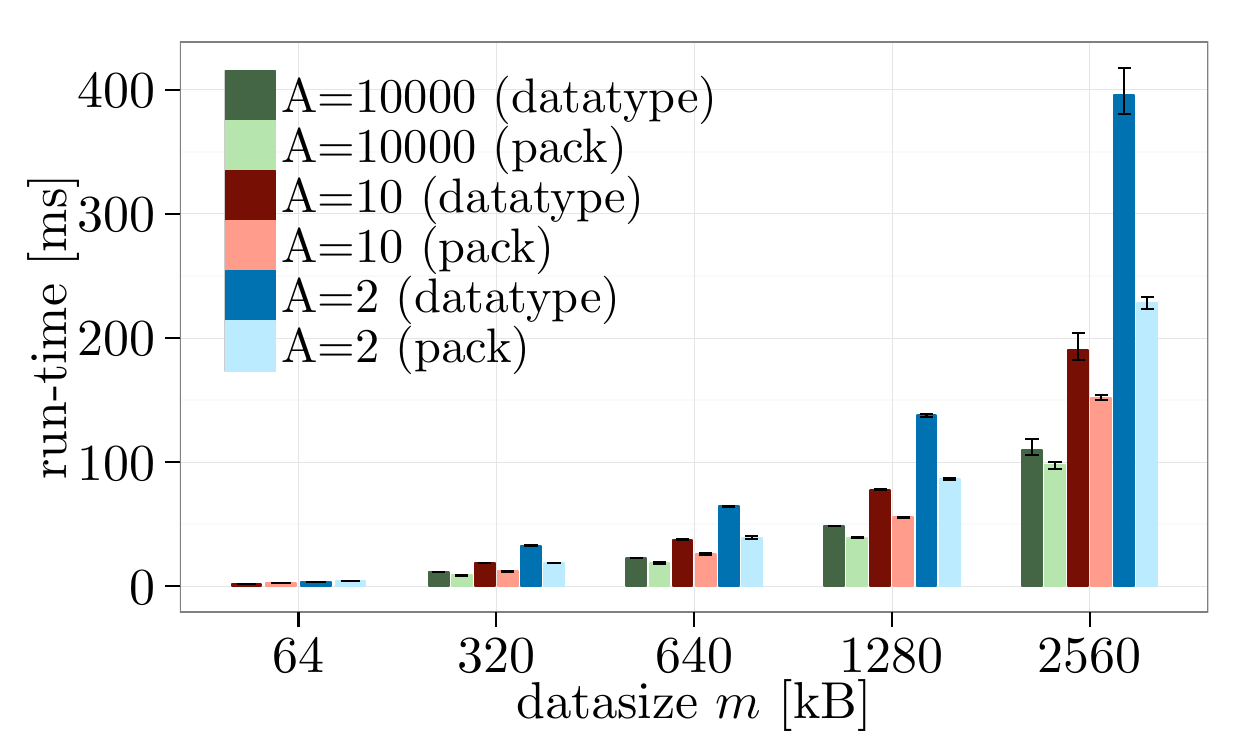}
\caption{%
\label{fig:exp:allgather-pack-bucket-32x1-nec}%
\dtbucket, \jupiternecmpi%
}%
\end{subfigure}%
\hfill%
\begin{subfigure}{.24\linewidth}
\centering
\includegraphics[width=\linewidth]{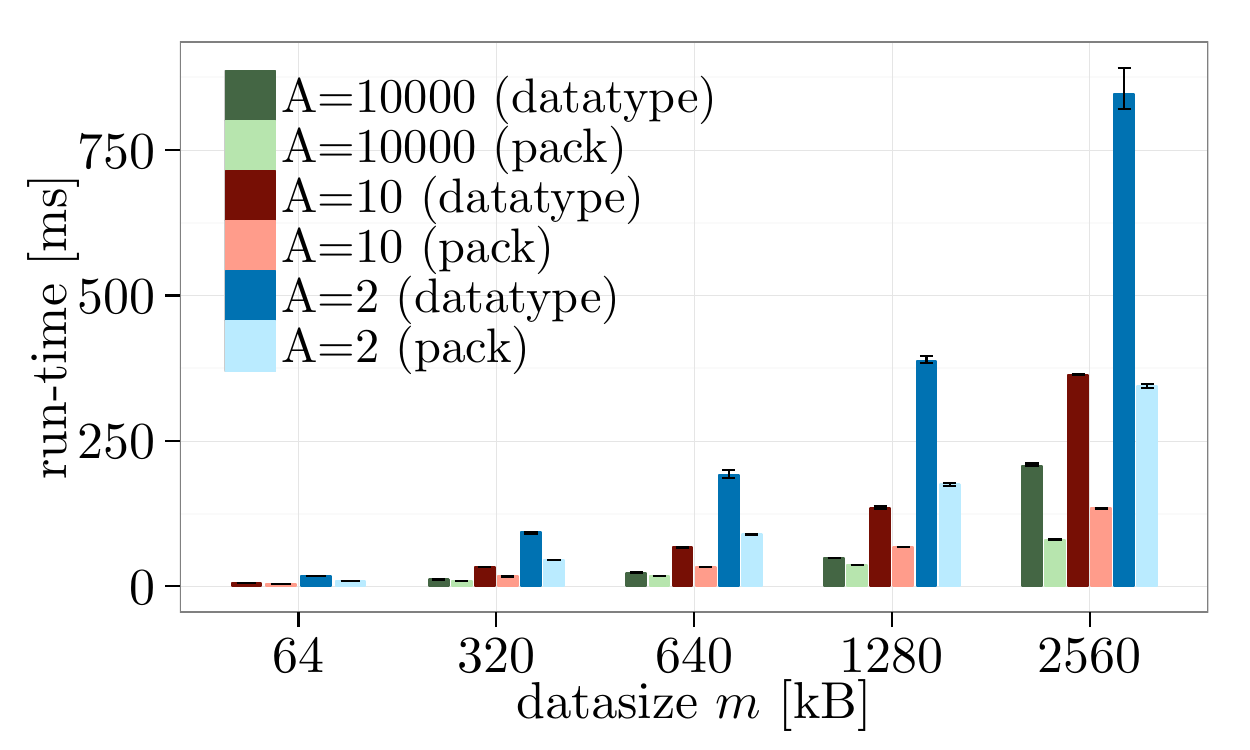}
\caption{%
\label{fig:exp:allgather-pack-bucket-32x1-mvapich}%
\dtbucket, \jupitermvapich%
}%
\end{subfigure}%
\caption{\label{fig:exp:allgather-pack-32x1}  Basic layouts \vs pack/unpack, element datatype: \mpiint, \num{32x1}~processes, \mpiallgather.}
\end{figure*}
\begin{figure}[t]
\centering
\begin{subfigure}{.49\linewidth}
\centering
\includegraphics[width=\linewidth]{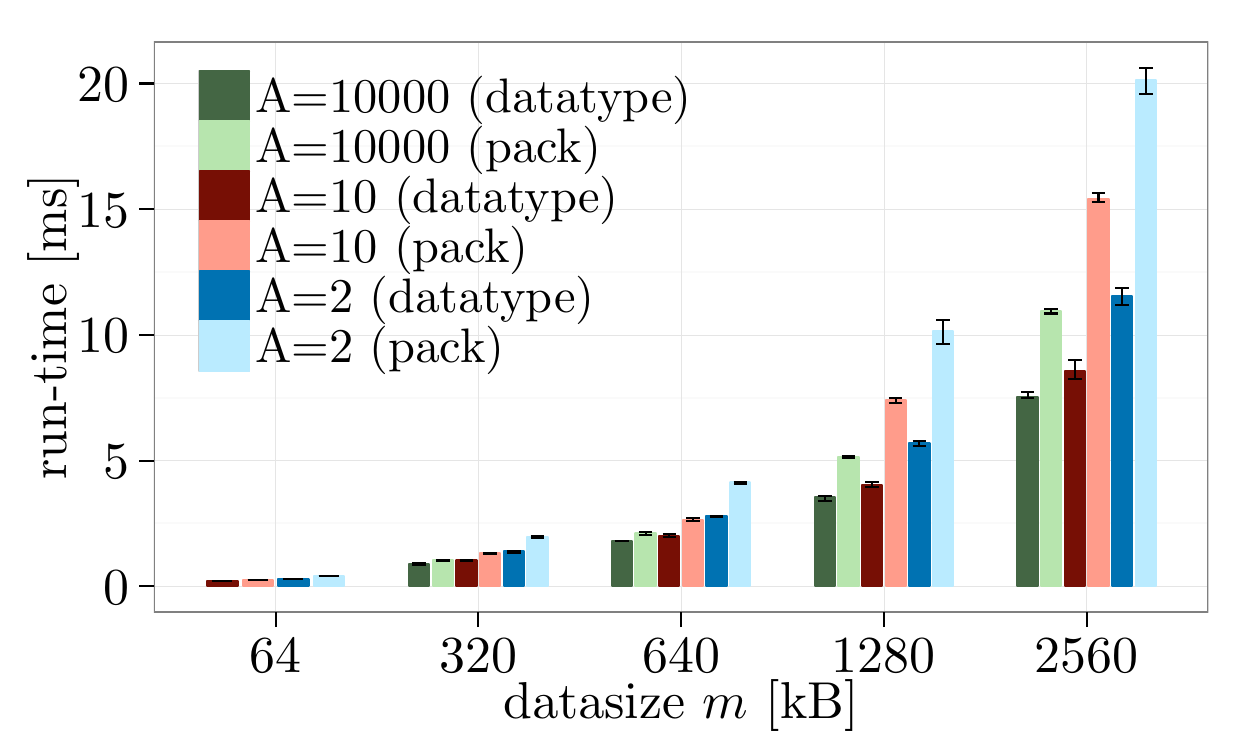}
\caption{%
\label{fig:exp:bcast-pack-alternating-onenode-nec}%
\jupiternecmpi%
}%
\end{subfigure}%
\hfill%
\begin{subfigure}{.49\linewidth}
\centering
\includegraphics[width=\linewidth]{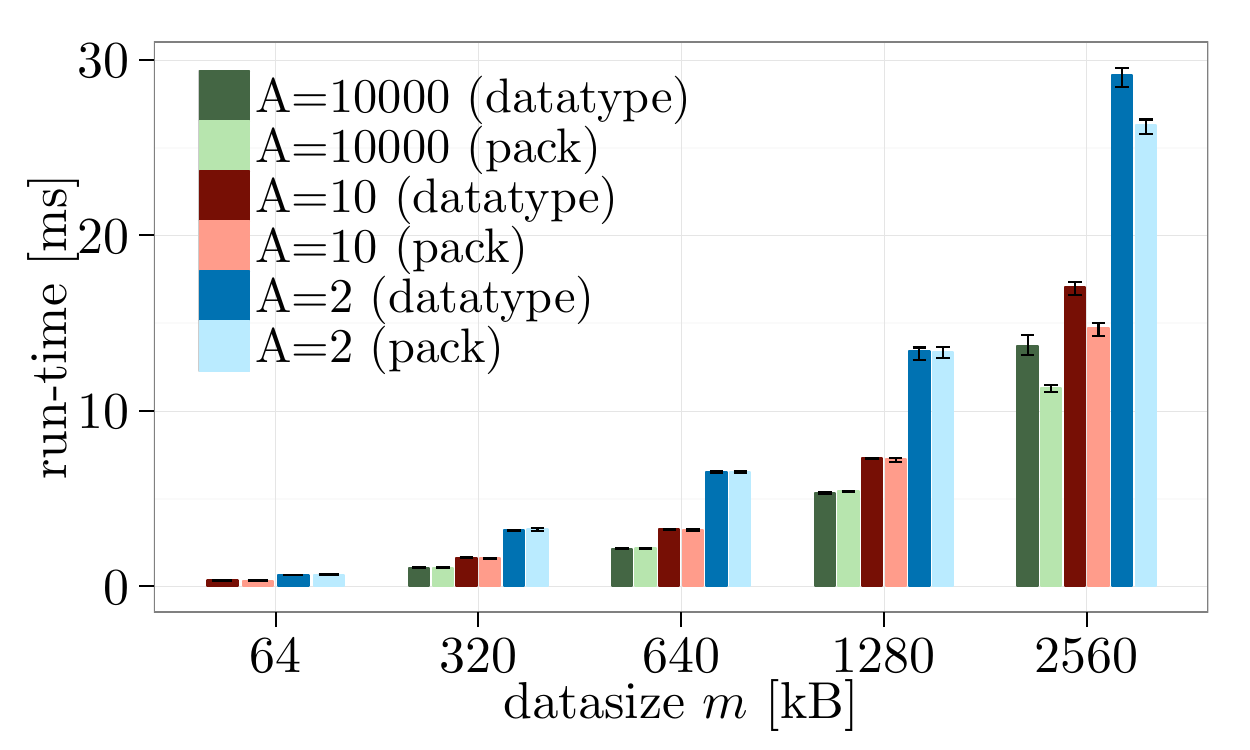}
\caption{%
\label{fig:exp:bcast-pack-alternating-onenode-mvapich}%
\jupitermvapich%
}%
\end{subfigure}%
\caption{\label{fig:exp:bcast-pack-onenode}  \dtalternating layout \vs pack/unpack, element datatype: \mpiint,  \num{1x16}~processes, \mpibcast.}
\end{figure}

Much to our surprise, we found many cases where the guidelines are
severely violated. For processes on different nodes, both the
\jupiternecmpi and \jupitermvapich libraries violate the guidelines
for all layouts, see \fig~\ref{fig:exp:allgather-pack-32x1} for two
examples. Especially with \jupitermvapich, the violations are severe
and amount to factors of two or more. 
With MPI processes on the same node, 
in many cases (with \jupiternecmpi) MPI derived datatypes performed
better than explicit packing and unpacking before and after
communication. Examples with the \mpibcast pattern are shown in
\fig~\ref{fig:exp:bcast-pack-onenode}.

\subexptest{exptest:contig}

As a sanity check for \Gl~(\ref{eqn:contig}), we create a contiguous
$n/k$-element datatype with the \mpicontig constructor for each of the
$k$-element datatypes of \Sec~\ref{sec:raw}. We compare the
performance of the two datatypes against each other for different
communication patterns.

\DEdescr
\begin{dtabular}{ll}
  \toprule
  Reference Layouts & \dtdtiled, \dtdblock\\
                   & \dtdbucket, \dtdalternating  \\
  Compared Layouts & \ddtcontig with subtypes: \\
                   & \dtdtiled, \dtdblock \\
                   &  \dtdbucket, \dtdalternating \\
  \midrule
  \blocksize~\VARblocksize  &  $\num{2}, \num{10}, \num{100}, \num{1000}, \num{1024}, \num{10000}$ \\
  \datasize~\VARdatasize & \SI{2000}{\Bytes}, \SI{2560000}{\Bytes}\\
  comm. patterns   & \pingpong \\
  \# of processes   & \num{2x1}, \num{1x2} \\
  \bottomrule
\end{dtabular}

\DEtypes %

The concrete \ddtcontig is shown in
\fig~\ref{fig:test_basic_vs_contig} with \dtdtiled as subtype. The
\dtdtiled subtype consists of units of $k=A$ elements, and in the
communication patterns, $n/k$ such units are communicated. In
contrast, the \ddtcontig contains all $n/k$ units in a single type,
so all $n$ elements are communicated using a count of one with this
datatype.

\begin{figure}[h!]
\centering
\begin{tikzpicture}[scale=0.4]
  \def\myyshift{-3pt}

  \draw (0,0) rectangle (11.5,1);
  \foreach \x in {0,.5,1,2,2.5,3,4,4.5,5,6,6.5,7,9.5,10.0,10.5}
  \draw[fill = lightgray] (\x,0) rectangle (\x+.5,1);
  \node at (8.5,0.5) {...};
  
  \def\toff{-0.1}

  \draw [decorate,decoration={brace,amplitude=5pt,mirror}]
  (0,\toff-.1) -- node[anchor = north west, xshift=-5, yshift=\myyshift]{\mpicontig}
  (1.5,\toff-.1);

  \draw [decorate,decoration={brace,amplitude=5pt,mirror}]
  (0,\toff-1.3-.1) -- node[anchor = north west, xshift=-5, yshift=\myyshift]{\mpiresized (\dttiled)}
  (2,\toff-1.3-.1);

  \draw (0,0) rectangle (11.5,1);
  \foreach \x in {0,.5,1,2,2.5,3,4,4.5,5,6,6.5,7.0,9.5,10.0,10.5}
  \draw[fill = lightgray] (\x,0) rectangle (\x+.5,1);
  \node at (8.5,0.5) {...};
  
  \def\toff{-4.5}
  \def\vpad{.0}

  \foreach \x in {0,2,4,9.5}
  \draw[fill=white] (\x,\toff) rectangle (\x+2,\toff+1) node[pos=.5] { $T_1$ };
  \draw[fill=white] (6,\toff) rectangle (9.5,\toff+1) node[pos=.5] {...};  

  \draw [decorate,decoration={brace,amplitude=5pt,mirror}]
  (0,\toff-.1) -- node[anchor = north, yshift=\myyshift, align=left]{\mpicontig}
  (11.5,\toff-.1);

  \foreach \x in {0,.5,1}
  \draw[fill = lightgray] (13.5+\x,\toff) rectangle (13.5+\x+.5,\toff+1);
  \draw (13.5,\toff) rectangle (15.5,\toff+1);
  \node at (13,\toff+.5)  (a) {$T_1$};
\end{tikzpicture}
\caption{Basic, static layouts (in this figure \dtdtiled, top)
  \vs \ddtcontig (bottom).\vspace{-5pt}}
\label{fig:test_basic_vs_contig}
\end{figure}

\newpage

\DEhypo
We expect no difference in performance between the reference and
compared layouts.

\DEresults

\begin{figure*}[!t]
\centering
\begin{subfigure}{.245\linewidth}
\centering
\includegraphics[width=\linewidth]{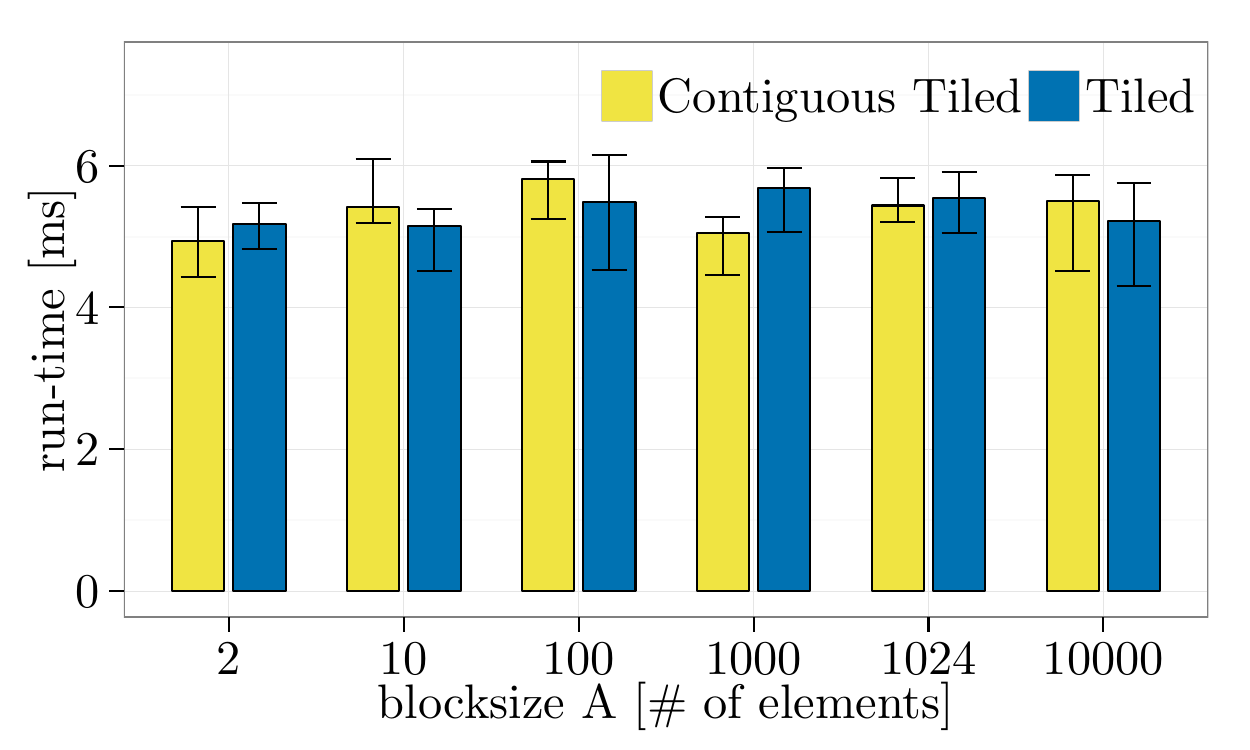}
\caption{%
\label{fig:exp:pingpong-contigtiled-2x1}%
\dttiled, \jupiternecmpi%
}%
\end{subfigure}%
\hfill%
\begin{subfigure}{.245\linewidth}
\centering
\includegraphics[width=\linewidth]{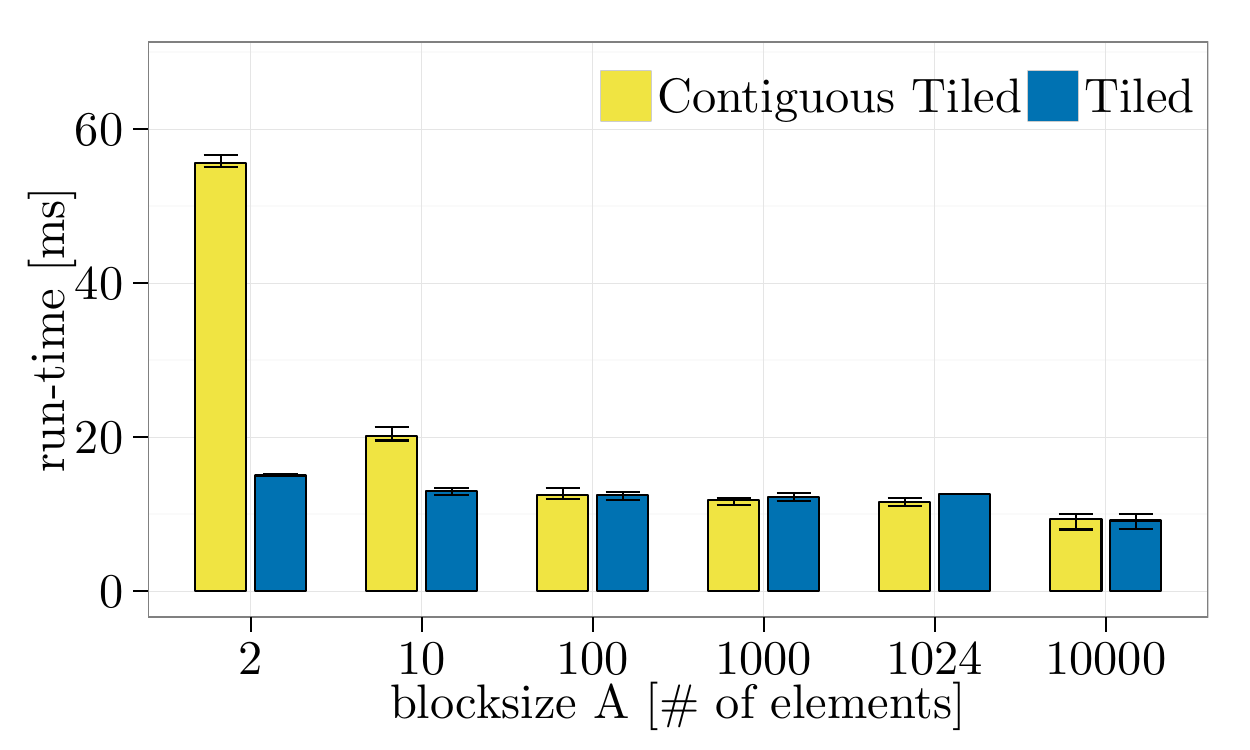}
\caption{%
\label{fig:exp:pingpong-contigtiled-2x1-mvapich}%
\dttiled, \jupitermvapich%
}%
\end{subfigure}%
\hfill%
\begin{subfigure}{.245\linewidth}
\centering
\includegraphics[width=\linewidth]{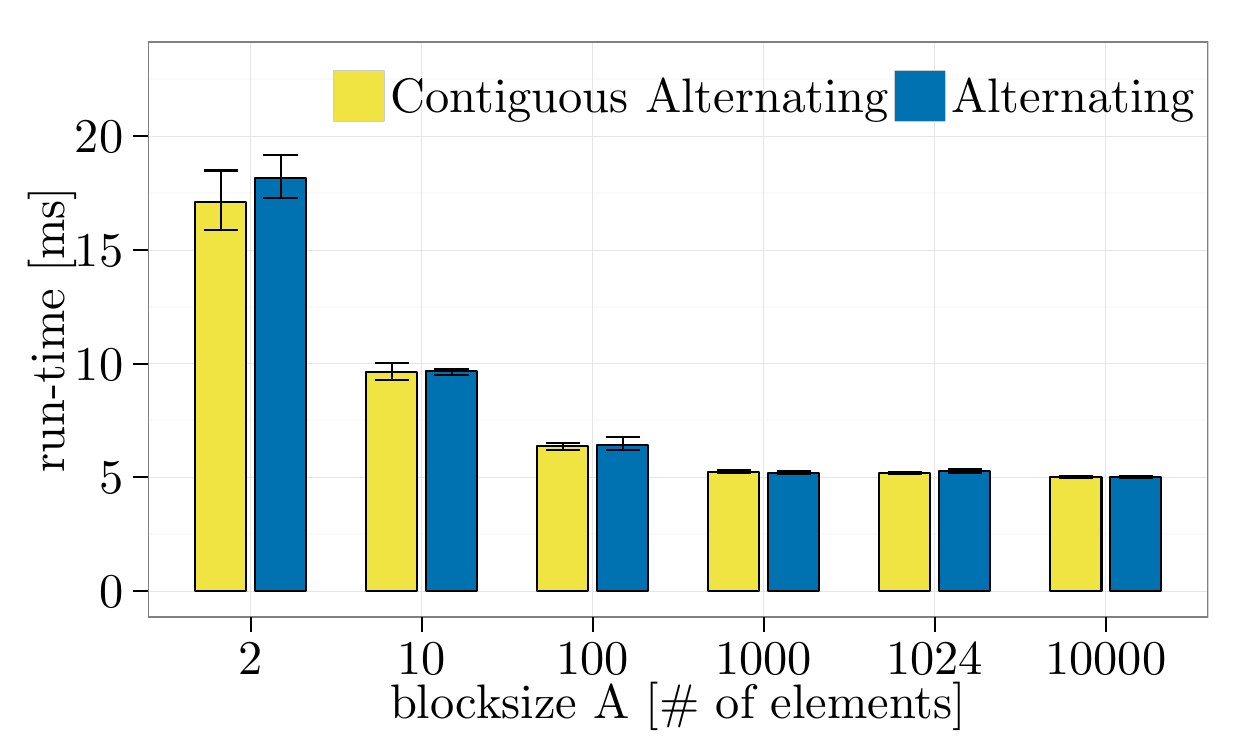}
\caption{%
\label{fig:exp:pingpong-contigalternating-2x1}%
\dtalternating, \jupiternecmpi%
}%
\end{subfigure}%
\hfill%
\begin{subfigure}{.245\linewidth}
\centering
\includegraphics[width=\linewidth]{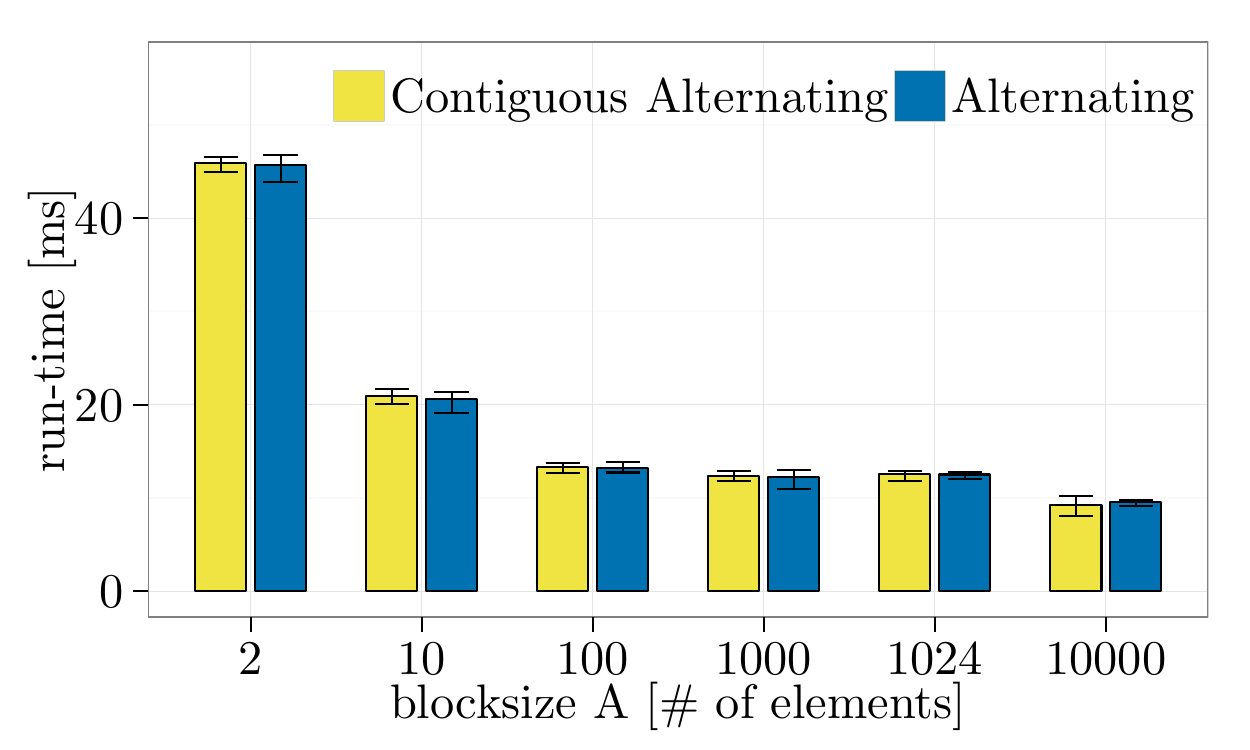}
\caption{%
\label{fig:exp:pingpong-contigalternating-2x1-mvapich}%
\dtalternating, \jupitermvapich%
}%
\end{subfigure}%
\caption{\label{fig:exp:pingpong-contig-largenbytes-2x1}  Basic layouts \vs \ddtcontig, $\VARdatasize=\SI{2.56}{\mega\byte}$, element datatype: \mpiint, \num{2x1}~processes, \pingpong.\vspace{-10pt}}
\end{figure*}

Again, surprisingly, the \jupitermvapich (and also \jupiteropenmpi,
not shown here) library violates the guideline for the \dtdtiled type,
with \ddtcontig being a factor of three slower with \dtdtiled as
subtype, see \fig~\ref{fig:exp:pingpong-contig-largenbytes-2x1}.  For
the other basic types \dtdblock, \dtdbucket, and \dtdalternating, no
significant performance difference between the left-hand and the
right-hand side of \Gl~(\ref{eqn:contig}) was detected.

\subexptest{exptest:tiled_struct}

The remaining expectation tests are concerned with verifying
\Gl~(\ref{eqn:normal}). In all these tests, we give different
descriptions of the same layout and study the communication
performance. 
We formulate our expectations (hypotheses) on the assumption that a
strong type normalization is not performed by the MPI libraries.

Our first experiment uses the regularly strided \dtdtiled layout, for
which we now know the baseline communication performance. We describe
this pattern as a larger block comprised of several \dtdtiled
subtypes.

\DEdescr
\begin{dtabular}{ll}
  \toprule
  Reference Layout & \dtdtiled \\
  Compared Layout & \dttiledstruct \\
  \midrule
  \blocksize~\VARblocksize  &  $\num{2}, \num{10}, \num{100}, \num{1000}, \num{1024}, \num{10000}$ \\
  \stride~\VARstride  & $A+2$ \\
  repetition counts $S_1, S_2$ & (a) $S_1=1$, $S_2=1$ \\
  				& (b) $S_1=2$, $S_2=3$\\
  \datasize~\VARdatasize & \SI{2000}{\Bytes}, \SI{2560000}{\Bytes}\\
  comm. patterns   & \pingpong \\
  \# of processes   & \num{2x1}, \num{1x2} \\
  \bottomrule
\end{dtabular}

\DEtypes 
The \dttiledstruct layout is a concatenation described with \mpistruct
of two smaller, contiguously strided layouts of $S_1$ and $S_2$ tiled
blocks.  The description is illustrated in
\fig~\ref{fig:test_tiled_vs_tiled_struct}. Each \dttiled sub-layout
has the same \blocksize $A$ and \stride $B$. The number of elements in
the structure is $(S_1+S_2)A$. We also allow the degenerated (fully
contiguous) case of $A=B$.
However, we only test the performance of the structured layout using
$A<B$, with $A$ and $B$ chosen as in \variantone (\cf
\tab~\ref{tab:test_cases_basic}).
\begin{figure}[!h]
\centering
\begin{tikzpicture}[scale=0.4]
  
  \def\toff{0.0}
  \def\vpad{.0}
  \def\myyshift{-3pt}

  \def\toneyoff{1.5}
  \def\tonexoff{0.0}

  \foreach \xoff in {0,2,4,6}
  \foreach \x in {0,.5,1} {
  \draw[fill = lightgray] (\tonexoff+\x+\xoff,\toneyoff) rectangle (\tonexoff+\x+\xoff+.5,\toneyoff+1);
  \draw (\tonexoff+\xoff+1.5,\toneyoff) rectangle (\tonexoff+\xoff+2,\toneyoff+1);
  }
  \node at (\tonexoff-.6,\toneyoff+.5)  (a) {$T_1$};

  \draw [<->]
  (0,\toneyoff+1.3) -- node[anchor = south, yshift=-.4*\myyshift]{$S_1$}
  (8,\toneyoff+1.3);

  \draw [decorate,decoration={brace,amplitude=5pt,mirror}]
  (0,\toneyoff-.1) -- node[anchor = north, yshift=\myyshift]{\dttiled}
  (2,\toneyoff-.1);

  \draw [decorate,decoration={brace,amplitude=5pt,mirror}]
  (0,\toneyoff-1.2) -- node[anchor = north, yshift=\myyshift]{\mpicontig}
  (8,\toneyoff-1.2);

  \def\toff{-2.0}

  \draw[fill=white] (0,\toff) rectangle (8,\toff+1) node[pos=.5] { $T_1$ };
  \draw[fill=white] (8,\toff) rectangle (11.5,\toff+1) node[pos=.5] { $T_2$ };  

  \draw [decorate,decoration={brace,amplitude=5pt,mirror}]
  (0,\toff-.1) -- node[anchor = north, yshift=\myyshift]{\mpistruct}
  (11.5,\toff-.1);
\end{tikzpicture}
\caption{\dttiled (top) \vs \dttiledstruct (bottom).}
\label{fig:test_tiled_vs_tiled_struct}
\end{figure}

\DEhypo

As explained, since the user can easily describe the \dttiled layout
with the \mpicontig and \mpiresized constructor, we would like to
expect that the MPI library can detect from the \dttiledstruct
description that the underlying pattern is a simple, tiled
pattern. That would require detecting that both sub-layouts of the
\mpistruct are indeed tiled (although with different repetition
counts) and have the same basetype. However, the heuristics used by
MPI libraries at \mpicommit time usually work
differently~\cite{KjolstadHoeflerSnir11,KjolstadHoeflerSnir12}. Therefore,
we actually expect to see cases where the \dttiledstruct description
performs worse than the reference layout.

\DEresults

\begin{figure}[t]
\centering
\begin{subfigure}{.49\linewidth}
\centering
\includegraphics[width=\linewidth]{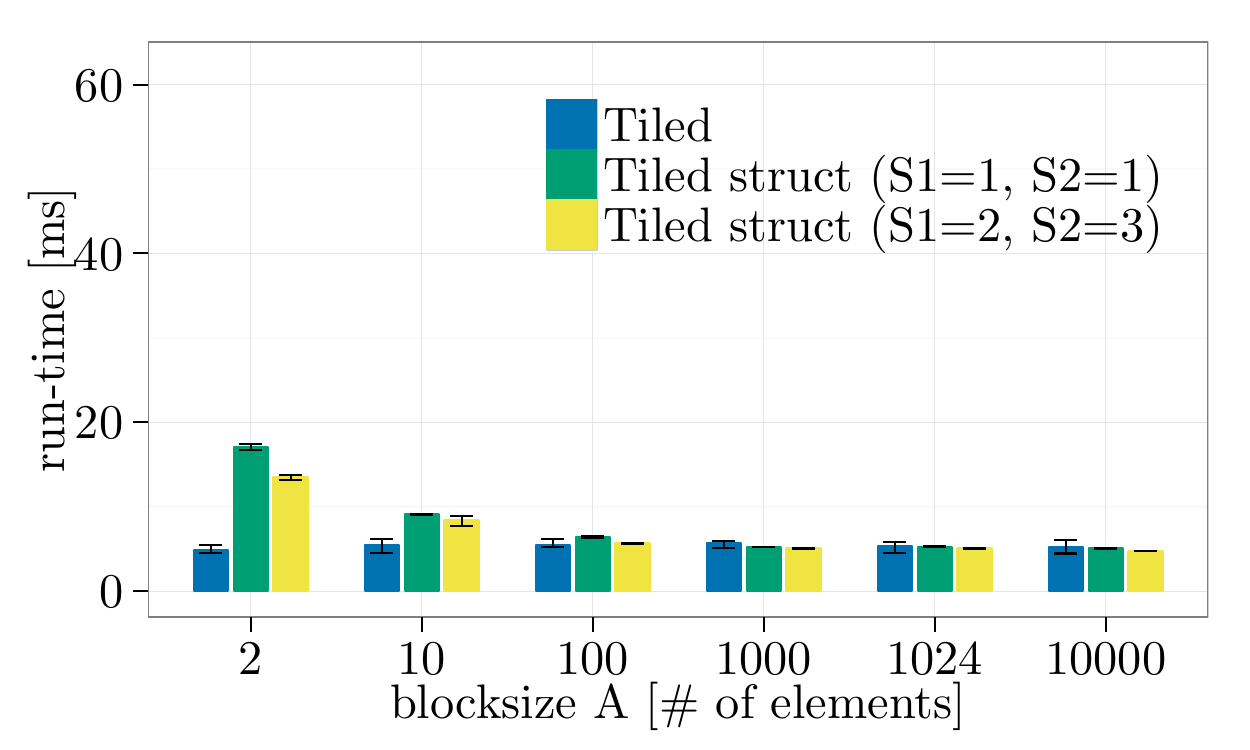}
\caption{%
\label{fig:exp:pingpong-tiledstruct-large-2x1}%
\jupiternecmpi
}%
\end{subfigure}%
\hfill%
\begin{subfigure}{.49\linewidth}
\centering
\includegraphics[width=\linewidth]{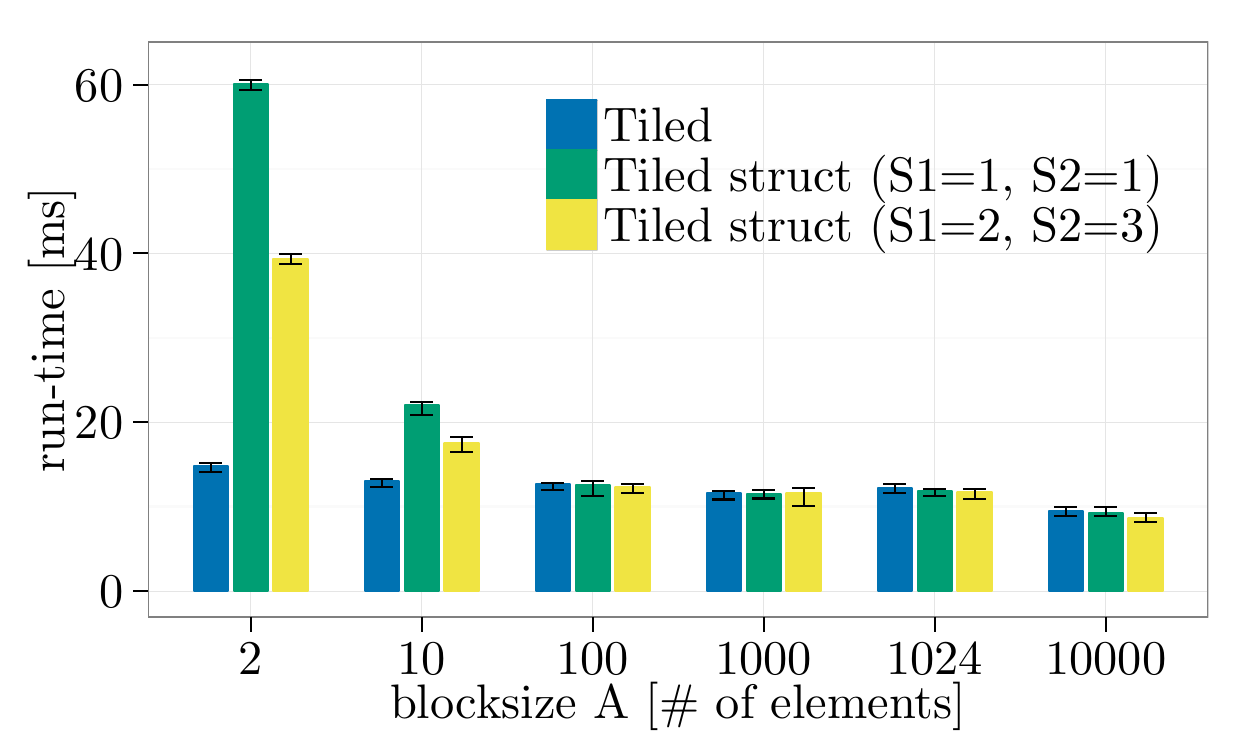}
\caption{%
\label{fig:exp:pingpong-tiledstruct-large-2x1-mvapich}%
\jupitermvapich
}%
\end{subfigure}%
\caption{\label{fig:exp:pingpong-tiledstruct} \dtdtiled \vs \dttiledstruct, element datatype: \mpiint, $\VARdatasize=\SI{2.56}{\mega\byte}$, \num{2}~nodes, \pingpong.\vspace{-5pt}}
\end{figure}

The results in \fig~\ref{fig:exp:pingpong-tiledstruct}, especially for
\jupitermvapich, show a large performance difference for $A=2$. Even
in the case of $S_1=S_2$, where the two subtypes are identically set
up, the \dttiledstruct performs several factors worse. This shows that
the normalization heuristics in the MPI libraries are insufficient to
identify the complex description of the simple, tiled layout and
to normalize accordingly.

\subexptest{exptest:tiled_vector}

The next experiment for \Gl~(\ref{eqn:normal}) is also a sanity check,
where we would expect no differences between two equally simple,
natural descriptions of a tiled layout. We now describe the \dtdtiled
layout as a vector of $n/A$ contiguous blocks of $A$ elements with
stride $B$. This is the ``most natural'' way in MPI to describe a
long, regularly strided layout, and is accomplished with
\mpivector. To have the same extent of the vector type, we resize the
extent of the vector to $n/A$ times the extent of \dtdtiled. The
\ddttiledvector datatype is our first example of a \emph{dynamically
  derived datatype} that can only be set up when the number of
elements $n$ to be communicated is known. As in all our experiments,
we do \emph{not} include the datatype setup time in the measured
\runtime.

\DEdescr
\begin{dtabular}{ll}
  \toprule
  Reference Layout & \dtdtiled \\
  Compared Layout & \ddttiledvector \\
  \midrule
  \blocksize~\VARblocksize  &  $\num{2}, \num{10}, \num{100}, \num{1000}, \num{1024}, \num{10000}$ \\
  \stride~\VARstride  & $A+2$ \\
  \datasize~\VARdatasize & \SI{2000}{\Bytes}, \SI{2560000}{\Bytes}\\
  comm. patterns   & \pingpong \\
  \# of processes   & \num{2x1}, \num{1x2} \\
  \bottomrule
\end{dtabular}

\DEtypes%
The two contrasted datatype layout descriptions \dtdtiled and
\ddttiledvector are illustrated in
\fig~\ref{fig:test_tiled_vs_tiled_vector}.

\begin{figure}[h!]
\centering
\begin{tikzpicture}[scale=0.4]
  \def\myyshift{-3pt}
  \def\toff{0.0}

  \draw (0,\toff) rectangle (11.5,\toff+1) node[pos=.5] {$T_1$};

  \foreach \x in {0,.5,1}
  \draw[fill = lightgray] (13+\x,\toff) rectangle (13+\x+.5,\toff+1);
  \draw (13.0,\toff) rectangle (15,\toff+1);
  \node at (12.5,\toff+.5)  (a) {$P$};

  \draw [decorate,decoration={brace,amplitude=5pt,mirror}]
  (0,\toff-.1) -- node[anchor = north, yshift=\myyshift]{\mpivector using pattern $P$}
  (11,\toff-.1);

  \draw [decorate,decoration={brace,amplitude=5pt,mirror}]
  (0,\toff-1.5-.1) -- node[anchor = north, yshift=\myyshift]{\mpiresized}
  (11.5,\toff-1.5-.1);
\end{tikzpicture}
\caption{\ddttiledvector defines the layout with \mpivector using the
  \dttiled pattern.}
\label{fig:test_tiled_vs_tiled_vector}
\end{figure}

\DEhypo 

We expect the performance of the two descriptions to match. 
The MPI internal representations of the two descriptions can be
expected to be similar, and concrete offsets for accessing the
elements in the layout can be computed easily by the datatype engine
given these representations. Since the \mpivector is a commonly used
datatype constructor, it may even have been specially optimized, such
that the description as \ddttiledvector might be slightly
advantageous.

\DEresults

\begin{figure}[htpb]
\centering
\begin{subfigure}{.49\linewidth}
\centering
\includegraphics[width=\linewidth]{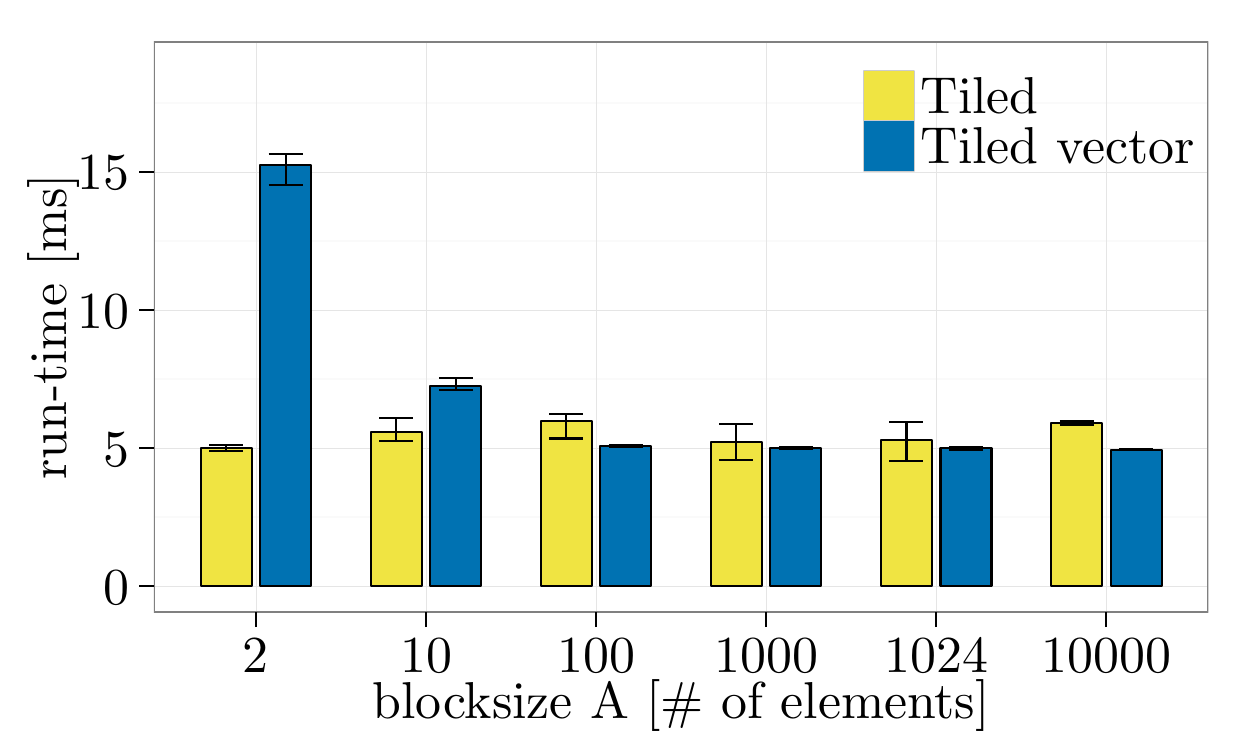}
\caption{%
\label{fig:exp:pingpong-tiledvec-large-2x1}%
\jupiternecmpi
}%
\end{subfigure}%
\hfill%
\begin{subfigure}{.49\linewidth}
\centering
\includegraphics[width=\linewidth]{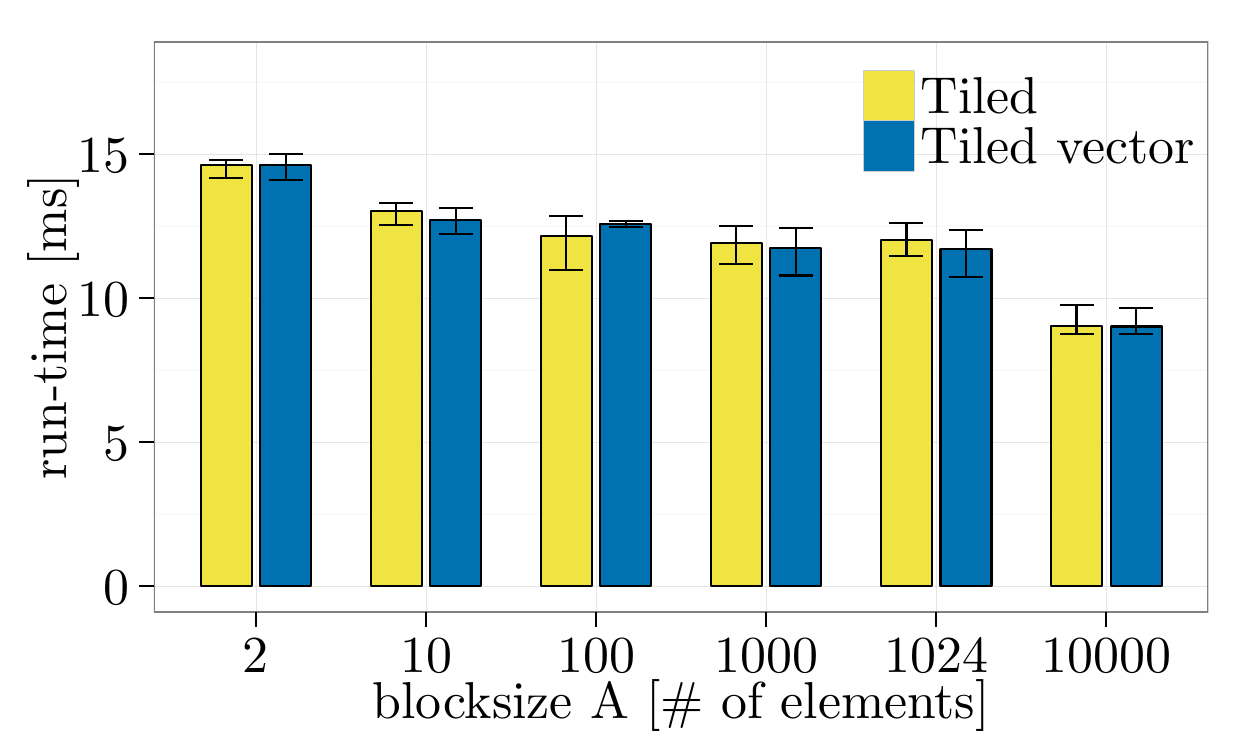}
\caption{%
\label{fig:exp:pingpong-tiledvec-large-2x1-mvapich}%
\jupitermvapich
}%
\end{subfigure}%
\caption{\label{fig:exp:pingpong-tiledvec}  \dtdtiled \vs \ddttiledvector, element datatype: \mpiint, $\VARdatasize=\SI{2.56}{\mega\byte}$, \num{2}~nodes, \pingpong.}
\end{figure}

As shown in~\fig~\ref{fig:exp:pingpong-tiledvec}, for the
\jupiternecmpi (and also for \jupiteropenmpi) library and small values
of $A$, the \ddttiledvector performs much worse than repeating the
\dtdtiled block. This is again surprising. However, the comparison of the
absolute performance between \jupiternecmpi and
\jupitermvapich shows that the bad performance of \ddttiledvector is
relative, in absolute terms it is on par with the performance in
\jupitermvapich for both descriptions of the layout. These findings
also illustrate that performance guidelines can only ensure
consistency in an MPI library. Guideline verification needs to be
complemented with benchmarking against hard baselines.

\subexptest{exptest:vector_tiled}

Our next description of the \dtdtiled layout is done using a nested
vector. We describe a larger block of a constant $S$ number of units
of $A$ elements and stride $B$ with the \mpivector constructor. On
this datatype, we build a dynamic vector of $n/(SA)$ blocks with a
stride of $SB$ elements. In order to express the stride correctly,
this vector has to be constructed with the \mpihvector constructor.

\DEdescr
\begin{dtabular}{ll}
  \toprule
  Reference Layout & \dtdtiled \\
  Compared Layout & \ddtvectortiled \\
  \midrule
  \blocksize~\VARblocksize  &  $\num{2}, \num{10}, \num{100}, \num{1000}, \num{1024}, \num{10000}$ \\
  \stride~\VARstride  & $A+2$ \\
  \datasize~\VARdatasize & \SI{2000}{\Bytes}, \SI{2560000}{\Bytes}\\
  comm. patterns   & \pingpong \\
  \# of processes   & \num{2x1}, \num{1x2} \\
  \bottomrule
\end{dtabular}

\DEtypes
The setup of the nested vector \ddtvectortiled \versus the basic layout 
\dtdtiled is illustrated in 
\fig~\ref{fig:test_tiled_vs_vector_tiled}.

\begin{figure}[h!]
\centering
\begin{tikzpicture}[scale=0.45]
  
  \def\toff{0.0}
  \def\vpad{.0}
  \def\myyshift{-3pt}

  \foreach \x in {0,2,4,9.5}
  \draw[fill=white] (\x,\toff) rectangle (\x+2,\toff+1) node[pos=.5] { $P$ };
  \draw[fill=white] (6,\toff) rectangle (9.5,\toff+1) node[pos=.5] {...};  
  \foreach \x in {2,4,6,9.5}
  \draw [line width=0.25mm] (\x, \toff-\vpad) -- (\x, \toff+1+\vpad);
  
  \foreach \x in {0,.5,1}
  \draw[fill = lightgray] (13.5+\x,\toff) rectangle (13.5+\x+.5,\toff+1);
  \draw (13.5,\toff) rectangle (15.5,\toff+1);
  \node at (13,\toff+.5)  (a) {$P$};

  \draw [<->]
  (0,\toff+1.2) -- node[anchor = south, xshift=0, yshift=0]{$S$}
  (4,\toff+1.2);

  \draw [decorate,decoration={brace,amplitude=5pt,mirror}]
  (0,\toff-.1) -- node[anchor = north west, xshift=-5, yshift=\myyshift]{\mpivector using pattern $P$}
  (4,\toff-.1);

  \draw [decorate,decoration={brace,amplitude=5pt,mirror}]
  (0,\toff-1.5) -- node[anchor = north, yshift=\myyshift]{\mpihvector}
  (11.5,\toff-1.5);

\end{tikzpicture}
\caption{\ddtvectortiled}
\label{fig:test_tiled_vs_vector_tiled}
\end{figure}

\DEhypo 
We expect that an MPI library will detect that the stride for the
outer vector is equal to $c$ times the stride of the inner vector,
such that the layout can also be described by a non-nested vector
constructor.  \mpicommit should perform the transformation. The
performance of the two descriptions should therefore be similar,
regardless of the communication pattern.

\DEresults

\begin{figure}[htpb]
\centering
\begin{subfigure}{.49\linewidth}
\centering
\includegraphics[width=\linewidth]{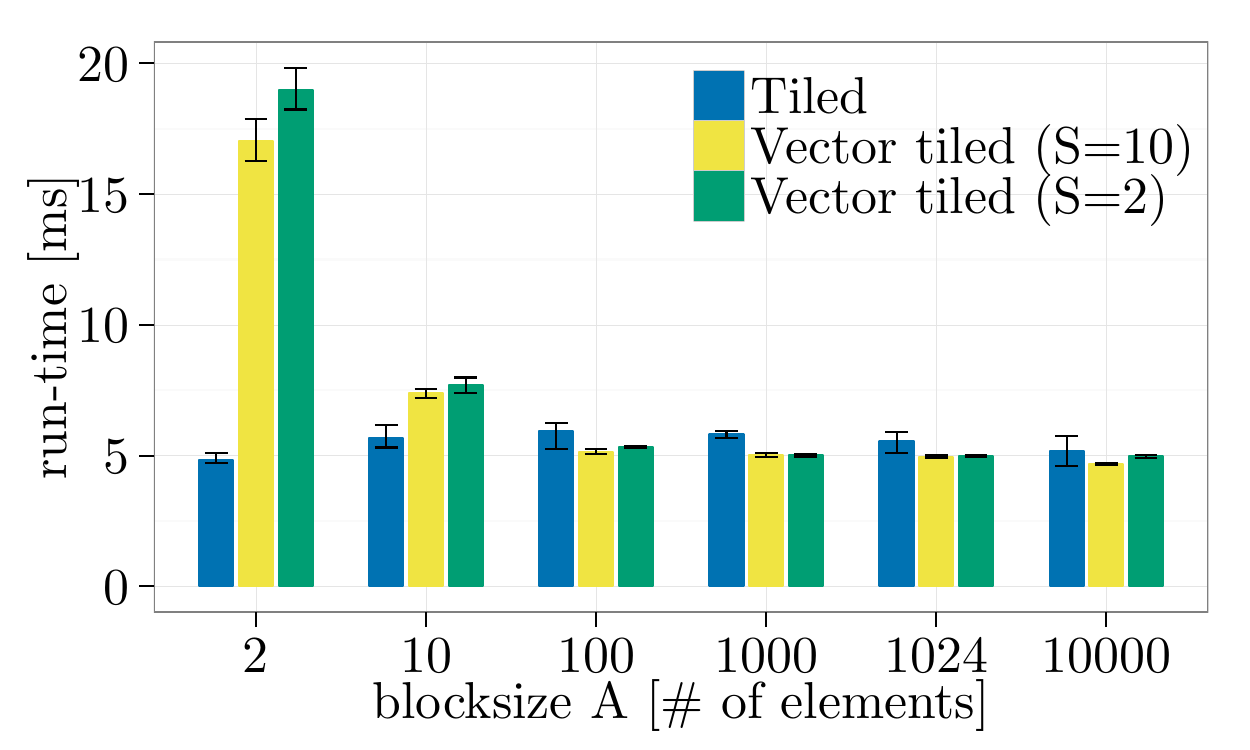}
\caption{%
\label{fig:exp:pingpong-vectortiled-large-2x1}%
\jupiternecmpi
}%
\end{subfigure}%
\hfill%
\begin{subfigure}{.49\linewidth}
\centering
\includegraphics[width=\linewidth]{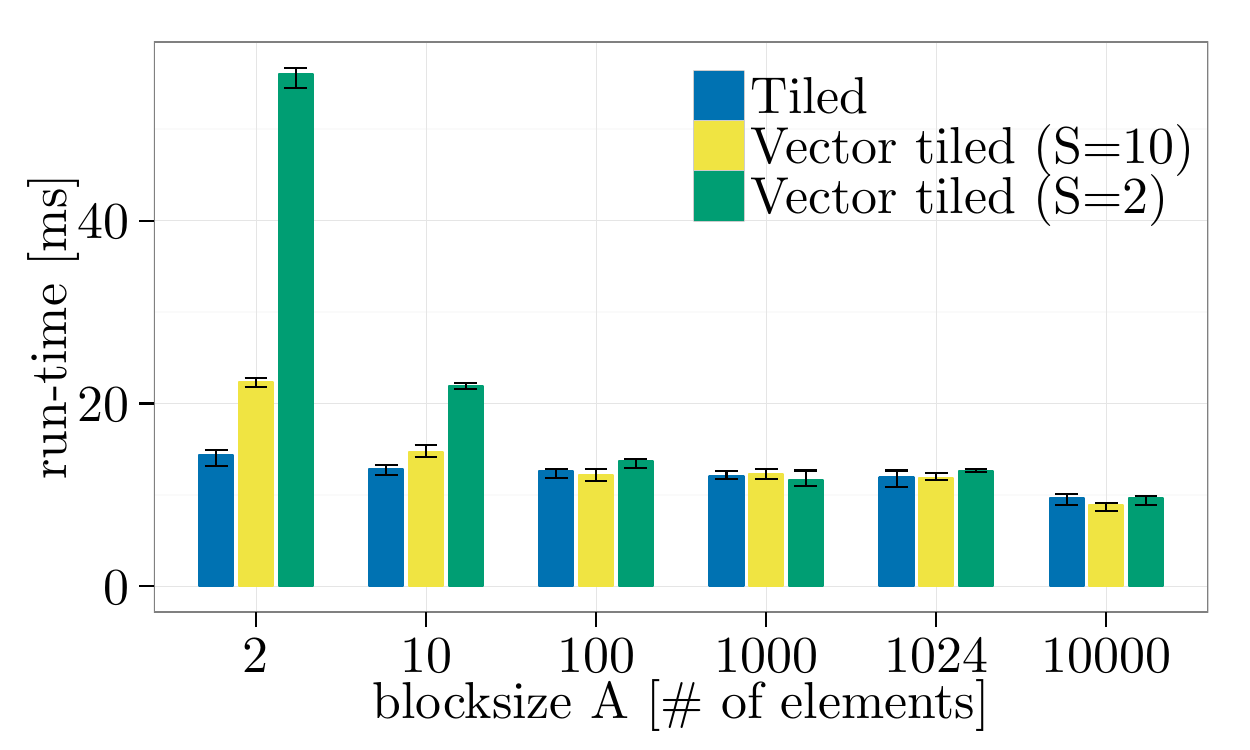}
\caption{%
\label{fig:exp:pingpong-vectortiled-large-2x1-mvapich}%
\jupitermvapich
}%
\end{subfigure}%
\caption{\label{fig:exp:pingpong-vectortiled} \dtdtiled \vs \ddtvectortiled, element datatype: \mpiint, $\VARdatasize=\SI{2.56}{\mega\byte}$, \num{2}~nodes, \pingpong.}
\end{figure}

To our surprise, apparently none of the MPI libraries normalizes the
two-level nested vector description into a better layout. For a small
unit of size $A=2$, the communication time with \ddtvectortiled
compared to the communication time with the simple \dtdtiled
description differ by factors from 2 to 4, as shown in
\fig~\ref{fig:exp:pingpong-vectortiled}.

\subexptest{exptest:block_indexed}

In this experiment, we look at different, explicit descriptions of the
more irregular layout \dtdblock, by explicitly listing the
displacements and number of elements in all $n/k$ blocks in the $n$
element layouts. The purpose of this experiment is to investigate the
relative penalty of having to traverse long, explicit lists of
displacements, \versus implicit, computed displacements.

\DEdescr
\begin{dtabular}{ll}
  \toprule
  Reference Layout & \dtdblock \\
  Compared Layout & \ddtblockindexed \\
  \midrule
  \blocksize~\VARblocksize  &  $\num{2}, \num{10}, \num{100}, \num{1000}, \num{1024}, \num{10000}$ \\
  \stride~\VARstride  & $B_1=A+1$,  $B_2=A+3$ \\
  \datasize~\VARdatasize & \SI{3200}{\Bytes}, \SI{2560000}{\Bytes}\\
  comm. patterns   & \pingpong \\
  \# of processes   & \num{2x1}, \num{1x2} \\
  \bottomrule
\end{dtabular}

\DEtypes%
The \ddtblockindexed layout uses the \textbf{\dtblock} layout with
given $A, B_1$, and $B_2$, described with the \mpiblock constructor
with $n/A$~indices and \blocksize~$A$. The block displacements can
easily be computed. This is illustrated in
\fig~\ref{fig:test_block_vs_block_indexed}.

\begin{figure}[h!]
\centering
\begin{tikzpicture}[scale=0.4]
  
  \def\myyshift{-3pt}

  \draw (0,0) rectangle (11.5,1);
  \foreach \x in {0,.5,2,2.5,3.5,4.0,5.5,6,7.0,7.5}
  \draw[fill = lightgray] (\x,0) rectangle (\x+.5,1);
  \foreach \x in {1,3,4.5,6.5}
  \draw[fill = white] (\x,0) rectangle (\x+.5,1);
  \node at (9.5,0.5) {...};
  
  \def\toff{-0.1}

  \draw [decorate,decoration={brace,amplitude=5pt,mirror}]
  (0,\toff-.1) -- node[anchor = north, yshift=\myyshift]{\mpiblock}
  (3.0,\toff-.1);

  \draw [decorate,decoration={brace,amplitude=5pt,mirror}]
  (0,\toff-1.3-.1) -- node[anchor = north, yshift=\myyshift]{\mpiresized (\dtblock)}
  (3.5,\toff-1.3-.1);

  \def\toff{-4.5}
  \def\vpad{.0}
 
  \foreach \x in {0,.5,2,2.5,3.5,4.0,5.5,6,7.0,7.5}
  \draw[fill = lightgray] (\x,\toff) rectangle (\x+.5,\toff+1);
  \foreach \x in {1,1.5,3,4.5,5,6.5}
  \draw[fill = white] (\x,\toff) rectangle (\x+.5,\toff+1);
  \draw[fill=white] (8,\toff) rectangle (11.5,\toff+1) node[pos=.5] {...};  
  
  \draw [decorate,decoration={brace,amplitude=5pt,mirror}]
  (0,\toff-.1) -- node[anchor = north, yshift=\myyshift]{\scriptsize\tt{block[0]}}
  (2,\toff-.1);

  \draw [decorate,decoration={brace,amplitude=5pt,mirror}]
  (2,\toff-.1) -- node[anchor = north, xshift=10pt, yshift=\myyshift]{\scriptsize\tt{block[1]}}
  (3.5,\toff-.1);

  \node at (5.3,\toff-.8) (b) {...};

  \draw [decorate,decoration={brace,amplitude=5pt,mirror}]
  (0,\toff-1-.1) -- node[anchor = north, yshift=\myyshift]{\mpiblock}
  (11.5,\toff-1-.1);

\end{tikzpicture}
\caption{Static \dtdblock (top) \vs\ dynamic \ddtblockindexed (bottom)
  description of the \dtdblock layout.}
\label{fig:test_block_vs_block_indexed}
\end{figure}

\DEhypo 

An MPI library should normalize both cases to the same internal
datatype representation with good performance. It is doubtful that
anything like that will happen. More importantly, it is not obvious
which of the two descriptions is better. A reasonable expectation is
that beyond some number of elements, the large array of displacements
(indices) in the \ddtblockindexed datatype will become expensive to
traverse, and that simple repetitions of the small, irregular
non-contiguous \dtblock pattern will perform better. 

\DEresults

For small \blocksizes of $A$, the \dtblock description is worse,
especially for the \jupiternecmpi library. Otherwise the performance
of the two descriptions looks similar. Nevertheless, the absolute
performance of the \jupiternecmpi library is still better. We do not
show the results here (see appendix).

\subexptest{exptest:alternating_indexed}

This experiment is similar to the previous one. Here, two descriptions of the
``most irregular'' of the four basic layouts, namely \dtdalternating are
contrasted.

\DEdescr
\begin{dtabular}{ll}
  \toprule
  Reference Layout & \dtdalternating \\
  Compared Layout & \ddtalternatingindexed \\
  \midrule
  \blocksize~\VARblocksize  &  $\num{2}, \num{10}, \num{100}, \num{1000}, \num{1024}, \num{10000}$ \\
  blocksizes $A_1$, $A_2$  & $A_1=A-1$,  $A_2=A+1$ \\
  \stride~\VARstride  & $B_1=A+1$,  $B_2=A+3$ \\
  \datasize~\VARdatasize & \SI{3200}{\Bytes}, \SI{2560000}{\Bytes}\\
  comm. patterns   & \pingpong \\
  \# of processes   & \num{2x1}, \num{1x2} \\
  \bottomrule
\end{dtabular}

\DEtypes%
The \ddtalternatingindexed datatype is based on the \dtalternating
layout with given $A_1, A_2, B_1$, and $B_2$, described with the
\mpiindex constructor with $n/(A_1+A_2)$ indices and blocksizes of
$A_1$ and $A_2$.  The data layout is illustrated in
\fig~\ref{fig:test_alternating_vs_alterning_indexed}.

\begin{figure}[h!]
\centering
\begin{tikzpicture}[scale=0.45]
  
  \def\myyshift{-3pt}

  \draw (0,0) rectangle (11.5,1);
  \foreach \x in {0,.5,2,3.5,4.0,5.5,7.0,7.5}
  \draw[fill = lightgray] (\x,0) rectangle (\x+.5,1);
  \foreach \x in {1,2.5,3,4.5,6,6.5}
  \draw[fill = white] (\x,0) rectangle (\x+.5,1);
  \node at (9.5,0.5) {...};
  
  \def\toff{-0.1}

  \draw [decorate,decoration={brace,amplitude=5pt,mirror}]
  (0,\toff-.1) -- node[anchor = north west, xshift=-5, yshift=\myyshift]{\mpiindex}
  (2.5,\toff-.1);

  \draw [decorate,decoration={brace,amplitude=5pt,mirror}]
  (0,\toff-1.3-.1) -- node[anchor = north west, xshift=-5, yshift=\myyshift]{\mpiresized (\dtalternating)}
  (3.5,\toff-1.3-.1);

  \def\toff{-4.5}
  \def\vpad{.0}
 
  \draw (0,\toff) rectangle (11.5,\toff+1);
  \foreach \x in {0,.5,2,3.5,4.0,5.5,7.0,7.5}
  \draw[fill = lightgray] (\x,\toff) rectangle (\x+.5,\toff+1);
  \foreach \x in {1,2.5,3,4.5,6,6.5}
  \draw[fill = white] (\x,\toff) rectangle (\x+.5,1+\toff);
  \draw[fill=white] (8,\toff) rectangle (11.5,\toff+1) node[pos=.5] {...};  
  
  \draw [decorate,decoration={brace,amplitude=5pt,mirror}]
  (0,\toff-.1) -- node[anchor = north, yshift=\myyshift]{\scriptsize\tt{block[0]}}
  (2,\toff-.1);

  \draw [decorate,decoration={brace,amplitude=5pt,mirror}]
  (2,\toff-.1) -- node[anchor = north, xshift=10pt, yshift=\myyshift]{\scriptsize\tt{block[1]}}
  (3.5,\toff-.1);

  \node at (5.3,\toff-.8) (b) {...};  

  \draw [decorate,decoration={brace,amplitude=5pt,mirror}]
  (0,\toff-1-.1) -- node[anchor = north, yshift=\myyshift]{\mpiindex}
  (11.5,\toff-1-.1);

\end{tikzpicture}
\caption{Static \dtdalternating (top) \vs\ dynamic
  \ddtalternatingindexed (bottom) description of the \dtdalternating
  layout.}
\label{fig:test_alternating_vs_alterning_indexed}
\end{figure}

\DEhypo

As for the previous experiment, it is not obvious which of the two
descriptions will perform better (under the pessimistic assumption
that no normalization takes place). Experimental results will give insight
on whether there is a penalty for large lists of displacements in the
\mpiindex constructor.

\DEresults

For larger values of $A$, the performance of the two descriptions
looks similar. For small values of $A$, the results are the opposite
of the previous experiment. Especially for the \jupiternecmpi, the
\ddtalternatingindexed description performs worse than the
\dtdalternating description. Results are not shown here (see
appendix).

\subexptest{exptest:alternating_repeated}

In our final test with the basic layouts, we look at an
\dtdalternating pattern where the stride $B_2$ of the second unit is
equal to the number of elements $A_2$ in the unit. This pattern can be
described as a repetition of small datatypes describing fixed
blocks. It can alternatively be formulated as a layout comprised of
(1) a first, small unit of $A_1$ elements, (2) a large, regularly
strided middle part, and (3) a last, small unit of $A_2$ elements. We
expect the second description to perform better, and we want to check
this hypothesis.

\DEdescr
\begin{dtabular}{ll}
  \toprule
  Compared Layouts & \ddtalternatingrepeated \\
				& \ddtalternatingstruct \\
  \midrule  
  \blocksize~\VARblocksize  &  $\num{2}, \num{10}, \num{100}, \num{1000}, \num{1024}, \num{10000}$ \\
  unit blocksizes $A_1$, $A_2$  & $A_1=A-1$,  $A_2=A+1$ \\
  \stride~\VARstride  & $A+1$ \\
  \datasize~\VARdatasize & \SI{3200}{\Bytes}, \SI{2560000}{\Bytes}\\
  comm. patterns   & \pingpong \\
  \# of processes   & \num{2x1}, \num{1x2} \\
  \bottomrule
\end{dtabular}

\DEtypes%
Here, two types describe an \textbf{alternating} layout with units of
$A_1$ and $A_2$ elements and strides $B_1=B$ and $B_2=A_2$,
respectively. The first datatype is called \ddtalternatingrepeated and
is defined as a fixed, alternating block (\fig~\ref{fig:twoviews}a).
The second is called \ddtalternatingstruct and is created with
\mpistruct using three subtypes: the first is a block of $A_1$
contiguous elements, the second subtype is a tiled vector, and the
third subtype is a contiguous block of $A_2$ elements
(\fig~\ref{fig:twoviews}b).

\begin{figure}[!h]
\centering
\begin{tikzpicture}[scale=0.45]

  \def\myyshift{-3pt}

  \draw (0,0) rectangle (11,1);
  \foreach \x in {0,1,1.5,2,3,3.5,4,5,5.5,6,8.5,9,9.5,10.5,11}
  \draw[fill = lightgray] (\x,0) rectangle (\x+.5,1);
  \node at (7.5,0.5) {...};
  
  \def\toff{-1.5}
  \def\vpad{.2}

  \foreach \x in {0,2,4,9.5}
  \draw[fill=white] (\x,\toff) rectangle (\x+2,\toff+1) node[pos=.5] { $T_1$ };
  \draw[fill=white] (6,\toff) rectangle (9.5,\toff+1) node[pos=.5] {...};  

  \foreach \x in {0,1,1.5}
  \draw[fill = lightgray] (13.5+\x,\toff) rectangle (13.5+\x+.5,\toff+1);
  \draw (13.5,\toff) rectangle (15.5,\toff+1);
  \node at (13,\toff+.5)  (a) {$T_1$};
  \node at (-.6,\toff+.5)   (b) {a)};

  \draw [decorate,decoration={brace,amplitude=5pt,mirror}]
  (0,\toff-.1) -- node[anchor = north, yshift=\myyshift]{\mpicontig}
  (11.5,\toff-.1);

  \draw [decorate,decoration={brace,amplitude=5pt,mirror}]
  (13.5,\toff-.1) -- node[anchor = north, yshift=\myyshift, align=left]{\mpiindex\\ (\dtalternating)}
  (15.5,\toff-.1);

  \def\toff{-5.5}

  \foreach \x in {0}
  \draw[fill=lightgray] (\x,\toff) rectangle (\x+.5,\toff+1); 
  \draw[fill=white] (.5,\toff) rectangle (1,\toff+1); 

  \draw[fill=white] (1,\toff) rectangle (10.5,\toff+1) node[pos=.5] {$T_2$}; 
  \draw [decorate,decoration={brace,amplitude=5pt,mirror}]
  (1,\toff-.1) -- node[anchor = north, yshift=\myyshift, align=left]{\mpivector using pattern $P$}
  (10.5,\toff-.1);

  \foreach \x in {10.5,11.0}
  \draw[fill=lightgray] (\x,\toff) rectangle (\x+.5,\toff+1); 

  \draw [decorate,decoration={brace,amplitude=5pt,mirror}]
  (0,\toff-1.5-.1) -- node[anchor = north, yshift=\myyshift, align=left]{\mpistruct}
  (11.5,\toff-1.5-.1);

  \foreach \x in {0,.5,1}
  \draw[fill = lightgray] (13.5+\x,\toff) rectangle (13.5+\x+.5,\toff+1);
  \draw (13.5,\toff) rectangle (15.5,\toff+1);
  \node at (13,\toff+.5)  (c) {$P$};
  \node at (-.6,\toff+.5)   (d) {b)};
\end{tikzpicture}
\caption{\ddtalternatingrepeated (top) \vs \ddtalternatingstruct
  (bottom).}
\label{fig:twoviews}
\end{figure}

\DEhypo 

With the description as an \ddtalternatingstruct, communication
performance should approach the performance of communicating a tiled
vector with \blocksizes of $A_1+A_2$ elements, when the total number
of elements $n$ goes up. Our previous measurements have given the
baseline performance for such \dtdtiled patterns, against which we can
compare. Since commonly used MPI normalization heuristics do probably
not change the description from \ddtalternatingrepeated to the
possibly better \ddtalternatingstruct, our expectation is that the
latter will perform better.

\DEresults
\begin{figure}[htpb]
\centering
\begin{subfigure}{.49\linewidth}
\centering
\includegraphics[width=\linewidth]{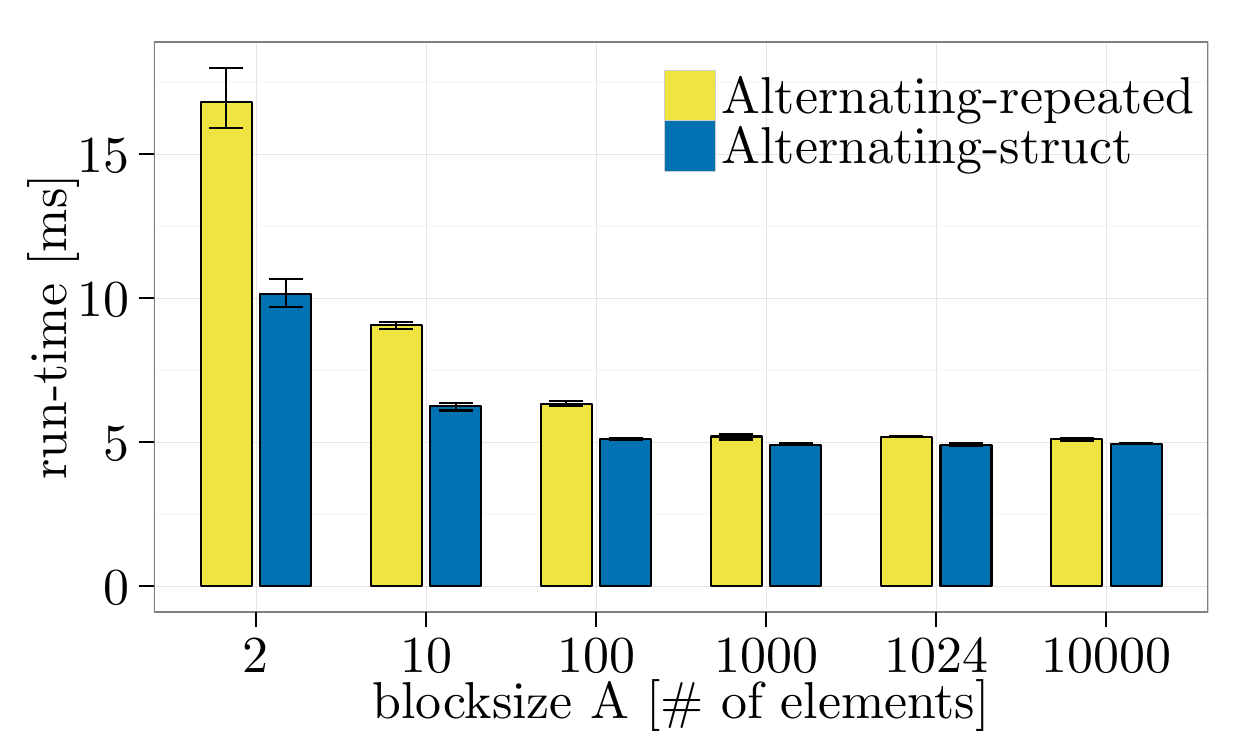}
\caption{%
\label{fig:exp:pingpong-alternatingrepeated-large-2x1}%
\jupiternecmpi%
}%
\end{subfigure}%
\hfill%
\begin{subfigure}{.49\linewidth}
\centering
\includegraphics[width=\linewidth]{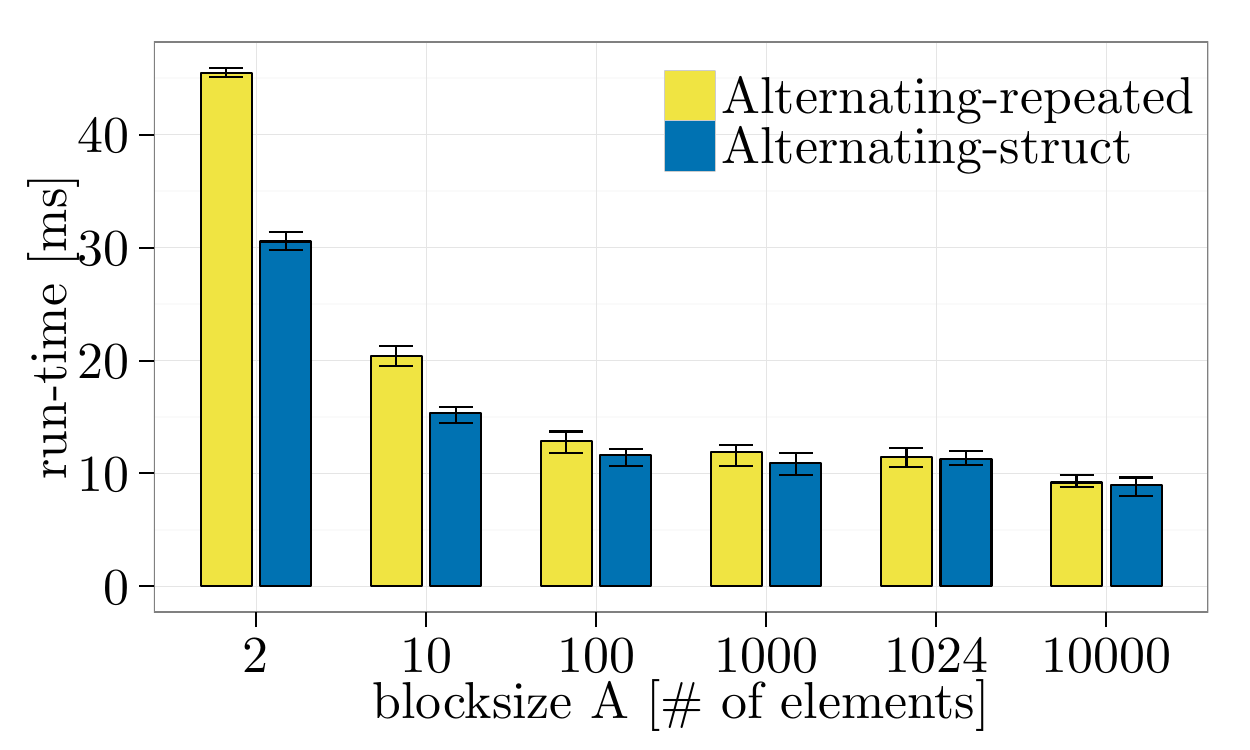}
\caption{%
\label{fig:exp:pingpong-alternatingrepeated-large-2x1-mvapich}%
\jupitermvapich%
}%
\end{subfigure}%
\caption{\label{fig:exp:pingpong-alternatingrepeated} \ddtalternatingrepeated \vs \ddtalternatingstruct, element datatype: \mpiint, $\VARdatasize=\SI{2.56}{\mega\byte}$, \num{2}~nodes, \pingpong.}
\end{figure}

For small values of $A$, the \ddtalternatingrepeated way of
communicating the pattern is indeed slower for all tested MPI
libraries; see \fig~\ref{fig:exp:pingpong-alternatingrepeated}. This
shows that a stronger type normalization (on the fly) than currently
performed by MPI libraries is needed to approach the baseline
performance. It also shows, and this is important, that the \mpistruct
constructor is needed for the best description even of homogeneous
layouts where all elements have the same basic type.

\newpage

\subexptest{exptest:rowcol}

For our final set of experiments, we use another layout.  Given an
$(n+1-A)\times A$ matrix, we want to communicate together the first
row of $A$ elements and the (remainder of the) first column of $n-A$
elements, for a total of $n$ elements. For examining
\Gl~(\ref{eqn:normal}), we again compare natural ways of describing
this layout, and compare the measured communication times.
A similar
example was used by Ganian~\etal~\cite{Traff16:alltypetree}.

\DEdescr
\begin{dtabular}{ll}
  \toprule
Compared Layouts& \ddtrowcolfullindexed \\
	 & \ddtrowcolcontiguousandindexed \\
	 & \ddtrowcolstruct \\
  \midrule
  \blocksize~\VARblocksize  & $\num{2},\num{10},\num{100},\num{128},\num{512},\num{1000},\num{1024},\num{5000},\num{10000}$\\
  \# of elements $n$  & $\num{100}, \num{10240}$ \\
  comm. patterns   & \pingpong \\
  \# of processes   & \num{2x1}, \num{1x2} \\
  \bottomrule
\end{dtabular}

\DEtypes

A row-column layout (submatrix of $(n+1-A)\times A$ matrix),
consisting of $A$ consecutive elements followed by $n-A$ elements in a
strided layout with stride $A$, can be described either by
\begin{myitemize}
\item using \mpiblock with $n$ indices (\ddtrowcolfullindexed),
\item using \mpiindex with $1+n-A$ indices
  (\ddtrowcolcontiguousandindexed), or
\item using \mpistruct consisting of a contiguous subtype of $A$
  elements, followed by a vector of $n-A$ blocks of one element with
  stride $A$ (\ddtrowcolstruct).
\end{myitemize}
The layout and the three possible descriptions as derived datatypes
are shown in \fig~\ref{fig:test_tiled_vs_rowcol}.

\begin{figure}[h!]
\centering
\begin{tikzpicture}[scale=0.4]
  
  \def\myyshift{-3pt}

  \def\tpicxoff{0.0}
  \def\tpicyoff{0.0}

  \foreach \x in {0,.5,1,1.5,2,2.5,3} {
    \draw[fill = lightgray] (\tpicxoff+\x,\tpicyoff) rectangle (\tpicxoff+\x+.5,\tpicyoff+1);
  }

  \foreach \y in {1,2.0,3.0,4.0} {
    \draw[fill = lightgray] (\tpicxoff,\tpicyoff-\y) rectangle (\tpicxoff+.5,\tpicyoff-\y+1);
  }

  \def\tyoff{-5.0}
  \def\txoff{1.0}
  \foreach \x in {0,.5,1,1.5} {
    \draw[fill = white] (\tpicxoff+\txoff+\x,\tpicyoff+\tyoff) rectangle (\tpicxoff+\txoff+\x+.5,\tpicyoff+\tyoff+1);
  }
  \draw[fill = white] (\tpicxoff+\txoff+2,\tpicyoff+\tyoff) rectangle (\tpicxoff+\txoff+4,\tpicyoff+\tyoff+1) node[pos=.5] {...};

  \draw node[anchor = north, yshift=\myyshift] at  (\tpicxoff+\txoff+2,\tpicyoff+\tyoff) {\scriptsize\mpiblock};

  \foreach \x in {0,.5,1,1.5} {
    \draw[->](\tpicxoff+\x+.25,\tpicyoff+.5)--(\tpicxoff+\x+1,\tpicyoff+-1)node{}--(\tpicxoff+\x+1.25,\tpicyoff+-4.5);
  }

  \def\tpicxoff{7.0}
  \def\tpicyoff{0.0}

  \foreach \x in {0,.5,1,1.5,2,2.5,3} {
    \draw[fill = lightgray] (\tpicxoff+\x,\tpicyoff) rectangle (\tpicxoff+\x+.5,\tpicyoff+1);
  }

  \foreach \y in {1,2.0,3.0,4.0} {
    \draw[fill = lightgray] (\tpicxoff,\tpicyoff-\y) rectangle (\tpicxoff+.5,\tpicyoff-\y+1);
  }

  \def\tyoff{-5.0}
  \def\txoff{1.0}
  \foreach \x in {0,.5,1,1.5} {
    \draw[fill = white] (\tpicxoff+\txoff+\x,\tpicyoff+\tyoff) rectangle (\tpicxoff+\txoff+\x+.5,\tpicyoff+\tyoff+1);
  }
  \draw[fill = white] (\tpicxoff+\txoff+2,\tpicyoff+\tyoff) rectangle (\tpicxoff+\txoff+4,\tpicyoff+\tyoff+1) node[pos=.5] {...};

  \draw node[anchor = north, yshift=\myyshift-10] at  (\tpicxoff+\txoff+2,\tpicyoff+\tyoff) {\scriptsize\mpiindex};

  \def\x{0.0}
  \foreach \y in {.5,1,1.5} {
    \draw[->](\tpicxoff+\x+.25,\tpicyoff-2*\y+.5)--(\tpicxoff+\y+1,\tpicyoff-2*\y+-1)node{}--(\tpicxoff+\y+1.25,\tpicyoff-4.5);
  }
  
  \draw [decorate,decoration={brace,amplitude=5pt,mirror}]
  (\tpicxoff,\tpicyoff) -- node[anchor = north, yshift=\myyshift]{block}
  (\tpicxoff+3.5,\tpicyoff-.1);

  \draw[->](\tpicxoff+1.5,\tpicyoff-1.2)--(\tpicxoff+2.5,\tpicyoff-2.5)node{}--(\tpicxoff+1.25,\tpicyoff-4.5);

  \def\tpicxoff{14.0}
  \def\tpicyoff{0.0}

  \foreach \x in {0,.5,1,1.5,2,2.5,3} {
    \draw[fill = lightgray] (\tpicxoff+\x,\tpicyoff) rectangle (\tpicxoff+\x+.5,\tpicyoff+1);
  }

  \foreach \y in {1,2.0,3.0,4.0} {
    \draw[fill = lightgray] (\tpicxoff,\tpicyoff-\y) rectangle (\tpicxoff+.5,\tpicyoff-\y+1);
  }

  \def\tyoff{-5.0}
  \def\txoff{1.0}
  \foreach \x in {0,.5} {
    \draw[fill = white] (\tpicxoff+\txoff+\x,\tpicyoff+\tyoff) rectangle (\tpicxoff+\txoff+\x+.5,\tpicyoff+\tyoff+1);
  }
  
  \draw [decorate,decoration={brace,amplitude=5pt,mirror}]
  (\tpicxoff,\tpicyoff) -- node[anchor = north, yshift=\myyshift]{block}
  (\tpicxoff+3.5,\tpicyoff-.1);

  \draw[->](\tpicxoff+1.5,\tpicyoff-1.2)--(\tpicxoff+2.5,\tpicyoff-2.5)node{}--(\tpicxoff+1.25,\tpicyoff-4.5);

  \draw [decorate,decoration={brace,amplitude=5pt,mirror}]
  (\tpicxoff,\tpicyoff) -- node[anchor = east,xshift=-10,yshift=40pt,rotate=90]{\scriptsize\mpivector}
  (\tpicxoff,\tpicyoff-4);

  \draw[->](\tpicxoff+.6,\tpicyoff-1.2)--(\tpicxoff+1.5,\tpicyoff-2.5)node{}--(\tpicxoff+1.75,\tpicyoff-4.5);

  \draw node[anchor = north, yshift=\myyshift] at  (\tpicxoff+\txoff+1,\tpicyoff+\tyoff) {\scriptsize\mpistruct};

\end{tikzpicture}
\caption{From left to right: \ddtrowcolfullindexed,
  \ddtrowcolcontiguousandindexed, \ddtrowcolstruct.}
\label{fig:test_tiled_vs_rowcol}
\end{figure}

\DEhypo

The latter representation is the most compact, and expectedly best
performing. This layout description also illustrates that the full
power of the \mpistruct constructor is needed, even for homogeneous
layouts of elements of the same basic type.  As we do not expect the
MPI libraries to perform a normalization into an efficient data
representation, our hypothesis is that the \ddtrowcolstruct datatype
will perform better than the two other types, and that
\ddtrowcolcontiguousandindexed may perform better than
\ddtrowcolfullindexed as the \blocksize~$A$ goes up.

\newpage

\DEresults

\begin{figure}[!t]
\centering
\begin{subfigure}{.49\linewidth}
\centering
\includegraphics[width=\linewidth]{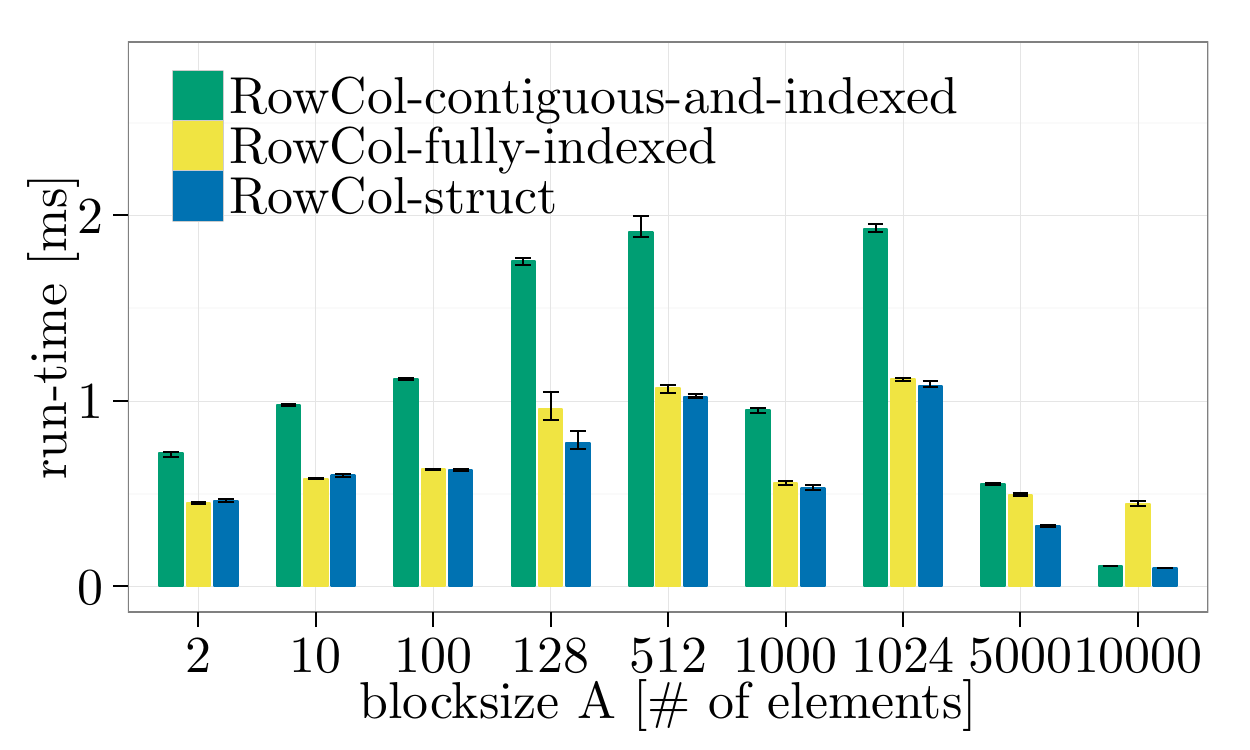}
\caption{%
\label{fig:exp:pingpong-rowcol-large-2x1}%
\jupiternecmpi%
}%
\end{subfigure}%
\hfill%
\begin{subfigure}{.49\linewidth}
\centering
\includegraphics[width=\linewidth]{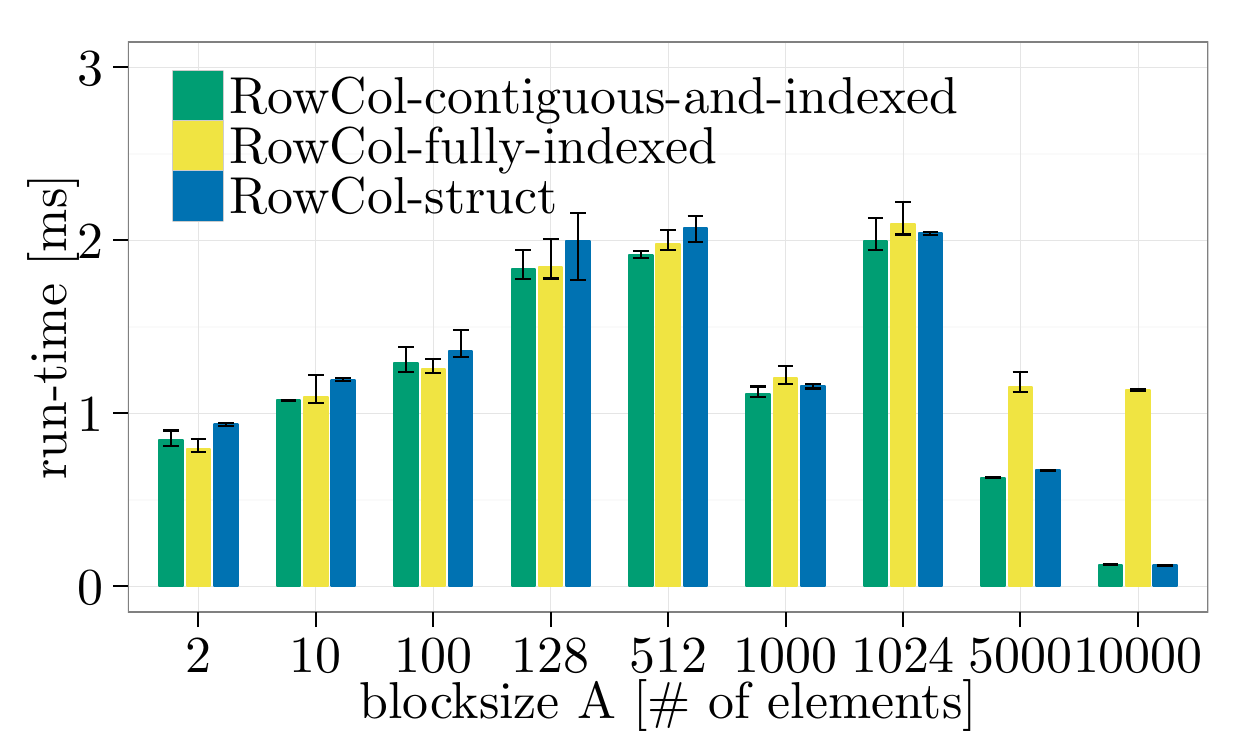}
\caption{%
\label{fig:exp:pingpong-rowcol-large-2x1-mvapich}%
 \jupitermvapich%
}%
\end{subfigure}%
\caption{\label{fig:exp:pingpong-rowcol} \ddtrowcolfullindexed, \ddtrowcolcontiguousandindexed, \ddtrowcolstruct, element datatype: \mpiint, $n=\num{10240}$, \extent increases with $A$,  \num{2}~nodes, \pingpong.}
\end{figure}

The results in \fig~\ref{fig:exp:pingpong-rowcol} confirm that for
\jupiternecmpi the compact description \ddtrowcolstruct gives the best
performance, closely followed by \ddtrowcolfullindexed.  For the
\jupitermvapich, the performance is worse (compared to \jupiternecmpi
as baseline) for all three descriptions, except for large values of
$A$.
The results for \jupiteropenmpi (see appendix) are as expected, with
the compact \ddtrowcolstruct description being close to a factor of
two faster than the other two (except for the large $A$ values).

\section{Summmary and Outlook}

We performed a large number of experiments to explore the performance
of communication with differently structured, non-contiguous data
layouts described by MPI derived datatypes. We focused on simple
tiled layouts parameterized by element counts and strides, and
structured the experiments as a set of expectation tests using MPI
performance guidelines.

The results were revealing and in many cases surprising and
disappointing.  For instance, it was unexpected that
\Gls~(\ref{eqn:pack}) and~(\ref{eqn:unpack}) would be violated, but we
found many cases (for all libraries) where these guidelines were
severely compromised. In such cases, the recommendation to use
datatypes is hard to justify. It is definitely important to look into
the reasons and improve the situation.  We also observe that the
communication performance with (non-trivial) derived datatypes is
quite different between the libraries. For example, the current
version of \mvapich does not handle derived datatypes as efficiently
as the other libraries. 

In addition, the simplest expectations concerning the use of the
\mpicontig constructor, captured in \Gl~(\ref{eqn:contig}), were
sometimes violated. We believe that these violations can and should be
repaired.

Our experiments around \Gl~(\ref{eqn:normal}) first and foremost show
that the way a given layout is described as a derived datatype matters
a lot. Or put differently, the heuristics employed by common MPI
libraries in \mpicommit are insufficient to find good internal
datatype representations. It is worthwhile to improve the situation,
since an application programmer currently needs a good intuition to
select an efficient derived datatype description. Simple rules of
thumb are not enough: our findings sometimes contradicted our own
intuitions and expectations. Furthermore, some experiments show that
datatype descriptions may be too localized to make a sufficiently good
normalization possible, namely, that both the repetition count and the
datatype are needed for computing the normalized type
description. However, normalization on the fly in each communication
call is not an option for a high-performance MPI library, especially
since (optimal) normalization may be very
expensive~\cite{Traff16:alltypetree}. Therefore, the performance
equivalence of the seemingly innocent \Gl~(\ref{eqn:contig}) cannot
hold when \mpicommit may normalize the contiguous type (right-hand
side of the guideline). MPI might need more query functionality for
the application programmer to explore and guide the normalization.

\section{Acknowledgments}

We thank Antoine Rougier and Felix Donatus L\"ubbe (TU Wien) for
helping us with some of the experiments.

\balance
\bibliographystyle{abbrv}
\bibliography{datatypeperf}

\onecolumn
\appendix

This appendix contains our currently full set of experiments, only some
of which are shown in the main text. The results are listed in the order
of the expectation tests.

\section{Experimental Results}

\appexp{exptest:basic_layouts} 

\begin{figure*}[htpb]
\centering
\begin{subfigure}{.33\linewidth}
\centering
\includegraphics[width=\linewidth]{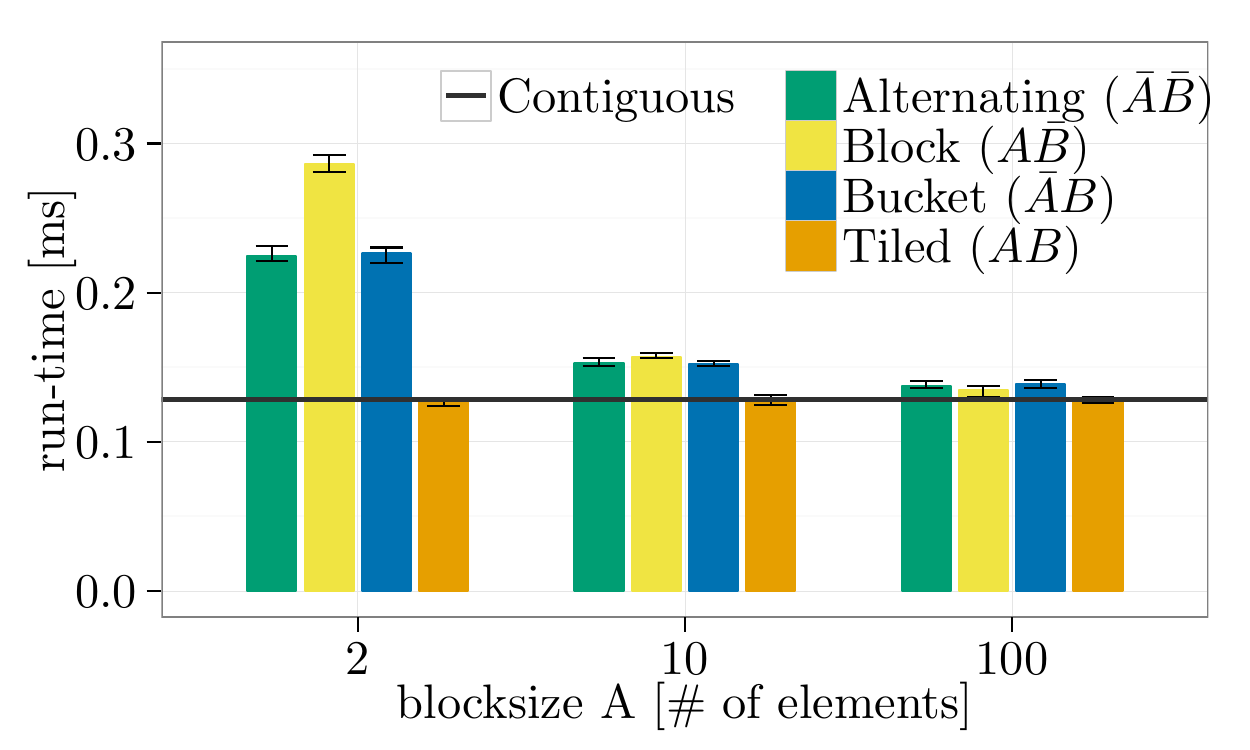}
\caption{%
\label{exp:allgather-nsmall-32p-nec}%
\mpiallgather%
}%
\end{subfigure}%
\hfill%
\begin{subfigure}{.33\linewidth}
\centering
\includegraphics[width=\linewidth]{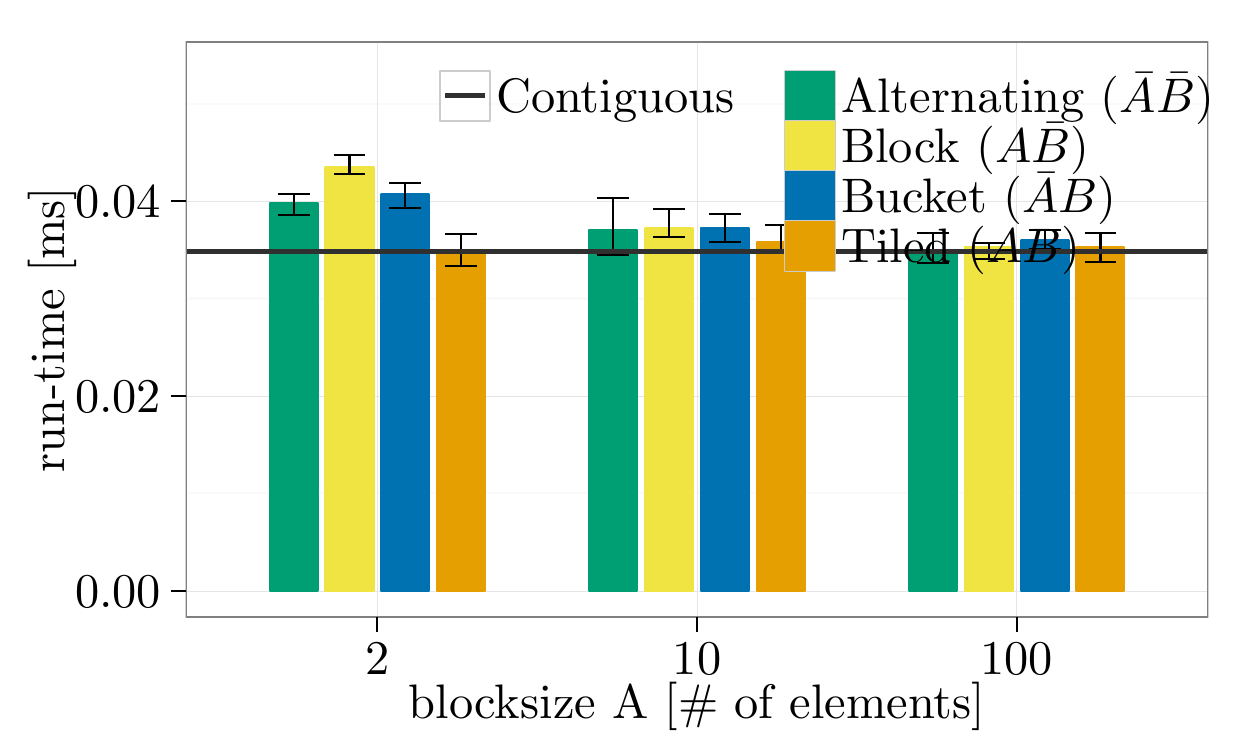}
\caption{%
\label{exp:bcast-nsmall-32p-nec}%
\mpibcast%
}%
\end{subfigure}%
\hfill%
\begin{subfigure}{.33\linewidth}
\centering
\includegraphics[width=\linewidth]{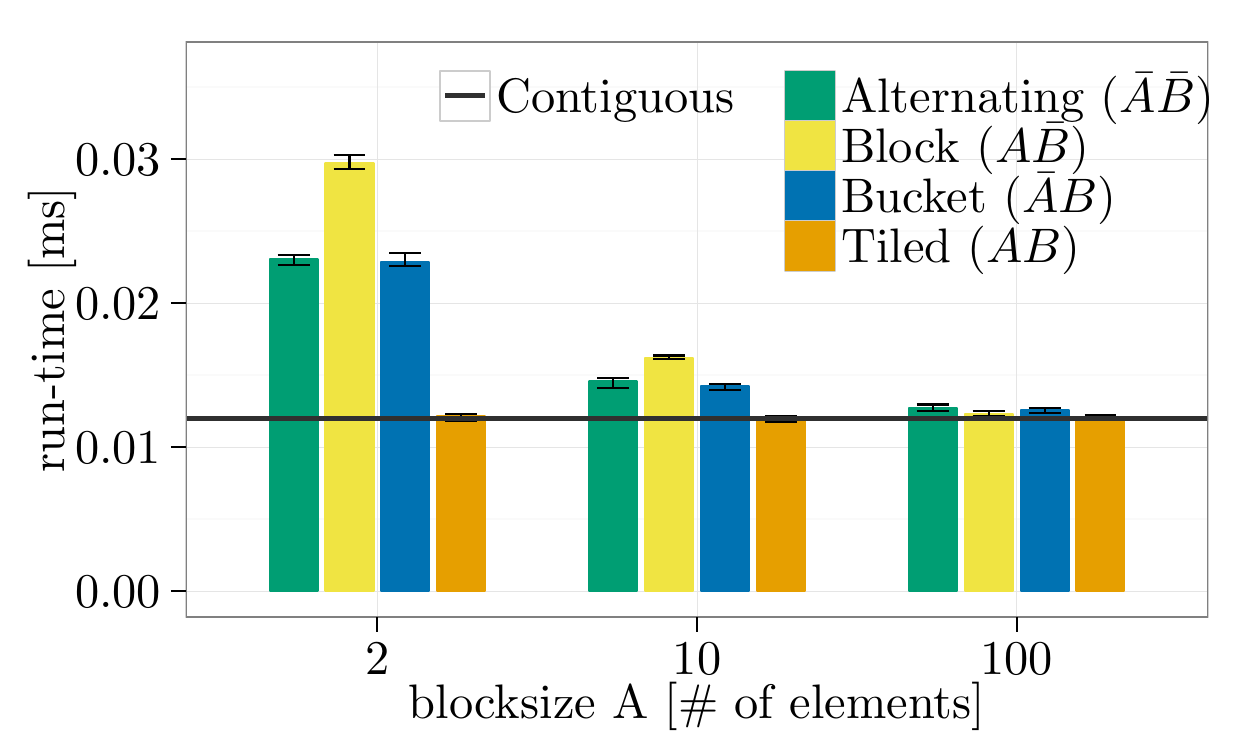}
\caption{%
\label{exp:pingpong-nsmall-32p-nec}%
\pingpong%
}%
\end{subfigure}%
\caption{\label{exp:layouts-nsmall-32p-nec}  Contiguous \vs typed,  $\VARdatasize=\SI{3.2}{\kilo\byte}$, element datatype: \mpiint, \num{32x1}~processes (\num{2x1} for \pingpong), \jupiternecmpi, \variantone.}
\end{figure*}

\begin{figure*}[htpb]
\centering
\begin{subfigure}{.33\linewidth}
\centering
\includegraphics[width=\linewidth]{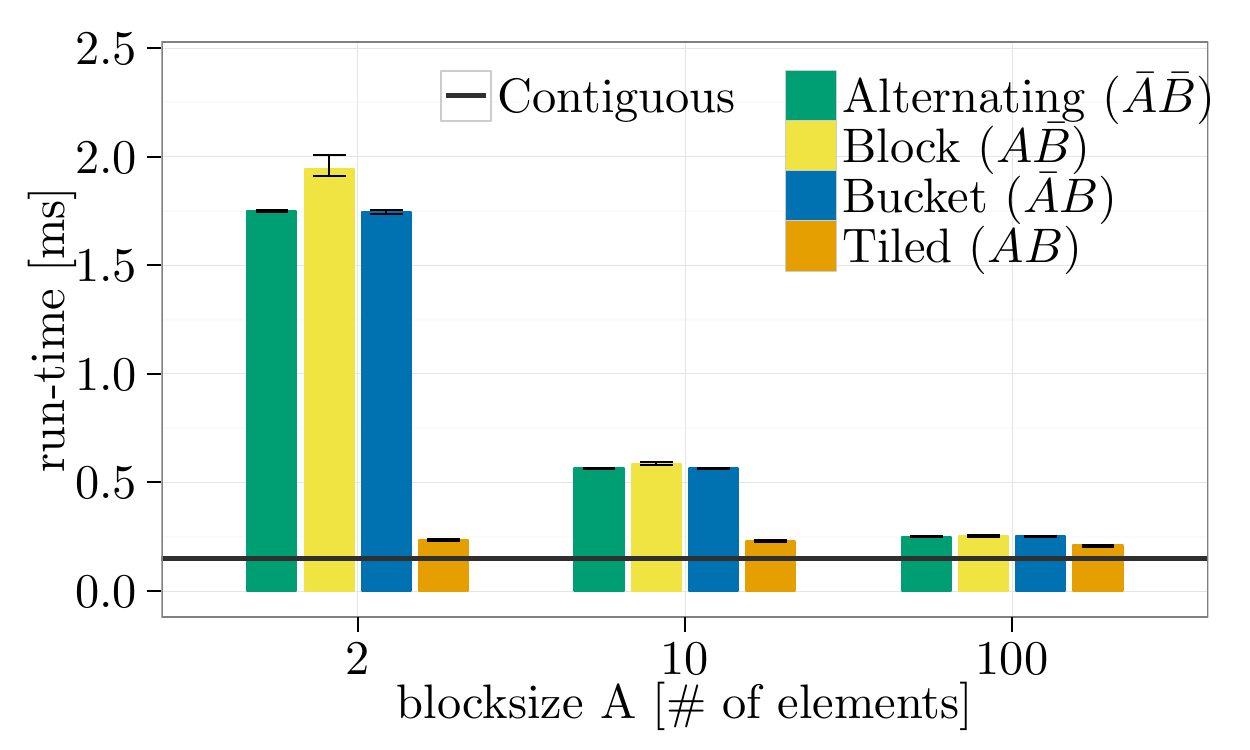}
\caption{%
\label{exp:allgather-nsmall-32p-mvapich}%
\mpiallgather%
}%
\end{subfigure}%
\hfill%
\begin{subfigure}{.33\linewidth}
\centering
\includegraphics[width=\linewidth]{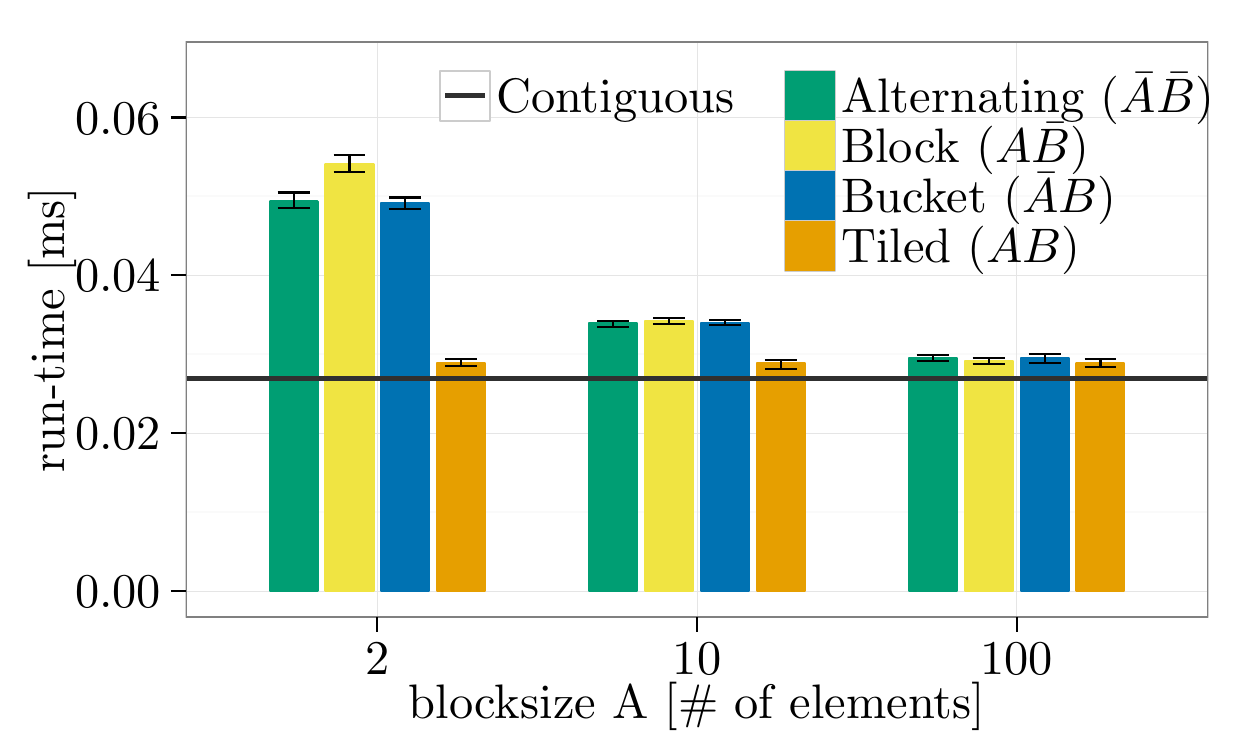}
\caption{%
\label{exp:bcast-nsmall-32p-mvapich}%
\mpibcast%
}%
\end{subfigure}%
\hfill%
\begin{subfigure}{.33\linewidth}
\centering
\includegraphics[width=\linewidth]{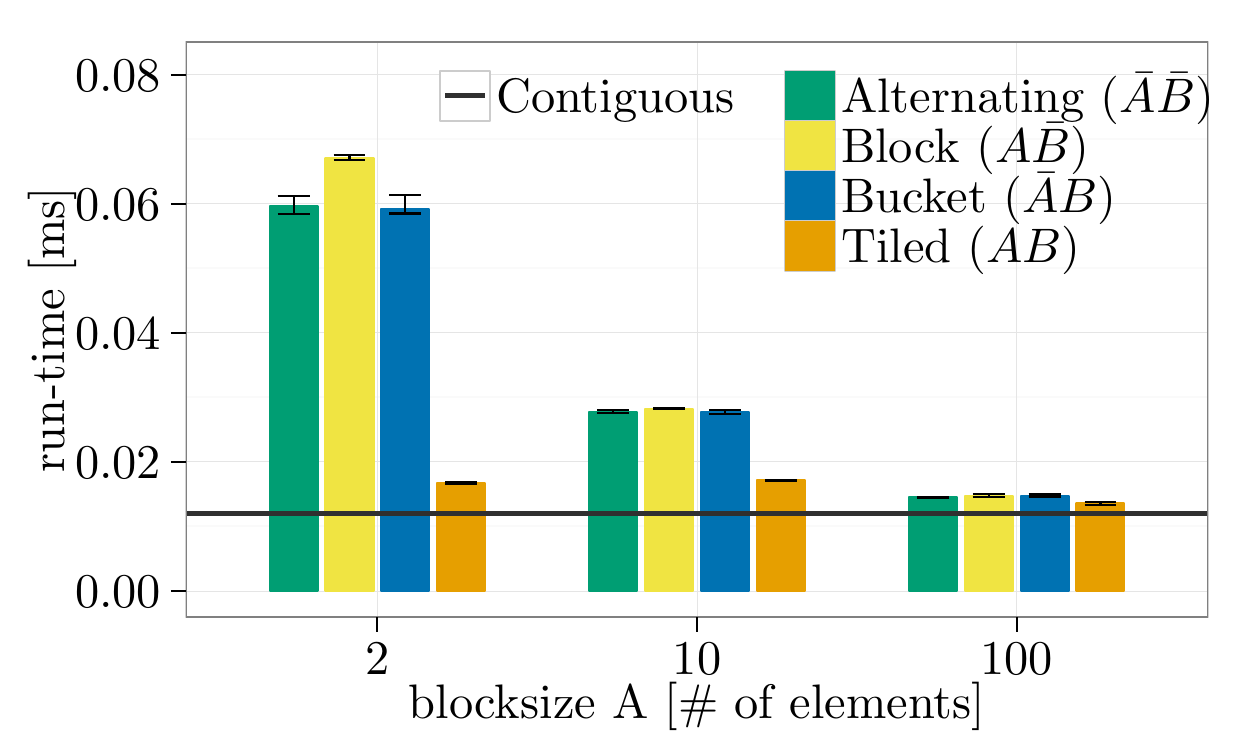}
\caption{%
\label{exp:pingpong-nsmall-32p-mvapich}%
\pingpong%
}%
\end{subfigure}%
\caption{\label{exp:layouts-nsmall-32p-mvapich}  Contiguous \vs typed, $\VARdatasize=\SI{3.2}{\kilo\byte}$, element datatype: \mpiint, \num{32x1}~processes (\num{2x1} for \pingpong), \jupitermvapich, \variantone.}
\end{figure*}

\begin{figure*}[htpb]
\centering
\begin{subfigure}{.33\linewidth}
\centering
\includegraphics[width=\linewidth]{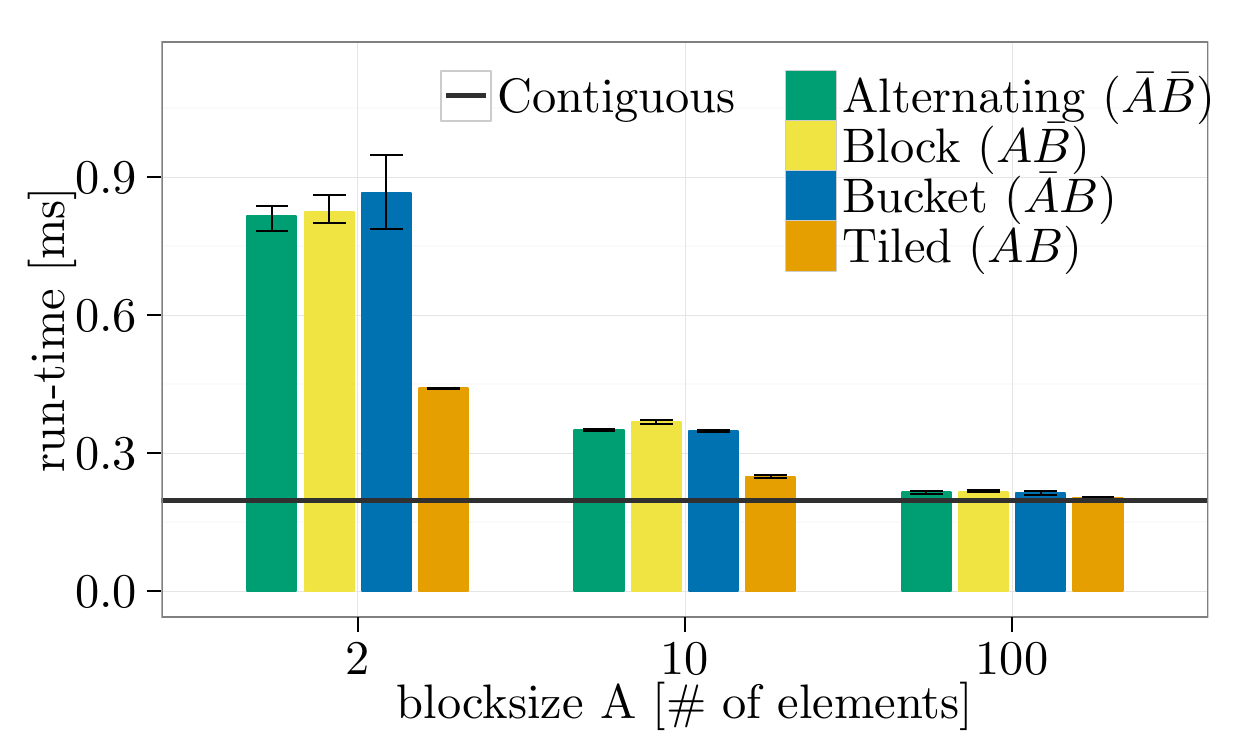}
\caption{%
\label{exp:allgather-nsmall-32p-openmpi}%
\mpiallgather%
}%
\end{subfigure}%
\hfill%
\begin{subfigure}{.33\linewidth}
\centering
\includegraphics[width=\linewidth]{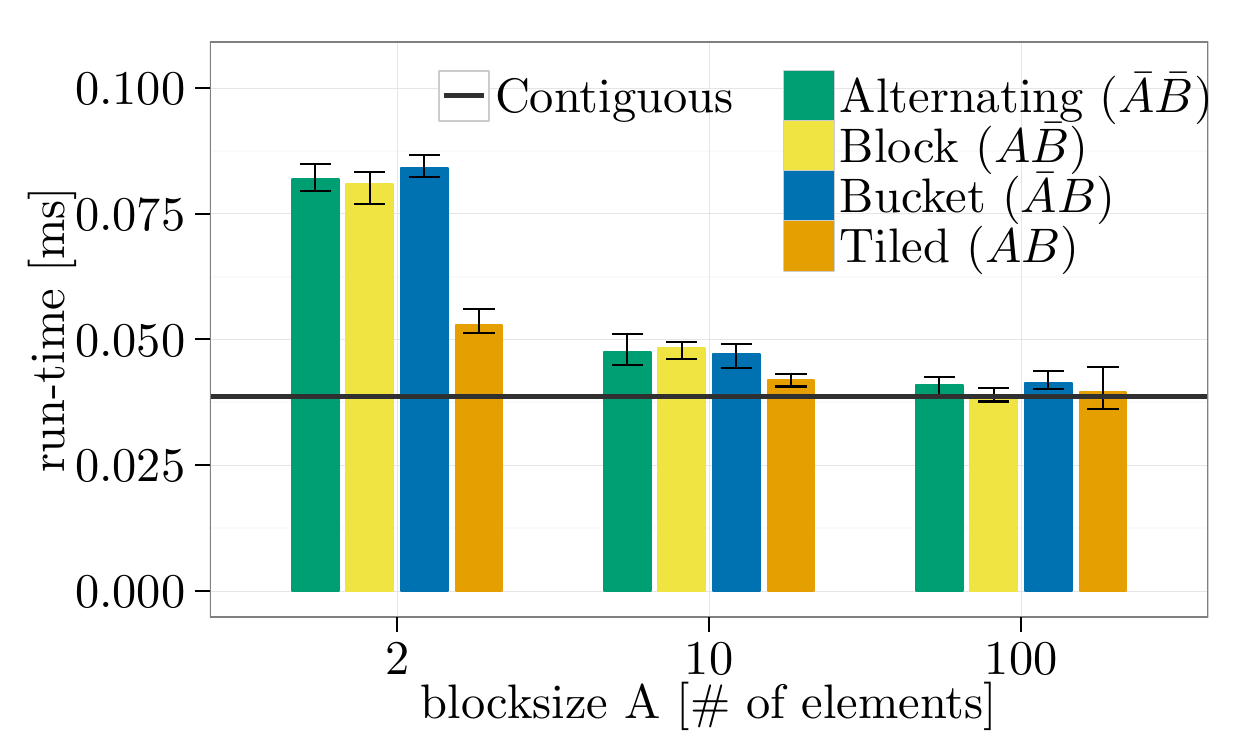}
\caption{%
\label{exp:bcast-nsmall-32p-openmpi}%
\mpibcast%
}%
\end{subfigure}%
\hfill%
\begin{subfigure}{.33\linewidth}
\centering
\includegraphics[width=\linewidth]{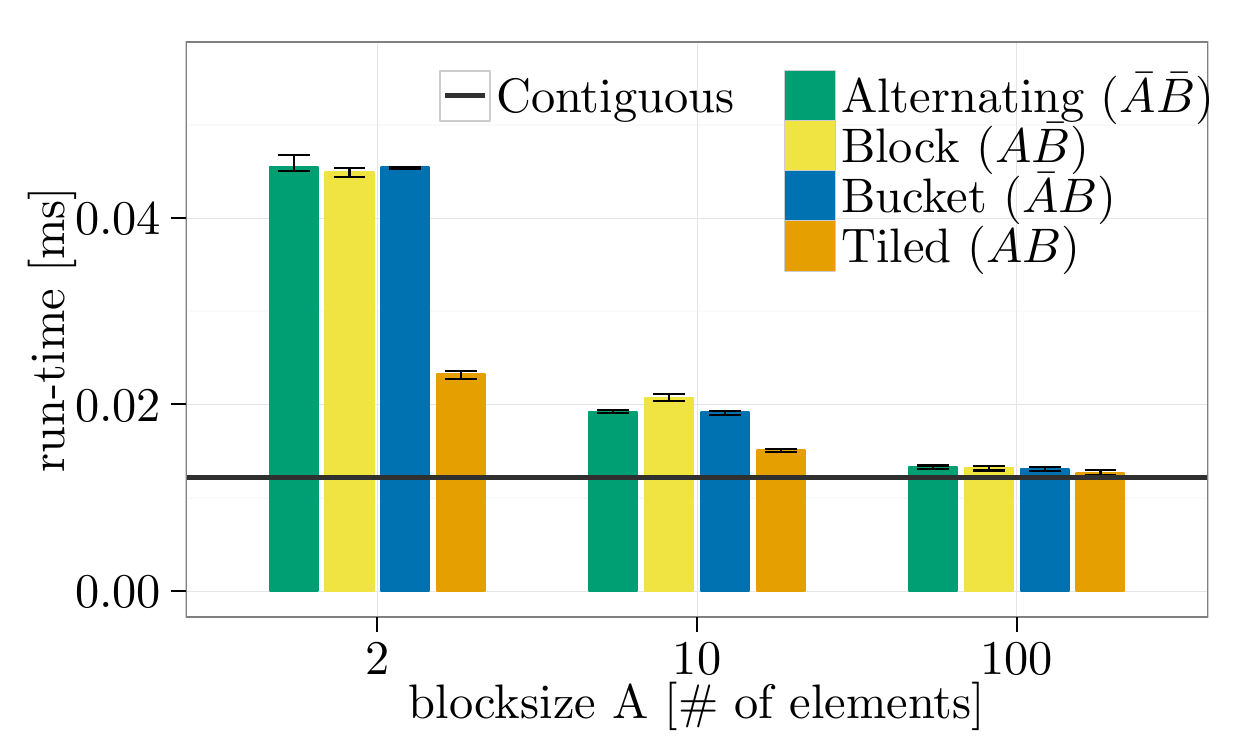}
\caption{%
\label{exp:pingpong-nsmall-32p-openmpi}%
\pingpong%
}%
\end{subfigure}%
\caption{\label{exp:layouts-nsmall-32p-openmpi}  Contiguous \vs typed, $\VARdatasize=\SI{3.2}{\kilo\byte}$, element datatype: \mpiint, \num{32x1}~processes (\num{2x1} for \pingpong), \jupiteropenmpi, \variantone.}
\end{figure*}

\begin{figure*}[htpb]
\centering
\begin{subfigure}{.33\linewidth}
\centering
\includegraphics[width=\linewidth]{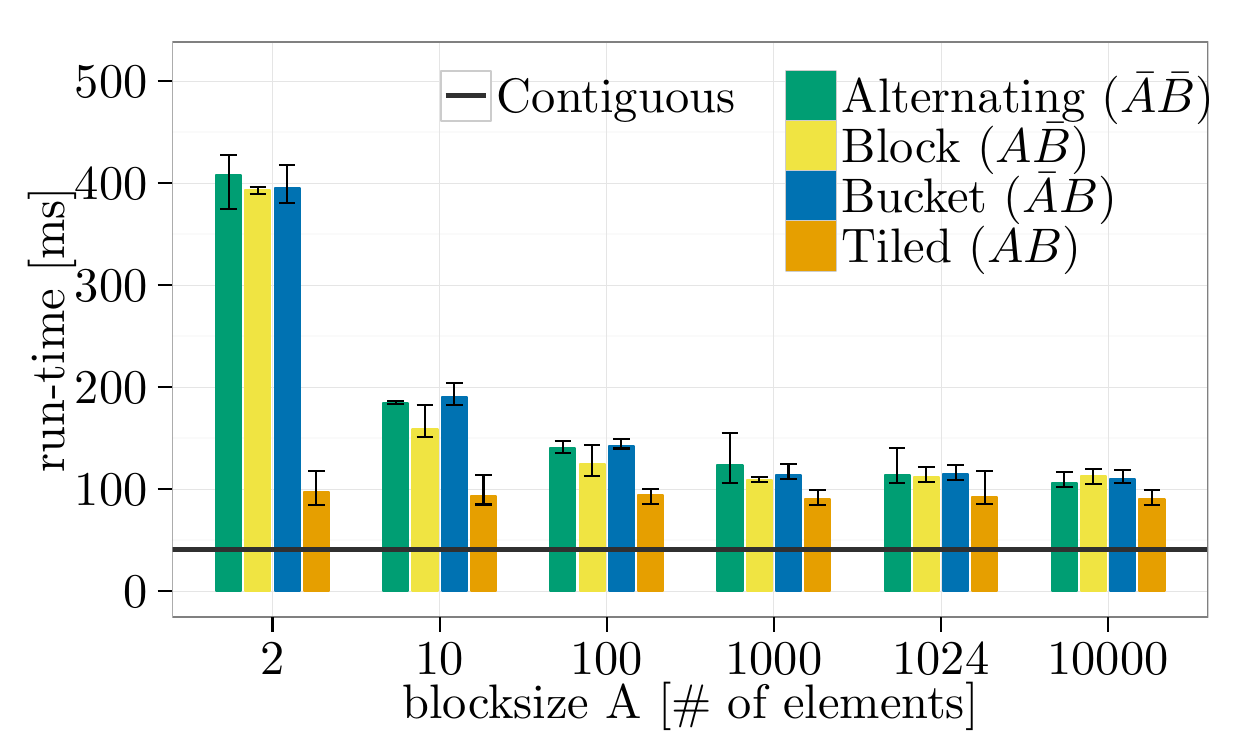}
\caption{%
\label{exp:allgather-nlarge-32p-nec}%
\mpiallgather%
}%
\end{subfigure}%
\hfill%
\begin{subfigure}{.33\linewidth}
\centering
\includegraphics[width=\linewidth]{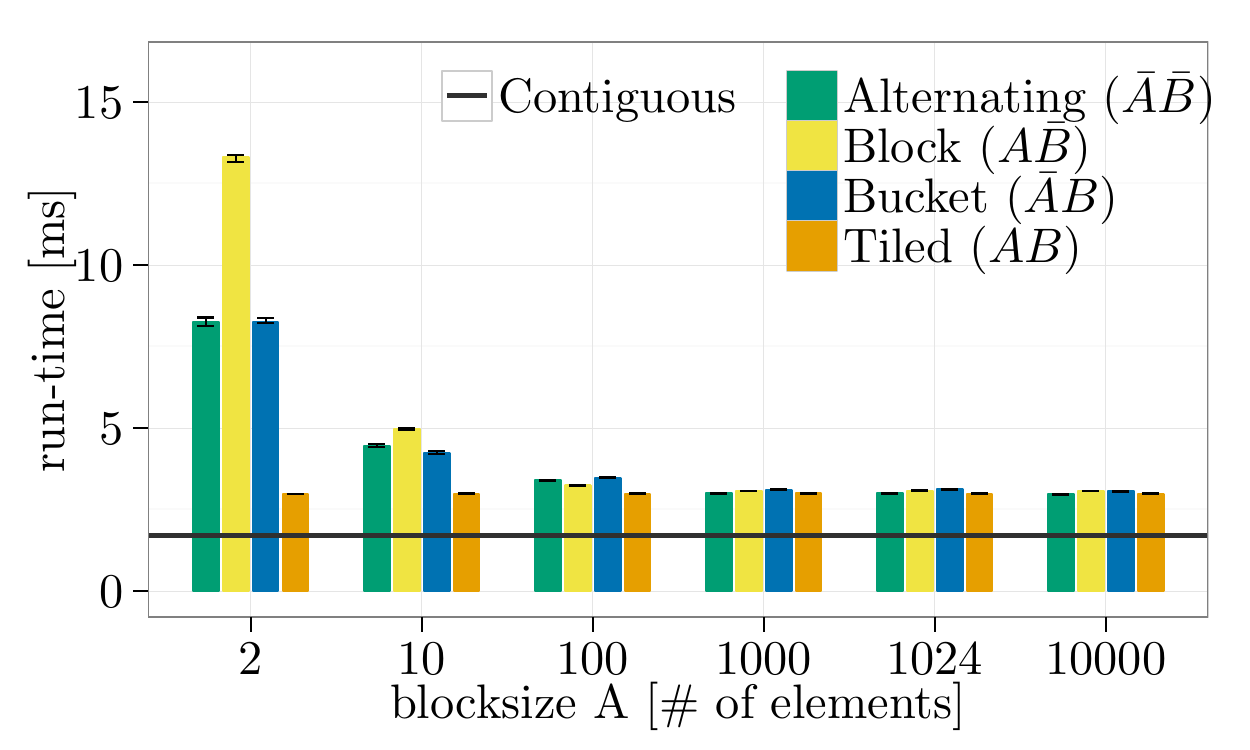}
\caption{%
\label{exp:bcast-nlarge-32p-nec}%
\mpibcast%
}%
\end{subfigure}%
\hfill%
\begin{subfigure}{.33\linewidth}
\centering
\includegraphics[width=\linewidth]{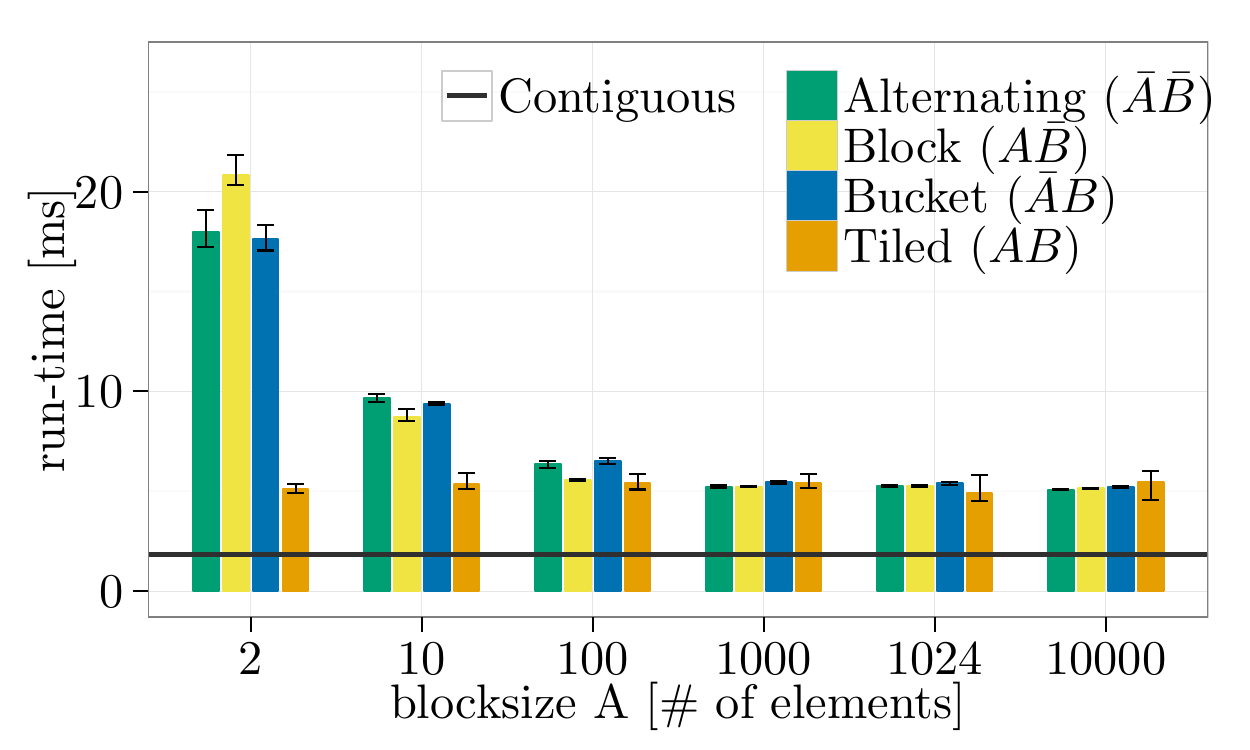}
\caption{%
\label{exp:pingpong-nlarge-32p-nec}%
\pingpong%
}%
\end{subfigure}%
\caption{\label{exp:layouts-nlarge-32p-nec}  Contiguous \vs typed,  $\VARdatasize=\SI{2.56}{\mega\byte}$, element datatype: \mpiint, \num{32x1}~processes (\num{2x1} for \pingpong), \jupiternecmpi, \variantone.}
\end{figure*}

\begin{figure*}[htpb]
\centering
\begin{subfigure}{.33\linewidth}
\centering
\includegraphics[width=\linewidth]{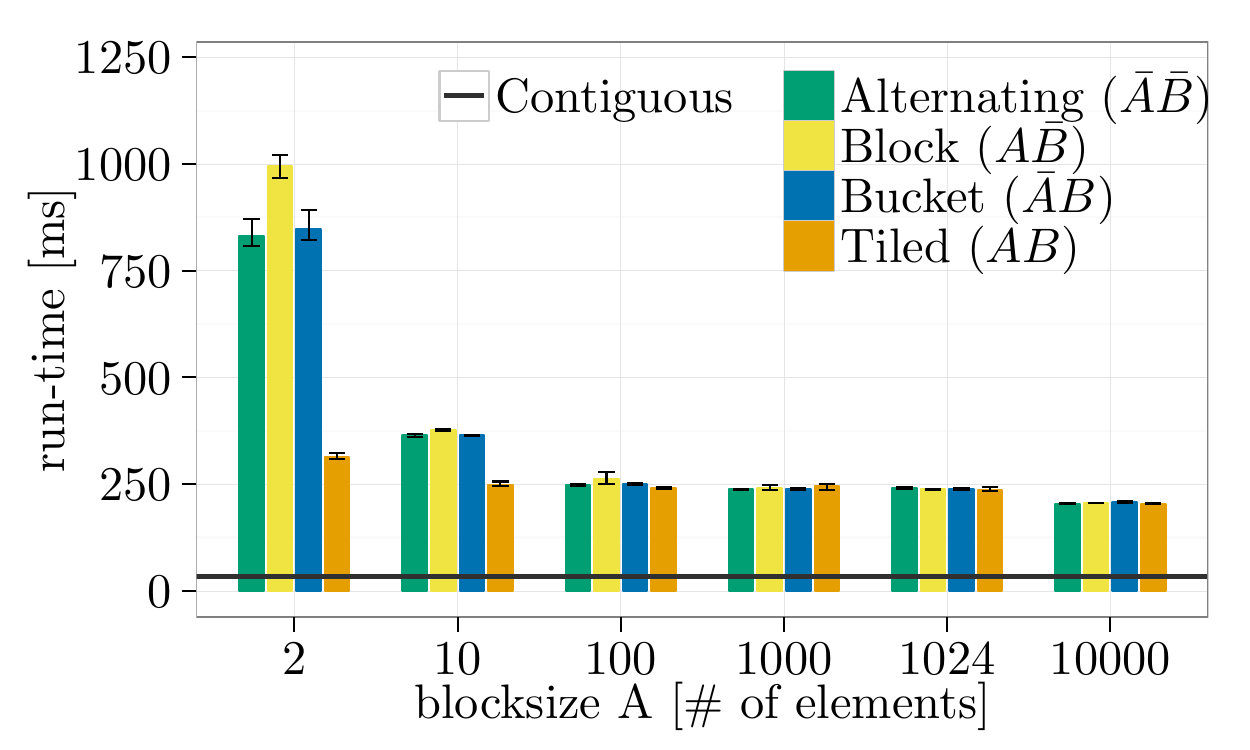}
\caption{%
\label{exp:allgather-nlarge-32p-mvapich}%
\mpiallgather%
}%
\end{subfigure}%
\hfill%
\begin{subfigure}{.33\linewidth}
\centering
\includegraphics[width=\linewidth]{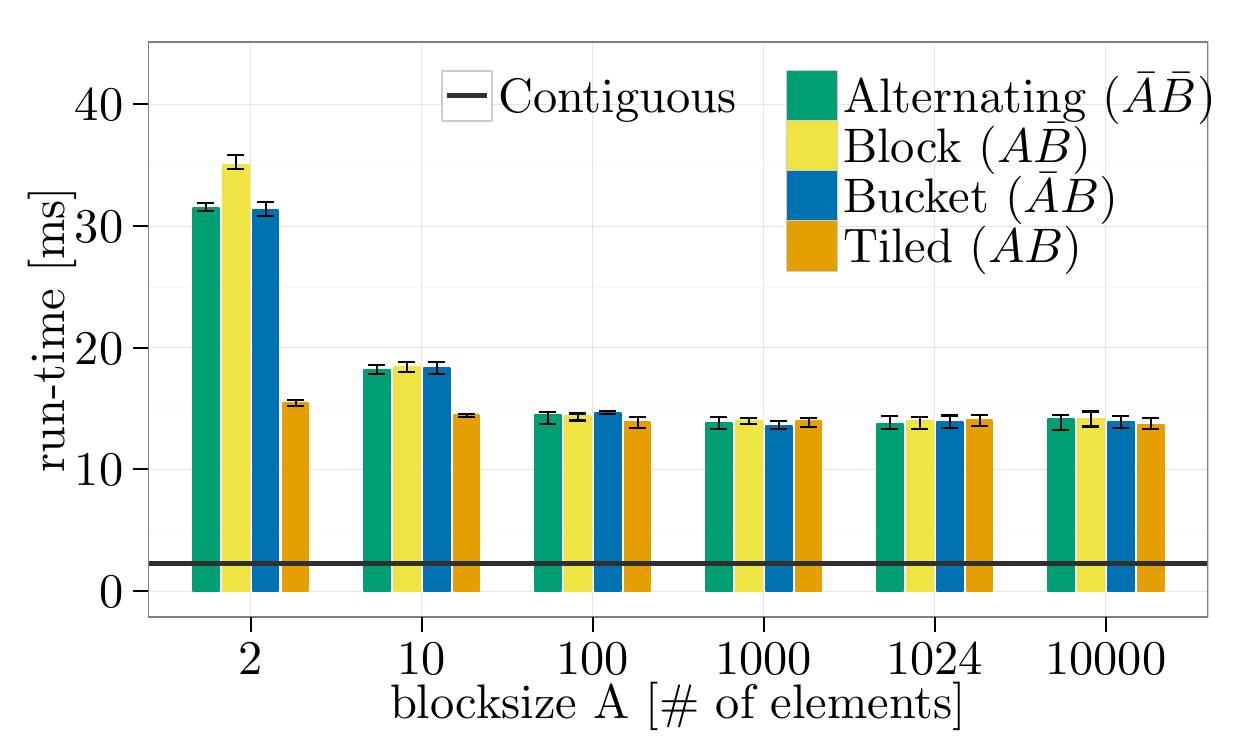}
\caption{%
\label{exp:bcast-nlarge-32p-mvapich}%
\mpibcast%
}%
\end{subfigure}%
\hfill%
\begin{subfigure}{.33\linewidth}
\centering
\includegraphics[width=\linewidth]{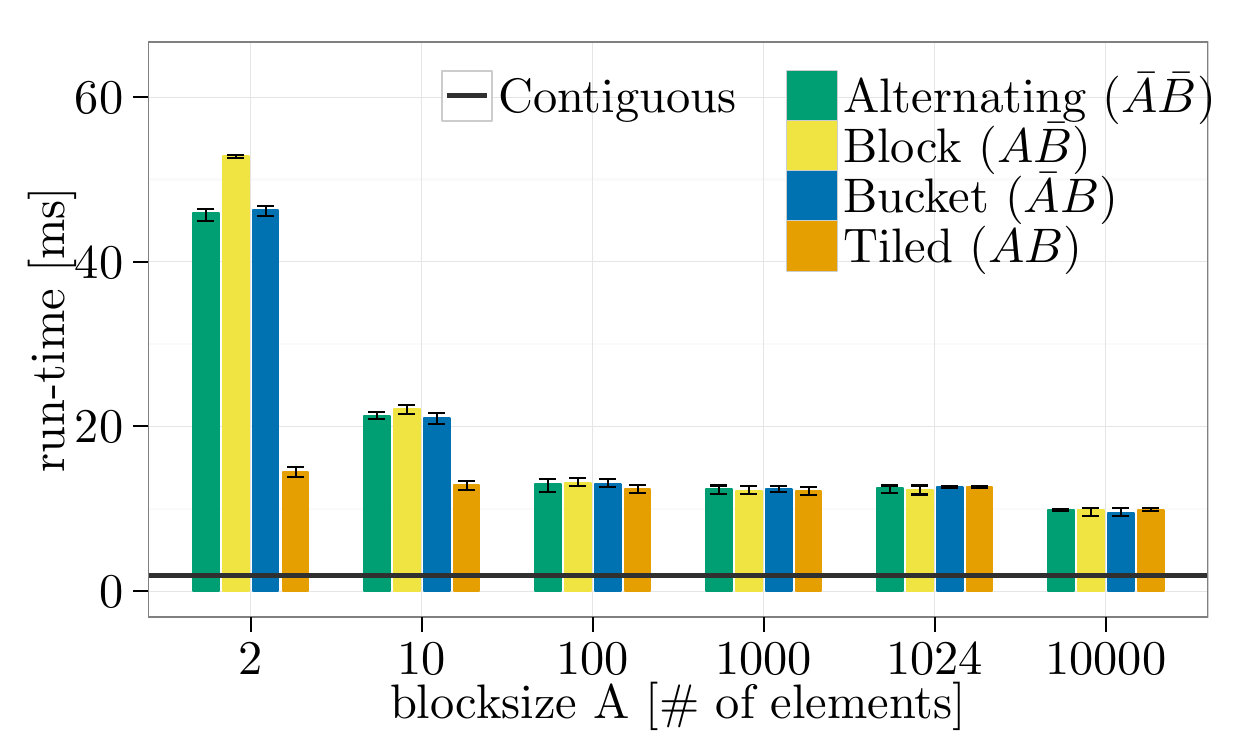}
\caption{%
\label{exp:pingpong-nlarge-32p-mvapich}%
\pingpong%
}%
\end{subfigure}%
\caption{\label{exp:layouts-nlarge-32p-mvapich}  Contiguous \vs typed,  $\VARdatasize=\SI{2.56}{\mega\byte}$, element datatype: \mpiint, \num{32x1}~processes (\num{2x1} for \pingpong), \jupitermvapich, \variantone.}
\end{figure*}

\begin{figure*}[htpb]
\centering
\begin{subfigure}{.33\linewidth}
\centering
\includegraphics[width=\linewidth]{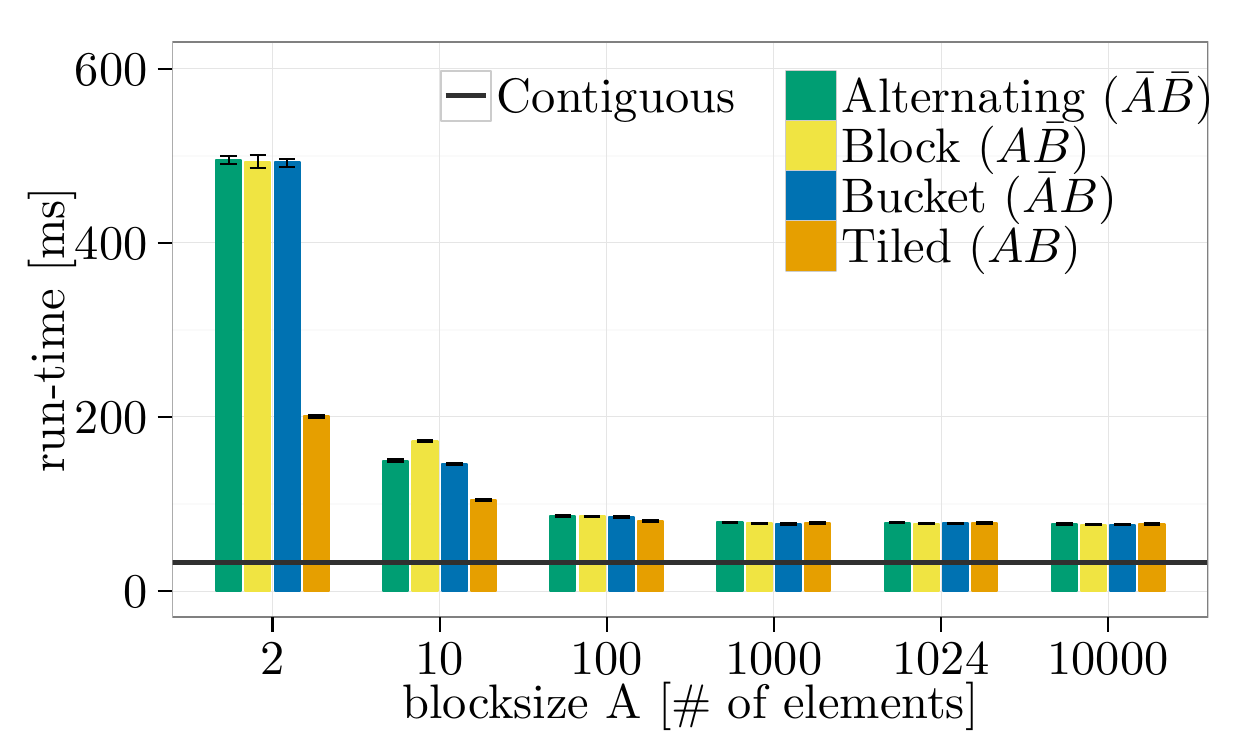}
\caption{%
\label{exp:allgather-nlarge-32p-openmpi}%
\mpiallgather%
}%
\end{subfigure}%
\hfill%
\begin{subfigure}{.33\linewidth}
\centering
\includegraphics[width=\linewidth]{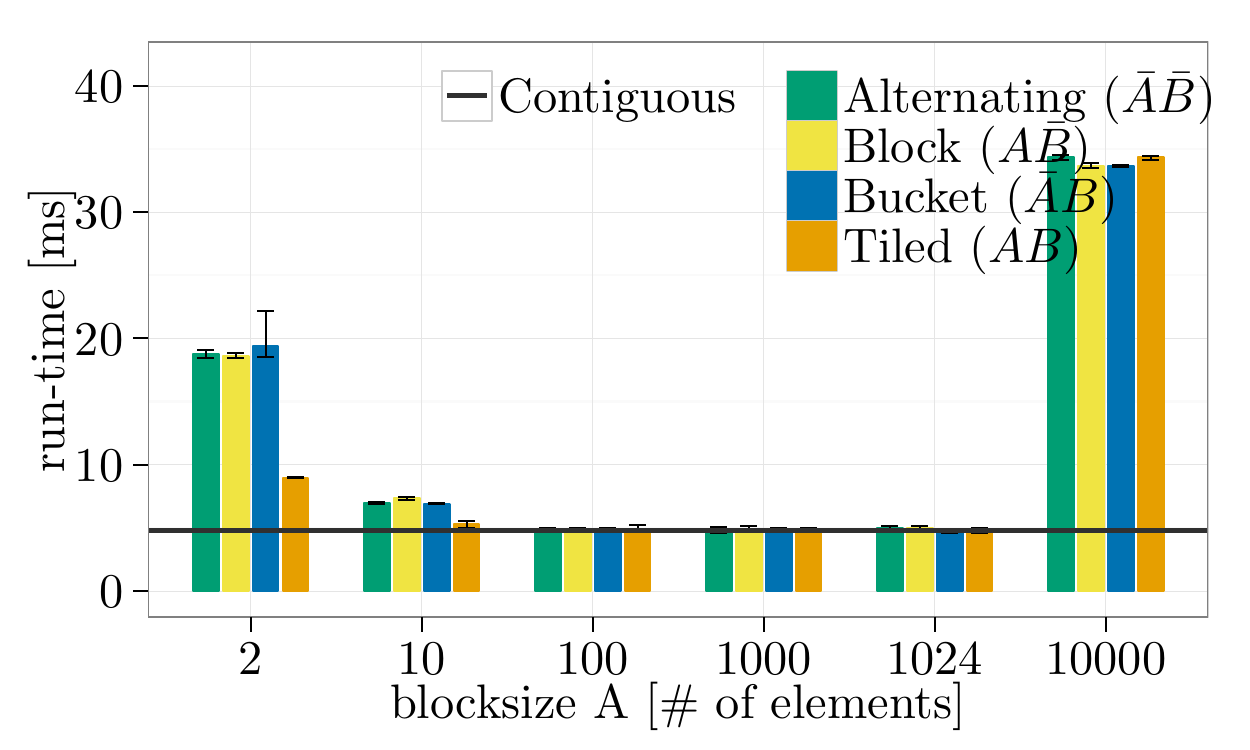}
\caption{%
\label{exp:bcast-nlarge-32p-openmpi}%
\mpibcast%
}%
\end{subfigure}%
\hfill%
\begin{subfigure}{.33\linewidth}
\centering
\includegraphics[width=\linewidth]{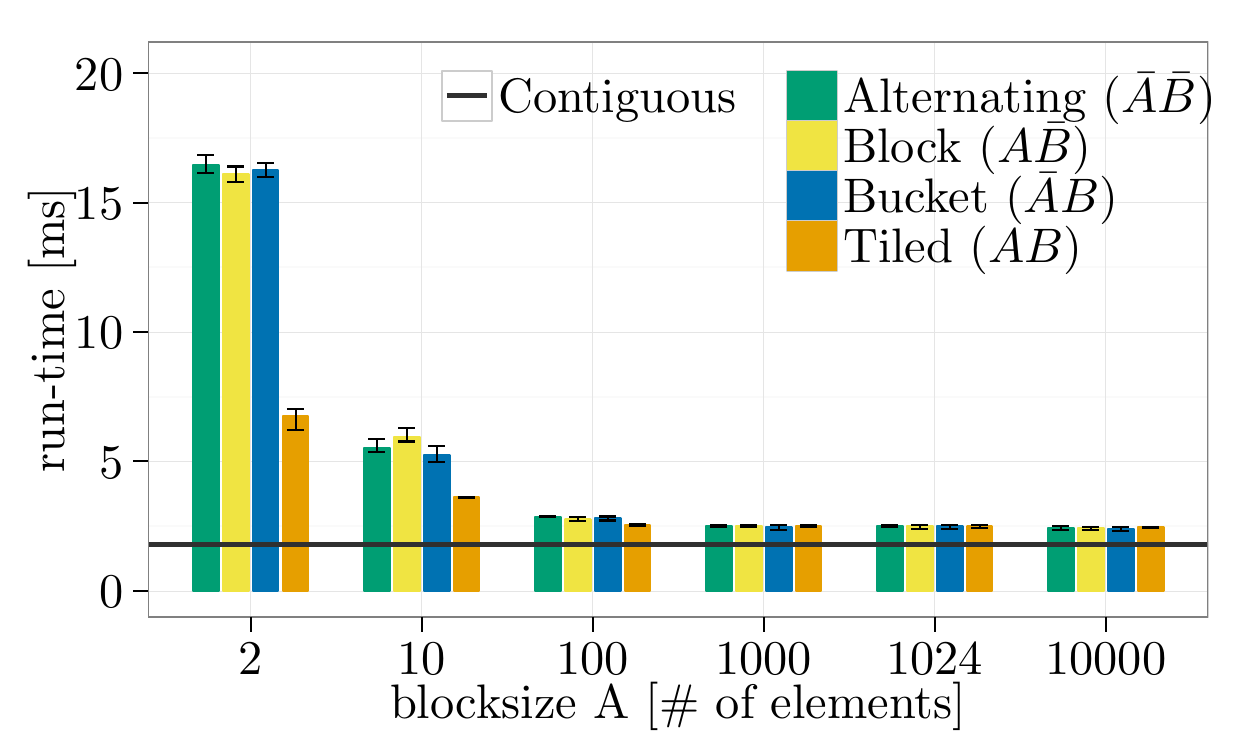}
\caption{%
\label{exp:pingpong-nlarge-32p-openmpi}%
\pingpong%
}%
\end{subfigure}%
\caption{\label{exp:layouts-nlarge-32p-openmpi}  Contiguous \vs typed,  $\VARdatasize=\SI{2.56}{\mega\byte}$, element datatype: \mpiint, \num{32x1}~processes (\num{2x1} for \pingpong), \jupiteropenmpi, \variantone.}
\end{figure*}

\begin{figure*}[htpb]
\centering
\begin{subfigure}{.33\linewidth}
\centering
\includegraphics[width=\linewidth]{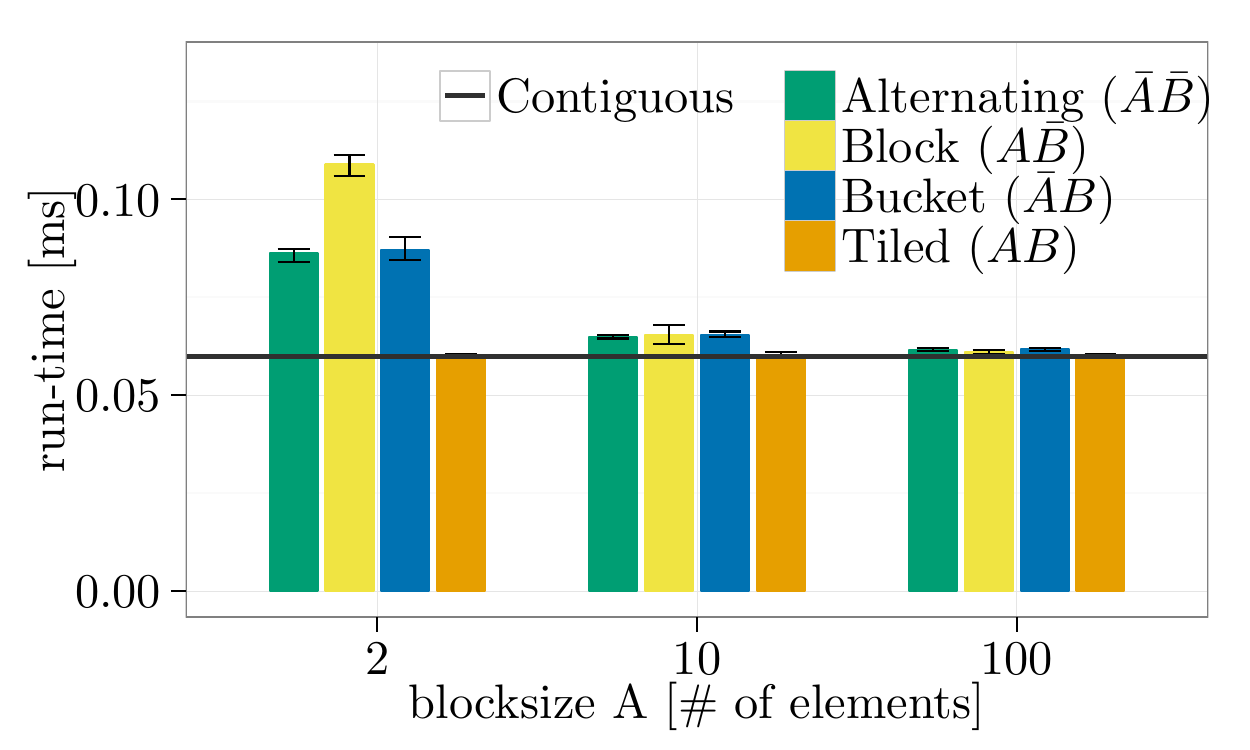}
\caption{%
\label{exp:allgather-nsmall-onenode-nec}%
\mpiallgather%
}%
\end{subfigure}%
\hfill%
\begin{subfigure}{.33\linewidth}
\centering
\includegraphics[width=\linewidth]{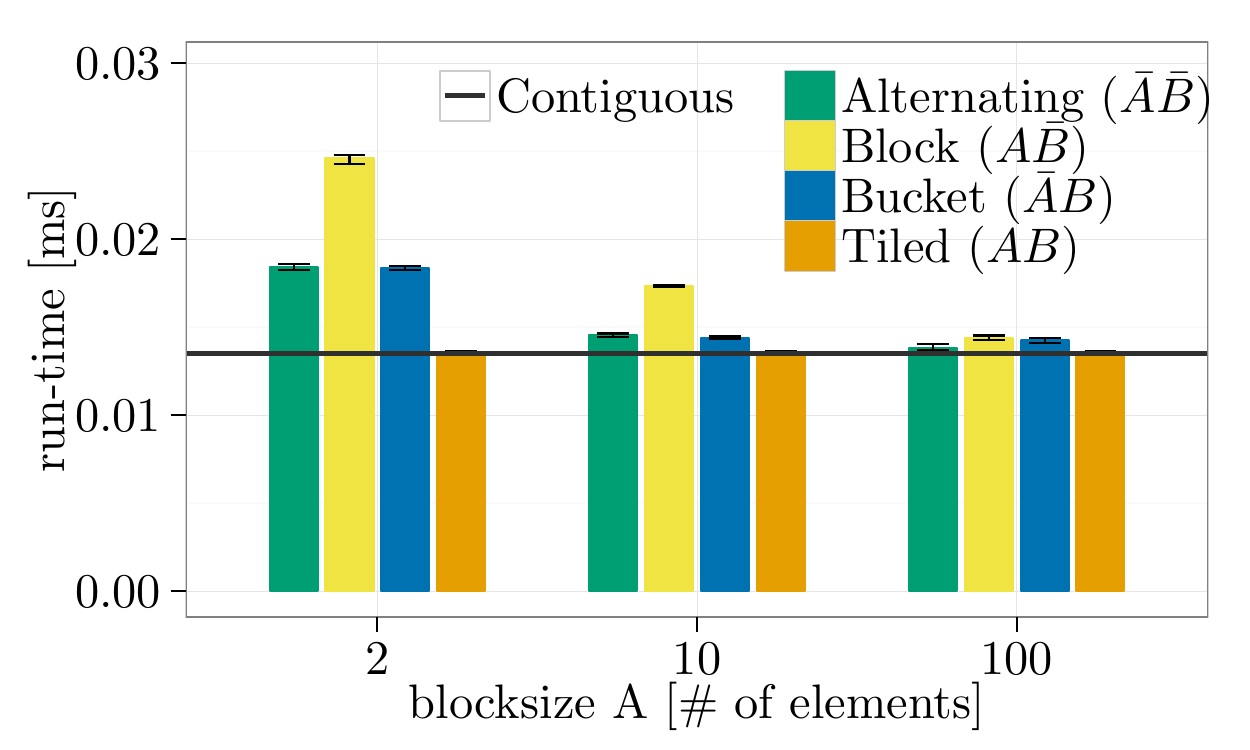}
\caption{%
\label{exp:bcast-nsmall-onenode-nec}%
\mpibcast%
}%
\end{subfigure}%
\hfill%
\begin{subfigure}{.33\linewidth}
\centering
\includegraphics[width=\linewidth]{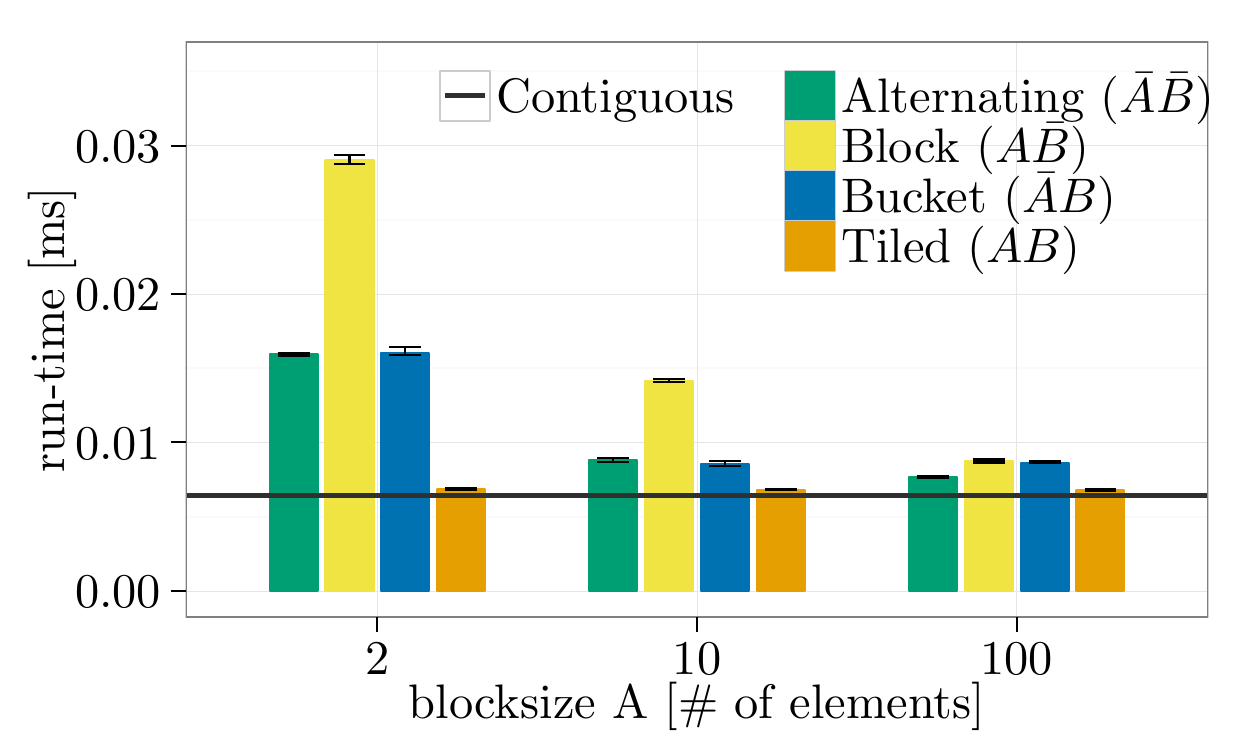}
\caption{%
\label{exp:pingpong-nsmall-onenode-nec}%
\pingpong%
}%
\end{subfigure}%
\caption{\label{exp:layouts-nsmall-onenode-nec}  Contiguous \vs typed, $\VARdatasize=\SI{3.2}{\kilo\byte}$, element datatype: \mpiint, one node, \num{16}~processes (\num{2} for \pingpong), \jupiternecmpi, \variantone.}
\end{figure*}

\begin{figure*}[htpb]
\centering
\begin{subfigure}{.33\linewidth}
\centering
\includegraphics[width=\linewidth]{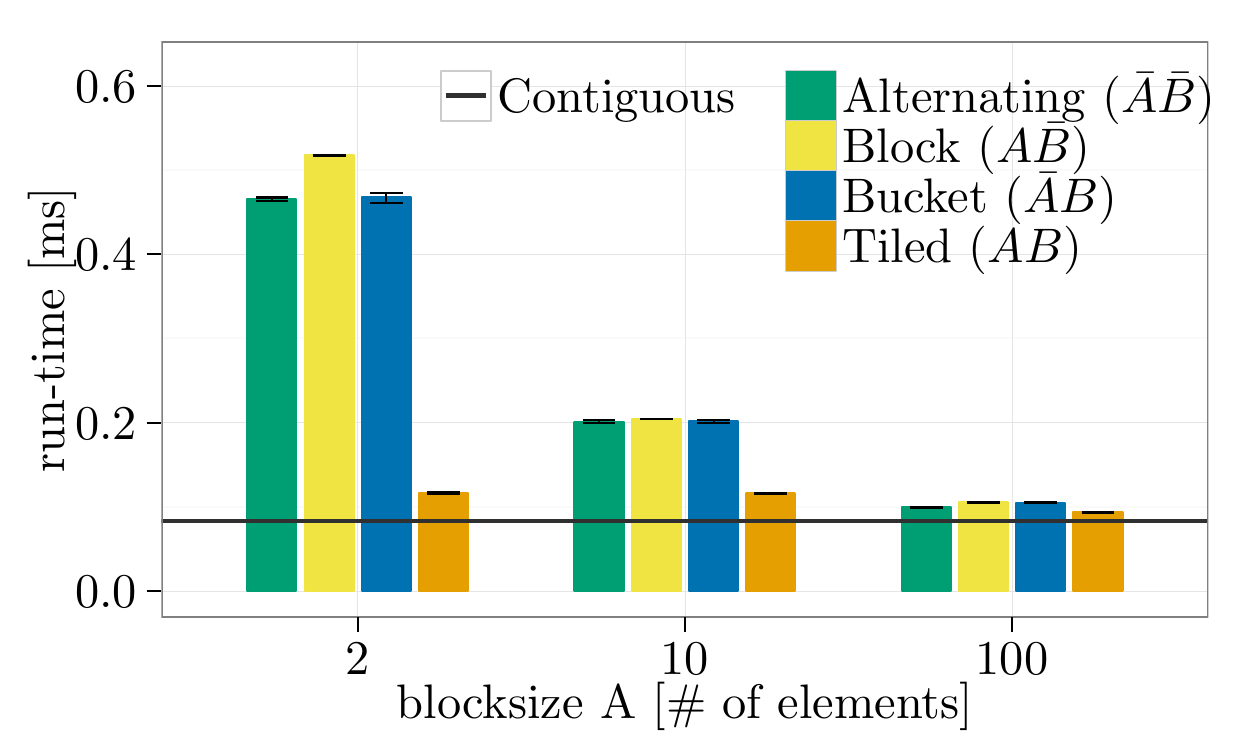}
\caption{%
\label{exp:allgather-nsmall-onenode-mvapich}%
\mpiallgather%
}%
\end{subfigure}%
\hfill%
\begin{subfigure}{.33\linewidth}
\centering
\includegraphics[width=\linewidth]{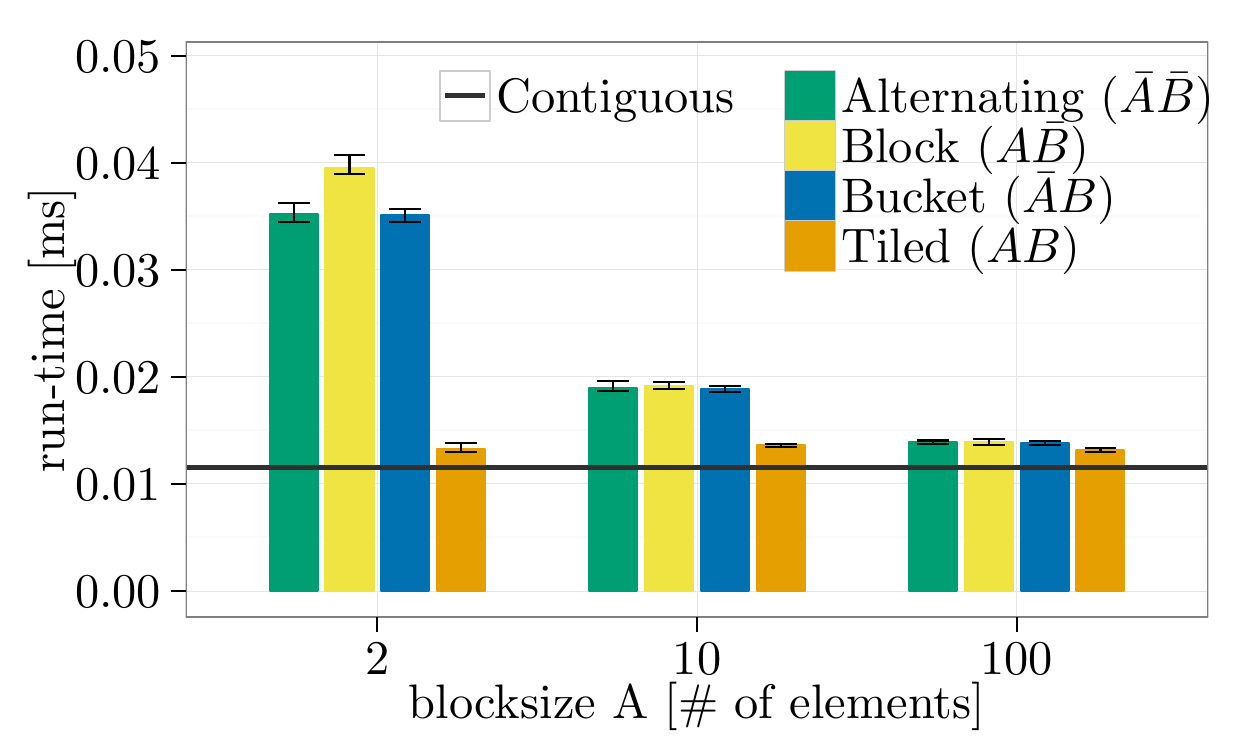}
\caption{%
\label{exp:bcast-nsmall-onenode-mvapich}%
\mpibcast%
}%
\end{subfigure}%
\hfill%
\begin{subfigure}{.33\linewidth}
\centering
\includegraphics[width=\linewidth]{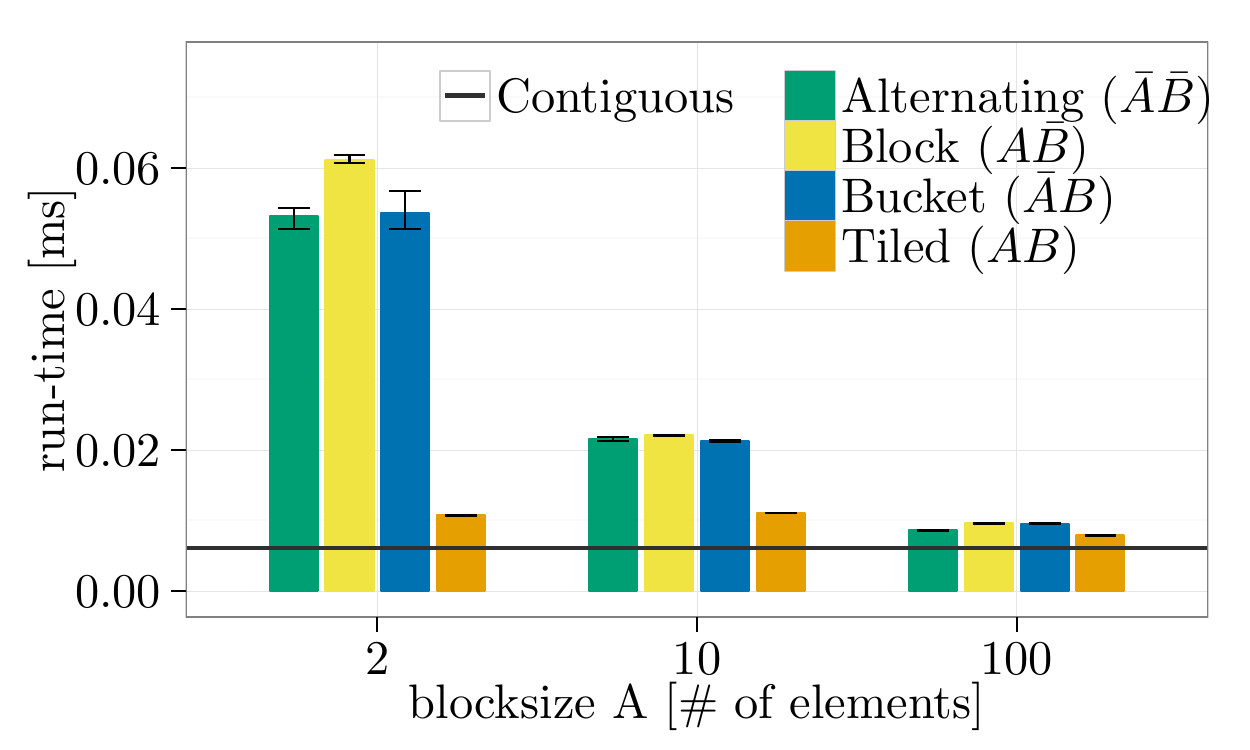}
\caption{%
\label{exp:pingpong-nsmall-onenode-mvapich}%
\pingpong%
}%
\end{subfigure}%
\caption{\label{exp:layouts-nsmall-onenode-mvapich}  Contiguous \vs typed, $\VARdatasize=\SI{3.2}{\kilo\byte}$, element datatype: \mpiint, one~node, \num{16}~processes (\num{2}~processes for \pingpong), \jupitermvapich, \variantone.}
\end{figure*}

\begin{figure*}[htpb]
\centering
\begin{subfigure}{.33\linewidth}
\centering
\includegraphics[width=\linewidth]{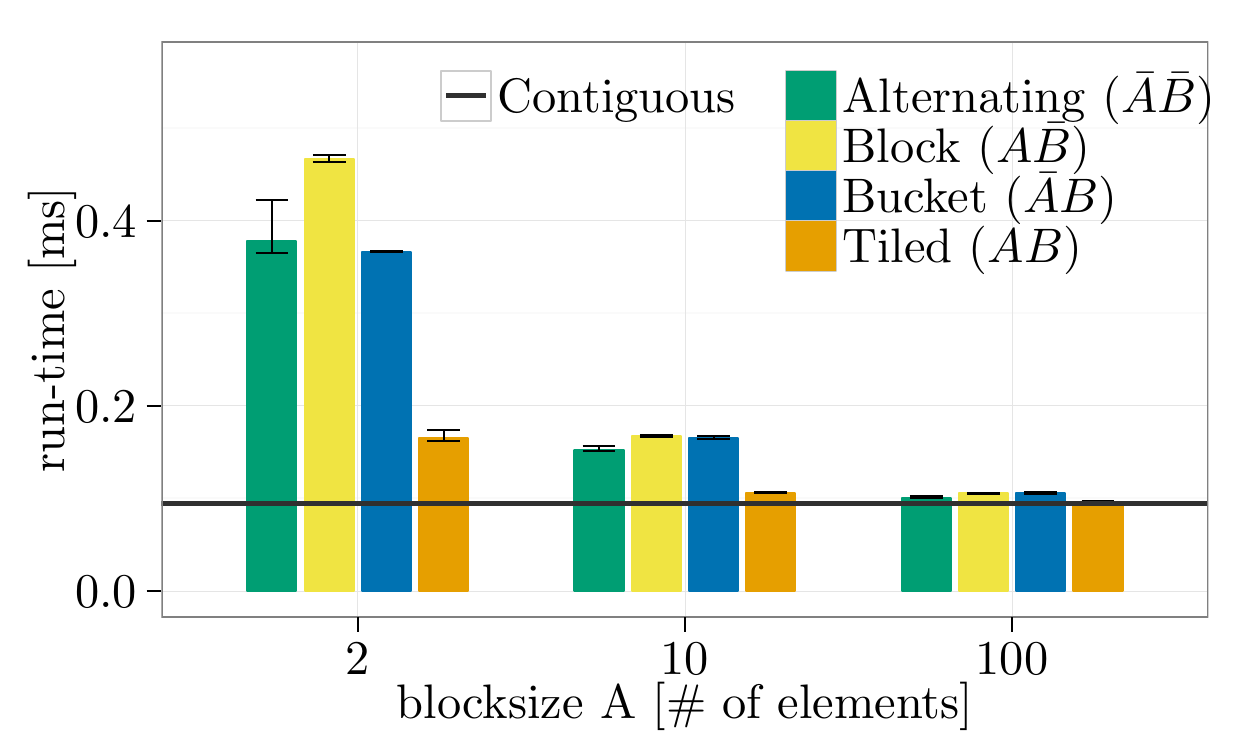}
\caption{%
\label{exp:allgather-nsmall-onenode-openmpi}%
\mpiallgather%
}%
\end{subfigure}%
\hfill%
\begin{subfigure}{.33\linewidth}
\centering
\includegraphics[width=\linewidth]{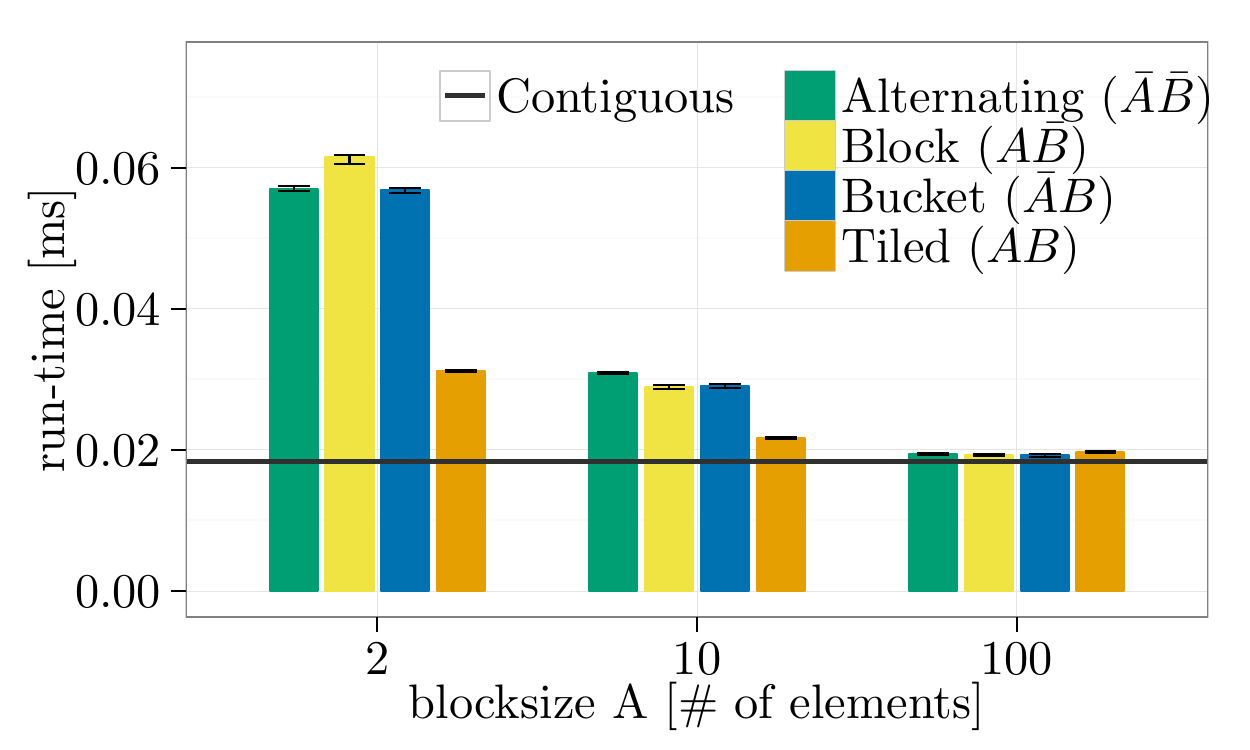}
\caption{%
\label{exp:bcast-nsmall-onenode-openmpi}%
\mpibcast%
}%
\end{subfigure}%
\hfill%
\begin{subfigure}{.33\linewidth}
\centering
\includegraphics[width=\linewidth]{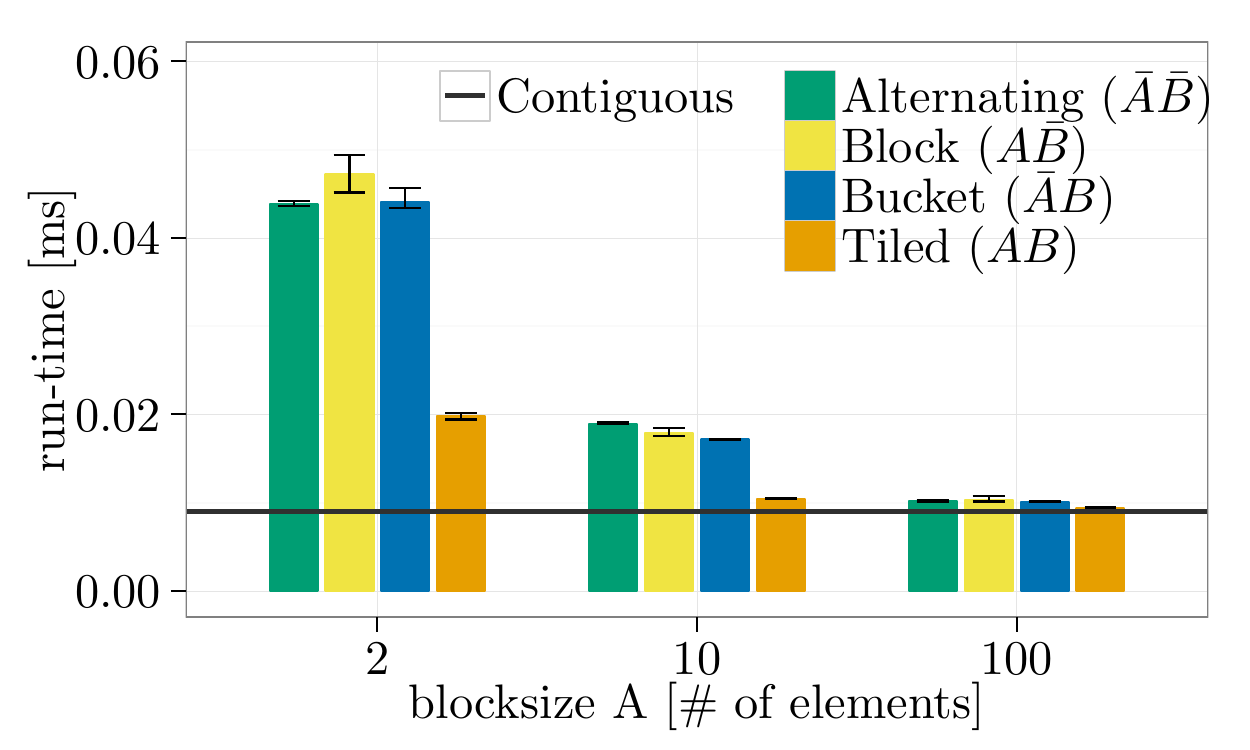}
\caption{%
\label{exp:pingpong-nsmall-onenode-openmpi}%
\pingpong%
}%
\end{subfigure}%
\caption{\label{exp:layouts-nsmall-onenode-openmpi}  Contiguous \vs typed, $\VARdatasize=\SI{3.2}{\kilo\byte}$, element datatype: \mpiint, \num{1}~node, \num{16}~processes (\num{2}~processes for \pingpong), \jupiteropenmpi, \variantone.}
\end{figure*}

\begin{figure*}[htpb]
\centering
\begin{subfigure}{.33\linewidth}
\centering
\includegraphics[width=\linewidth]{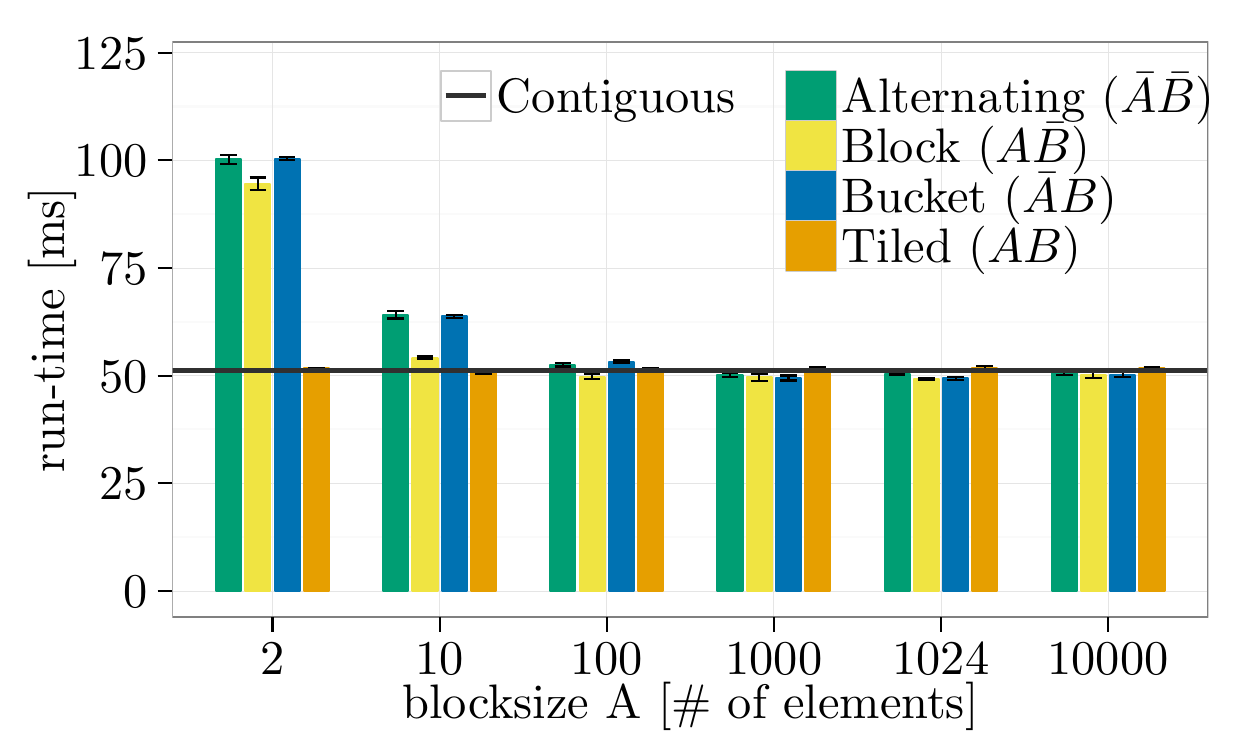}
\caption{%
\label{exp:allgather-nlarge-onenode-nec}%
\mpiallgather%
}%
\end{subfigure}%
\hfill%
\begin{subfigure}{.33\linewidth}
\centering
\includegraphics[width=\linewidth]{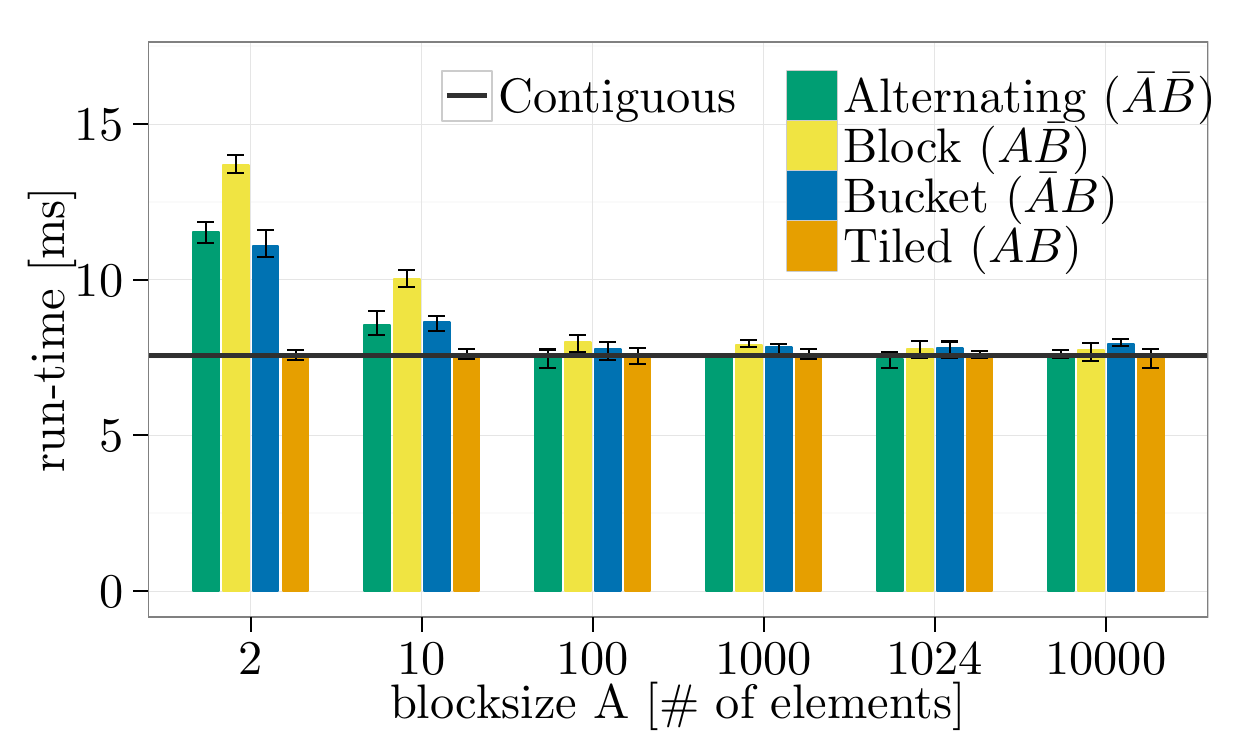}
\caption{%
\label{exp:bcast-nlarge-onenode-nec}%
\mpibcast%
}%
\end{subfigure}%
\hfill%
\begin{subfigure}{.33\linewidth}
\centering
\includegraphics[width=\linewidth]{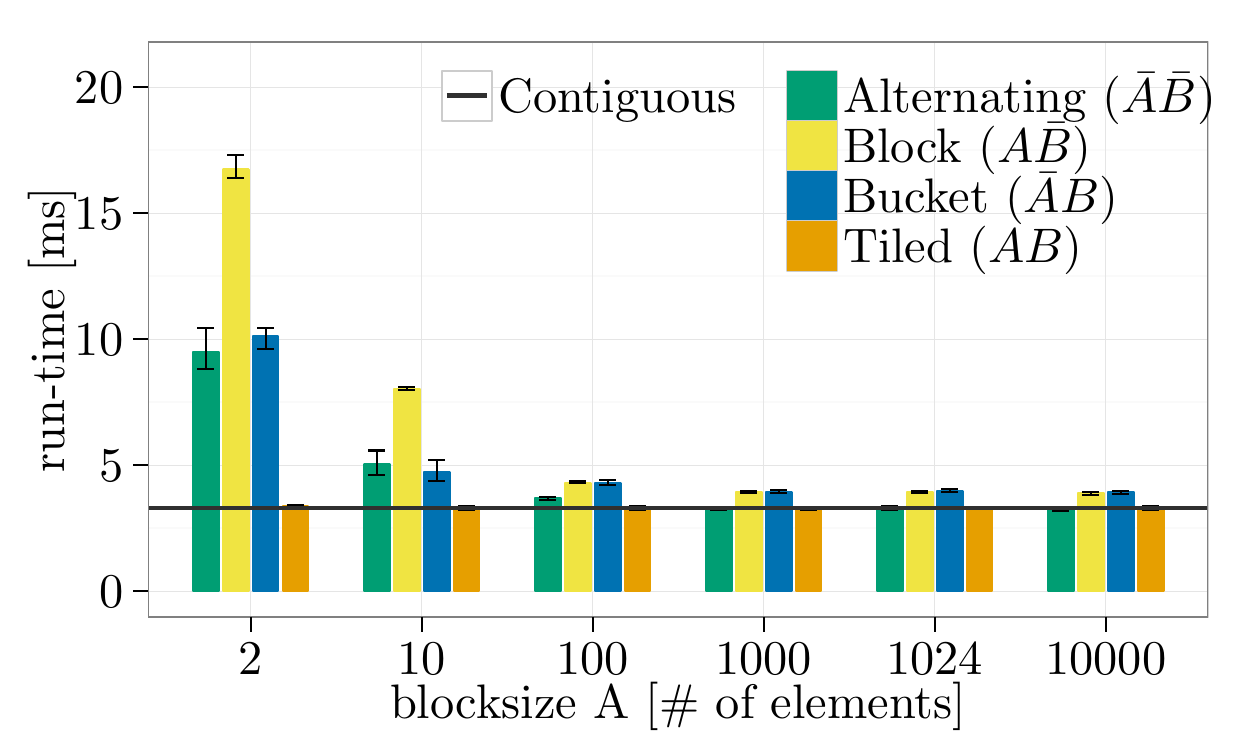}
\caption{%
\label{exp:pingpong-nlarge-onenode-nec}%
\pingpong%
}%
\end{subfigure}%
\caption{\label{exp:layouts-nlarge-onenode-nec}  Contiguous \vs typed,  $\VARdatasize=\SI{2.56}{\mega\byte}$, element datatype: \mpiint, one~node, \num{16}~processes (\num{2}~processes for \pingpong), \jupiternecmpi, \variantone.}
\end{figure*}

\begin{figure*}[htpb]
\centering
\begin{subfigure}{.33\linewidth}
\centering
\includegraphics[width=\linewidth]{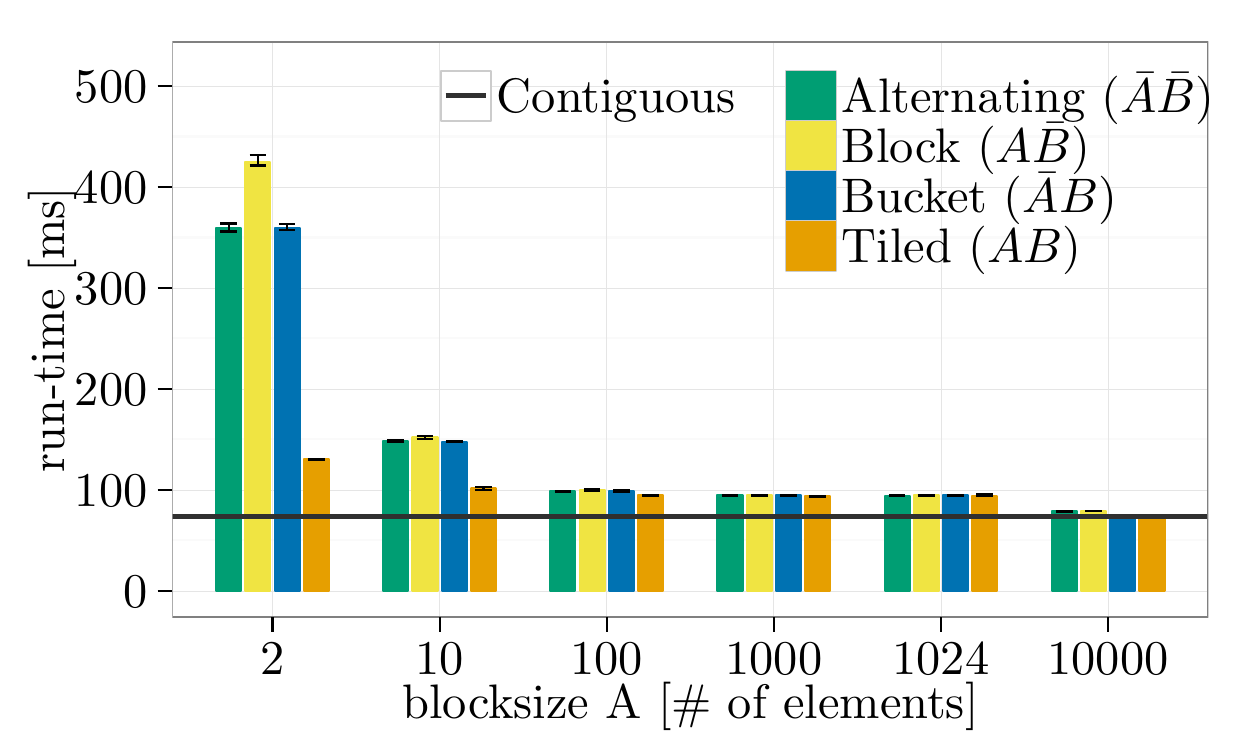}
\caption{%
\label{exp:allgather-nlarge-onenode-mvapich}%
\mpiallgather%
}%
\end{subfigure}%
\hfill%
\begin{subfigure}{.33\linewidth}
\centering
\includegraphics[width=\linewidth]{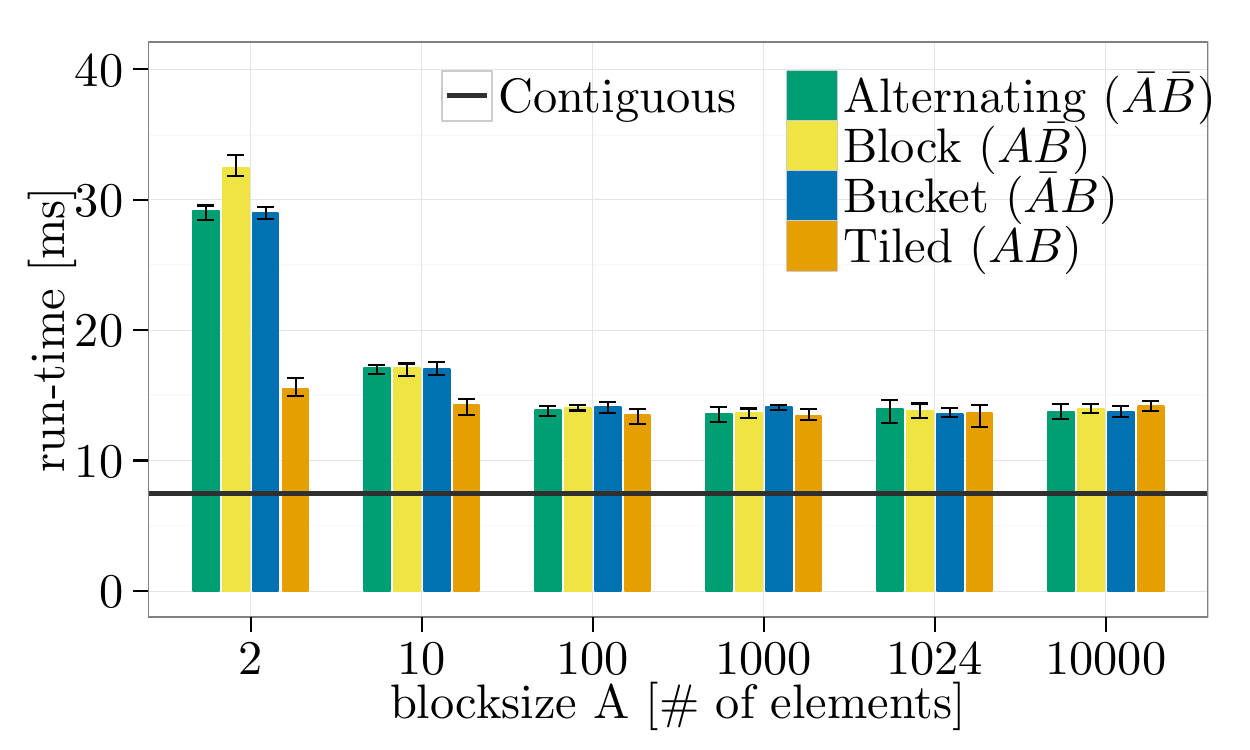}
\caption{%
\label{exp:bcast-nlarge-onenode-mvapich}%
\mpibcast%
}%
\end{subfigure}%
\hfill%
\begin{subfigure}{.33\linewidth}
\centering
\includegraphics[width=\linewidth]{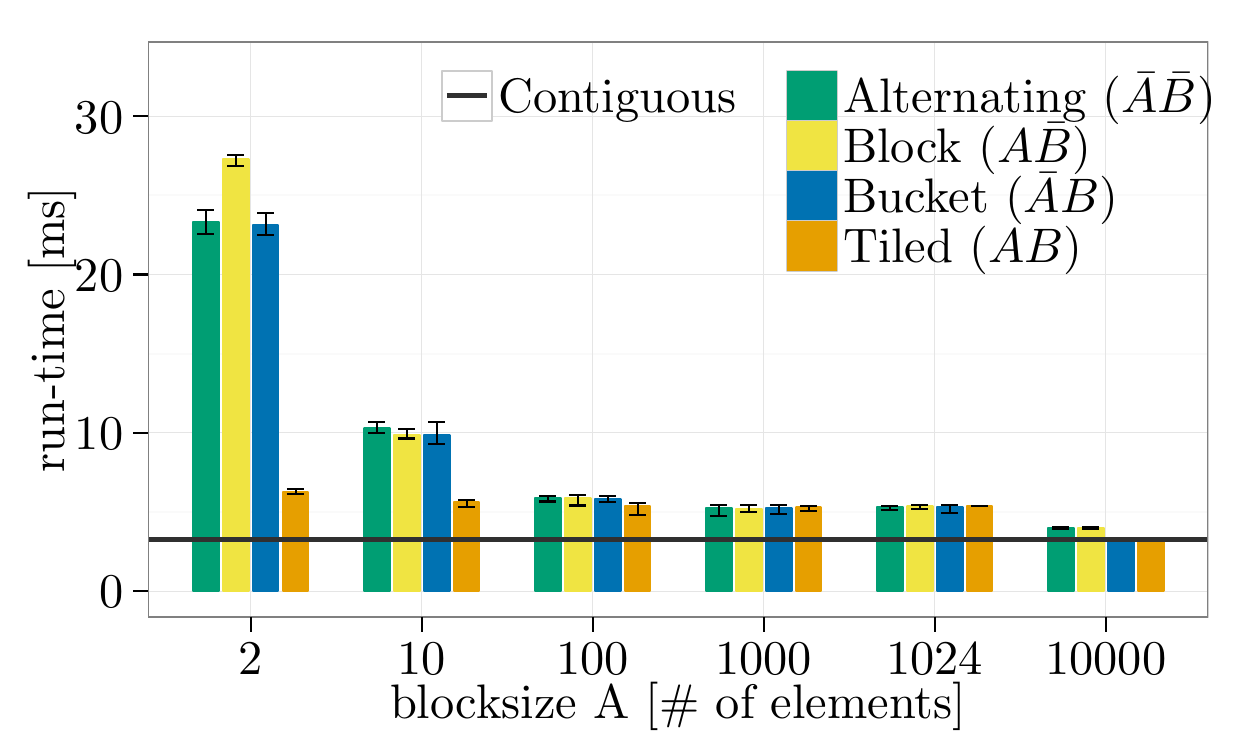}
\caption{%
\label{exp:pingpong-nlarge-onenode-mvapich}%
\pingpong%
}%
\end{subfigure}%
\caption{\label{exp:layouts-nlarge-onenode-mvapich}  Contiguous \vs typed,  $\VARdatasize=\SI{2.56}{\mega\byte}$, element datatype: \mpiint,  one~node, \num{16}~processes (\num{2}~processes for \pingpong), \jupitermvapich, \variantone.}
\end{figure*}

\begin{figure*}[htpb]
\centering
\begin{subfigure}{.33\linewidth}
\centering
\includegraphics[width=\linewidth]{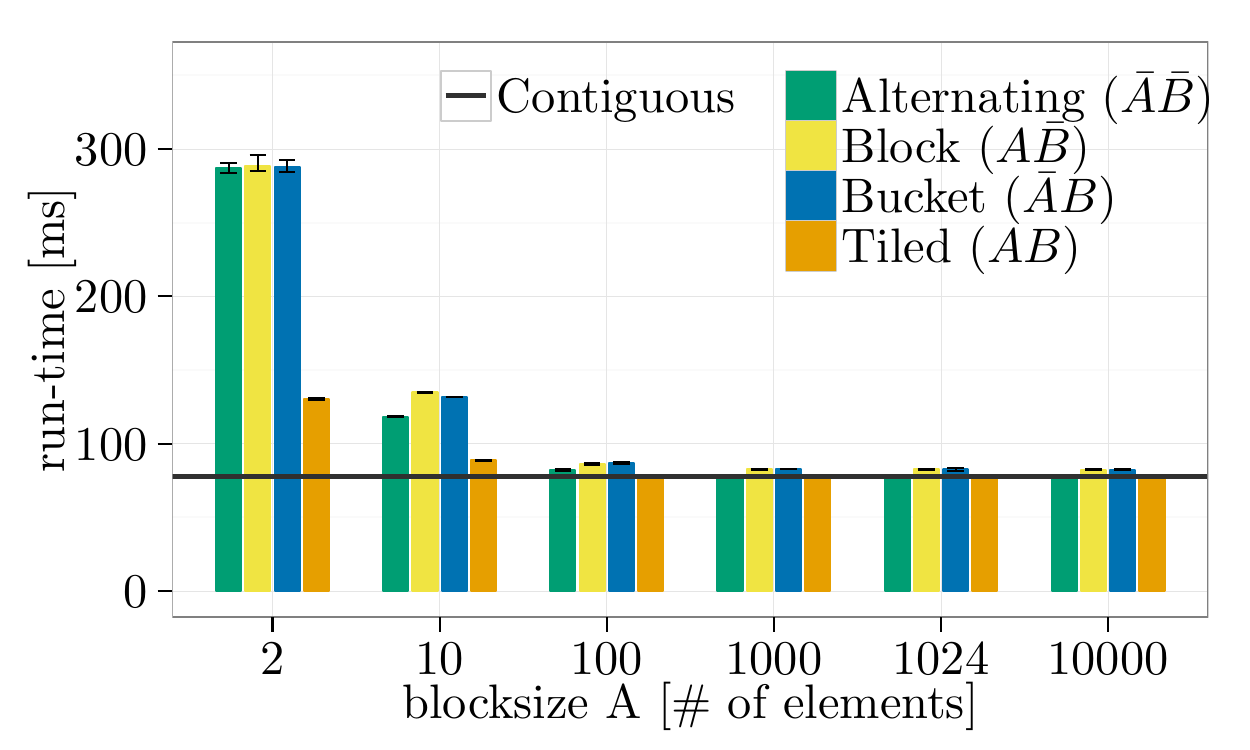}
\caption{%
\label{exp:allgather-nlarge-onenode-openmpi}%
\mpiallgather%
}%
\end{subfigure}%
\hfill%
\begin{subfigure}{.33\linewidth}
\centering
\includegraphics[width=\linewidth]{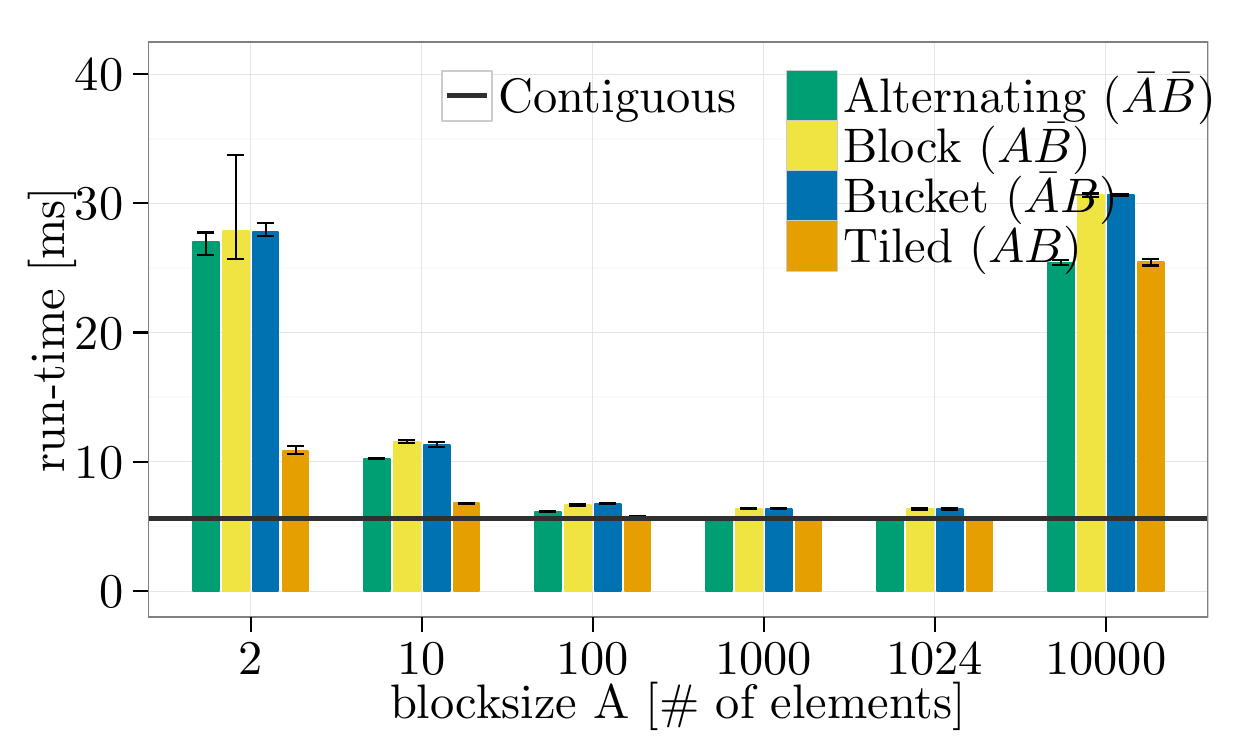}
\caption{%
\label{exp:bcast-nlarge-onenode-openmpi}%
\mpibcast%
}%
\end{subfigure}%
\hfill%
\begin{subfigure}{.33\linewidth}
\centering
\includegraphics[width=\linewidth]{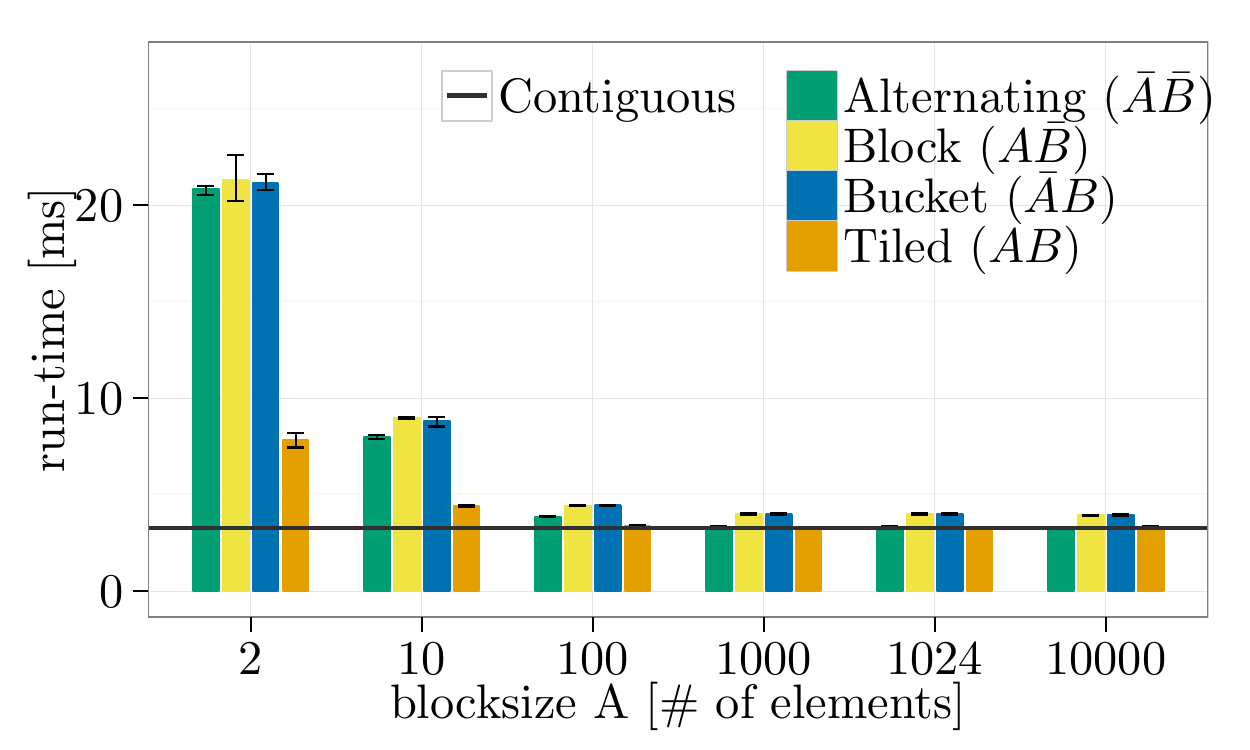}
\caption{%
\label{exp:pingpong-nlarge-onenode-openmpi}%
\pingpong%
}%
\end{subfigure}%
\caption{\label{exp:layouts-nlarge-onenode-openmpi}  Contiguous \vs typed,  $\VARdatasize=\SI{2.56}{\mega\byte}$, element datatype: \mpiint,  \num{1}~node, \num{16}~processes (\num{2}~processes for \pingpong), \jupiteropenmpi, \variantone.}
\end{figure*}

\begin{figure*}[htpb]
\centering
\begin{subfigure}{.33\linewidth}
\centering
\includegraphics[width=\linewidth]{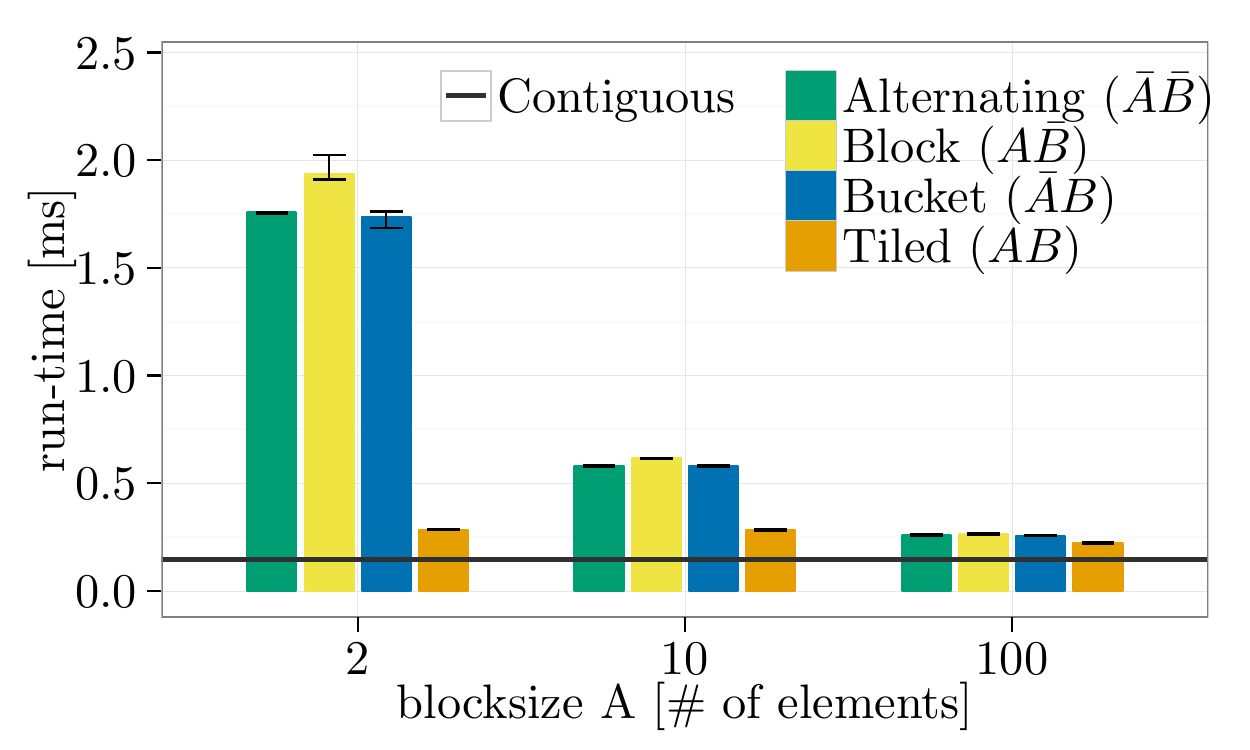}
\caption{%
\label{exp:allgather-nsmall-mvapich-vartwo}%
\mpiallgather%
}%
\end{subfigure}%
\hfill%
\begin{subfigure}{.33\linewidth}
\centering
\includegraphics[width=\linewidth]{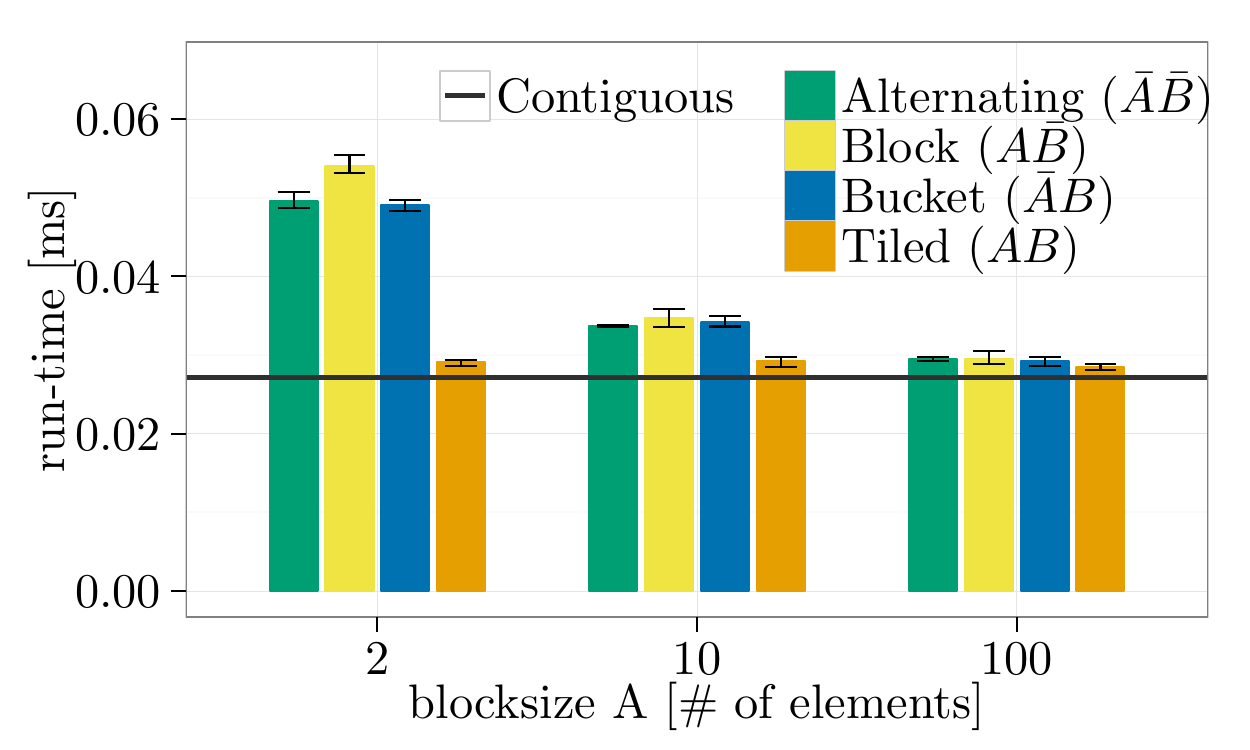}
\caption{%
\label{exp:bcast-nsmall-mvapich-vartwo}%
\mpibcast%
}%
\end{subfigure}%
\hfill%
\begin{subfigure}{.33\linewidth}
\centering
\includegraphics[width=\linewidth]{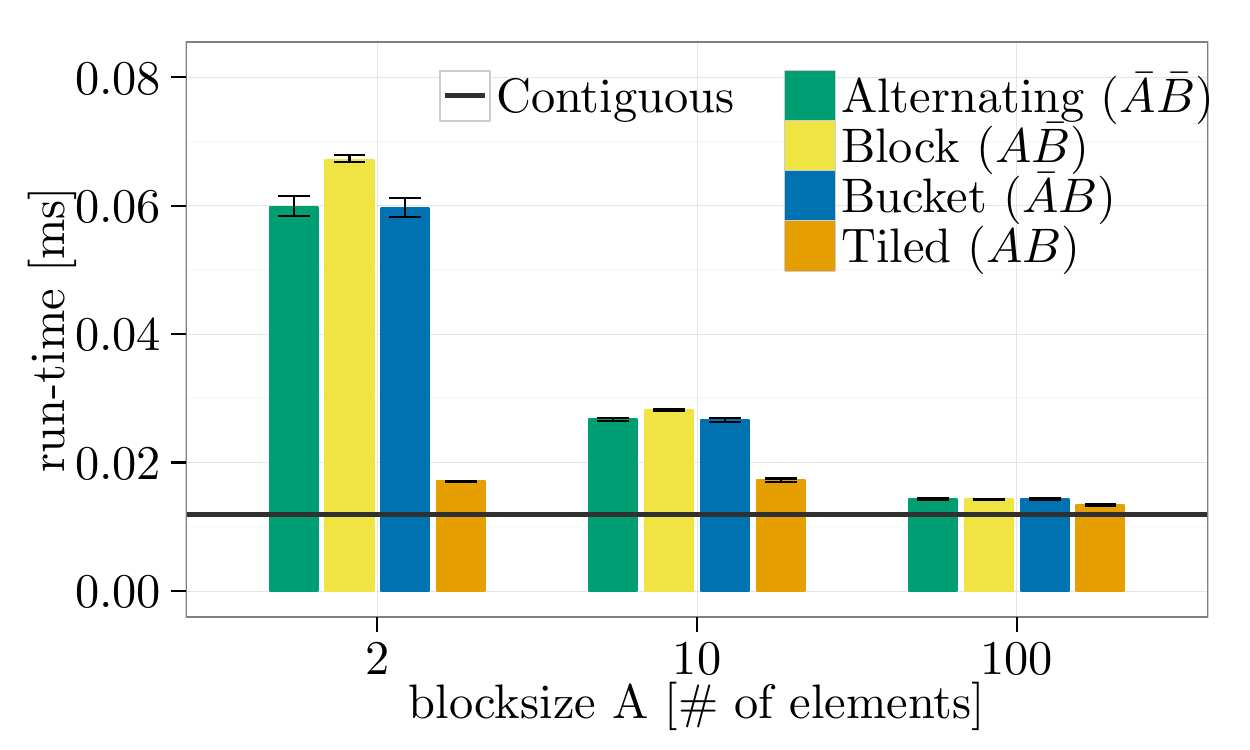}
\caption{%
\label{exp:pingpong-nsmall-mvapich-vartwo}%
\pingpong%
}%
\end{subfigure}%
\caption{\label{exp:layouts-small-32p-mvapich-vartwo}  Contiguous \vs typed, $\VARdatasize=\SI{3.2}{\kilo\byte}$, element datatype: \mpiint, \num{32x1}~processes (\num{2x1} for \pingpong), \jupitermvapich, \varianttwo.}
\end{figure*}

\begin{figure*}[htpb]
\centering
\begin{subfigure}{.33\linewidth}
\centering
\includegraphics[width=\linewidth]{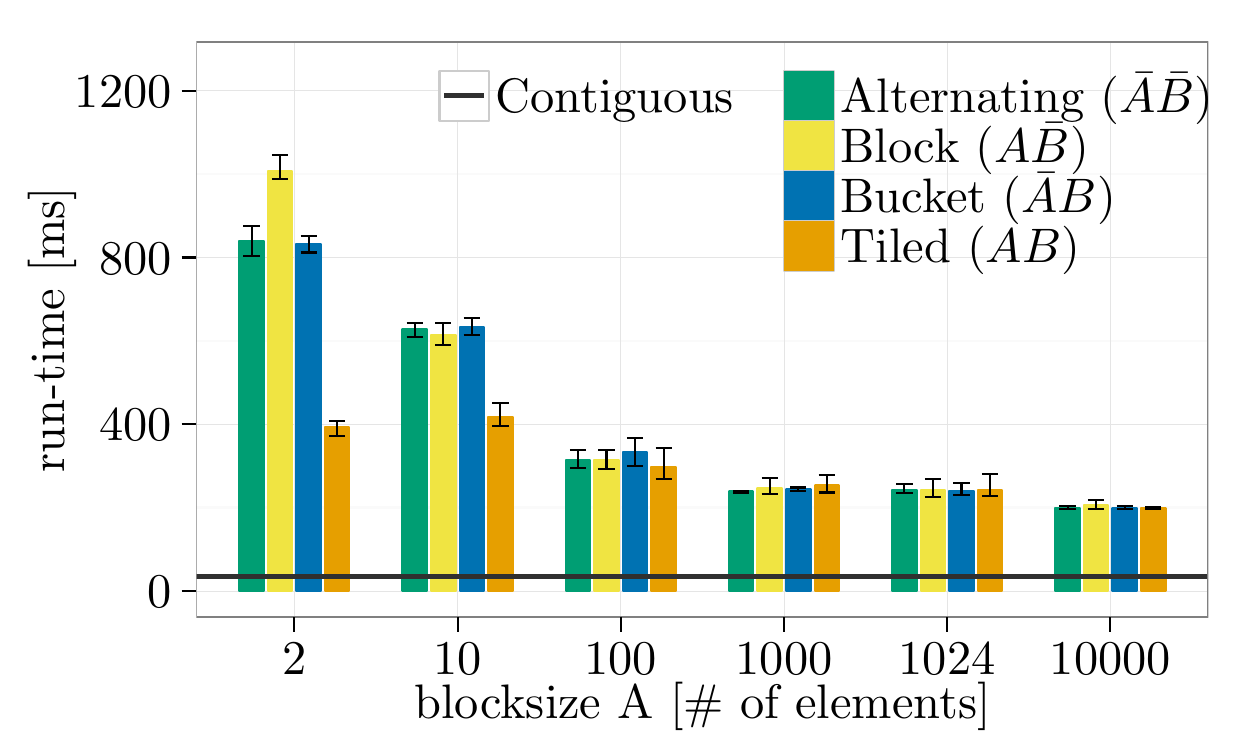}
\caption{%
\label{exp:allgather-nlarge-mvapich-vartwo}%
\mpiallgather%
}%
\end{subfigure}%
\hfill%
\begin{subfigure}{.33\linewidth}
\centering
\includegraphics[width=\linewidth]{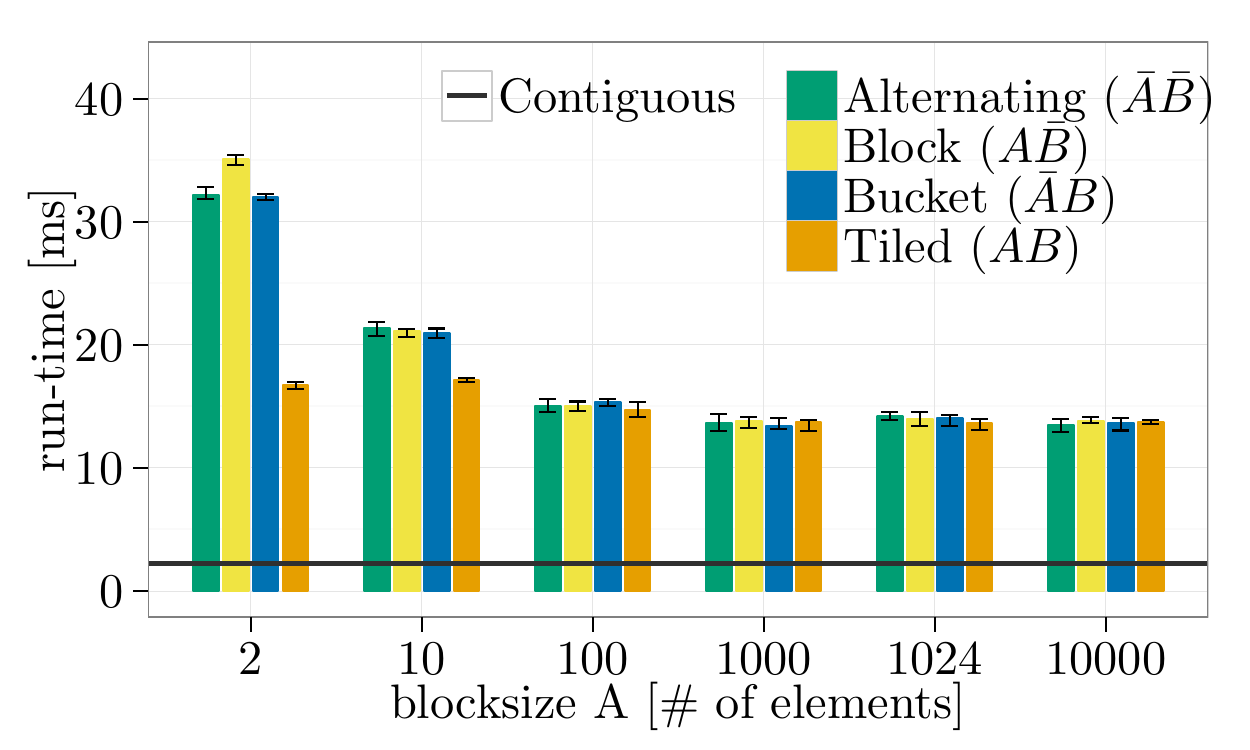}
\caption{%
\label{exp:bcast-nlarge-mvapich-vartwo}%
\mpibcast%
}%
\end{subfigure}%
\hfill%
\begin{subfigure}{.33\linewidth}
\centering
\includegraphics[width=\linewidth]{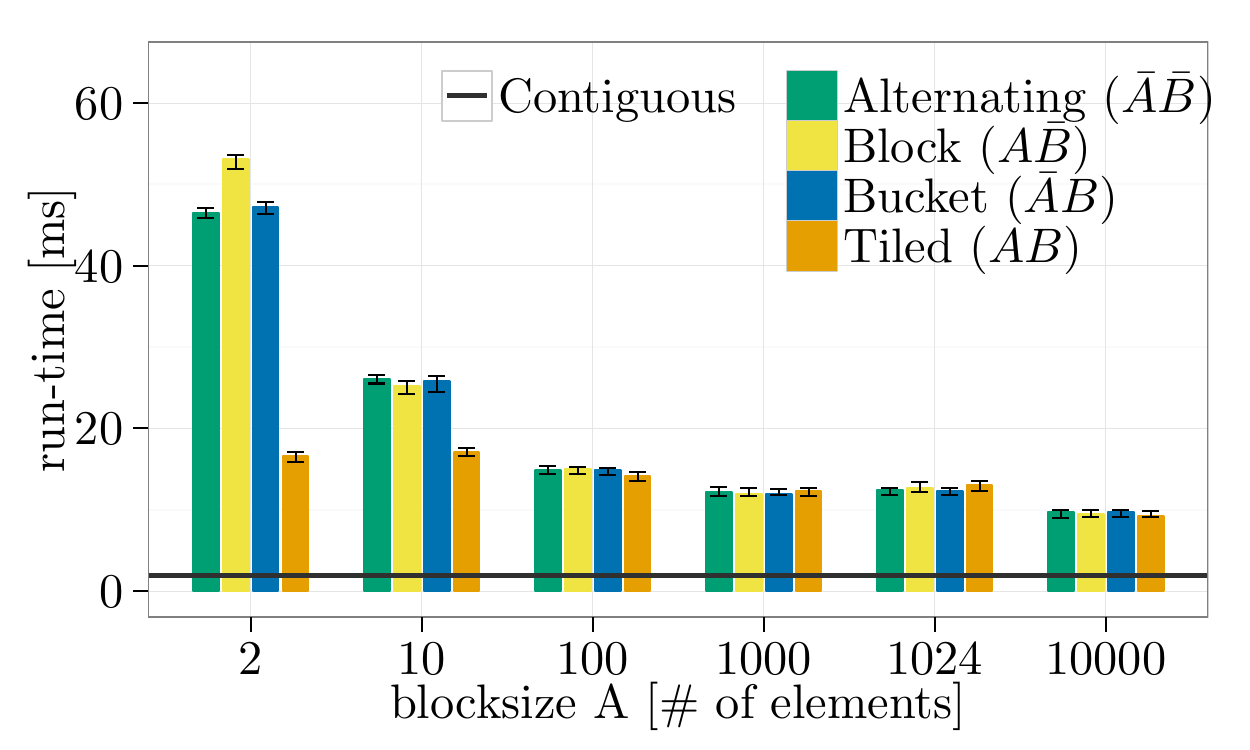}
\caption{%
\label{exp:pingpong-nlarge-mvapich-vartwo}%
\pingpong%
}%
\end{subfigure}%
\caption{\label{exp:layouts-large-32p-mvapich-vartwo} Contiguous \vs typed, $\VARdatasize=\SI{2.56}{\mega\byte}$, element datatype: \mpiint, \num{32x1}~processes (\num{2x1} for \pingpong), \jupitermvapich, \varianttwo.}
\end{figure*}

\begin{figure*}[htpb]
\centering
\begin{subfigure}{.33\linewidth}
\centering
\includegraphics[width=\linewidth]{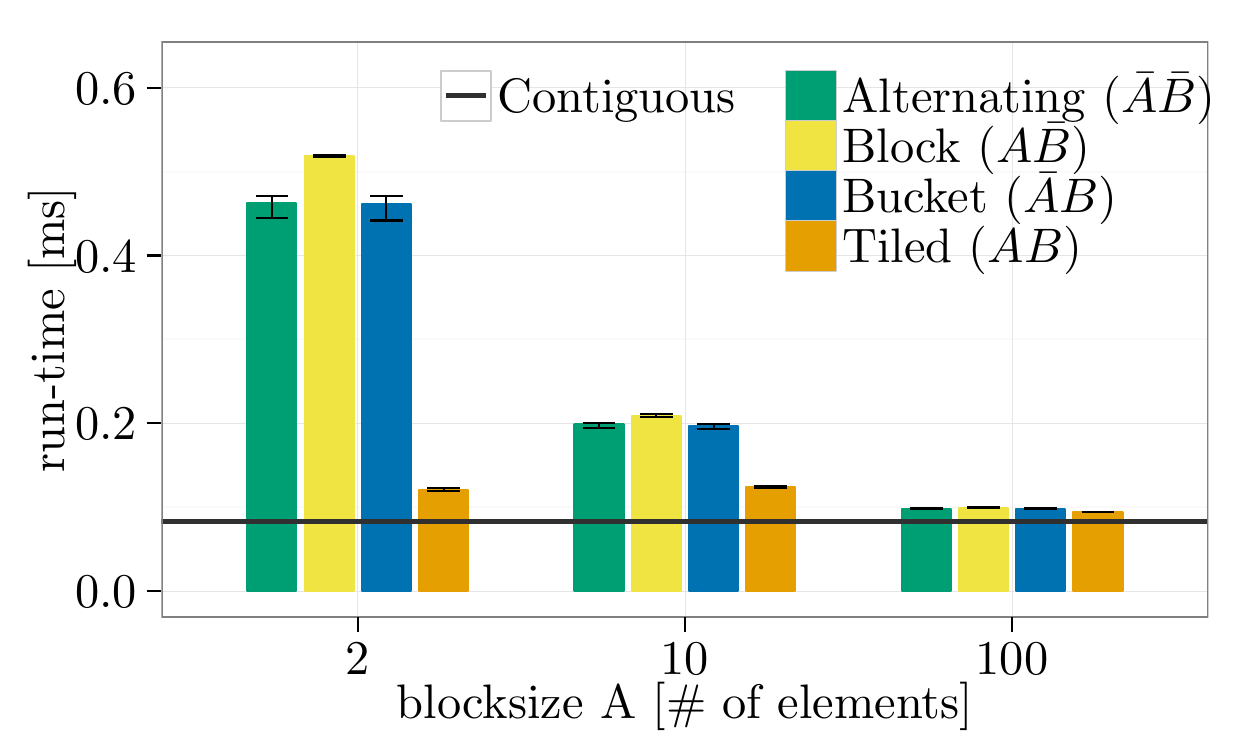}
\caption{%
\label{exp:allgather-nsmall-onenode-mvapich-vartwo}%
\mpiallgather%
}%
\end{subfigure}%
\hfill%
\begin{subfigure}{.33\linewidth}
\centering
\includegraphics[width=\linewidth]{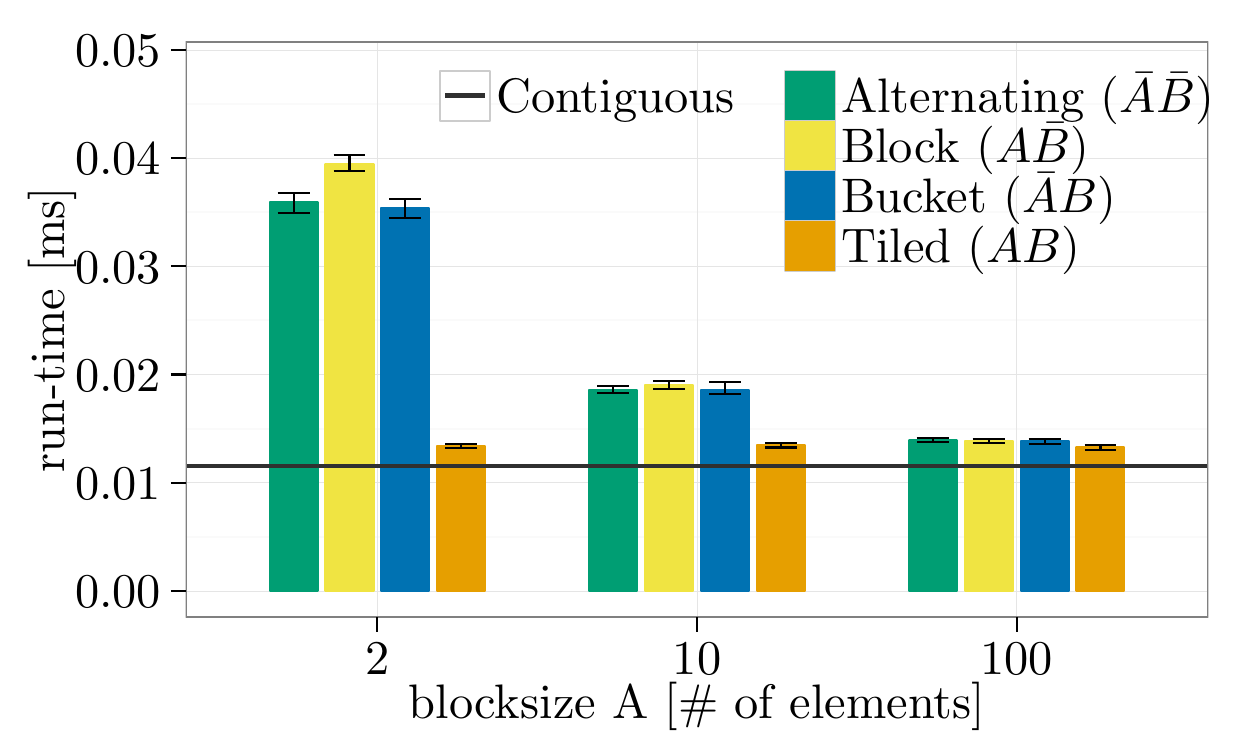}
\caption{%
\label{exp:bcast-nsmall-onenode-mvapich-vartwo}%
\mpibcast%
}%
\end{subfigure}%
\hfill%
\begin{subfigure}{.33\linewidth}
\centering
\includegraphics[width=\linewidth]{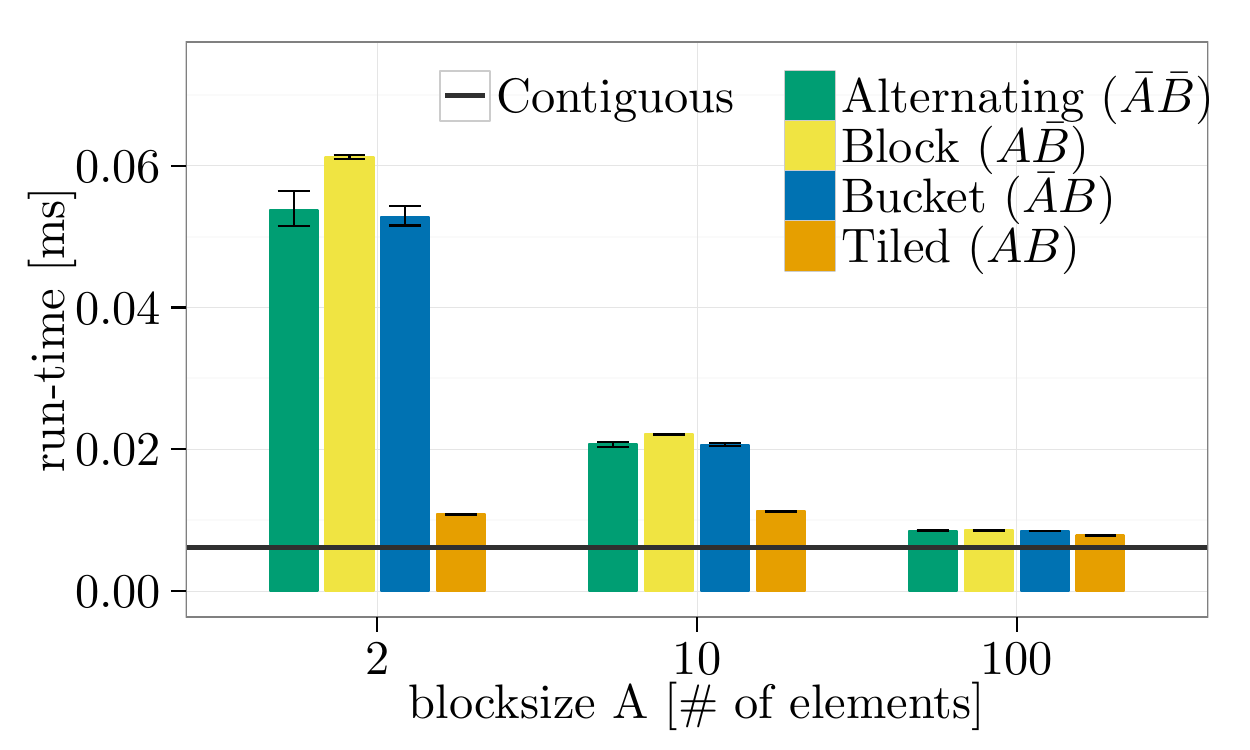}
\caption{%
\label{exp:pingpong-nsmall-onenode-mvapich-vartwo}%
\pingpong%
}%
\end{subfigure}%
\caption{\label{exp:layouts-small-onenode-mvapich-vartwo}  Contiguous \vs typed, $\VARdatasize=\SI{3.2}{\kilo\byte}$, element datatype: \mpiint, one node, \num{16}~processes (\num{2} for \pingpong), \jupitermvapich, \varianttwo.}
\end{figure*}

\begin{figure*}[htpb]
\centering
\begin{subfigure}{.33\linewidth}
\centering
\includegraphics[width=\linewidth]{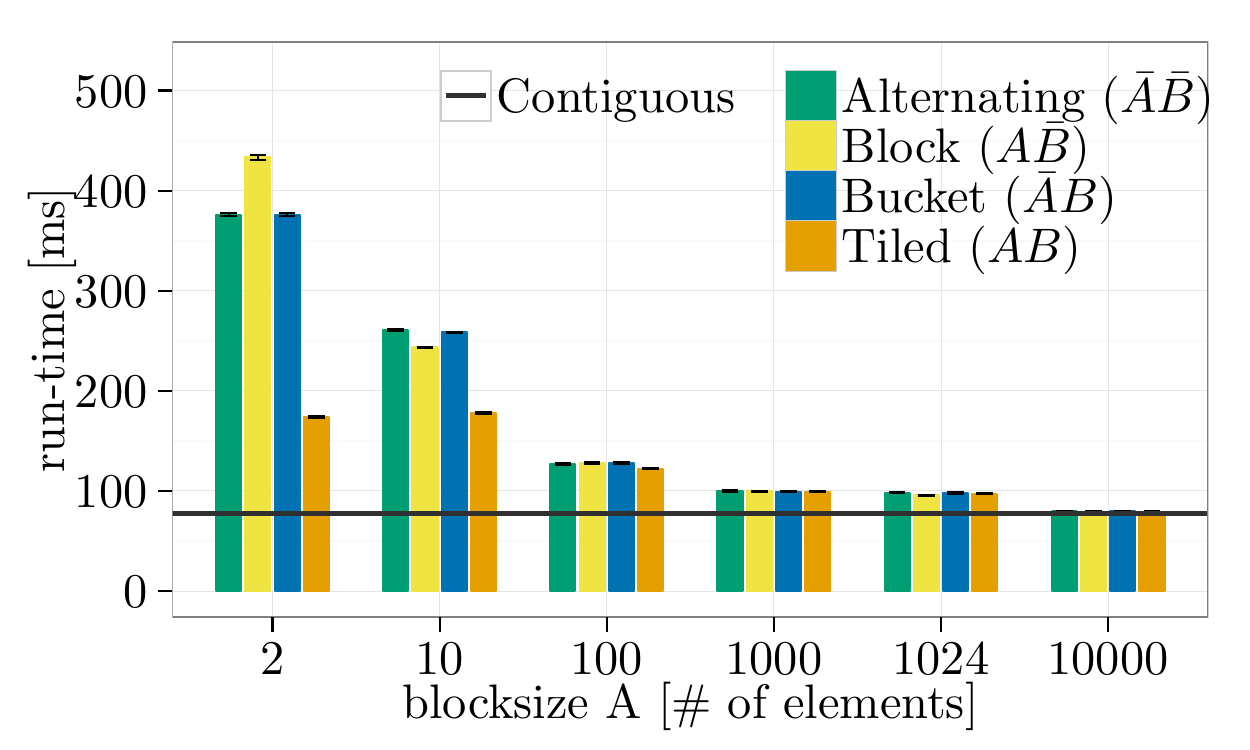}
\caption{%
\label{exp:allgather-nlarge-onenode-mvapich-vartwo}%
\mpiallgather%
}%
\end{subfigure}%
\hfill%
\begin{subfigure}{.33\linewidth}
\centering
\includegraphics[width=\linewidth]{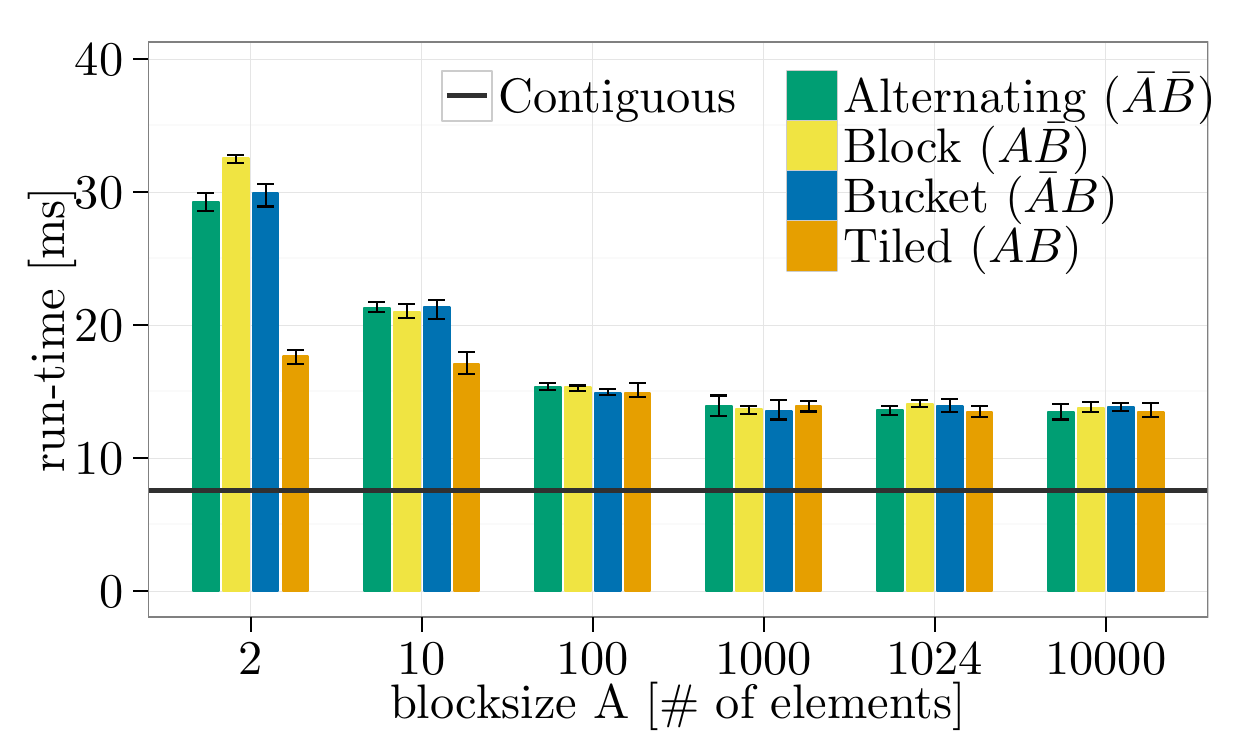}
\caption{%
\label{exp:bcast-nlarge-onenode-mvapich-vartwo}%
\mpibcast%
}%
\end{subfigure}%
\hfill%
\begin{subfigure}{.33\linewidth}
\centering
\includegraphics[width=\linewidth]{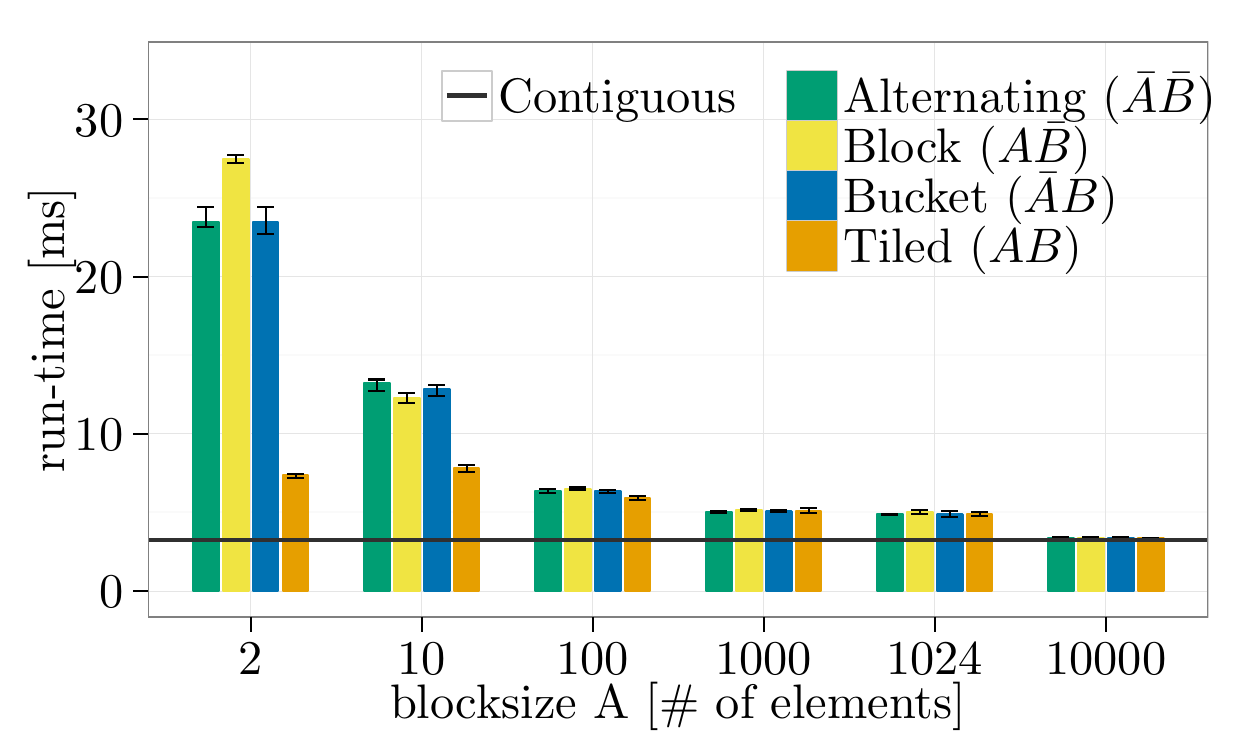}
\caption{%
\label{exp:pingpong-nlarge-onenode-mvapich-vartwo}%
\pingpong%
}%
\end{subfigure}%
\caption{\label{exp:layouts-large-onenode-mvapich-vartwo} Contiguous \vs typed, $\VARdatasize=\SI{2.56}{\mega\byte}$, element datatype: \mpiint, one node, \num{16}~processes (\num{2} for \pingpong), \jupitermvapich, \varianttwo.}
\end{figure*}

\appexpdesc{
  \begin{expitemize}
    \item \dtcontig, \dtdtiled, \dtdblock, \dtdbucket, \dtdalternating
    \item \pingpong, \mpibcast, \mpiallgather
  \end{expitemize}
}{
  \begin{expitemize}
    \item \expparam{\jupiternecmpi, small \datasize, \variantone}{\fig~\ref{exp:layouts-nsmall-32p-nec}}
    \item \expparam{\jupitermvapich, small \datasize, \variantone}{\fig~\ref{exp:layouts-nsmall-32p-mvapich}}
    \item \expparam{\jupiteropenmpi, small \datasize, \variantone}{\fig~\ref{exp:layouts-nsmall-32p-openmpi}}
    \item \expparam{\jupiternecmpi, large \datasize, \variantone}{\fig~\ref{exp:layouts-nlarge-32p-nec}}
    \item \expparam{\jupitermvapich, large \datasize, \variantone}{\fig~\ref{exp:layouts-nlarge-32p-mvapich}}
    \item \expparam{\jupiteropenmpi, large \datasize, \variantone, one node}{\fig~\ref{exp:layouts-nlarge-32p-openmpi}}
    \item \expparam{\jupiternecmpi, small \datasize, \variantone, one node}{\fig~\ref{exp:layouts-nsmall-onenode-nec}}
    \item \expparam{\jupitermvapich, small \datasize, \variantone, one node}{\fig~\ref{exp:layouts-nsmall-onenode-mvapich}}
    \item \expparam{\jupiteropenmpi, small \datasize, \variantone, one node}{\fig~\ref{exp:layouts-nsmall-onenode-openmpi}}
    \item \expparam{\jupiternecmpi, large \datasize, \variantone, one node}{\fig~\ref{exp:layouts-nlarge-onenode-nec}}
    \item \expparam{\jupitermvapich, large \datasize, \variantone, one node}{\fig~\ref{exp:layouts-nlarge-onenode-mvapich}}
    \item \expparam{\jupiteropenmpi, large \datasize, \variantone, one node}{\fig~\ref{exp:layouts-nlarge-onenode-openmpi}}
    \item \expparam{\jupitermvapich, small \datasize, \varianttwo}{\fig~\ref{exp:layouts-small-32p-mvapich-vartwo}}
    \item \expparam{\jupitermvapich, large \datasize, \varianttwo}{\fig~\ref{exp:layouts-large-32p-mvapich-vartwo}}
    \item \expparam{\jupitermvapich, small \datasize, \varianttwo, one node}{\fig~\ref{exp:layouts-small-onenode-mvapich-vartwo}}
    \item \expparam{\jupitermvapich, large \datasize, \varianttwo, one node}{\fig~\ref{exp:layouts-large-onenode-mvapich-vartwo}}
  \end{expitemize}  
}

\FloatBarrier

\appexp{exptest:tiled_het}

\begin{figure*}[htpb]
\centering
\begin{subfigure}{.24\linewidth}
\centering
\includegraphics[width=\linewidth]{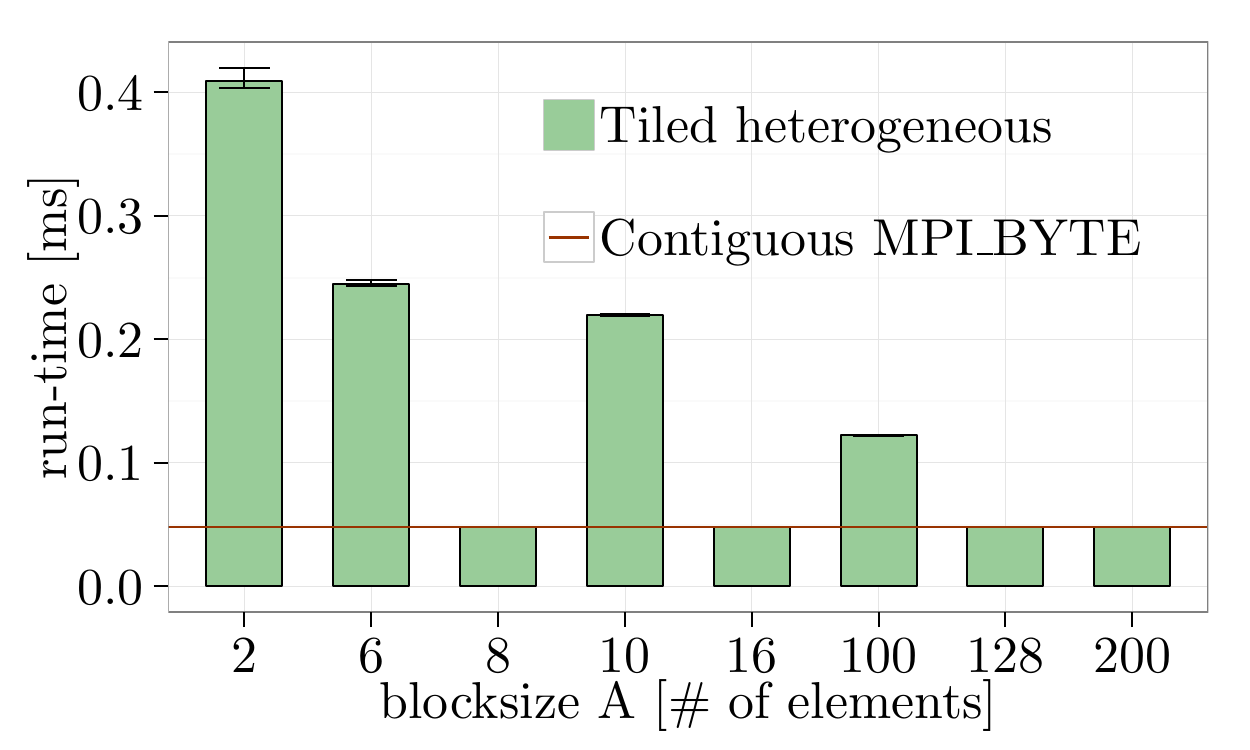}
\caption{%
\label{exp:pingpong-heterog-small-2x1-nec}%
$\VARdatasize = \SI{48}{\kilo\byte}$, \num{2}~nodes%
}%
\end{subfigure}%
\hfill%
\begin{subfigure}{.24\linewidth}
\centering
\includegraphics[width=\linewidth]{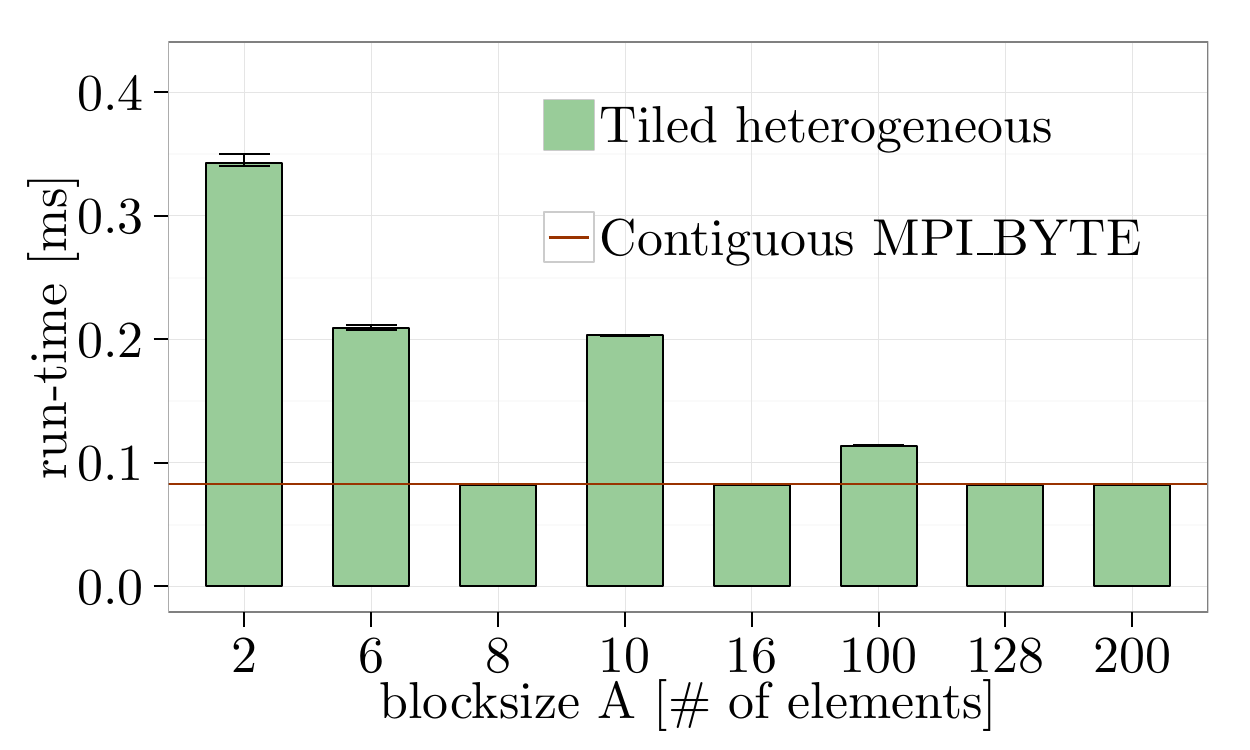}
\caption{%
\label{exp:pingpong-heterog-small-1x2-nec}%
$\VARdatasize = \SI{48}{\kilo\byte}$, same node%
}%
\end{subfigure}%
\hfill%
\begin{subfigure}{.24\linewidth}
\centering
\includegraphics[width=\linewidth]{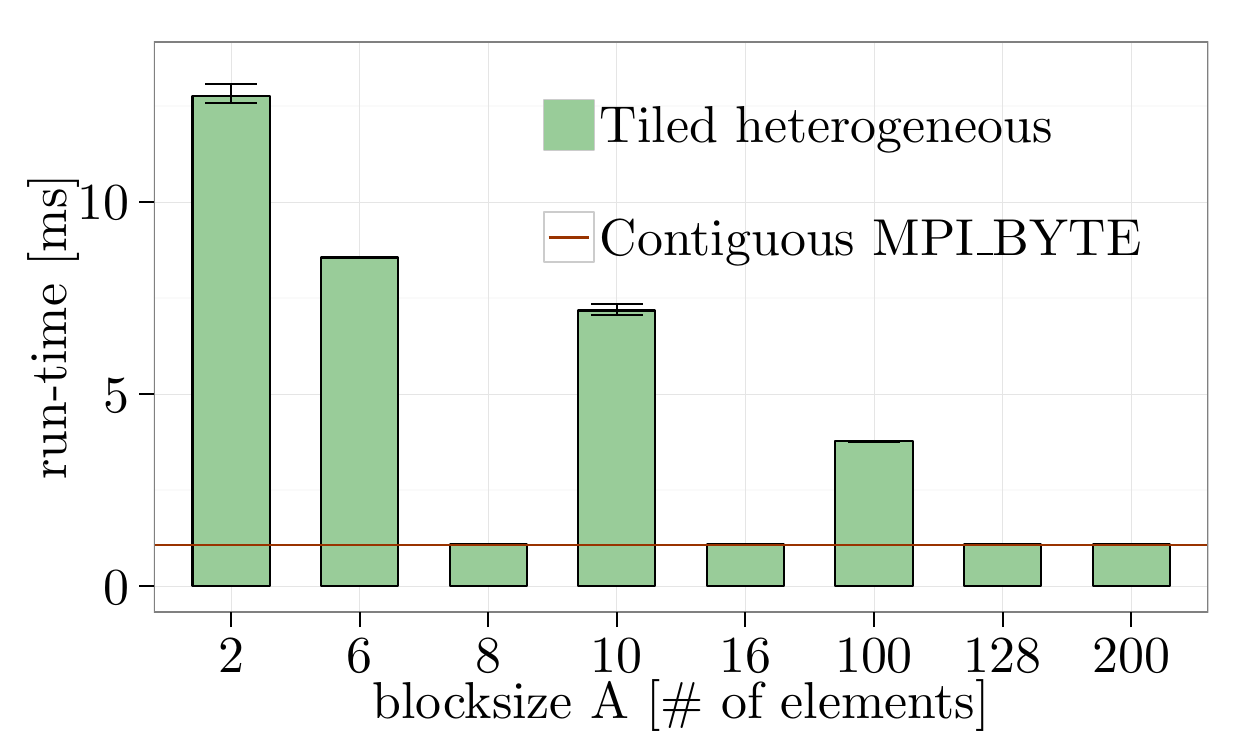}
\caption{%
\label{exp:pingpong-heterog-large-2x1-nec}%
$\VARdatasize = \SI{1.5}{\mega\byte}$, \num{2}~nodes%
}%
\end{subfigure}%
\hfill%
\begin{subfigure}{.24\linewidth}
\centering
\includegraphics[width=\linewidth]{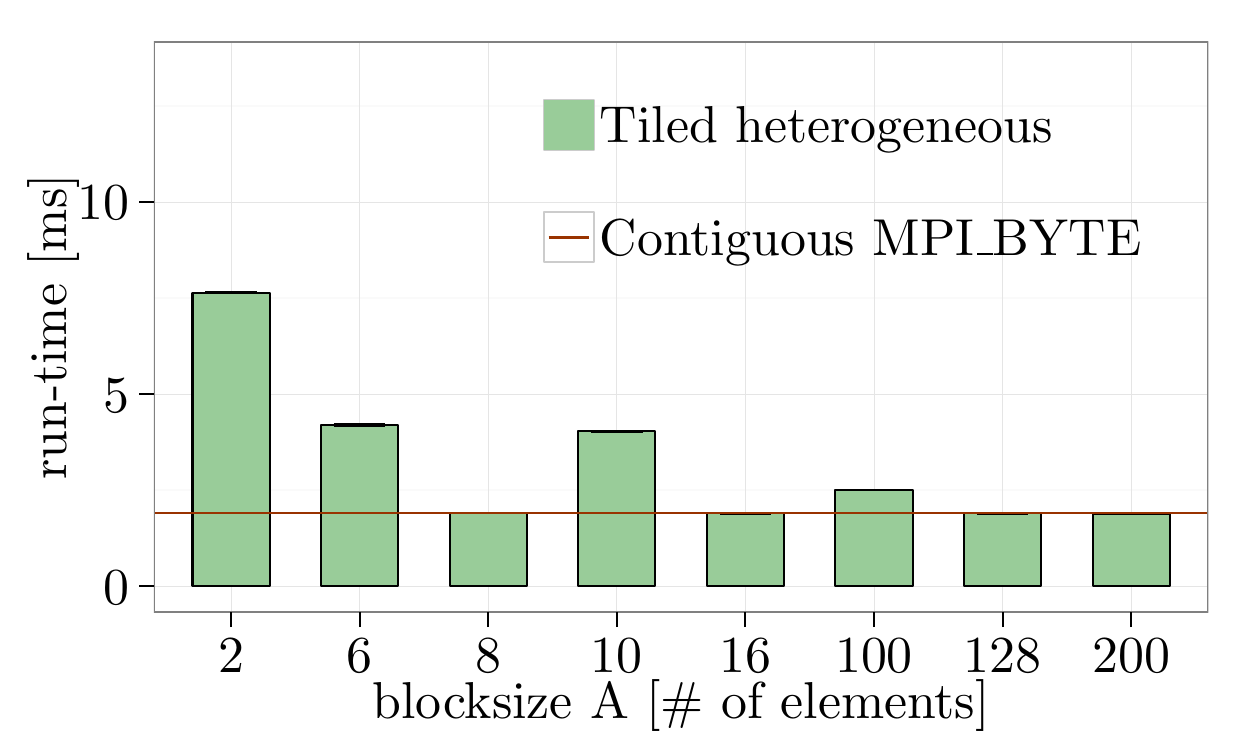}
\caption{%
\label{exp:pingpong-heterog-large-1x2-nec}%
$\VARdatasize = \SI{1.5}{\mega\byte}$, same node%
}%
\end{subfigure}%
\caption{\label{exp:pingpong-heterog-nec} \dtcontig \vs \dtdtiledhet $A$=$B$, element datatype: \mpiint, \pingpong, \jupiternecmpi.}
\end{figure*}

\begin{figure*}[htpb]
\centering
\begin{subfigure}{.24\linewidth}
\centering
\includegraphics[width=\linewidth]{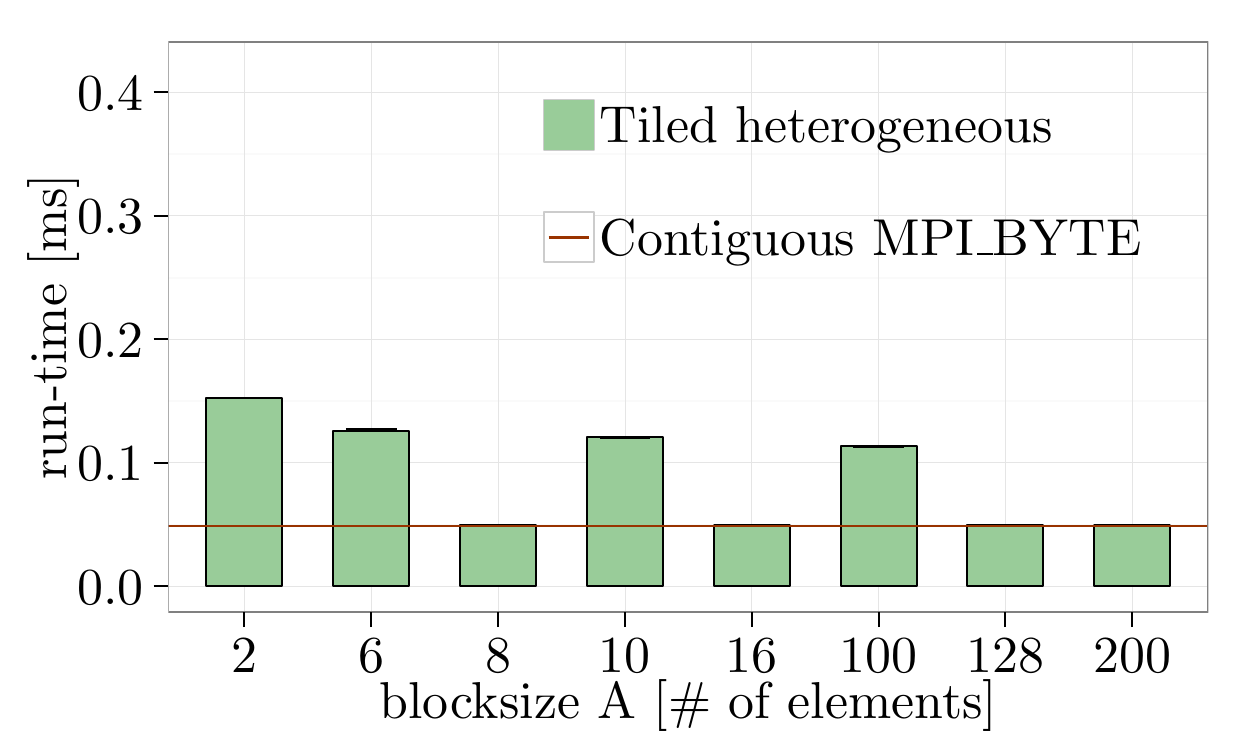}
\caption{%
\label{exp:pingpong-heterog-small-2x1-mvapich}%
$\VARdatasize = \SI{48}{\kilo\byte}$, \num{2}~nodes%
}%
\end{subfigure}%
\hfill%
\begin{subfigure}{.24\linewidth}
\centering
\includegraphics[width=\linewidth]{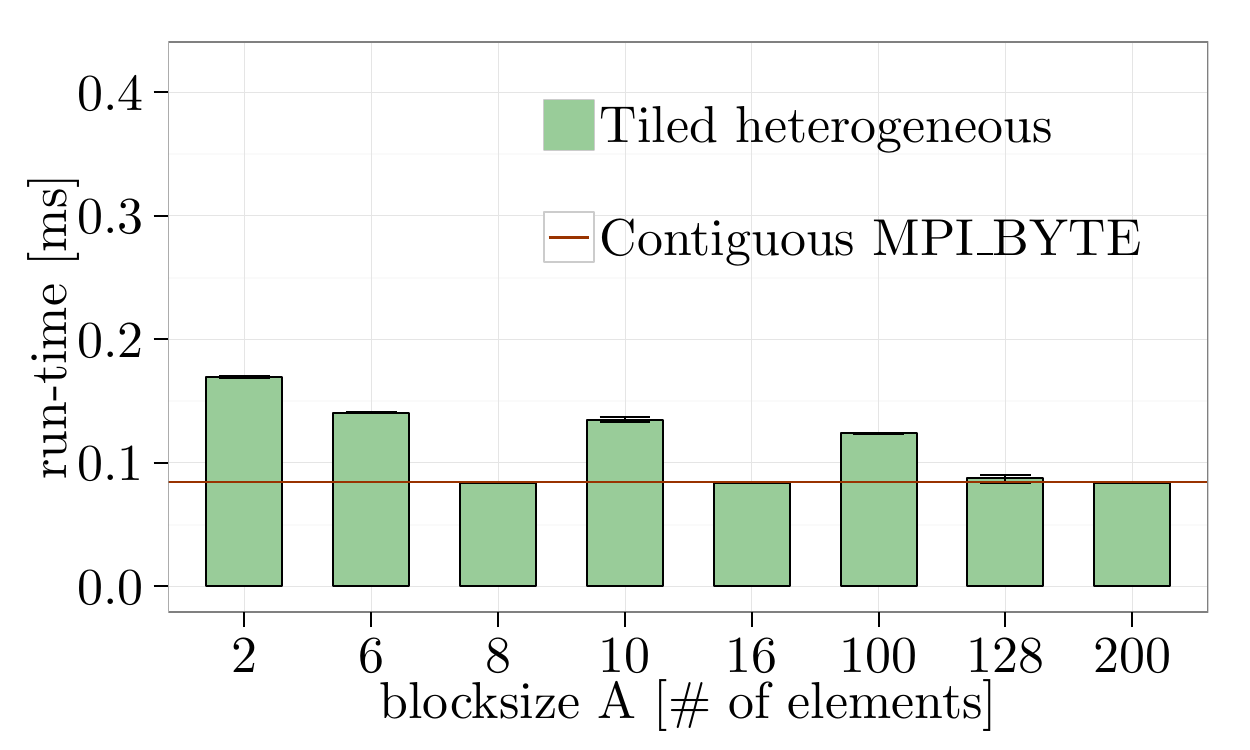}
\caption{%
\label{exp:pingpong-heterog-small-1x2-mvapich}%
$\VARdatasize = \SI{48}{\kilo\byte}$, same node%
}%
\end{subfigure}%
\hfill%
\begin{subfigure}{.24\linewidth}
\centering
\includegraphics[width=\linewidth]{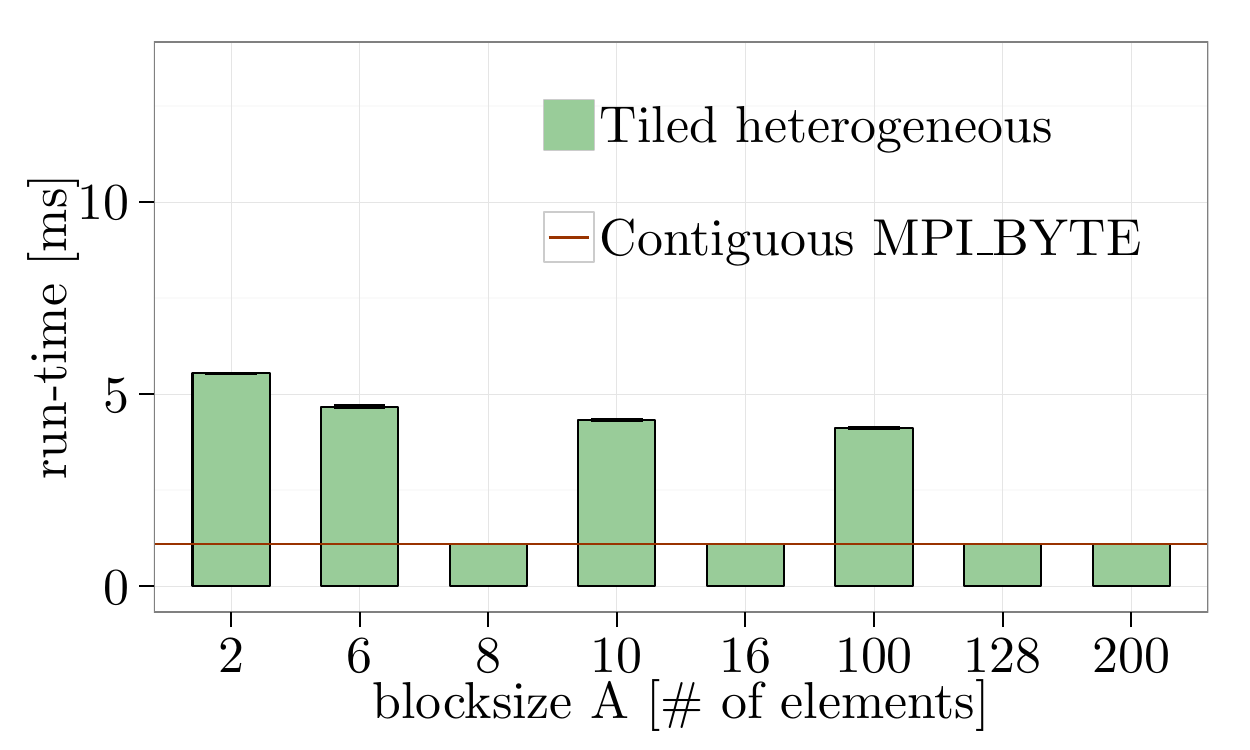}
\caption{%
\label{exp:pingpong-heterog-large-2x1-mvapich}%
$\VARdatasize = \SI{1.5}{\mega\byte}$, \num{2}~nodes%
}%
\end{subfigure}%
\hfill%
\begin{subfigure}{.24\linewidth}
\centering
\includegraphics[width=\linewidth]{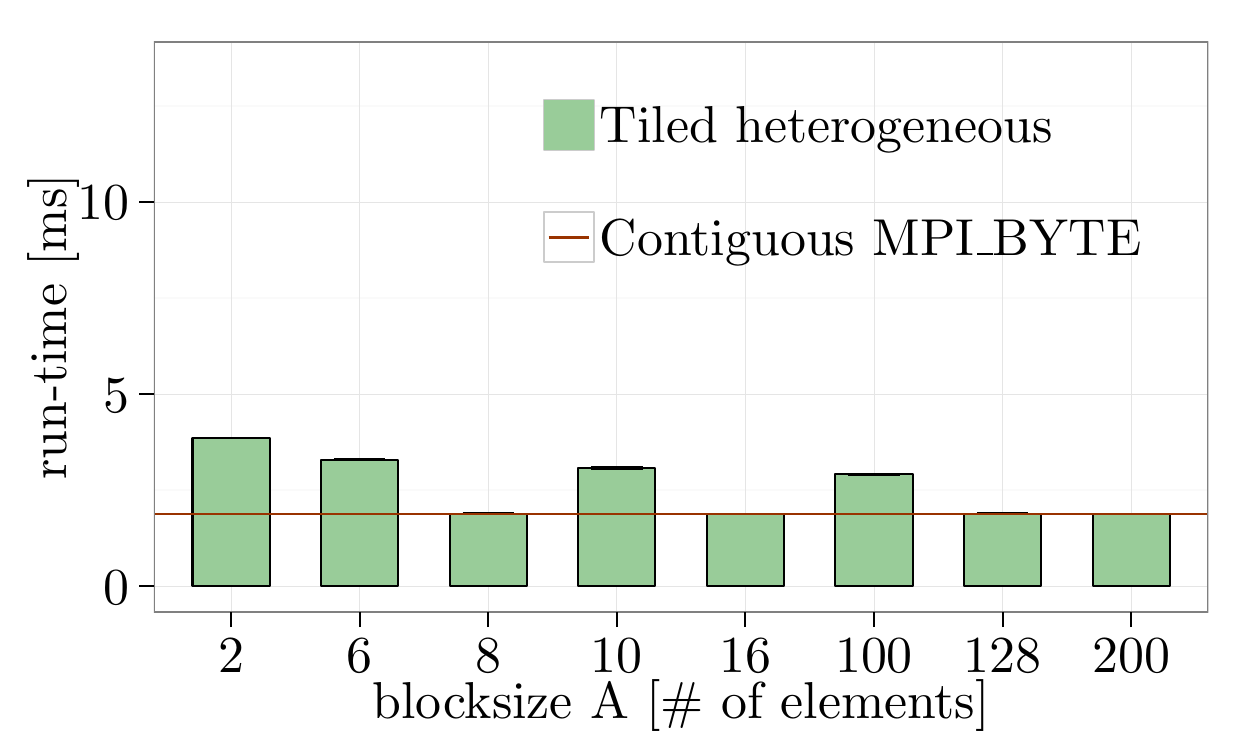}
\caption{%
\label{exp:pingpong-heterog-large-1x2-mvapich}%
$\VARdatasize = \SI{1.5}{\mega\byte}$, same node%
}%
\end{subfigure}%
\caption{\label{exp:pingpong-heterog-mvapich}  \dtcontig \vs \dtdtiledhet $A$=$B$, element datatype: \mpiint, \pingpong, \jupitermvapich.}
\end{figure*}

\begin{figure*}[htpb]
\centering
\begin{subfigure}{.24\linewidth}
\centering
\includegraphics[width=\linewidth]{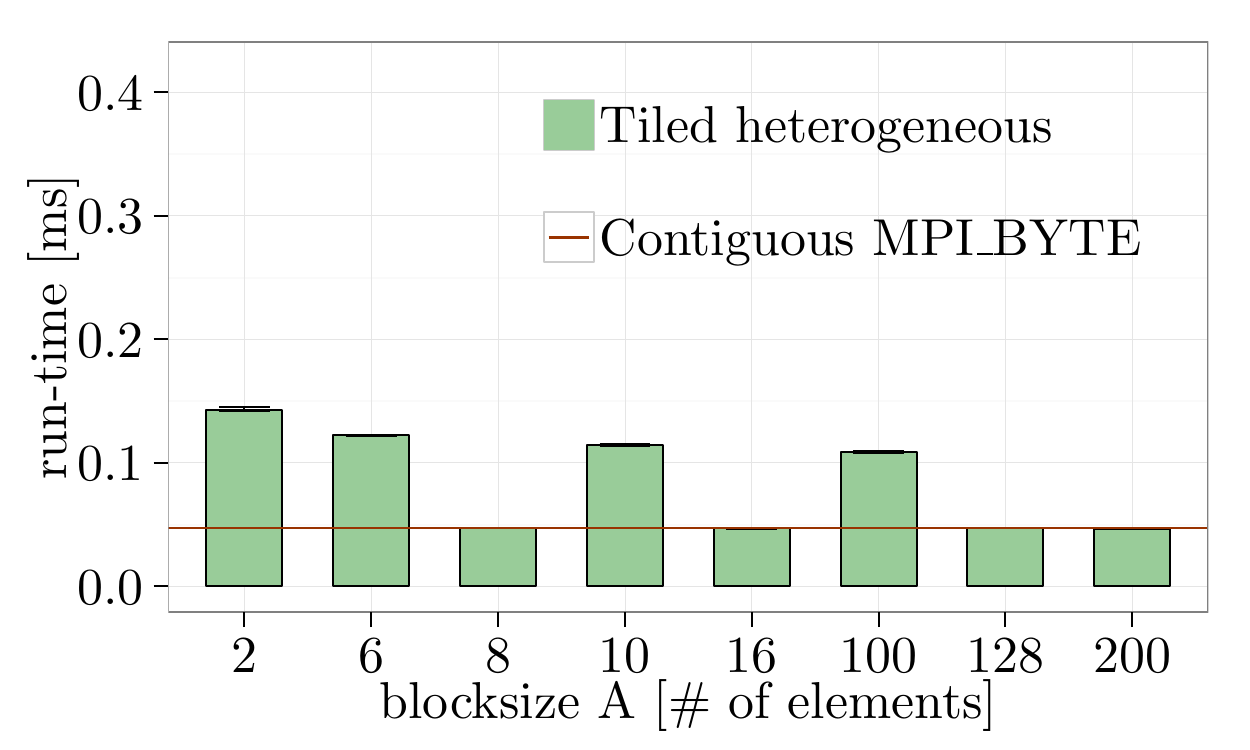}
\caption{%
\label{exp:pingpong-heterog-small-2x1-openmpi}%
$\VARdatasize = \SI{48}{\kilo\byte}$, \num{2}~nodes%
}%
\end{subfigure}%
\hfill%
\begin{subfigure}{.24\linewidth}
\centering
\includegraphics[width=\linewidth]{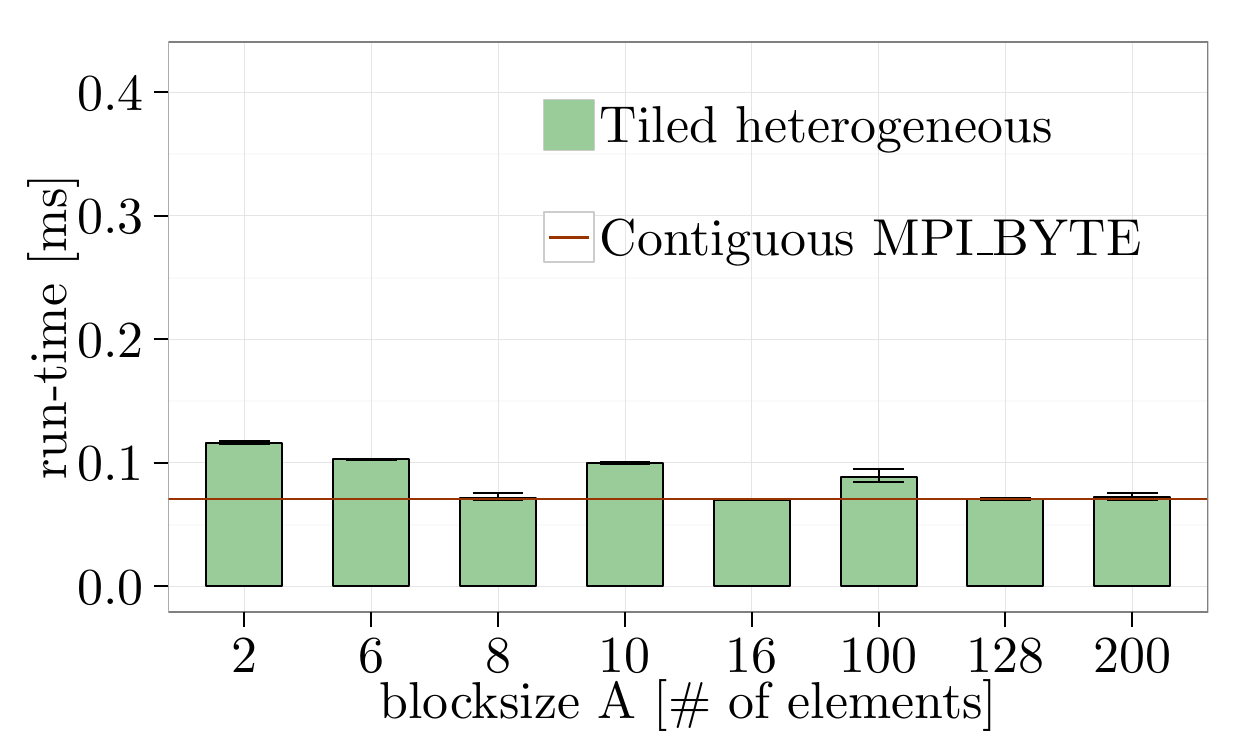}
\caption{%
\label{exp:pingpong-heterog-small-1x2-openmpi}%
$\VARdatasize = \SI{48}{\kilo\byte}$, same node%
}%
\end{subfigure}%
\hfill%
\begin{subfigure}{.24\linewidth}
\centering
\includegraphics[width=\linewidth]{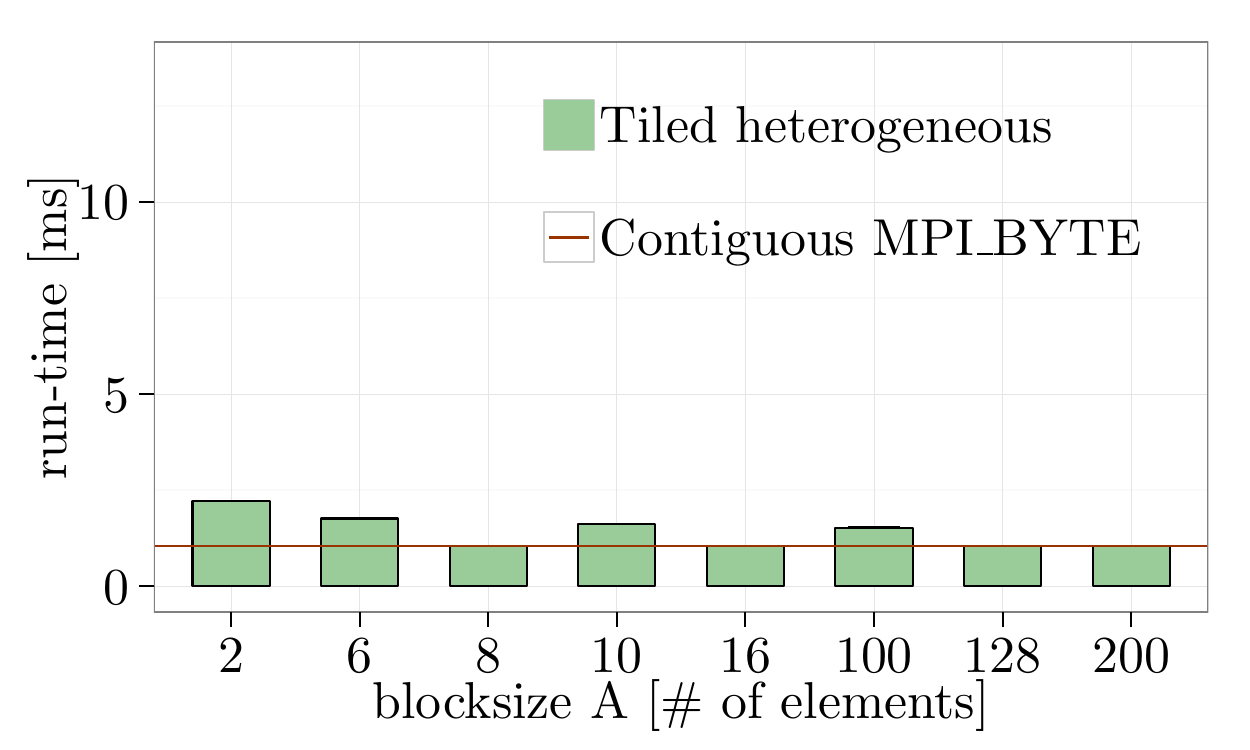}
\caption{%
\label{exp:pingpong-heterog-large-2x1-openmpi}%
$\VARdatasize = \SI{1.5}{\mega\byte}$, \num{2}~nodes%
}%
\end{subfigure}%
\hfill%
\begin{subfigure}{.24\linewidth}
\centering
\includegraphics[width=\linewidth]{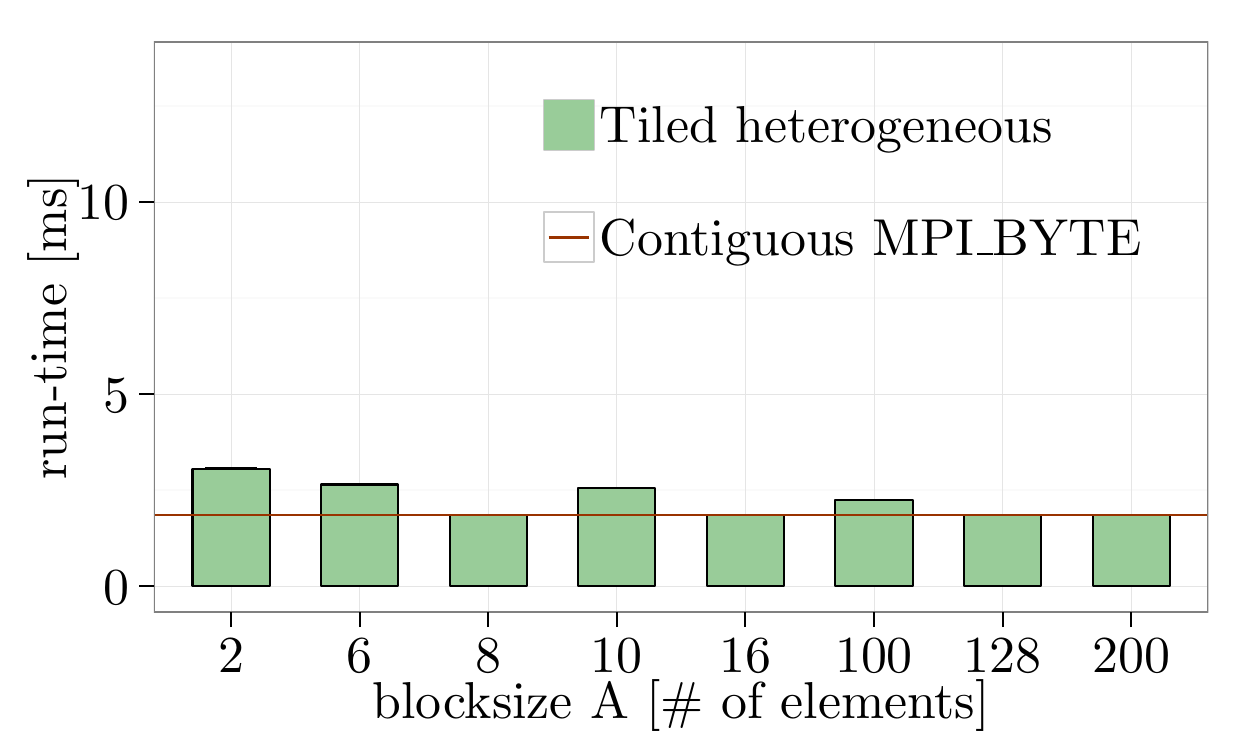}
\caption{%
\label{exp:pingpong-heterog-large-1x2-openmpi}%
$\VARdatasize = \SI{1.5}{\mega\byte}$, same node%
}%
\end{subfigure}%
\caption{\label{exp:pingpong-heterog-openmpi}  \dtcontig \vs \dtdtiledhet $A$=$B$, element datatype: \mpiint, \pingpong, \jupiteropenmpi.}
\end{figure*}

\appexpdesc{
  \begin{expitemize}
    \item \dtcontig, \dtdtiledhet
    \item \pingpong
  \end{expitemize}
}{
  \begin{expitemize}
    \item \expparam{\jupiternecmpi}{\fig~\ref{exp:pingpong-heterog-nec}}
    \item \expparam{\jupitermvapich}{\fig~\ref{exp:pingpong-heterog-mvapich}}
    \item \expparam{\jupiteropenmpi}{\fig~\ref{exp:pingpong-heterog-openmpi}}
  \end{expitemize}  
}

\FloatBarrier

\appexp{exptest:pack_unpack}

\begin{figure*}[htpb]
\centering
\begin{subfigure}{.24\linewidth}
\centering
\includegraphics[width=\linewidth]{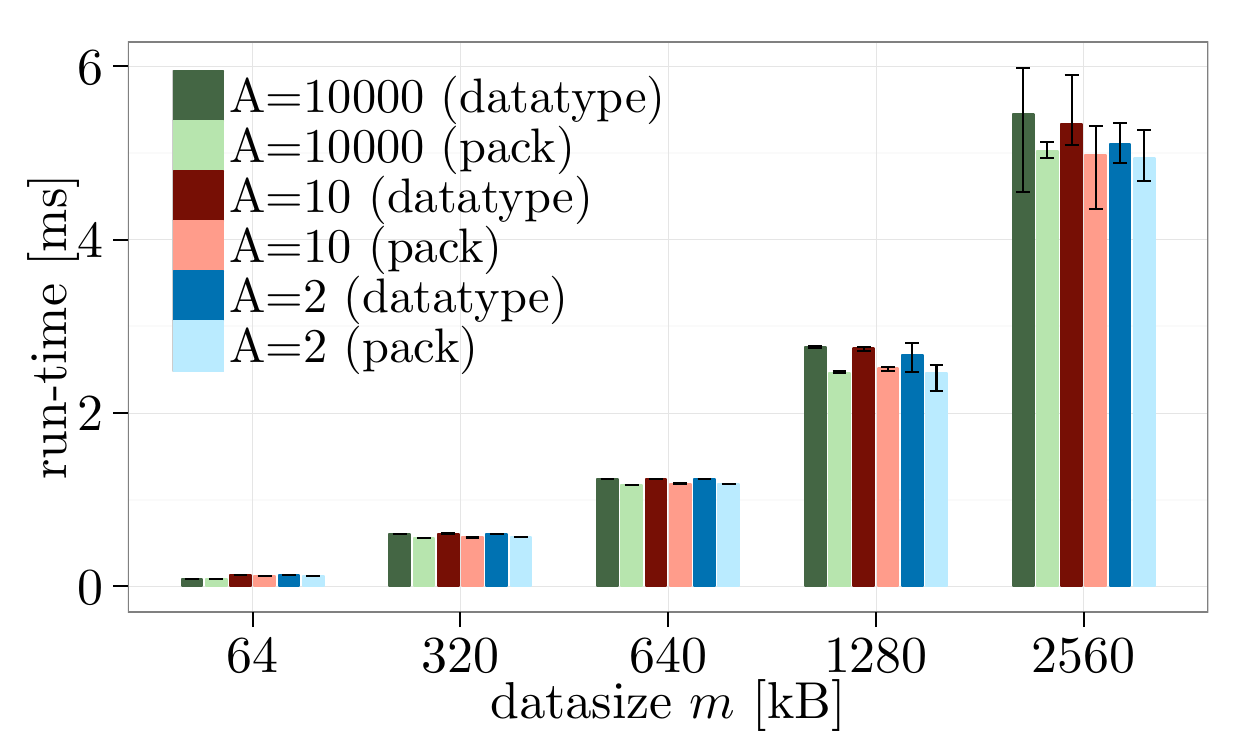}
\caption{%
\label{exp:pingpong-pack-tiled-2x1-nec}%
\dttiled%
}%
\end{subfigure}%
\hfill%
\begin{subfigure}{.24\linewidth}
\centering
\includegraphics[width=\linewidth]{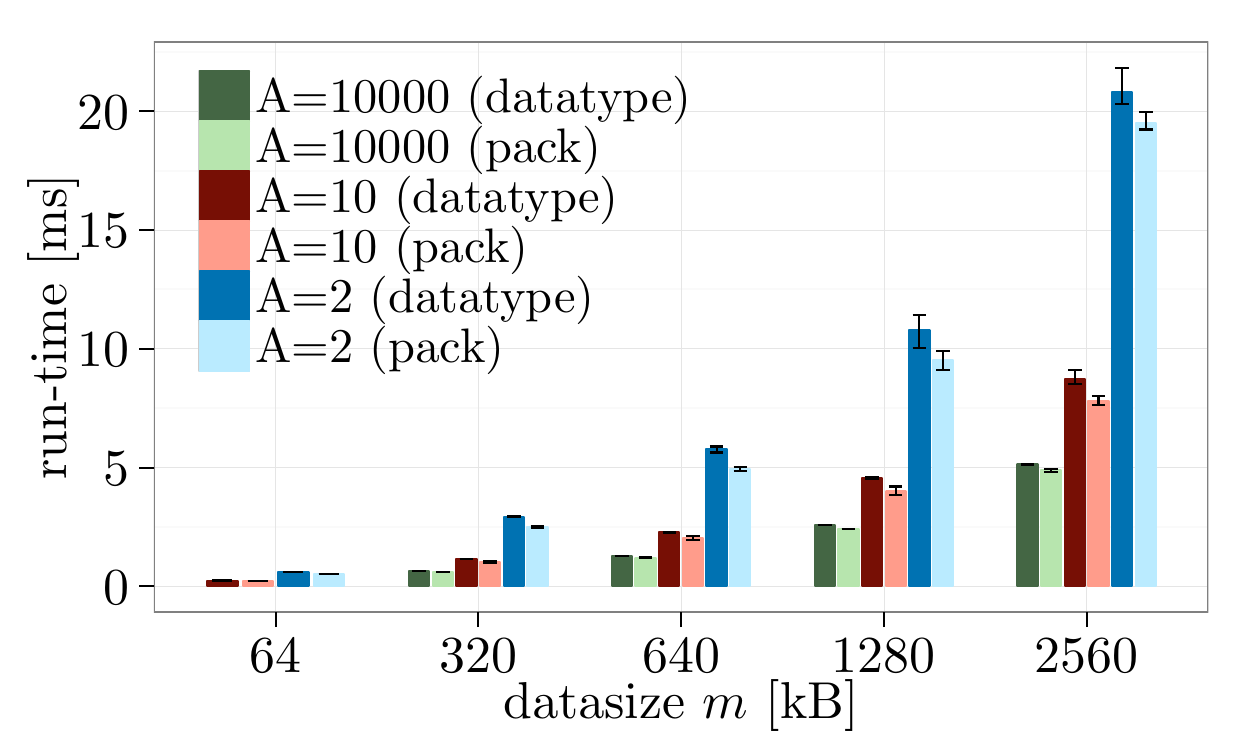}
\caption{%
\label{exp:pingpong-pack-block-2x1-nec}%
\dtblock%
}%
\end{subfigure}%
\hfill%
\begin{subfigure}{.24\linewidth}
\centering
\includegraphics[width=\linewidth]{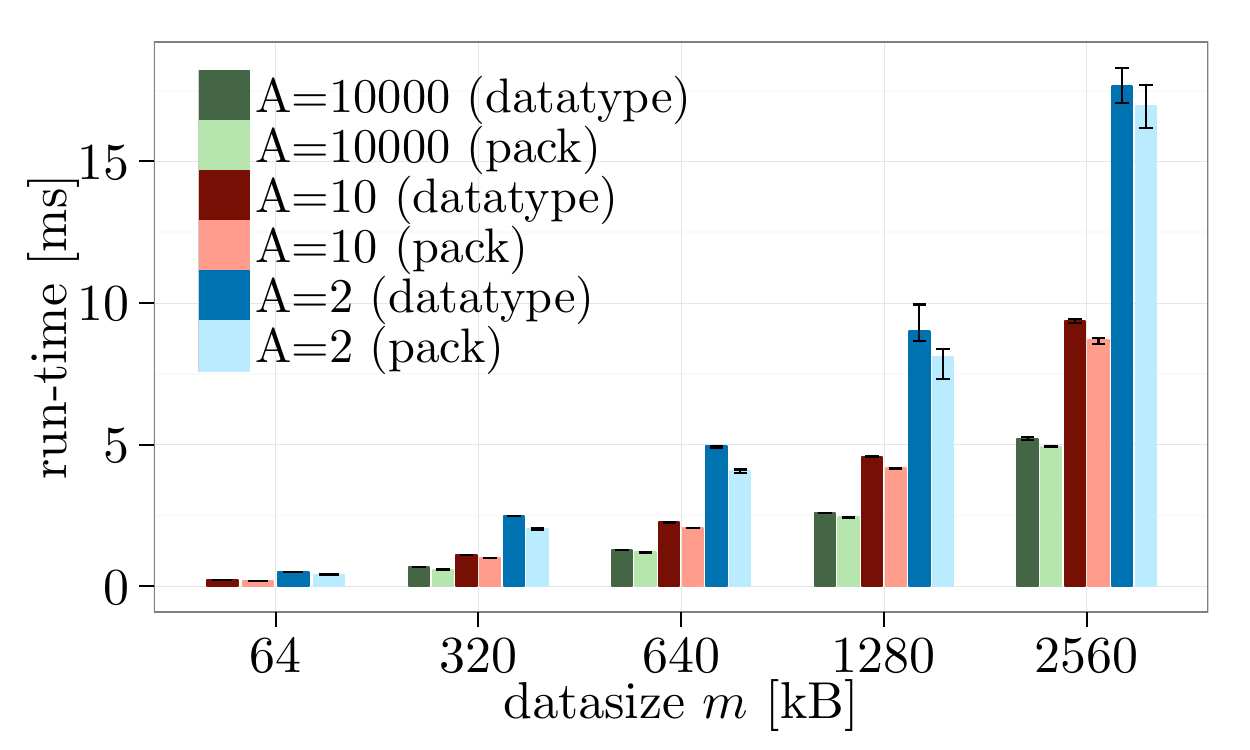}
\caption{%
\label{exp:pingpong-pack-bucket-2x1-nec}%
\dtbucket%
}%
\end{subfigure}%
\hfill%
\begin{subfigure}{.24\linewidth}
\centering
\includegraphics[width=\linewidth]{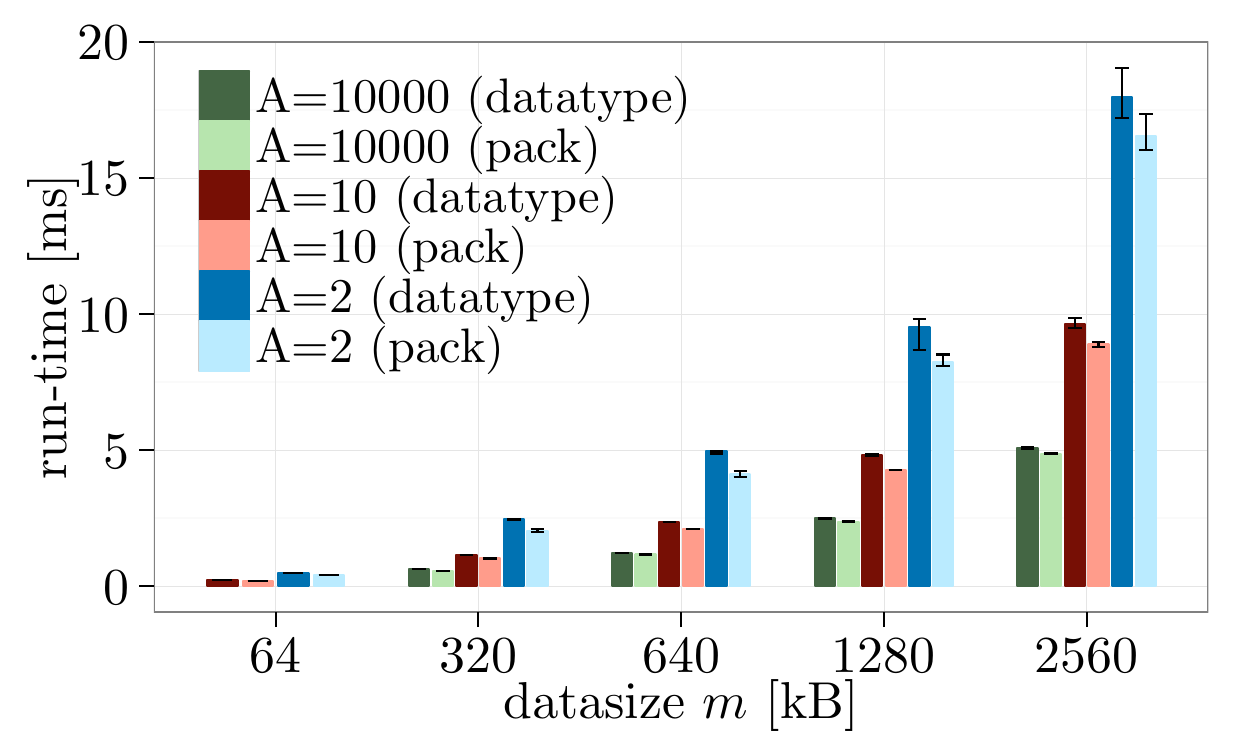}
\caption{%
\label{exp:pingpong-pack-alternating-2x1-nec}%
\dtalternating%
}%
\end{subfigure}%
\caption{\label{exp:pingpong-pack-2x1-nec}  Basic layouts \vs pack/unpack, element datatype: \mpiint, \num{2x1}~processes, \pingpong, \jupiternecmpi.}
\end{figure*}

\begin{figure*}[htpb]
\centering
\begin{subfigure}{.24\linewidth}
\centering
\includegraphics[width=\linewidth]{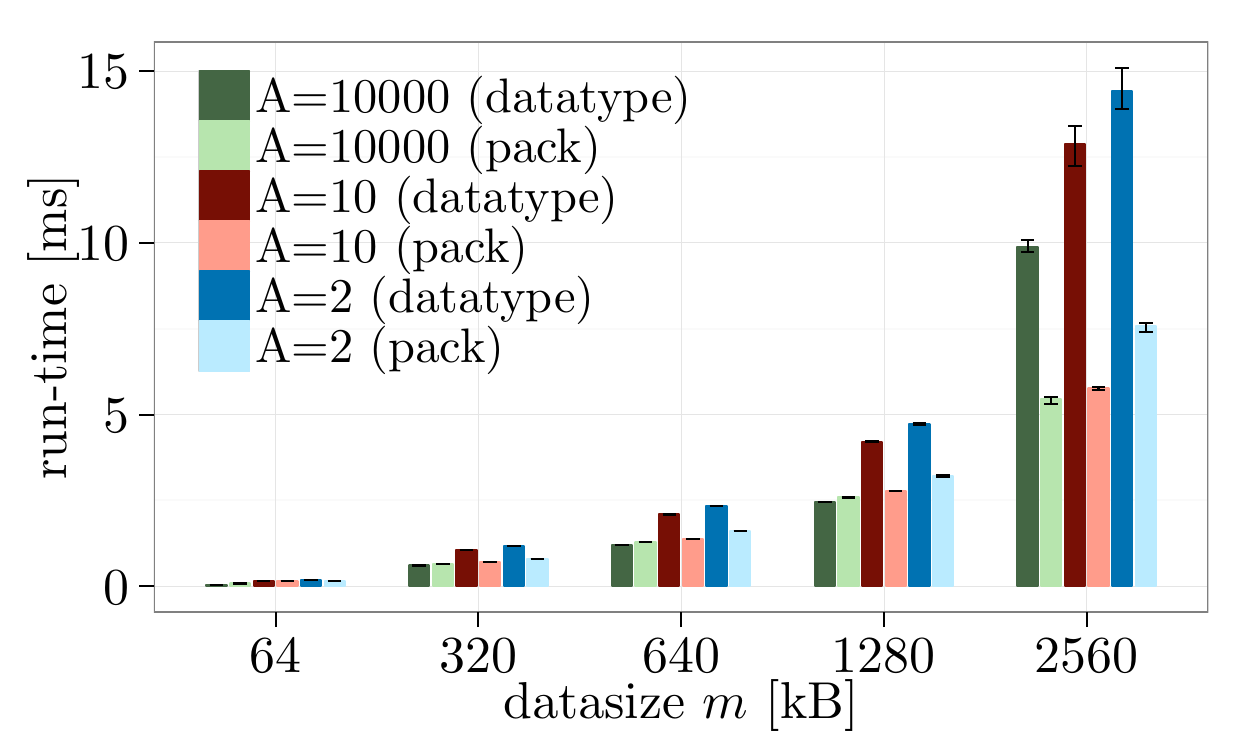}
\caption{%
\label{exp:pingpong-pack-tiled-2x1-mvapich}%
\dttiled%
}%
\end{subfigure}%
\hfill%
\begin{subfigure}{.24\linewidth}
\centering
\includegraphics[width=\linewidth]{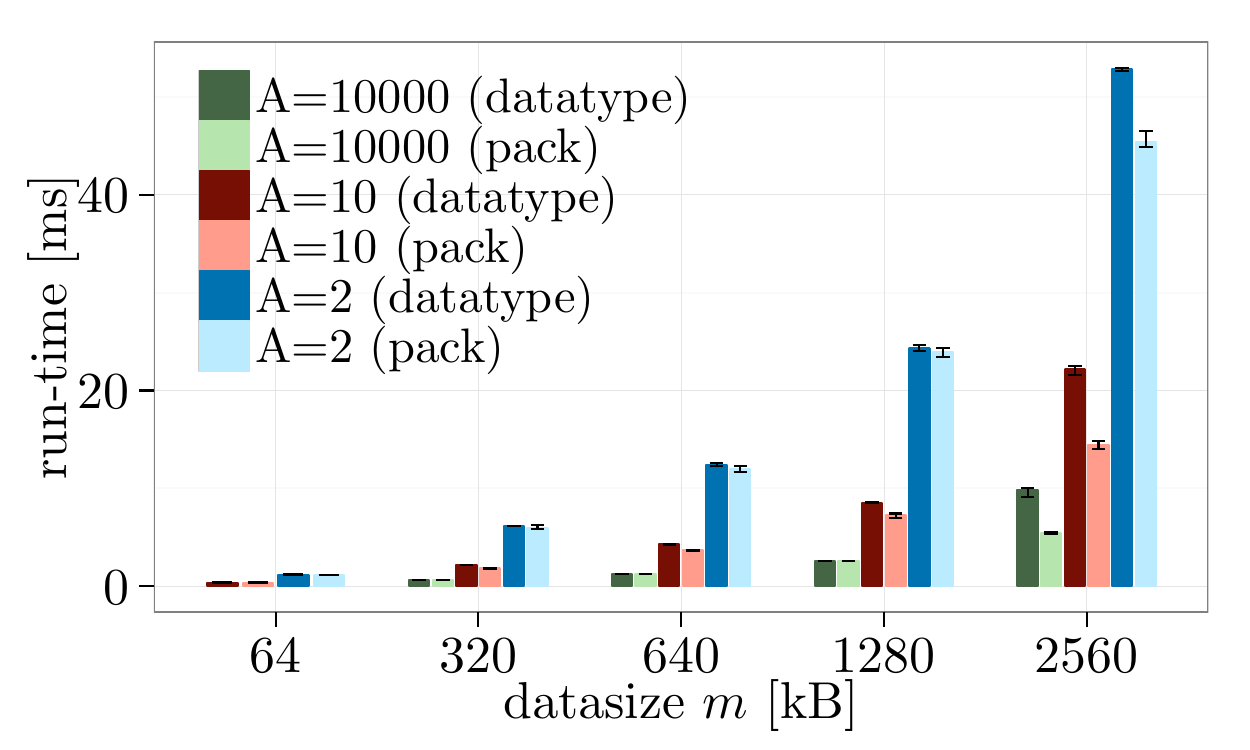}
\caption{%
\label{exp:pingpong-pack-block-2x1-mvapich}%
\dtblock%
}%
\end{subfigure}%
\hfill%
\begin{subfigure}{.24\linewidth}
\centering
\includegraphics[width=\linewidth]{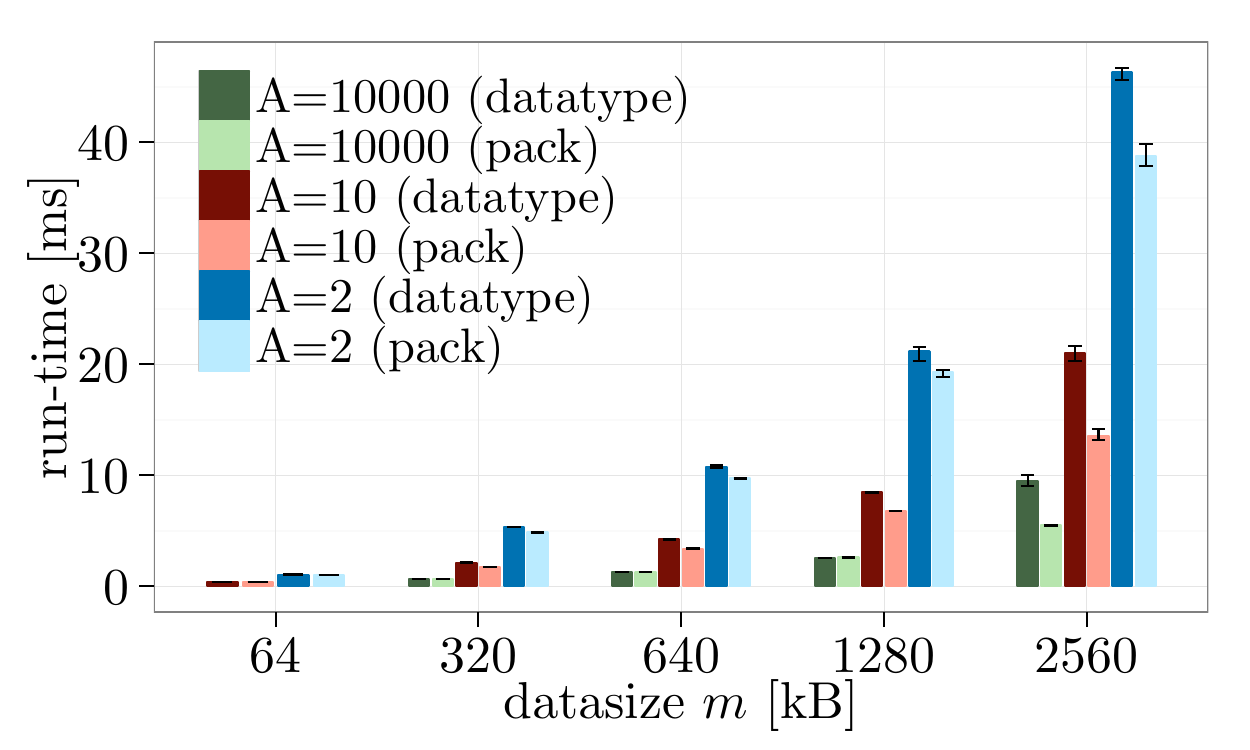}
\caption{%
\label{exp:pingpong-pack-bucket-2x1-mvapich}%
\dtbucket%
}%
\end{subfigure}%
\hfill%
\begin{subfigure}{.24\linewidth}
\centering
\includegraphics[width=\linewidth]{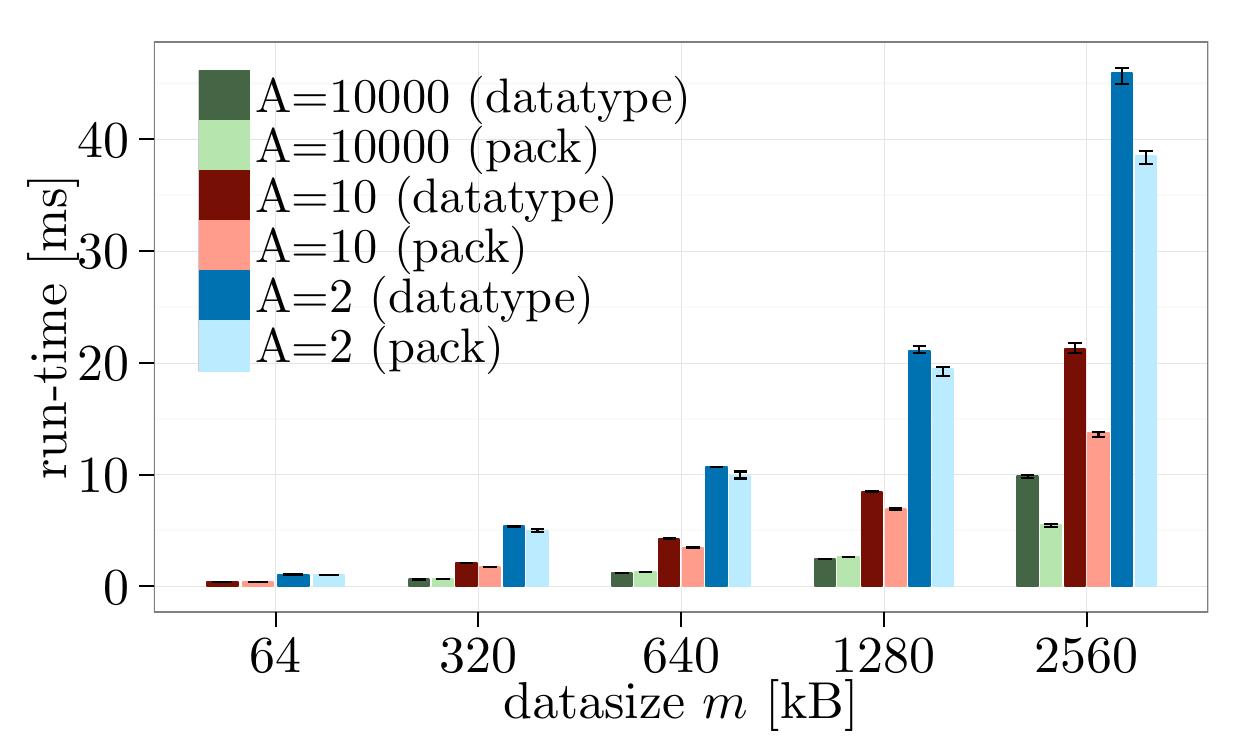}
\caption{%
\label{exp:pingpong-pack-alternating-2x1-mvapich}%
\dtalternating%
}%
\end{subfigure}%
\caption{\label{exp:pingpong-pack-2x1-mvapich}  Basic layouts \vs pack/unpack, element datatype: \mpiint, \num{2x1}~processes, \pingpong, \jupitermvapich.}
\end{figure*}

\begin{figure*}[htpb]
\centering
\begin{subfigure}{.24\linewidth}
\centering
\includegraphics[width=\linewidth]{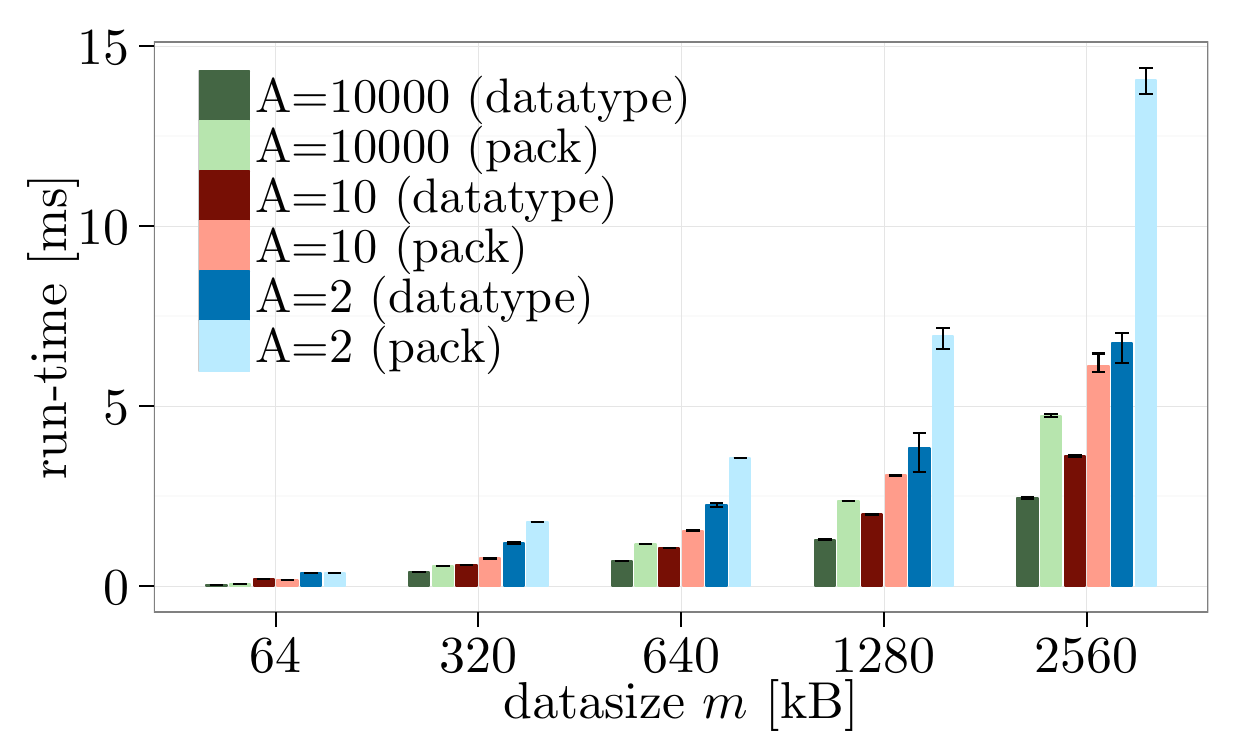}
\caption{%
\label{exp:pingpong-pack-tiled-2x1-openmpi}%
\dttiled%
}%
\end{subfigure}%
\hfill%
\begin{subfigure}{.24\linewidth}
\centering
\includegraphics[width=\linewidth]{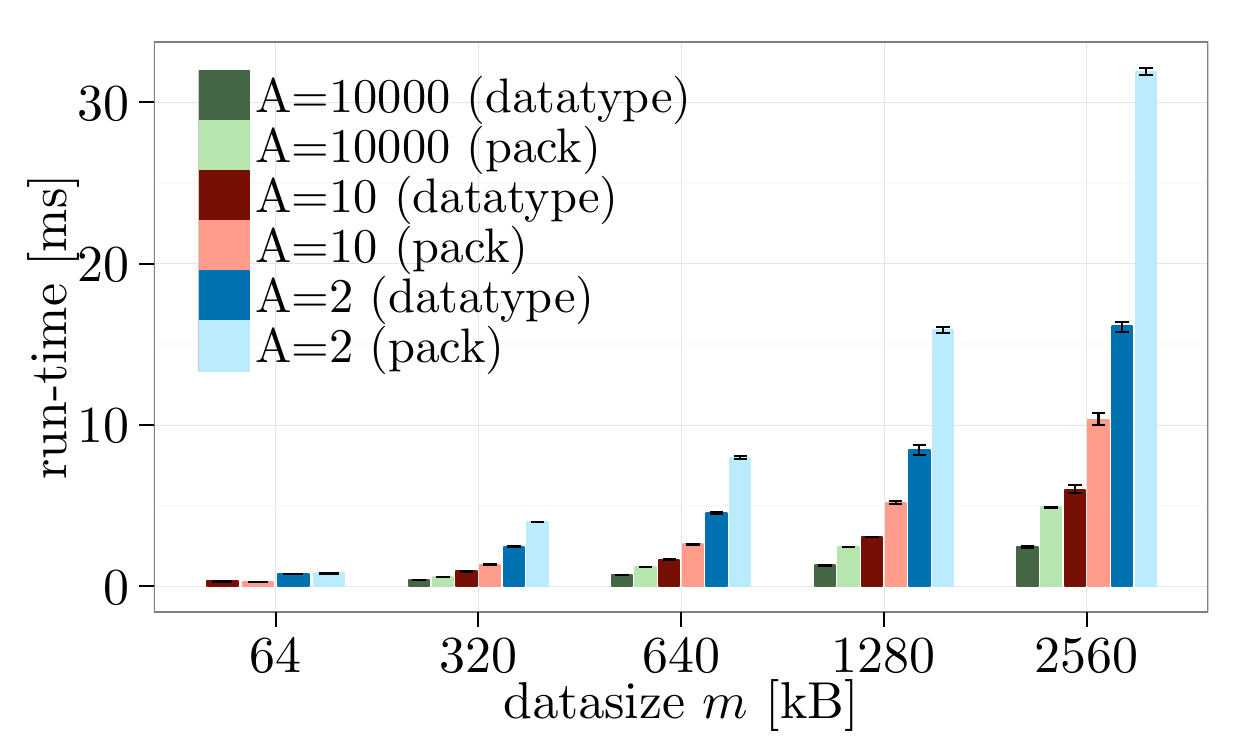}
\caption{%
\label{exp:pingpong-pack-block-2x1-openmpi}%
\dtblock%
}%
\end{subfigure}%
\hfill%
\begin{subfigure}{.24\linewidth}
\centering
\includegraphics[width=\linewidth]{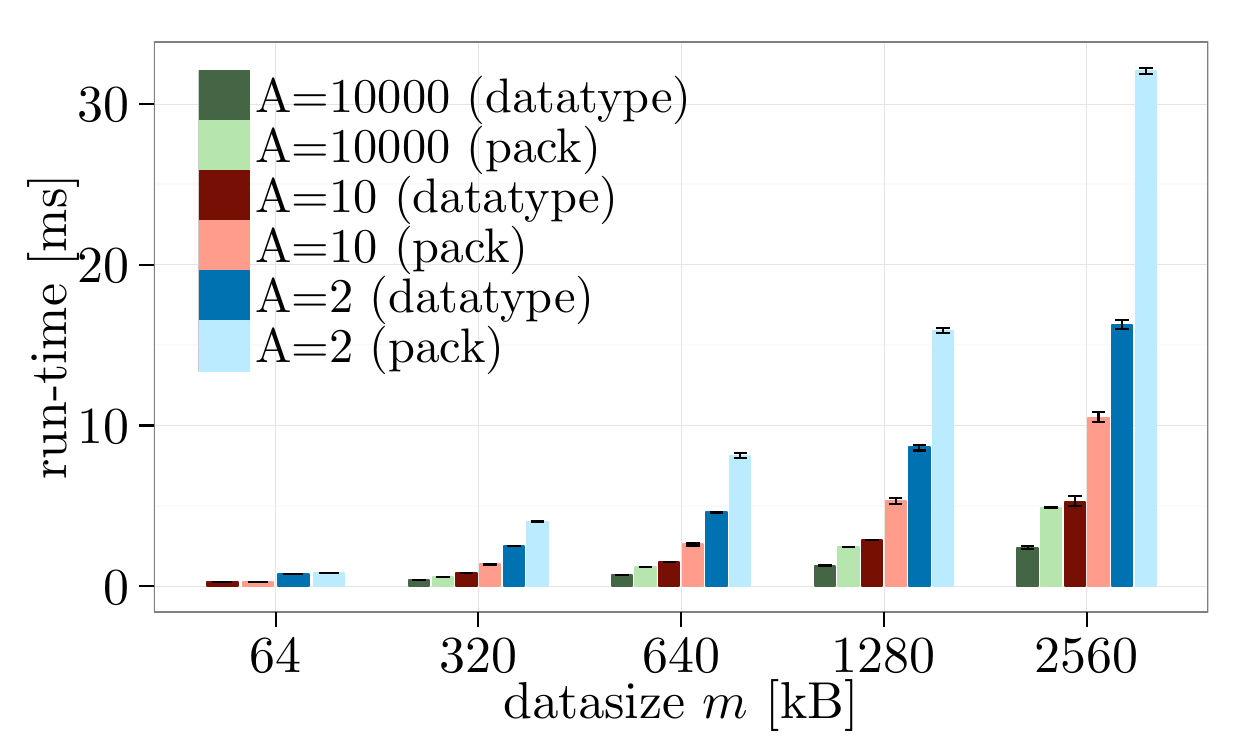}
\caption{%
\label{exp:pingpong-pack-bucket-2x1-openmpi}%
\dtbucket%
}%
\end{subfigure}%
\hfill%
\begin{subfigure}{.24\linewidth}
\centering
\includegraphics[width=\linewidth]{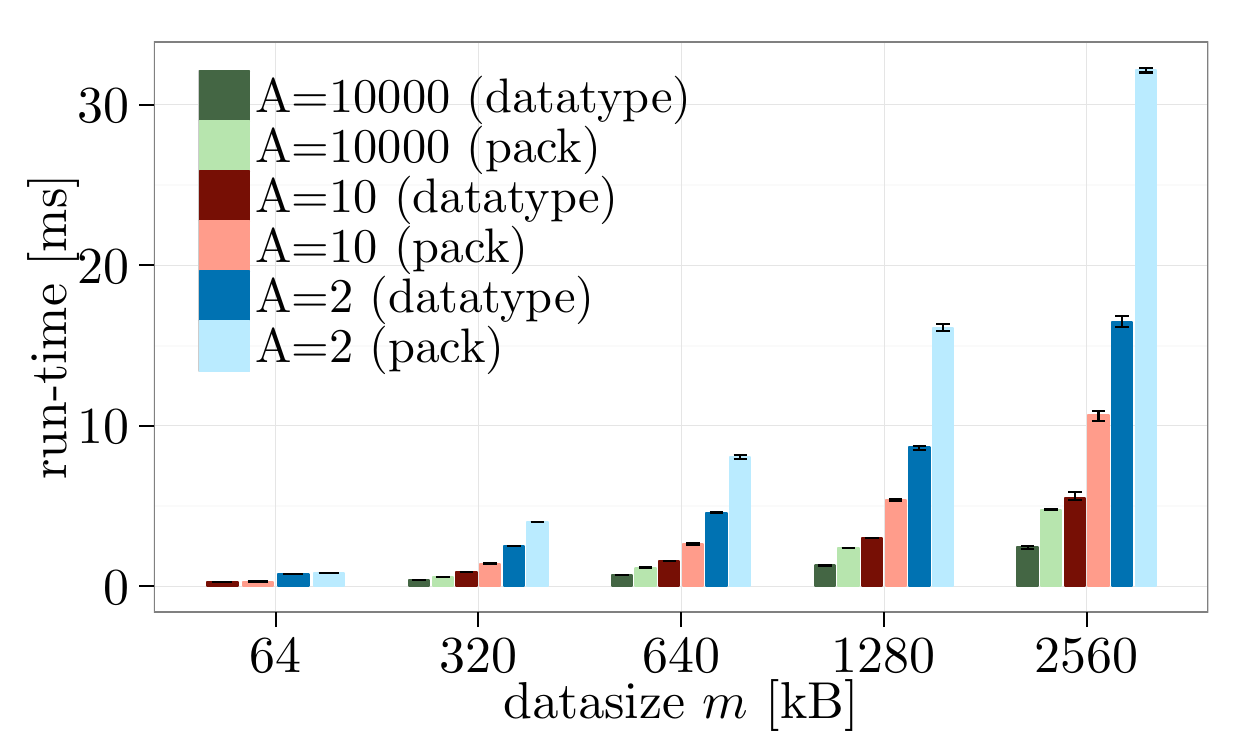}
\caption{%
\label{exp:pingpong-pack-alternating-2x1-openmpi}%
\dtalternating%
}%
\end{subfigure}%
\caption{\label{exp:pingpong-pack-2x1-openmpi}  Basic layouts \vs pack/unpack, element datatype: \mpiint, \num{2x1}~processes, \pingpong, \jupiteropenmpi.}
\end{figure*}

\begin{figure*}[htpb]
\centering
\begin{subfigure}{.24\linewidth}
\centering
\includegraphics[width=\linewidth]{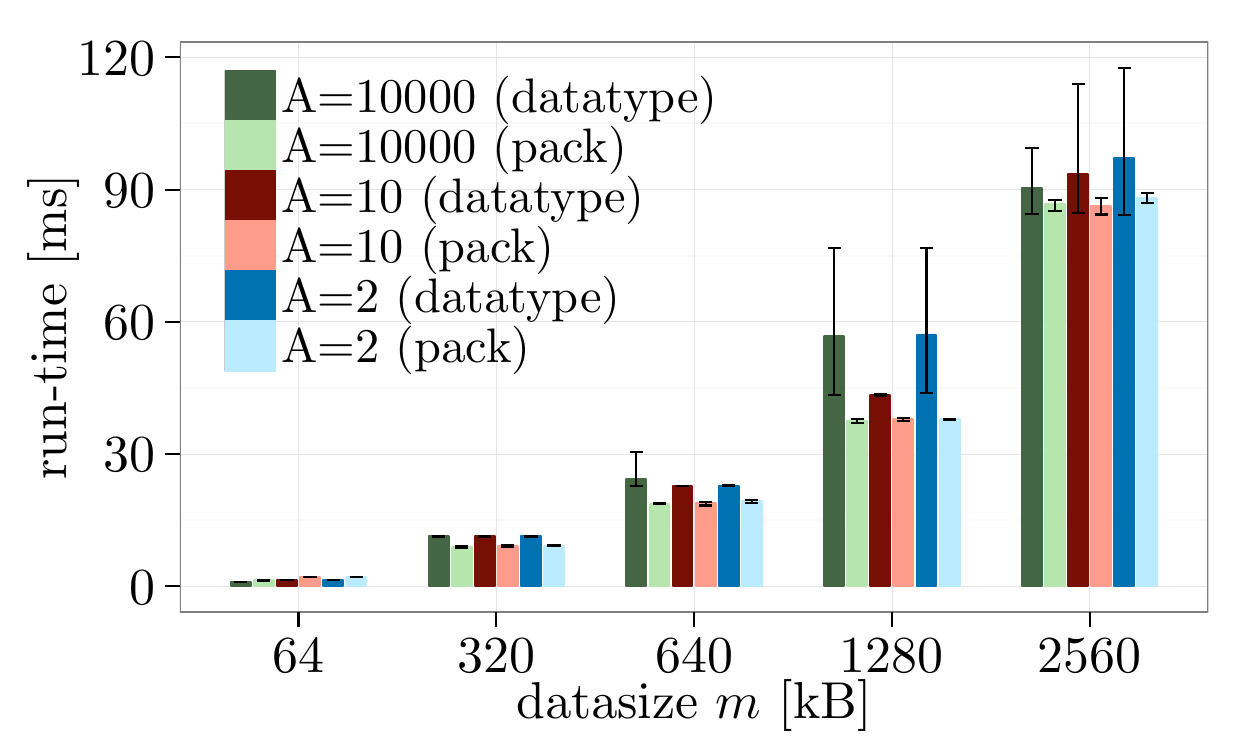}
\caption{%
\label{exp:allgather-pack-tiled-32x1-nec}%
\dttiled%
}%
\end{subfigure}%
\hfill%
\begin{subfigure}{.24\linewidth}
\centering
\includegraphics[width=\linewidth]{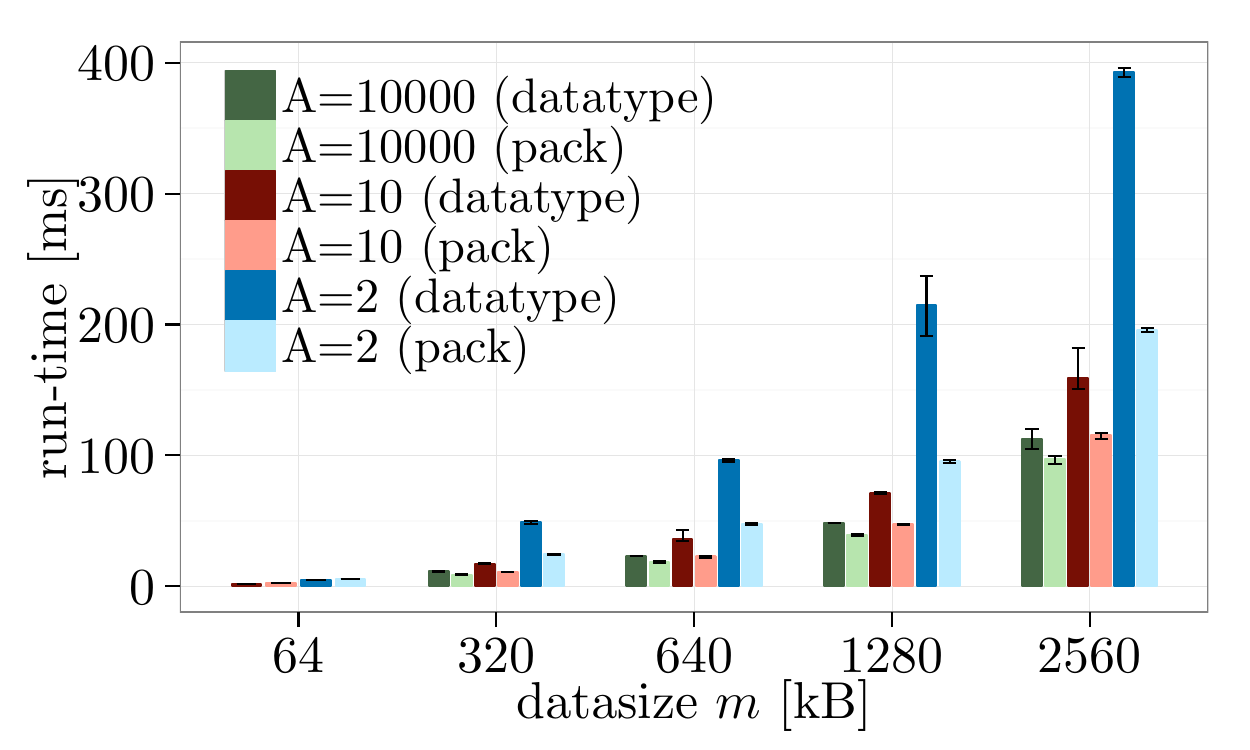}
\caption{%
\label{exp:allgather-pack-block-32x1-nec}%
\dtblock%
}%
\end{subfigure}%
\hfill%
\begin{subfigure}{.24\linewidth}
\centering
\includegraphics[width=\linewidth]{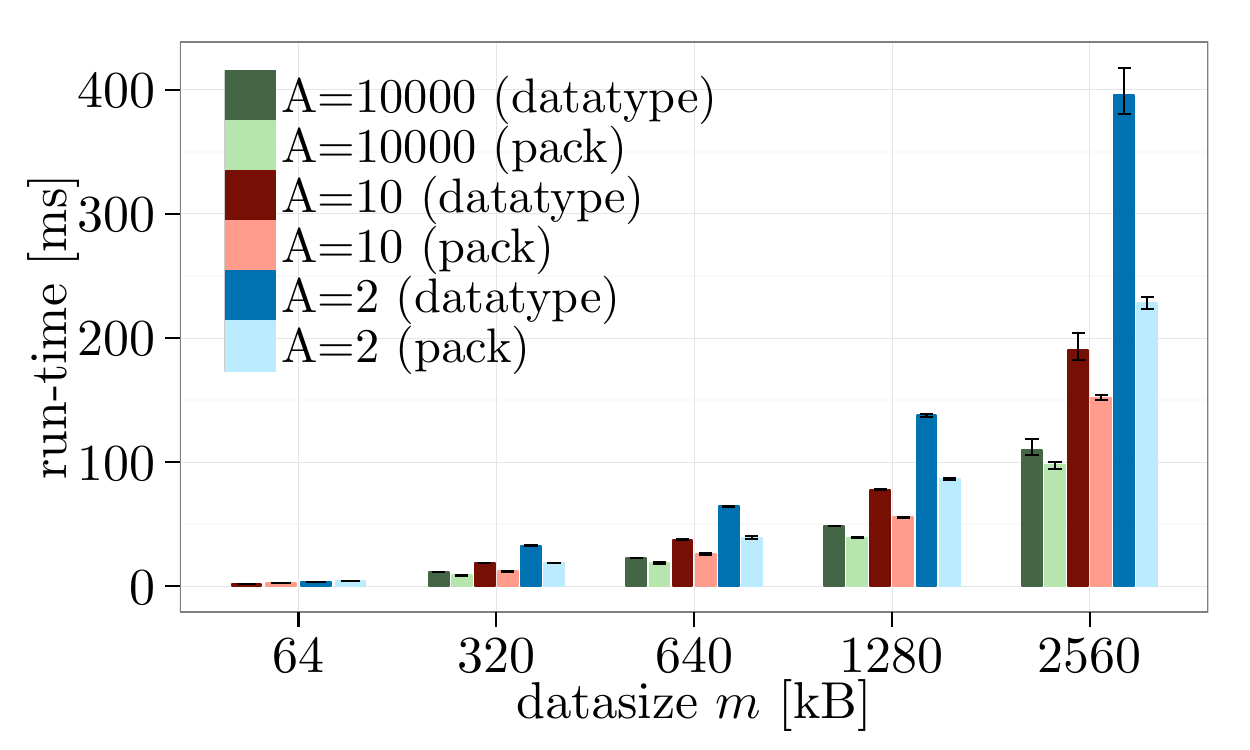}
\caption{%
\label{exp:allgather-pack-bucket-32x1-nec}%
\dtbucket%
}%
\end{subfigure}%
\hfill%
\begin{subfigure}{.24\linewidth}
\centering
\includegraphics[width=\linewidth]{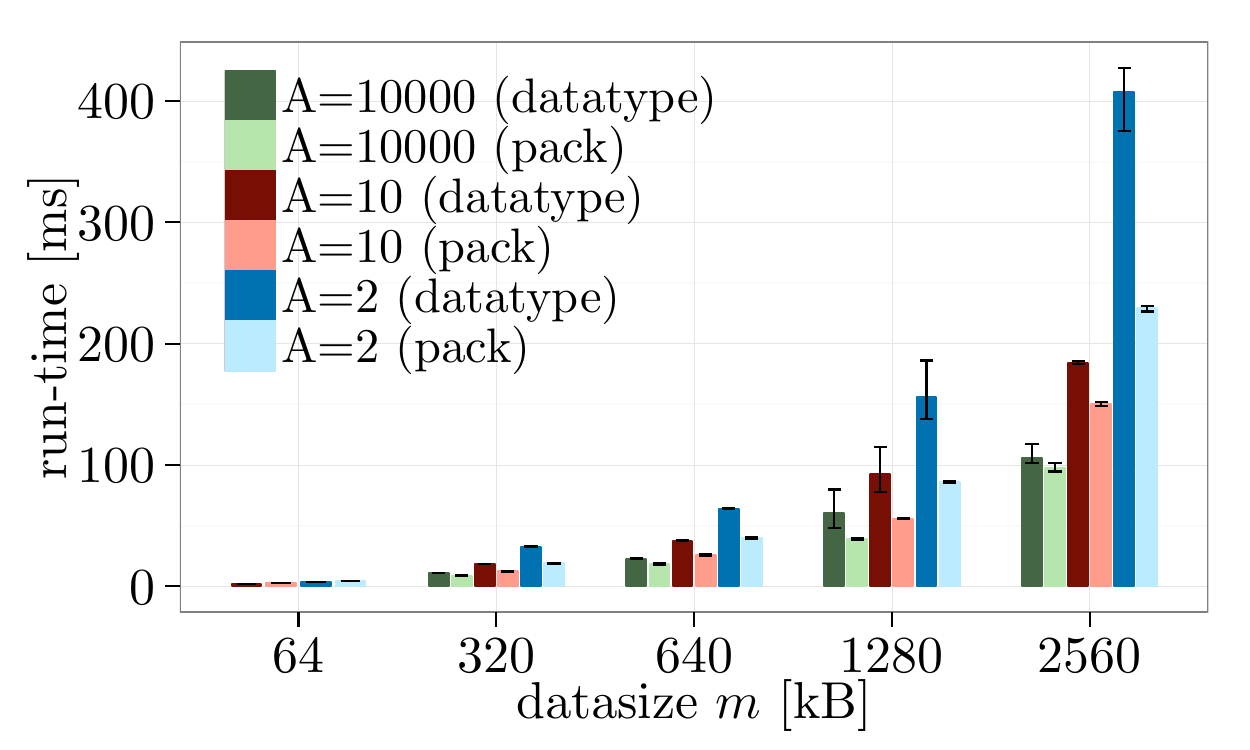}
\caption{%
\label{exp:allgather-pack-alternating-32x1-nec}%
\dtalternating%
}%
\end{subfigure}%
\caption{\label{exp:allgather-pack-32x1-nec}  Basic layouts \vs pack/unpack, element datatype: \mpiint, \num{32x1}~processes, \mpiallgather, \jupiternecmpi.}
\end{figure*}

\begin{figure*}[htpb]
\centering
\begin{subfigure}{.24\linewidth}
\centering
\includegraphics[width=\linewidth]{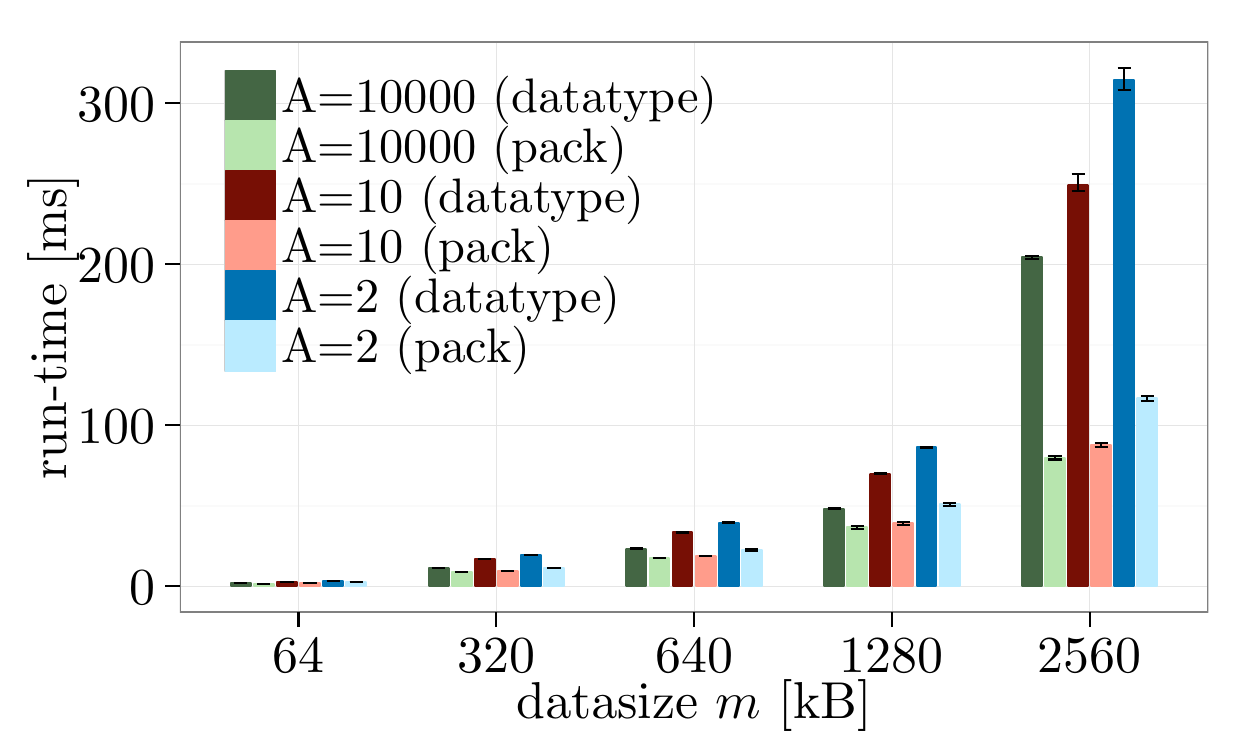}
\caption{%
\label{exp:allgather-pack-tiled-32x1-mvapich}%
\dttiled%
}%
\end{subfigure}%
\hfill%
\begin{subfigure}{.24\linewidth}
\centering
\includegraphics[width=\linewidth]{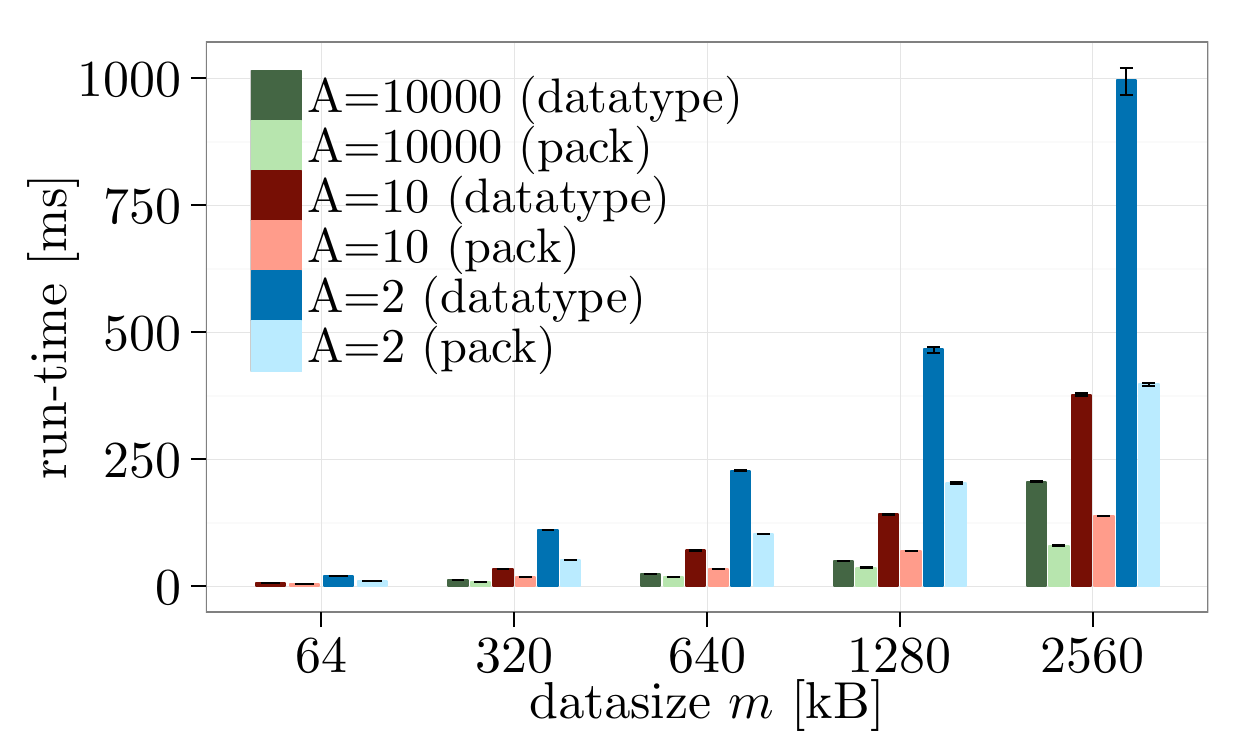}
\caption{%
\label{exp:allgather-pack-block-32x1-mvapich}%
\dtblock%
}%
\end{subfigure}%
\hfill%
\begin{subfigure}{.24\linewidth}
\centering
\includegraphics[width=\linewidth]{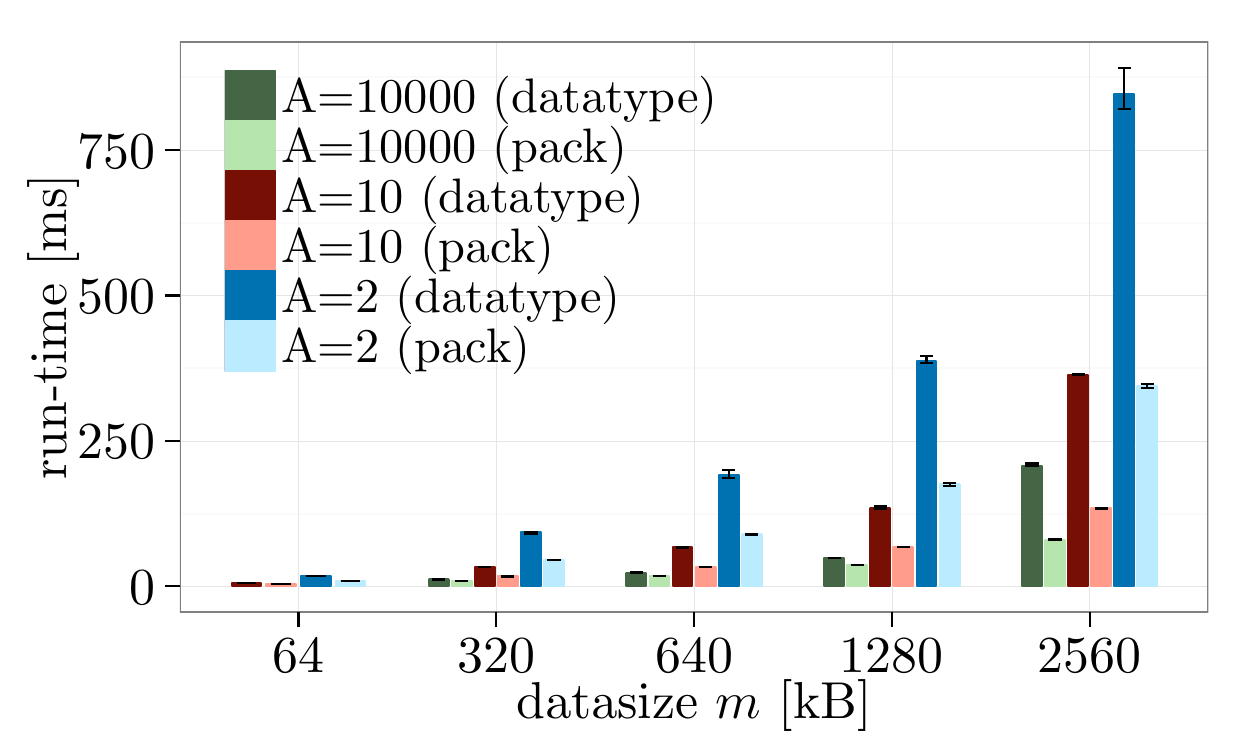}
\caption{%
\label{exp:allgather-pack-bucket-32x1-mvapich}%
\dtbucket%
}%
\end{subfigure}%
\hfill%
\begin{subfigure}{.24\linewidth}
\centering
\includegraphics[width=\linewidth]{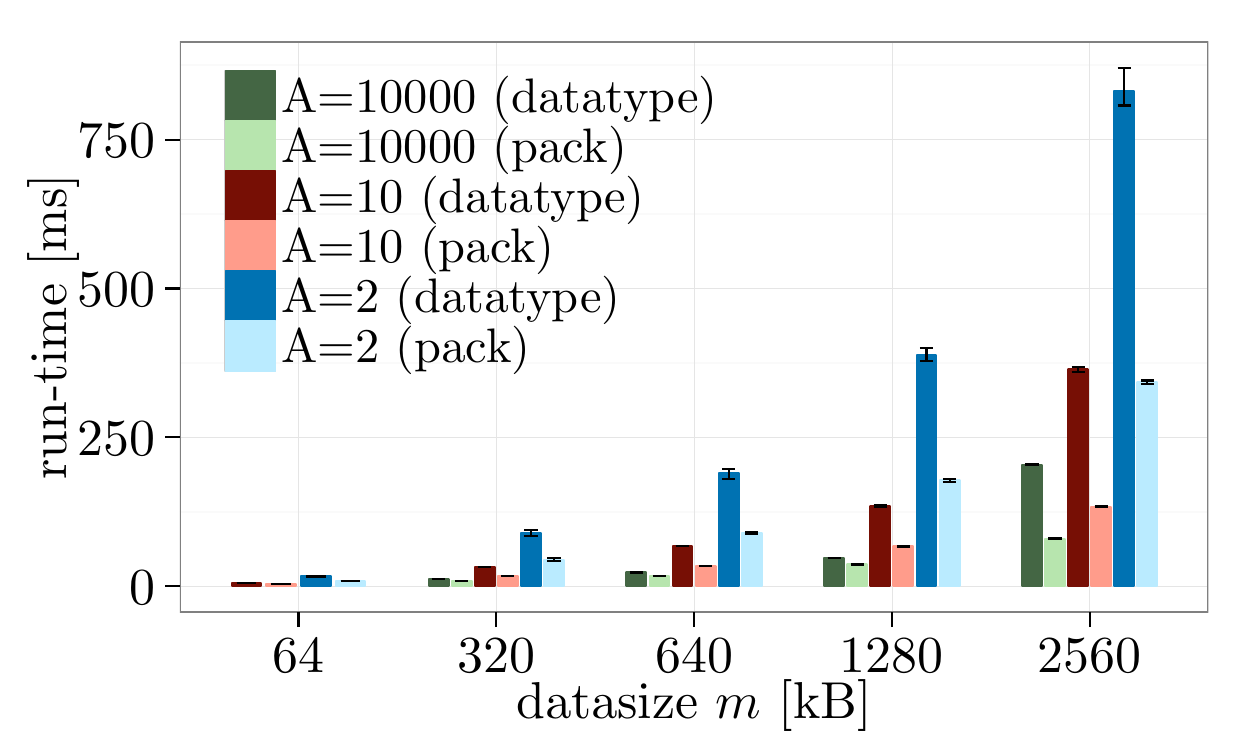}
\caption{%
\label{exp:allgather-pack-alternating-32x1-mvapich}%
\dtalternating%
}%
\end{subfigure}%
\caption{\label{exp:allgather-pack-32x1-mvapich}  Basic layouts \vs pack/unpack, element datatype: \mpiint, \num{32x1}~processes, \mpiallgather, \jupitermvapich.}
\end{figure*}

\begin{figure*}[htpb]
\centering
\begin{subfigure}{.24\linewidth}
\centering
\includegraphics[width=\linewidth]{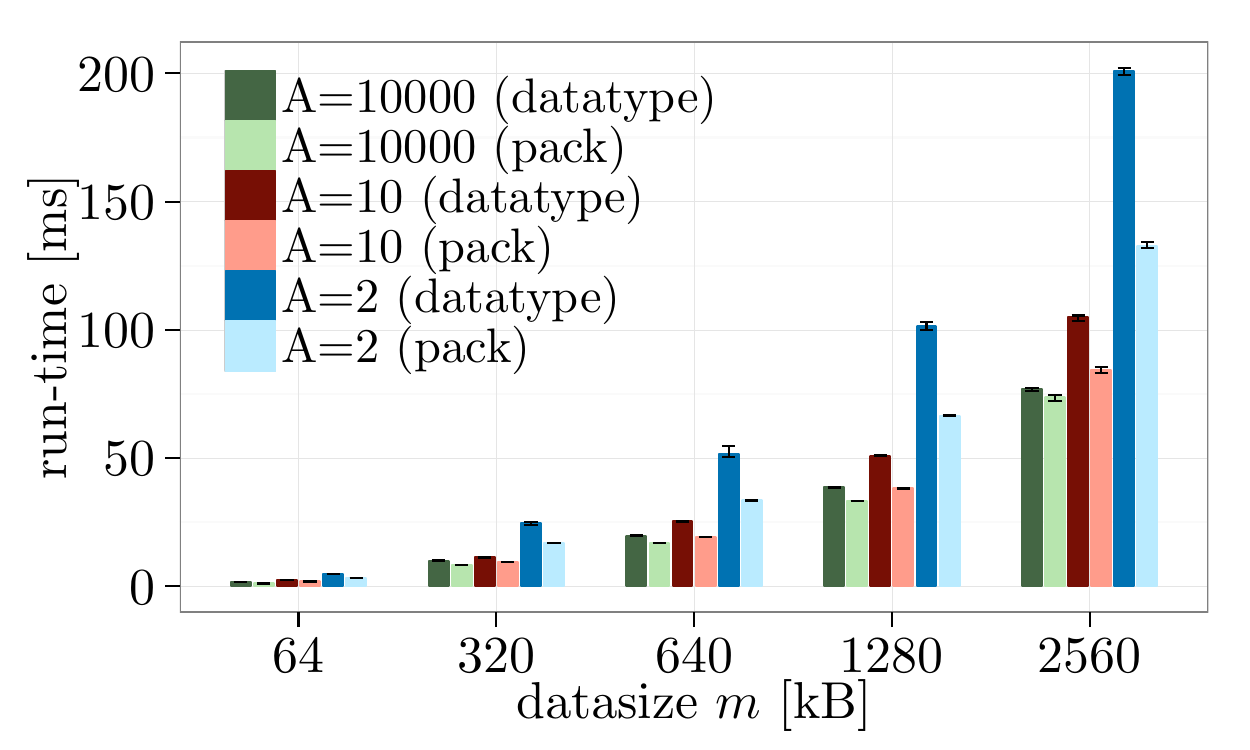}
\caption{%
\label{exp:allgather-pack-tiled-32x1-openmpi}%
\dttiled%
}%
\end{subfigure}%
\hfill%
\begin{subfigure}{.24\linewidth}
\centering
\includegraphics[width=\linewidth]{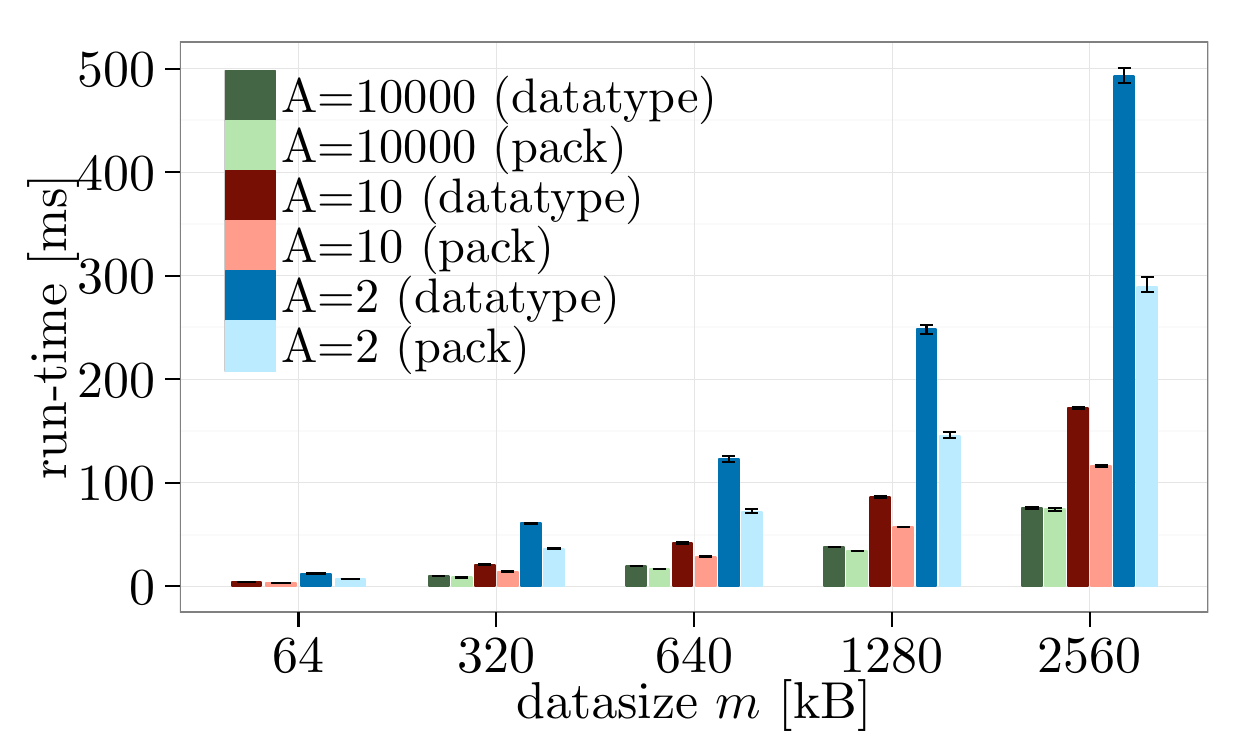}
\caption{%
\label{exp:allgather-pack-block-32x1-openmpi}%
\dtblock%
}%
\end{subfigure}%
\hfill%
\begin{subfigure}{.24\linewidth}
\centering
\includegraphics[width=\linewidth]{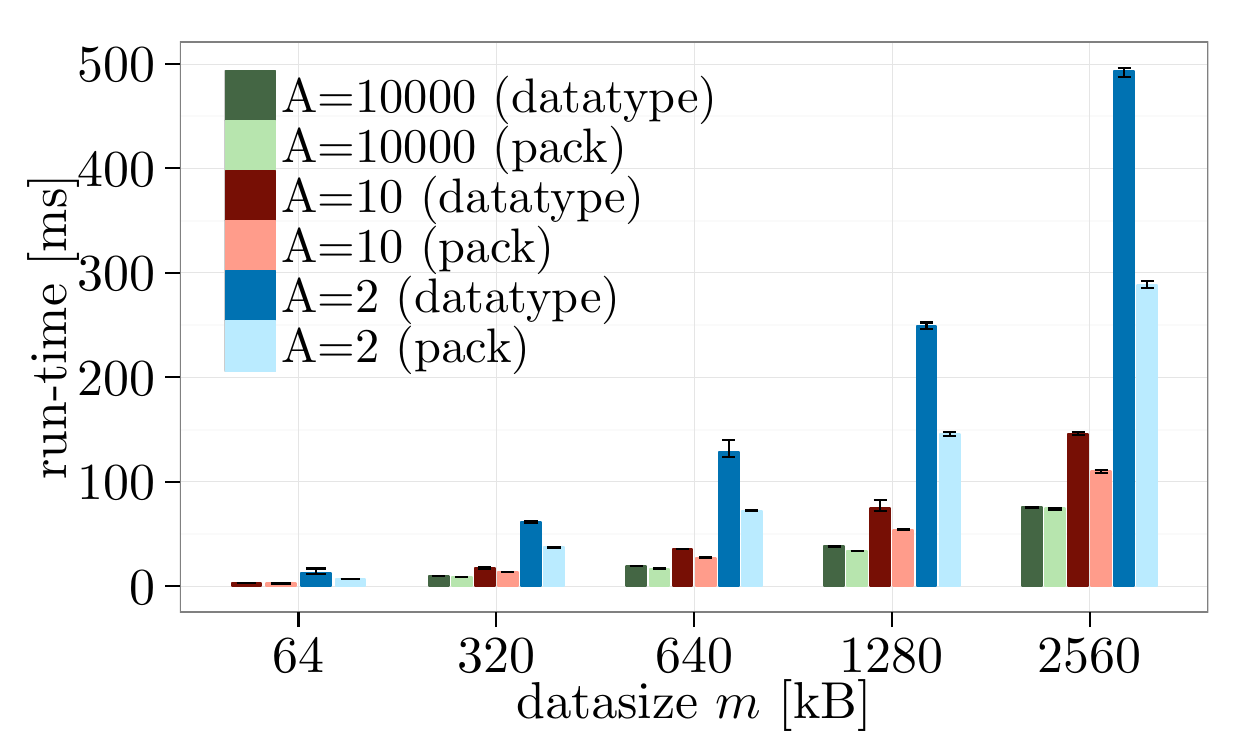}
\caption{%
\label{exp:allgather-pack-bucket-32x1-openmpi}%
\dtbucket%
}%
\end{subfigure}%
\hfill%
\begin{subfigure}{.24\linewidth}
\centering
\includegraphics[width=\linewidth]{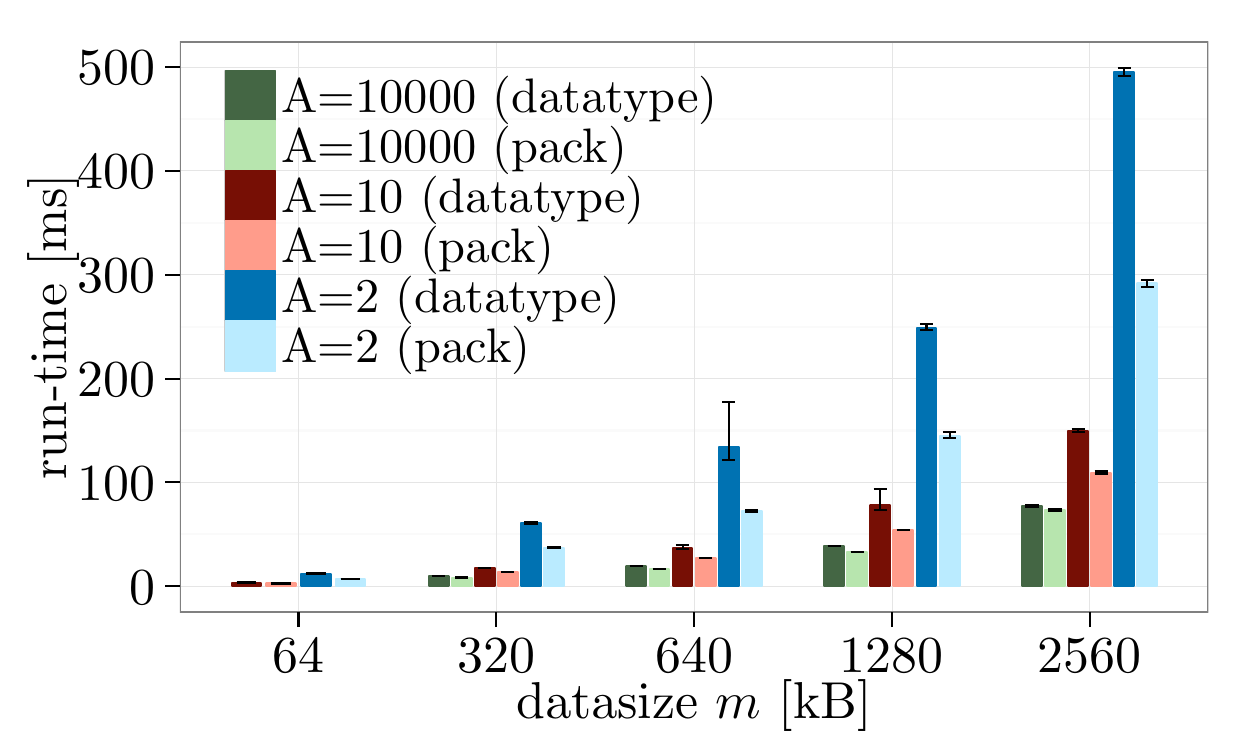}
\caption{%
\label{exp:allgather-pack-alternating-32x1-openmpi}%
\dtalternating%
}%
\end{subfigure}%
\caption{\label{exp:allgather-pack-32x1-openmpi}  Basic layouts \vs pack/unpack, element datatype: \mpiint, \num{32x1}~processes, \mpiallgather, \jupiteropenmpi.}
\end{figure*}

\begin{figure*}[htpb]
\centering
\begin{subfigure}{.24\linewidth}
\centering
\includegraphics[width=\linewidth]{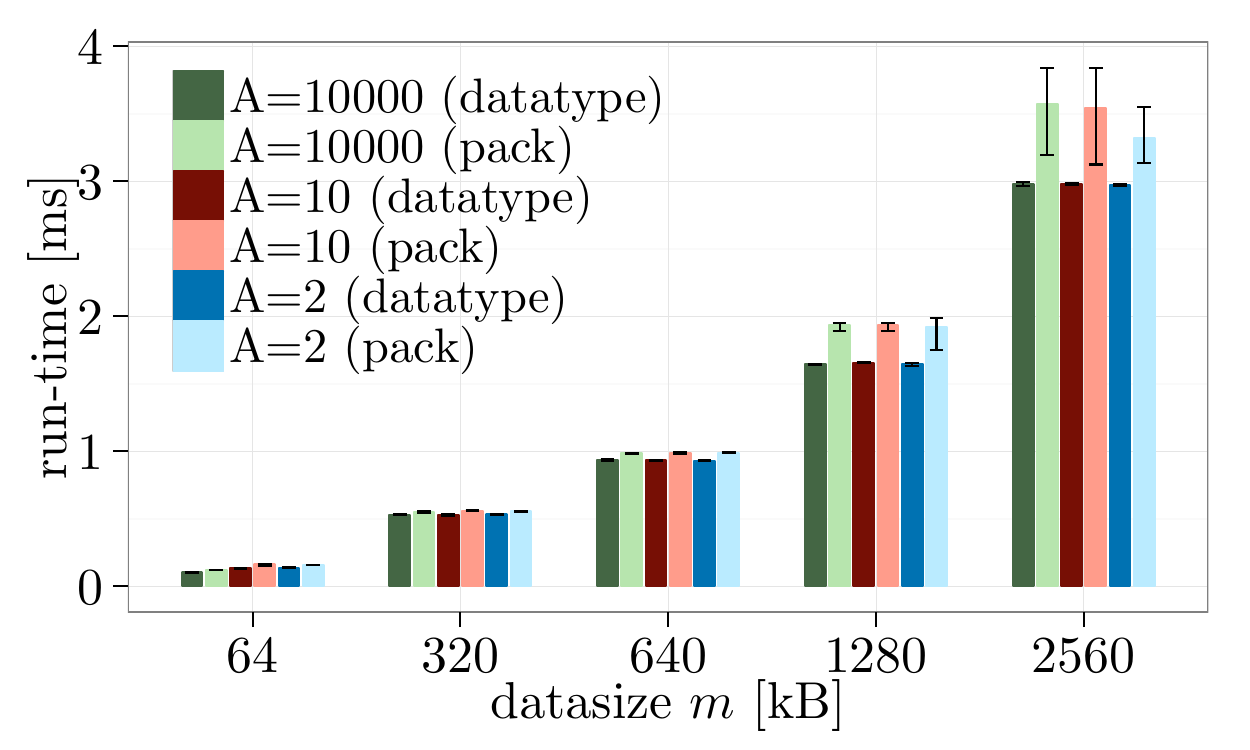}
\caption{%
\label{exp:bcast-pack-tiled-32x1-nec}%
\dttiled%
}%
\end{subfigure}%
\hfill%
\begin{subfigure}{.24\linewidth}
\centering
\includegraphics[width=\linewidth]{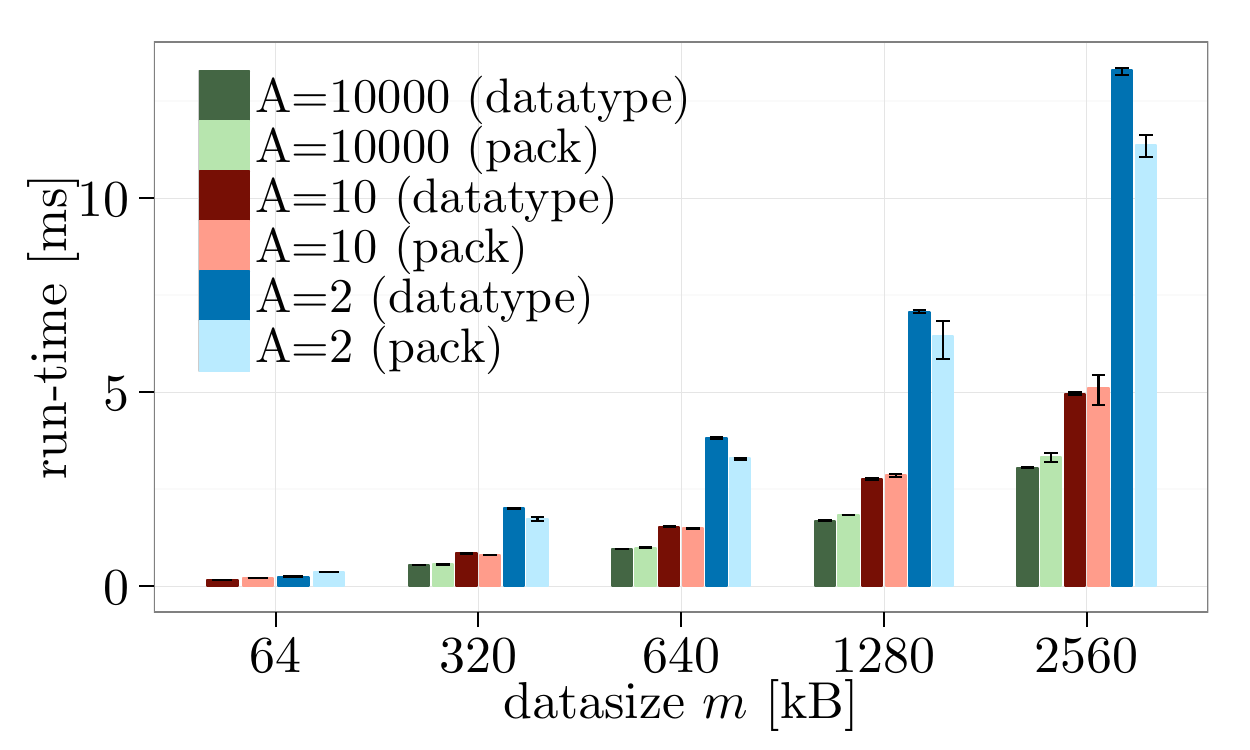}
\caption{%
\label{exp:bcast-pack-block-32x1-nec}%
\dtblock%
}%
\end{subfigure}%
\hfill%
\begin{subfigure}{.24\linewidth}
\centering
\includegraphics[width=\linewidth]{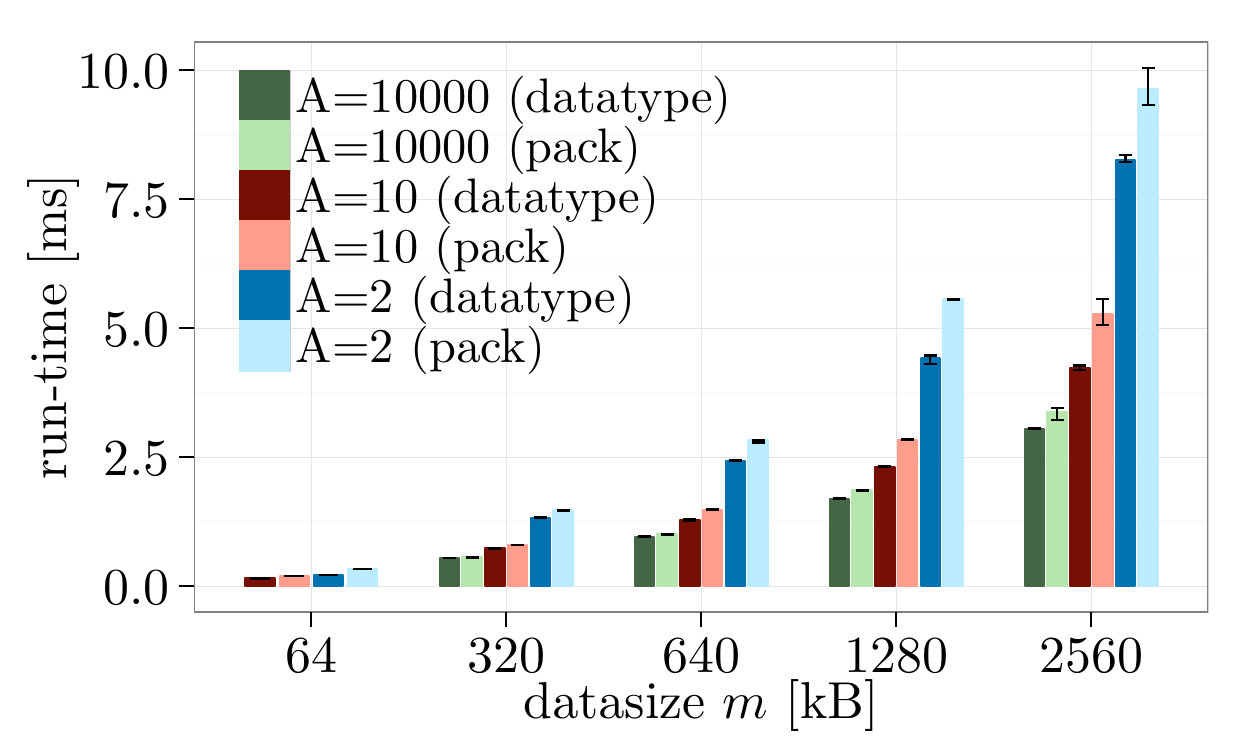}
\caption{%
\label{exp:bcast-pack-bucket-32x1-nec}%
\dtbucket%
}%
\end{subfigure}%
\hfill%
\begin{subfigure}{.24\linewidth}
\centering
\includegraphics[width=\linewidth]{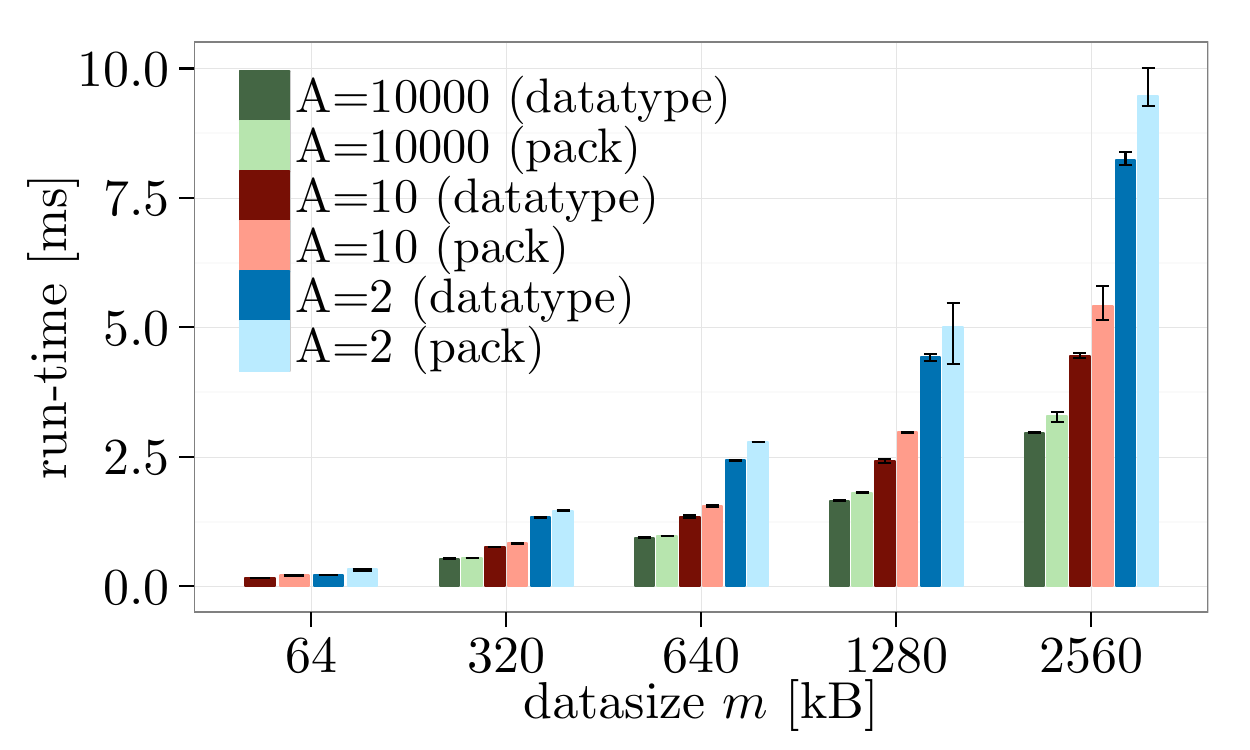}
\caption{%
\label{exp:bcast-pack-alternating-32x1-nec}%
\dtalternating%
}%
\end{subfigure}%
\caption{\label{exp:bcast-pack-32x1-nec}  Basic layouts \vs pack/unpack, element datatype: \mpiint, \num{32x1}~processes, \mpibcast, \jupiternecmpi.}
\end{figure*}

\begin{figure*}[htpb]
\centering
\begin{subfigure}{.24\linewidth}
\centering
\includegraphics[width=\linewidth]{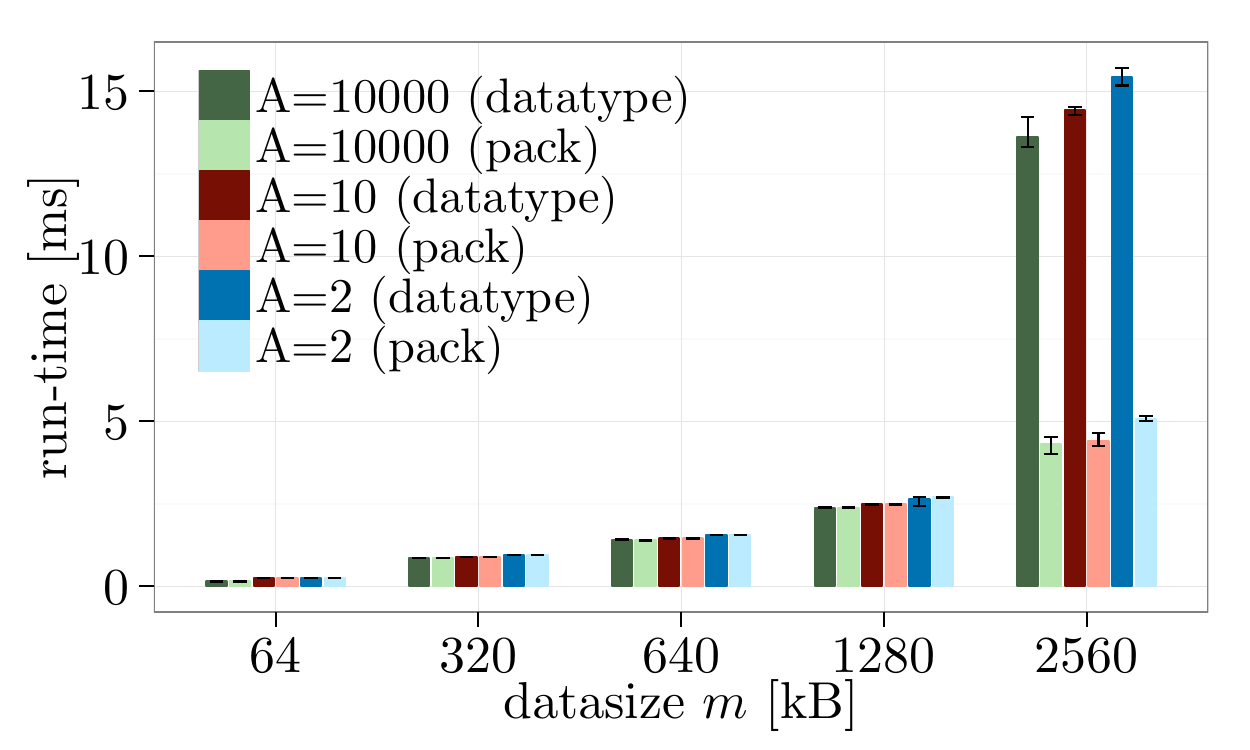}
\caption{%
\label{exp:bcast-pack-tiled-32x1-mvapich}%
\dttiled%
}%
\end{subfigure}%
\hfill%
\begin{subfigure}{.24\linewidth}
\centering
\includegraphics[width=\linewidth]{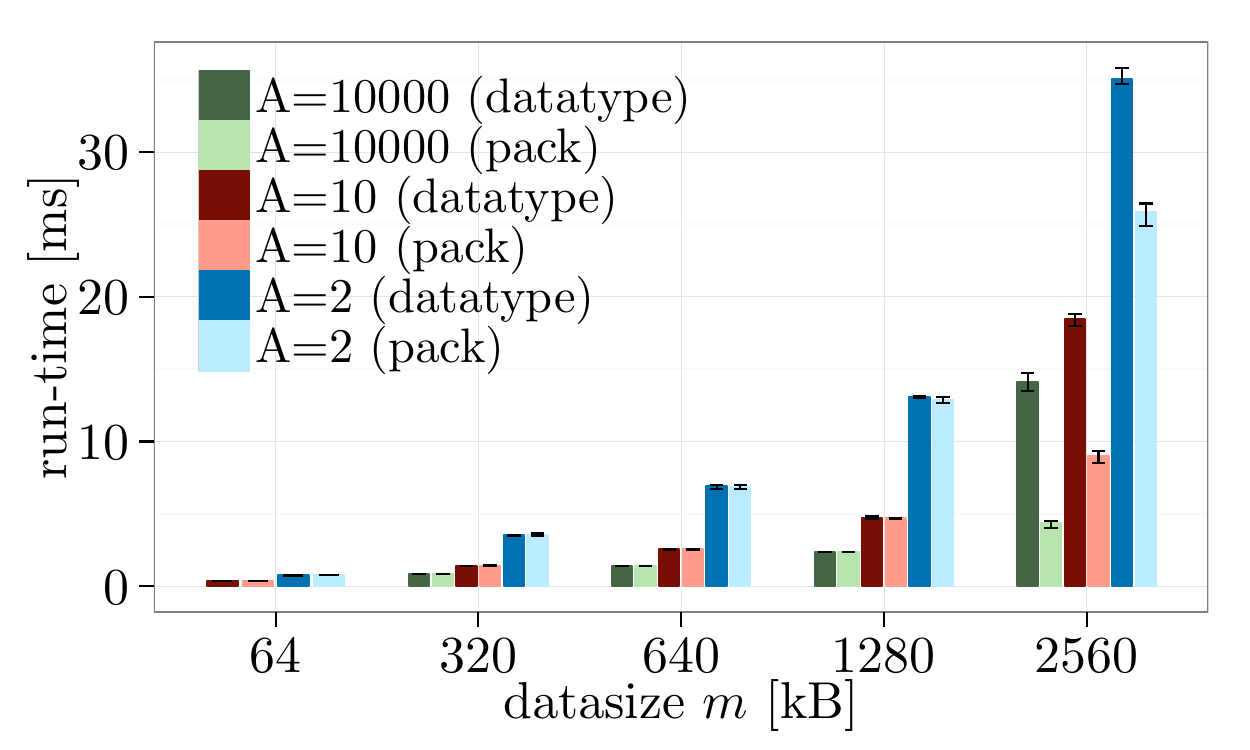}
\caption{%
\label{exp:bcast-pack-block-32x1-mvapich}%
\dtblock%
}%
\end{subfigure}%
\hfill%
\begin{subfigure}{.24\linewidth}
\centering
\includegraphics[width=\linewidth]{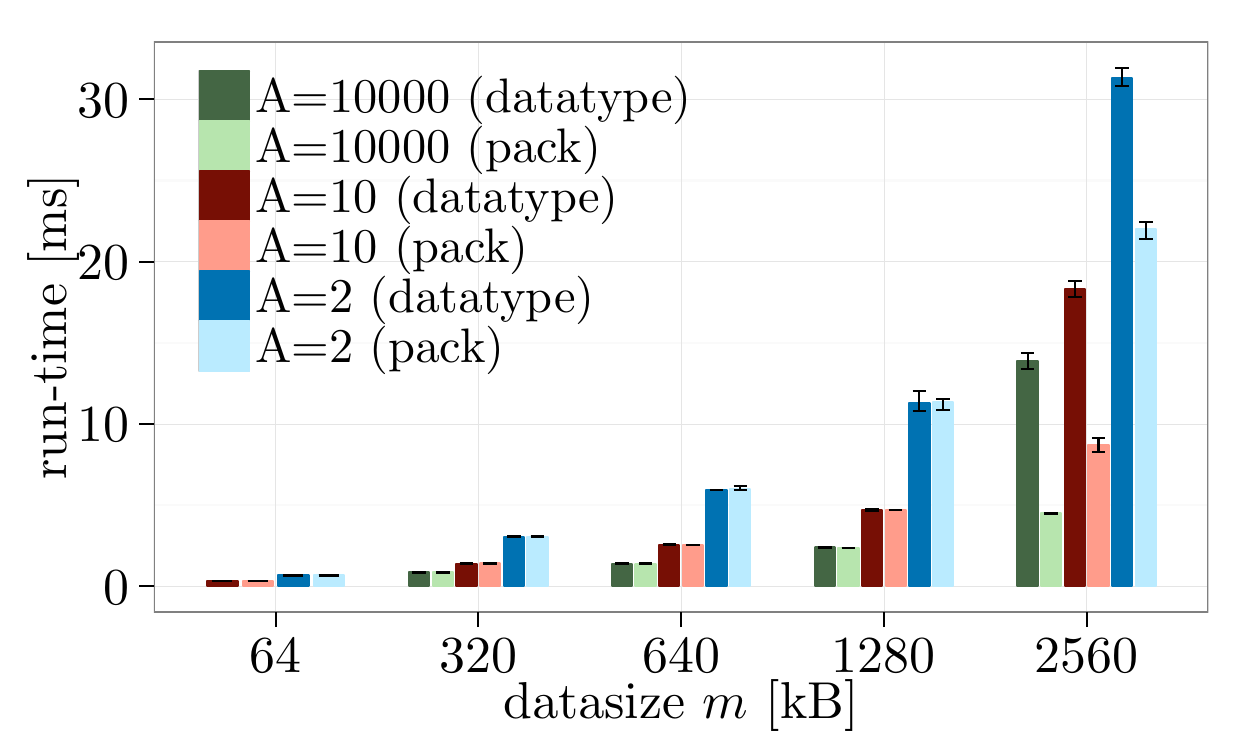}
\caption{%
\label{exp:bcast-pack-bucket-32x1-mvapich}%
\dtbucket%
}%
\end{subfigure}%
\hfill%
\begin{subfigure}{.24\linewidth}
\centering
\includegraphics[width=\linewidth]{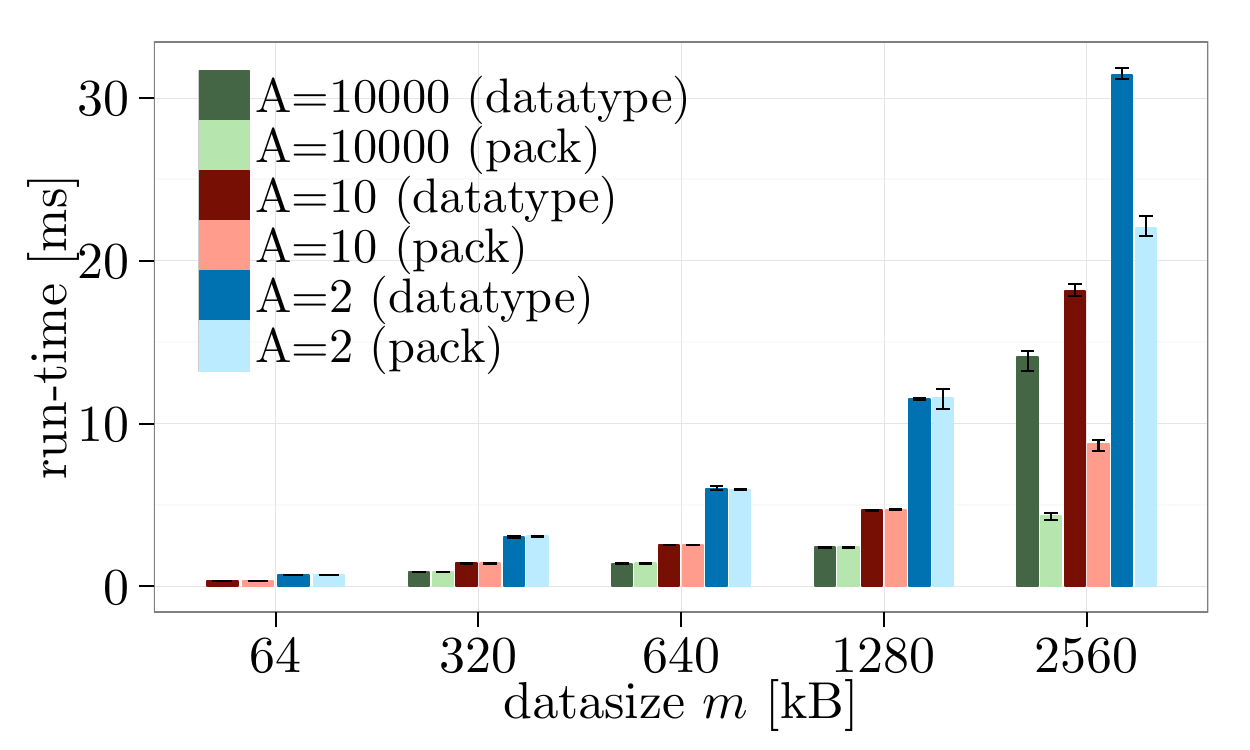}
\caption{%
\label{exp:bcast-pack-alternating-32x1-mvapich}%
\dtalternating%
}%
\end{subfigure}%
\caption{\label{exp:bcast-pack-32x1-mvapich}  Basic layouts \vs pack/unpack, element datatype: \mpiint, \num{32x1}~processes, \mpibcast, \jupitermvapich.}
\end{figure*}

\begin{figure*}[htpb]
\centering
\begin{subfigure}{.24\linewidth}
\centering
\includegraphics[width=\linewidth]{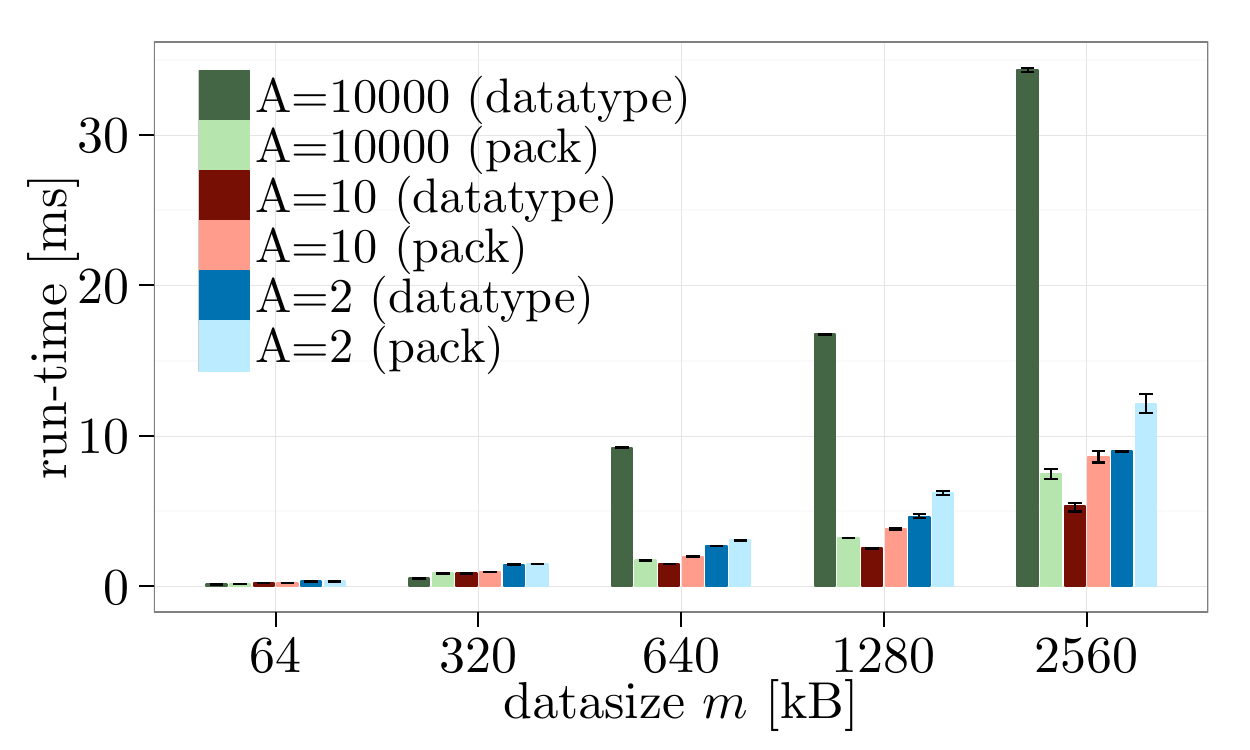}
\caption{%
\label{exp:bcast-pack-tiled-32x1-openmpi}%
\dttiled%
}%
\end{subfigure}%
\hfill%
\begin{subfigure}{.24\linewidth}
\centering
\includegraphics[width=\linewidth]{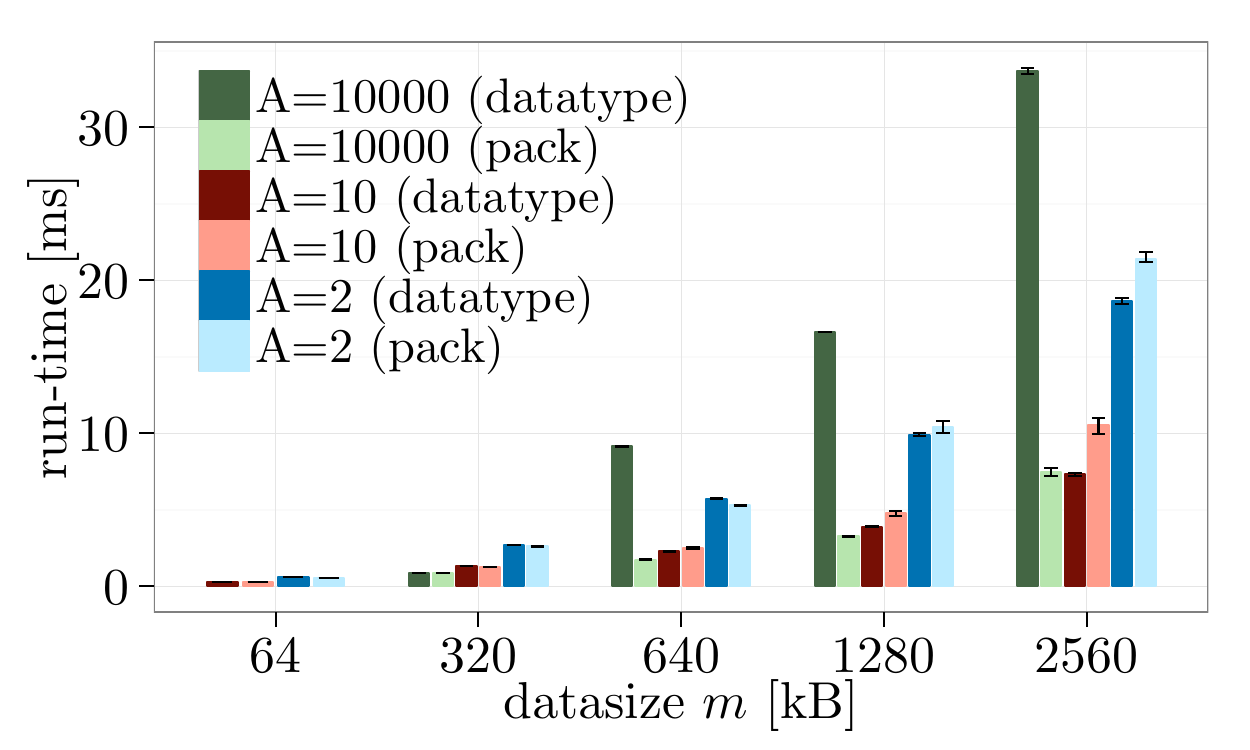}
\caption{%
\label{exp:bcast-pack-block-32x1-openmpi}%
\dtblock%
}%
\end{subfigure}%
\hfill%
\begin{subfigure}{.24\linewidth}
\centering
\includegraphics[width=\linewidth]{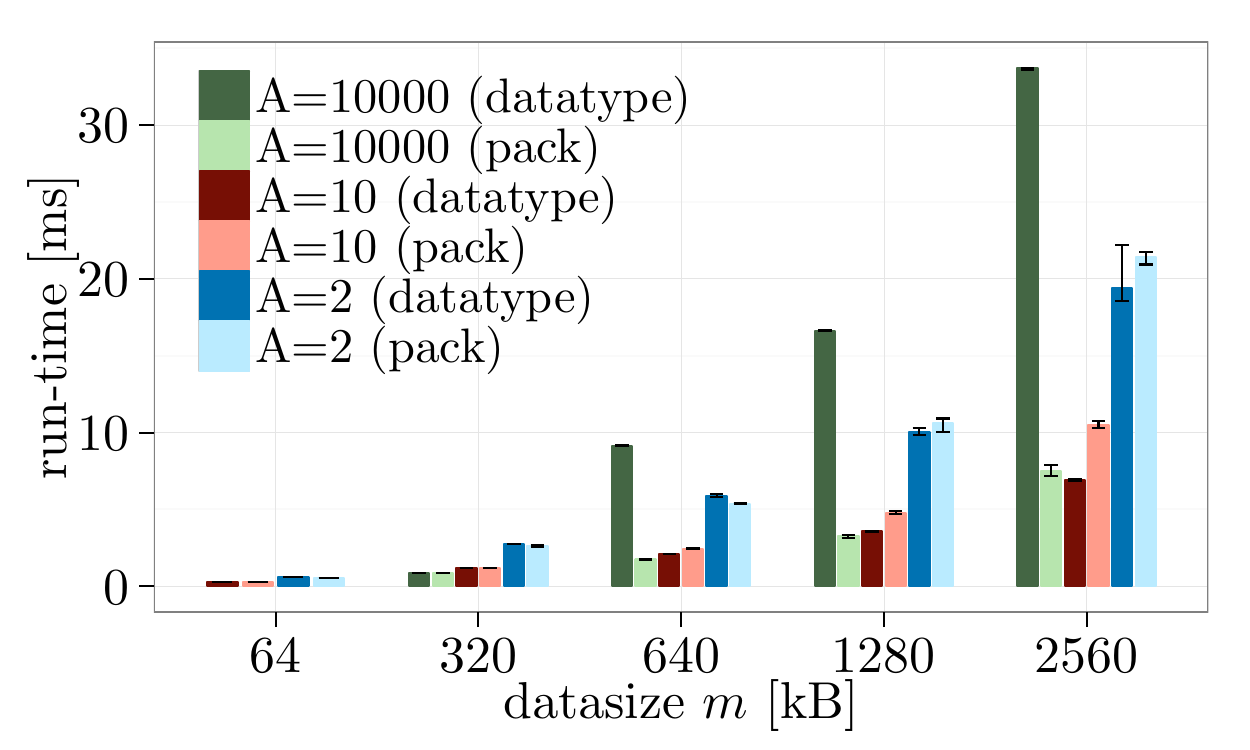}
\caption{%
\label{exp:bcast-pack-bucket-32x1-openmpi}%
\dtbucket%
}%
\end{subfigure}%
\hfill%
\begin{subfigure}{.24\linewidth}
\centering
\includegraphics[width=\linewidth]{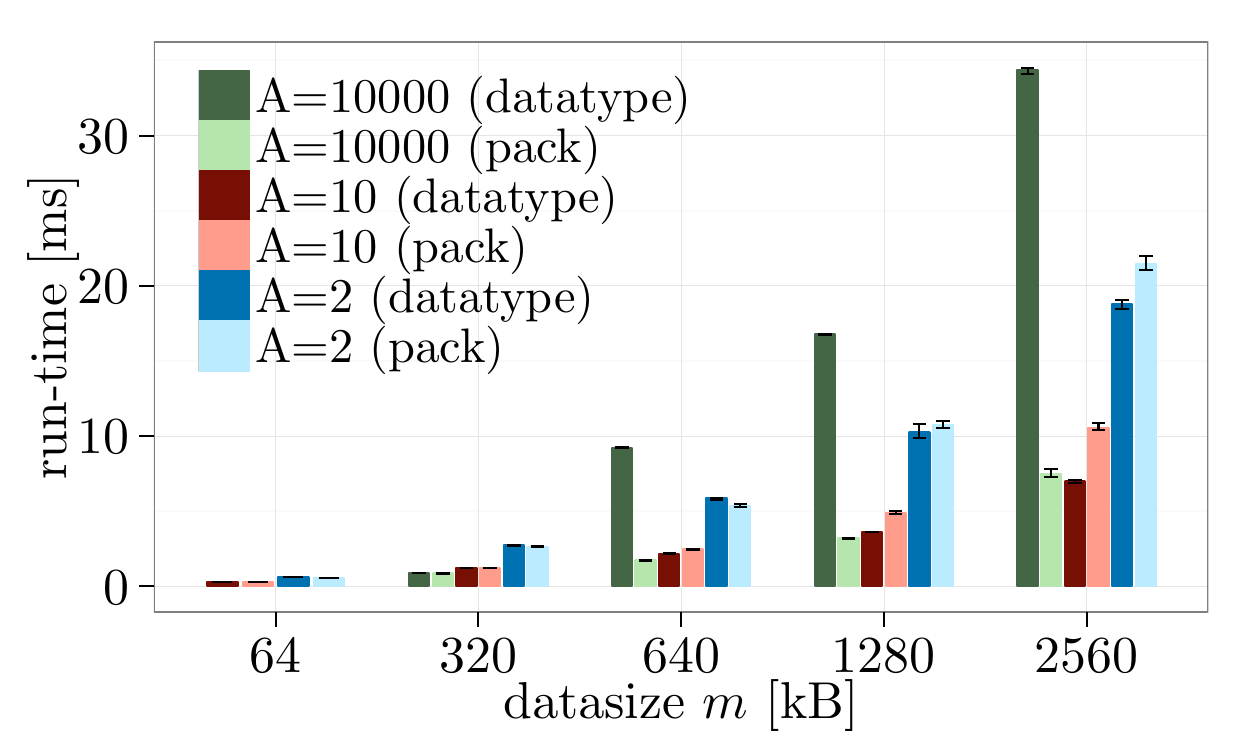}
\caption{%
\label{exp:bcast-pack-alternating-32x1-openmpi}%
\dtalternating%
}%
\end{subfigure}%
\caption{\label{exp:bcast-pack-32x1-openmpi}  Basic layouts \vs pack/unpack, element datatype: \mpiint, \num{32x1}~processes, \mpibcast, \jupiteropenmpi.}
\end{figure*}

\begin{figure*}[htpb]
\centering
\begin{subfigure}{.24\linewidth}
\centering
\includegraphics[width=\linewidth]{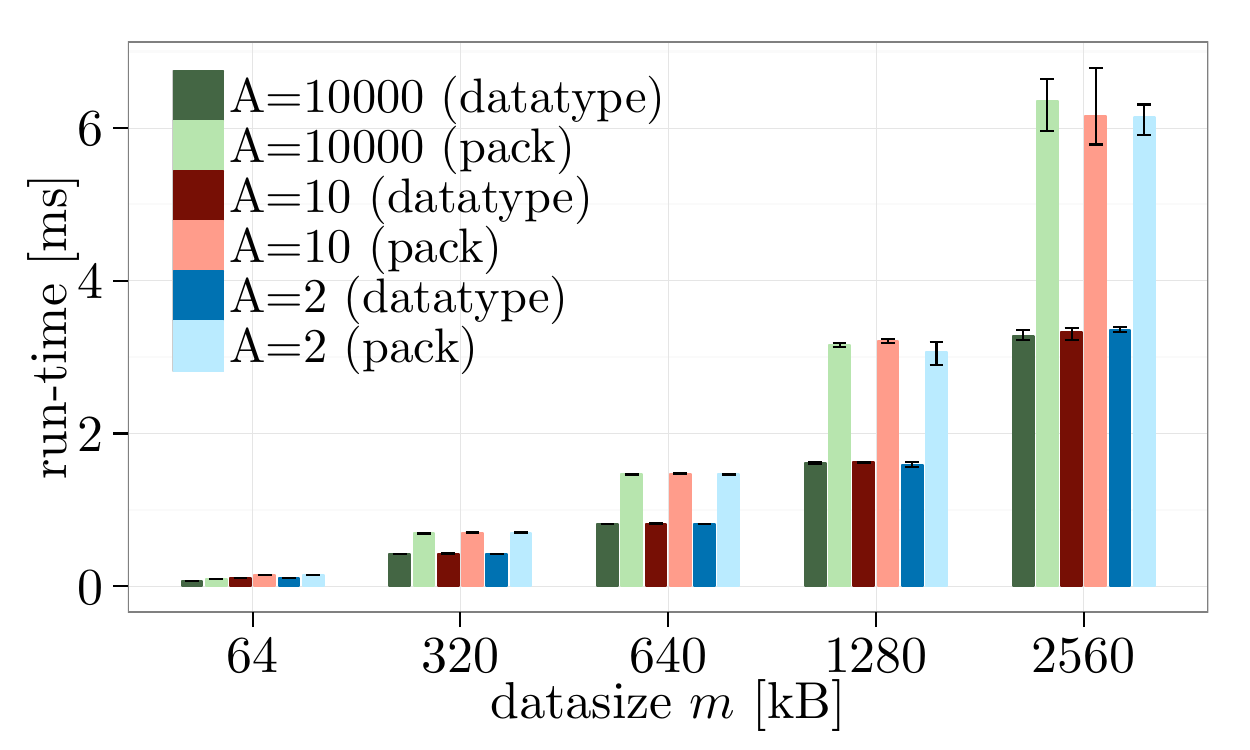}
\caption{%
\label{exp:pingpong-pack-tiled-onenode-nec}%
\dttiled%
}%
\end{subfigure}%
\hfill%
\begin{subfigure}{.24\linewidth}
\centering
\includegraphics[width=\linewidth]{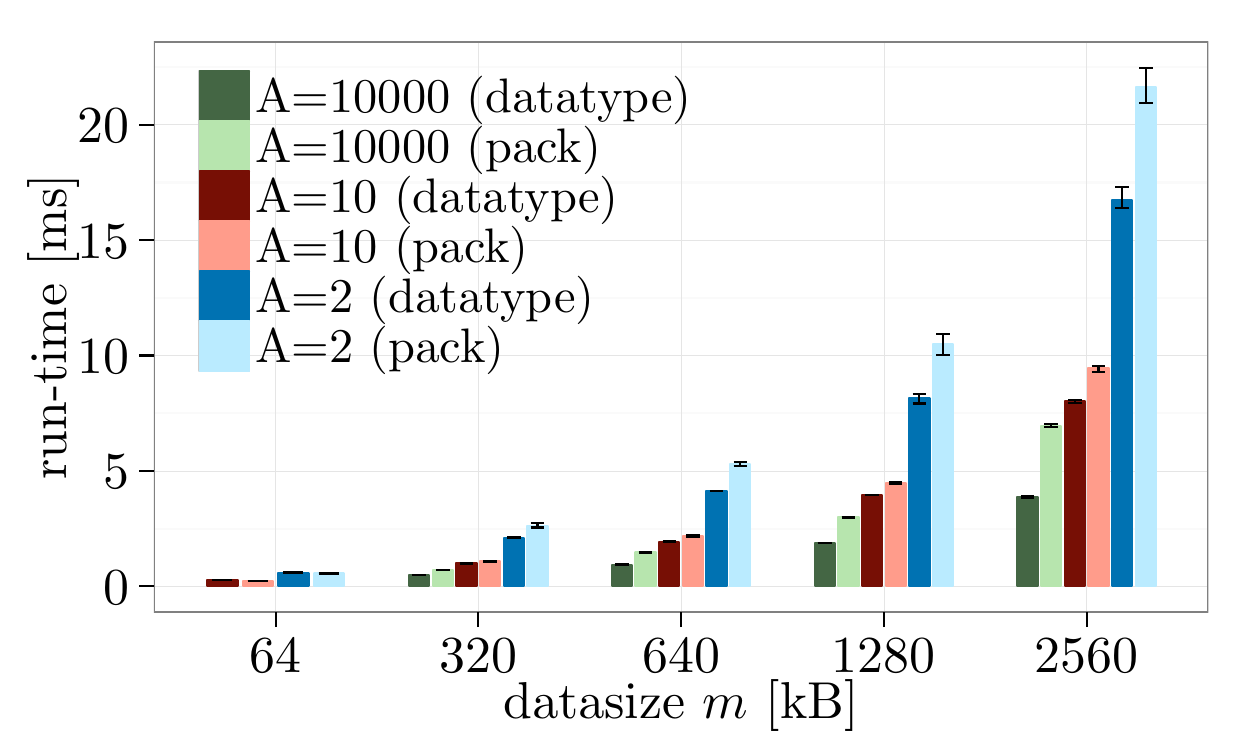}
\caption{%
\label{exp:pingpong-pack-block-onenode-nec}%
\dtblock%
}%
\end{subfigure}%
\hfill%
\begin{subfigure}{.24\linewidth}
\centering
\includegraphics[width=\linewidth]{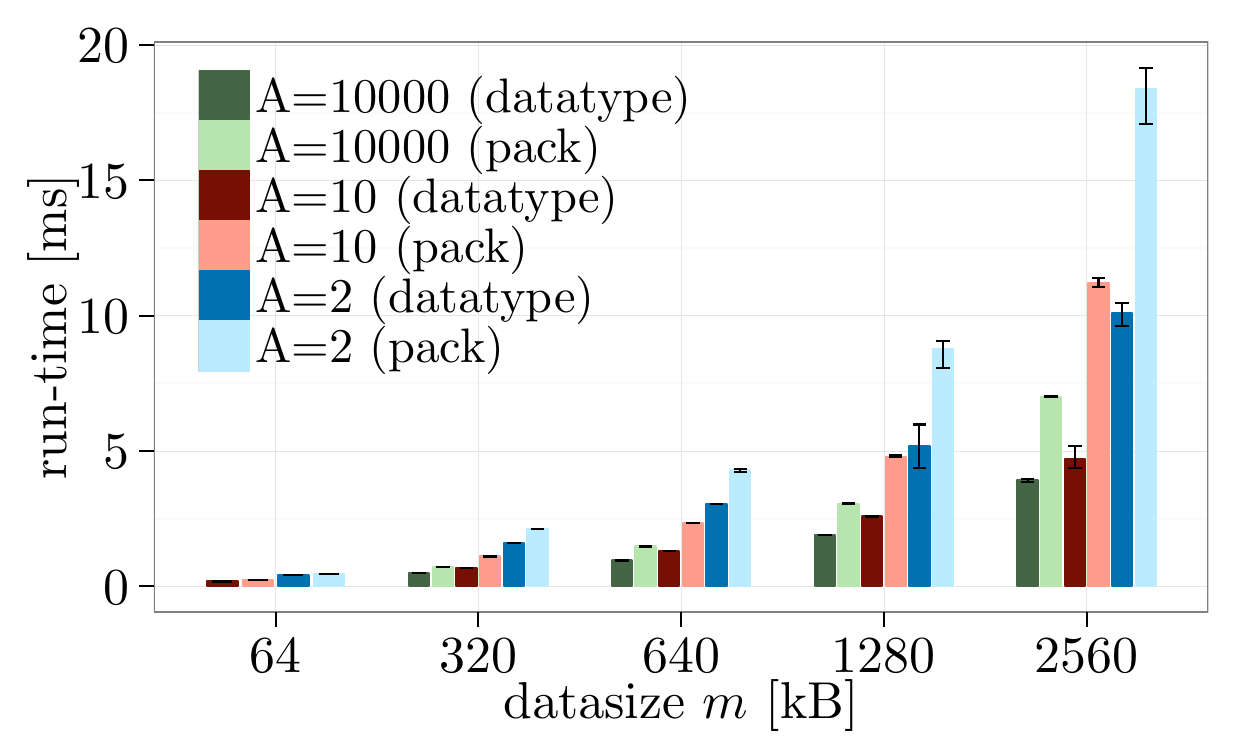}
\caption{%
\label{exp:pingpong-pack-bucket-onenode-nec}%
\dtbucket%
}%
\end{subfigure}%
\hfill%
\begin{subfigure}{.24\linewidth}
\centering
\includegraphics[width=\linewidth]{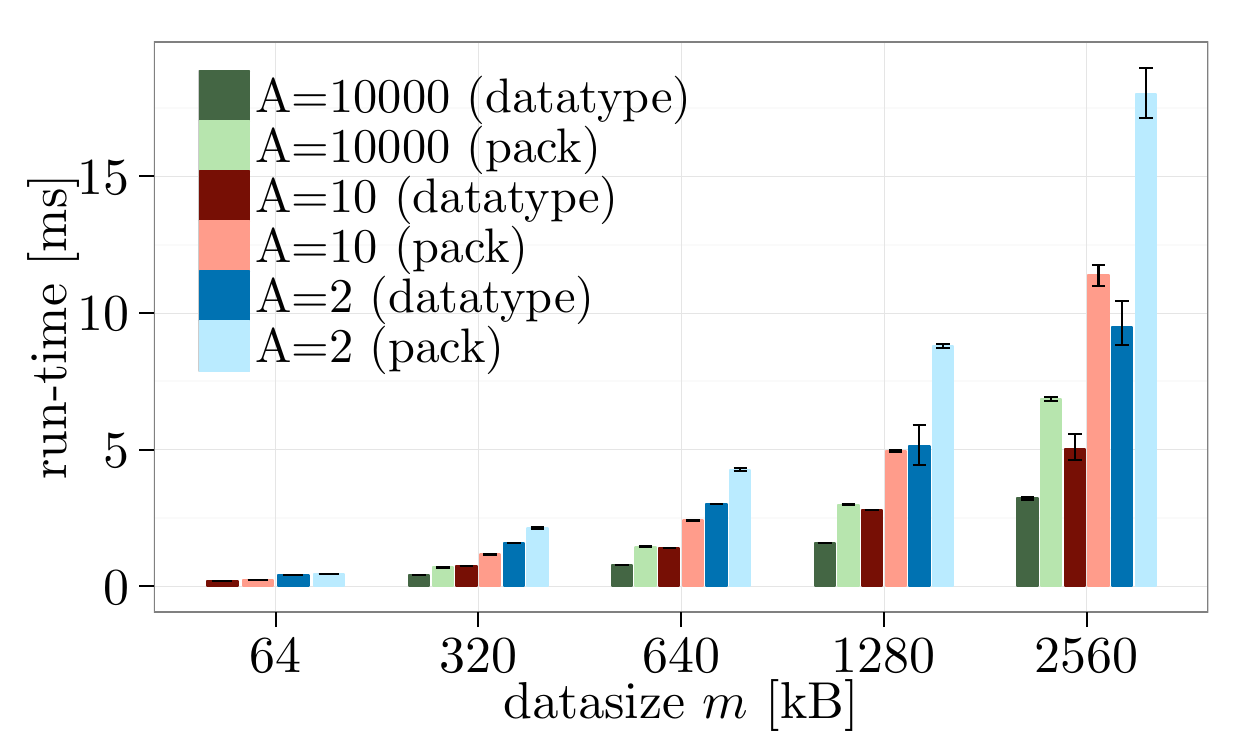}
\caption{%
\label{exp:pingpong-pack-alternating-onenode-nec}%
\dtalternating%
}%
\end{subfigure}%
\caption{\label{exp:pingpong-pack-onenode-nec}  Basic layouts \vs pack/unpack, element datatype: \mpiint, one~node, \num{2}~processes, \pingpong, \jupiternecmpi.}
\end{figure*}

\begin{figure*}[htpb]
\centering
\begin{subfigure}{.24\linewidth}
\centering
\includegraphics[width=\linewidth]{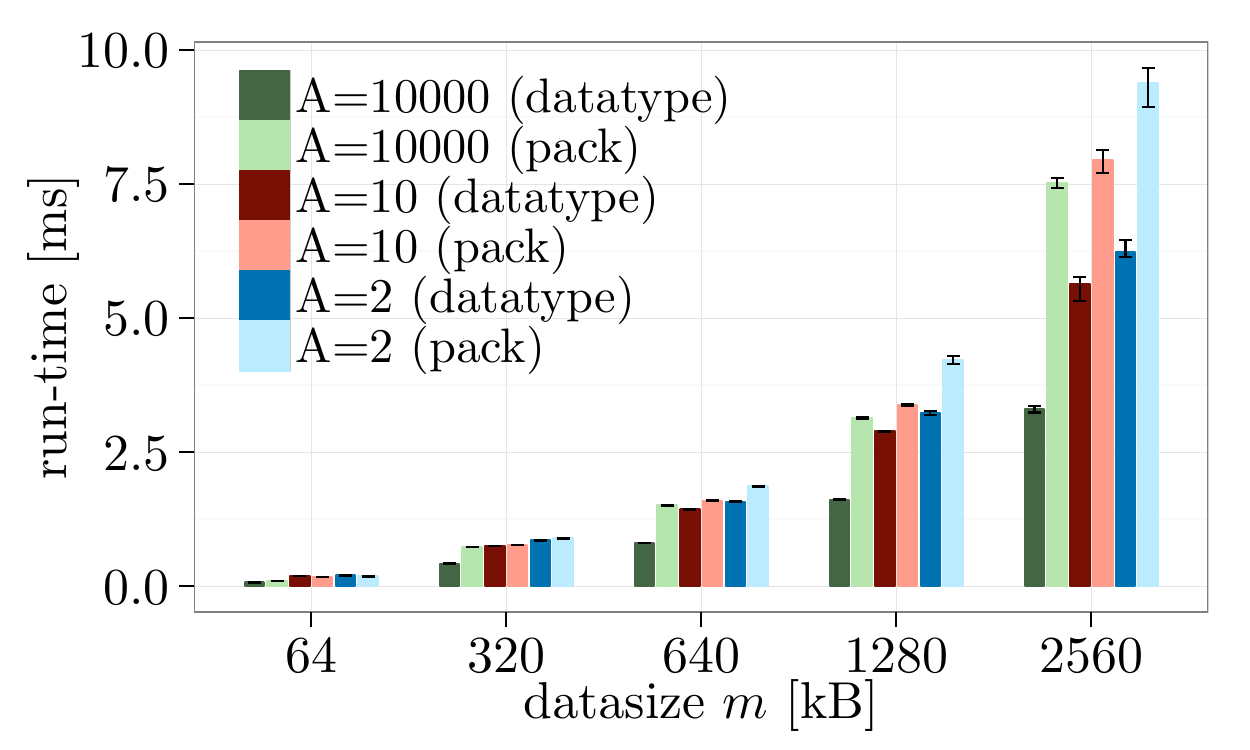}
\caption{%
\label{exp:pingpong-pack-tiled-onenode-mvapich}%
\dttiled%
}%
\end{subfigure}%
\hfill%
\begin{subfigure}{.24\linewidth}
\centering
\includegraphics[width=\linewidth]{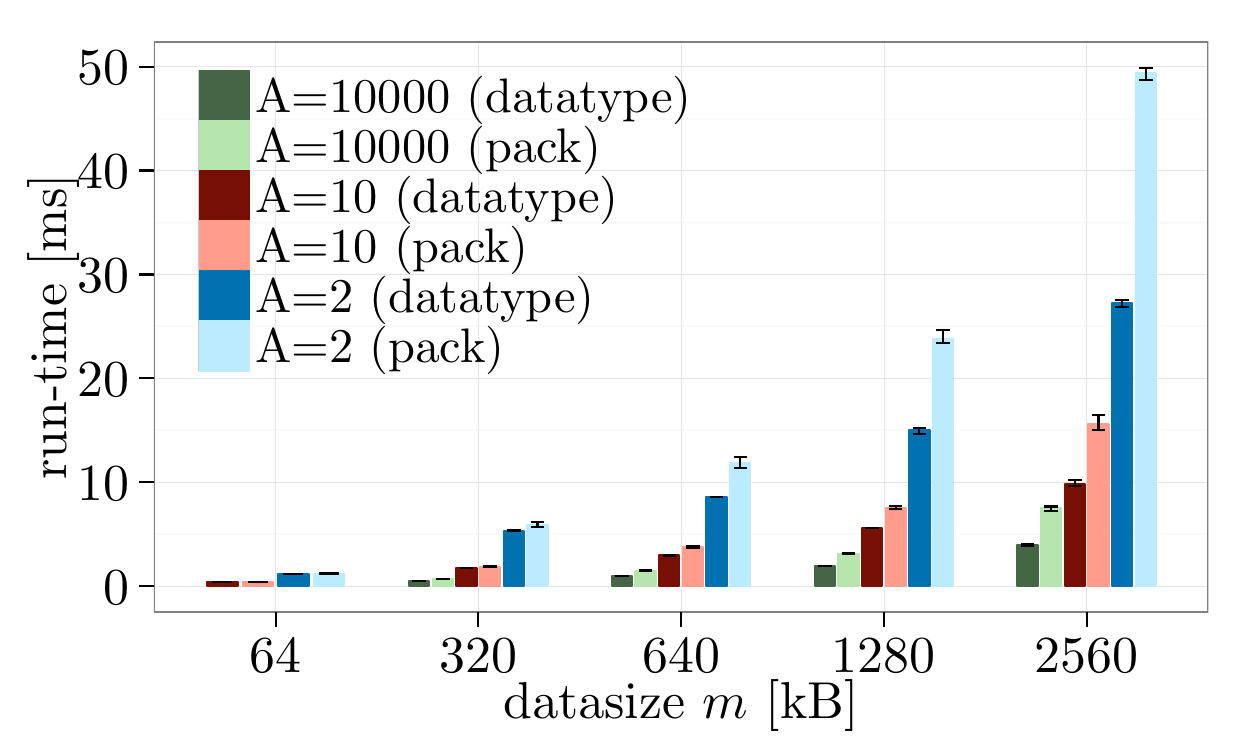}
\caption{%
\label{exp:pingpong-pack-block-onenode-mvapich}%
\dtblock%
}%
\end{subfigure}%
\hfill%
\begin{subfigure}{.24\linewidth}
\centering
\includegraphics[width=\linewidth]{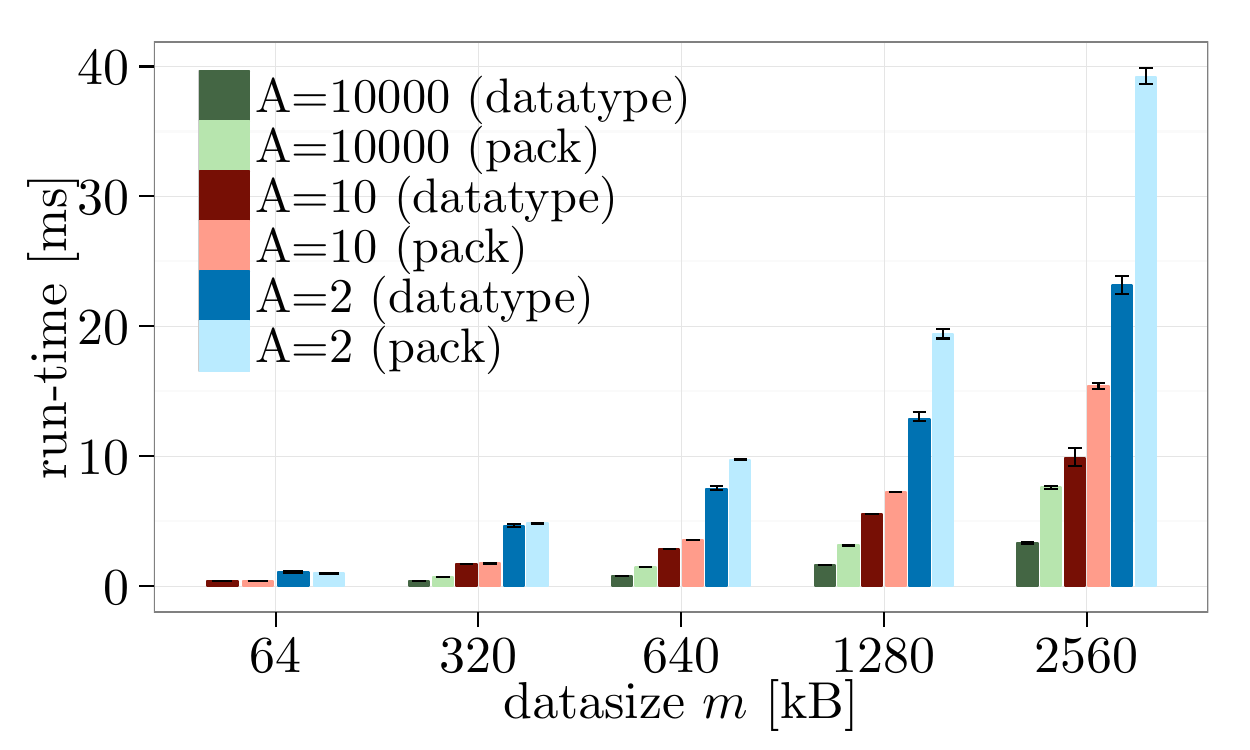}
\caption{%
\label{exp:pingpong-pack-bucket-onenode-mvapich}%
\dtbucket%
}%
\end{subfigure}%
\hfill%
\begin{subfigure}{.24\linewidth}
\centering
\includegraphics[width=\linewidth]{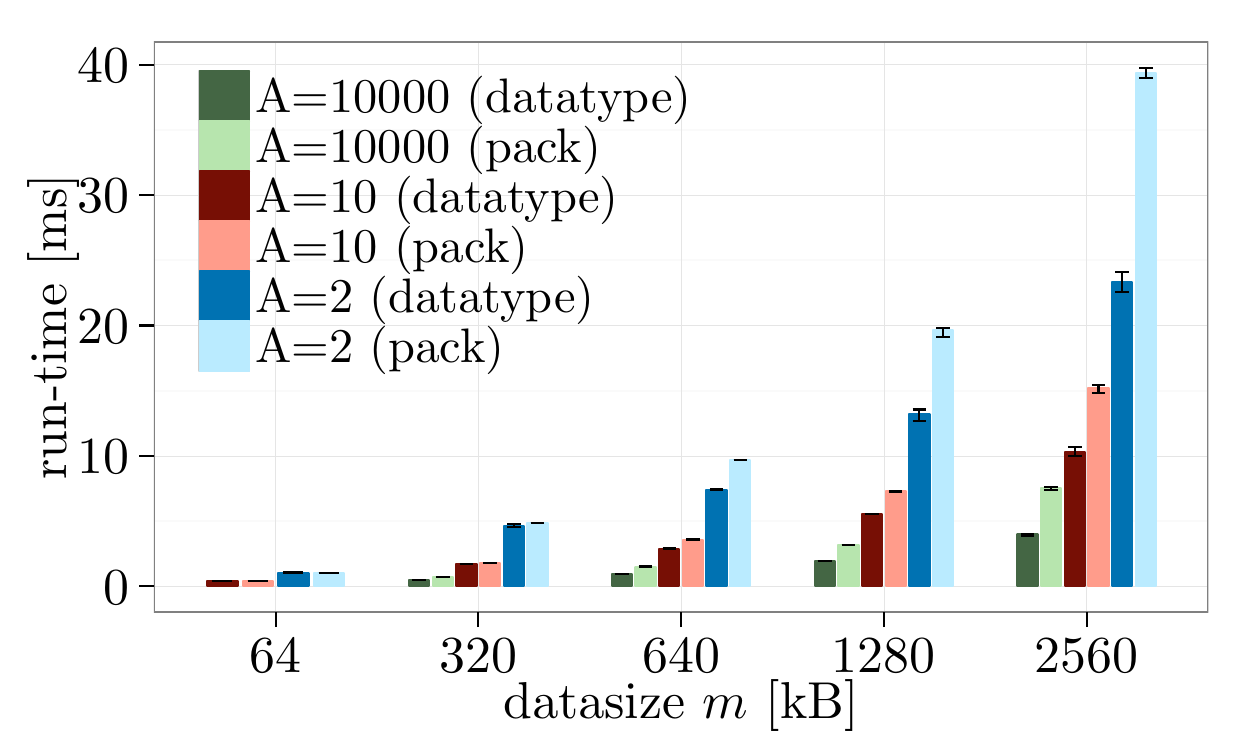}
\caption{%
\label{exp:pingpong-pack-alternating-onenode-mvapich}%
\dtalternating%
}%
\end{subfigure}%
\caption{\label{exp:pingpong-pack-onenode-mvapich}  Basic layouts \vs pack/unpack, element datatype: \mpiint, one~node, \num{2}~processes, \pingpong, \jupitermvapich.}
\end{figure*}

\begin{figure*}[htpb]
\centering
\begin{subfigure}{.24\linewidth}
\centering
\includegraphics[width=\linewidth]{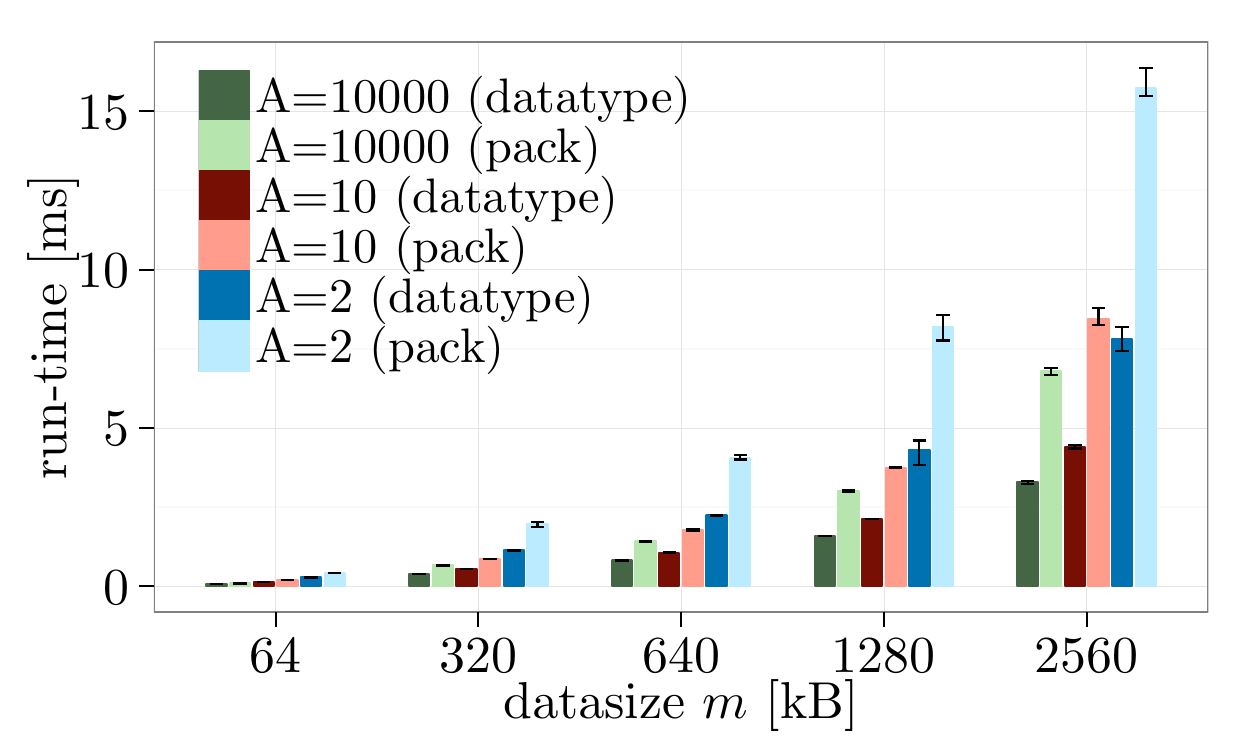}
\caption{%
\label{exp:pingpong-pack-tiled-onenode-openmpi}%
\dttiled%
}%
\end{subfigure}%
\hfill%
\begin{subfigure}{.24\linewidth}
\centering
\includegraphics[width=\linewidth]{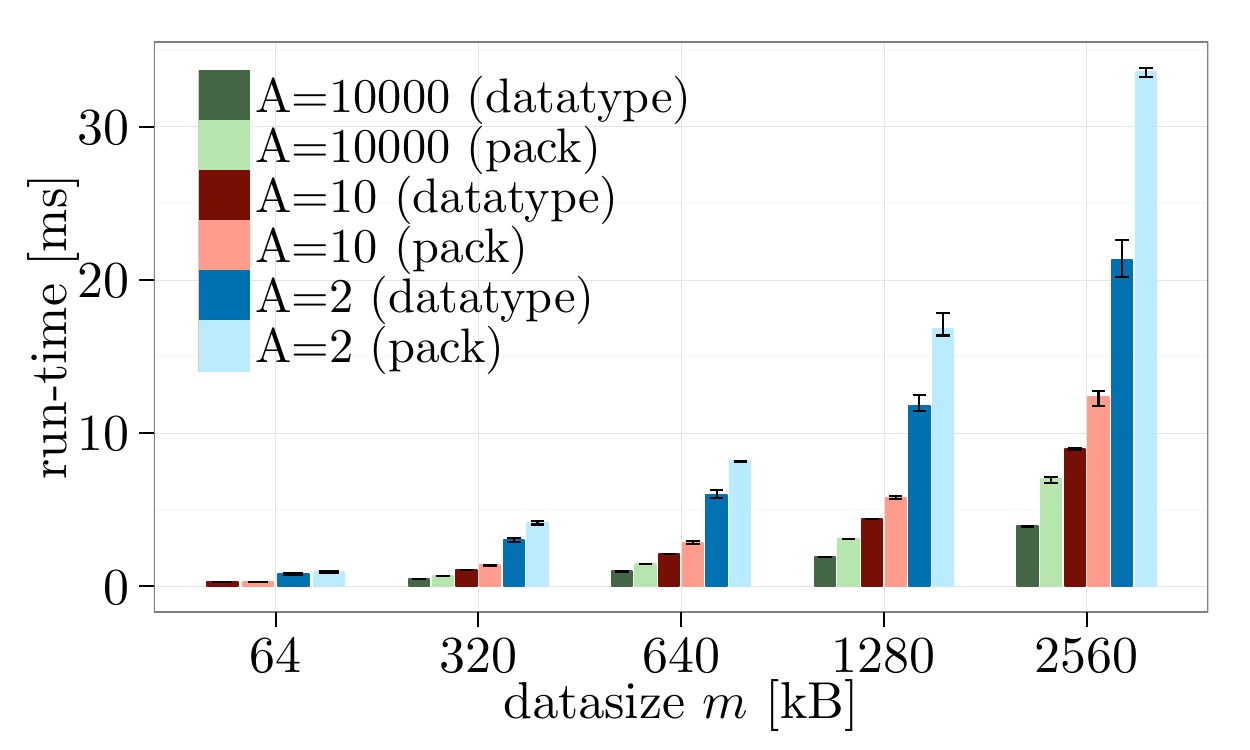}
\caption{%
\label{exp:pingpong-pack-block-onenode-openmpi}%
\dtblock%
}%
\end{subfigure}%
\hfill%
\begin{subfigure}{.24\linewidth}
\centering
\includegraphics[width=\linewidth]{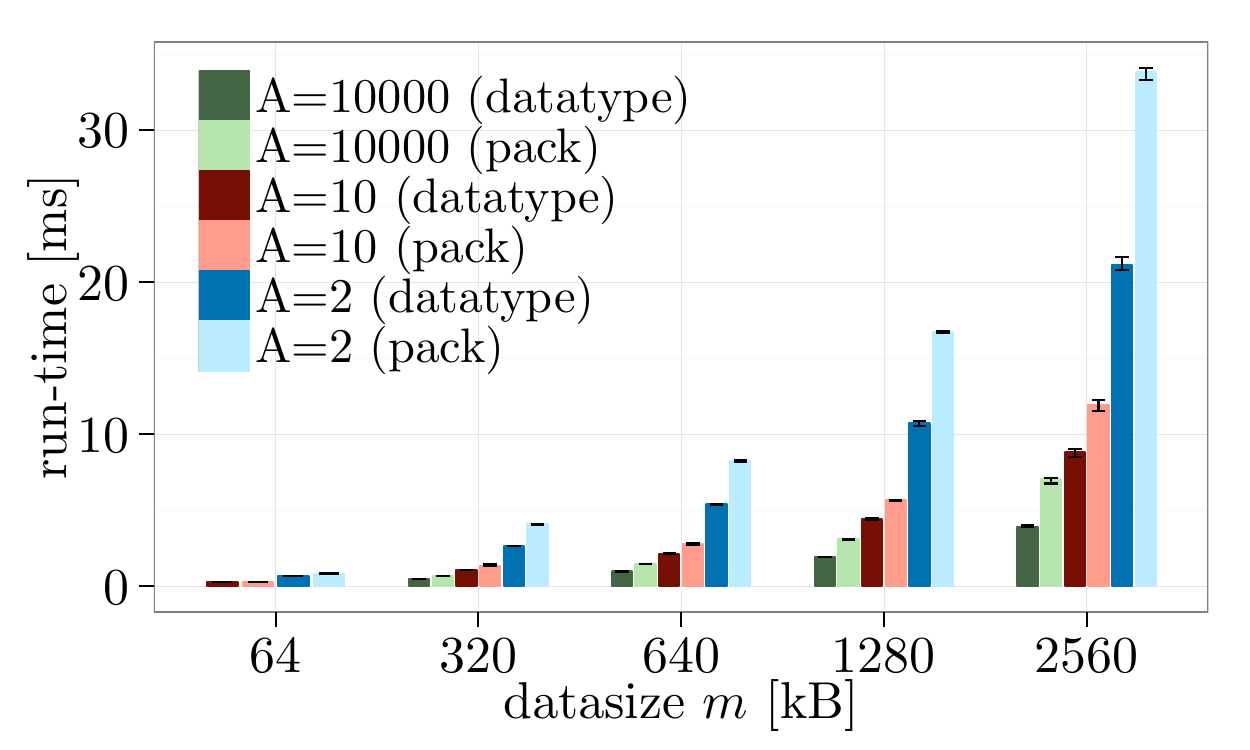}
\caption{%
\label{exp:pingpong-pack-bucket-onenode-openmpi}%
\dtbucket%
}%
\end{subfigure}%
\hfill%
\begin{subfigure}{.24\linewidth}
\centering
\includegraphics[width=\linewidth]{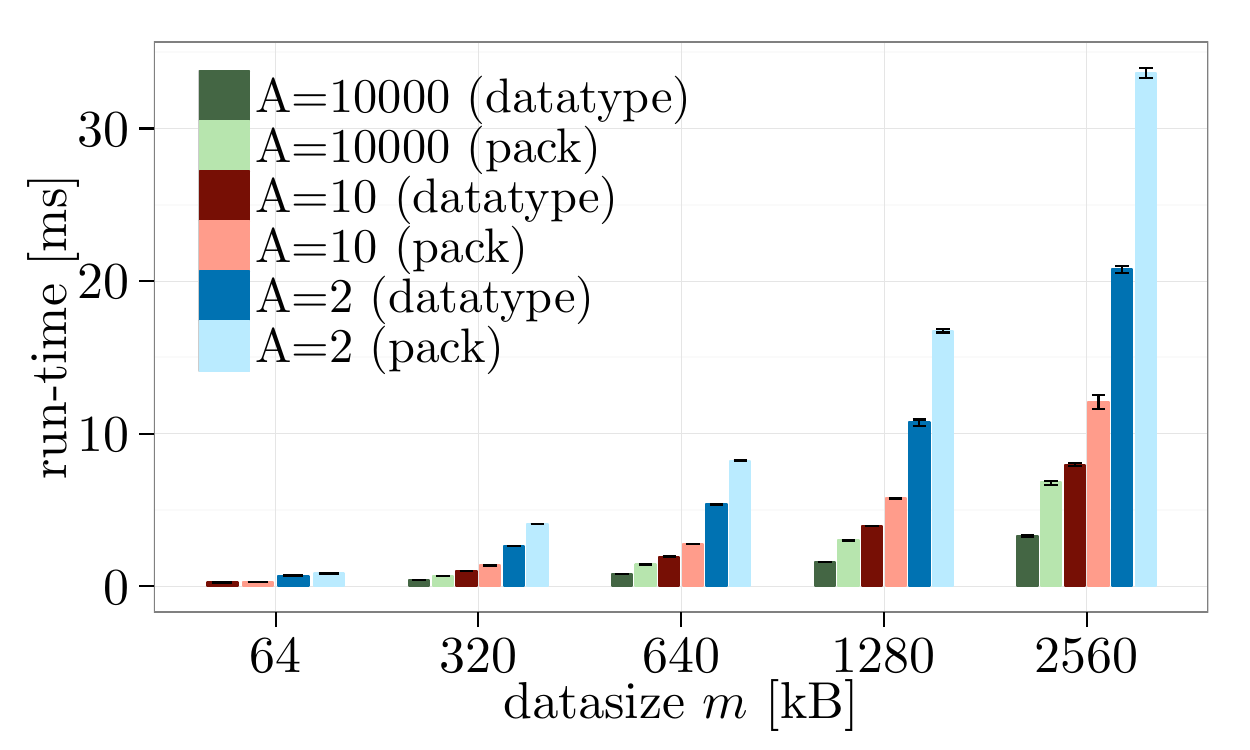}
\caption{%
\label{exp:pingpong-pack-alternating-onenode-openmpi}%
\dtalternating%
}%
\end{subfigure}%
\caption{\label{exp:pingpong-pack-onenode-openmpi}  Basic layouts \vs pack/unpack, element datatype: \mpiint, one~node, \num{2}~processes, \pingpong, \jupiteropenmpi.}
\end{figure*}

\begin{figure*}[htpb]
\centering
\begin{subfigure}{.24\linewidth}
\centering
\includegraphics[width=\linewidth]{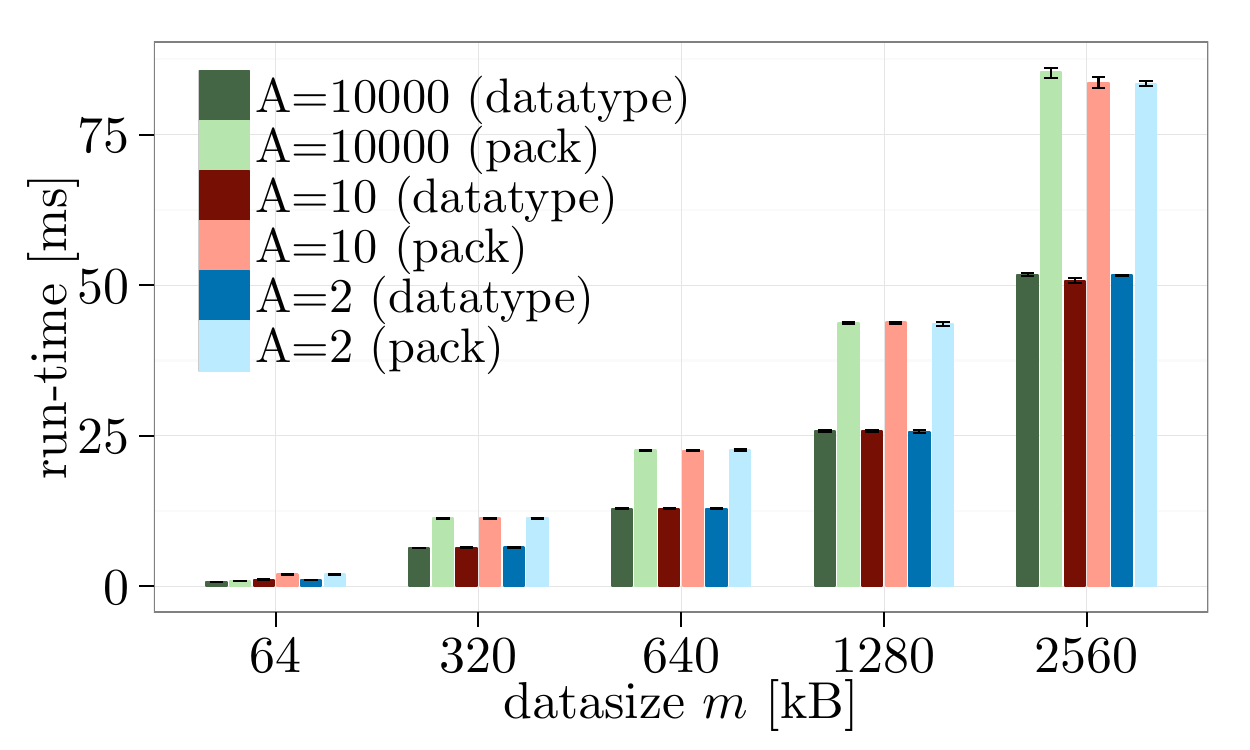}
\caption{%
\label{exp:allgather-pack-tiled-onenode-nec}%
\dttiled%
}%
\end{subfigure}%
\hfill%
\begin{subfigure}{.24\linewidth}
\centering
\includegraphics[width=\linewidth]{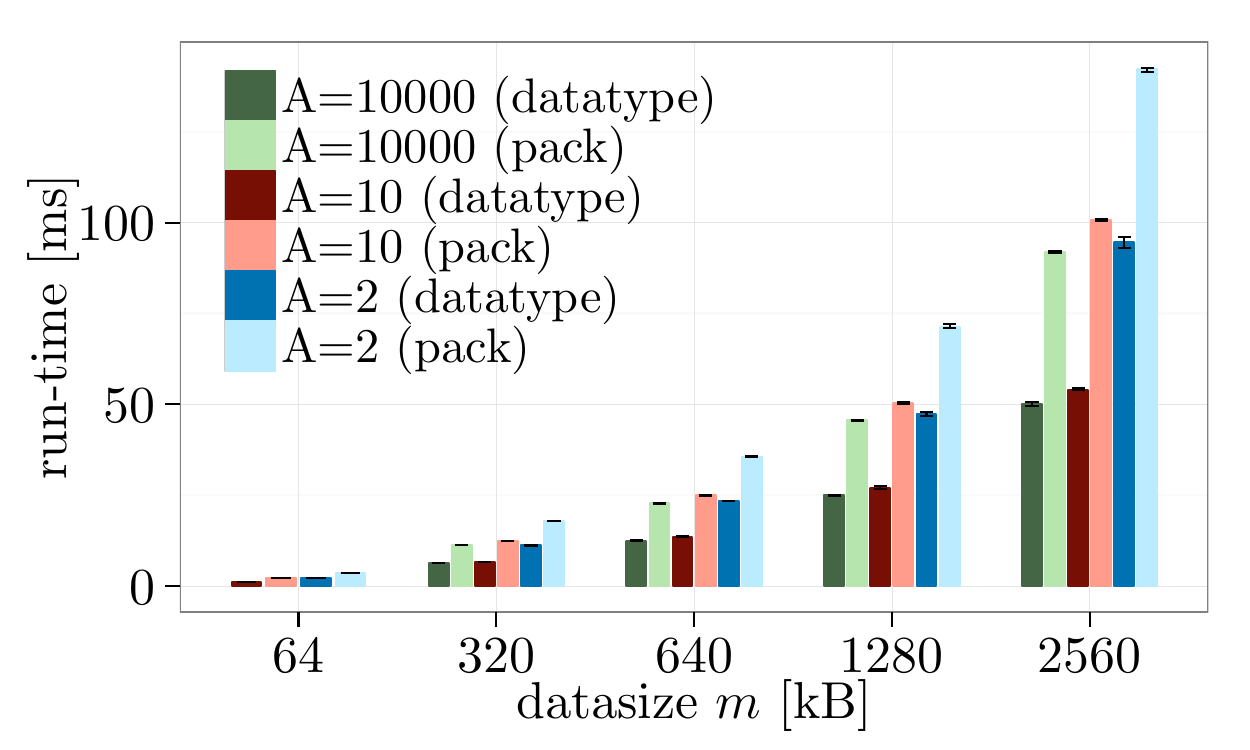}
\caption{%
\label{exp:allgather-pack-block-onenode-nec}%
\dtblock%
}%
\end{subfigure}%
\hfill%
\begin{subfigure}{.24\linewidth}
\centering
\includegraphics[width=\linewidth]{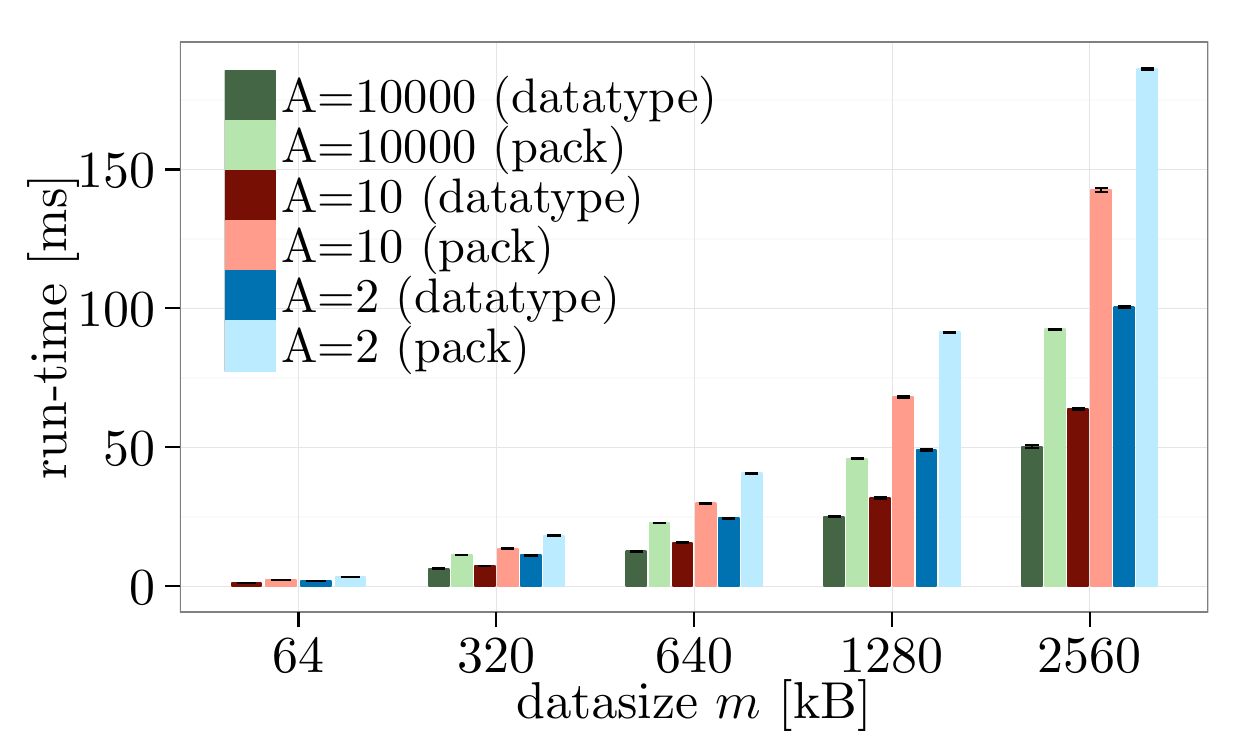}
\caption{%
\label{exp:allgather-pack-bucket-onenode-nec}%
\dtbucket%
}%
\end{subfigure}%
\hfill%
\begin{subfigure}{.24\linewidth}
\centering
\includegraphics[width=\linewidth]{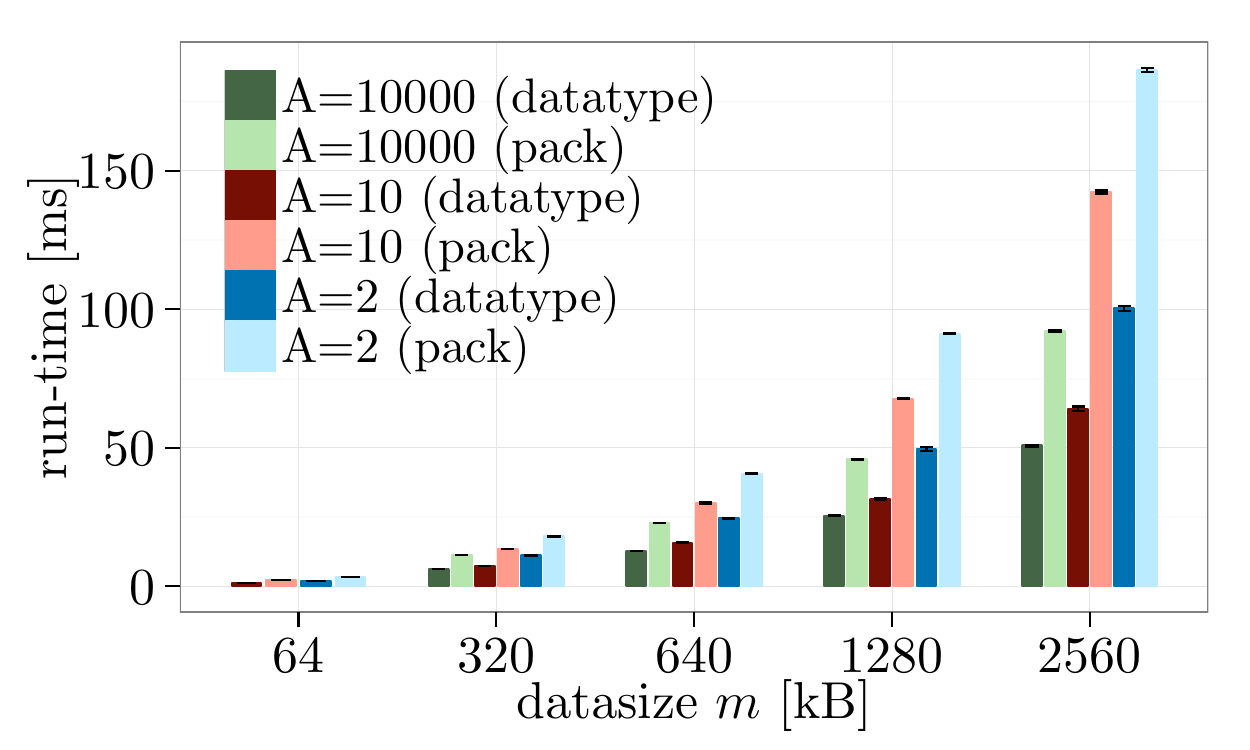}
\caption{%
\label{exp:allgather-pack-alternating-onenode-nec}%
\dtalternating%
}%
\end{subfigure}%
\caption{\label{exp:allgather-pack-onenode-nec}  Basic layouts \vs pack/unpack, element datatype: \mpiint, one~node, \num{16}~processes, \mpiallgather, \jupiternecmpi.}
\end{figure*}

\begin{figure*}[htpb]
\centering
\begin{subfigure}{.24\linewidth}
\centering
\includegraphics[width=\linewidth]{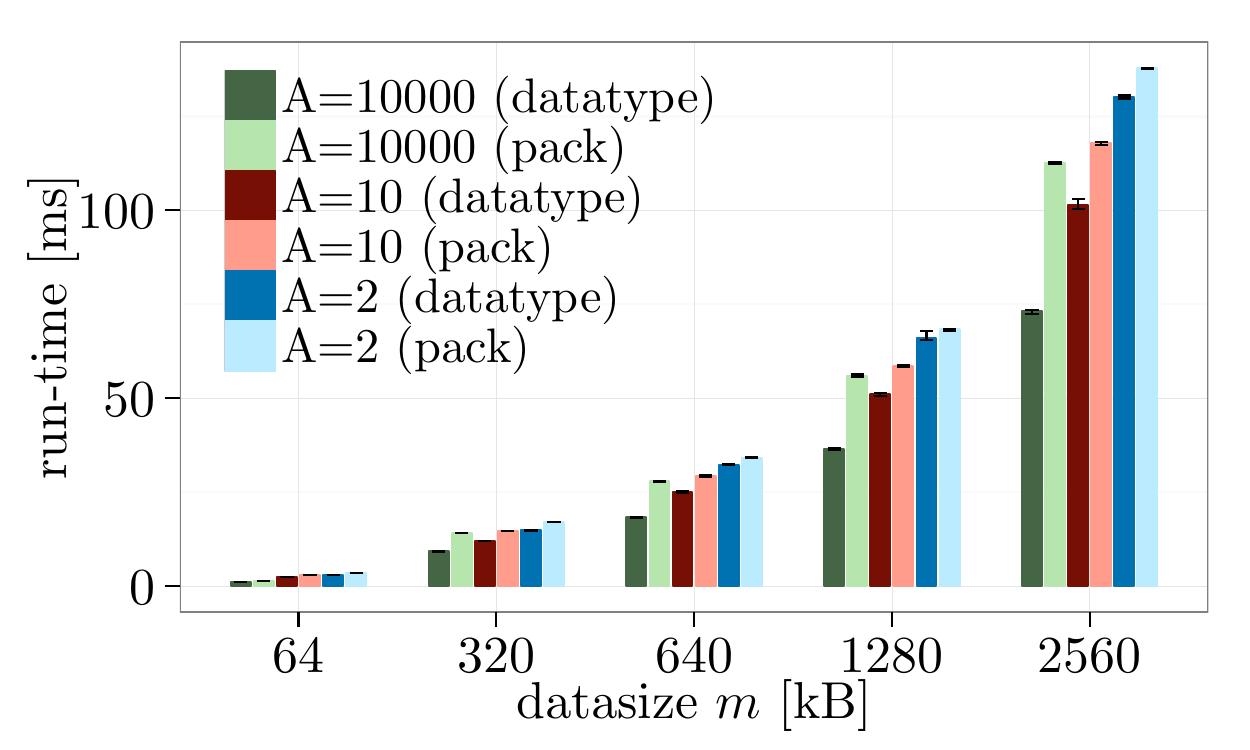}
\caption{%
\label{exp:allgather-pack-tiled-onenode-mvapich}%
\dttiled%
}%
\end{subfigure}%
\hfill%
\begin{subfigure}{.24\linewidth}
\centering
\includegraphics[width=\linewidth]{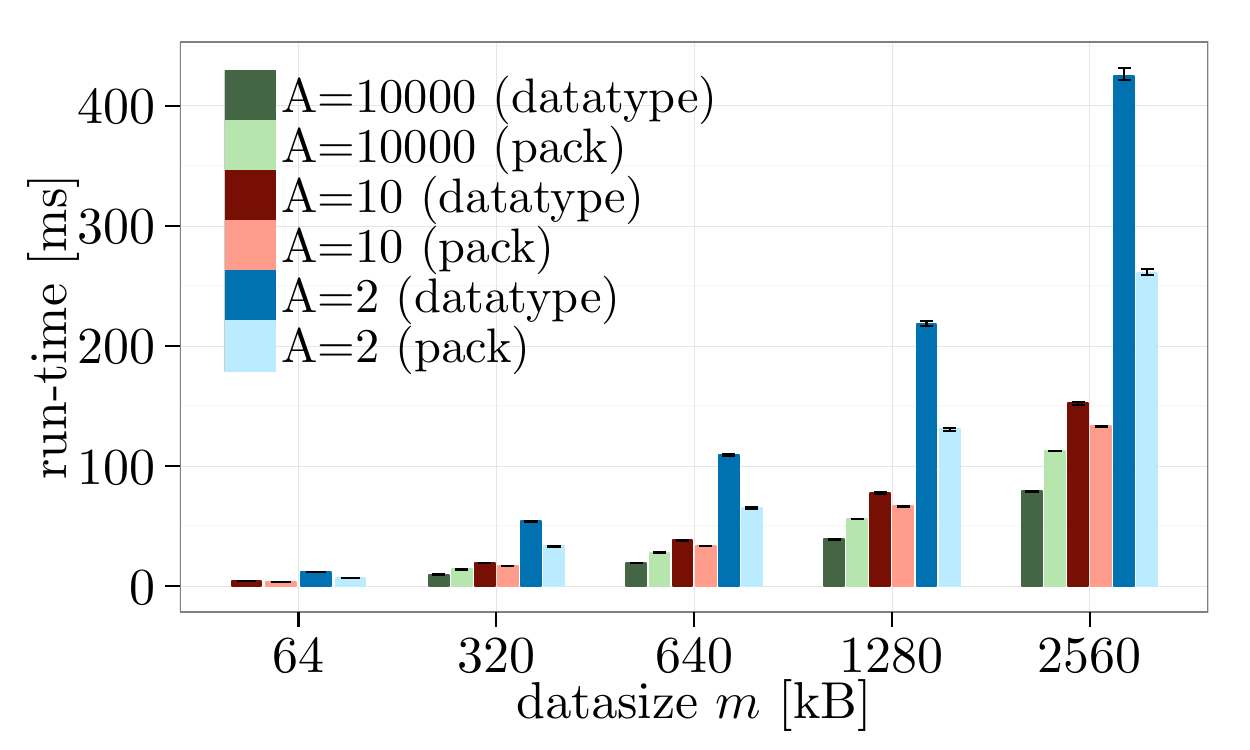}
\caption{%
\label{exp:allgather-pack-block-onenode-mvapich}%
\dtblock%
}%
\end{subfigure}%
\hfill%
\begin{subfigure}{.24\linewidth}
\centering
\includegraphics[width=\linewidth]{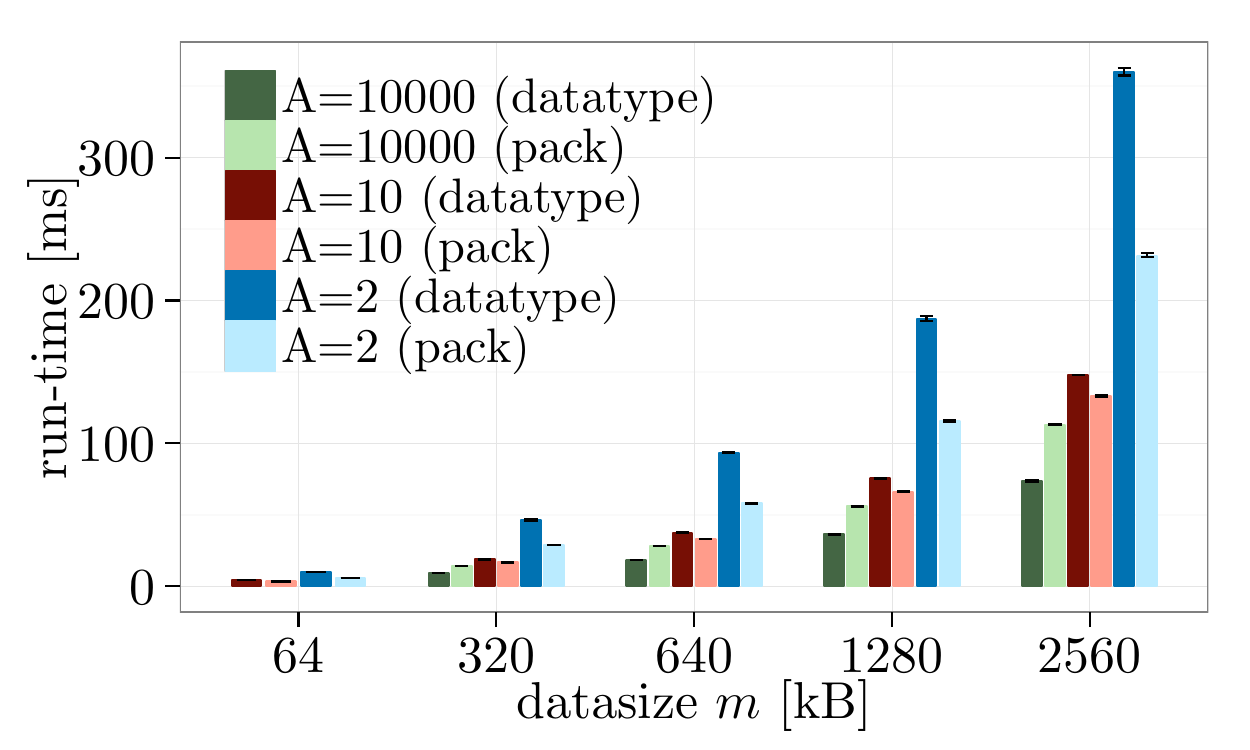}
\caption{%
\label{exp:allgather-pack-bucket-onenode-mvapich}%
\dtbucket%
}%
\end{subfigure}%
\hfill%
\begin{subfigure}{.24\linewidth}
\centering
\includegraphics[width=\linewidth]{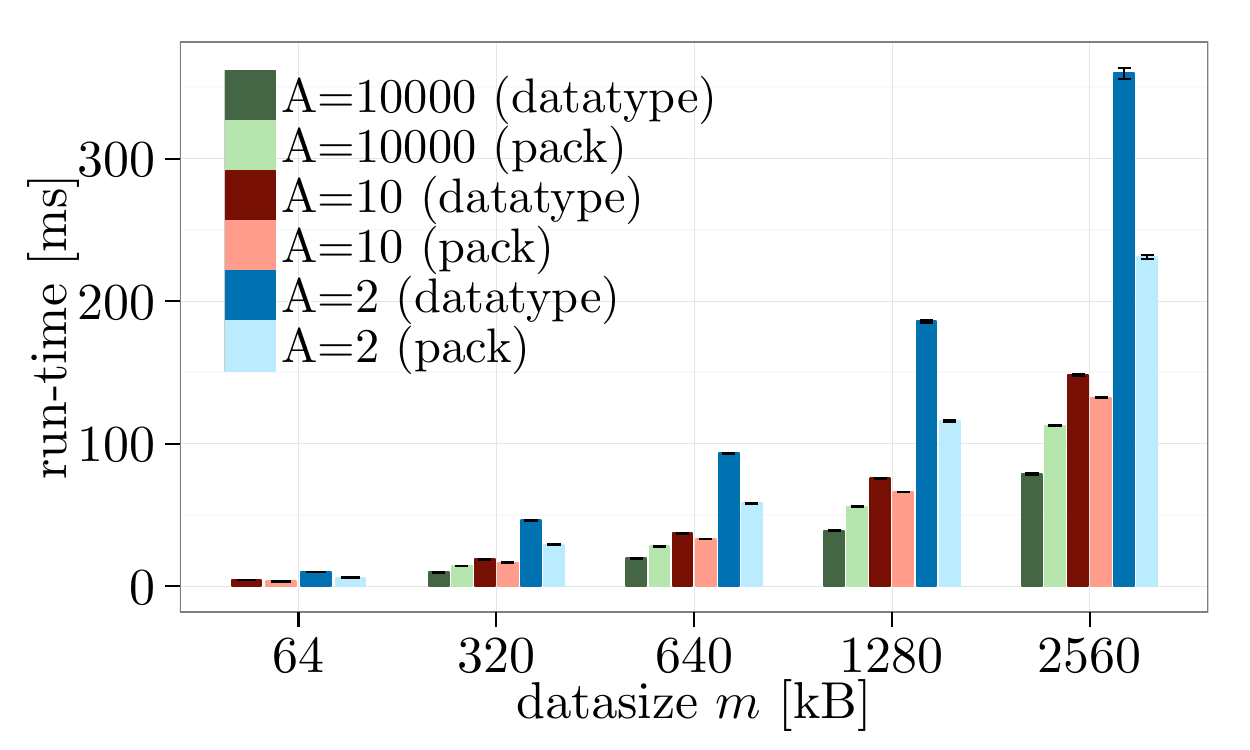}
\caption{%
\label{exp:allgather-pack-alternating-onenode-mvapich}%
\dtalternating%
}%
\end{subfigure}%
\caption{\label{exp:allgather-pack-onenode-mvapich}  Basic layouts \vs pack/unpack, element datatype: \mpiint, one~node, \num{16}~processes, \mpiallgather, \jupitermvapich.}
\end{figure*}

\begin{figure*}[htpb]
\centering
\begin{subfigure}{.24\linewidth}
\centering
\includegraphics[width=\linewidth]{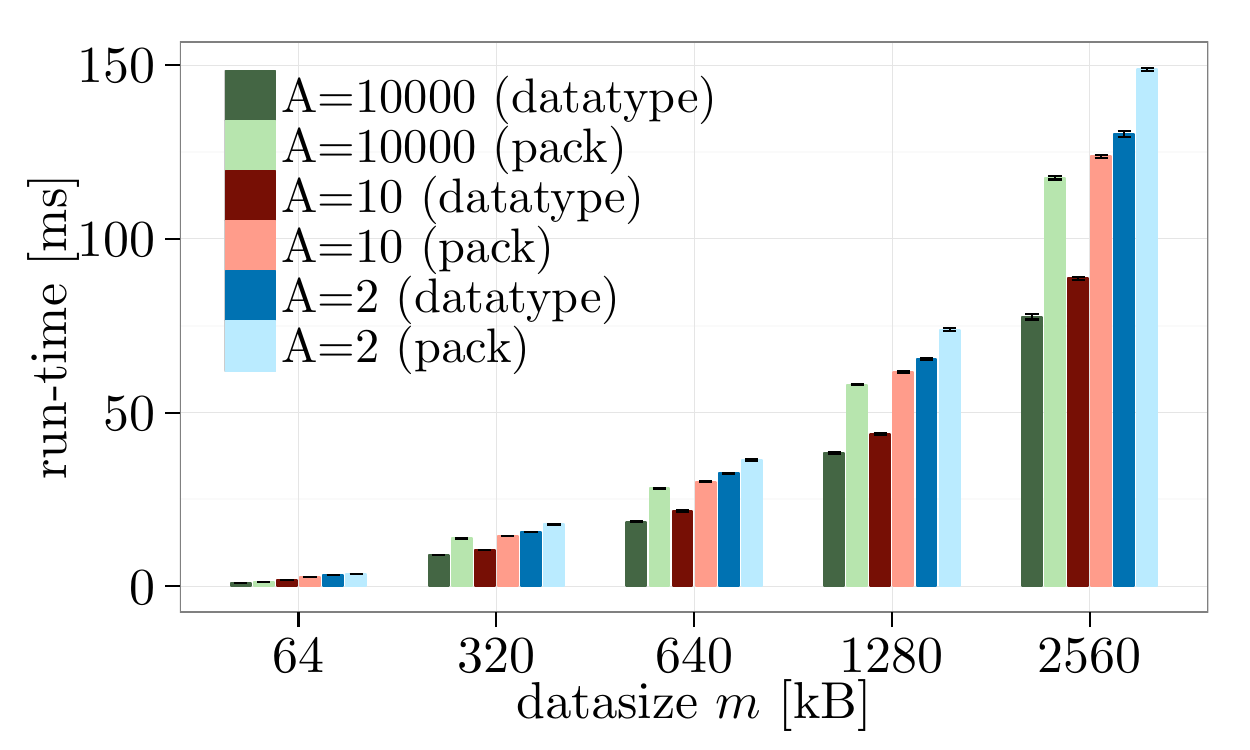}
\caption{%
\label{exp:allgather-pack-tiled-onenode-openmpi}%
\dttiled%
}%
\end{subfigure}%
\hfill%
\begin{subfigure}{.24\linewidth}
\centering
\includegraphics[width=\linewidth]{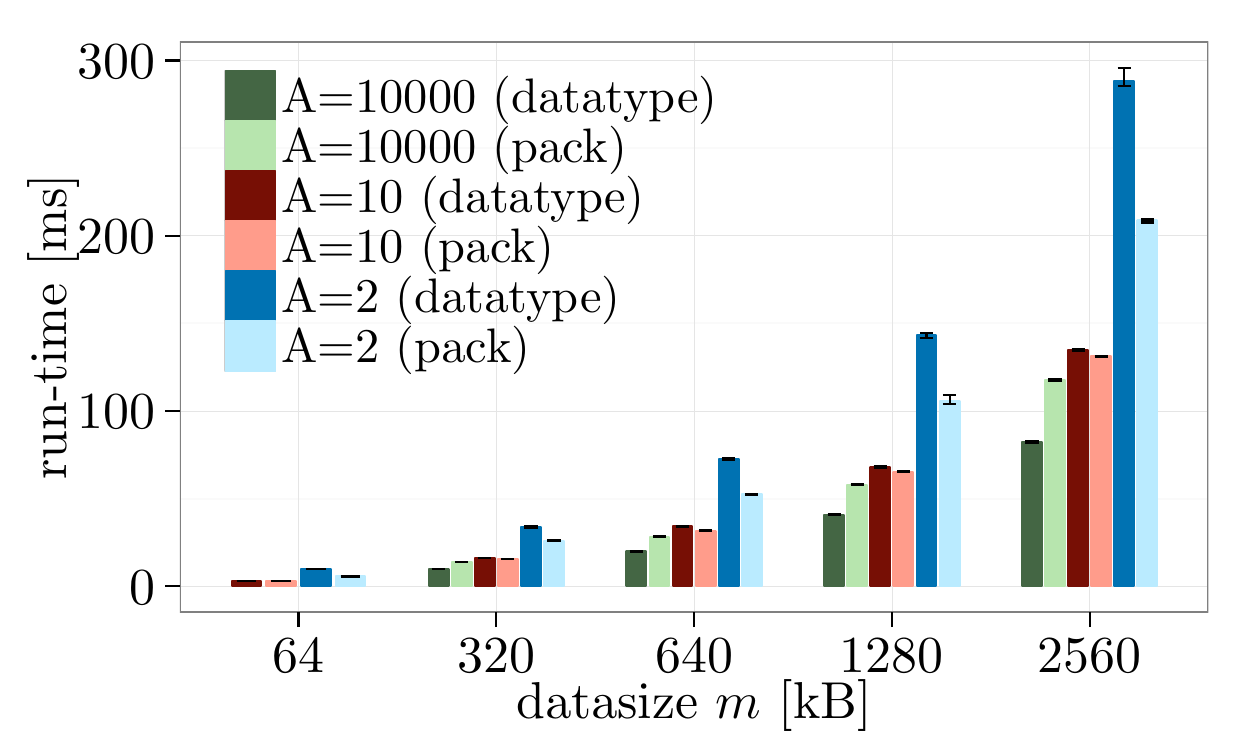}
\caption{%
\label{exp:allgather-pack-block-onenode-openmpi}%
\dtblock%
}%
\end{subfigure}%
\hfill%
\begin{subfigure}{.24\linewidth}
\centering
\includegraphics[width=\linewidth]{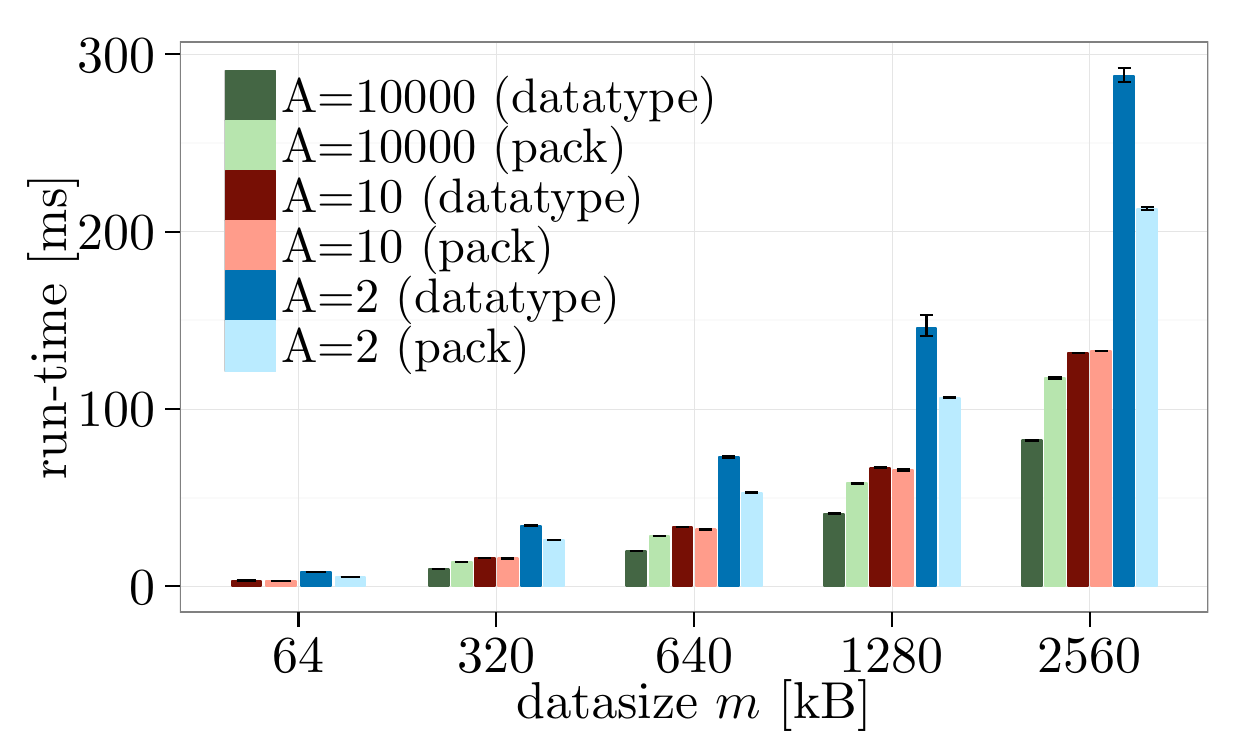}
\caption{%
\label{exp:allgather-pack-bucket-onenode-openmpi}%
\dtbucket%
}%
\end{subfigure}%
\hfill%
\begin{subfigure}{.24\linewidth}
\centering
\includegraphics[width=\linewidth]{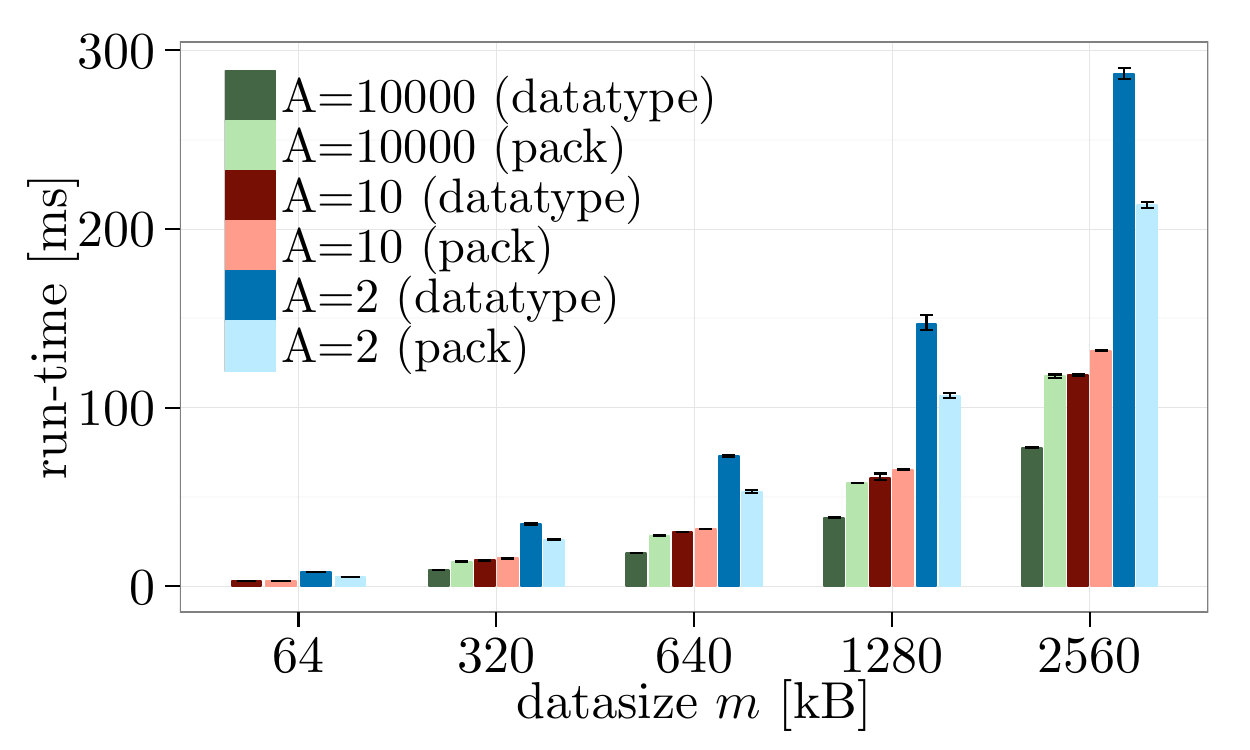}
\caption{%
\label{exp:allgather-pack-alternating-onenode-openmpi}%
\dtalternating%
}%
\end{subfigure}%
\caption{\label{exp:allgather-pack-onenode-openmpi}  Basic layouts \vs pack/unpack, element datatype: \mpiint, one~node, \num{16}~processes, \mpiallgather, \jupiteropenmpi.}
\end{figure*}

\begin{figure*}[htpb]
\centering
\begin{subfigure}{.24\linewidth}
\centering
\includegraphics[width=\linewidth]{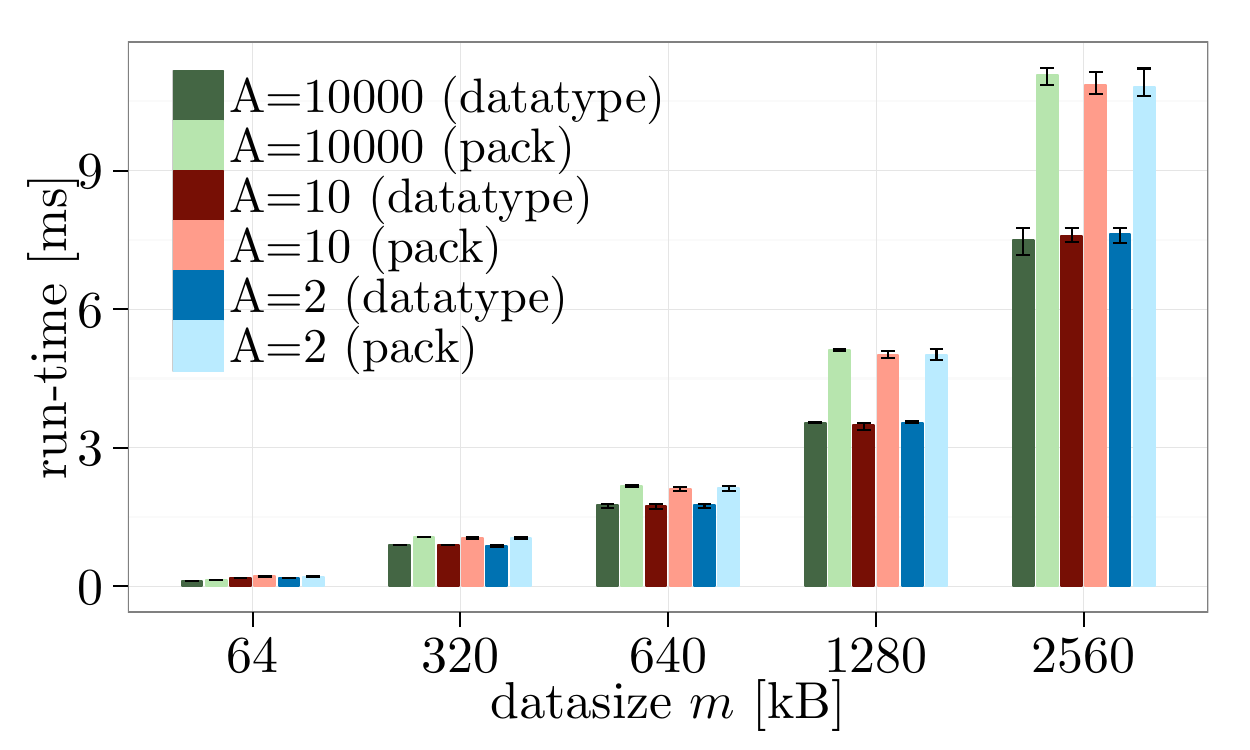}
\caption{%
\label{exp:bcast-pack-tiled-onenode-nec}%
\dttiled%
}%
\end{subfigure}%
\hfill%
\begin{subfigure}{.24\linewidth}
\centering
\includegraphics[width=\linewidth]{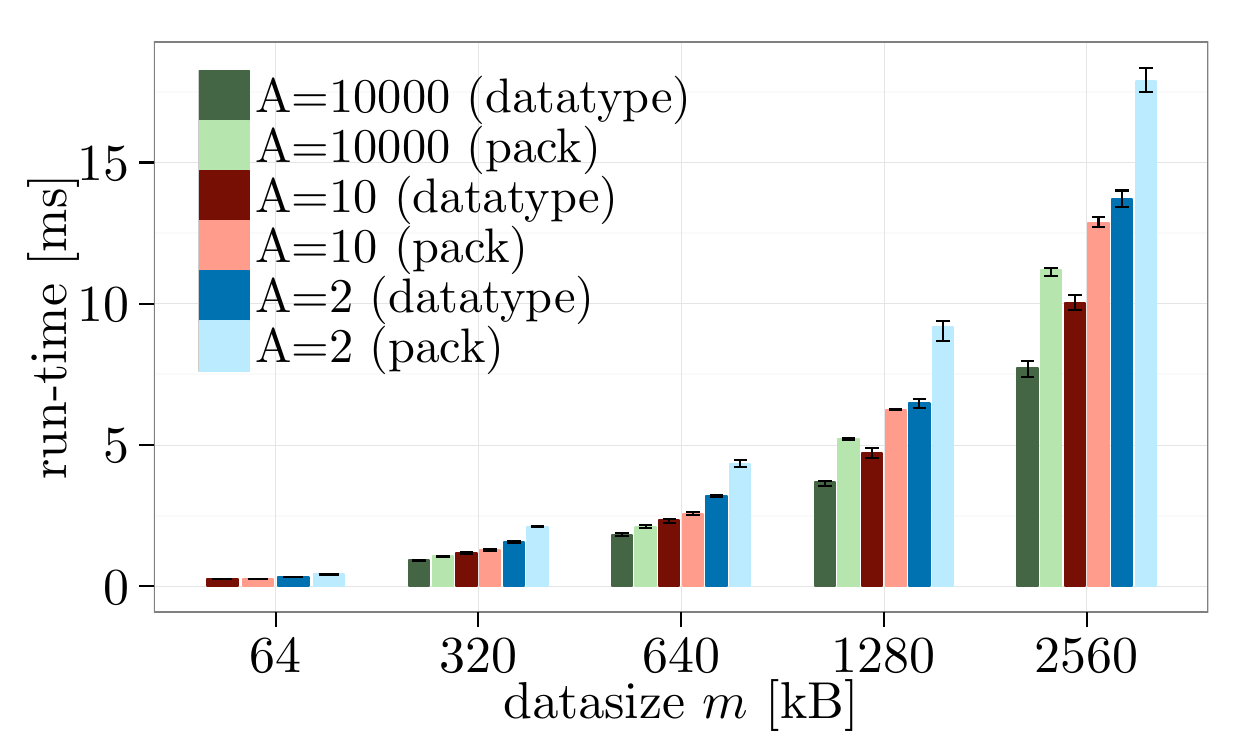}
\caption{%
\label{exp:bcast-pack-block-onenode-nec}%
\dtblock%
}%
\end{subfigure}%
\hfill%
\begin{subfigure}{.24\linewidth}
\centering
\includegraphics[width=\linewidth]{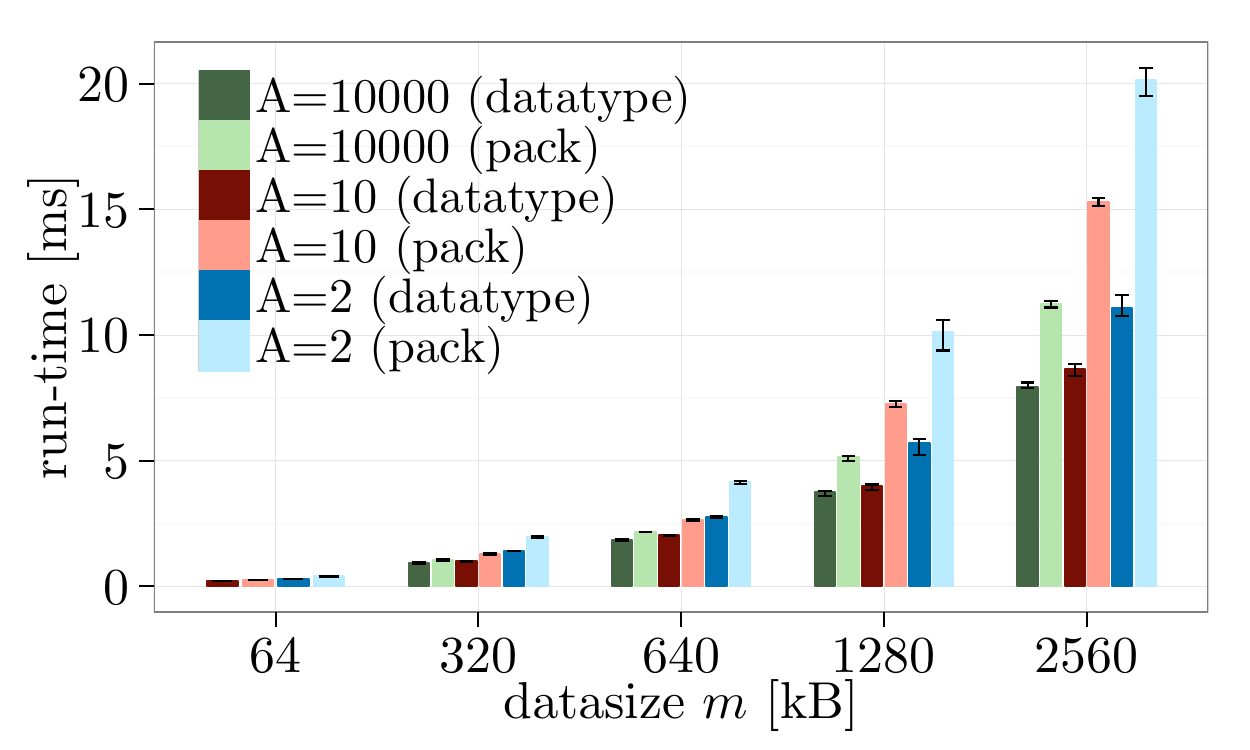}
\caption{%
\label{exp:bcast-pack-bucket-onenode-nec}%
\dtbucket%
}%
\end{subfigure}%
\hfill%
\begin{subfigure}{.24\linewidth}
\centering
\includegraphics[width=\linewidth]{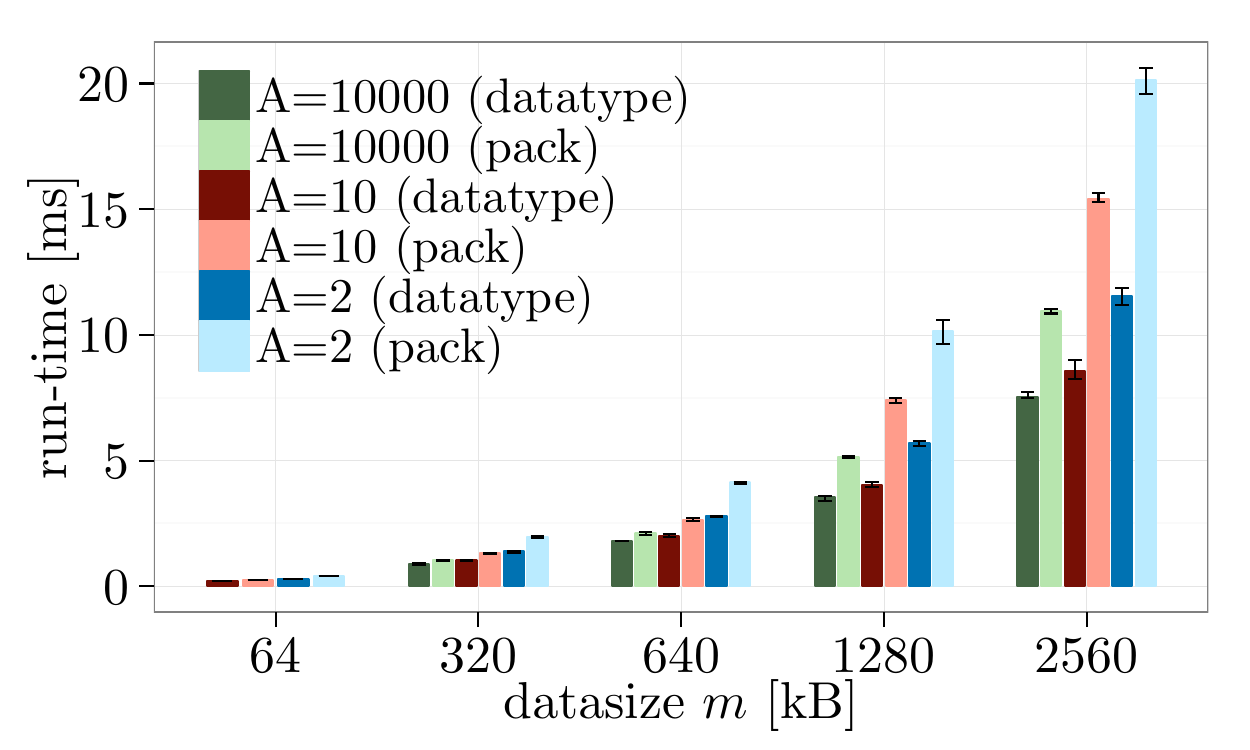}
\caption{%
\label{exp:bcast-pack-alternating-onenode-nec}%
\dtalternating%
}%
\end{subfigure}%
\caption{\label{exp:bcast-pack-onenode-nec}  Basic layouts \vs pack/unpack, element datatype: \mpiint, one~node, \num{16}~processes, \mpibcast, \jupiternecmpi.}
\end{figure*}

\begin{figure*}[htpb]
\centering
\begin{subfigure}{.24\linewidth}
\centering
\includegraphics[width=\linewidth]{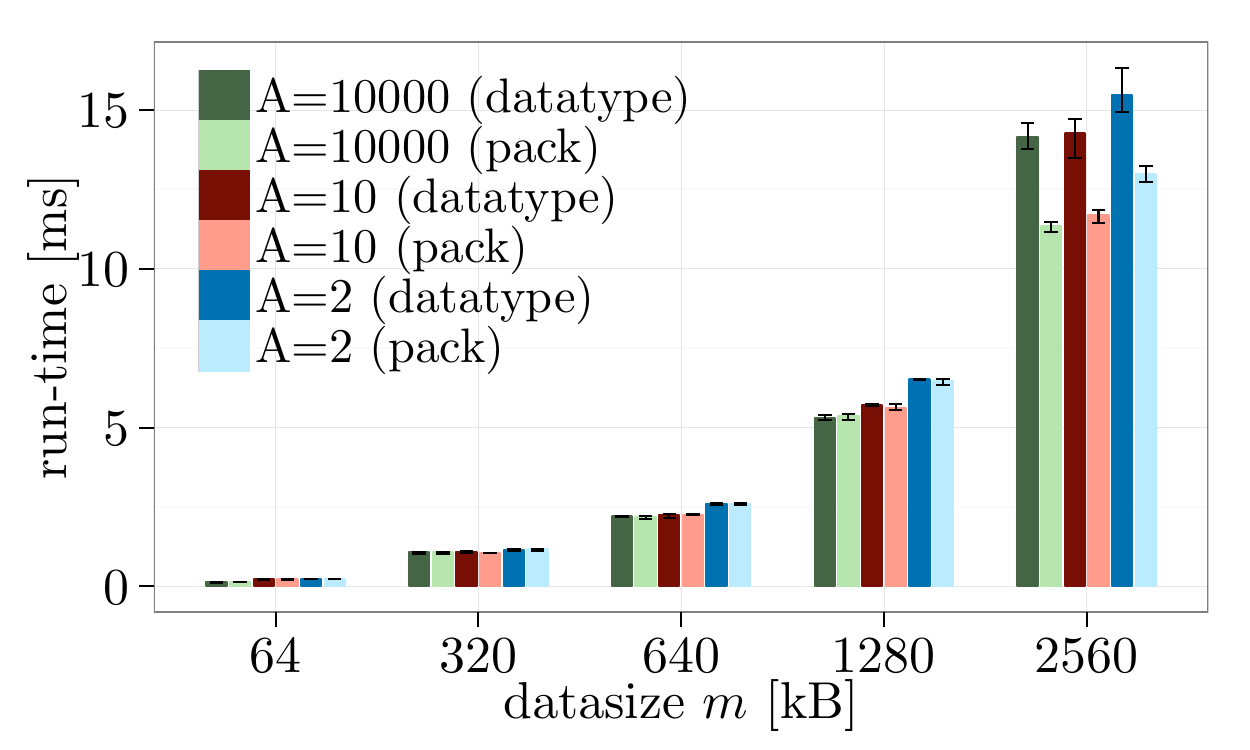}
\caption{%
\label{exp:bcast-pack-tiled-onenode-mvapich}%
\dttiled%
}%
\end{subfigure}%
\hfill%
\begin{subfigure}{.24\linewidth}
\centering
\includegraphics[width=\linewidth]{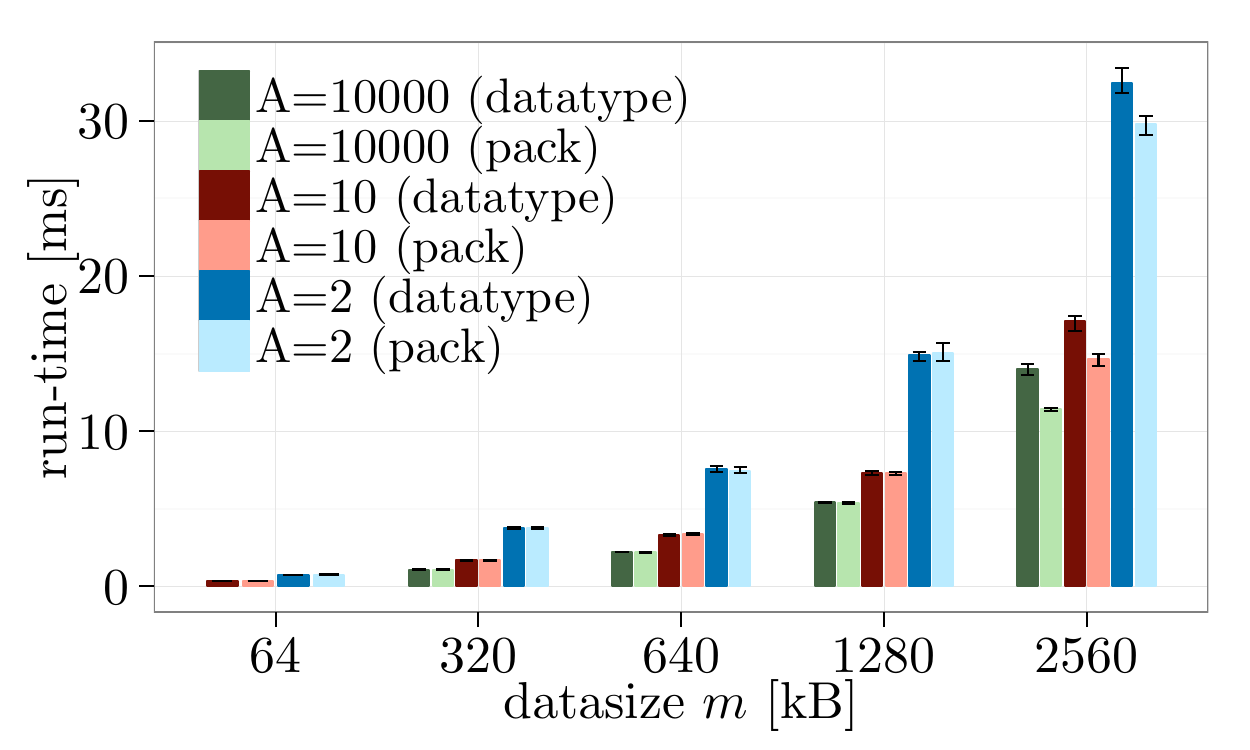}
\caption{%
\label{exp:bcast-pack-block-onenode-mvapich}%
\dtblock%
}%
\end{subfigure}%
\hfill%
\begin{subfigure}{.24\linewidth}
\centering
\includegraphics[width=\linewidth]{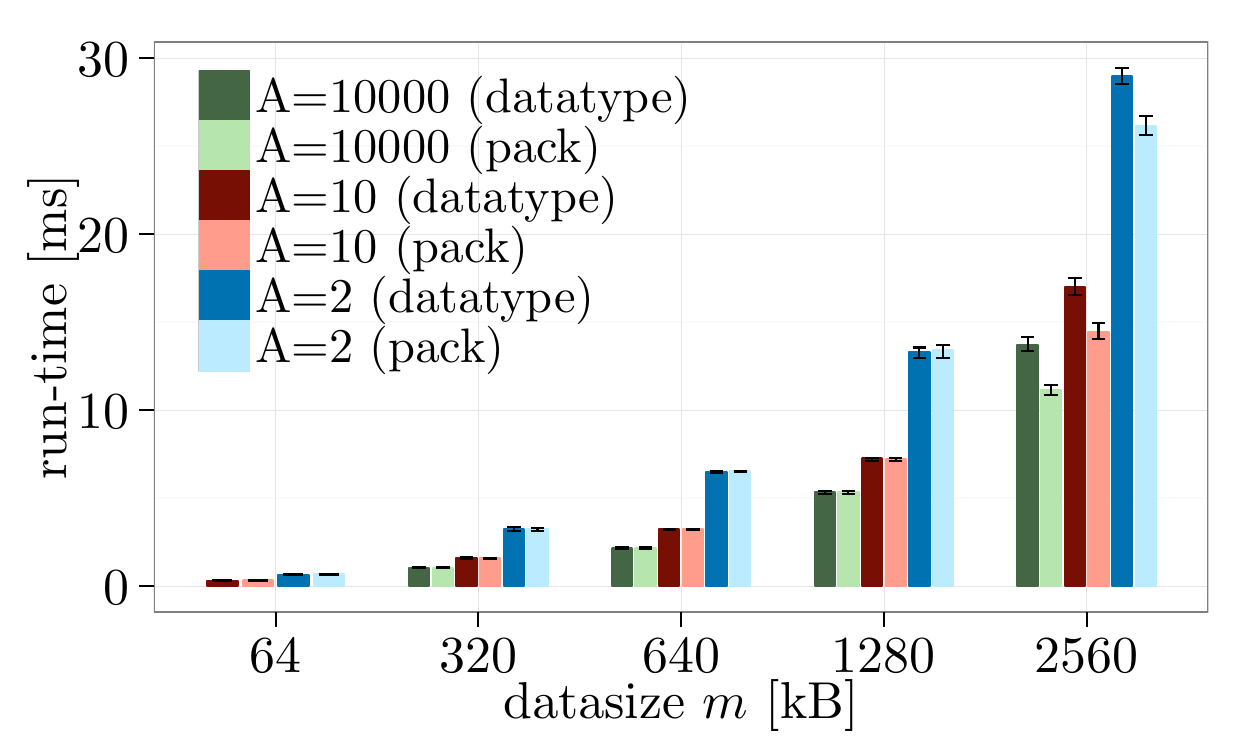}
\caption{%
\label{exp:bcast-pack-bucket-onenode-mvapich}%
\dtbucket%
}%
\end{subfigure}%
\hfill%
\begin{subfigure}{.24\linewidth}
\centering
\includegraphics[width=\linewidth]{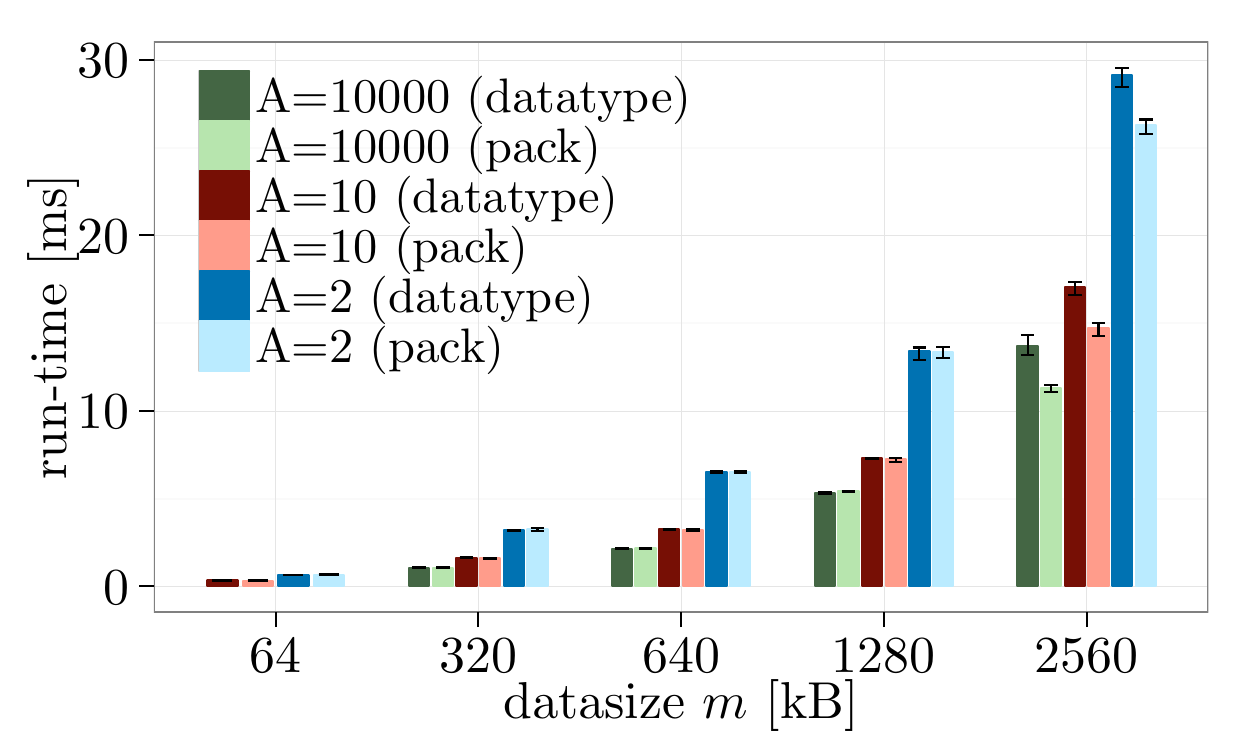}
\caption{%
\label{exp:bcast-pack-alternating-onenode-mvapich}%
\dtalternating%
}%
\end{subfigure}%
\caption{\label{exp:bcast-pack-onenode-mvapich}  Basic layouts \vs pack/unpack, element datatype: \mpiint, one~node, \num{16}~processes, \mpibcast, \jupitermvapich.}
\end{figure*}

\begin{figure*}[htpb]
\centering
\begin{subfigure}{.24\linewidth}
\centering
\includegraphics[width=\linewidth]{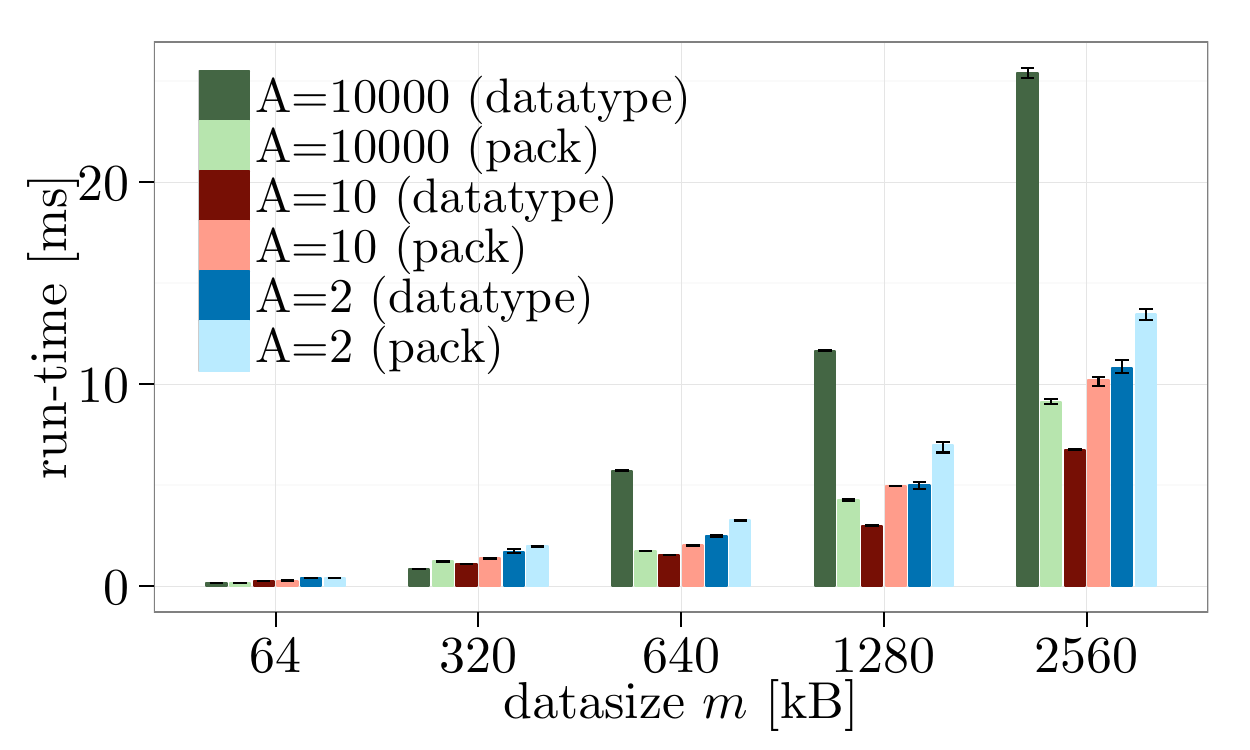}
\caption{%
\label{exp:bcast-pack-tiled-onenode-openmpi}%
\dttiled%
}%
\end{subfigure}%
\hfill%
\begin{subfigure}{.24\linewidth}
\centering
\includegraphics[width=\linewidth]{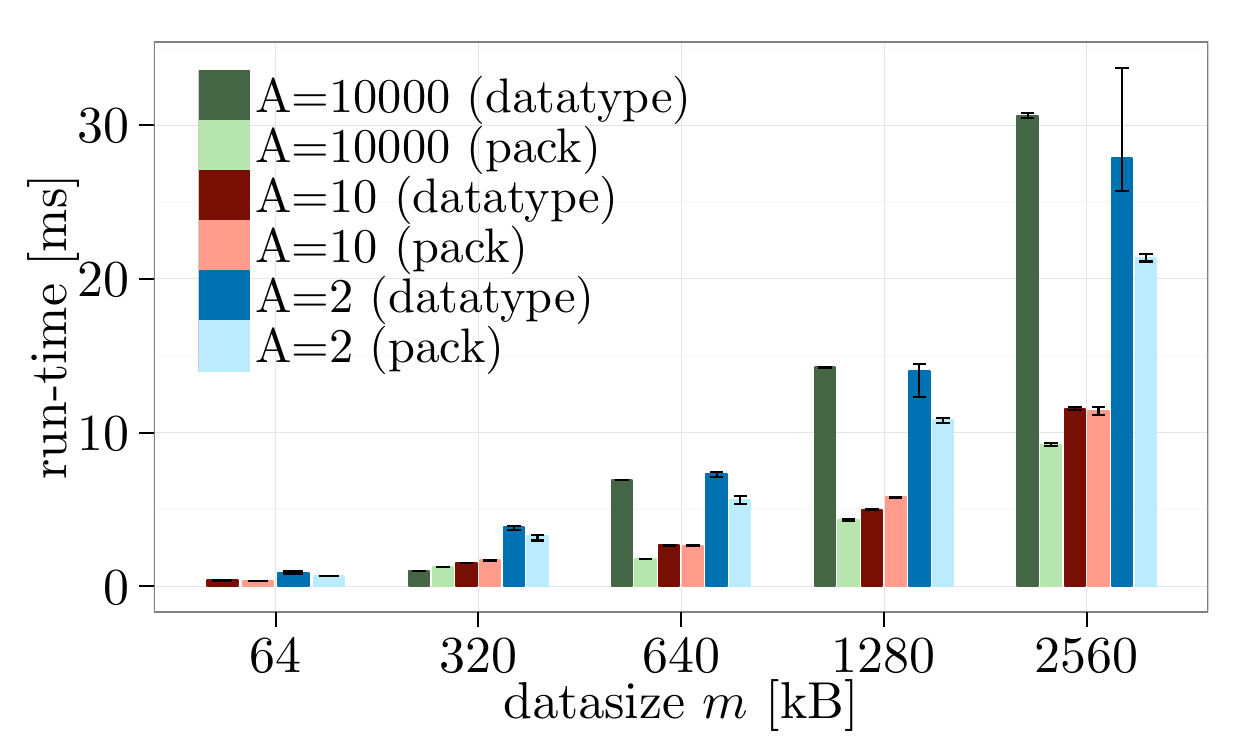}
\caption{%
\label{exp:bcast-pack-block-onenode-openmpi}%
\dtblock%
}%
\end{subfigure}%
\hfill%
\begin{subfigure}{.24\linewidth}
\centering
\includegraphics[width=\linewidth]{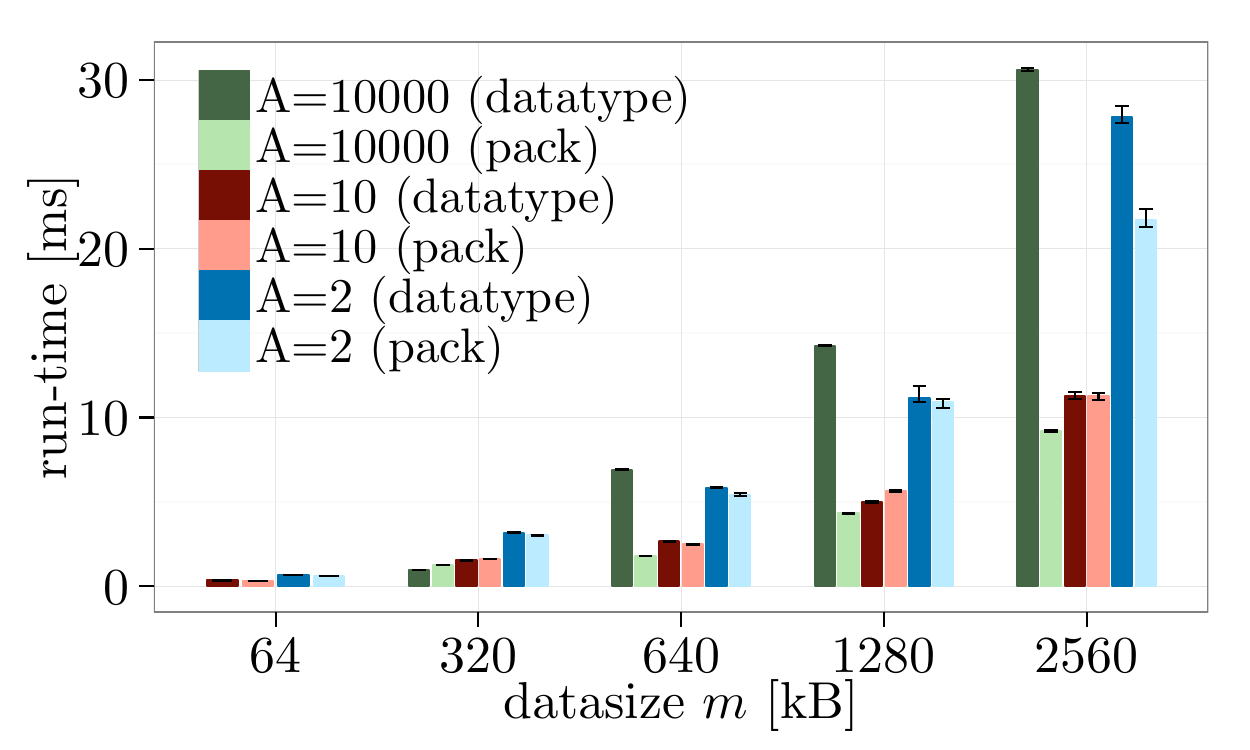}
\caption{%
\label{exp:bcast-pack-bucket-onenode-openmpi}%
\dtbucket%
}%
\end{subfigure}%
\hfill%
\begin{subfigure}{.24\linewidth}
\centering
\includegraphics[width=\linewidth]{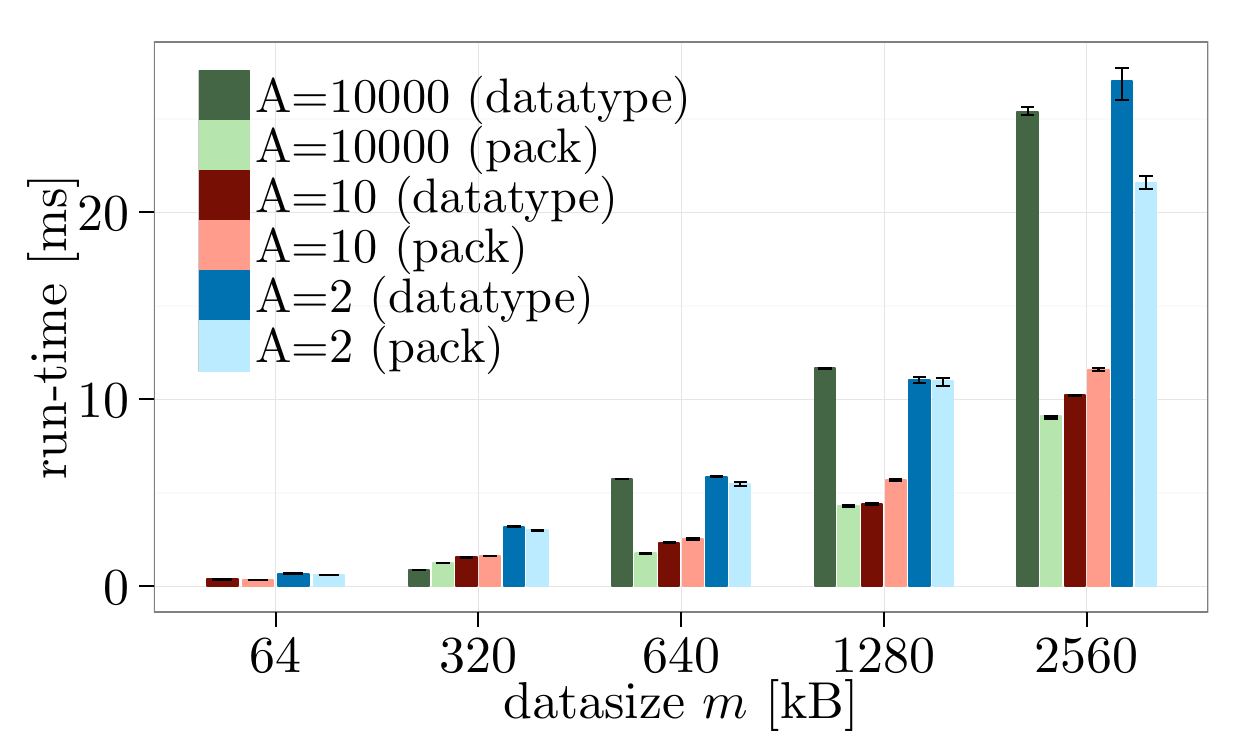}
\caption{%
\label{exp:bcast-pack-alternating-onenode-openmpi}%
\dtalternating%
}%
\end{subfigure}%
\caption{\label{exp:bcast-pack-onenode-openmpi}  Basic layouts \vs pack/unpack, element datatype: \mpiint, one~node, \num{16}~processes, \mpibcast, \jupiteropenmpi.}
\end{figure*}

\appexpdesc{
  \begin{expitemize}
    \item pack \vs unpack for basic layouts (\dttiled, \dtblock, \dtbucket, \dtalternating)
    \item \mpiallgather, \mpibcast, \pingpong
  \end{expitemize}
}{
  \begin{expitemize}
    \item \expparam{\jupiternecmpi, \pingpong, \num{2x1}~processes}{\fig~\ref{exp:pingpong-pack-2x1-nec}}
    \item \expparam{\jupitermvapich, \pingpong, \num{2x1}~processes}{\fig~\ref{exp:pingpong-pack-2x1-mvapich}}
    \item \expparam{\jupiteropenmpi, \pingpong, \num{2x1}~processes}{\fig~\ref{exp:pingpong-pack-2x1-openmpi}}
    \item \expparam{\jupiternecmpi, \mpiallgather, \num{32x1}~processes}{\fig~\ref{exp:allgather-pack-32x1-nec}}
    \item \expparam{\jupitermvapich, \mpiallgather, \num{32x1}~processes}{\fig~\ref{exp:allgather-pack-32x1-mvapich}}
    \item \expparam{\jupiteropenmpi, \mpiallgather, \num{32x1}~processes}{\fig~\ref{exp:allgather-pack-32x1-openmpi}}
    \item \expparam{\jupiternecmpi, \mpibcast, \num{32x1}~processes}{\fig~\ref{exp:bcast-pack-32x1-nec}}
    \item \expparam{\jupitermvapich, \mpibcast, \num{32x1}~processes}{\fig~\ref{exp:bcast-pack-32x1-mvapich}}
    \item \expparam{\jupiteropenmpi, \mpibcast, \num{32x1}~processes}{\fig~\ref{exp:bcast-pack-32x1-openmpi}}
    \item \expparam{\jupiternecmpi, \pingpong, one~node, \num{2}~processes}{\fig~\ref{exp:pingpong-pack-onenode-nec}}
    \item \expparam{\jupitermvapich, \pingpong, one~node, \num{2}~processes}{\fig~\ref{exp:pingpong-pack-onenode-mvapich}}
    \item \expparam{\jupiteropenmpi, \pingpong, one~node, \num{2}~processes}{\fig~\ref{exp:pingpong-pack-onenode-openmpi}}
    \item \expparam{\jupiternecmpi, \mpiallgather, one~node, \num{16}~processes}{\fig~\ref{exp:allgather-pack-onenode-nec}}
    \item \expparam{\jupitermvapich, \mpiallgather, one~node, \num{16}~processes}{\fig~\ref{exp:allgather-pack-onenode-mvapich}}
    \item \expparam{\jupiteropenmpi, \mpiallgather, one~node, \num{16}~processes}{\fig~\ref{exp:allgather-pack-onenode-openmpi}}
    \item \expparam{\jupiternecmpi, \mpibcast, one~node, \num{16}~processes}{\fig~\ref{exp:bcast-pack-onenode-nec}}
    \item \expparam{\jupitermvapich, \mpibcast, one~node, \num{16}~processes}{\fig~\ref{exp:bcast-pack-onenode-mvapich}}
    \item \expparam{\jupiteropenmpi, \mpibcast, one~node, \num{16}~processes}{\fig~\ref{exp:bcast-pack-onenode-openmpi}}
  \end{expitemize}  
}

\FloatBarrier
\clearpage

\FloatBarrier
\clearpage

\appexp{exptest:contig}

\begin{figure*}[htpb]
\centering
\begin{subfigure}{.24\linewidth}
\centering
\includegraphics[width=\linewidth]{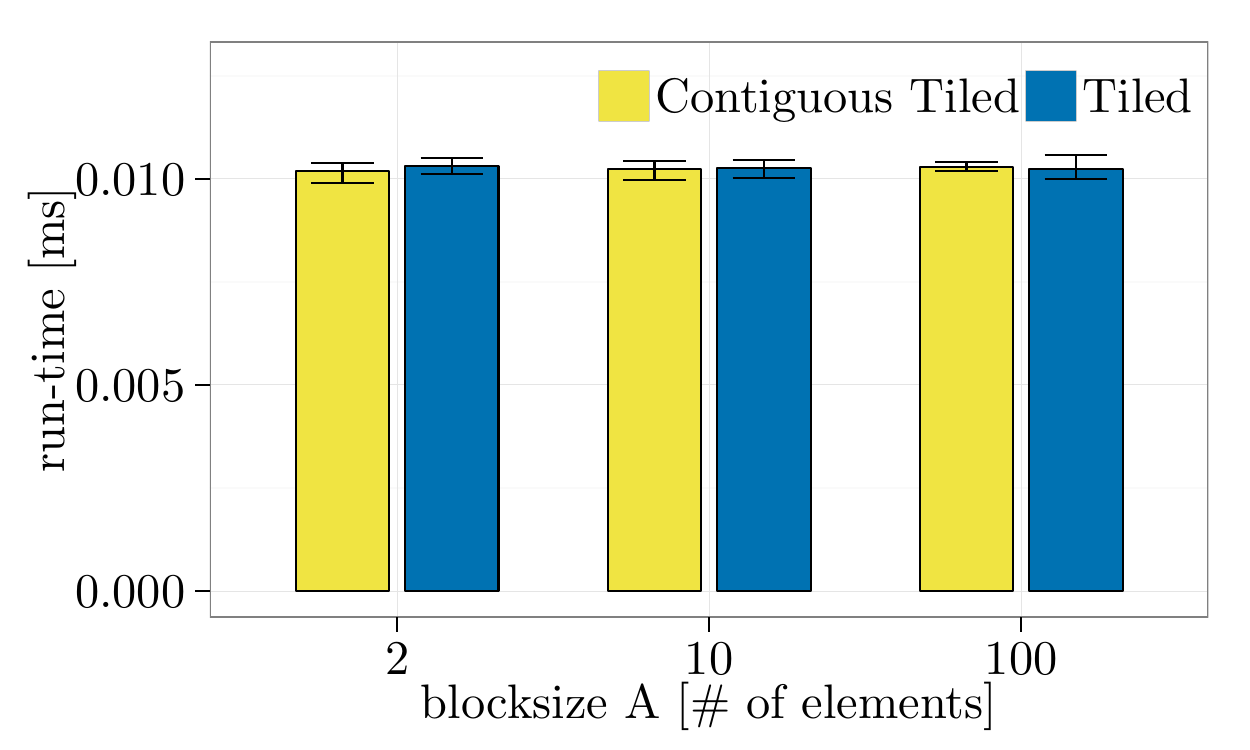}
\caption{%
\label{exp:pingpong-contigtiled-smallnbytes-2x1}%
\dttiled%
}%
\end{subfigure}%
\hfill%
\begin{subfigure}{.24\linewidth}
\centering
\includegraphics[width=\linewidth]{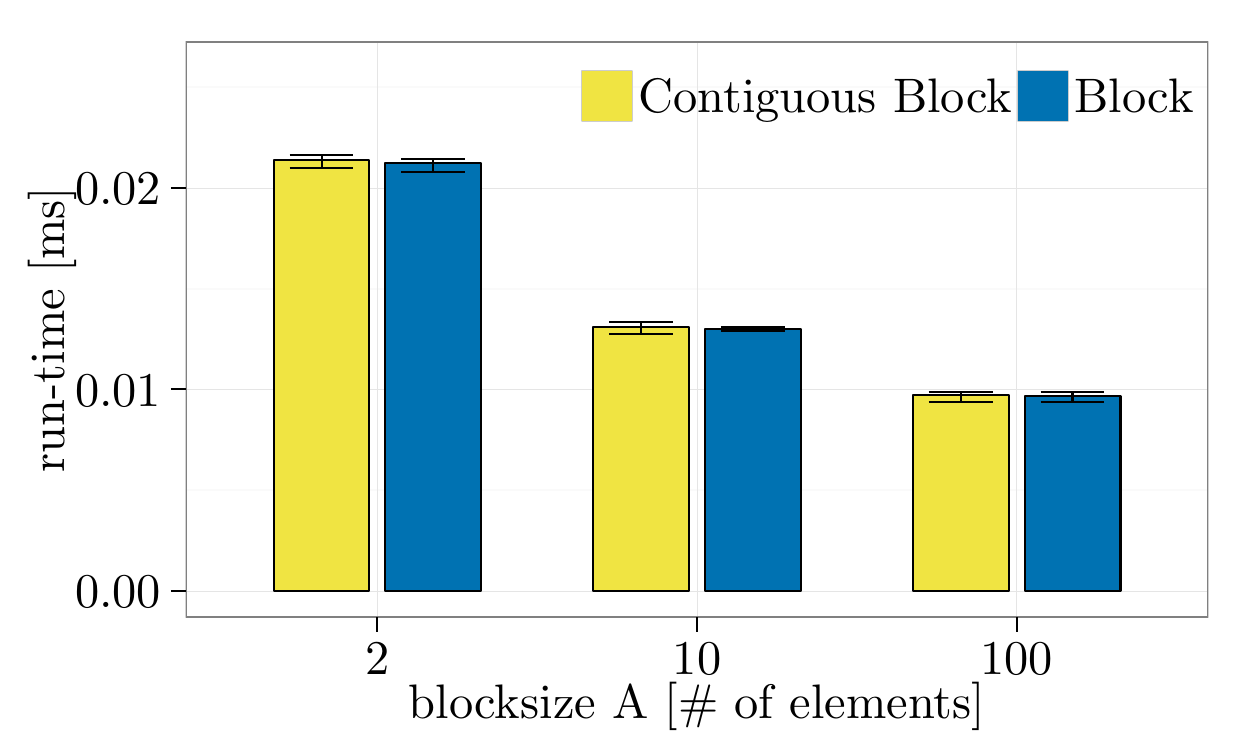}
\caption{%
\label{exp:pingpong-contigblock-smallnbytes-2x1}%
\dtblock%
}%
\end{subfigure}%
\hfill%
\begin{subfigure}{.24\linewidth}
\centering
\includegraphics[width=\linewidth]{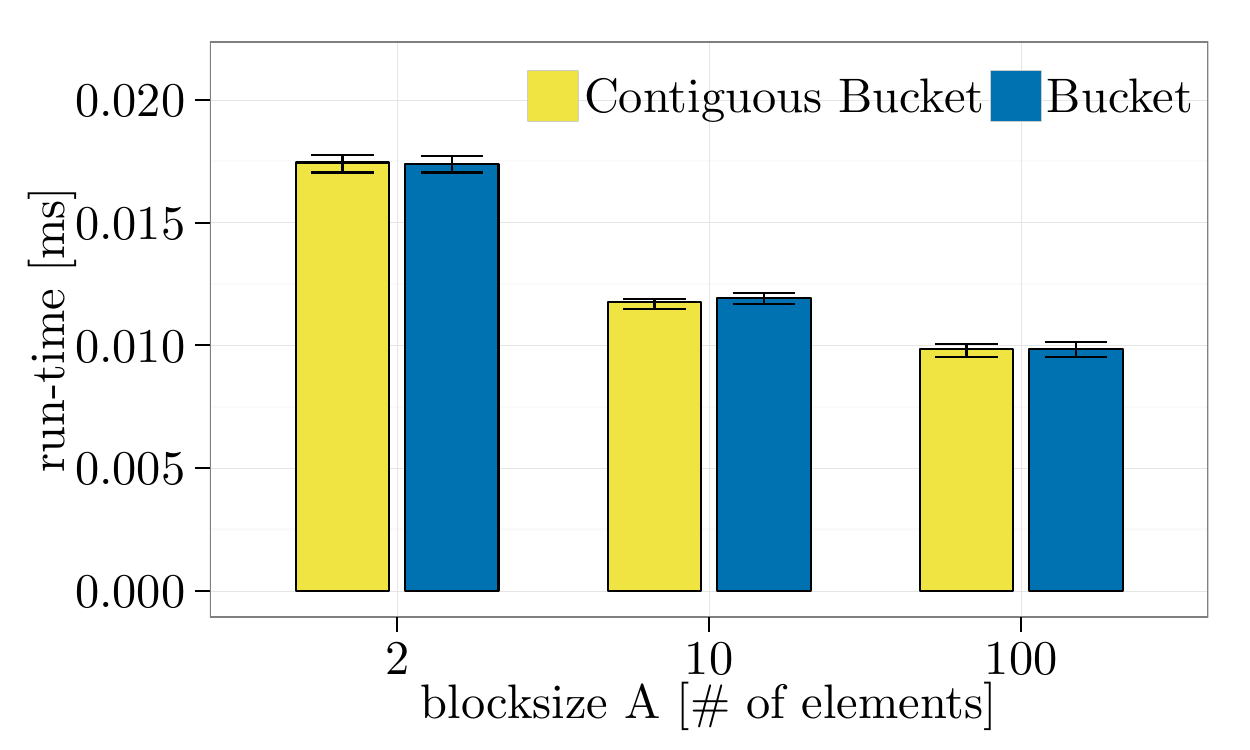}
\caption{%
\label{exp:pingpong-contigbucket-smallnbytes-2x1}%
\dtbucket%
}%
\end{subfigure}%
\hfill%
\begin{subfigure}{.24\linewidth}
\centering
\includegraphics[width=\linewidth]{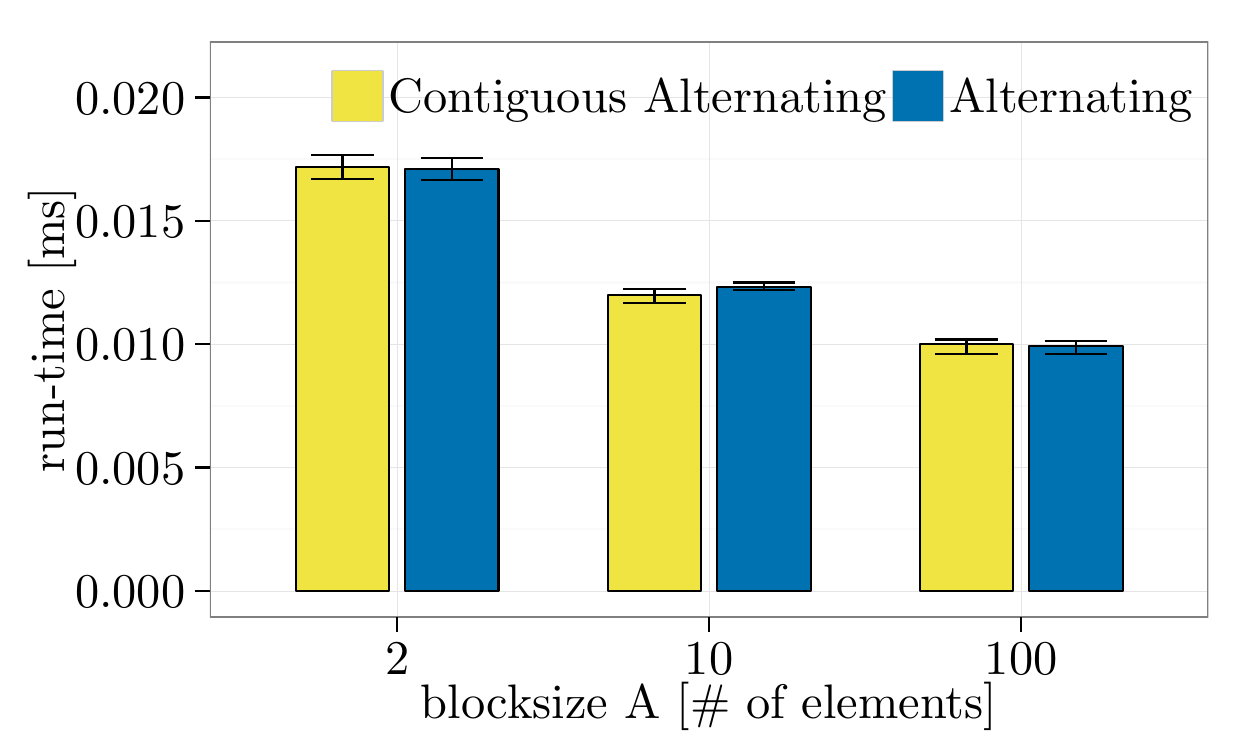}
\caption{%
\label{exp:pingpong-contigalternating-smallnbytes-2x1}%
\dtalternating%
}%
\end{subfigure}%
\caption{\label{exp:pingpong-contig-smallnbytes-2x1}  Basic layouts \vs \ddtcontig, $\VARdatasize=\SI{2}{\kilo\byte}$, element datatype: \mpiint, \num{2x1}~processes, \pingpong, \jupiternecmpi (similar results for \num{1x2}~processes).}
\end{figure*}

\begin{figure*}[htpb]
\centering
\begin{subfigure}{.24\linewidth}
\centering
\includegraphics[width=\linewidth]{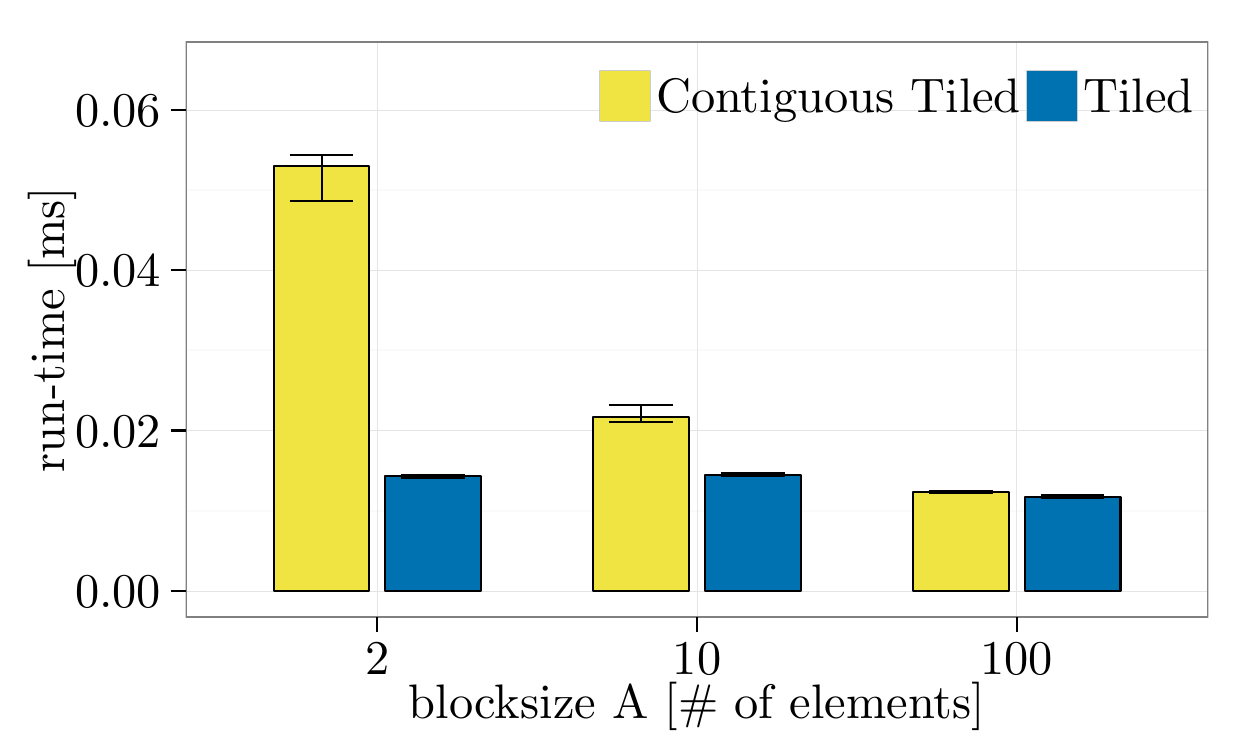}
\caption{%
\label{exp:pingpong-contigtiled-smallnbytes-2x1-mvapich}%
\dttiled%
}%
\end{subfigure}%
\hfill%
\begin{subfigure}{.24\linewidth}
\centering
\includegraphics[width=\linewidth]{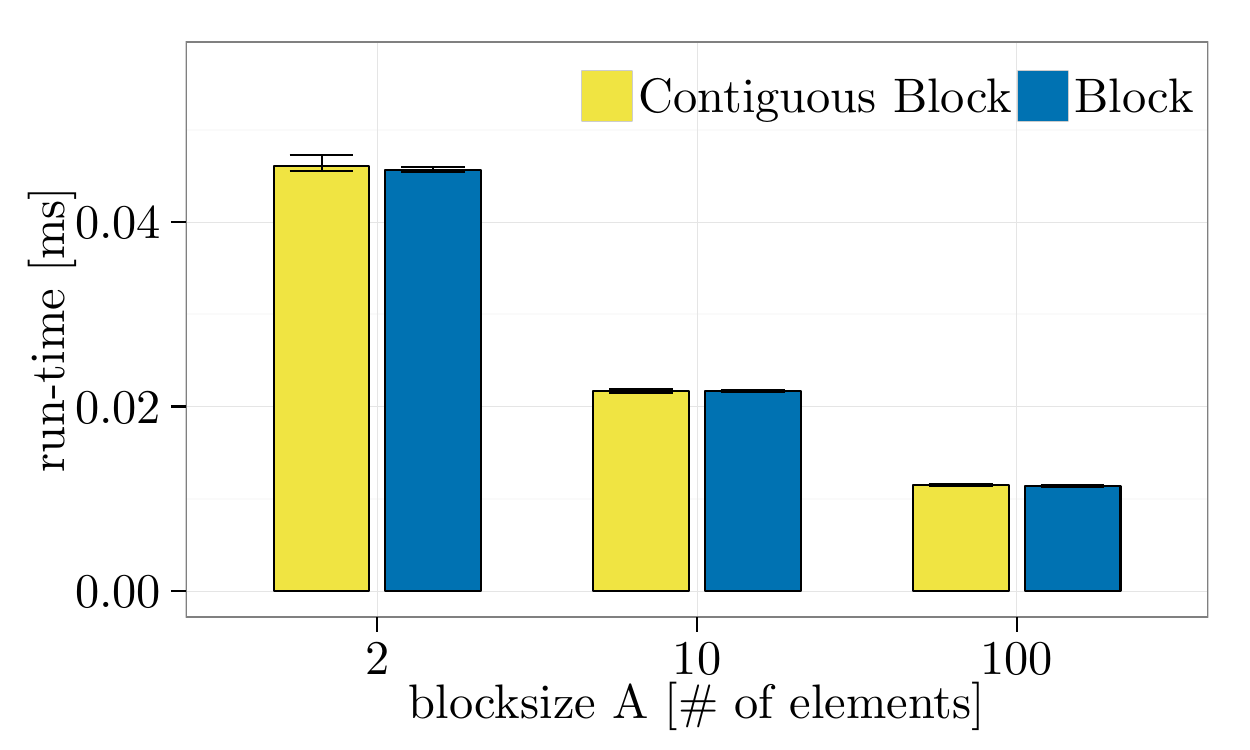}
\caption{%
\label{exp:pingpong-contigblock-smallnbytes-2x1-mvapich}%
\dtblock%
}%
\end{subfigure}%
\hfill%
\begin{subfigure}{.24\linewidth}
\centering
\includegraphics[width=\linewidth]{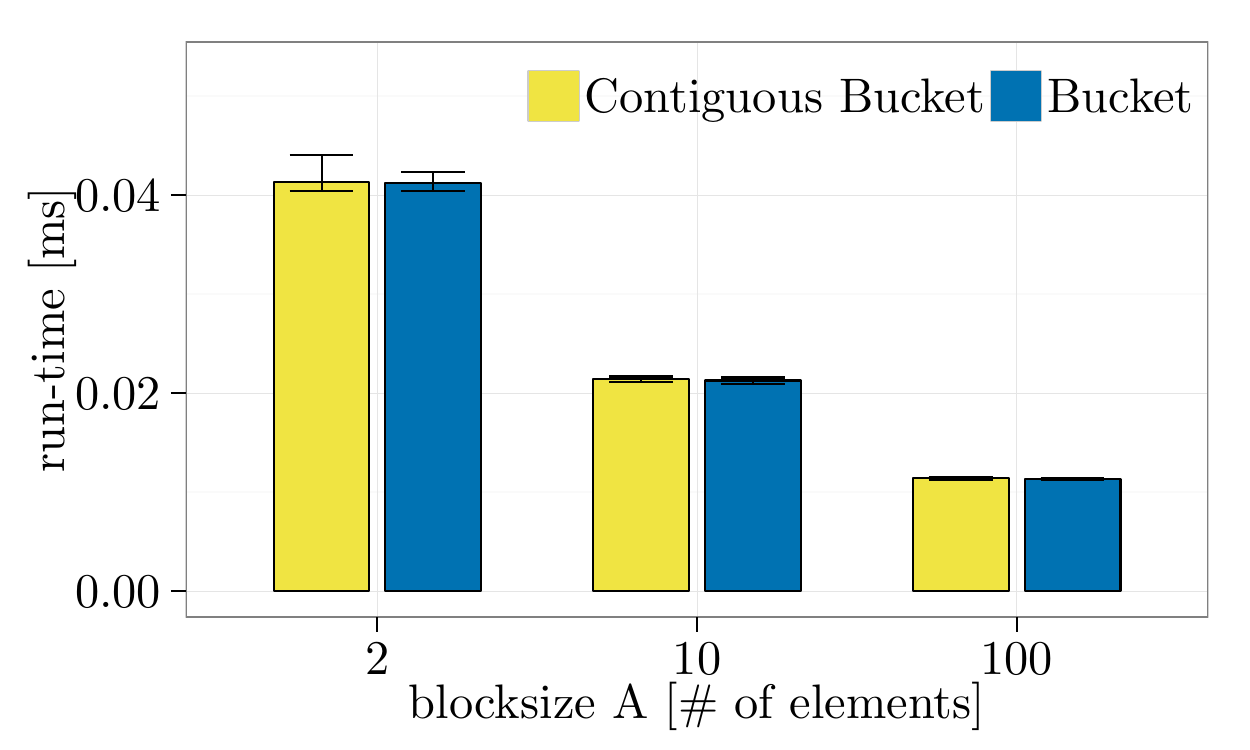}
\caption{%
\label{exp:pingpong-contigbucket-smallnbytes-2x1-mvapich}%
\dtbucket%
}%
\end{subfigure}%
\hfill%
\begin{subfigure}{.24\linewidth}
\centering
\includegraphics[width=\linewidth]{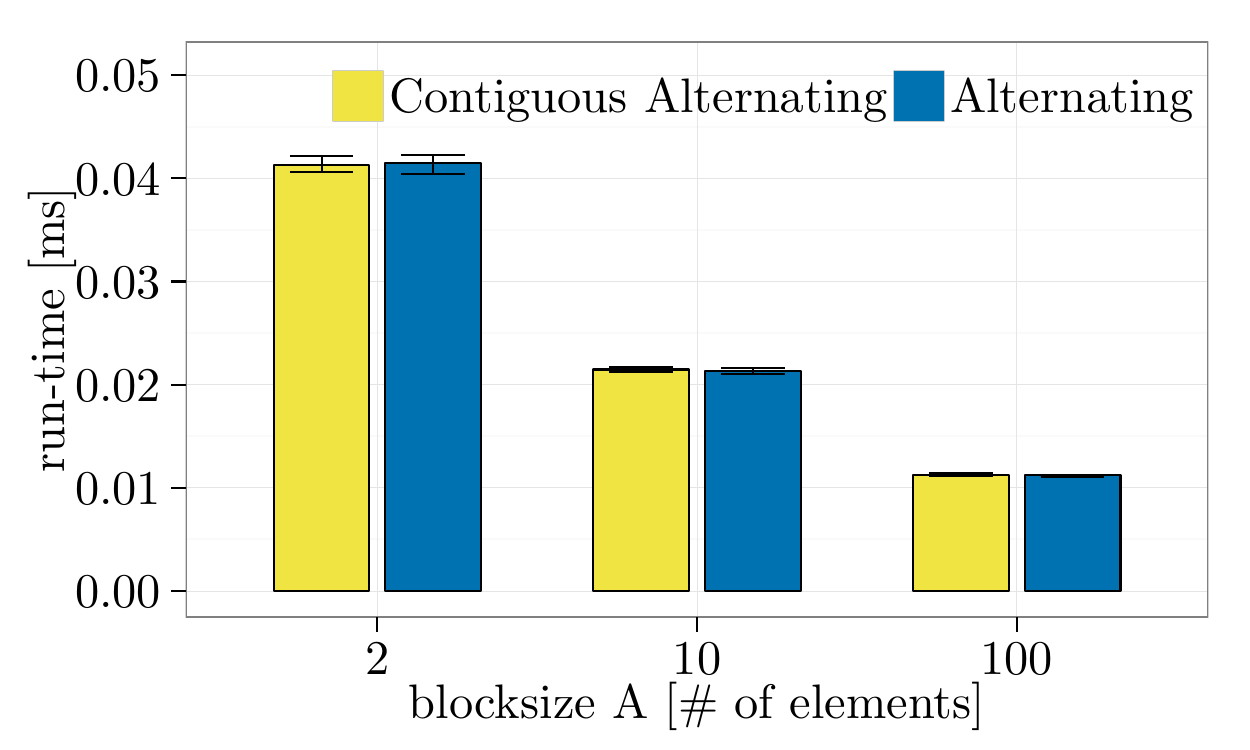}
\caption{%
\label{exp:pingpong-contigalternating-smallnbytes-2x1-mvapich}%
\dtalternating%
}%
\end{subfigure}%
\caption{\label{exp:pingpong-contig-smallnbytes-2x1-mvapich}  Basic layouts \vs \ddtcontig, $\VARdatasize=\SI{2}{\kilo\byte}$, element datatype: \mpiint, \num{2x1}~processes, \pingpong, \jupitermvapich (similar results for \num{1x2}~processes).}
\end{figure*}

\begin{figure*}[htpb]
\centering
\begin{subfigure}{.24\linewidth}
\centering
\includegraphics[width=\linewidth]{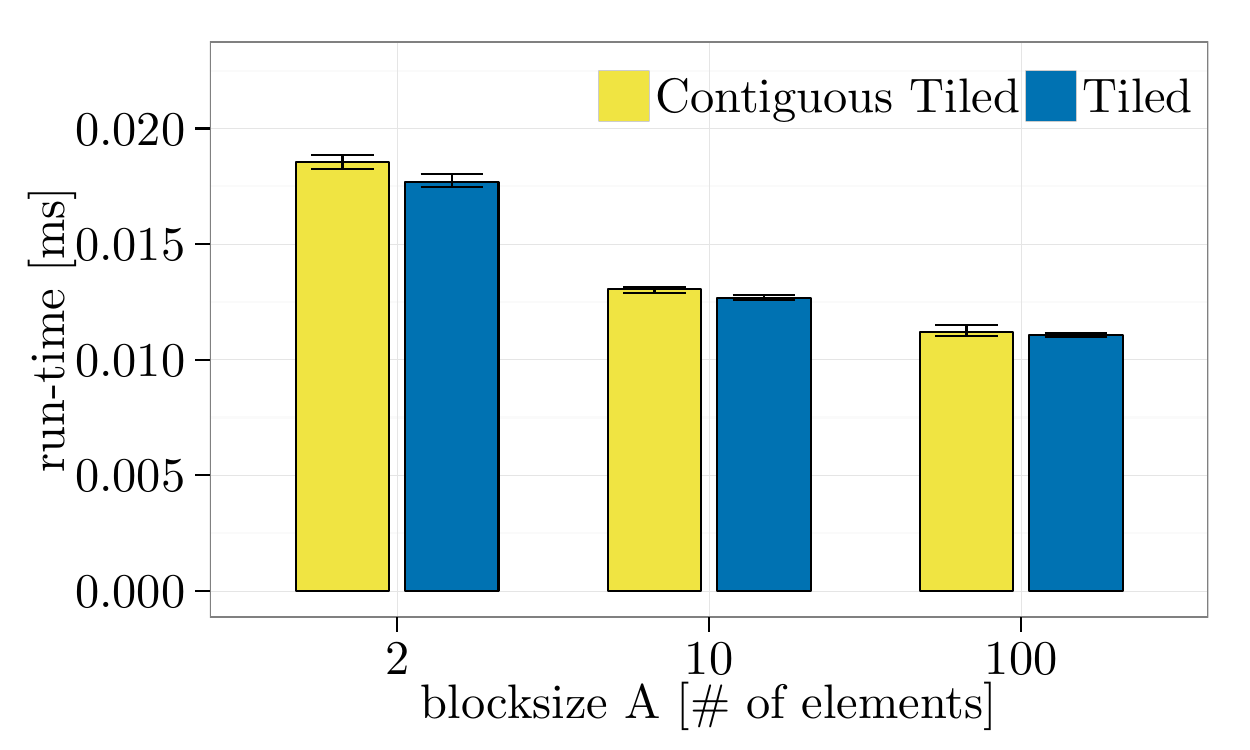}
\caption{%
\label{exp:pingpong-contigtiled-smallnbytes-2x1-openmpi}%
\dttiled%
}%
\end{subfigure}%
\hfill%
\begin{subfigure}{.24\linewidth}
\centering
\includegraphics[width=\linewidth]{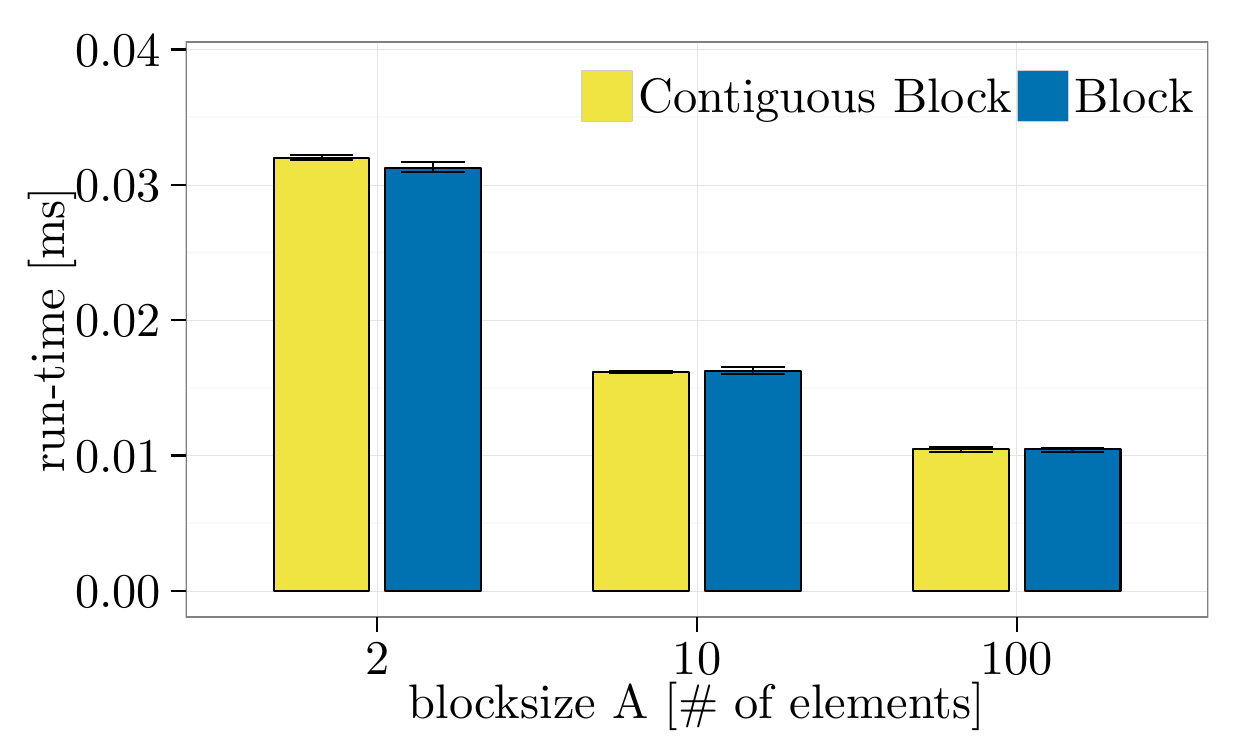}
\caption{%
\label{exp:pingpong-contigblock-smallnbytes-2x1-openmpi}%
\dtblock%
}%
\end{subfigure}%
\hfill%
\begin{subfigure}{.24\linewidth}
\centering
\includegraphics[width=\linewidth]{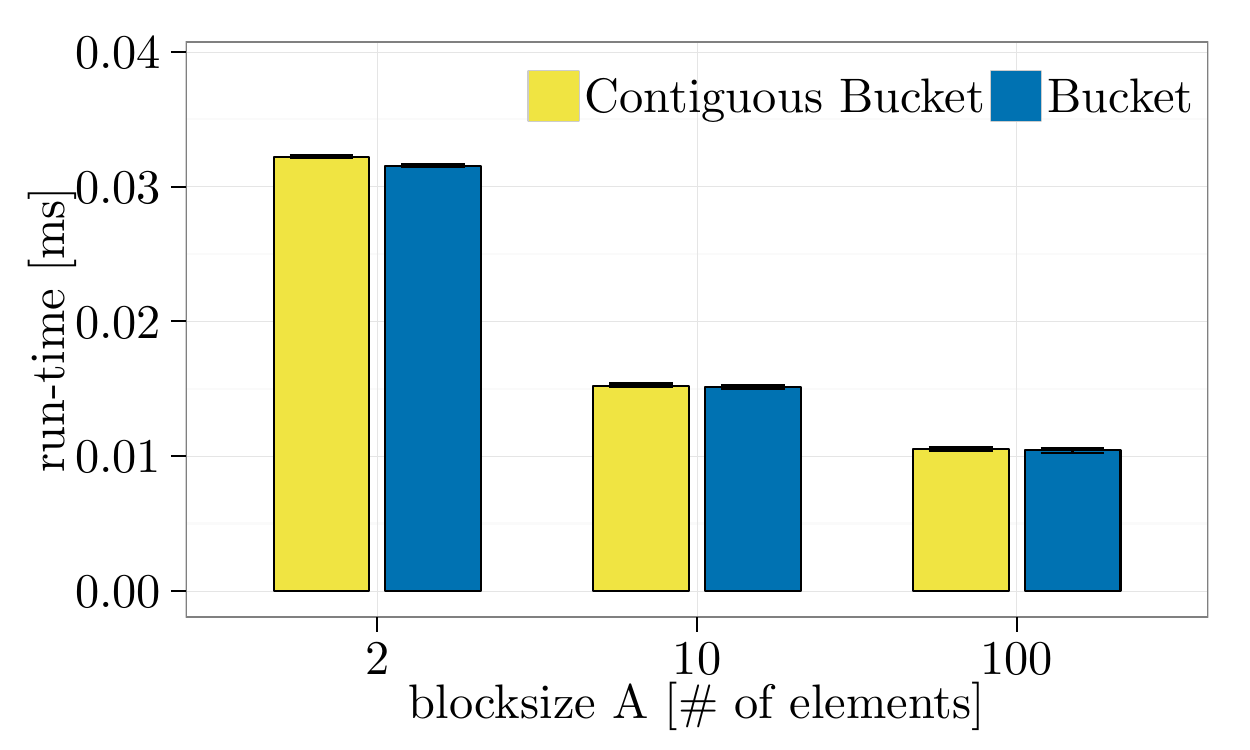}
\caption{%
\label{exp:pingpong-contigbucket-smallnbytes-2x1-openmpi}%
\dtbucket%
}%
\end{subfigure}%
\hfill%
\begin{subfigure}{.24\linewidth}
\centering
\includegraphics[width=\linewidth]{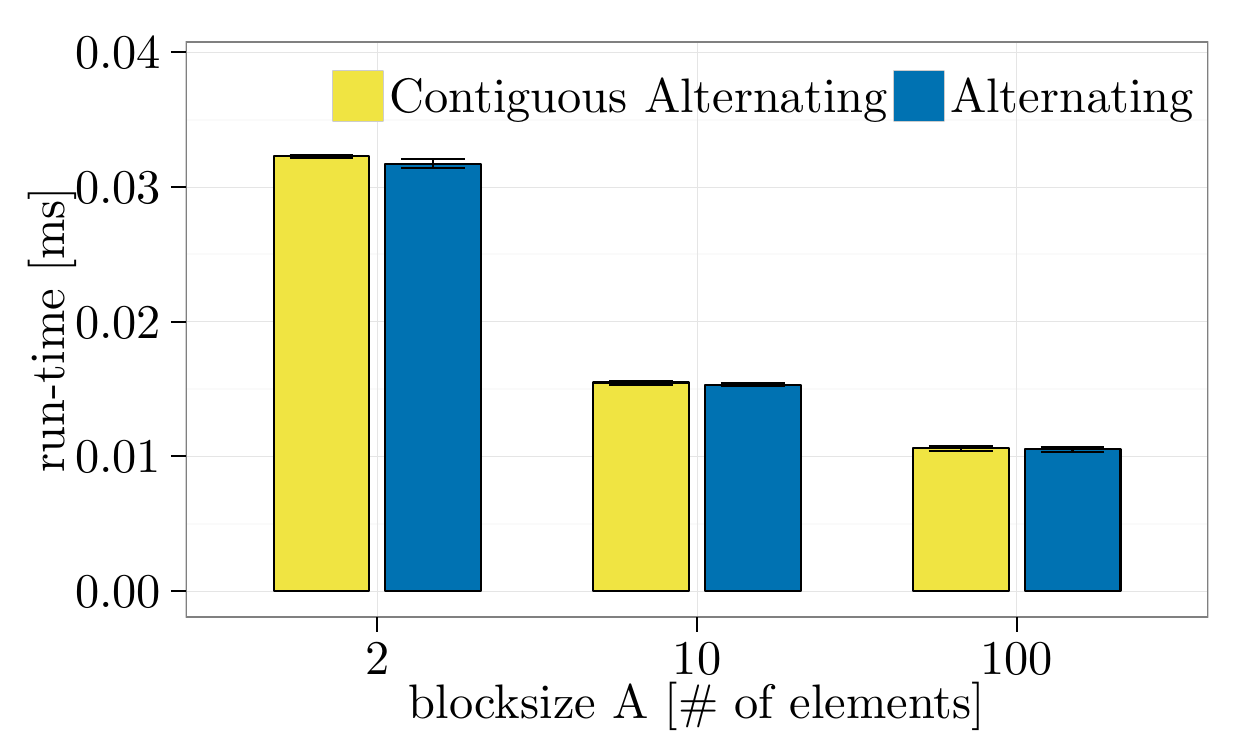}
\caption{%
\label{exp:pingpong-contigalternating-smallnbytes-2x1-openmpi}%
\dtalternating%
}%
\end{subfigure}%
\caption{\label{exp:pingpong-contig-smallnbytes-2x1-openmpi}  Basic layouts \vs \ddtcontig, $\VARdatasize=\SI{2}{\kilo\byte}$, element datatype: \mpiint, \num{2x1}~processes, \pingpong, \jupiteropenmpi (similar results for \num{1x2}~processes).}
\end{figure*}

\begin{figure*}[htpb]
\centering
\begin{subfigure}{.24\linewidth}
\centering
\includegraphics[width=\linewidth]{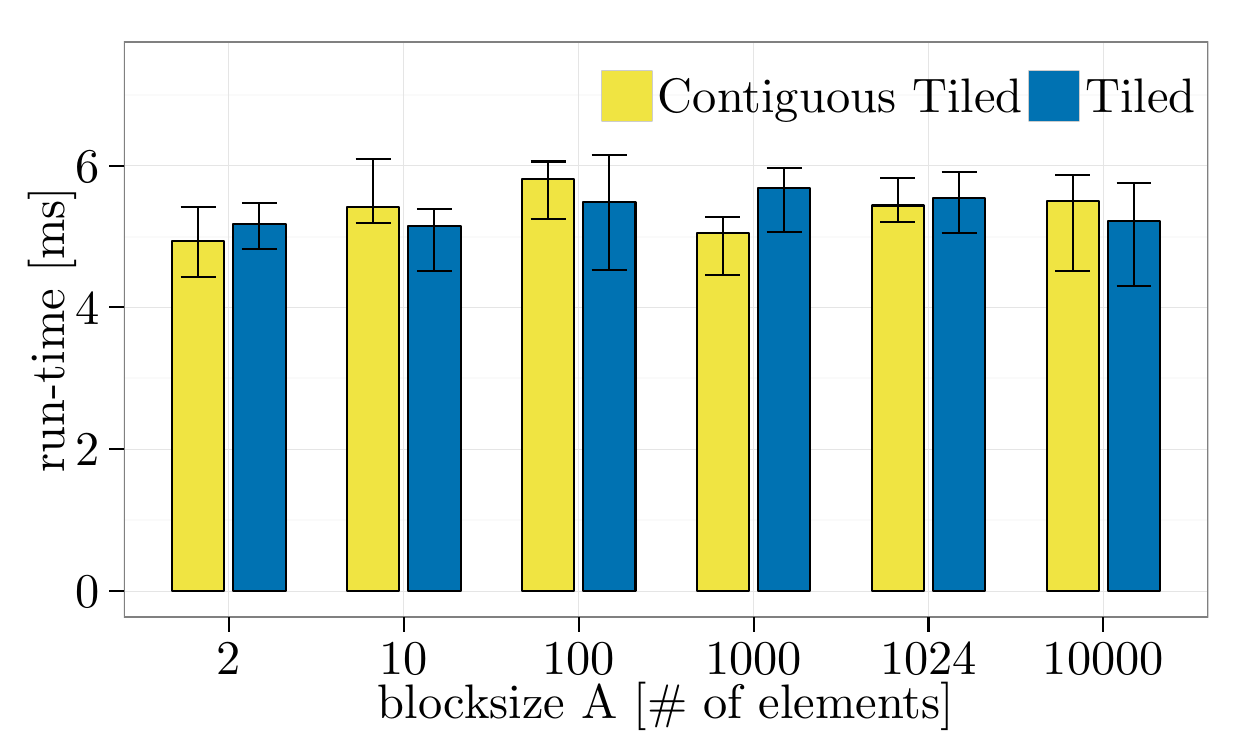}
\caption{%
\label{exp:pingpong-contigtiled-largenbytes-2x1}%
\dttiled%
}%
\end{subfigure}%
\hfill%
\begin{subfigure}{.24\linewidth}
\centering
\includegraphics[width=\linewidth]{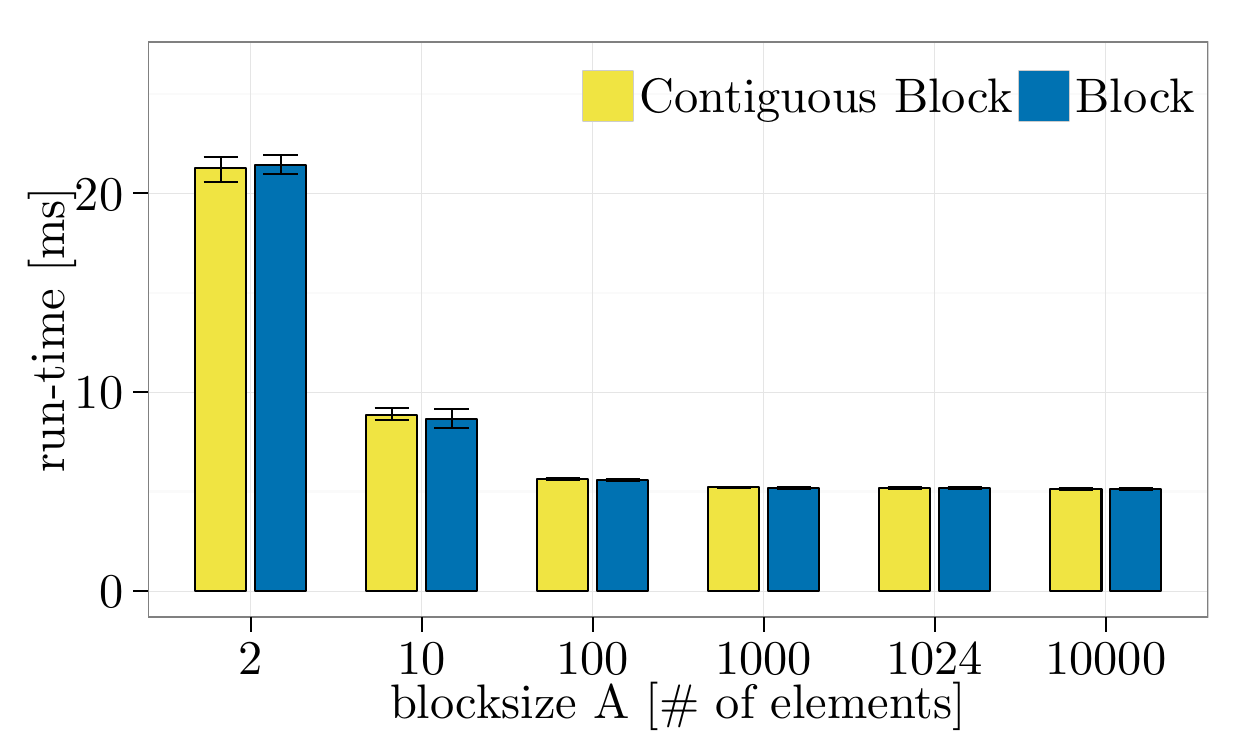}
\caption{%
\label{exp:pingpong-contigblock-largenbytes-2x1}%
\dtblock%
}%
\end{subfigure}%
\hfill%
\begin{subfigure}{.24\linewidth}
\centering
\includegraphics[width=\linewidth]{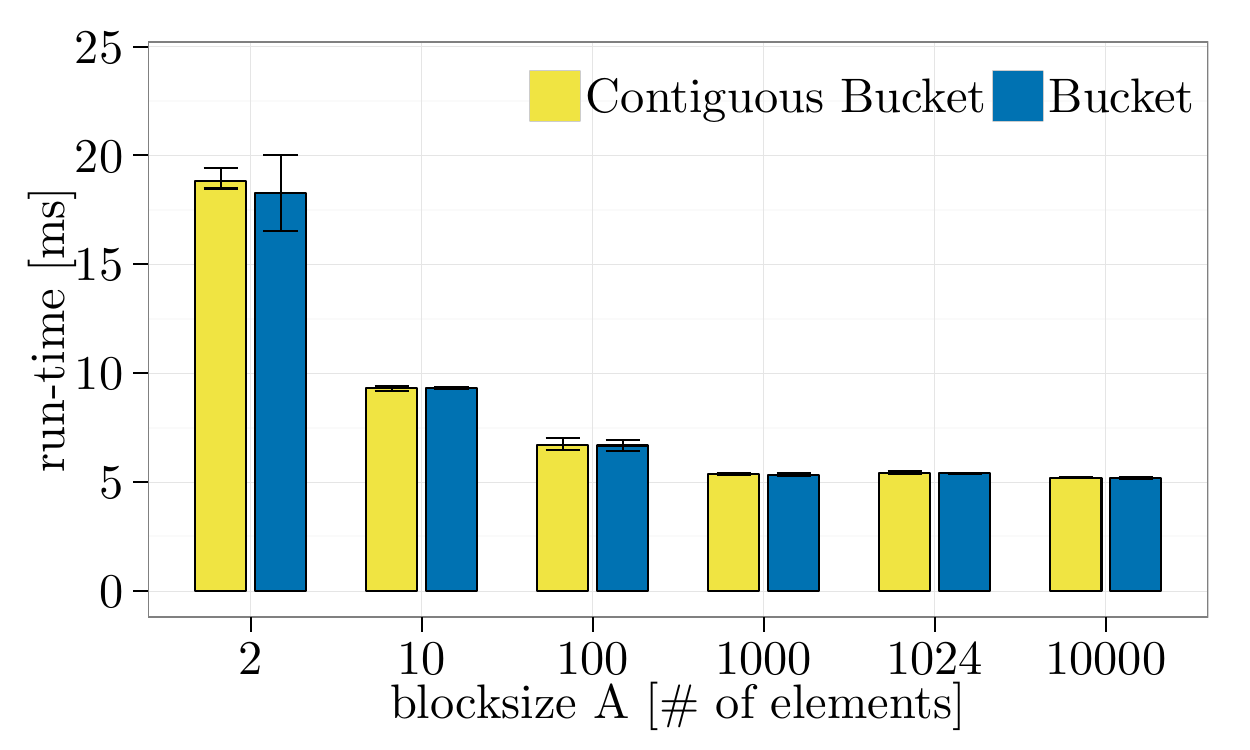}
\caption{%
\label{exp:pingpong-contigbucket-largenbytes-2x1}%
\dtbucket%
}%
\end{subfigure}%
\hfill%
\begin{subfigure}{.24\linewidth}
\centering
\includegraphics[width=\linewidth]{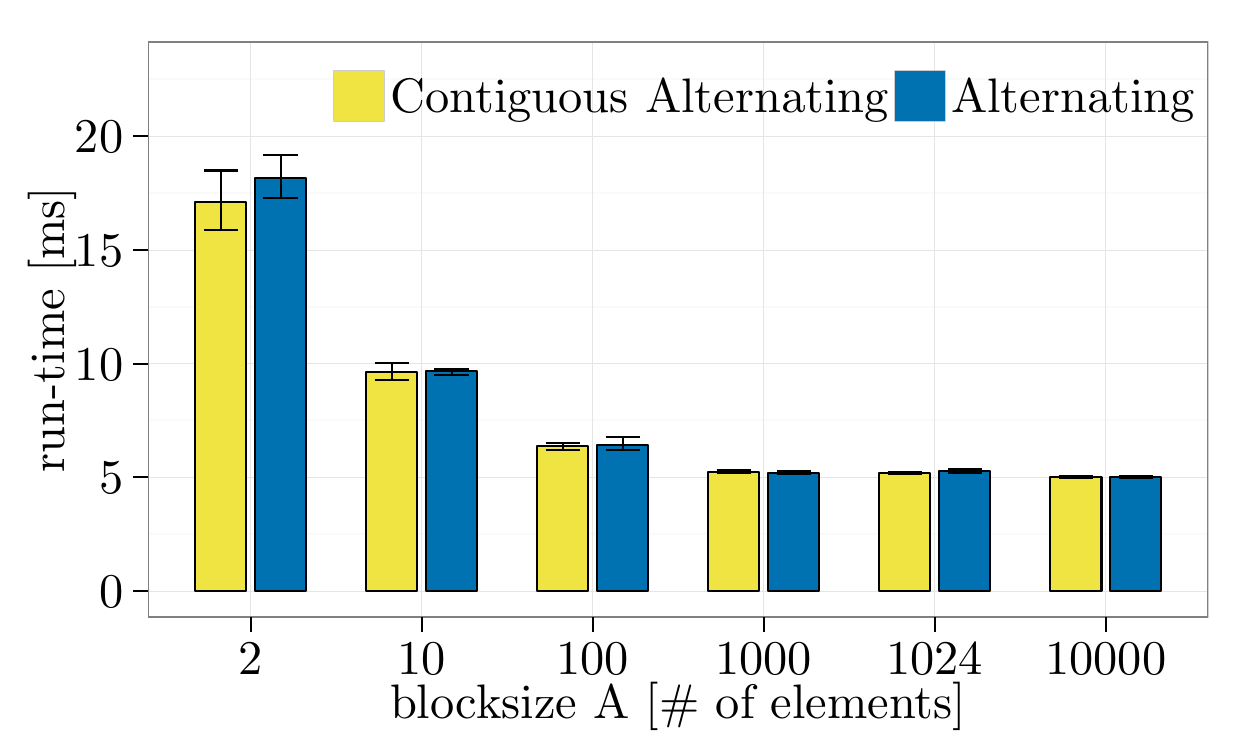}
\caption{%
\label{exp:pingpong-contigalternating-largenbytes-2x1}%
\dtalternating%
}%
\end{subfigure}%
\caption{\label{exp:pingpong-contig-largenbytes-2x1}  Basic layouts \vs \ddtcontig, $\VARdatasize=\SI{2.56}{\mega\byte}$, element datatype: \mpiint, \num{2x1}~processes, \pingpong, \jupiternecmpi (similar results for \num{1x2}~processes).}
\end{figure*}

\begin{figure*}[htpb]
\centering
\begin{subfigure}{.24\linewidth}
\centering
\includegraphics[width=\linewidth]{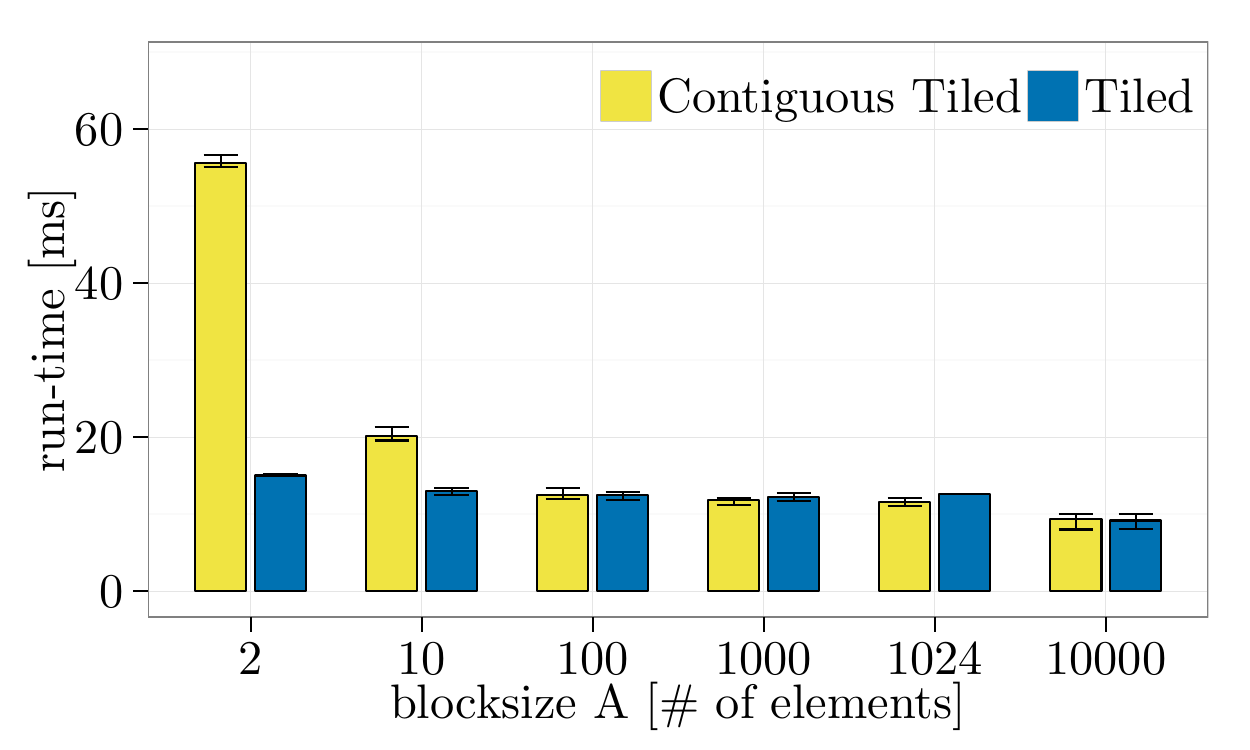}
\caption{%
\label{exp:pingpong-contigtiled-2x1-mvapich}%
\dttiled%
}%
\end{subfigure}%
\hfill%
\begin{subfigure}{.24\linewidth}
\centering
\includegraphics[width=\linewidth]{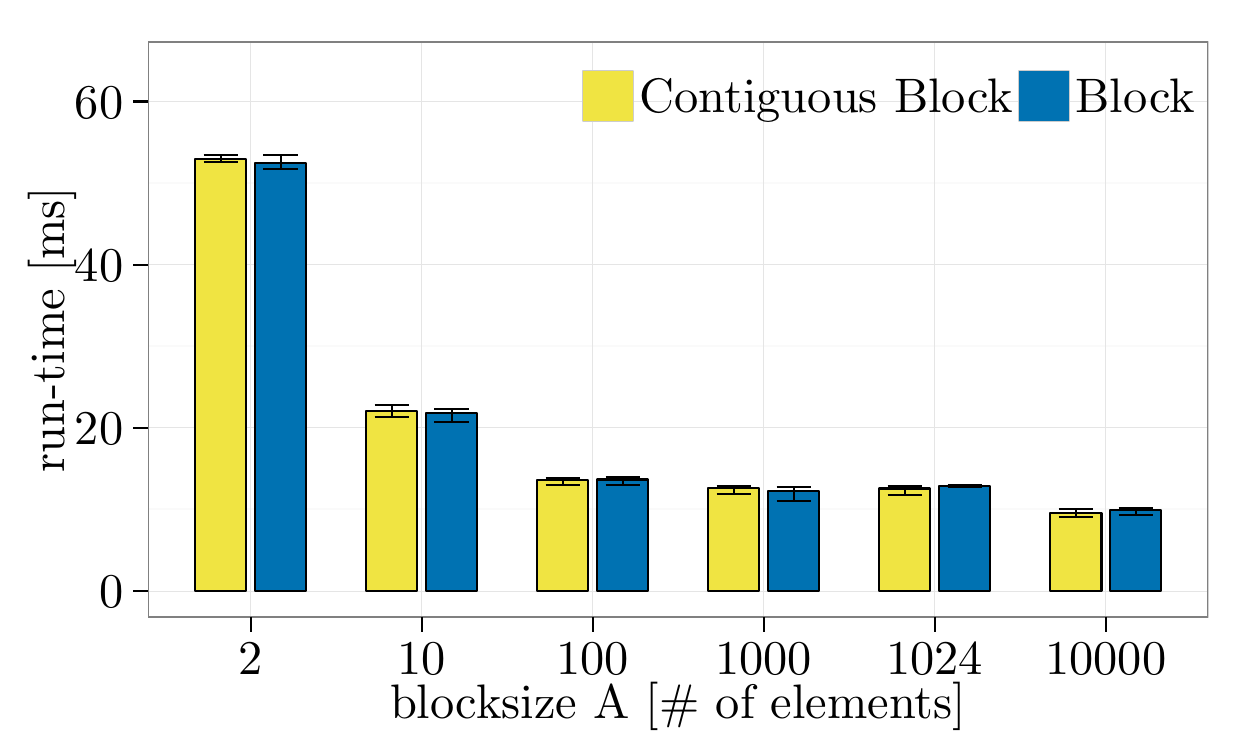}
\caption{%
\label{exp:pingpong-contigblock-2x1-mvapich}%
\dtblock%
}%
\end{subfigure}%
\hfill%
\begin{subfigure}{.24\linewidth}
\centering
\includegraphics[width=\linewidth]{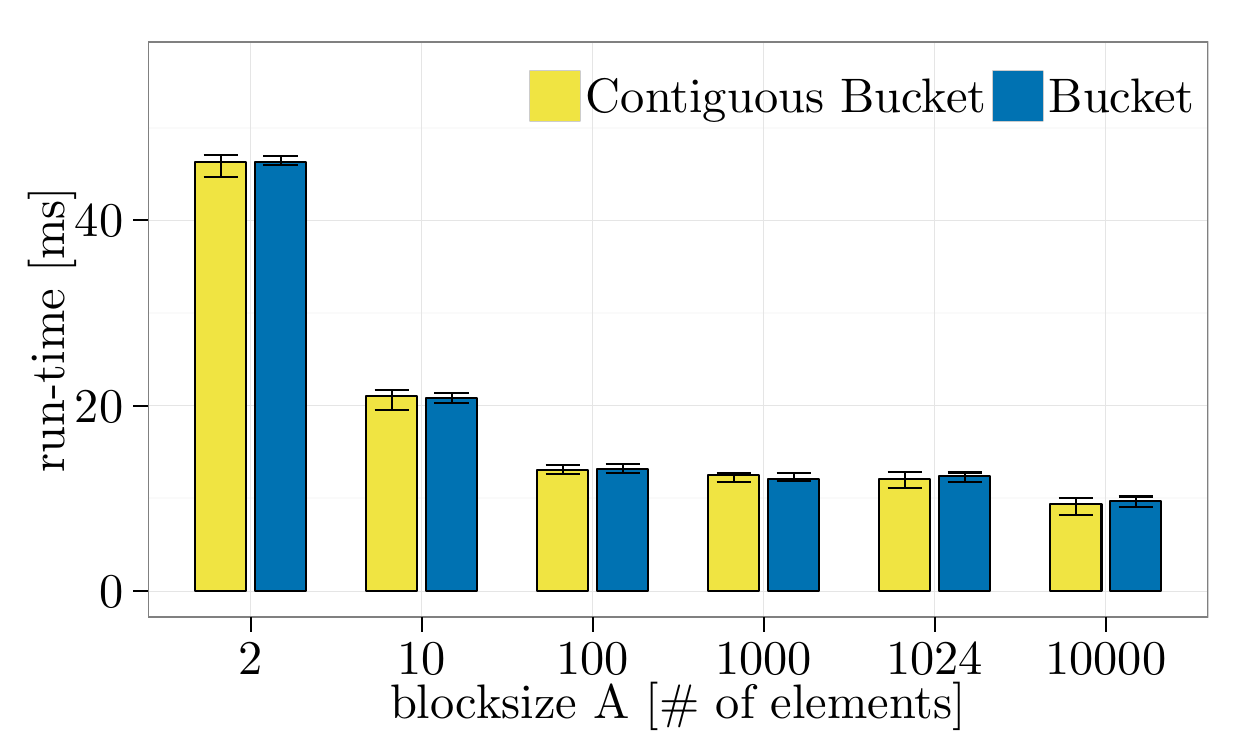}
\caption{%
\label{exp:pingpong-contigbucket-2x1-mvapich}%
\dtbucket%
}%
\end{subfigure}%
\hfill%
\begin{subfigure}{.24\linewidth}
\centering
\includegraphics[width=\linewidth]{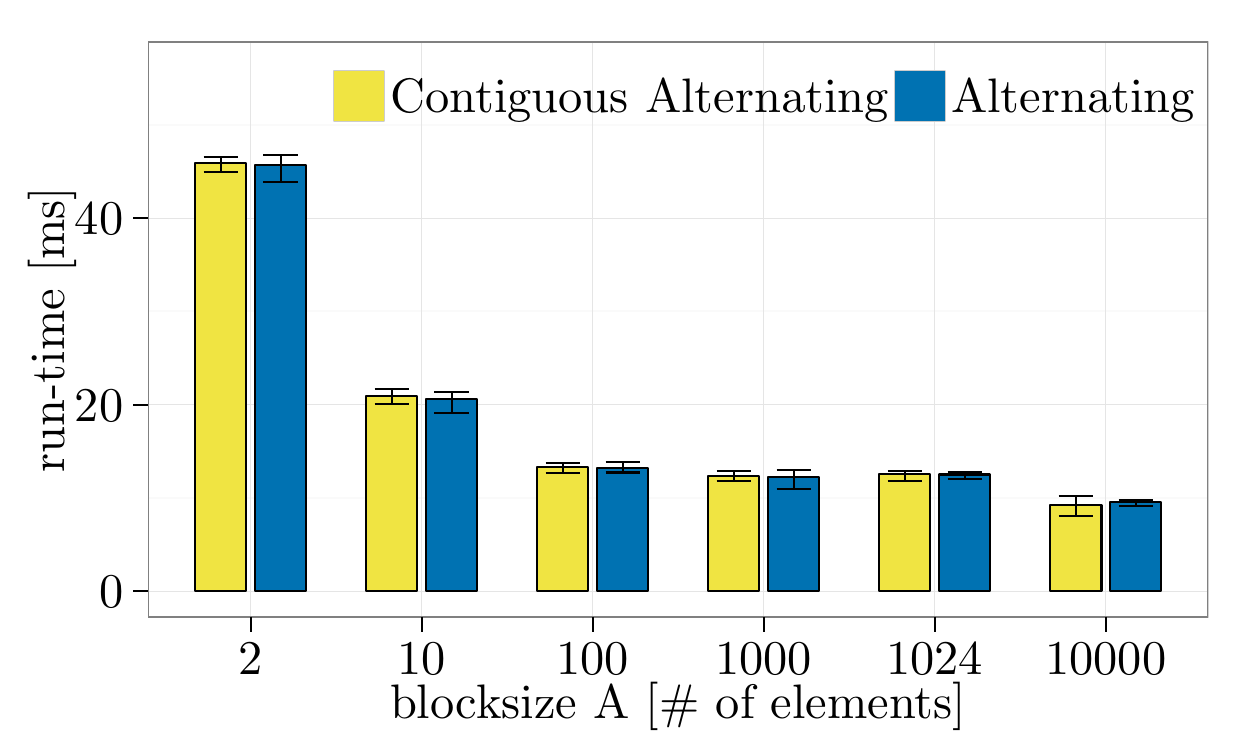}
\caption{%
\label{exp:pingpong-contigalternating-2x1-mvapich}%
\dtalternating%
}%
\end{subfigure}%
\caption{\label{exp:pingpong-contig-largenbytes-2x1-mvapich}  Basic layouts \vs \ddtcontig, $\VARdatasize=\SI{2.56}{\mega\byte}$, element datatype: \mpiint, \num{2x1}~processes, \pingpong, \jupitermvapich (similar results for \num{1x2}~processes).}
\end{figure*}

\begin{figure*}[htpb]
\centering
\begin{subfigure}{.24\linewidth}
\centering
\includegraphics[width=\linewidth]{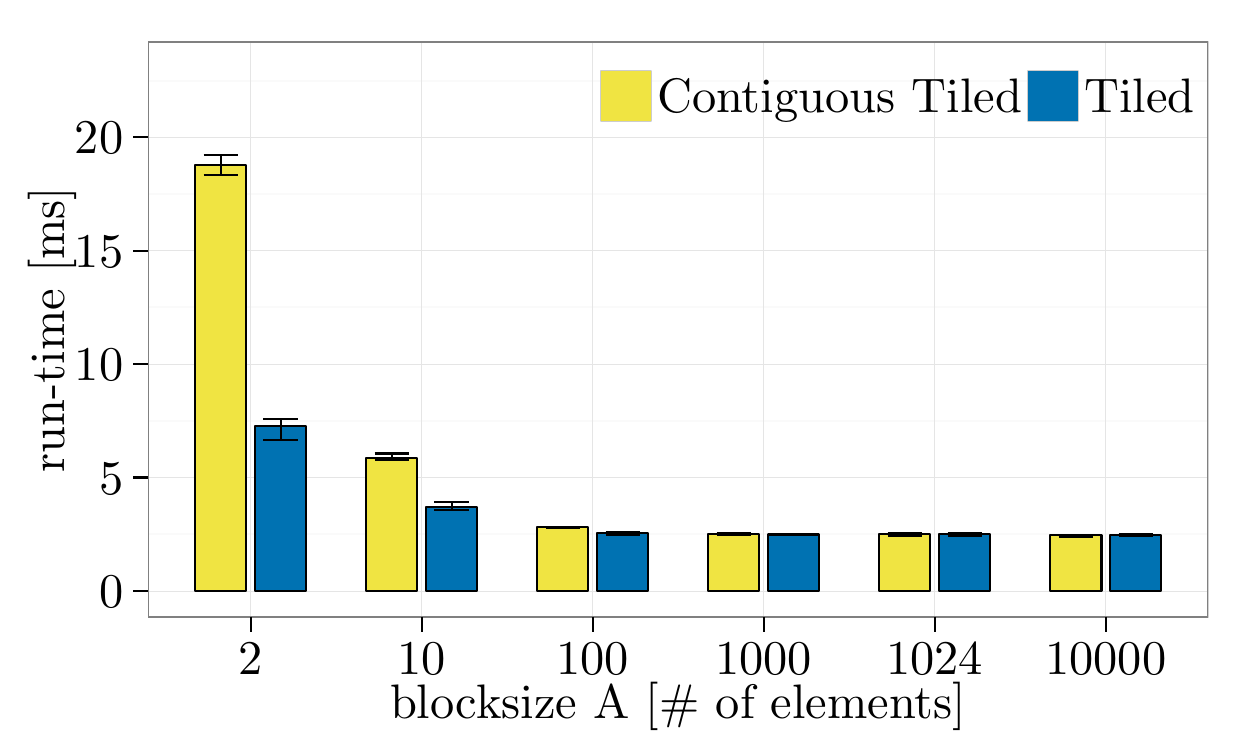}
\caption{%
\label{exp:pingpong-contigtiled-largenbytes-2x1-openmpi}%
\dttiled%
}%
\end{subfigure}%
\hfill%
\begin{subfigure}{.24\linewidth}
\centering
\includegraphics[width=\linewidth]{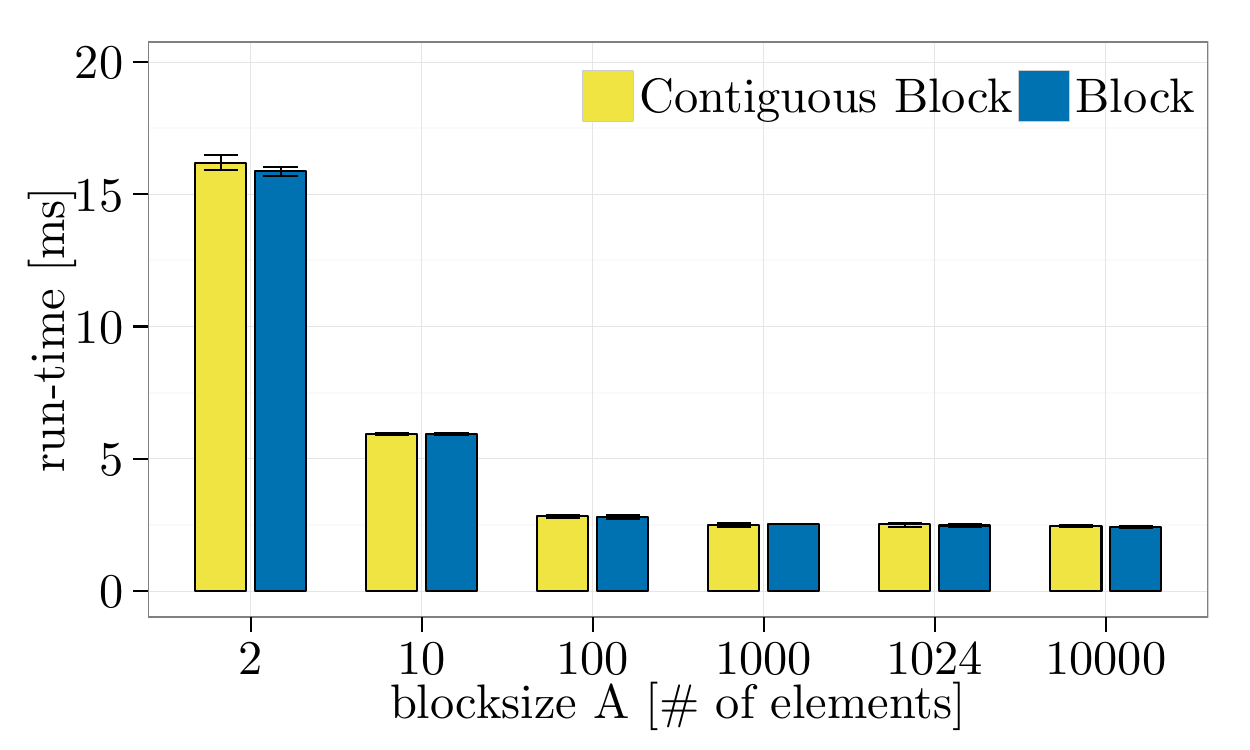}
\caption{%
\label{exp:pingpong-contigblock-largenbytes-2x1-openmpi}%
\dtblock%
}%
\end{subfigure}%
\hfill%
\begin{subfigure}{.24\linewidth}
\centering
\includegraphics[width=\linewidth]{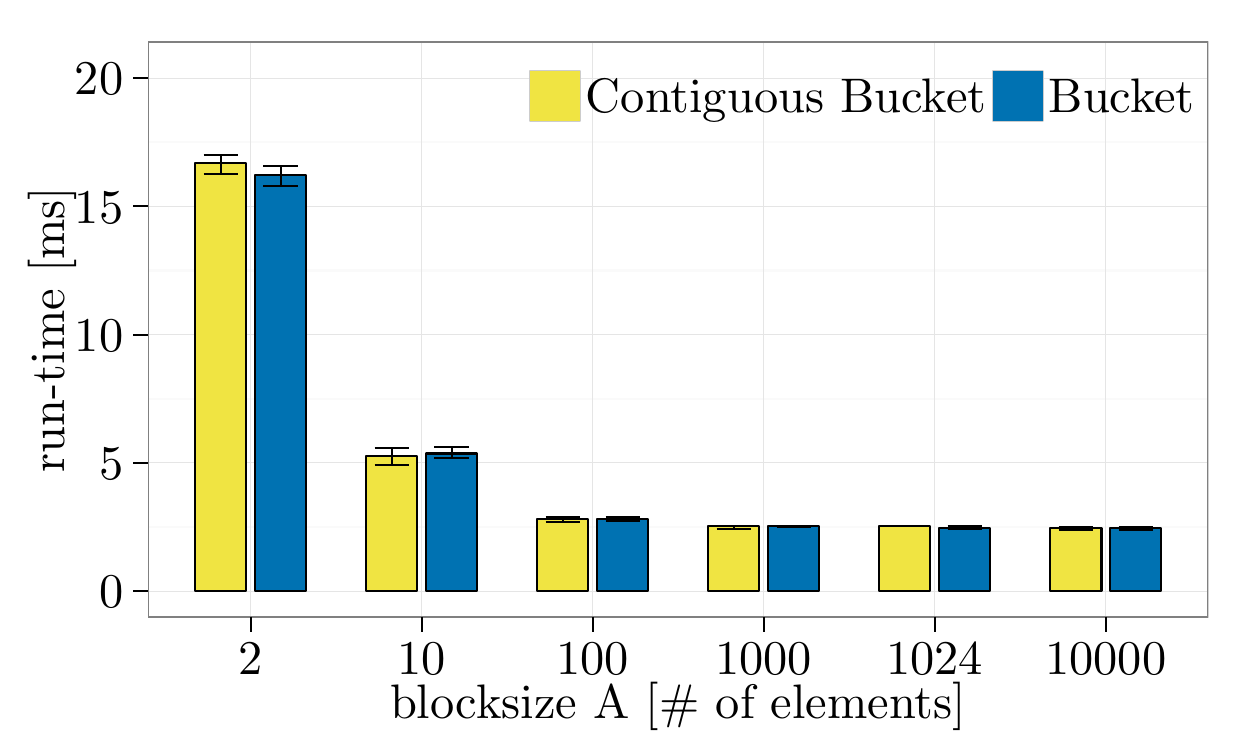}
\caption{%
\label{exp:pingpong-contigbucket-largenbytes-2x1-openmpi}%
\dtbucket%
}%
\end{subfigure}%
\hfill%
\begin{subfigure}{.24\linewidth}
\centering
\includegraphics[width=\linewidth]{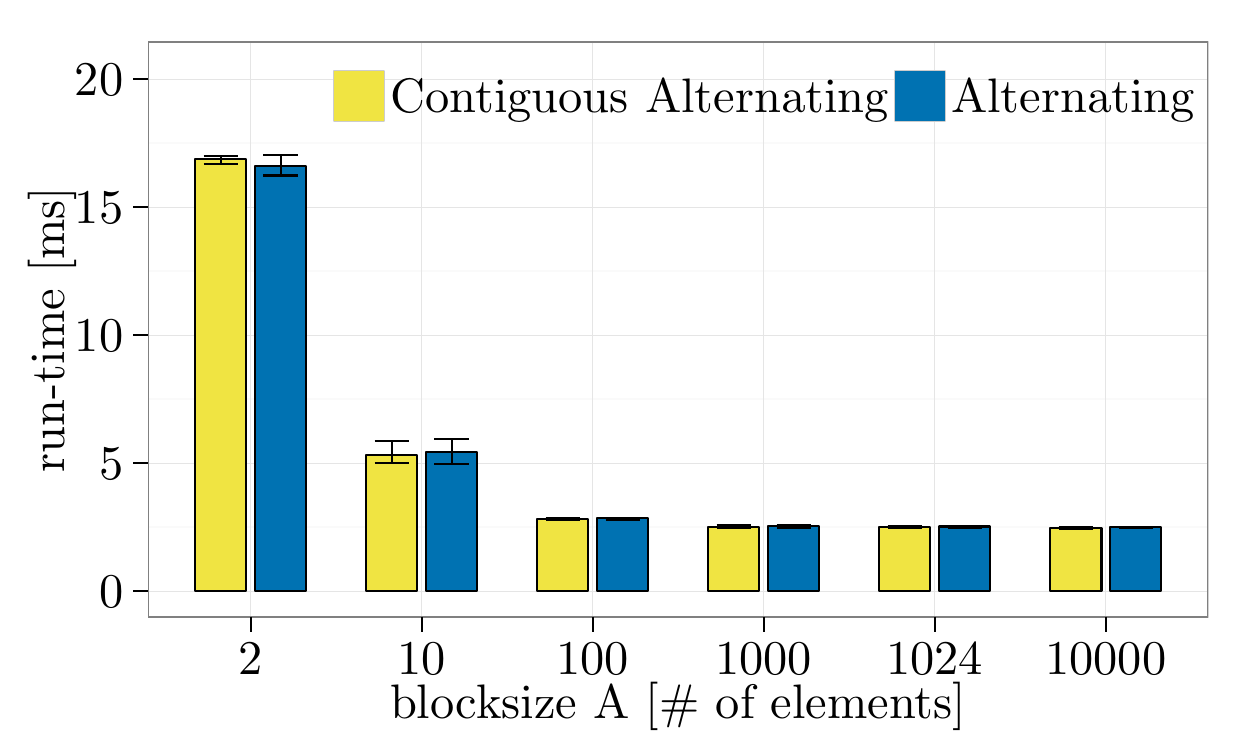}
\caption{%
\label{exp:pingpong-contigalternating-largenbytes-2x1-openmpi}%
\dtalternating%
}%
\end{subfigure}%
\caption{\label{exp:pingpong-contig-largenbytes-2x1-openmpi}  Basic layouts \vs \ddtcontig, $\VARdatasize=\SI{2.56}{\mega\byte}$, element datatype: \mpiint, \num{2x1}~processes, \pingpong, \jupiteropenmpi (similar results for \num{1x2}~processes).}
\end{figure*}

\appexpdesc{
  \begin{expitemize}
    \item \ddtcontig \vs basic layouts (\dttiled, \dtblock, \dtbucket, \dtalternating)
    \item \pingpong
  \end{expitemize}
}{
  \begin{expitemize}
    \item \expparam{\jupiternecmpi, small \datasize}{\fig~\ref{exp:pingpong-contig-smallnbytes-2x1}}
    \item \expparam{\jupitermvapich, small \datasize}{\fig~\ref{exp:pingpong-contig-smallnbytes-2x1-mvapich}}
    \item \expparam{\jupiteropenmpi, small \datasize}{\fig~\ref{exp:pingpong-contig-smallnbytes-2x1-openmpi}}
    \item \expparam{\jupiternecmpi, large \datasize}{\fig~\ref{exp:pingpong-contig-largenbytes-2x1}}
    \item \expparam{\jupitermvapich, large \datasize}{\fig~\ref{exp:pingpong-contig-largenbytes-2x1-mvapich}}
    \item \expparam{\jupiteropenmpi, large \datasize}{\fig~\ref{exp:pingpong-contig-largenbytes-2x1-openmpi}}
  \end{expitemize}  
}

\FloatBarrier
\clearpage

\appexp{exptest:tiled_struct}

\begin{figure*}[htpb]
\centering
\begin{subfigure}{.24\linewidth}
\centering
\includegraphics[width=\linewidth]{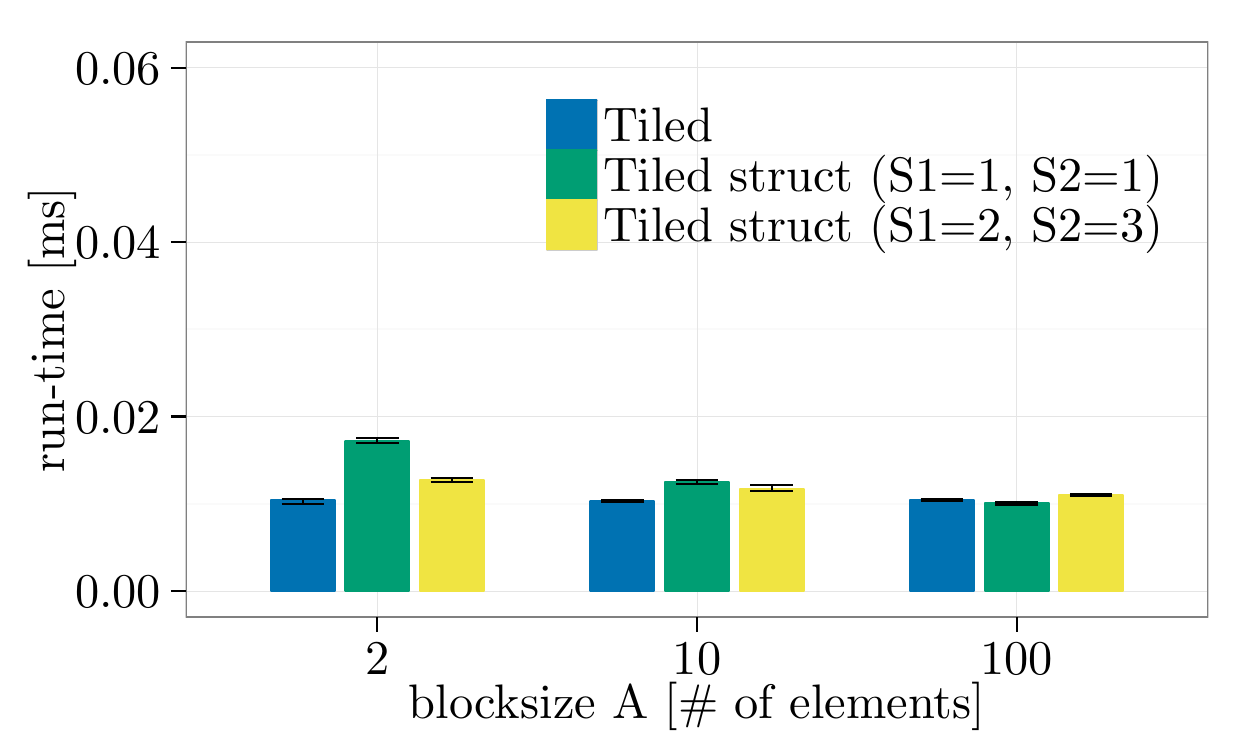}
\caption{%
\label{exp:pingpong-tiledstruct-small-2x1}%
$\VARdatasize=\SI{2}{\kilo\byte}$, \num{2}~nodes%
}%
\end{subfigure}%
\hfill%
\begin{subfigure}{.24\linewidth}
\centering
\includegraphics[width=\linewidth]{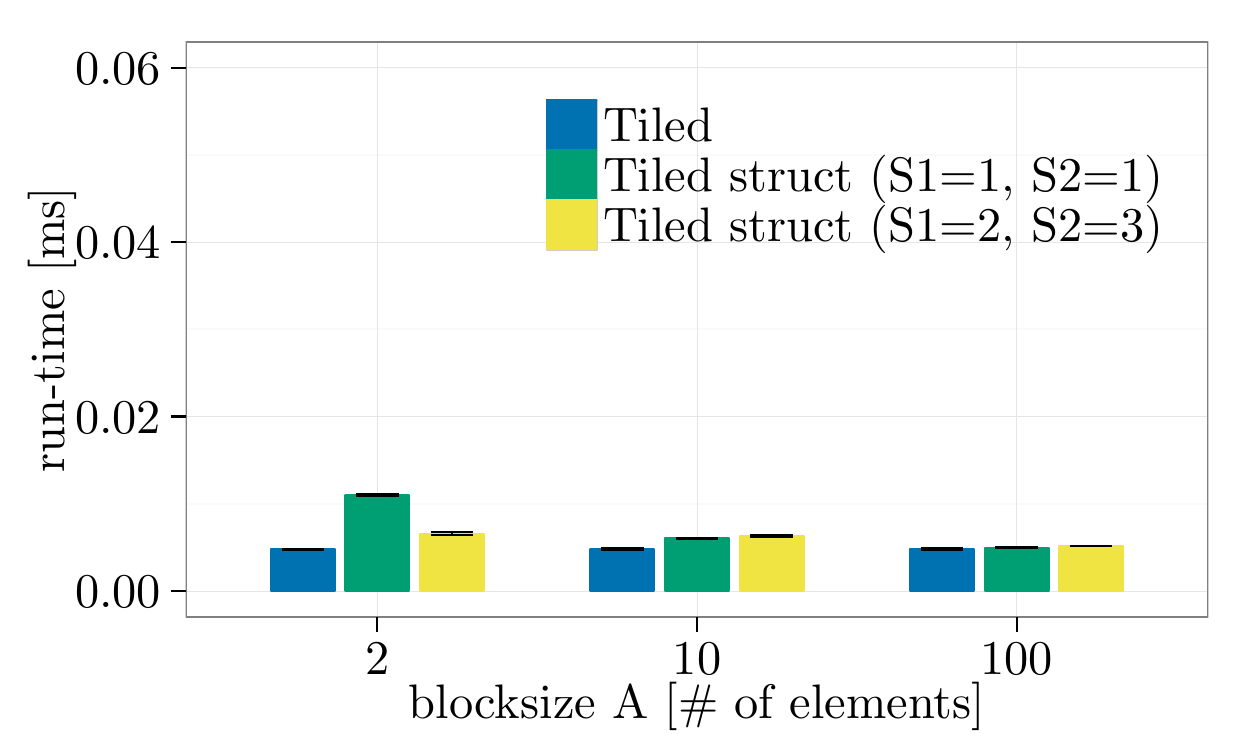}
\caption{%
\label{exp:pingpong-tiledstruct-small-1x2}%
$\VARdatasize=\SI{2}{\kilo\byte}$, same node%
}%
\end{subfigure}%
\hfill%
\begin{subfigure}{.24\linewidth}
\centering
\includegraphics[width=\linewidth]{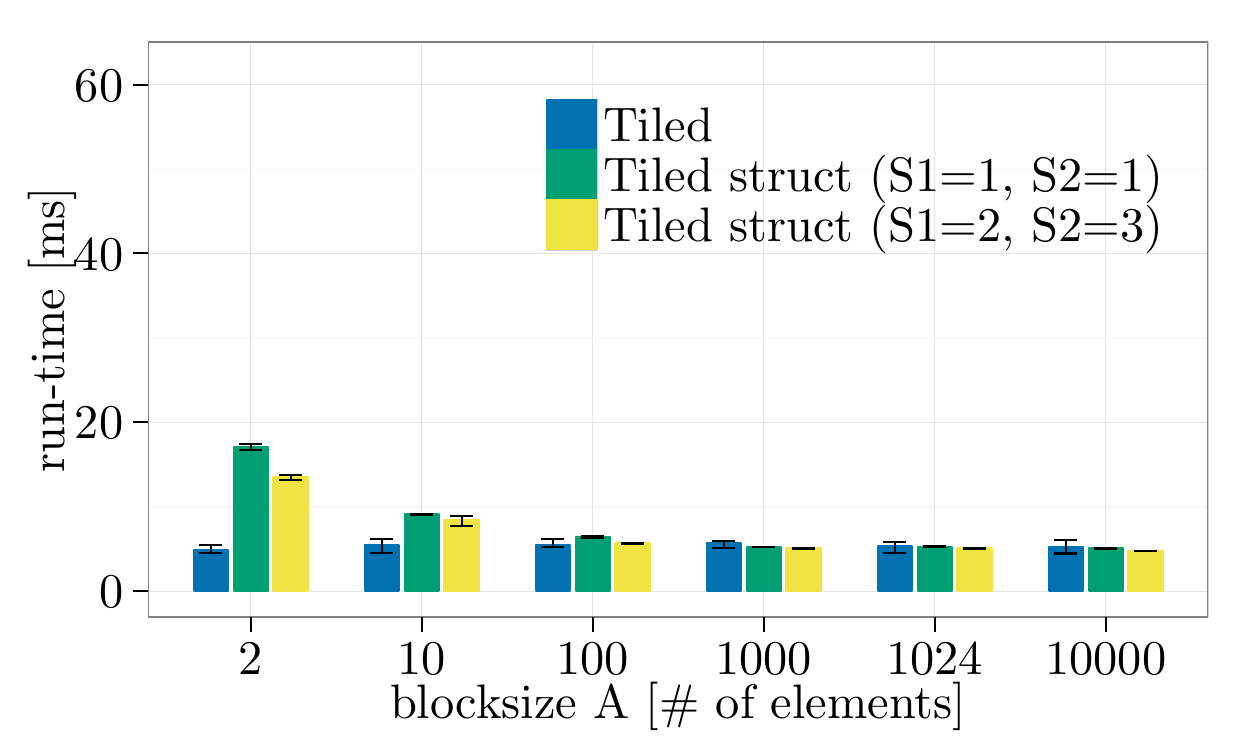}
\caption{%
\label{exp:pingpong-tiledstruct-large-2x1}%
$\VARdatasize=\SI{2.56}{\mega\byte}$, \num{2}~nodes%
}%
\end{subfigure}%
\hfill%
\begin{subfigure}{.24\linewidth}
\centering
\includegraphics[width=\linewidth]{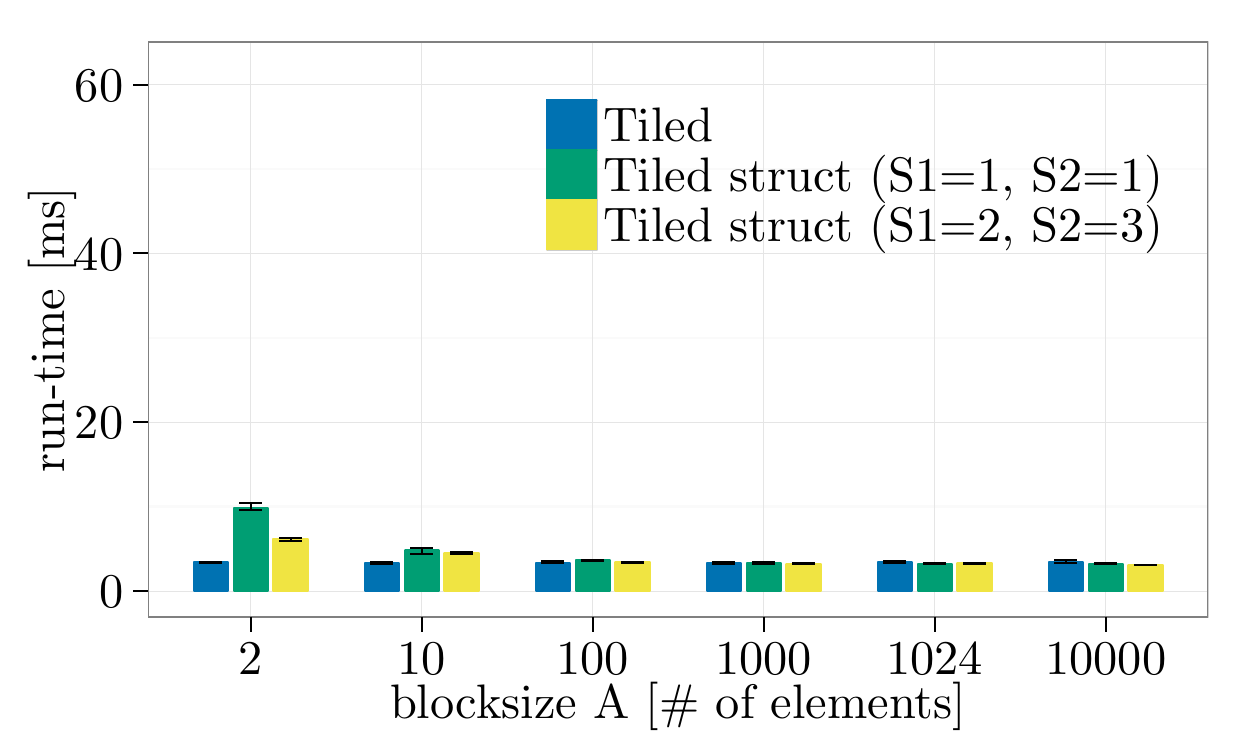}
\caption{%
\label{exp:pingpong-tiledstruct-large-1x2}%
$\VARdatasize=\SI{2.56}{\mega\byte}$, same node%
}%
\end{subfigure}%
\caption{\label{exp:pingpong-tiledstruct-nec} \dtdtiled \vs \dttiledstruct, element datatype: \mpiint, \pingpong, \jupiternecmpi.}
\end{figure*}

\begin{figure*}[htpb]
\centering
\begin{subfigure}{.24\linewidth}
\centering
\includegraphics[width=\linewidth]{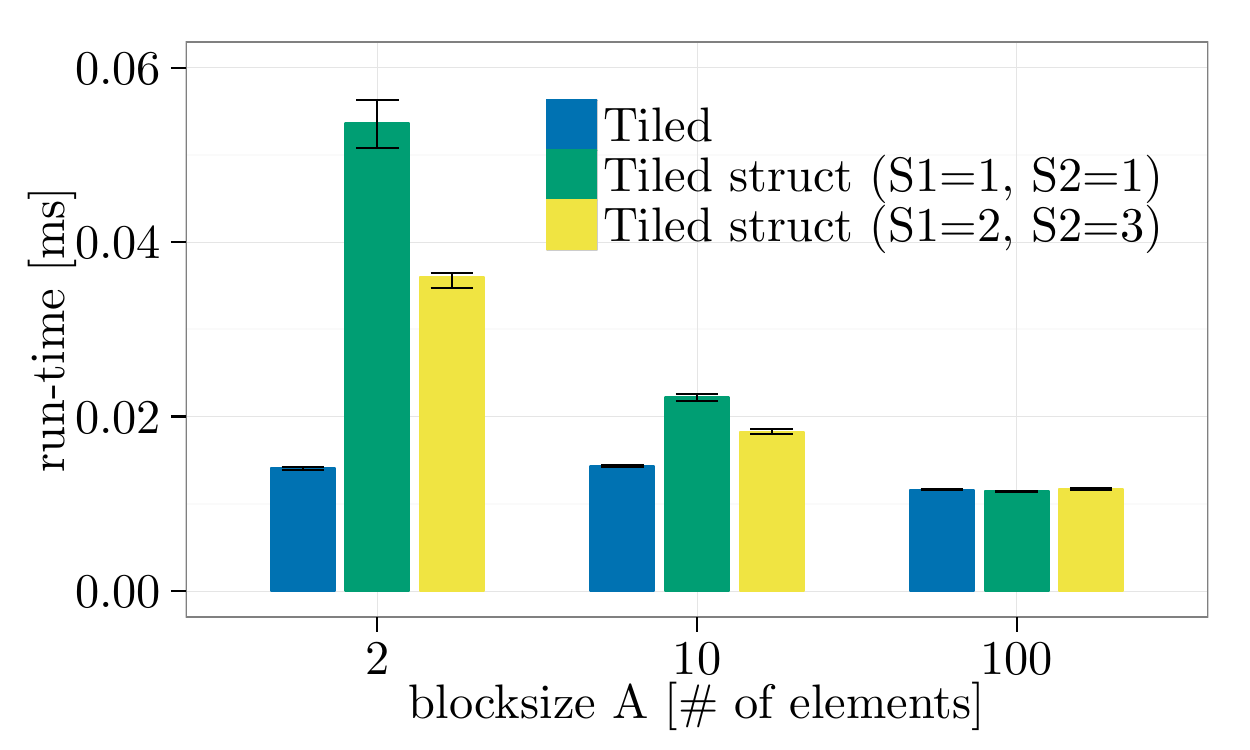}
\caption{%
\label{exp:pingpong-tiledstruct-small-2x1-mvapich}%
$\VARdatasize=\SI{2}{\kilo\byte}$, \num{2}~nodes%
}%
\end{subfigure}%
\hfill%
\begin{subfigure}{.24\linewidth}
\centering
\includegraphics[width=\linewidth]{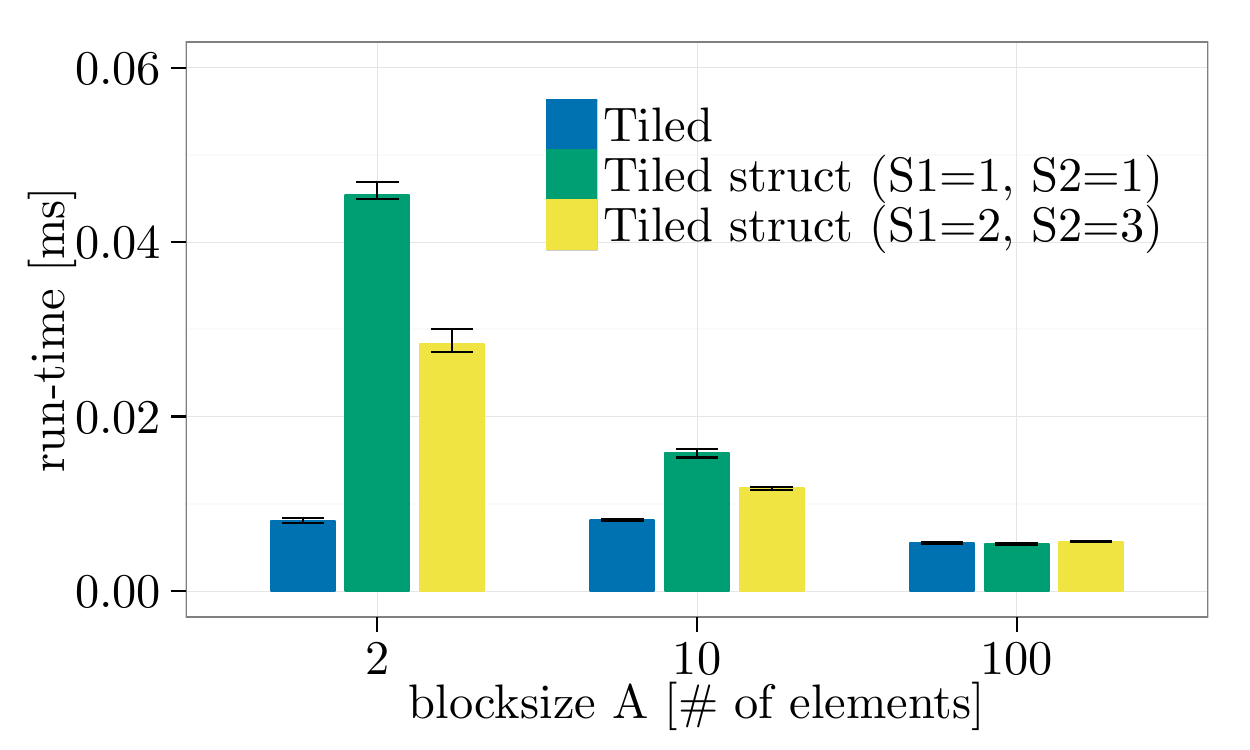}
\caption{%
\label{exp:pingpong-tiledstruct-small-1x2-mvapich}%
$\VARdatasize=\SI{2}{\kilo\byte}$, same node%
}%
\end{subfigure}%
\hfill%
\begin{subfigure}{.24\linewidth}
\centering
\includegraphics[width=\linewidth]{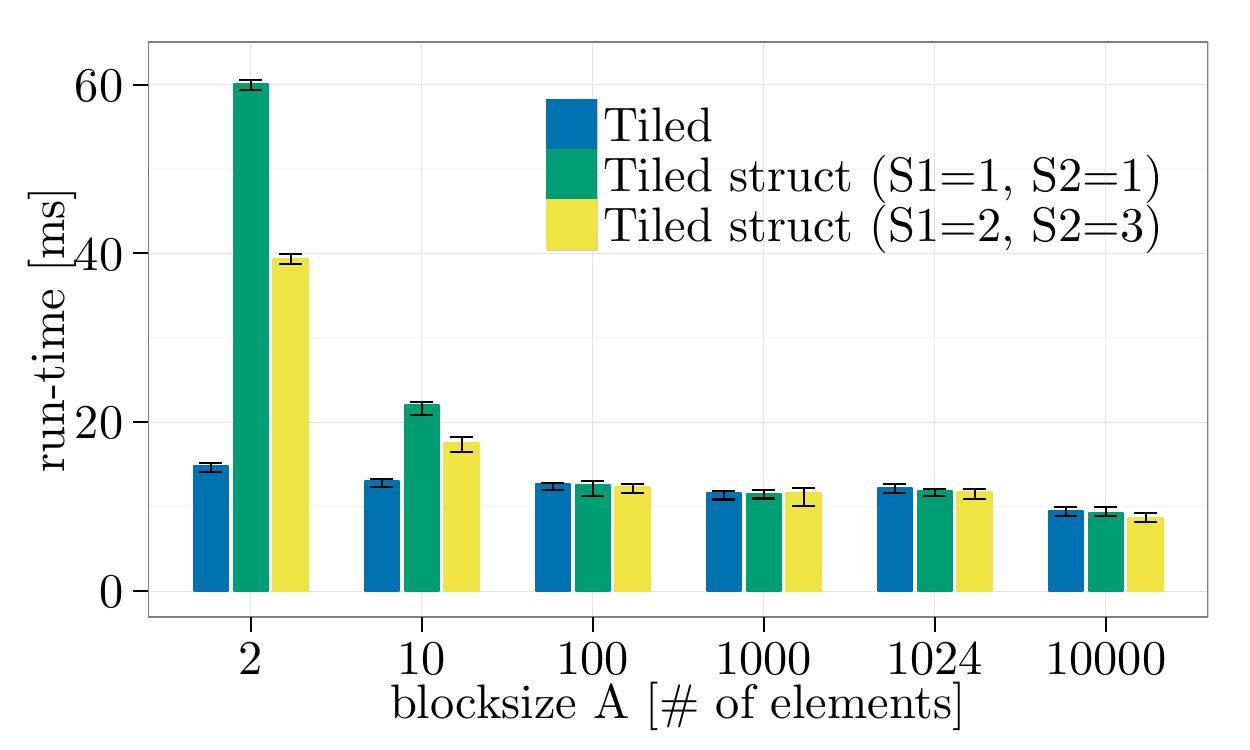}
\caption{%
\label{exp:pingpong-tiledstruct-large-2x1-mvapich}%
$\VARdatasize=\SI{2.56}{\mega\byte}$, \num{2}~nodes%
}%
\end{subfigure}%
\hfill%
\begin{subfigure}{.24\linewidth}
\centering
\includegraphics[width=\linewidth]{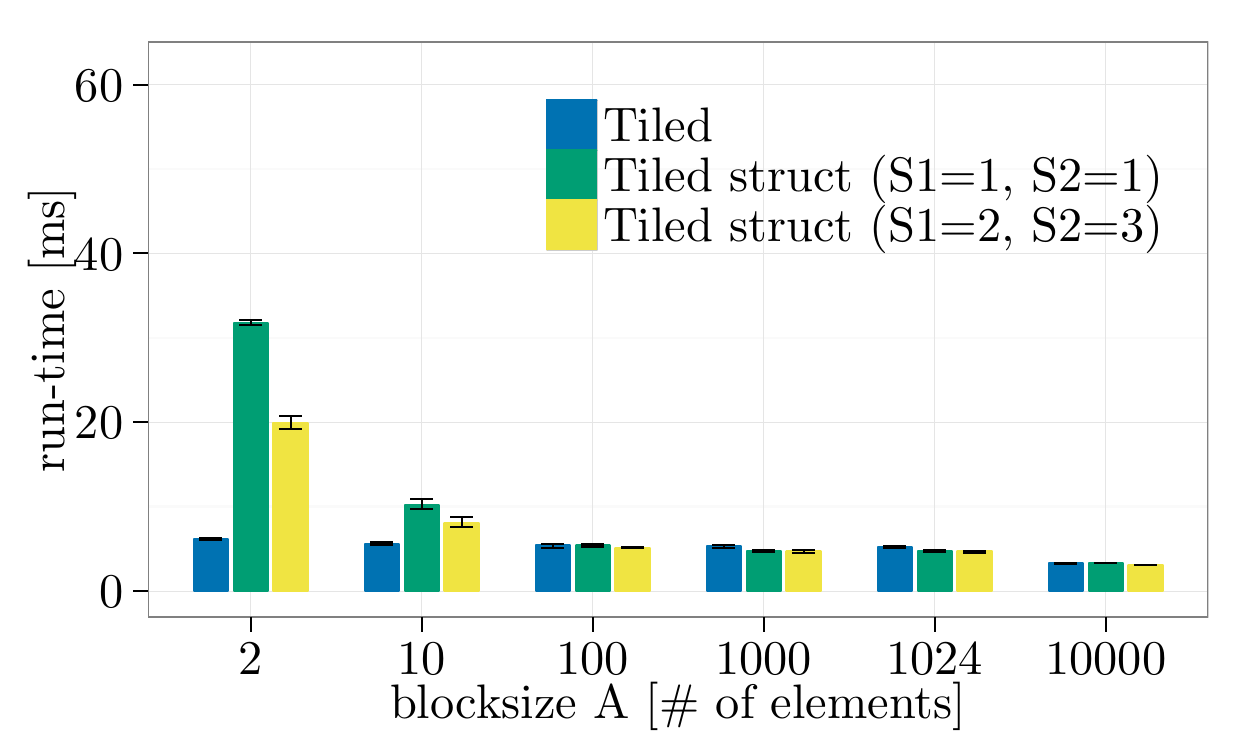}
\caption{%
\label{exp:pingpong-tiledstruct-large-1x2-mvapich}%
$\VARdatasize=\SI{2.56}{\mega\byte}$, same node%
}%
\end{subfigure}%
\caption{\label{exp:pingpong-tiledstruct-mvapich} \dtdtiled \vs \dttiledstruct, element datatype: \mpiint, \pingpong, \jupitermvapich.}
\end{figure*}

\begin{figure*}[htpb]
\centering
\begin{subfigure}{.24\linewidth}
\centering
\includegraphics[width=\linewidth]{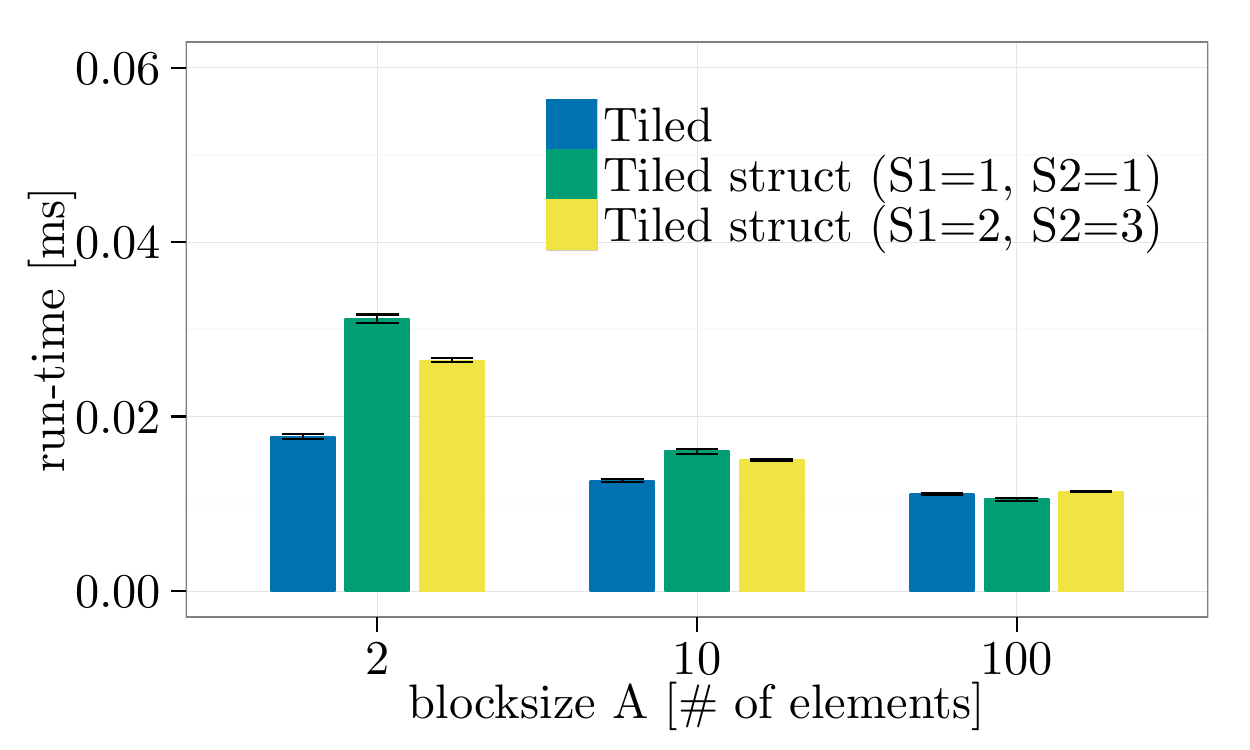}
\caption{%
\label{exp:pingpong-tiledstruct-small-2x1-openmpi}%
$\VARdatasize=\SI{2}{\kilo\byte}$, \num{2}~nodes%
}%
\end{subfigure}%
\hfill%
\begin{subfigure}{.24\linewidth}
\centering
\includegraphics[width=\linewidth]{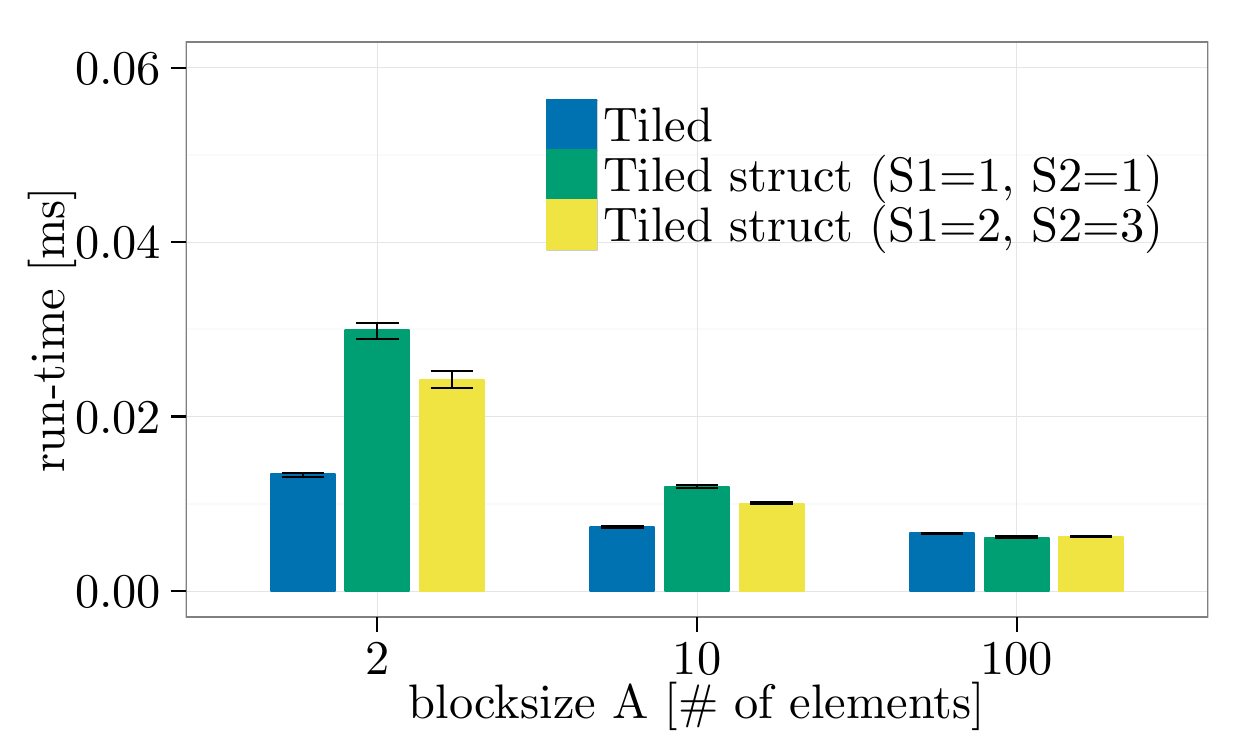}
\caption{%
\label{exp:pingpong-tiledstruct-small-1x2-openmpi}%
$\VARdatasize=\SI{2}{\kilo\byte}$, same node%
}%
\end{subfigure}%
\hfill%
\begin{subfigure}{.24\linewidth}
\centering
\includegraphics[width=\linewidth]{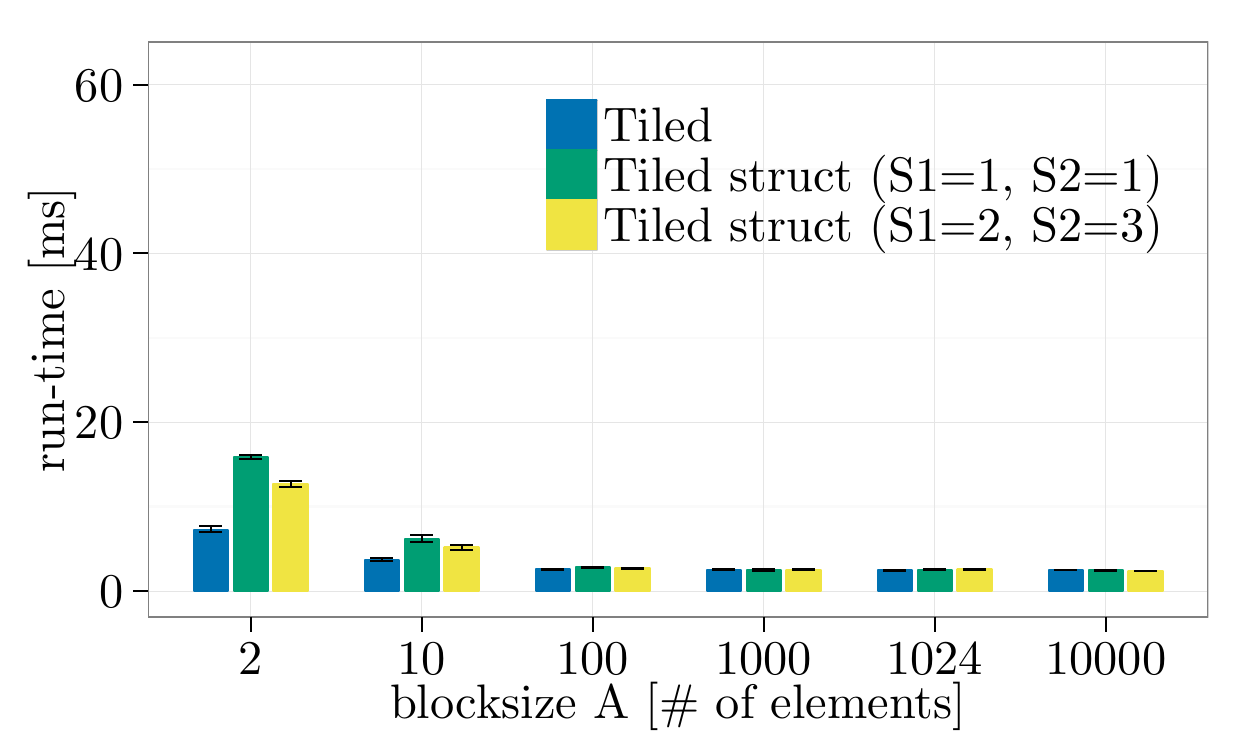}
\caption{%
\label{exp:pingpong-tiledstruct-large-2x1-openmpi}%
$\VARdatasize=\SI{2.56}{\mega\byte}$, \num{2}~nodes%
}%
\end{subfigure}%
\hfill%
\begin{subfigure}{.24\linewidth}
\centering
\includegraphics[width=\linewidth]{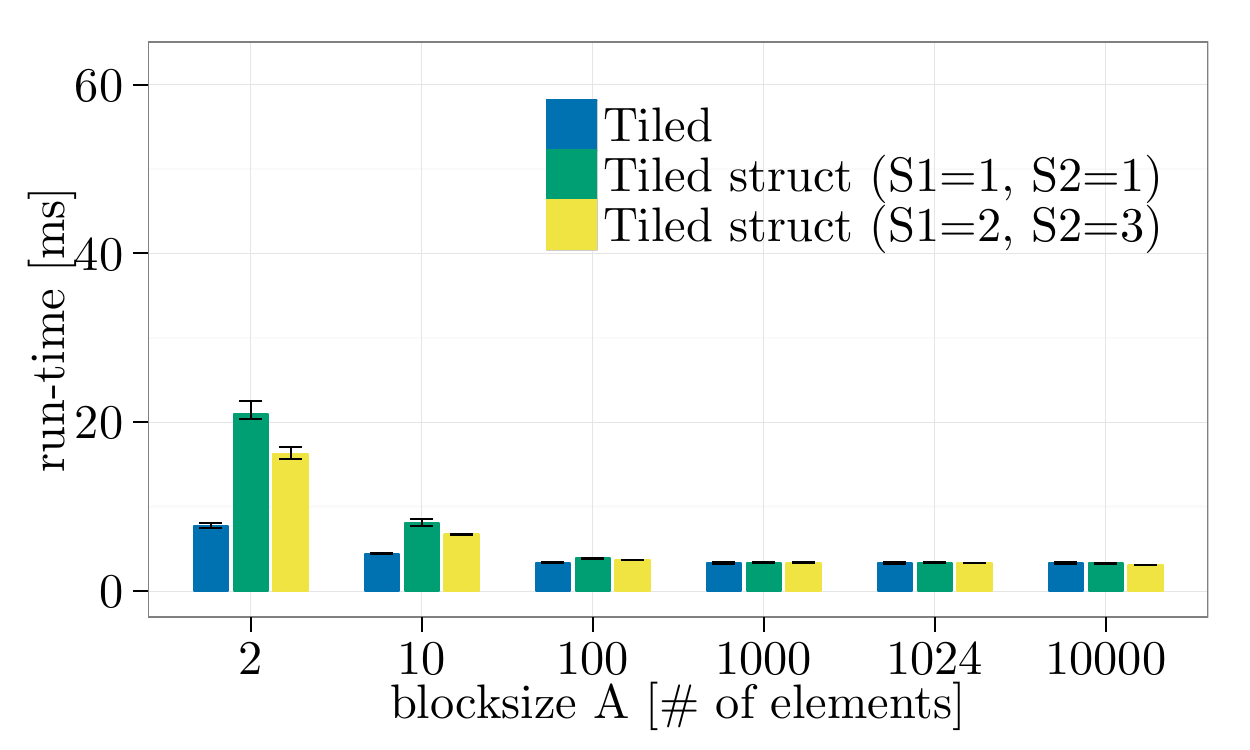}
\caption{%
\label{exp:pingpong-tiledstruct-large-1x2-openmpi}%
$\VARdatasize=\SI{2.56}{\mega\byte}$, same node%
}%
\end{subfigure}%
\caption{\label{exp:pingpong-tiledstruct-openmpi} \dtdtiled \vs \dttiledstruct, element datatype: \mpiint, \pingpong, \jupiteropenmpi.}
\end{figure*}

\appexpdesc{
  \begin{expitemize}
    \item \dtdtiled, \dttiledstruct
    \item \pingpong
  \end{expitemize}
}{
  \begin{expitemize}
    \item \expparam{\jupiternecmpi}{\fig~\ref{exp:pingpong-tiledstruct-nec}}
    \item \expparam{\jupitermvapich}{\fig~\ref{exp:pingpong-tiledstruct-mvapich}}
    \item \expparam{\jupiteropenmpi}{\fig~\ref{exp:pingpong-tiledstruct-openmpi}}
  \end{expitemize}  
}

\FloatBarrier
\clearpage

\appexp{exptest:tiled_vector}

\begin{figure*}[htpb]
\centering
\begin{subfigure}{.24\linewidth}
\centering
\includegraphics[width=\linewidth]{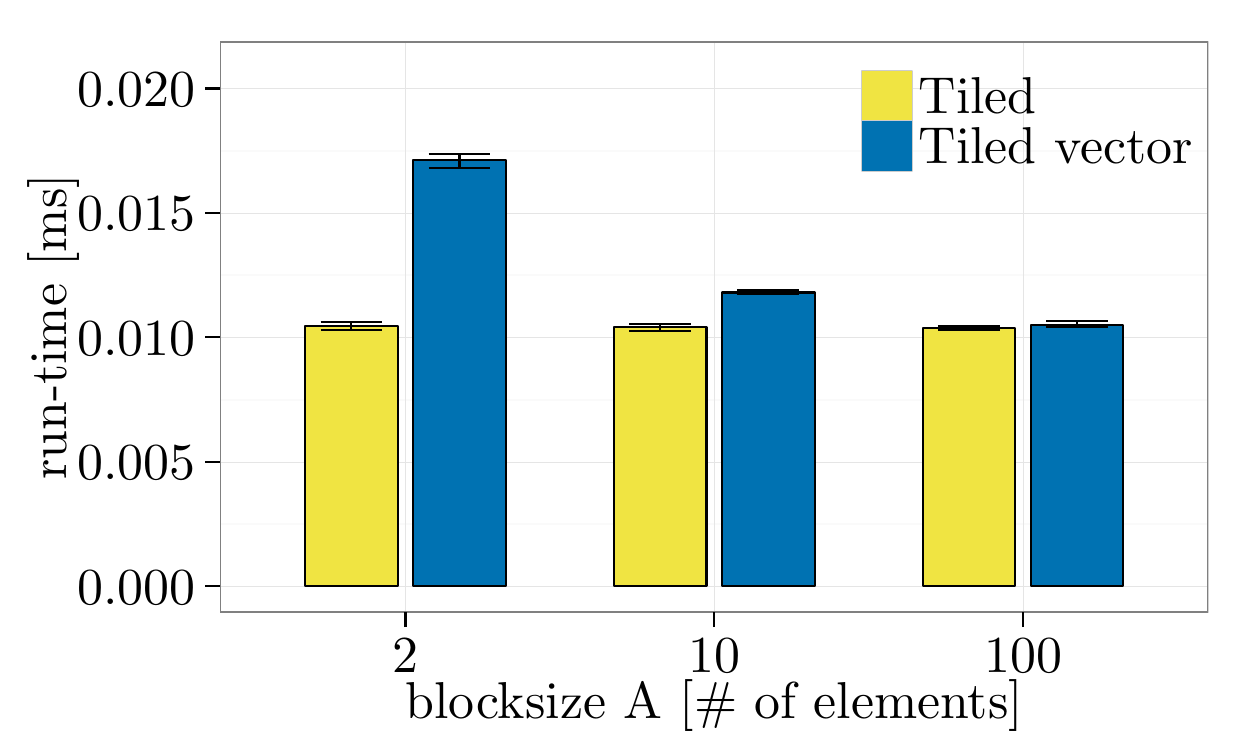}
\caption{%
\label{exp:pingpong-tiledvec-small-2x1}%
$\VARdatasize=\SI{2}{\kilo\byte}$, \num{2}~nodes%
}%
\end{subfigure}%
\hfill%
\begin{subfigure}{.24\linewidth}
\centering
\includegraphics[width=\linewidth]{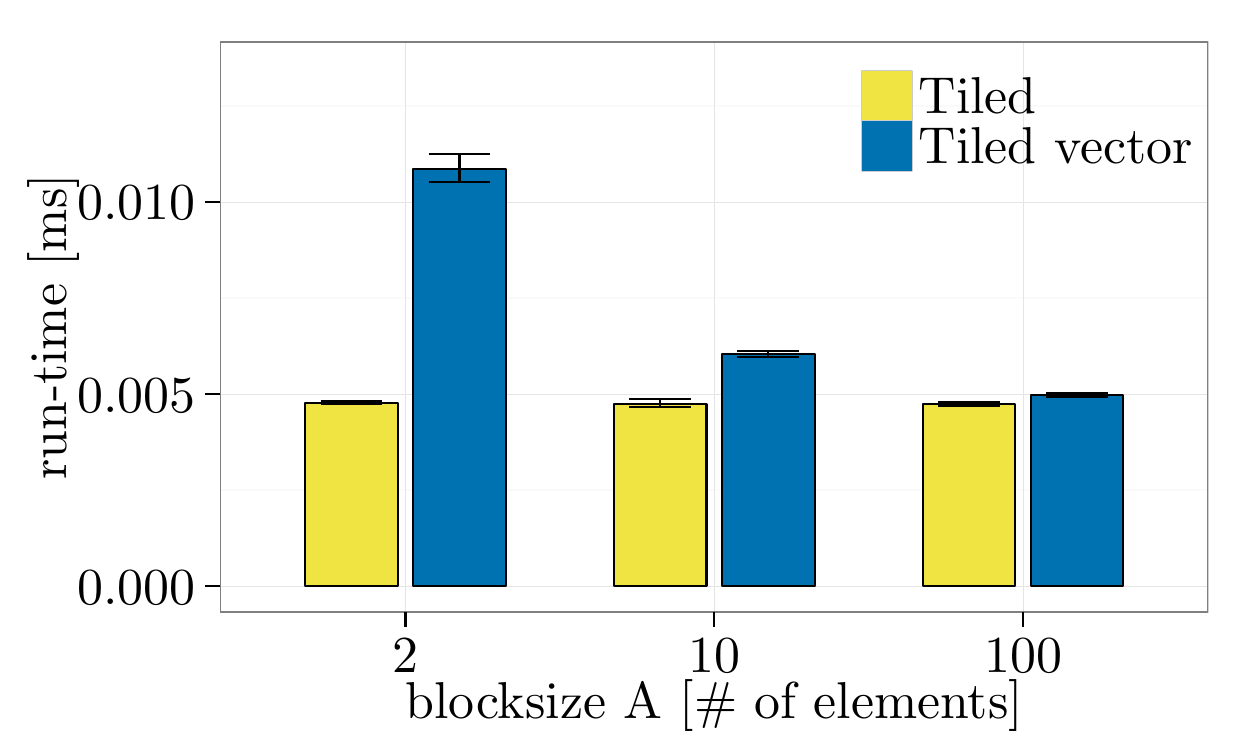}
\caption{%
\label{exp:pingpong-tiledvec-small-1x2}%
$\VARdatasize=\SI{2}{\kilo\byte}$, same node%
}%
\end{subfigure}%
\hfill%
\begin{subfigure}{.24\linewidth}
\centering
\includegraphics[width=\linewidth]{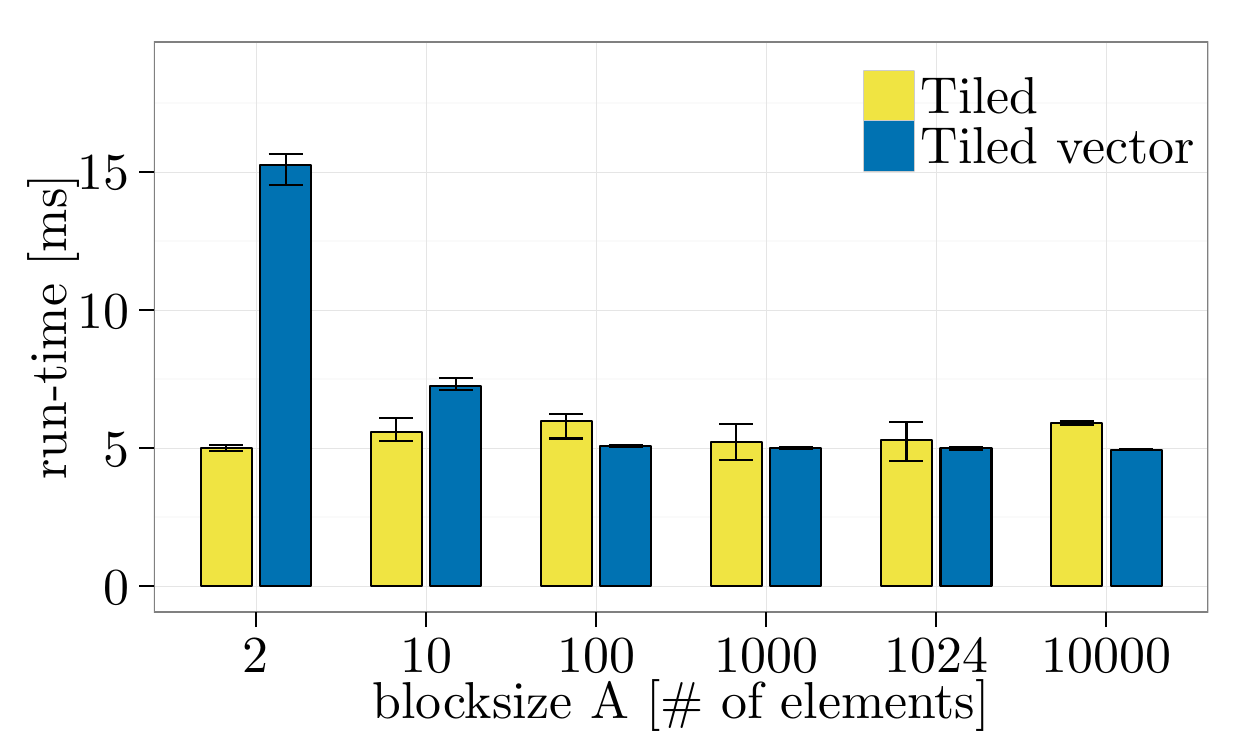}
\caption{%
\label{exp:pingpong-tiledvec-large-2x1}%
$\VARdatasize=\SI{2.56}{\mega\byte}$, \num{2}~nodes%
}%
\end{subfigure}%
\hfill%
\begin{subfigure}{.24\linewidth}
\centering
\includegraphics[width=\linewidth]{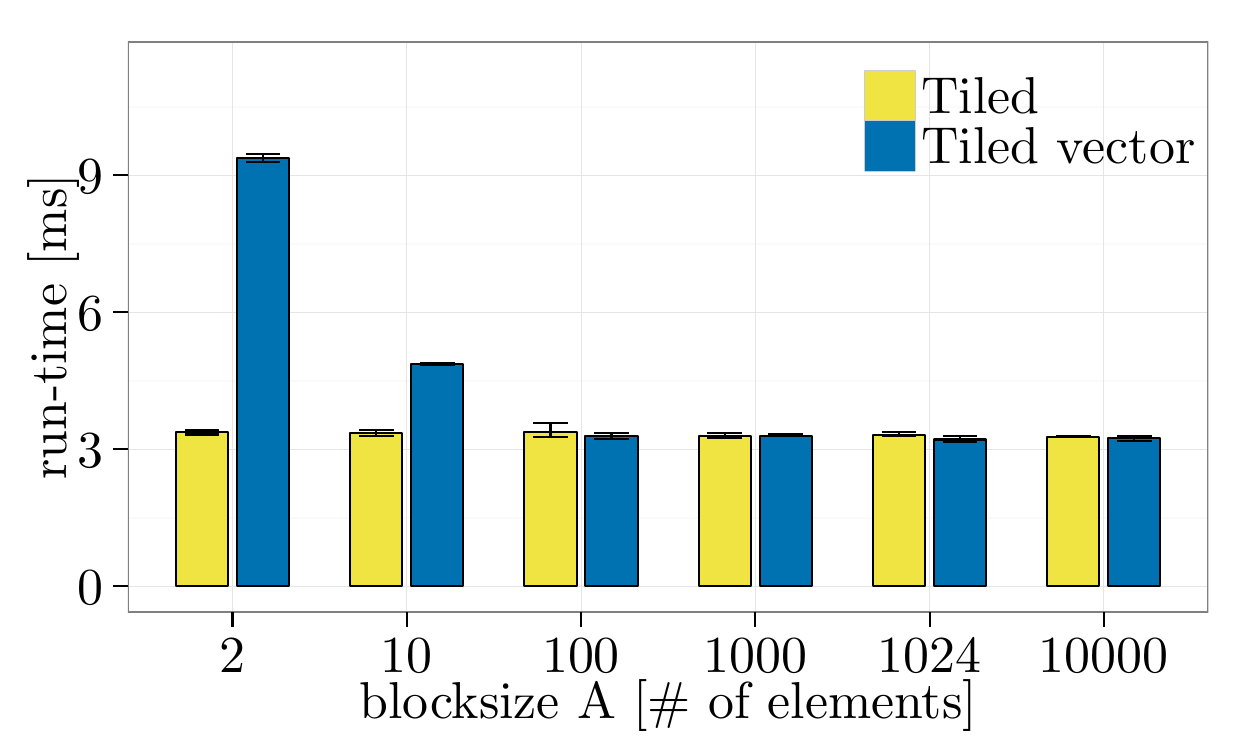}
\caption{%
\label{exp:pingpong-tiledvec-large-1x2}%
$\VARdatasize=\SI{2.56}{\mega\byte}$, same node%
}%
\end{subfigure}%
\caption{\label{exp:pingpong-tiledvec-nec} \dtdtiled \vs \ddttiledvector, element datatype: \mpiint, \pingpong, \jupiternecmpi.}
\end{figure*}

\begin{figure*}[htpb]
\centering
\begin{subfigure}{.24\linewidth}
\centering
\includegraphics[width=\linewidth]{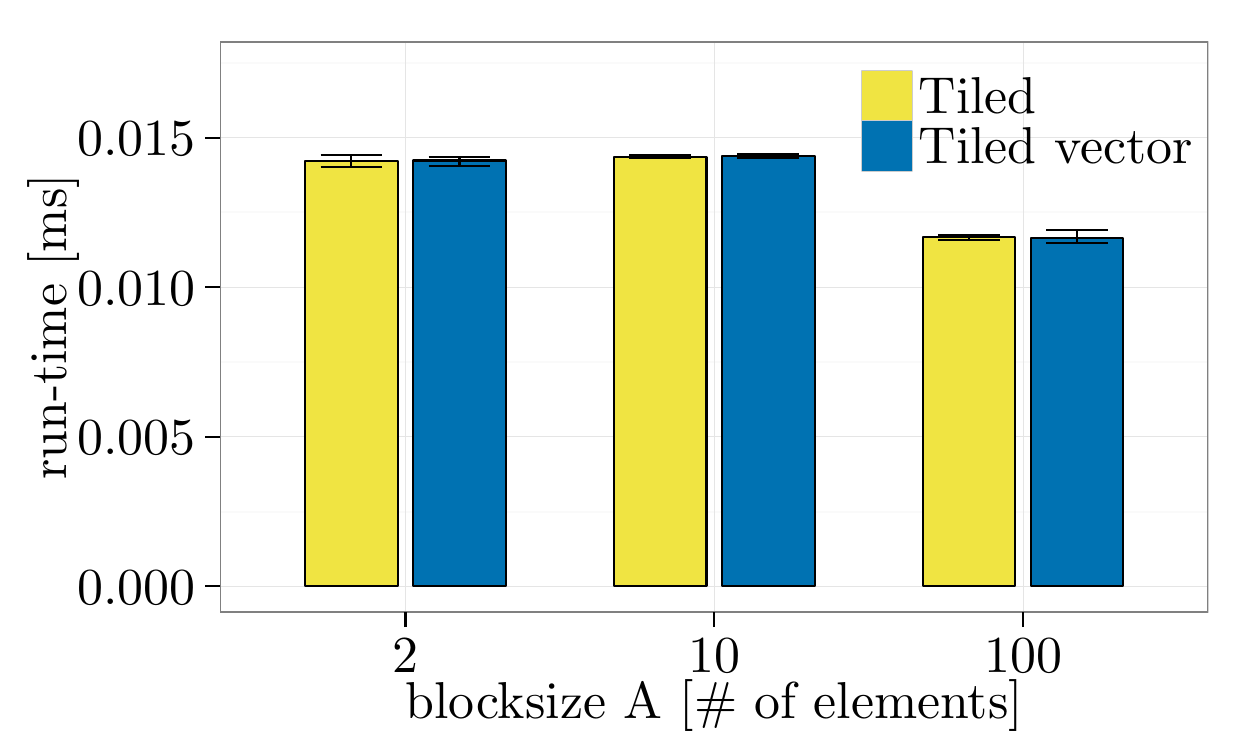}
\caption{%
\label{exp:pingpong-tiledvec-small-2x1-mvapich}%
$\VARdatasize=\SI{2}{\kilo\byte}$, \num{2}~nodes%
}%
\end{subfigure}%
\hfill%
\begin{subfigure}{.24\linewidth}
\centering
\includegraphics[width=\linewidth]{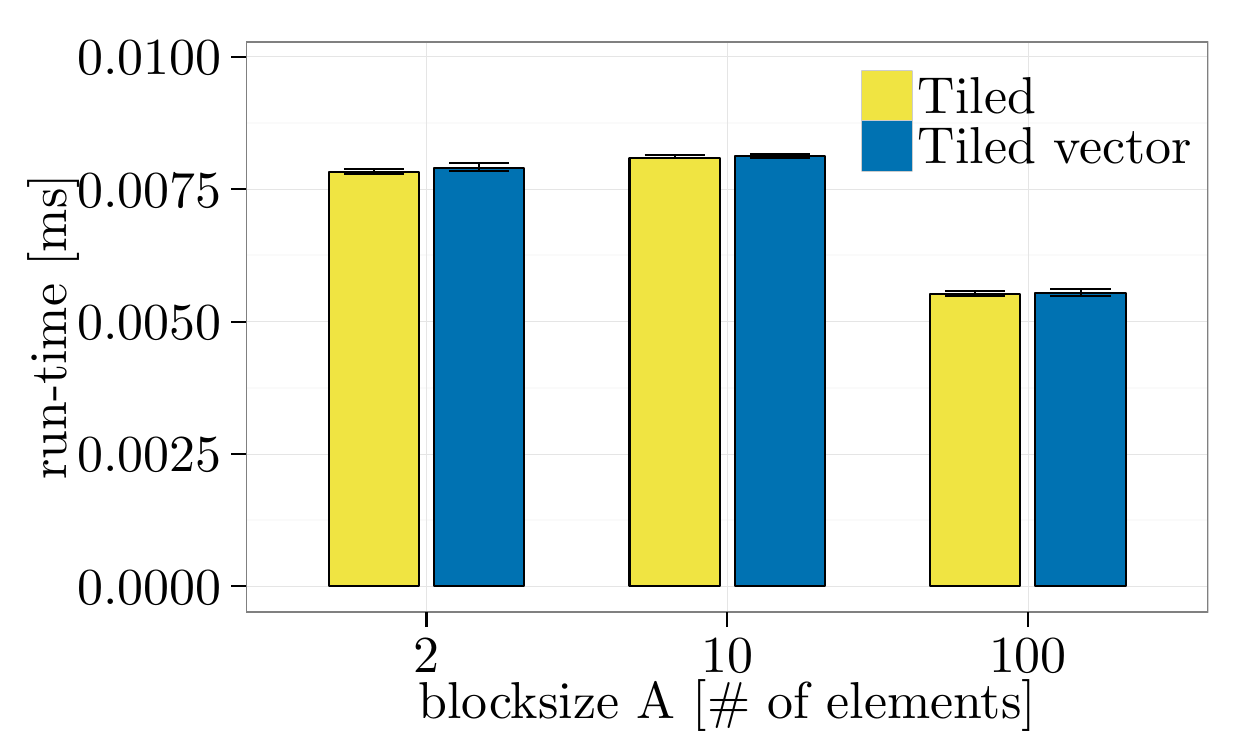}
\caption{%
\label{exp:pingpong-tiledvec-small-1x2-mvapich}%
$\VARdatasize=\SI{2}{\kilo\byte}$, same node%
}%
\end{subfigure}%
\hfill%
\begin{subfigure}{.24\linewidth}
\centering
\includegraphics[width=\linewidth]{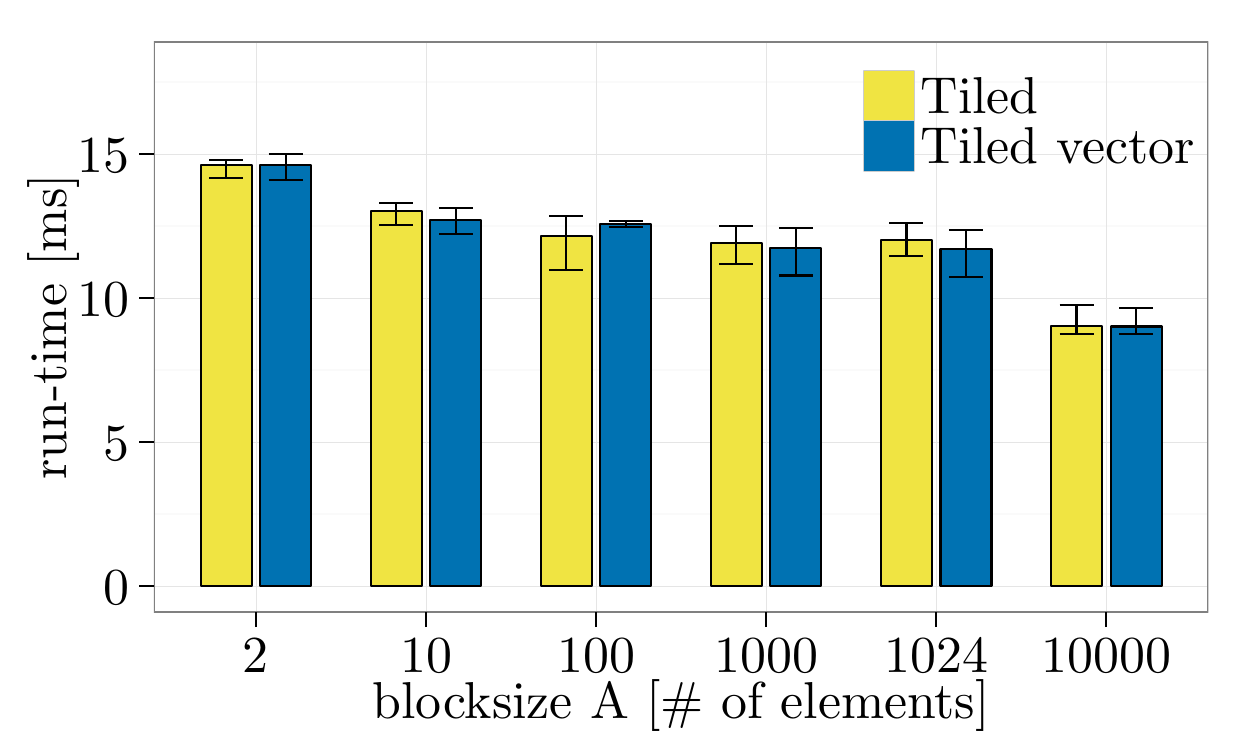}
\caption{%
\label{exp:pingpong-tiledvec-large-2x1-mvapich}%
$\VARdatasize=\SI{2.56}{\mega\byte}$, \num{2}~nodes%
}%
\end{subfigure}%
\hfill%
\begin{subfigure}{.24\linewidth}
\centering
\includegraphics[width=\linewidth]{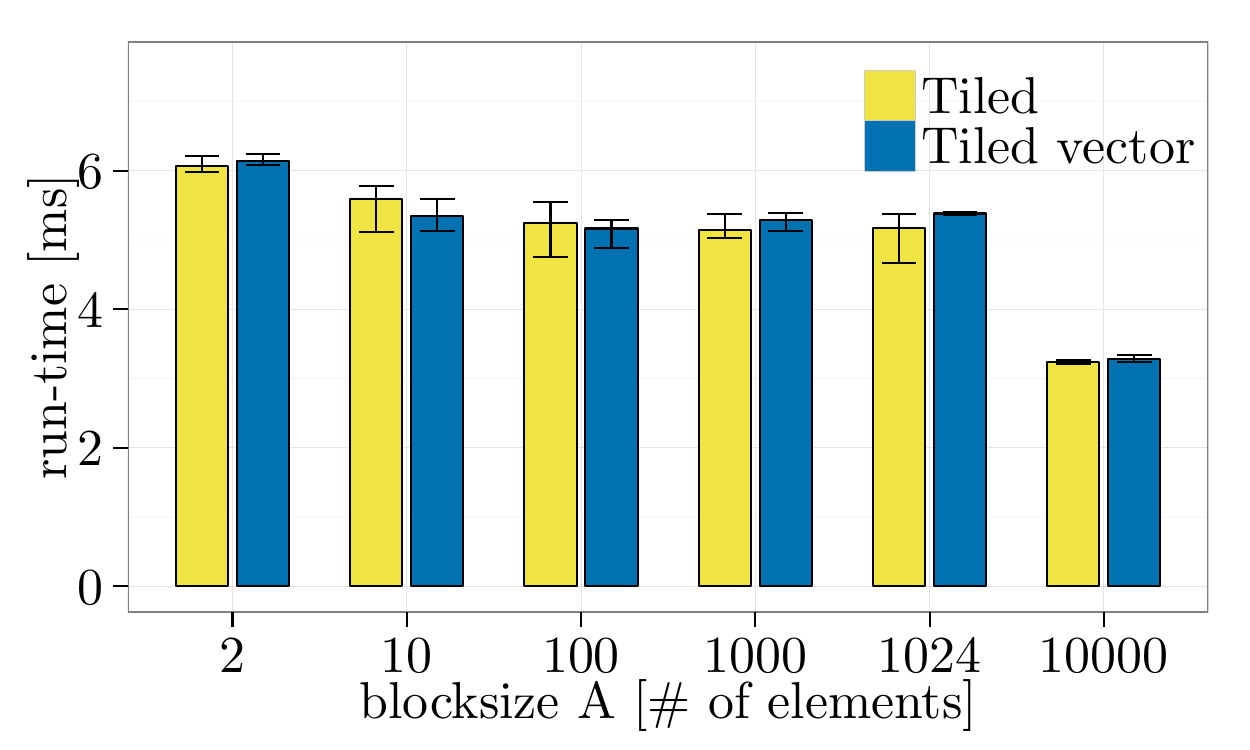}
\caption{%
\label{exp:pingpong-tiledvec-large-1x2-mvapich}%
$\VARdatasize=\SI{2.56}{\mega\byte}$, same node%
}%
\end{subfigure}%
\caption{\label{exp:pingpong-tiledvec-mvapich}  \dtdtiled \vs \ddttiledvector, element datatype: \mpiint, \pingpong, \jupitermvapich.}
\end{figure*}

\begin{figure*}[htpb]
\centering
\begin{subfigure}{.24\linewidth}
\centering
\includegraphics[width=\linewidth]{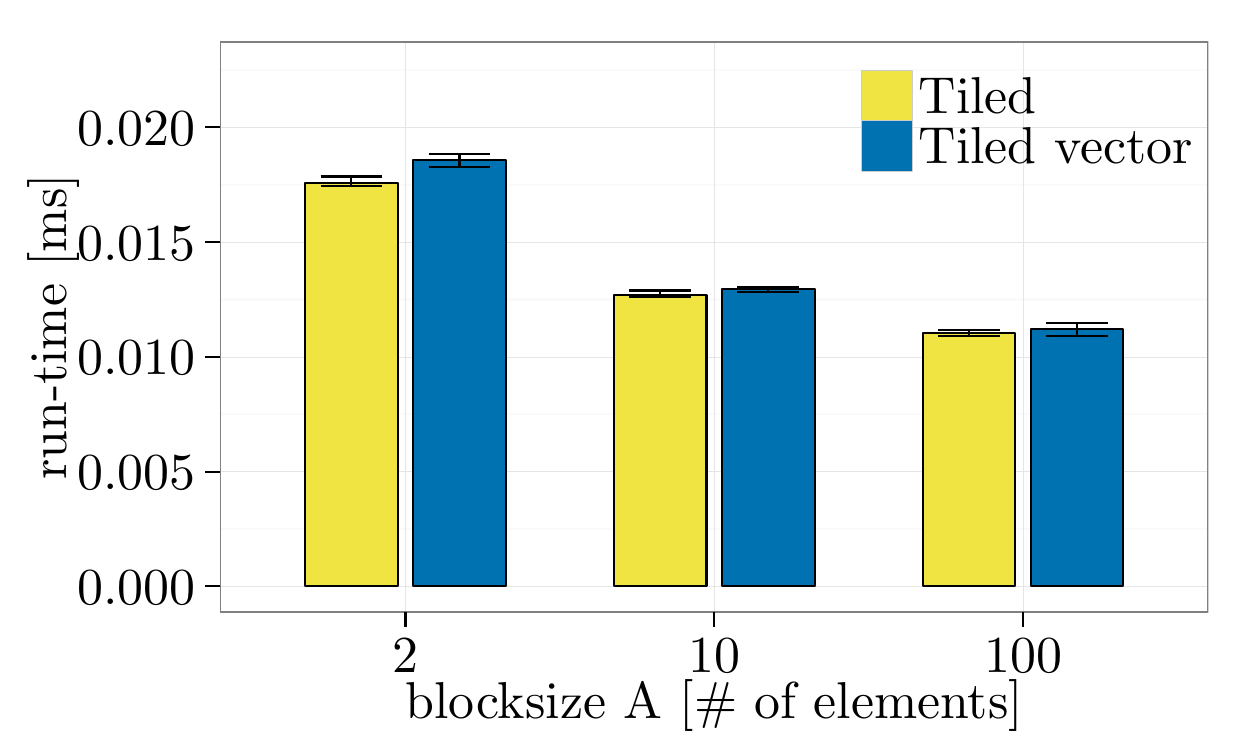}
\caption{%
\label{exp:pingpong-tiledvec-small-2x1-openmpi}%
$\VARdatasize=\SI{2}{\kilo\byte}$, \num{2}~nodes%
}%
\end{subfigure}%
\hfill%
\begin{subfigure}{.24\linewidth}
\centering
\includegraphics[width=\linewidth]{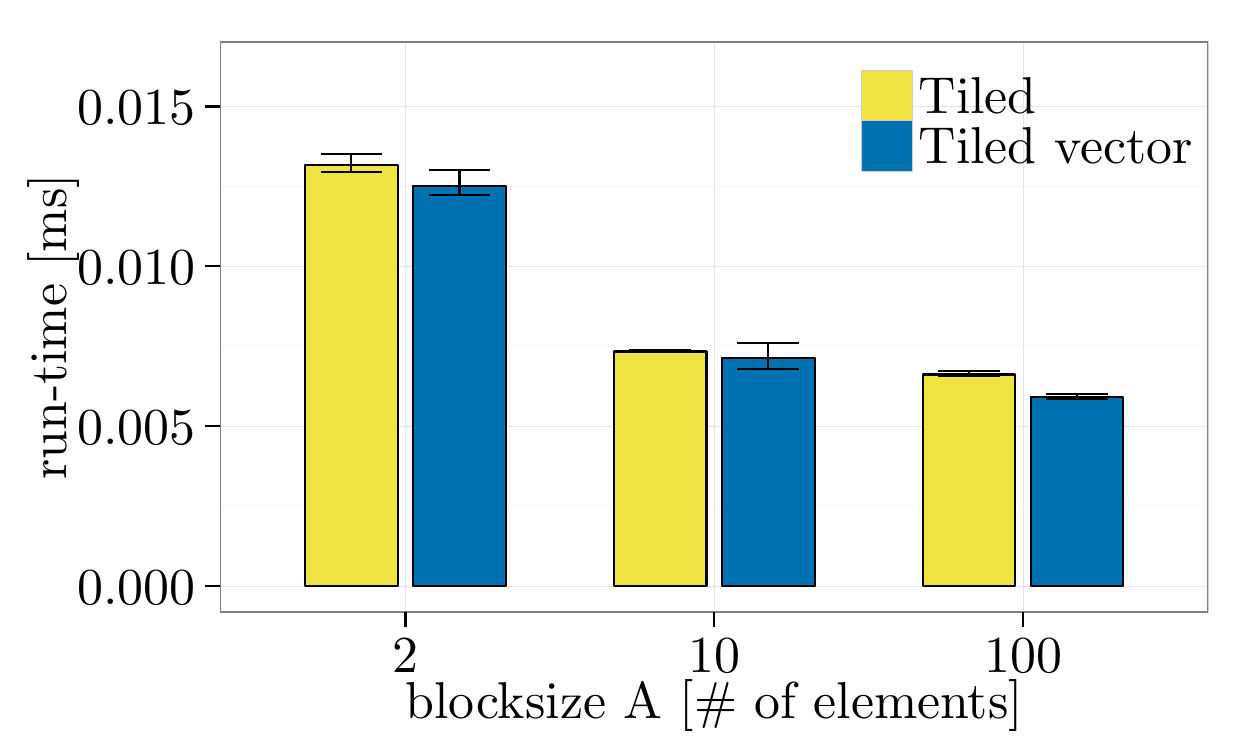}
\caption{%
\label{exp:pingpong-tiledvec-small-1x2-openmpi}%
$\VARdatasize=\SI{2}{\kilo\byte}$, same node%
}%
\end{subfigure}%
\hfill%
\begin{subfigure}{.24\linewidth}
\centering
\includegraphics[width=\linewidth]{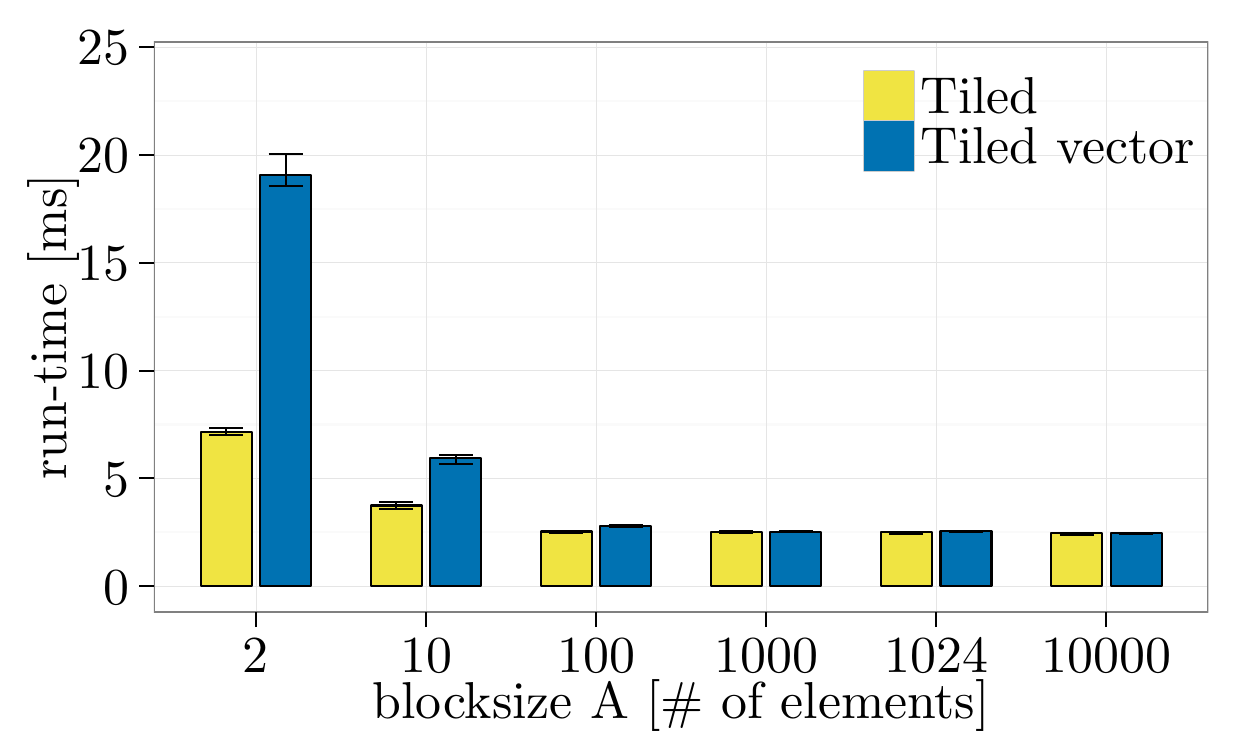}
\caption{%
\label{exp:pingpong-tiledvec-large-2x1-openmpi}%
$\VARdatasize=\SI{2.56}{\mega\byte}$, \num{2}~nodes%
}%
\end{subfigure}%
\hfill%
\begin{subfigure}{.24\linewidth}
\centering
\includegraphics[width=\linewidth]{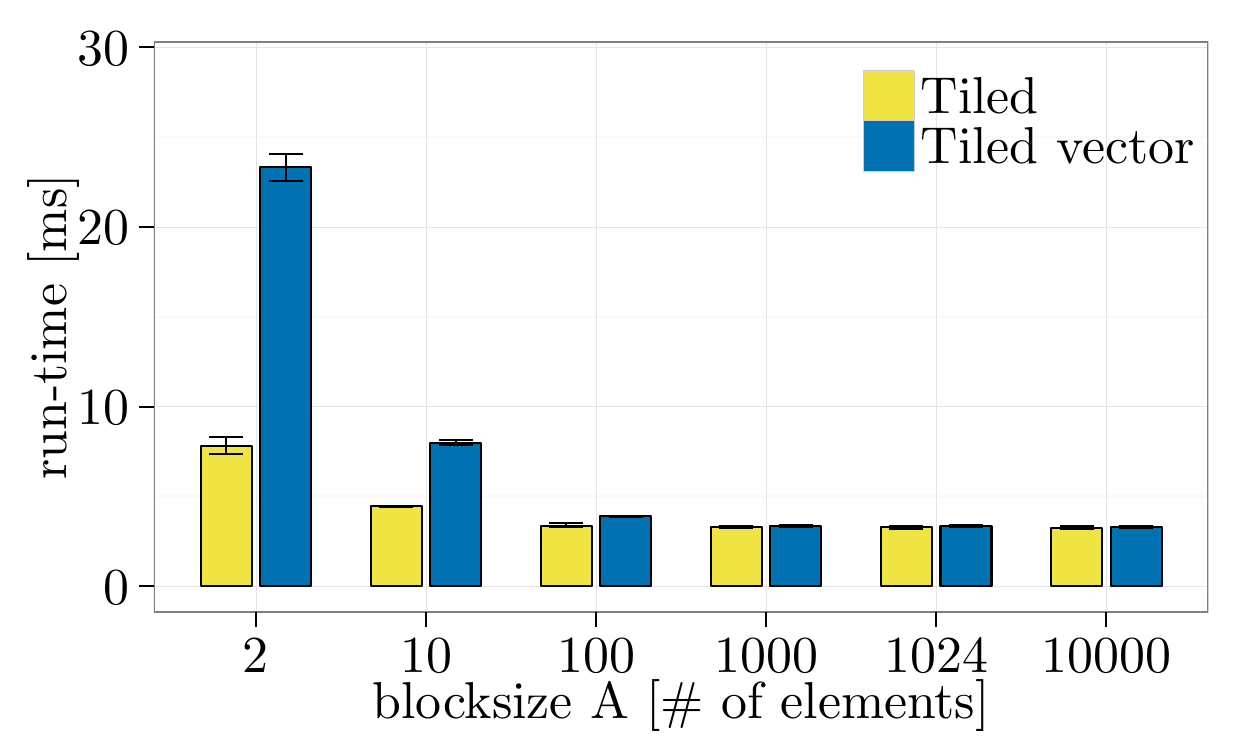}
\caption{%
\label{exp:pingpong-tiledvec-large-1x2-openmpi}%
$\VARdatasize=\SI{2.56}{\mega\byte}$, same node%
}%
\end{subfigure}%
\caption{\label{exp:pingpong-tiledvec-openmpi}  \dtdtiled \vs \ddttiledvector, element datatype: \mpiint, \pingpong, \jupiteropenmpi.}
\end{figure*}

\appexpdesc{
  \begin{expitemize}
    \item \dtdtiled, \ddttiledvector
    \item \pingpong
  \end{expitemize}
}{
  \begin{expitemize}
    \item \expparam{\jupiternecmpi}{\fig~\ref{exp:pingpong-tiledvec-nec}}
    \item \expparam{\jupitermvapich}{\fig~\ref{exp:pingpong-tiledvec-mvapich}}
    \item \expparam{\jupiteropenmpi}{\fig~\ref{exp:pingpong-tiledvec-openmpi}}
  \end{expitemize}  
}

\FloatBarrier
\clearpage

\appexp{exptest:vector_tiled}

\begin{figure*}[htpb]
\centering
\begin{subfigure}{.24\linewidth}
\centering
\includegraphics[width=\linewidth]{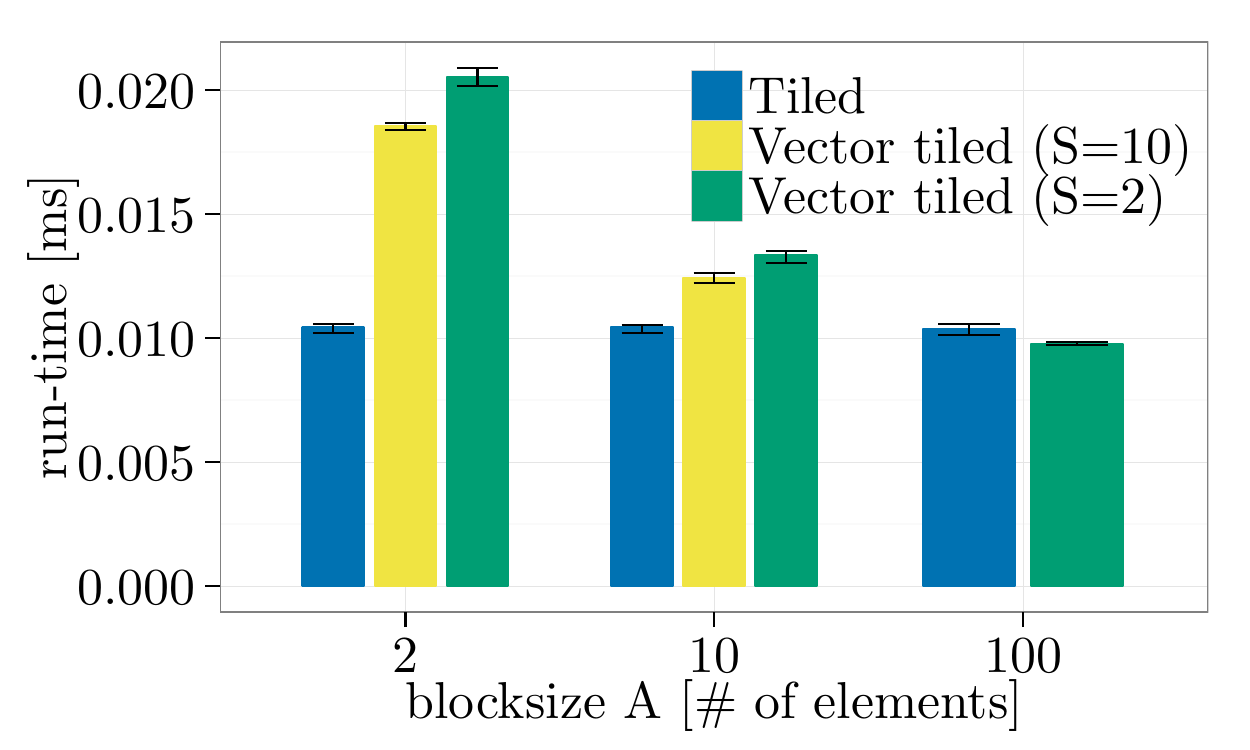}
\caption{%
\label{exp:pingpong-vectortiled-small-2x1}%
$\VARdatasize=\SI{2}{\kilo\byte}$, \num{2}~nodes%
}%
\end{subfigure}%
\hfill%
\begin{subfigure}{.24\linewidth}
\centering
\includegraphics[width=\linewidth]{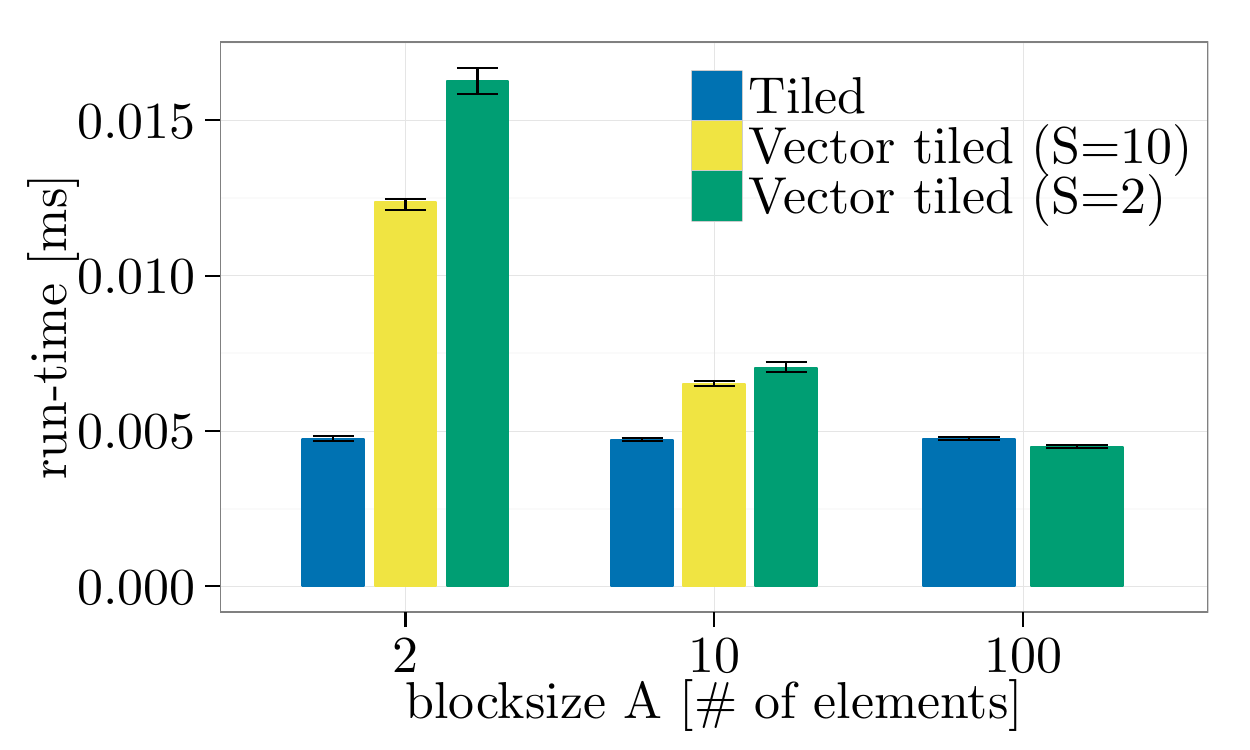}
\caption{%
\label{exp:pingpong-vectortiled-small-1x2}%
$\VARdatasize=\SI{2}{\kilo\byte}$, same node%
}%
\end{subfigure}%
\hfill%
\begin{subfigure}{.24\linewidth}
\centering
\includegraphics[width=\linewidth]{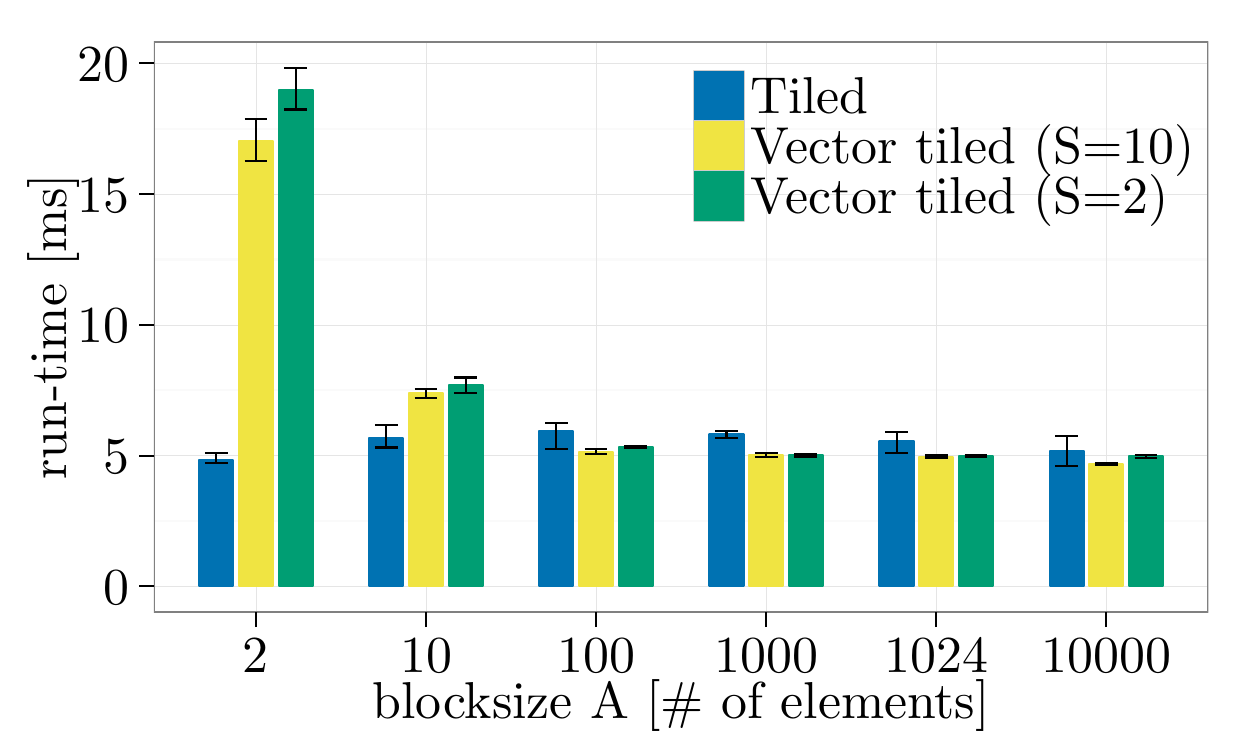}
\caption{%
\label{exp:pingpong-vectortiled-large-2x1}%
$\VARdatasize=\SI{2.56}{\mega\byte}$, \num{2}~nodes%
}%
\end{subfigure}%
\hfill%
\begin{subfigure}{.24\linewidth}
\centering
\includegraphics[width=\linewidth]{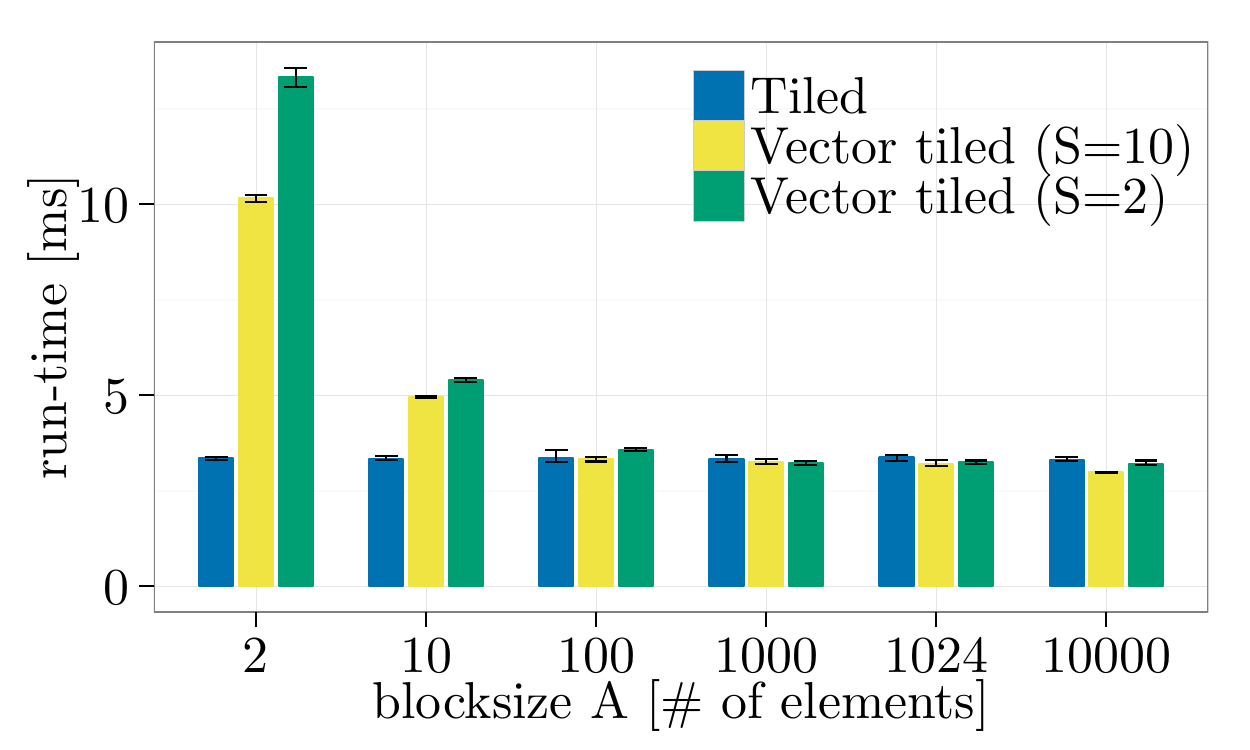}
\caption{%
\label{exp:pingpong-vectortiled-large-1x2}%
$\VARdatasize=\SI{2.56}{\mega\byte}$, same node%
}%
\end{subfigure}%
\caption{\label{exp:pingpong-vectortiled-nec} \dtdtiled \vs \ddtvectortiled, element datatype: \mpiint, \pingpong, \jupiternecmpi.}
\end{figure*}

\begin{figure*}[htpb]
\centering
\begin{subfigure}{.24\linewidth}
\centering
\includegraphics[width=\linewidth]{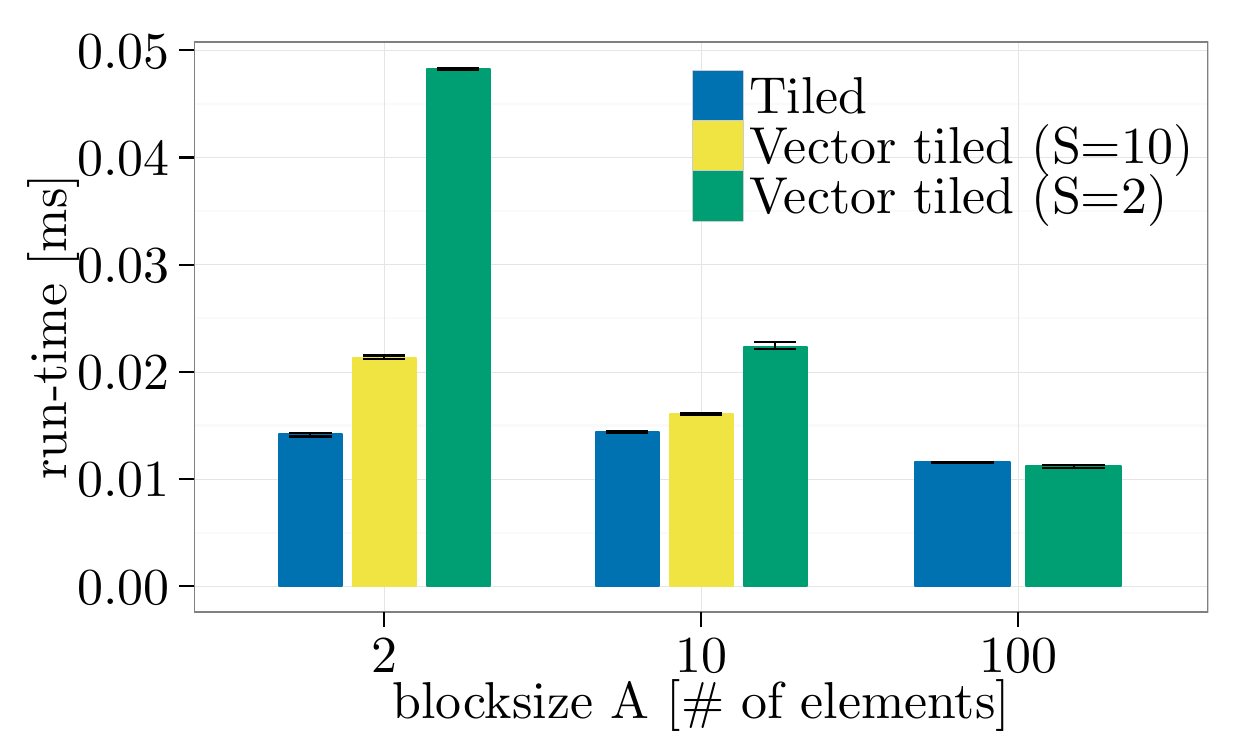}
\caption{%
\label{exp:pingpong-vectortiled-small-2x1-mvapich}%
$\VARdatasize=\SI{2}{\kilo\byte}$, \num{2}~nodes%
}%
\end{subfigure}%
\hfill%
\begin{subfigure}{.24\linewidth}
\centering
\includegraphics[width=\linewidth]{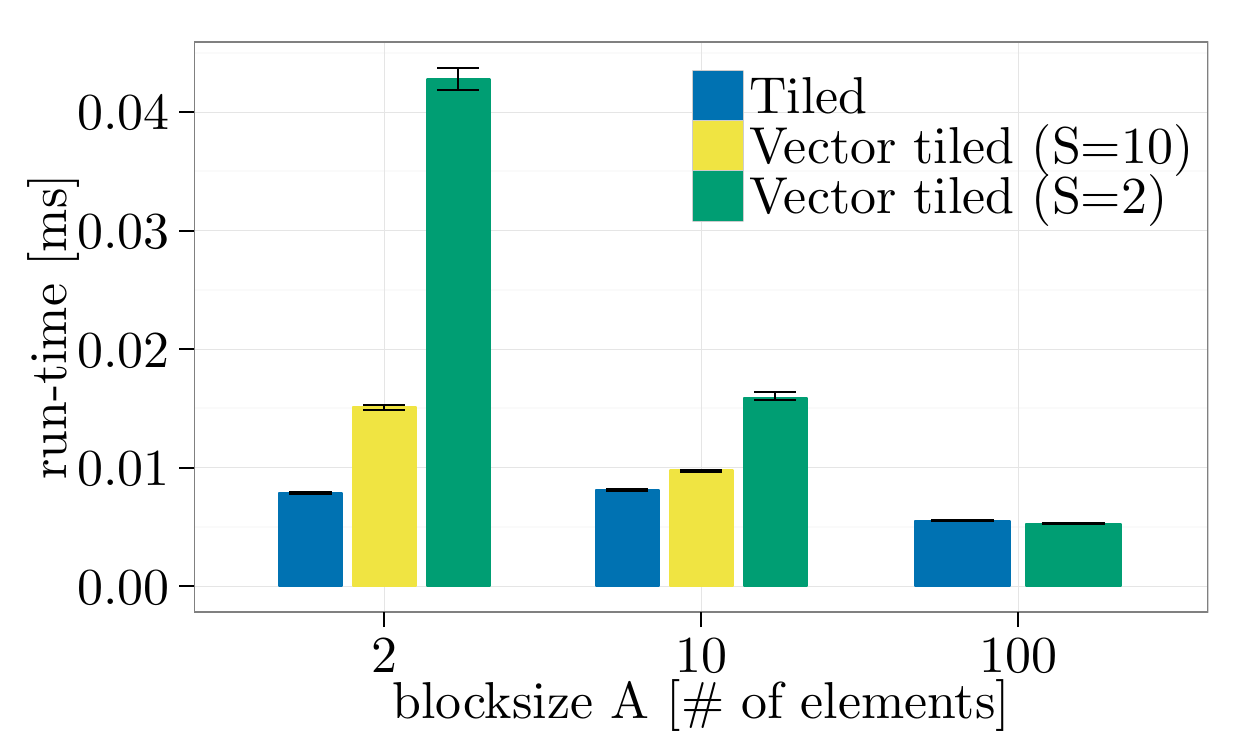}
\caption{%
\label{exp:pingpong-vectortiled-small-1x2-mvapich}%
$\VARdatasize=\SI{2}{\kilo\byte}$, same node%
}%
\end{subfigure}%
\hfill%
\begin{subfigure}{.24\linewidth}
\centering
\includegraphics[width=\linewidth]{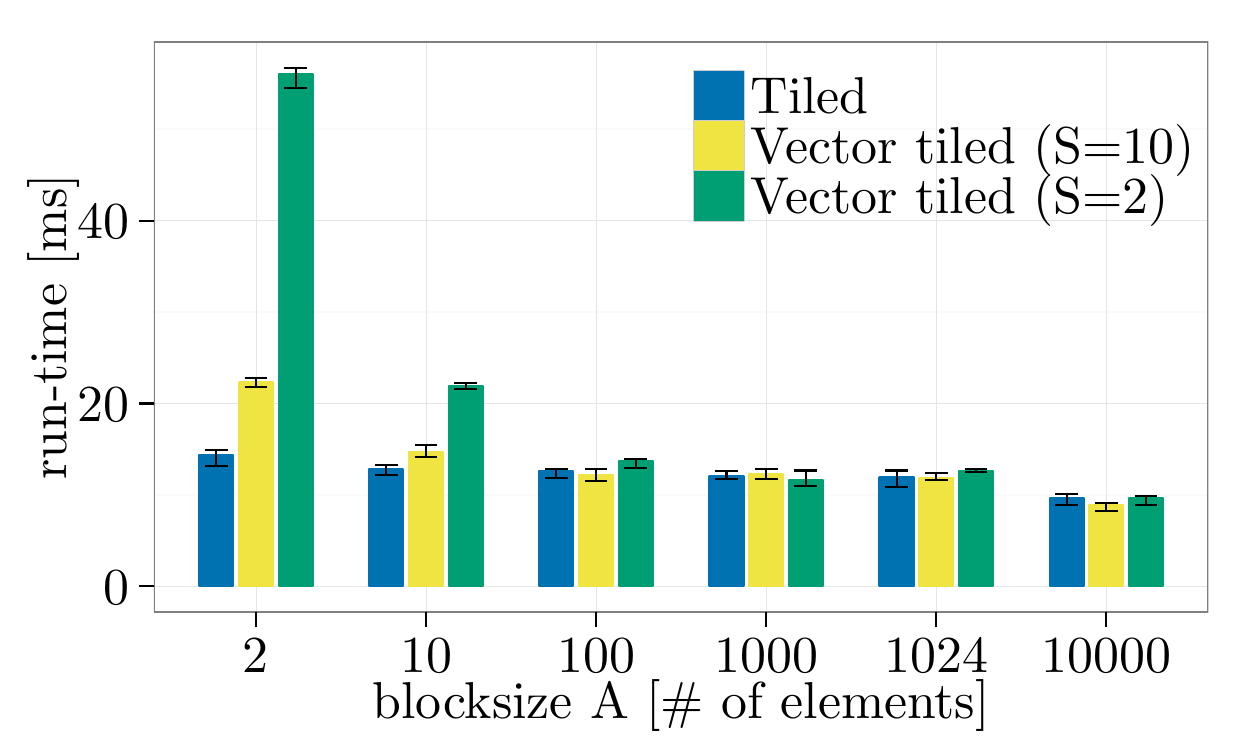}
\caption{%
\label{exp:pingpong-vectortiled-large-2x1-mvapich}%
$\VARdatasize=\SI{2.56}{\mega\byte}$, \num{2}~nodes%
}%
\end{subfigure}%
\hfill%
\begin{subfigure}{.24\linewidth}
\centering
\includegraphics[width=\linewidth]{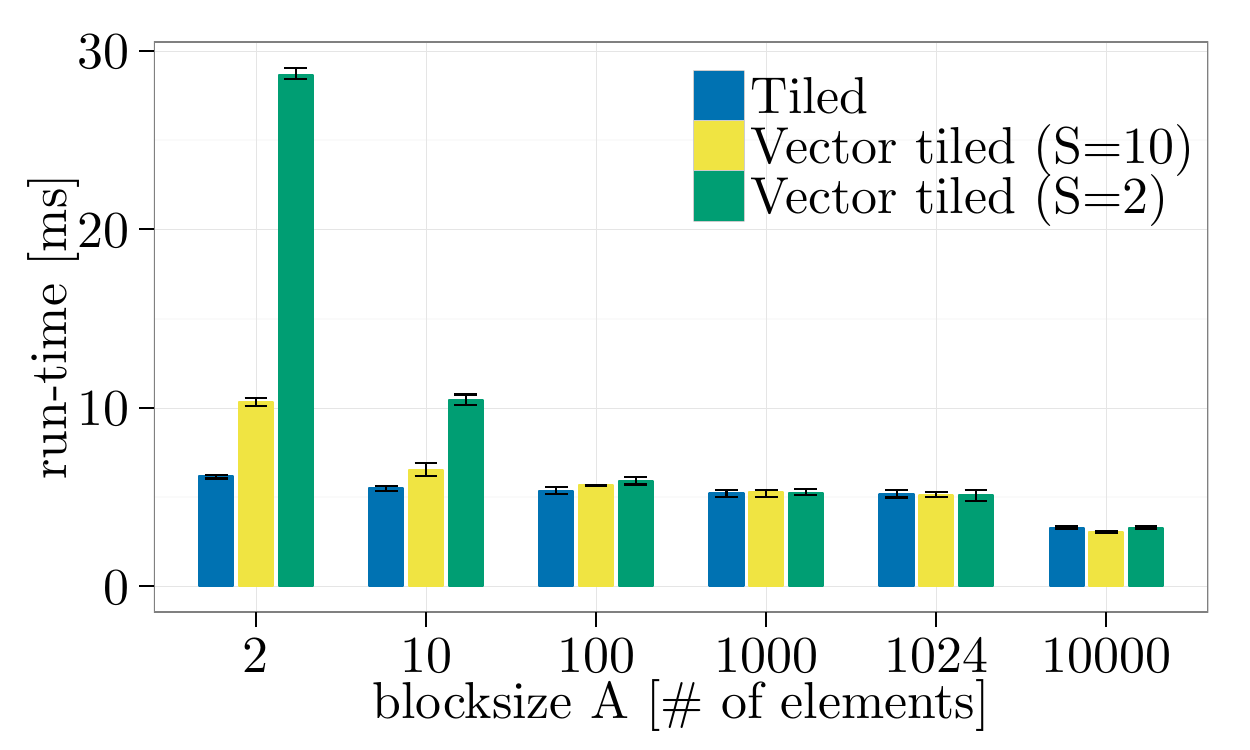}
\caption{%
\label{exp:pingpong-vectortiled-large-1x2-mvapich}%
$\VARdatasize=\SI{2.56}{\mega\byte}$, same node%
}%
\end{subfigure}%
\caption{\label{exp:pingpong-vectortiled-mvapich} \dtdtiled \vs \ddtvectortiled, element datatype: \mpiint, \pingpong, \jupitermvapich.}
\end{figure*}

\begin{figure*}[htpb]
\centering
\begin{subfigure}{.24\linewidth}
\centering
\includegraphics[width=\linewidth]{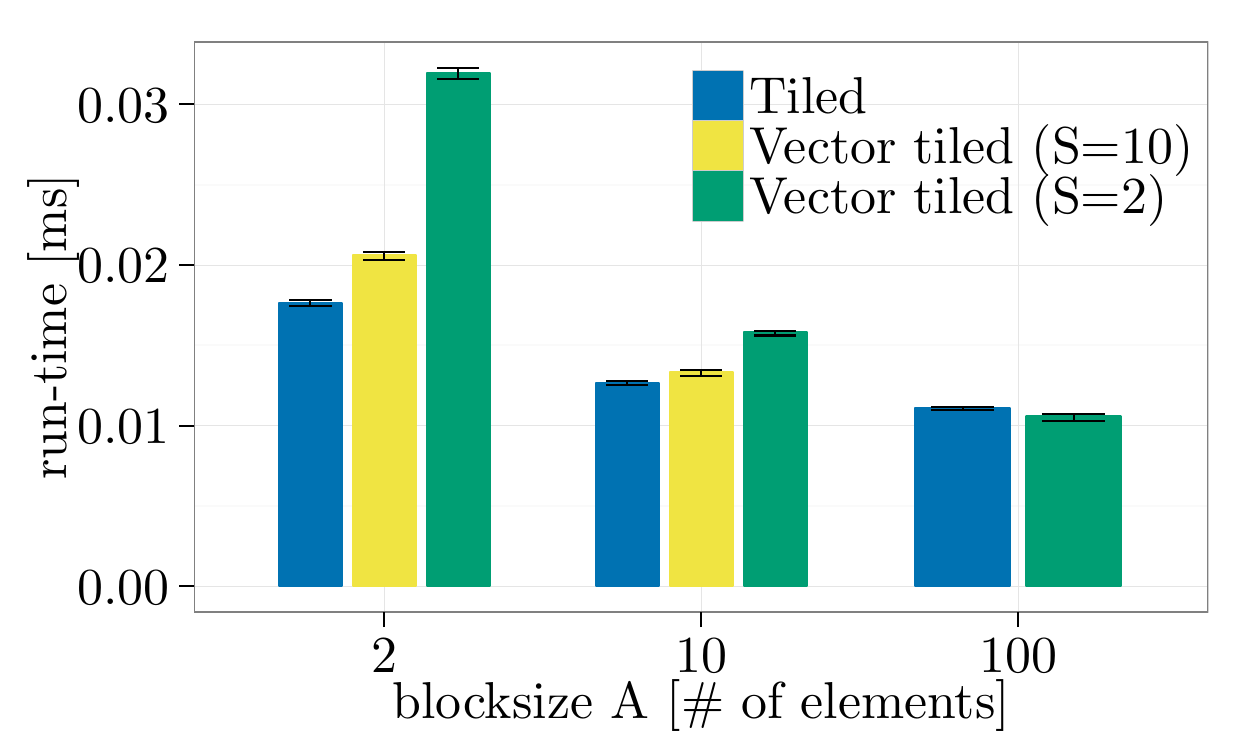}
\caption{%
\label{exp:pingpong-vectortiled-small-2x1-openmpi}%
$\VARdatasize=\SI{2}{\kilo\byte}$, \num{2}~nodes%
}%
\end{subfigure}%
\hfill%
\begin{subfigure}{.24\linewidth}
\centering
\includegraphics[width=\linewidth]{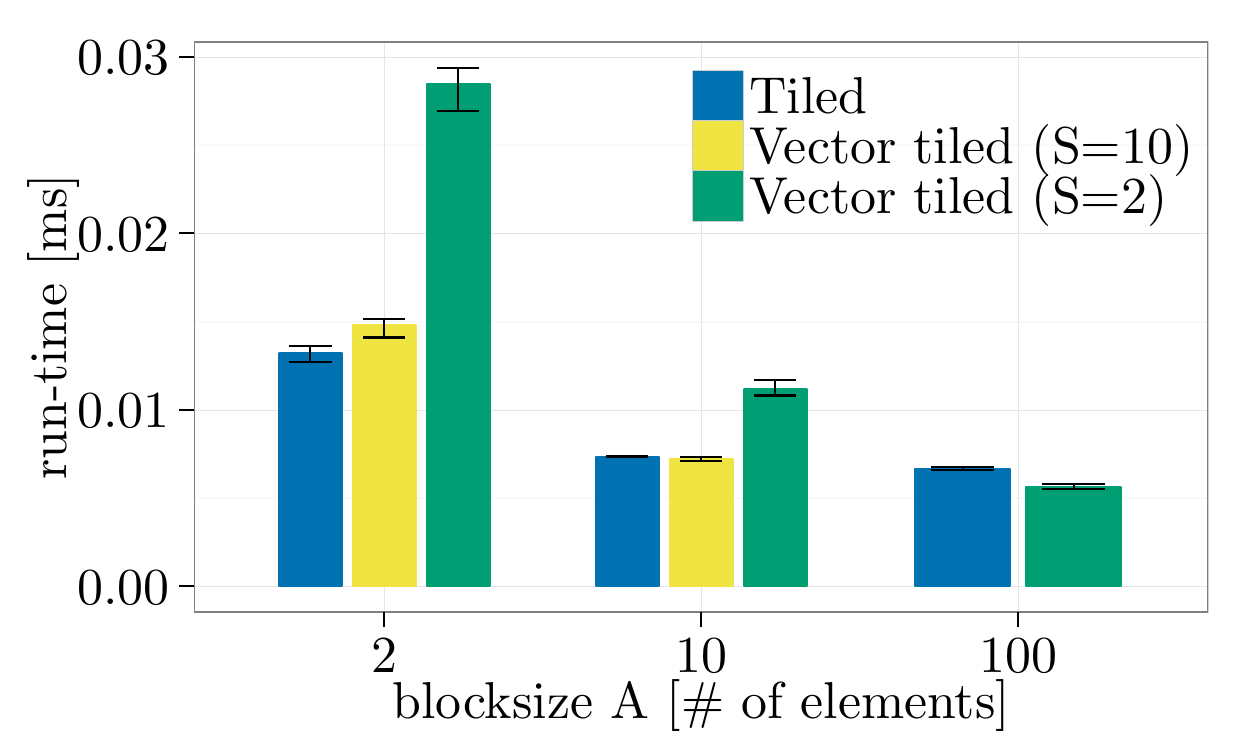}
\caption{%
\label{exp:pingpong-vectortiled-small-1x2-openmpi}%
$\VARdatasize=\SI{2}{\kilo\byte}$, same node%
}%
\end{subfigure}%
\hfill%
\begin{subfigure}{.24\linewidth}
\centering
\includegraphics[width=\linewidth]{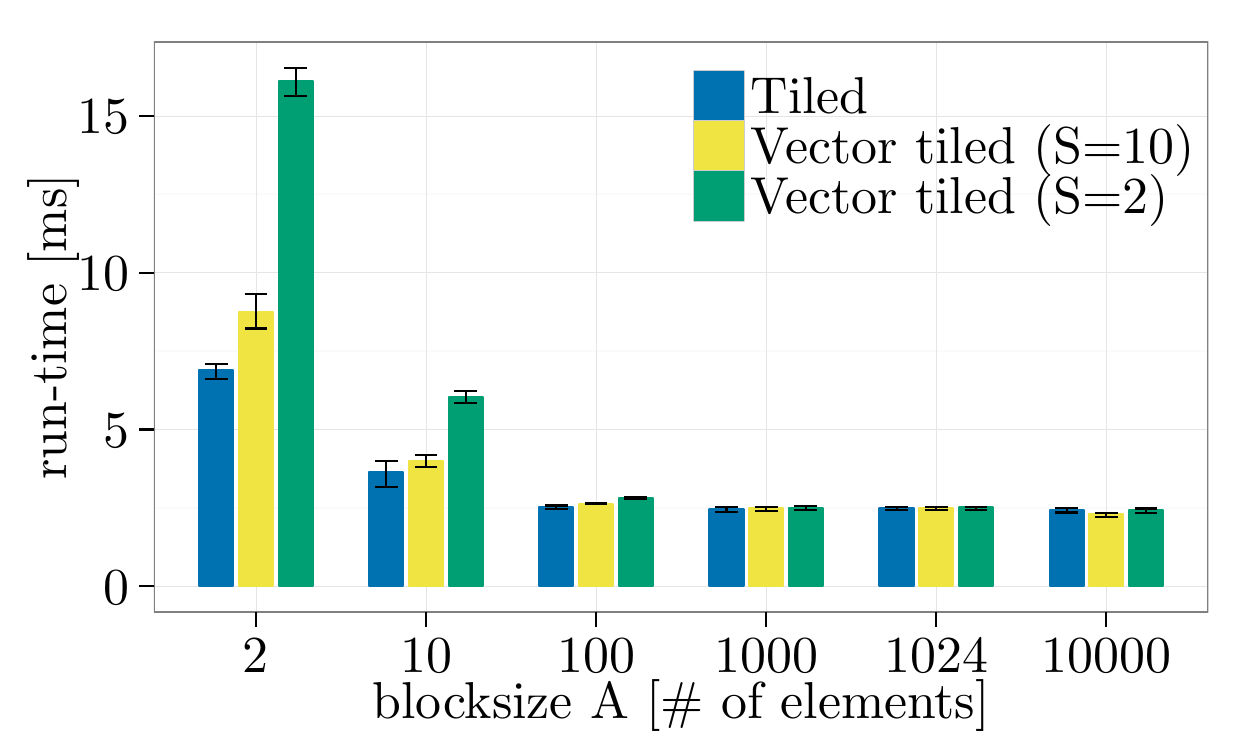}
\caption{%
\label{exp:pingpong-vectortiled-large-2x1-openmpi}%
$\VARdatasize=\SI{2.56}{\mega\byte}$, \num{2}~nodes%
}%
\end{subfigure}%
\hfill%
\begin{subfigure}{.24\linewidth}
\centering
\includegraphics[width=\linewidth]{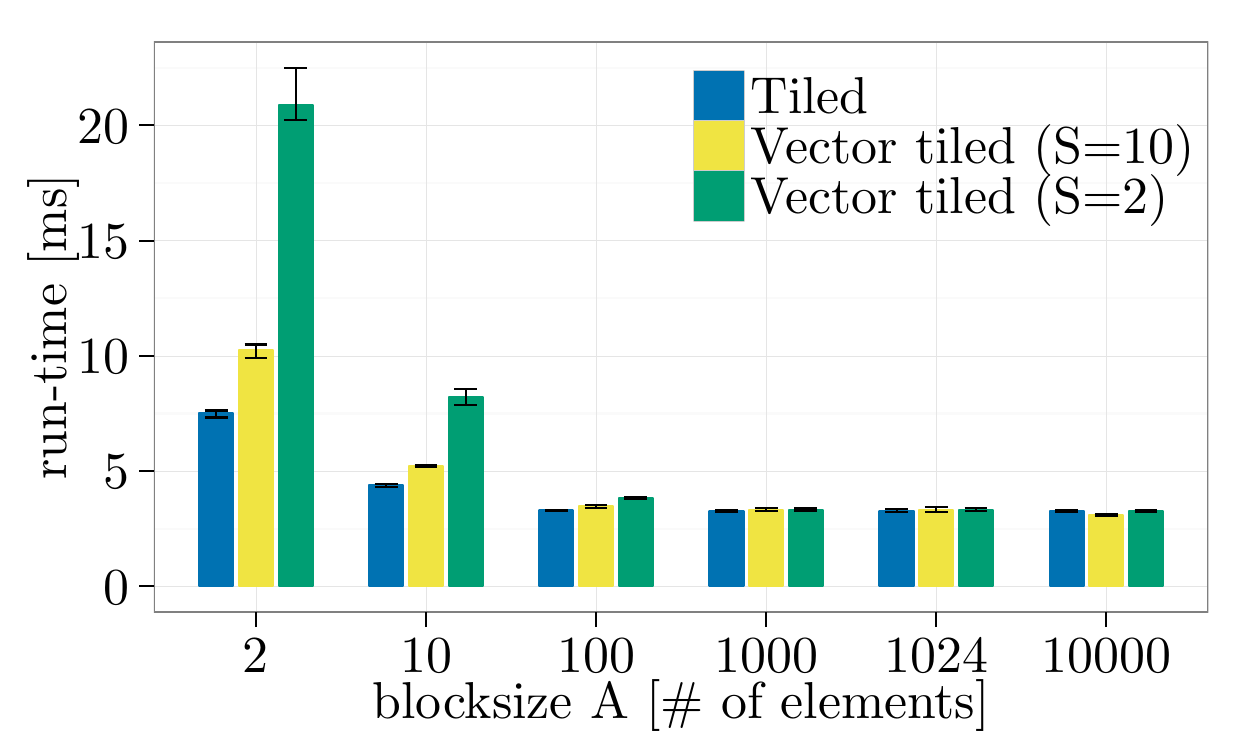}
\caption{%
\label{exp:pingpong-vectortiled-large-1x2-openmpi}%
$\VARdatasize=\SI{2.56}{\mega\byte}$, same node%
}%
\end{subfigure}%
\caption{\label{exp:pingpong-vectortiled-openmpi} \dtdtiled \vs \ddtvectortiled, element datatype: \mpiint, \pingpong, \jupiteropenmpi.}
\end{figure*}

\appexpdesc{
  \begin{expitemize}
    \item \dtdtiled, \ddtvectortiled
    \item \pingpong
  \end{expitemize}
}{
  \begin{expitemize}
    \item \expparam{\jupiternecmpi}{\fig~\ref{exp:pingpong-vectortiled-nec}}
    \item \expparam{\jupitermvapich}{\fig~\ref{exp:pingpong-vectortiled-mvapich}}
    \item \expparam{\jupiteropenmpi}{\fig~\ref{exp:pingpong-vectortiled-openmpi}}
  \end{expitemize}  
}

\FloatBarrier
\clearpage

\appexp{exptest:block_indexed}

\begin{figure*}[htpb]
\centering
\begin{subfigure}{.24\linewidth}
\centering
\includegraphics[width=\linewidth]{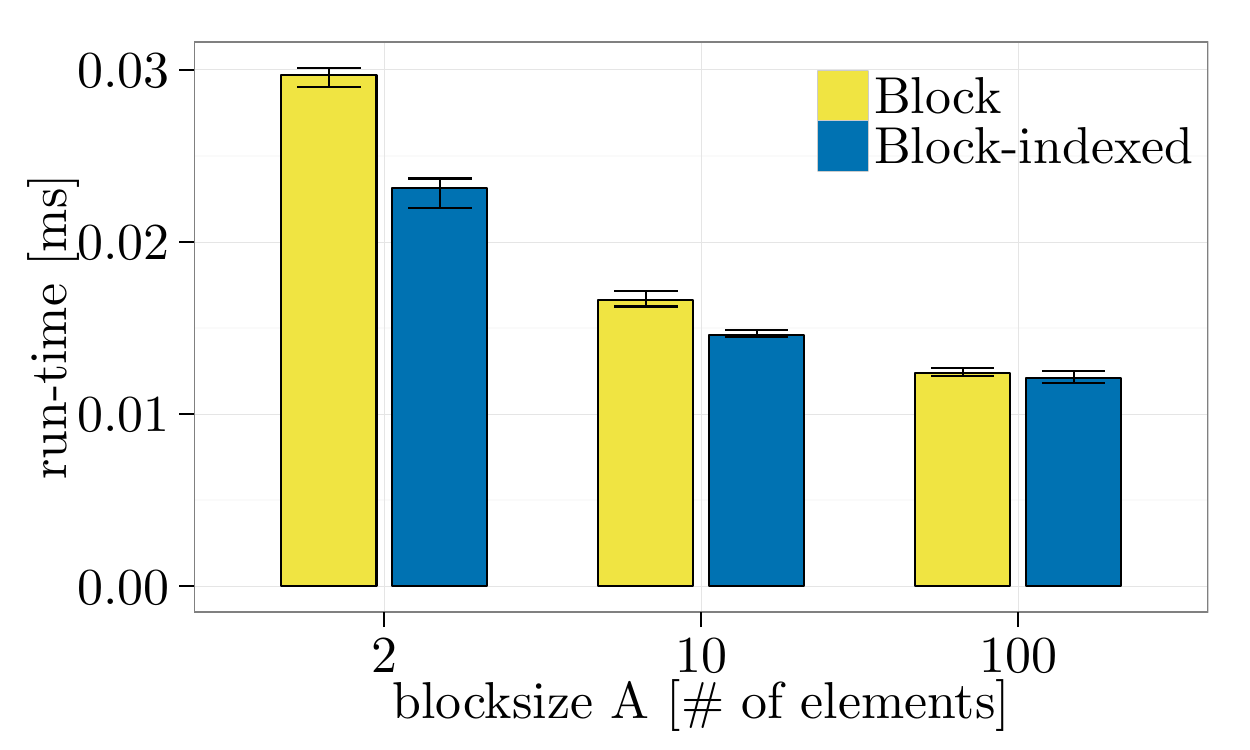}
\caption{%
\label{exp:pingpong-blockindexed-small-2x1}%
$\VARdatasize=\SI{3.2}{\kilo\byte}$, \num{2}~nodes%
}%
\end{subfigure}%
\hfill%
\begin{subfigure}{.24\linewidth}
\centering
\includegraphics[width=\linewidth]{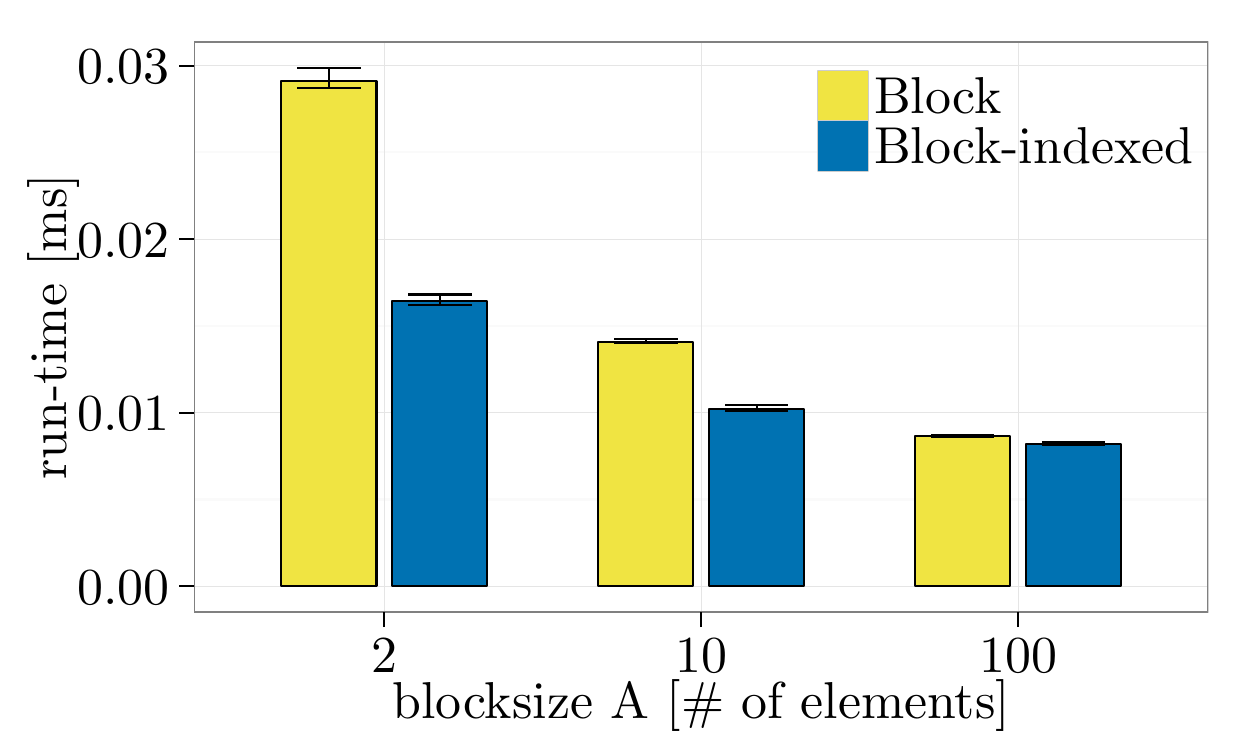}
\caption{%
\label{exp:pingpong-blockindexed-small-1x2}%
$\VARdatasize=\SI{3.2}{\kilo\byte}$, same node%
}%
\end{subfigure}%
\hfill%
\begin{subfigure}{.24\linewidth}
\centering
\includegraphics[width=\linewidth]{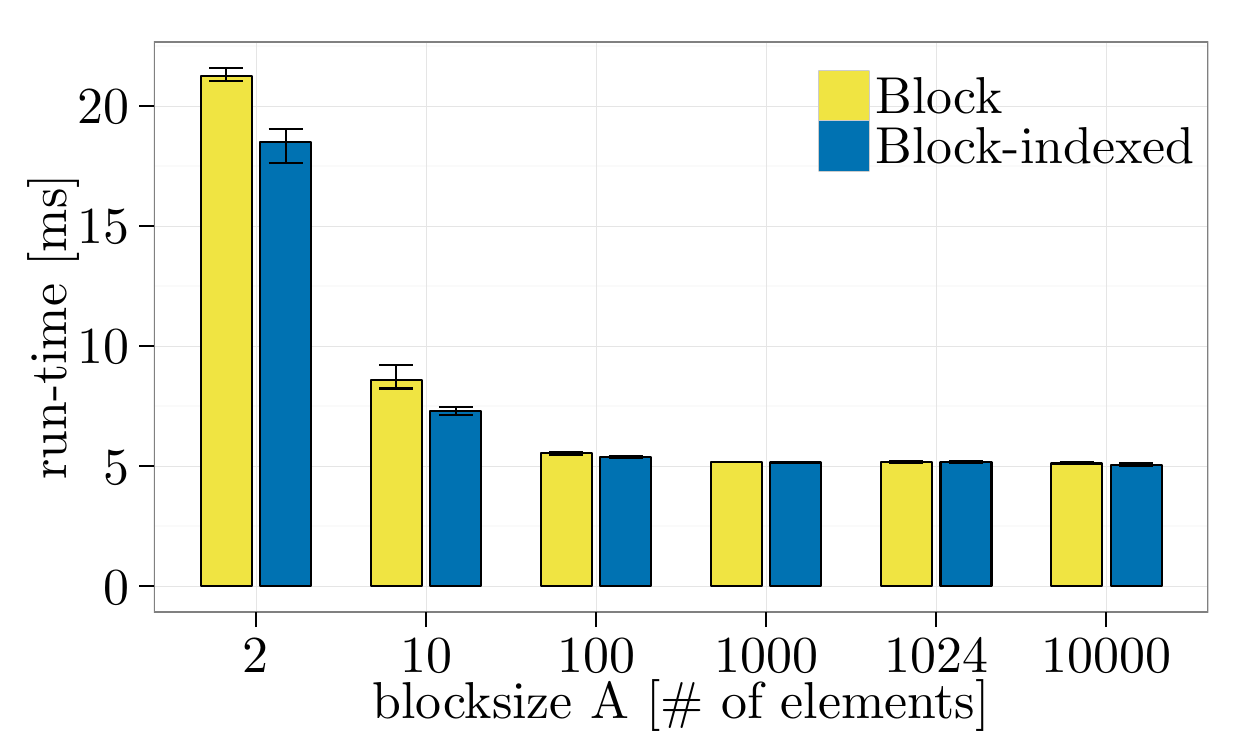}
\caption{%
\label{exp:pingpong-blockindexed-large-2x1}%
$\VARdatasize=\SI{2.56}{\mega\byte}$, \num{2}~nodes%
}%
\end{subfigure}%
\hfill%
\begin{subfigure}{.24\linewidth}
\centering
\includegraphics[width=\linewidth]{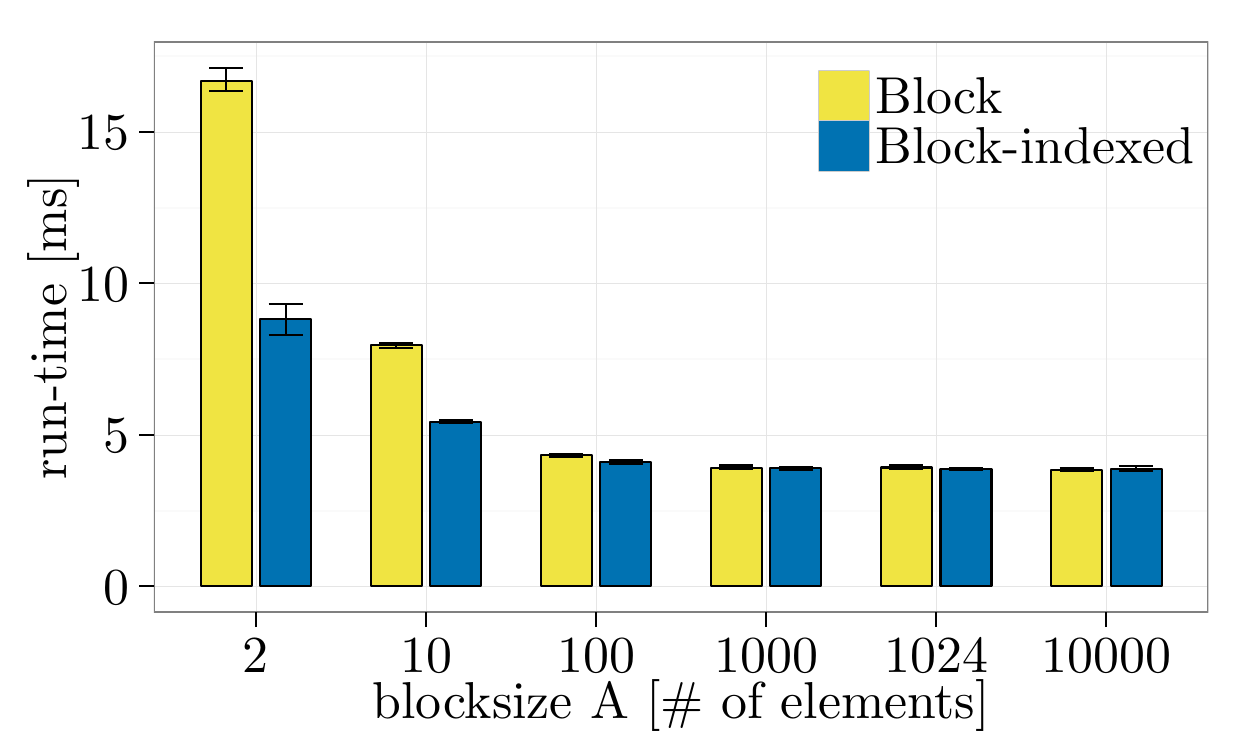}
\caption{%
\label{exp:pingpong-blockindexed-large-1x2}%
$\VARdatasize=\SI{2.56}{\mega\byte}$, same node%
}%
\end{subfigure}%
\caption{\label{exp:pingpong-blockindexed-nec} \dtblock \vs \ddtblockindexed, element datatype: \mpiint, \pingpong, \jupiternecmpi.}
\end{figure*}

\begin{figure*}[htpb]
\centering
\begin{subfigure}{.24\linewidth}
\centering
\includegraphics[width=\linewidth]{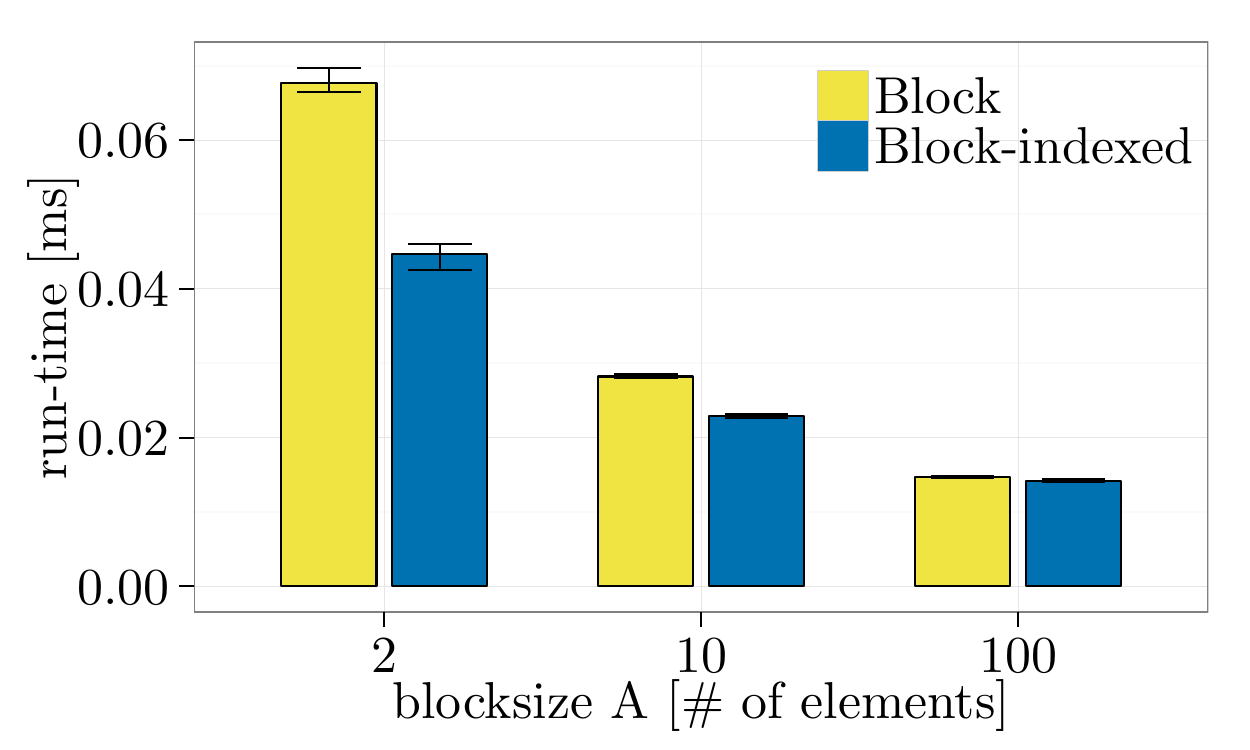}
\caption{%
\label{exp:pingpong-blockindexed-small-2x1-mvapich}%
$\VARdatasize=\SI{3.2}{\kilo\byte}$, \num{2}~nodes%
}%
\end{subfigure}%
\hfill%
\begin{subfigure}{.24\linewidth}
\centering
\includegraphics[width=\linewidth]{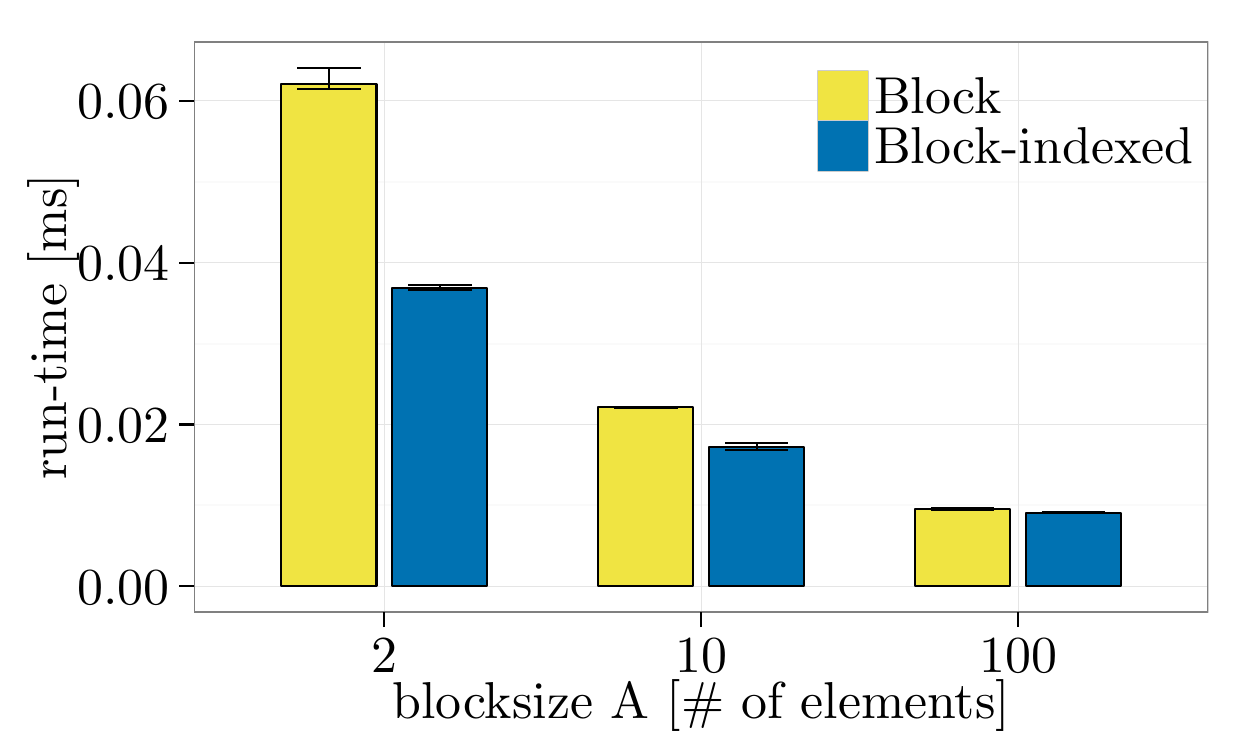}
\caption{%
\label{exp:pingpong-blockindexed-small-1x2-mvapich}%
$\VARdatasize=\SI{3.2}{\kilo\byte}$, same node%
}%
\end{subfigure}%
\hfill%
\begin{subfigure}{.24\linewidth}
\centering
\includegraphics[width=\linewidth]{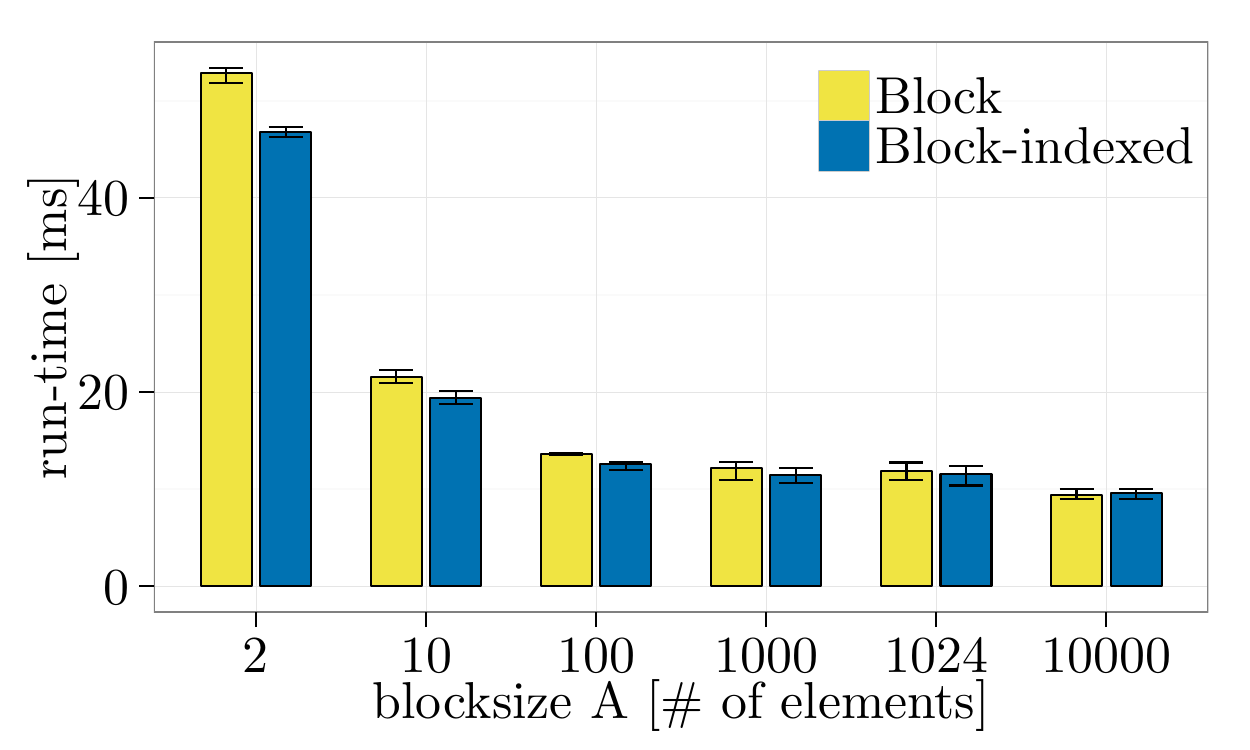}
\caption{%
\label{exp:pingpong-blockindexed-large-2x1-mvapich}%
$\VARdatasize=\SI{2.56}{\mega\byte}$, \num{2}~nodes%
}%
\end{subfigure}%
\hfill%
\begin{subfigure}{.24\linewidth}
\centering
\includegraphics[width=\linewidth]{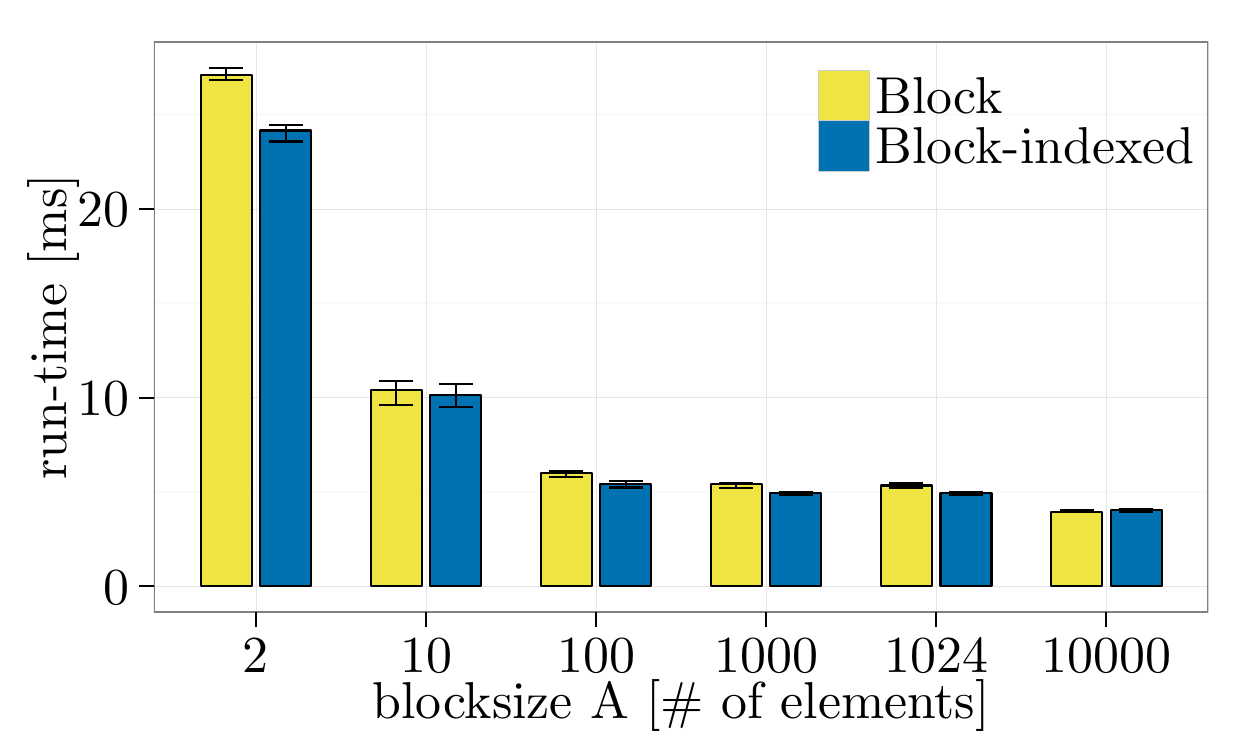}
\caption{%
\label{exp:pingpong-blockindexed-large-1x2-mvapich}%
$\VARdatasize=\SI{2.56}{\mega\byte}$, same node%
}%
\end{subfigure}%
\caption{\label{exp:pingpong-blockindexed-mvapich} \dtblock \vs \ddtblockindexed, element datatype: \mpiint, \pingpong, \jupitermvapich.}
\end{figure*}

\begin{figure*}[htpb]
\centering
\begin{subfigure}{.24\linewidth}
\centering
\includegraphics[width=\linewidth]{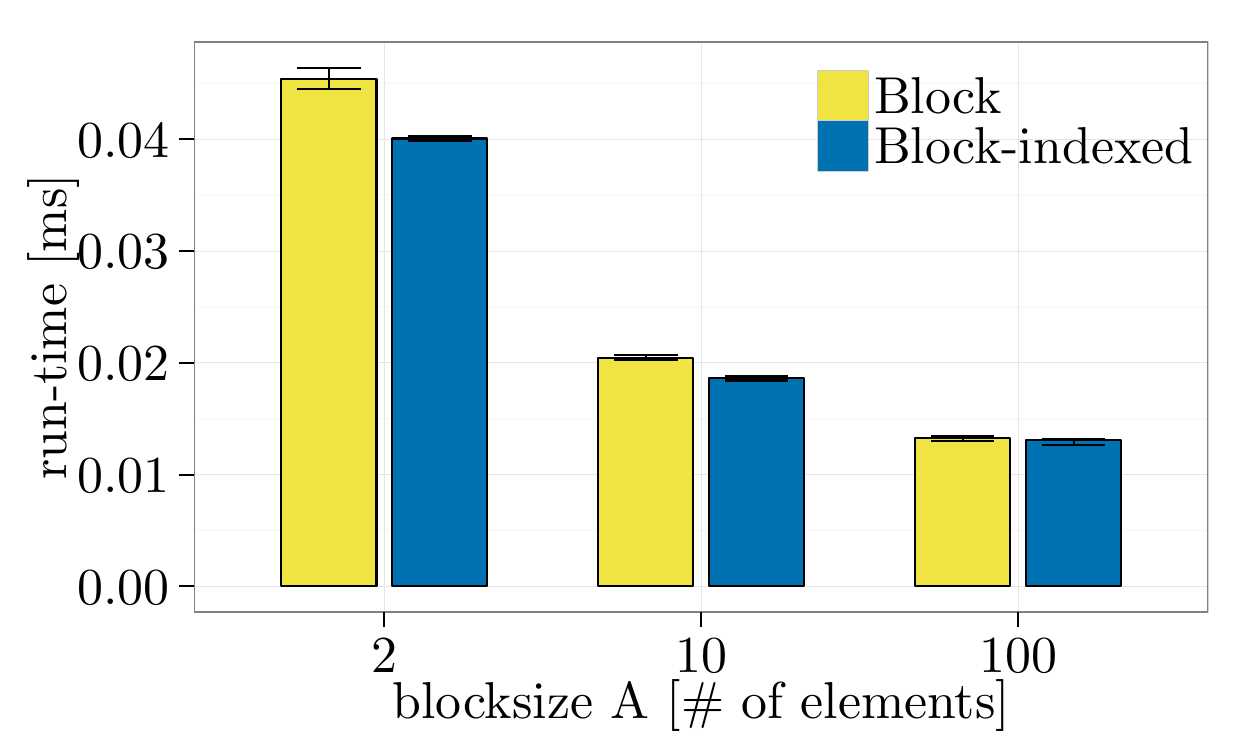}
\caption{%
\label{exp:pingpong-blockindexed-small-2x1-openmpi}%
$\VARdatasize=\SI{3.2}{\kilo\byte}$, \num{2}~nodes%
}%
\end{subfigure}%
\hfill%
\begin{subfigure}{.24\linewidth}
\centering
\includegraphics[width=\linewidth]{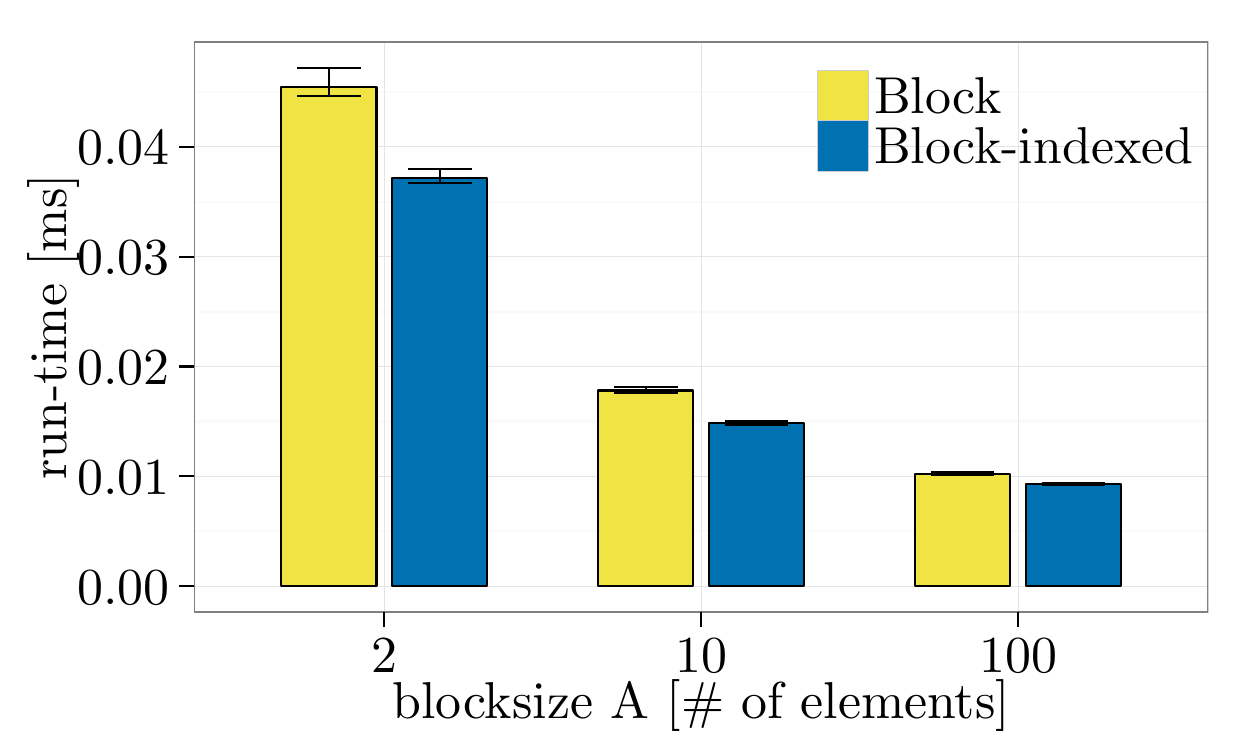}
\caption{%
\label{exp:pingpong-blockindexed-small-1x2-openmpi}%
$\VARdatasize=\SI{3.2}{\kilo\byte}$, same node%
}%
\end{subfigure}%
\hfill%
\begin{subfigure}{.24\linewidth}
\centering
\includegraphics[width=\linewidth]{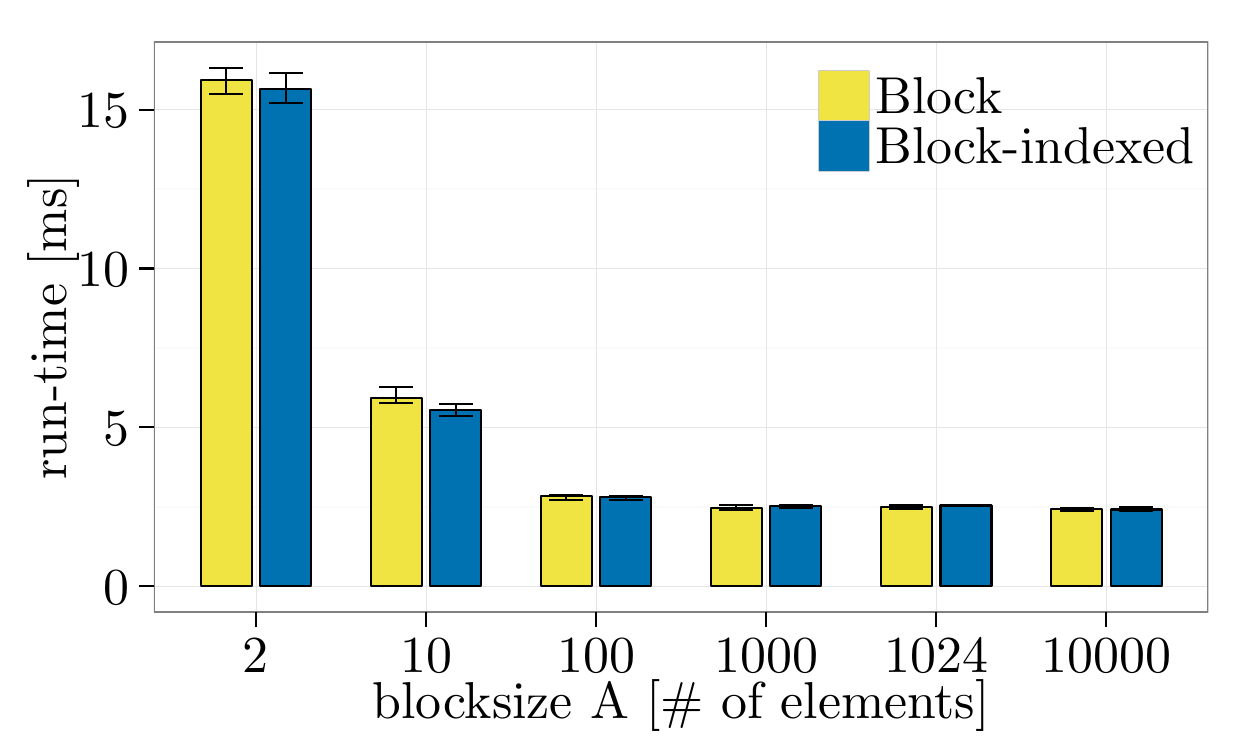}
\caption{%
\label{exp:pingpong-blockindexed-large-2x1-openmpi}%
$\VARdatasize=\SI{2.56}{\mega\byte}$, \num{2}~nodes%
}%
\end{subfigure}%
\hfill%
\begin{subfigure}{.24\linewidth}
\centering
\includegraphics[width=\linewidth]{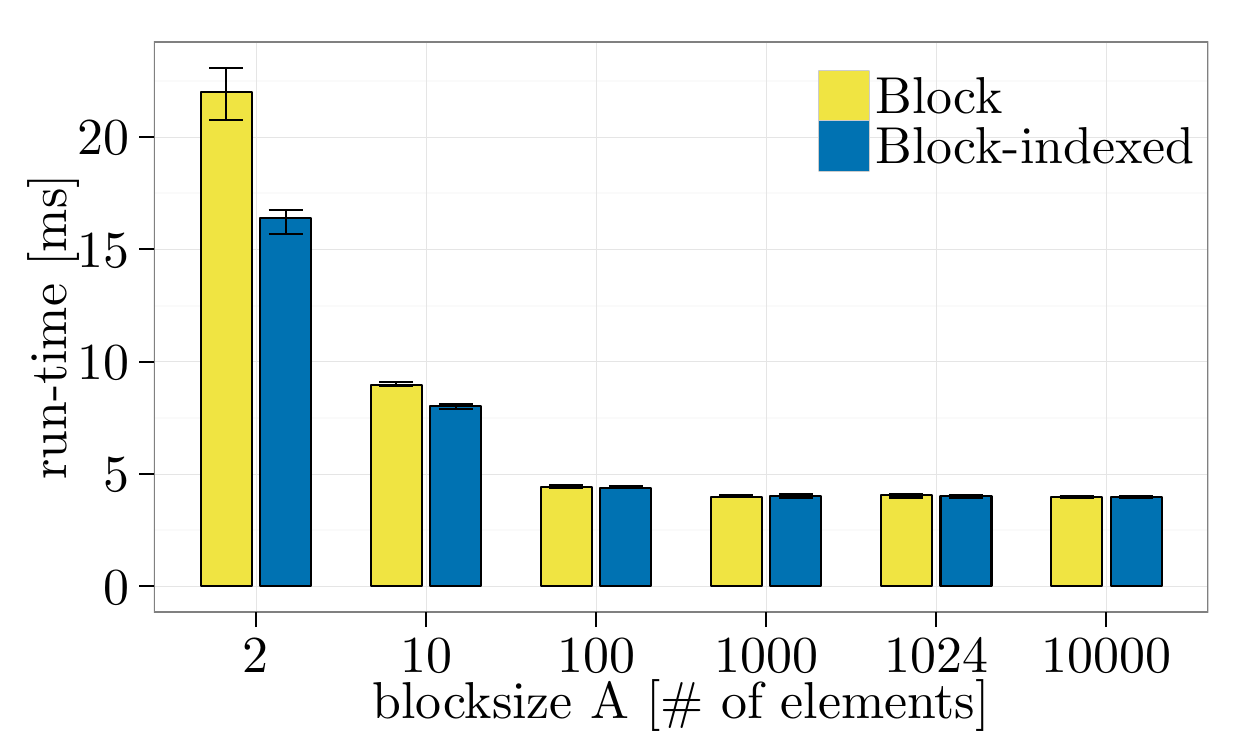}
\caption{%
\label{exp:pingpong-blockindexed-large-1x2-openmpi}%
$\VARdatasize=\SI{2.56}{\mega\byte}$, same node%
}%
\end{subfigure}%
\caption{\label{exp:pingpong-blockindexed-openmpi} \dtblock \vs \ddtblockindexed, element datatype: \mpiint, \pingpong, \jupiteropenmpi.}
\end{figure*}

\appexpdesc{
  \begin{expitemize}
    \item \dtblock, \ddtblockindexed
    \item \pingpong
  \end{expitemize}
}{
  \begin{expitemize}
    \item \expparam{\jupiternecmpi}{\fig~\ref{exp:pingpong-blockindexed-nec}}
    \item \expparam{\jupitermvapich}{\fig~\ref{exp:pingpong-blockindexed-mvapich}}
    \item \expparam{\jupiteropenmpi}{\fig~\ref{exp:pingpong-blockindexed-openmpi}}
  \end{expitemize}  
}

\FloatBarrier
\clearpage

\appexp{exptest:alternating_indexed}

\begin{figure*}[htpb]
\centering
\begin{subfigure}{.24\linewidth}
\centering
\includegraphics[width=\linewidth]{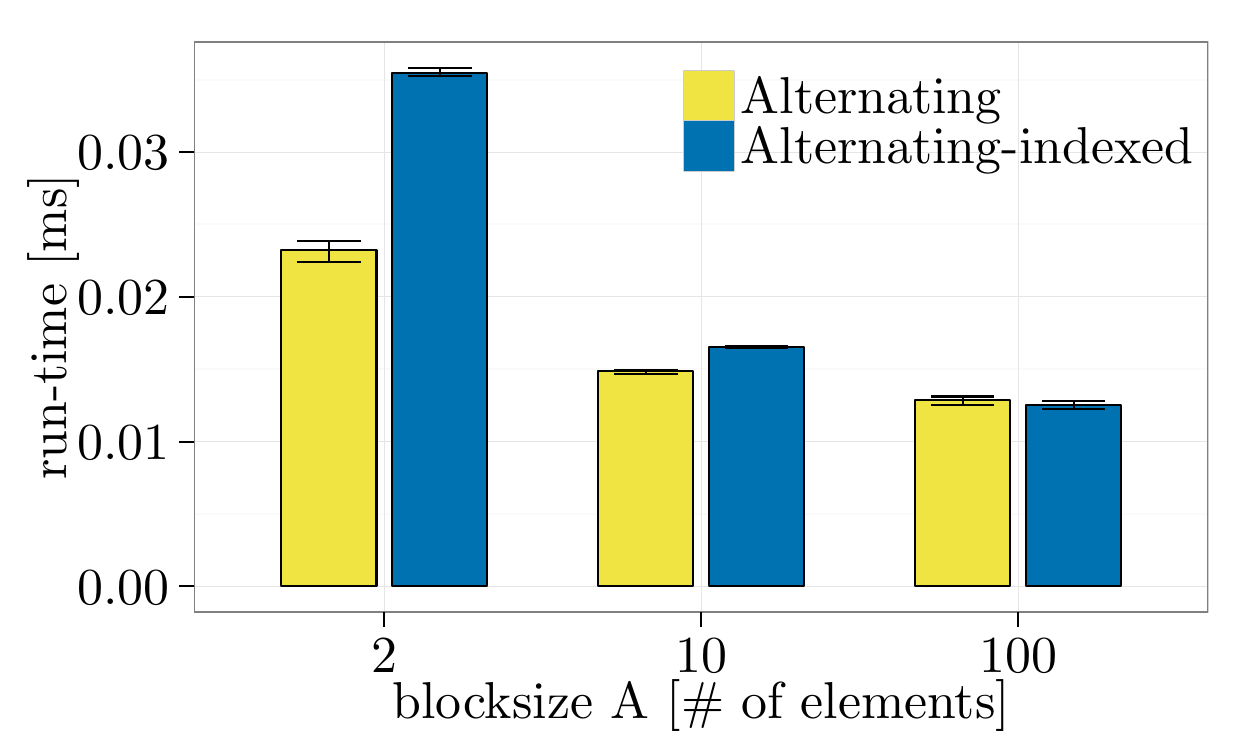}
\caption{%
\label{exp:pingpong-alternatingindexed-small-2x1}%
$\VARdatasize=\SI{3.2}{\kilo\byte}$, \num{2}~nodes%
}%
\end{subfigure}%
\hfill%
\begin{subfigure}{.24\linewidth}
\centering
\includegraphics[width=\linewidth]{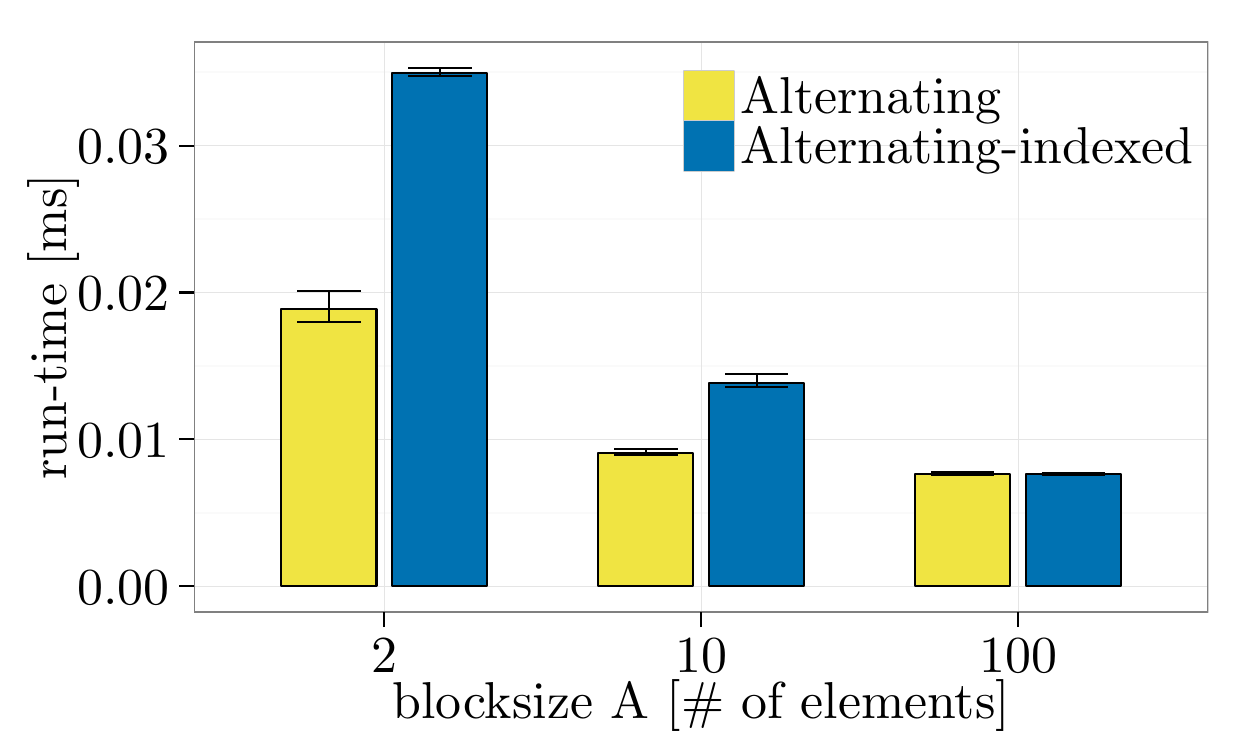}
\caption{%
\label{exp:pingpong-alternatingindexed-small-1x2}%
$\VARdatasize=\SI{3.2}{\kilo\byte}$, same node%
}%
\end{subfigure}%
\hfill%
\begin{subfigure}{.24\linewidth}
\centering
\includegraphics[width=\linewidth]{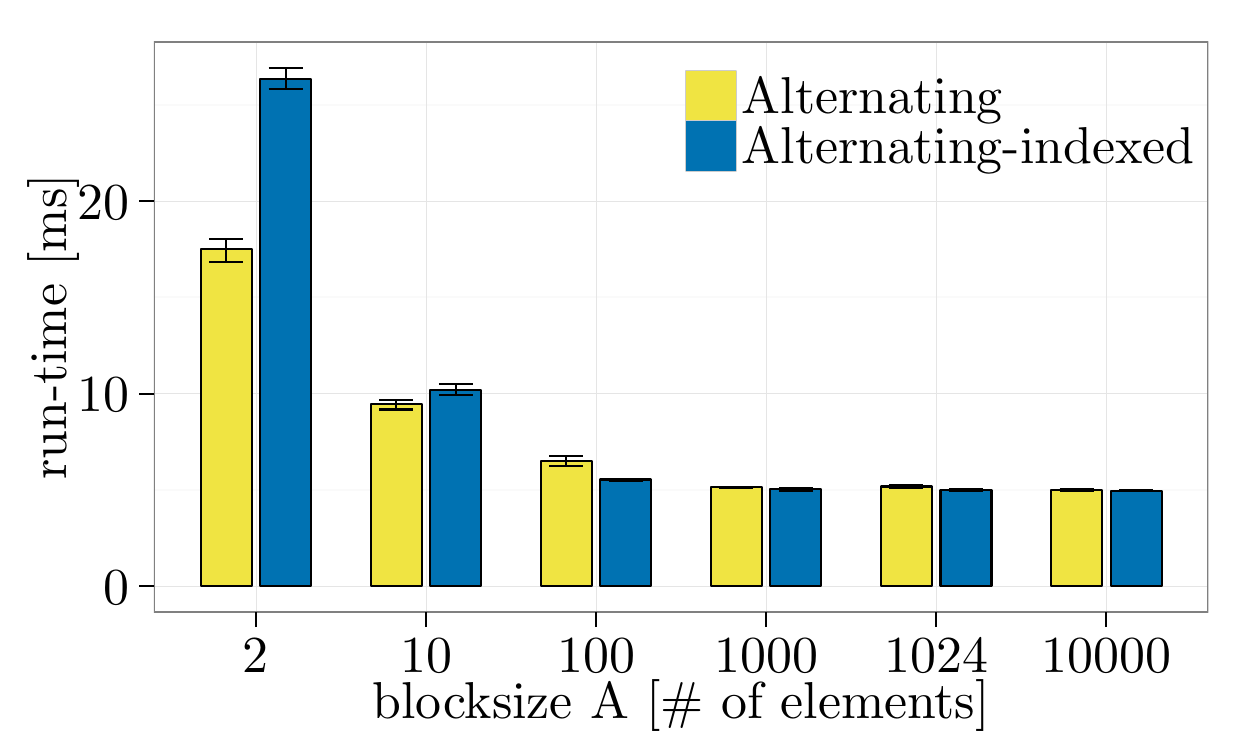}
\caption{%
\label{exp:pingpong-alternatingindexed-large-2x1}%
$\VARdatasize=\SI{2.56}{\mega\byte}$, \num{2}~nodes%
}%
\end{subfigure}%
\hfill%
\begin{subfigure}{.24\linewidth}
\centering
\includegraphics[width=\linewidth]{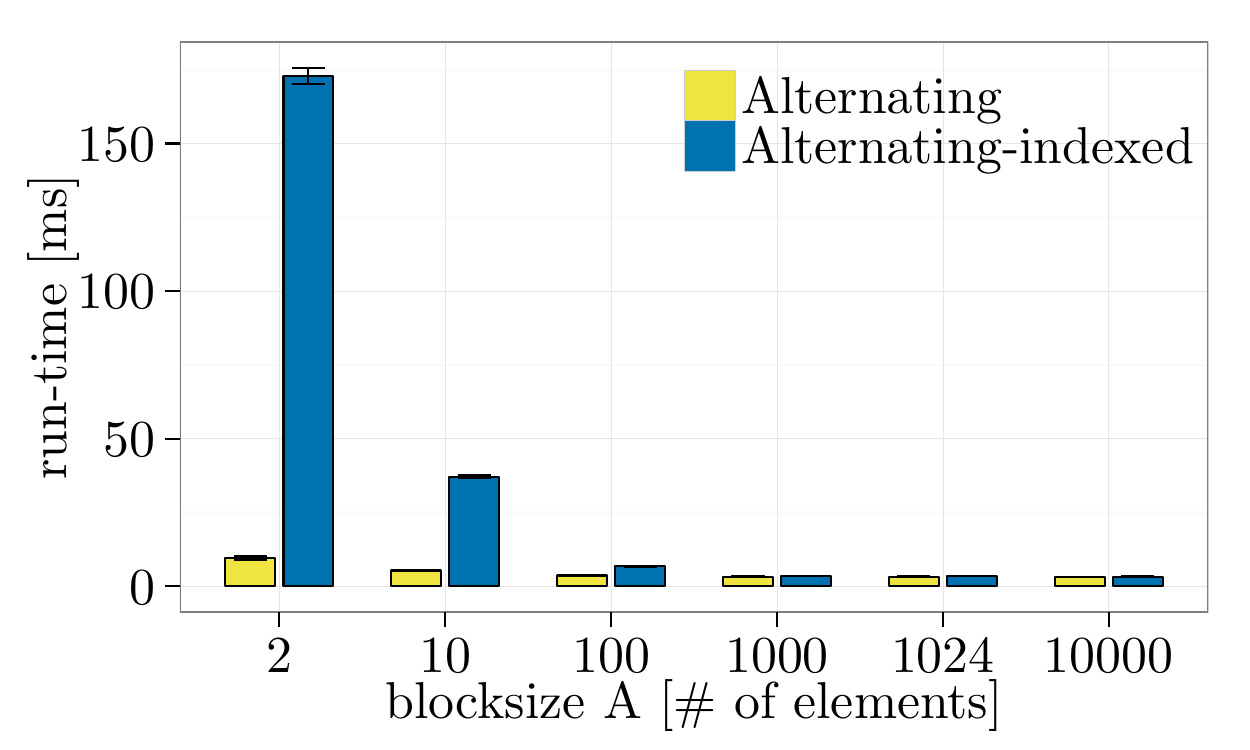}
\caption{%
\label{exp:pingpong-alternatingindexed-large-1x2}%
$\VARdatasize=\SI{2.56}{\mega\byte}$, same node%
}%
\end{subfigure}%
\caption{\label{exp:pingpong-alternatingindexed-nec} \dtalternating \vs \ddtalternatingindexed, element datatype: \mpiint, \pingpong, \jupiternecmpi.}
\end{figure*}

\begin{figure*}[htpb]
\centering
\begin{subfigure}{.24\linewidth}
\centering
\includegraphics[width=\linewidth]{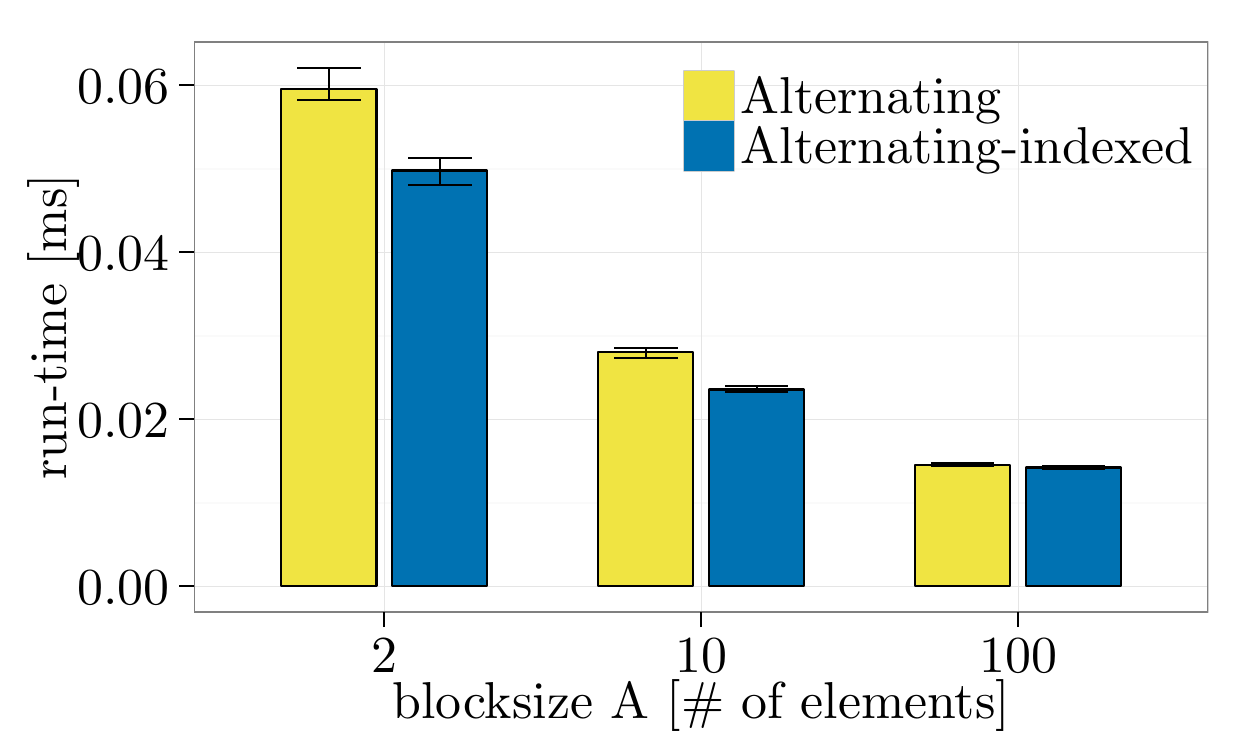}
\caption{%
\label{exp:pingpong-alternatingindexed-small-2x1-mvapich}%
$\VARdatasize=\SI{3.2}{\kilo\byte}$, \num{2}~nodes%
}%
\end{subfigure}%
\hfill%
\begin{subfigure}{.24\linewidth}
\centering
\includegraphics[width=\linewidth]{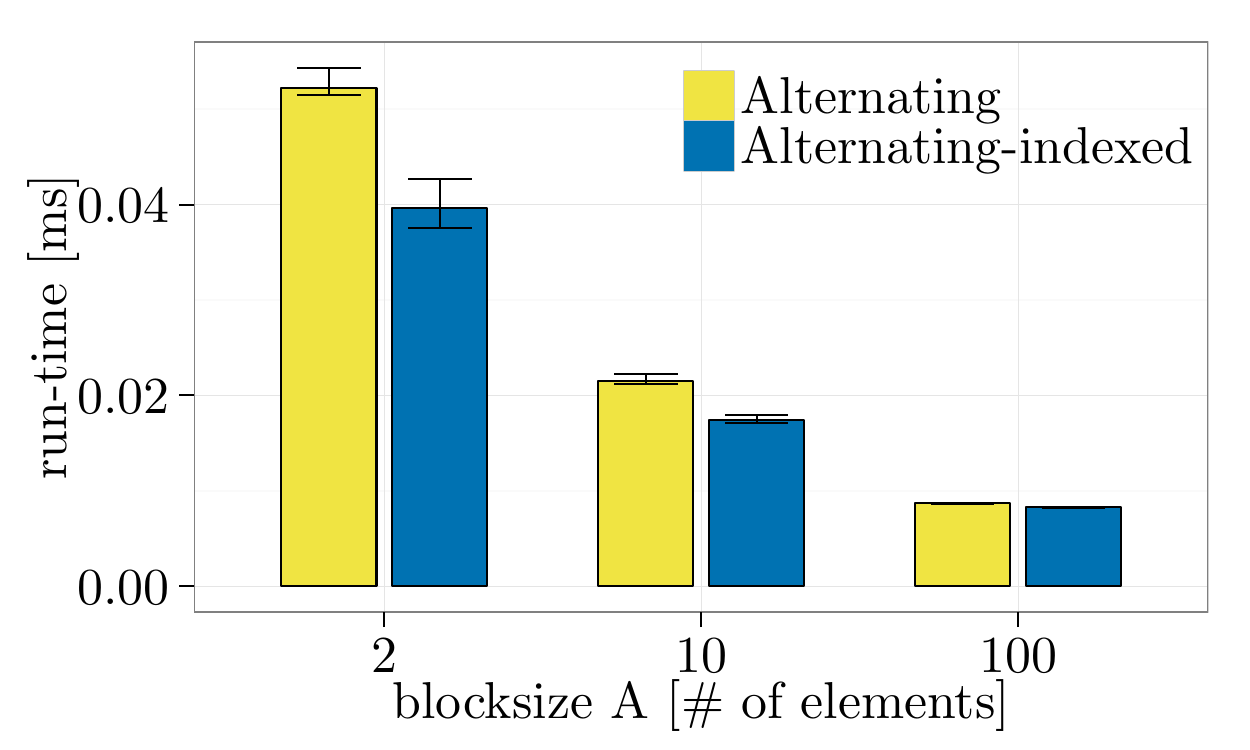}
\caption{%
\label{exp:pingpong-alternatingindexed-small-1x2-mvapich}%
$\VARdatasize=\SI{3.2}{\kilo\byte}$, same node%
}%
\end{subfigure}%
\hfill%
\begin{subfigure}{.24\linewidth}
\centering
\includegraphics[width=\linewidth]{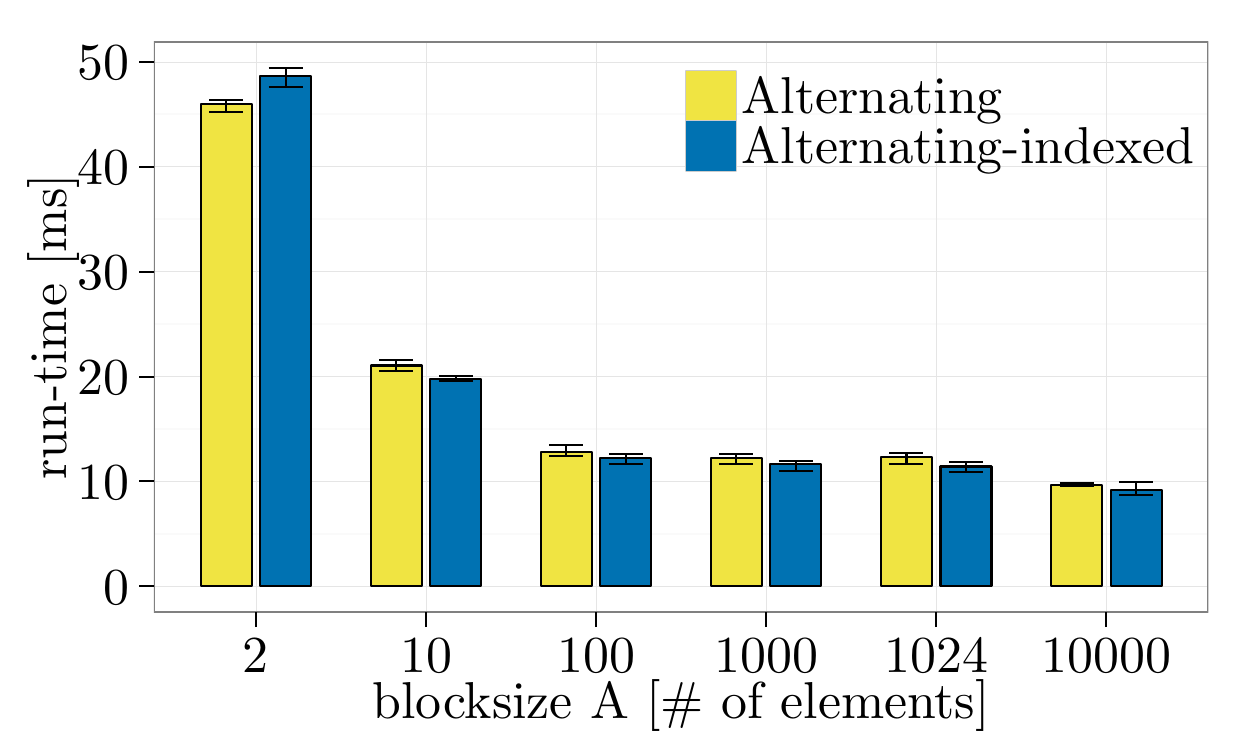}
\caption{%
\label{exp:pingpong-alternatingindexed-large-2x1-mvapich}%
$\VARdatasize=\SI{2.56}{\mega\byte}$, \num{2}~nodes%
}%
\end{subfigure}%
\hfill%
\begin{subfigure}{.24\linewidth}
\centering
\includegraphics[width=\linewidth]{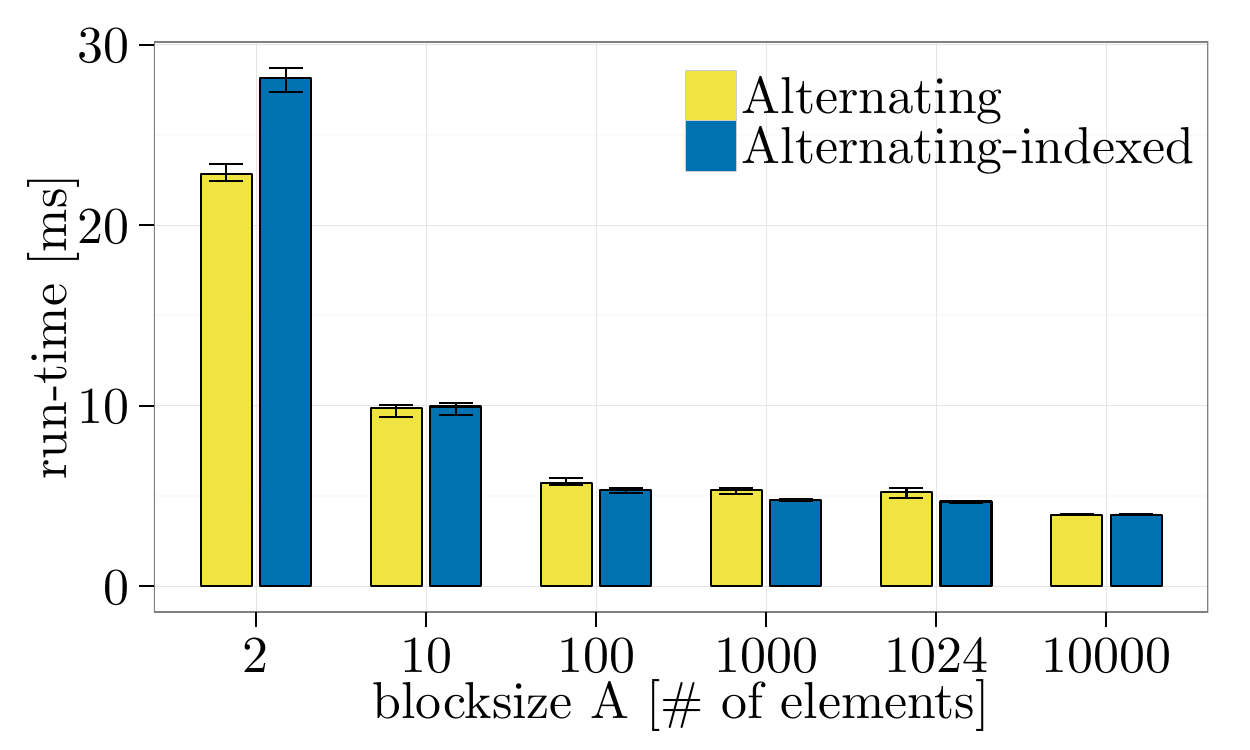}
\caption{%
\label{exp:pingpong-alternatingindexed-large-1x2-mvapich}%
$\VARdatasize=\SI{2.56}{\mega\byte}$, same node%
}%
\end{subfigure}%
\caption{\label{exp:pingpong-alternatingindexed-mvapich} \dtalternating \vs \ddtalternatingindexed, element datatype: \mpiint, \pingpong, \jupitermvapich.}
\end{figure*}

\begin{figure*}[htpb]
\centering
\begin{subfigure}{.24\linewidth}
\centering
\includegraphics[width=\linewidth]{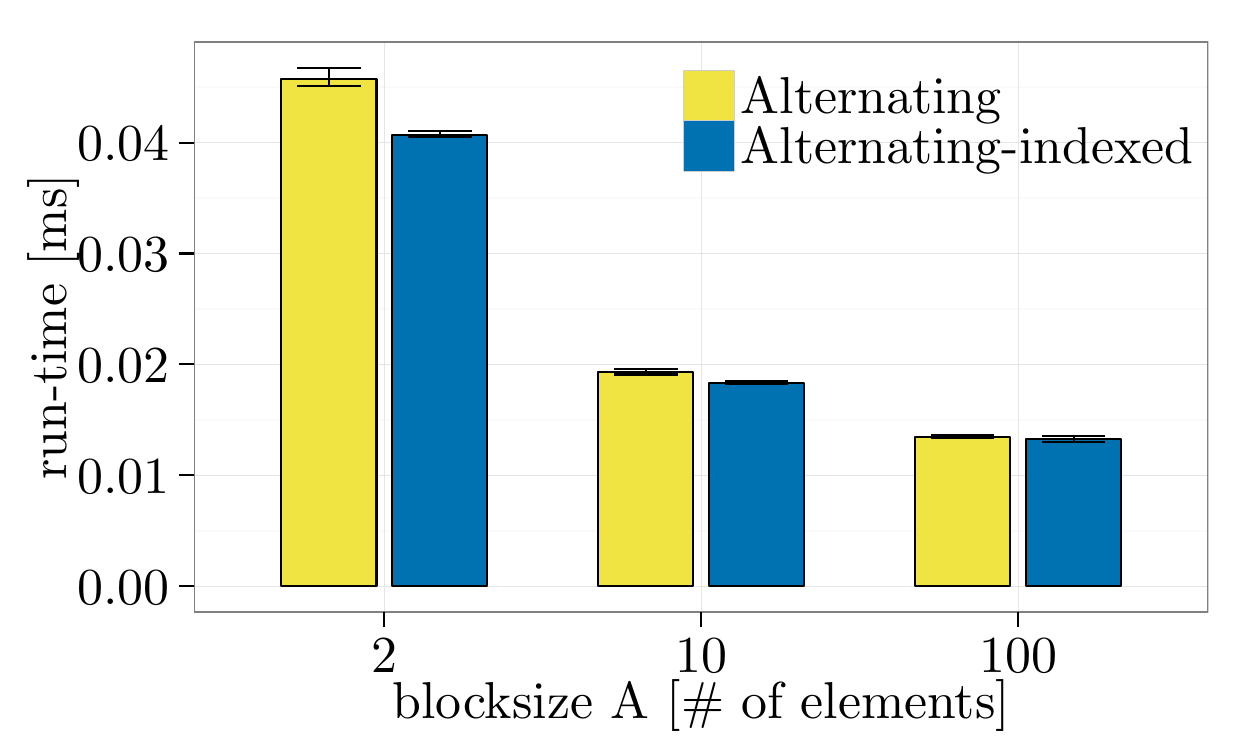}
\caption{%
\label{exp:pingpong-alternatingindexed-small-2x1-openmpi}%
$\VARdatasize=\SI{3.2}{\kilo\byte}$, \num{2}~nodes%
}%
\end{subfigure}%
\hfill%
\begin{subfigure}{.24\linewidth}
\centering
\includegraphics[width=\linewidth]{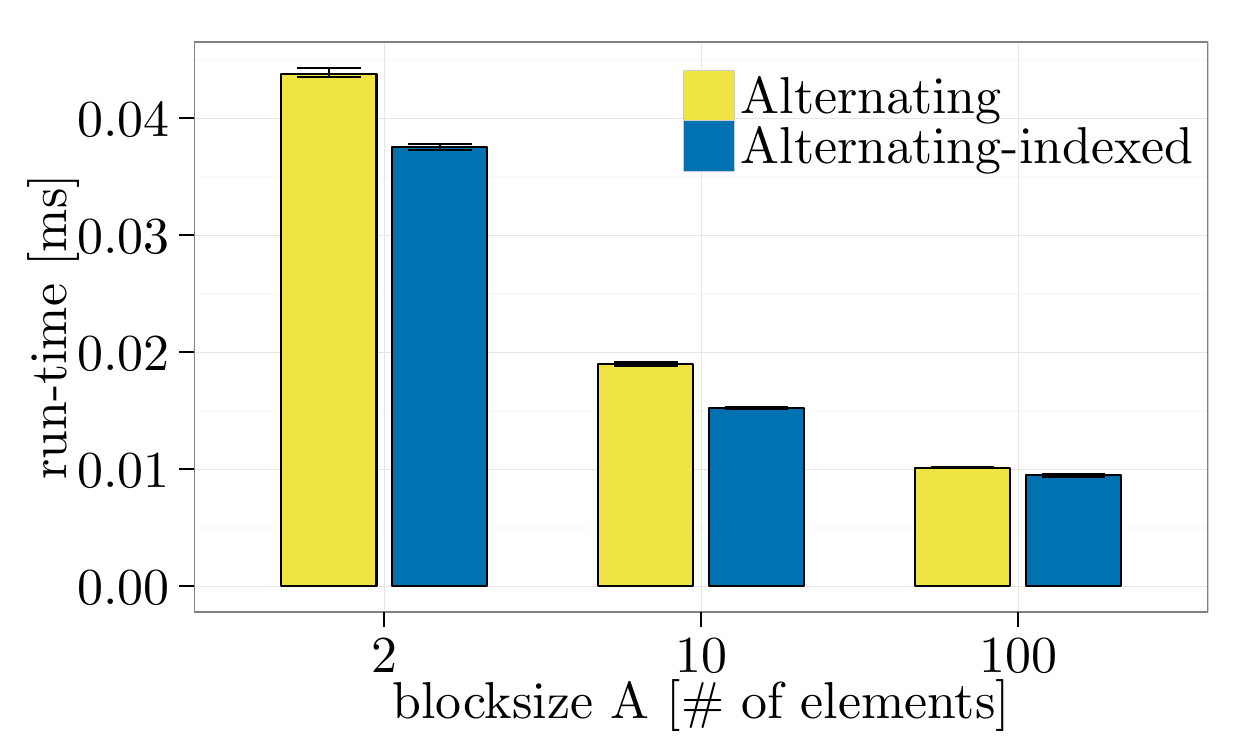}
\caption{%
\label{exp:pingpong-alternatingindexed-small-1x2-openmpi}%
$\VARdatasize=\SI{3.2}{\kilo\byte}$, same node%
}%
\end{subfigure}%
\hfill%
\begin{subfigure}{.24\linewidth}
\centering
\includegraphics[width=\linewidth]{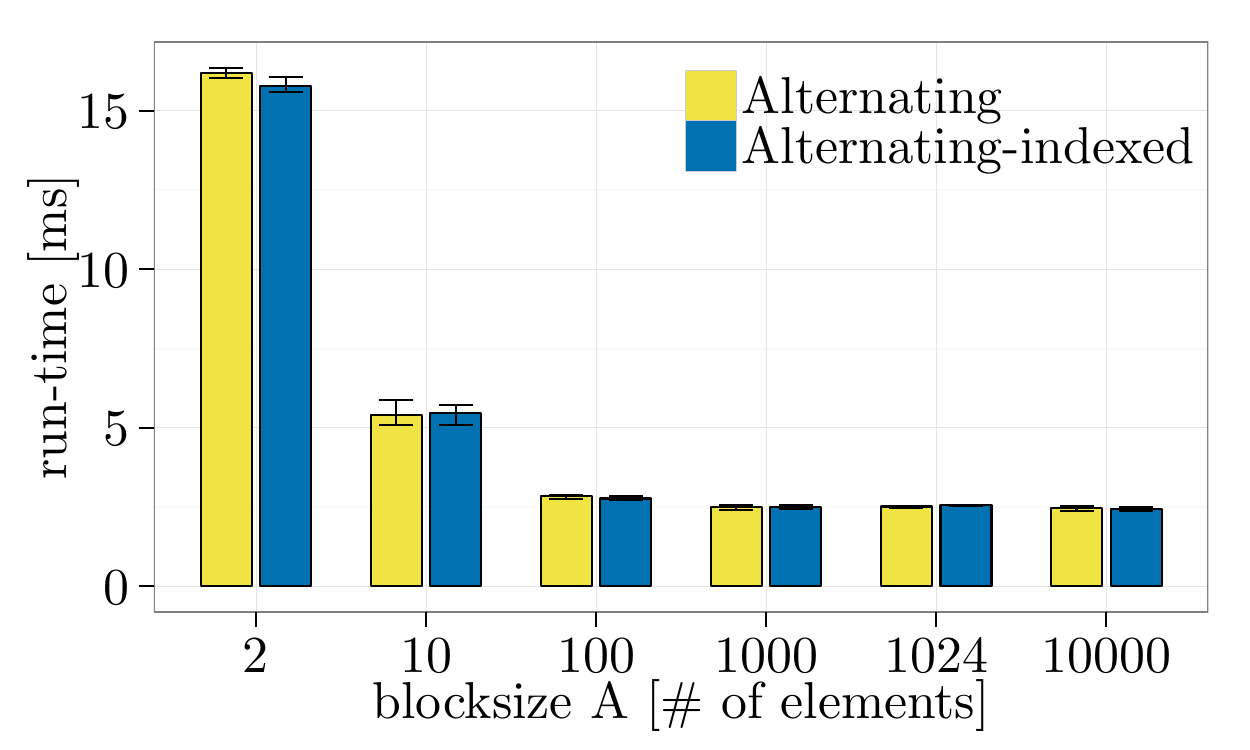}
\caption{%
\label{exp:pingpong-alternatingindexed-large-2x1-openmpi}%
$\VARdatasize=\SI{2.56}{\mega\byte}$, \num{2}~nodes%
}%
\end{subfigure}%
\hfill%
\begin{subfigure}{.24\linewidth}
\centering
\includegraphics[width=\linewidth]{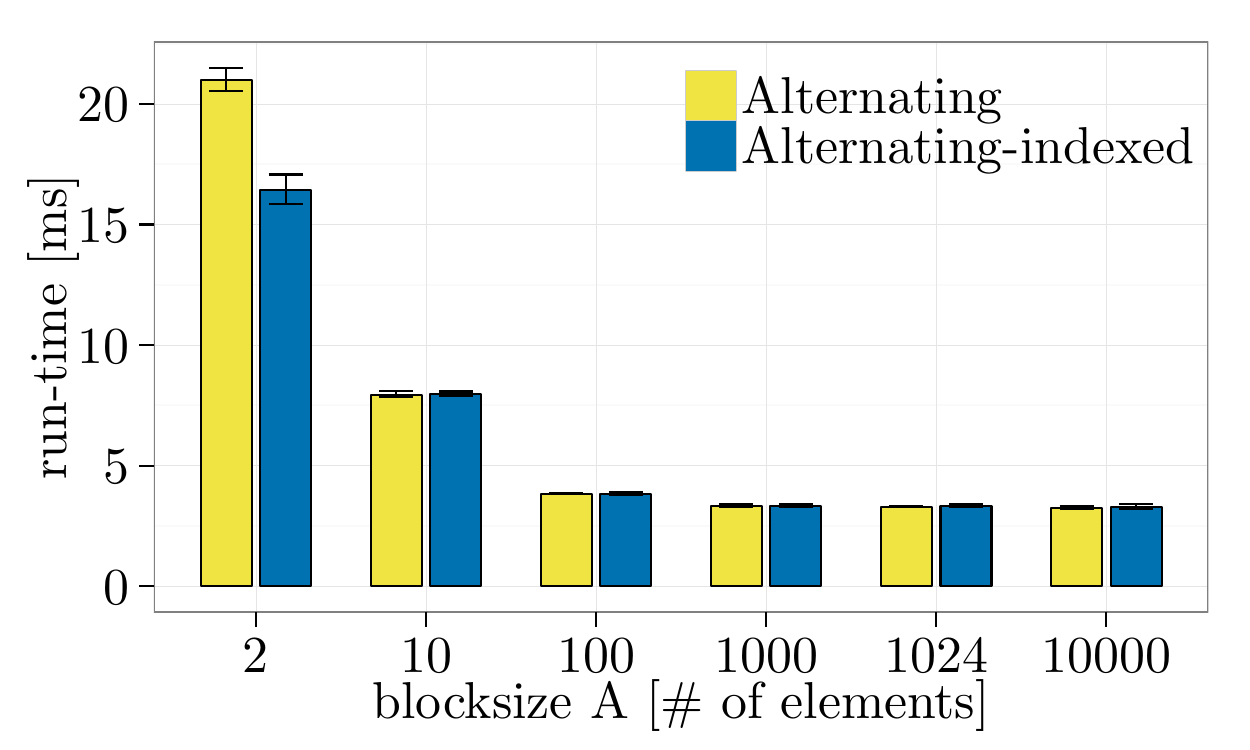}
\caption{%
\label{exp:pingpong-alternatingindexed-large-1x2-openmpi}%
$\VARdatasize=\SI{2.56}{\mega\byte}$, same node%
}%
\end{subfigure}%
\caption{\label{exp:pingpong-alternatingindexed-openmpi} \dtalternating \vs \ddtalternatingindexed, element datatype: \mpiint, \pingpong, \jupiteropenmpi.}
\end{figure*}

\appexpdesc{
  \begin{expitemize}
    \item \dtalternating, \ddtalternatingindexed
    \item \pingpong
  \end{expitemize}
}{
  \begin{expitemize}
    \item \expparam{\jupiternecmpi}{\fig~\ref{exp:pingpong-alternatingindexed-nec}}
    \item \expparam{\jupitermvapich}{\fig~\ref{exp:pingpong-alternatingindexed-mvapich}}
    \item \expparam{\jupiteropenmpi}{\fig~\ref{exp:pingpong-alternatingindexed-openmpi}}
  \end{expitemize}  
}

\FloatBarrier
\clearpage

\begin{figure*}[htpb]
\centering
\begin{subfigure}{.24\linewidth}
\centering
\includegraphics[width=\linewidth]{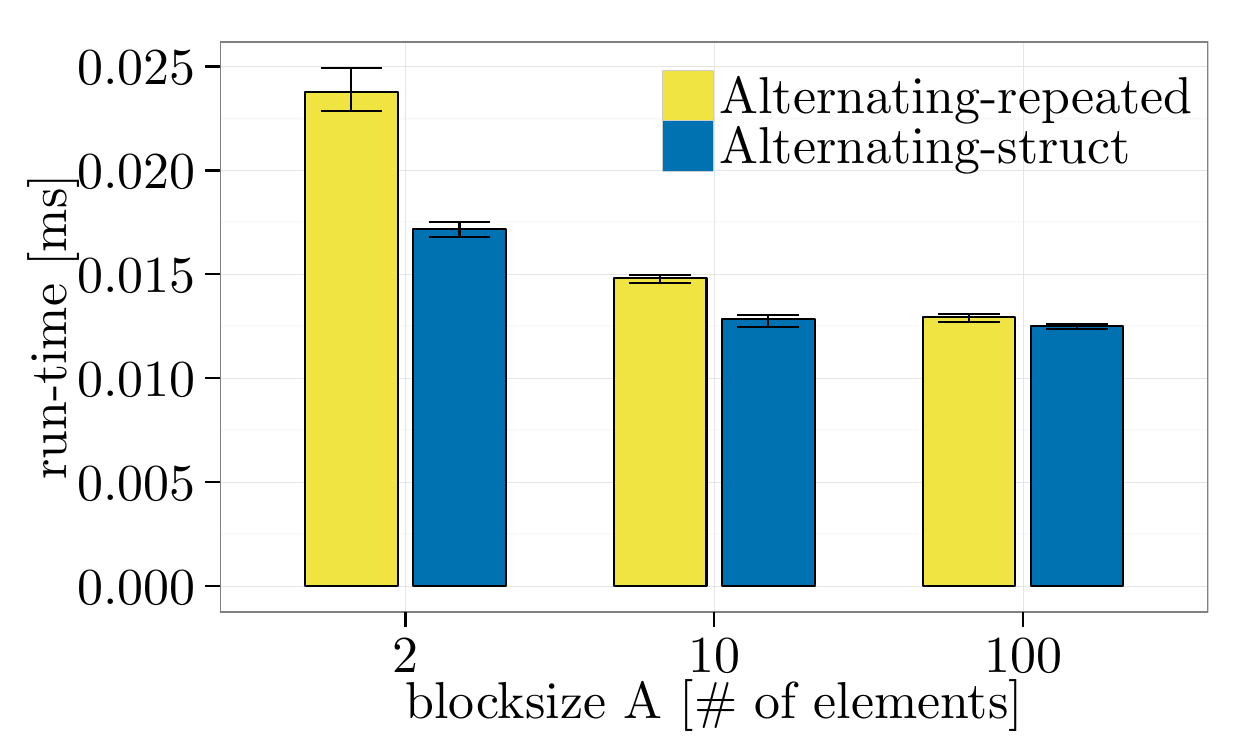}
\caption{%
\label{exp:pingpong-alternatingrepeated-small-2x1}%
$\VARdatasize=\SI{3.2}{\kilo\byte}$, \num{2}~nodes%
}%
\end{subfigure}%
\hfill%
\begin{subfigure}{.24\linewidth}
\centering
\includegraphics[width=\linewidth]{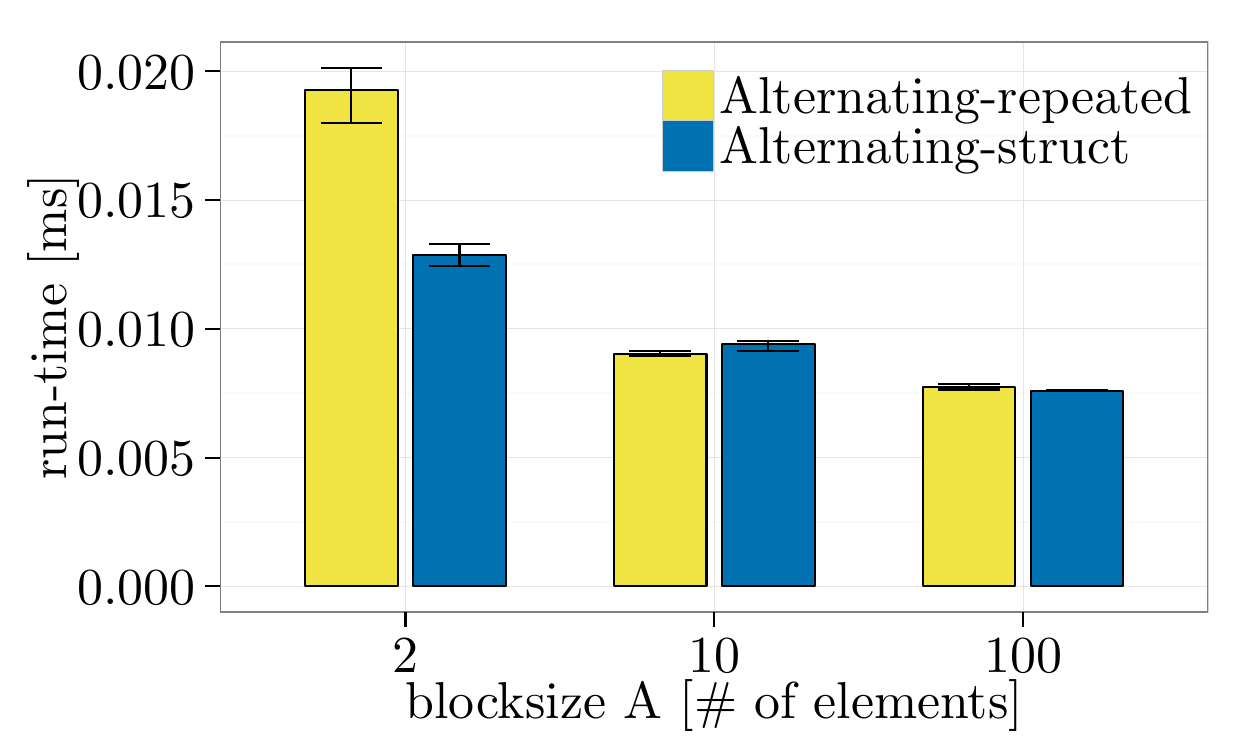}
\caption{%
\label{exp:pingpong-alternatingrepeated-small-1x2}%
$\VARdatasize=\SI{3.2}{\kilo\byte}$, same node%
}%
\end{subfigure}%
\hfill%
\begin{subfigure}{.24\linewidth}
\centering
\includegraphics[width=\linewidth]{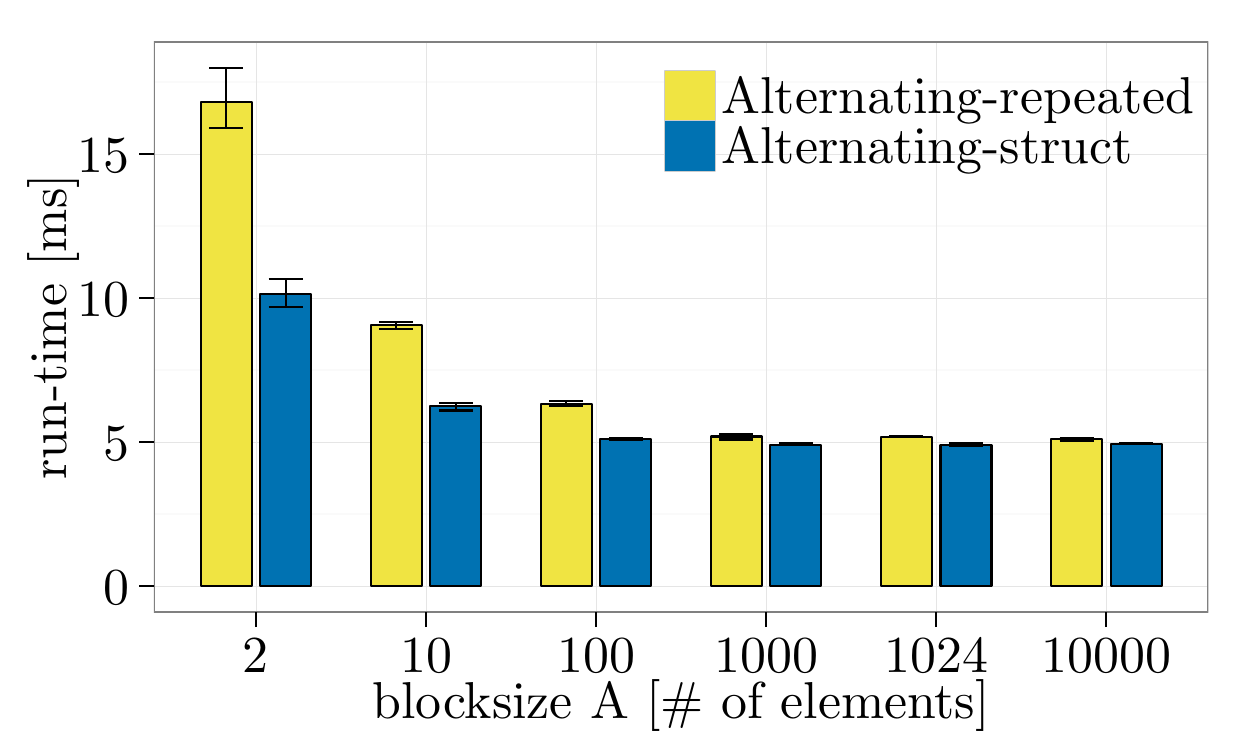}
\caption{%
\label{exp:pingpong-alternatingrepeated-large-2x1}%
$\VARdatasize=\SI{2.56}{\mega\byte}$, \num{2}~nodes%
}%
\end{subfigure}%
\hfill%
\begin{subfigure}{.24\linewidth}
\centering
\includegraphics[width=\linewidth]{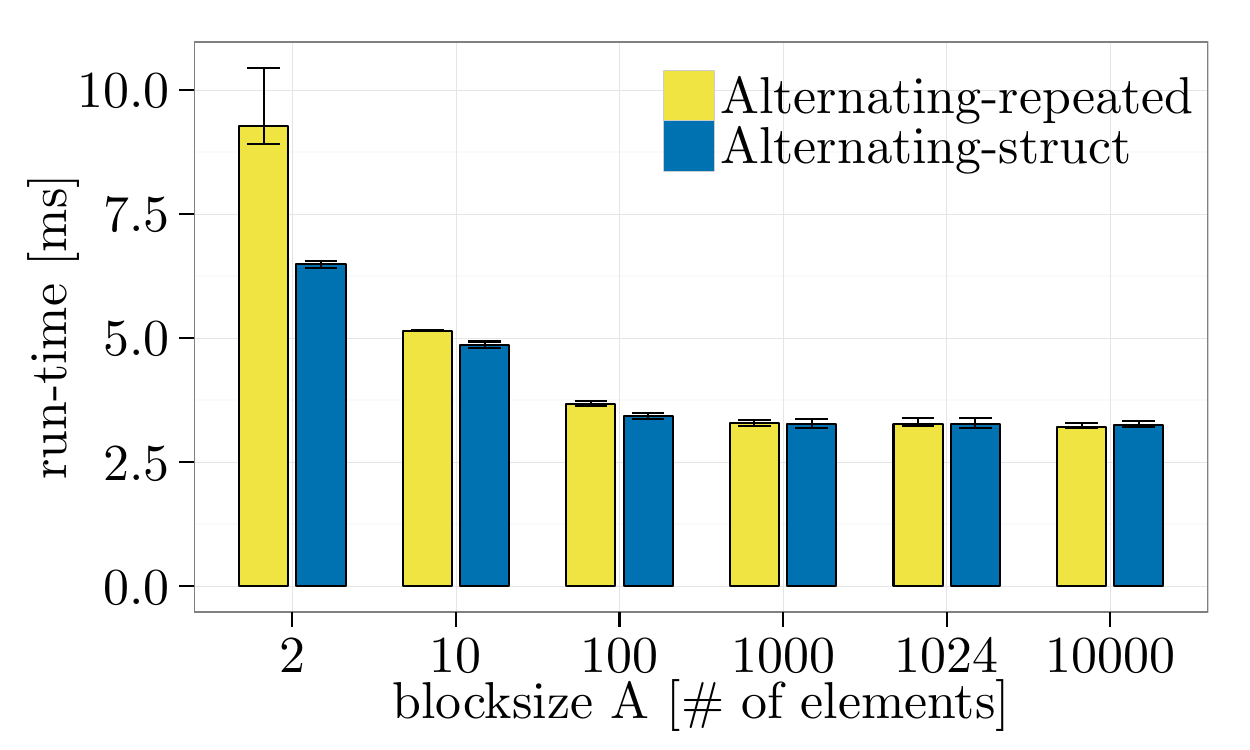}
\caption{%
\label{exp:pingpong-alternatingrepeated-large-1x2}%
$\VARdatasize=\SI{2.56}{\mega\byte}$, same node%
}%
\end{subfigure}%
\caption{\label{exp:pingpong-alternatingrepeated-nec} \ddtalternatingrepeated \vs \ddtalternatingstruct, element datatype: \mpiint, \pingpong, \jupiternecmpi.}
\end{figure*}

\begin{figure*}[htpb]
\centering
\begin{subfigure}{.24\linewidth}
\centering
\includegraphics[width=\linewidth]{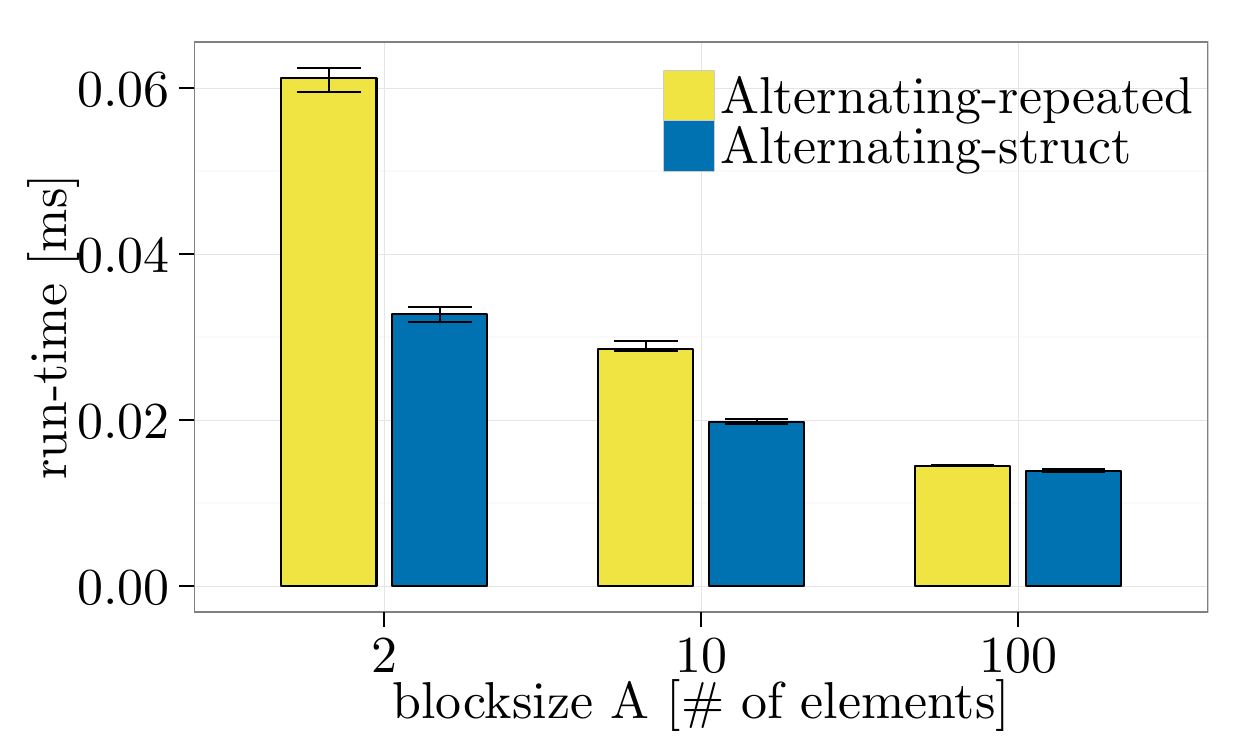}
\caption{%
\label{exp:pingpong-alternatingrepeated-small-2x1-mvapich}%
$\VARdatasize=\SI{3.2}{\kilo\byte}$, \num{2}~nodes%
}%
\end{subfigure}%
\hfill%
\begin{subfigure}{.24\linewidth}
\centering
\includegraphics[width=\linewidth]{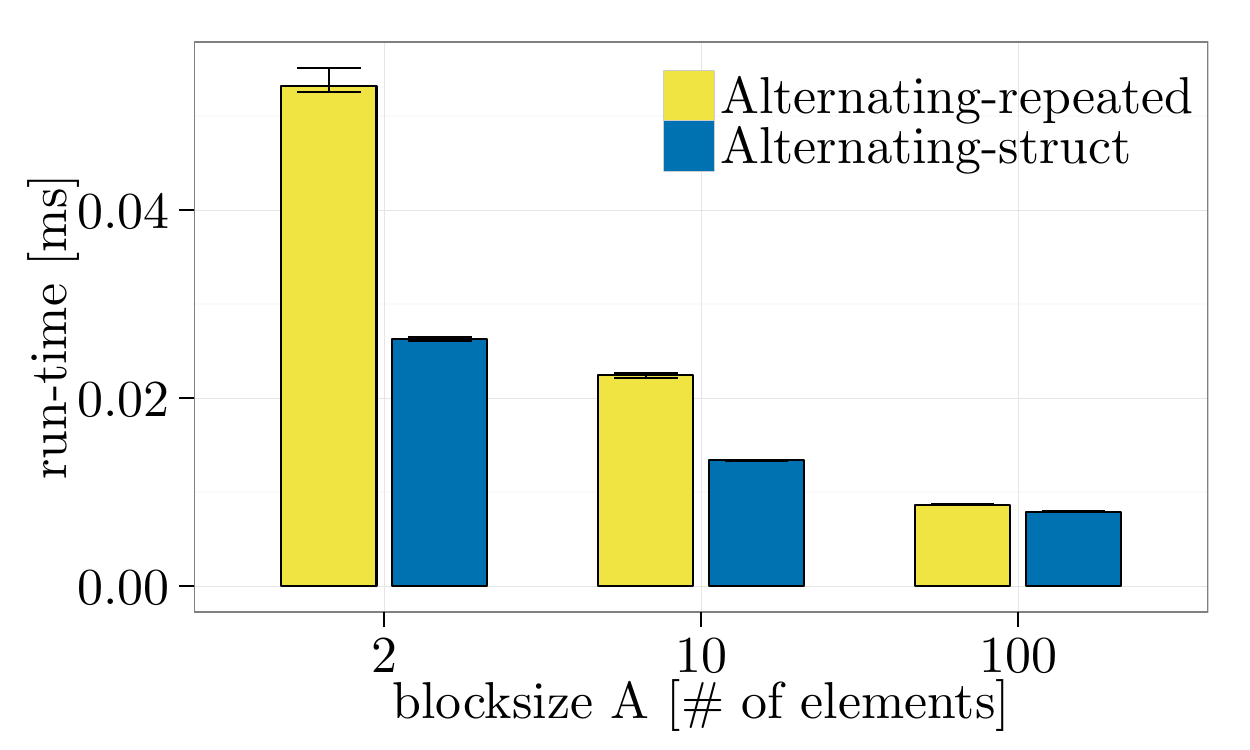}
\caption{%
\label{exp:pingpong-alternatingrepeated-small-1x2-mvapich}%
$\VARdatasize=\SI{3.2}{\kilo\byte}$, same node%
}%
\end{subfigure}%
\hfill%
\begin{subfigure}{.24\linewidth}
\centering
\includegraphics[width=\linewidth]{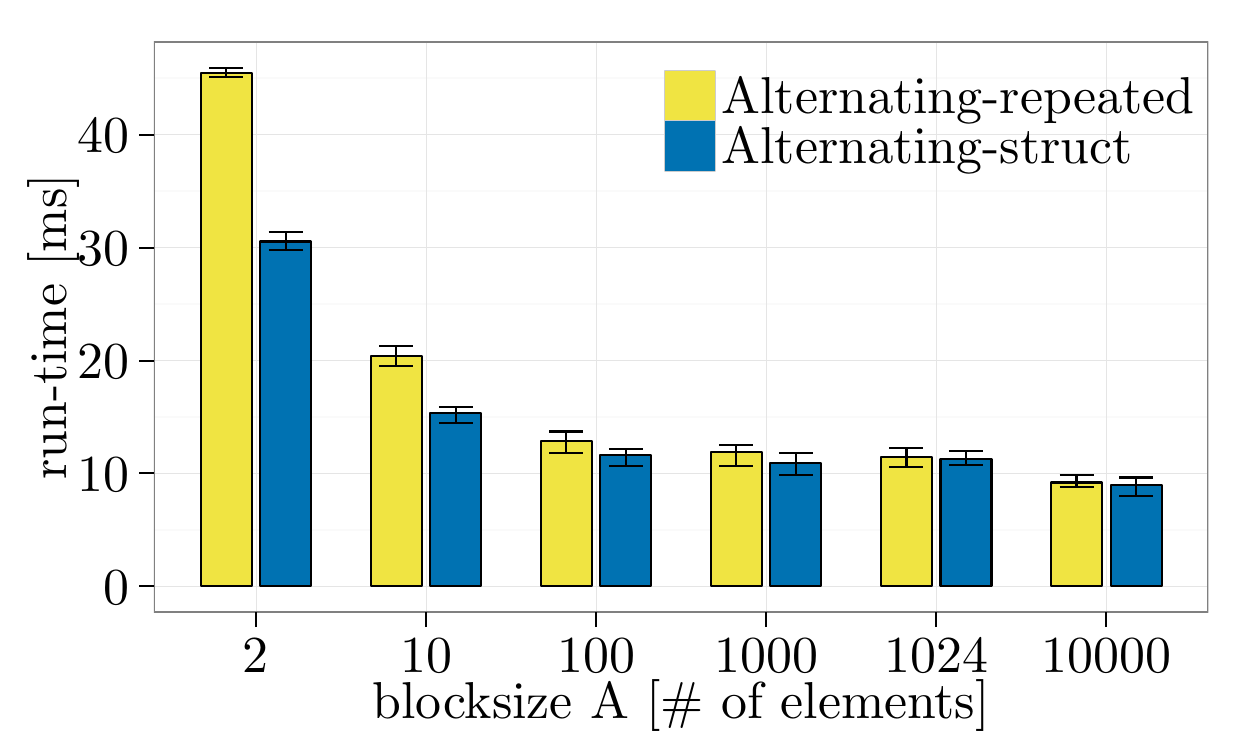}
\caption{%
\label{exp:pingpong-alternatingrepeated-large-2x1-mvapich}%
$\VARdatasize=\SI{2.56}{\mega\byte}$, \num{2}~nodes%
}%
\end{subfigure}%
\hfill%
\begin{subfigure}{.24\linewidth}
\centering
\includegraphics[width=\linewidth]{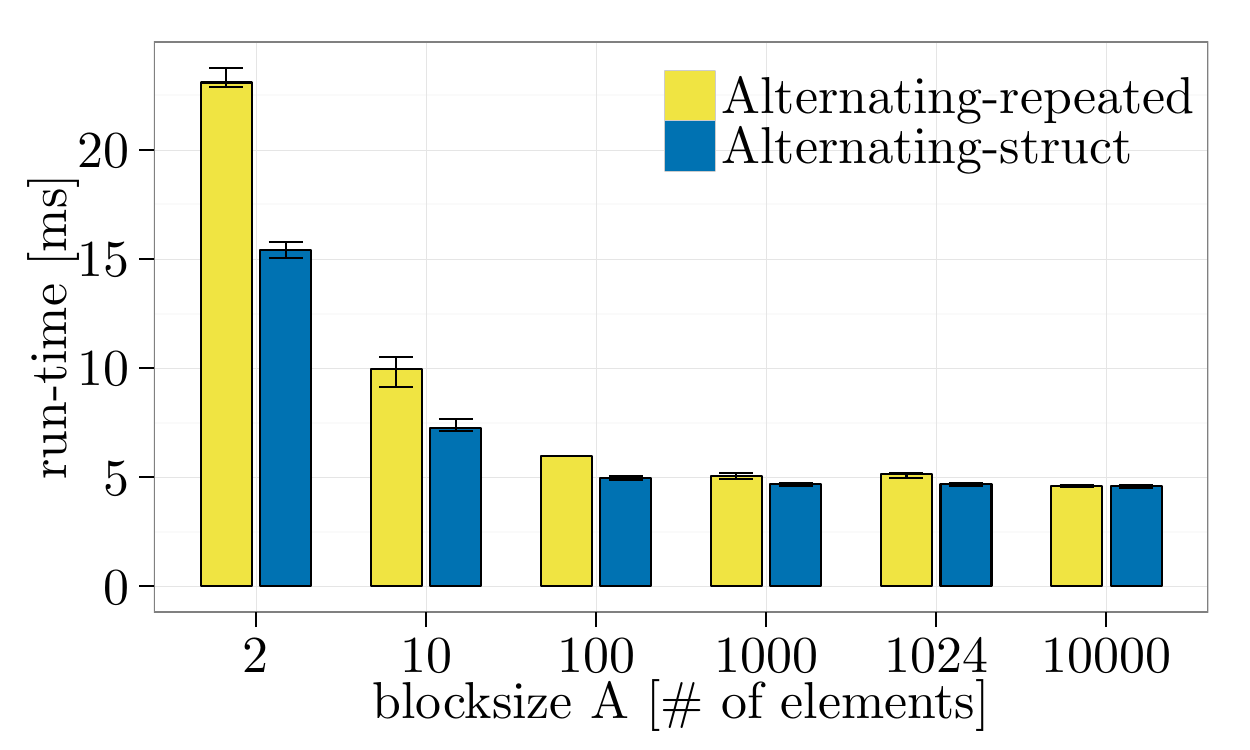}
\caption{%
\label{exp:pingpong-alternatingrepeated-large-1x2-mvapich}%
$\VARdatasize=\SI{2.56}{\mega\byte}$, same node%
}%
\end{subfigure}%
\caption{\label{exp:pingpong-alternatingrepeated-mvapich} \ddtalternatingrepeated \vs \ddtalternatingstruct, element datatype: \mpiint, \pingpong, \jupitermvapich.}
\end{figure*}

\begin{figure*}[htpb]
\centering
\begin{subfigure}{.24\linewidth}
\centering
\includegraphics[width=\linewidth]{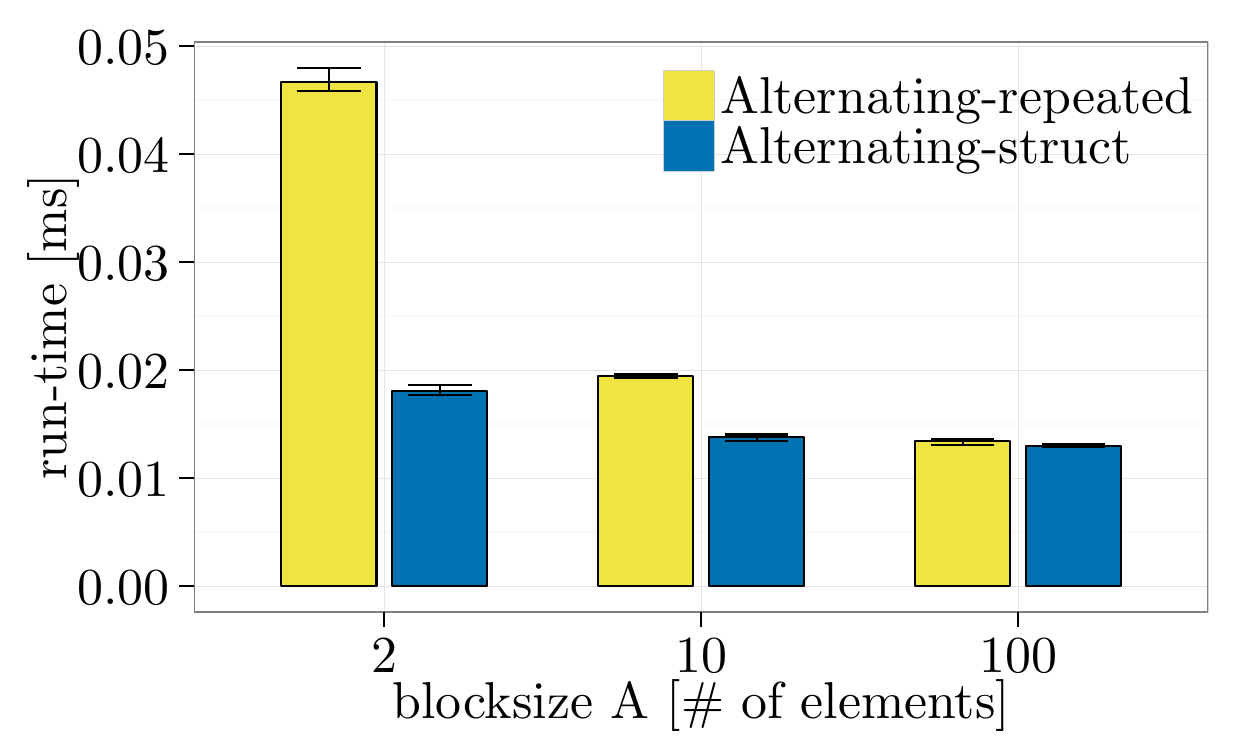}
\caption{%
\label{exp:pingpong-alternatingrepeated-small-2x1-openmpi}%
$\VARdatasize=\SI{3.2}{\kilo\byte}$, \num{2}~nodes%
}%
\end{subfigure}%
\hfill%
\begin{subfigure}{.24\linewidth}
\centering
\includegraphics[width=\linewidth]{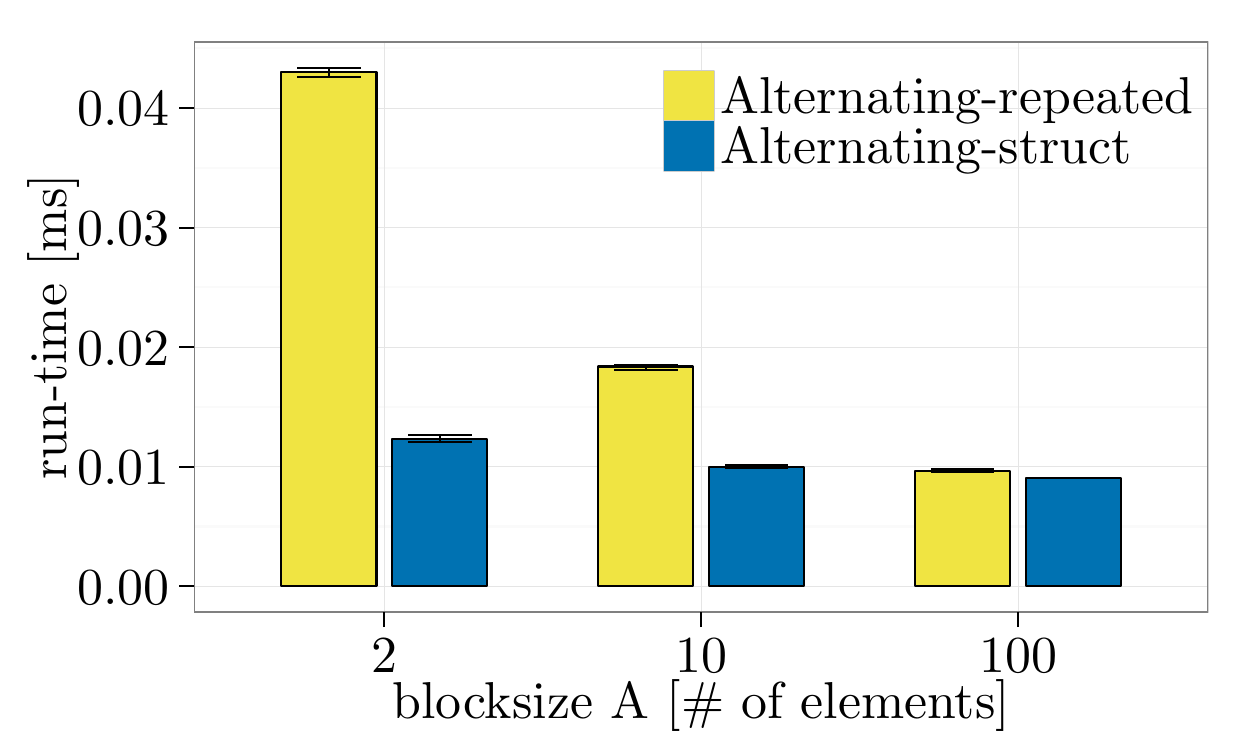}
\caption{%
\label{exp:pingpong-alternatingrepeated-small-1x2-openmpi}%
$\VARdatasize=\SI{3.2}{\kilo\byte}$, same node%
}%
\end{subfigure}%
\hfill%
\begin{subfigure}{.24\linewidth}
\centering
\includegraphics[width=\linewidth]{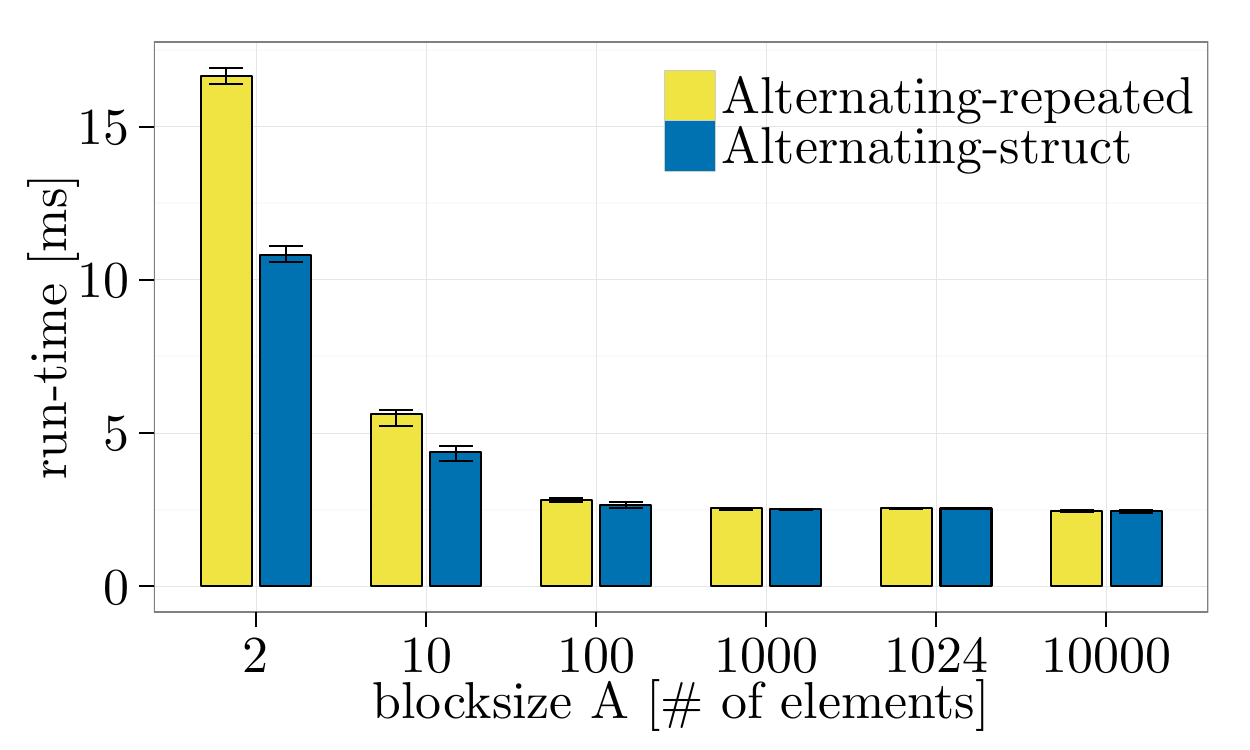}
\caption{%
\label{exp:pingpong-alternatingrepeated-large-2x1-openmpi}%
$\VARdatasize=\SI{2.56}{\mega\byte}$, \num{2}~nodes%
}%
\end{subfigure}%
\hfill%
\begin{subfigure}{.24\linewidth}
\centering
\includegraphics[width=\linewidth]{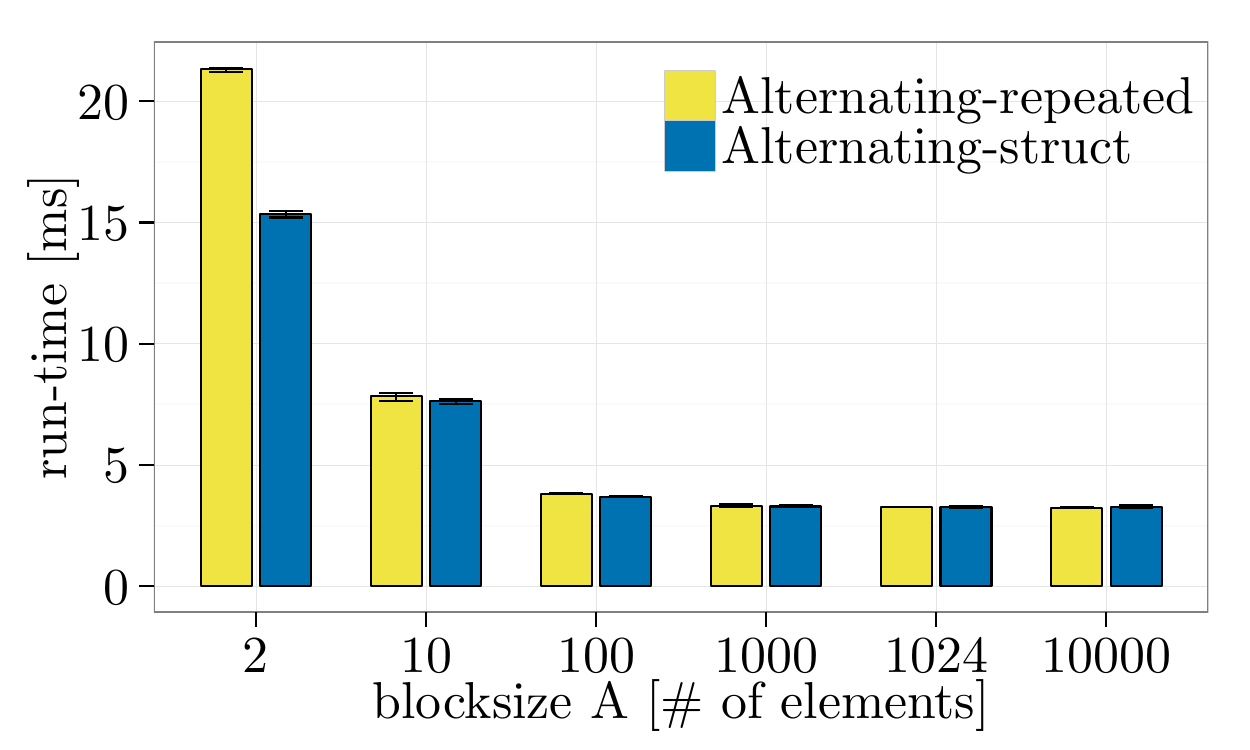}
\caption{%
\label{exp:pingpong-alternatingrepeated-large-1x2-openmpi}%
$\VARdatasize=\SI{2.56}{\mega\byte}$, same node%
}%
\end{subfigure}%
\caption{\label{exp:pingpong-alternatingrepeated-openmpi} \ddtalternatingrepeated \vs \ddtalternatingstruct, element datatype: \mpiint, \pingpong, \jupiteropenmpi.}
\end{figure*}

\appexp{exptest:alternating_repeated}
\appexpdesc{
  \begin{expitemize}
    \item \ddtalternatingrepeated, \ddtalternatingstruct
    \item \pingpong
  \end{expitemize}
}{
  \begin{expitemize}
    \item \expparam{\jupiternecmpi}{\fig~\ref{exp:pingpong-alternatingrepeated-nec}}
    \item \expparam{\jupitermvapich}{\fig~\ref{exp:pingpong-alternatingrepeated-mvapich}}
    \item \expparam{\jupiteropenmpi}{\fig~\ref{exp:pingpong-alternatingrepeated-openmpi}}
  \end{expitemize}  
}

\FloatBarrier
\clearpage

\begin{figure*}[htpb]
\centering
\begin{subfigure}{.24\linewidth}
\centering
\includegraphics[width=\linewidth]{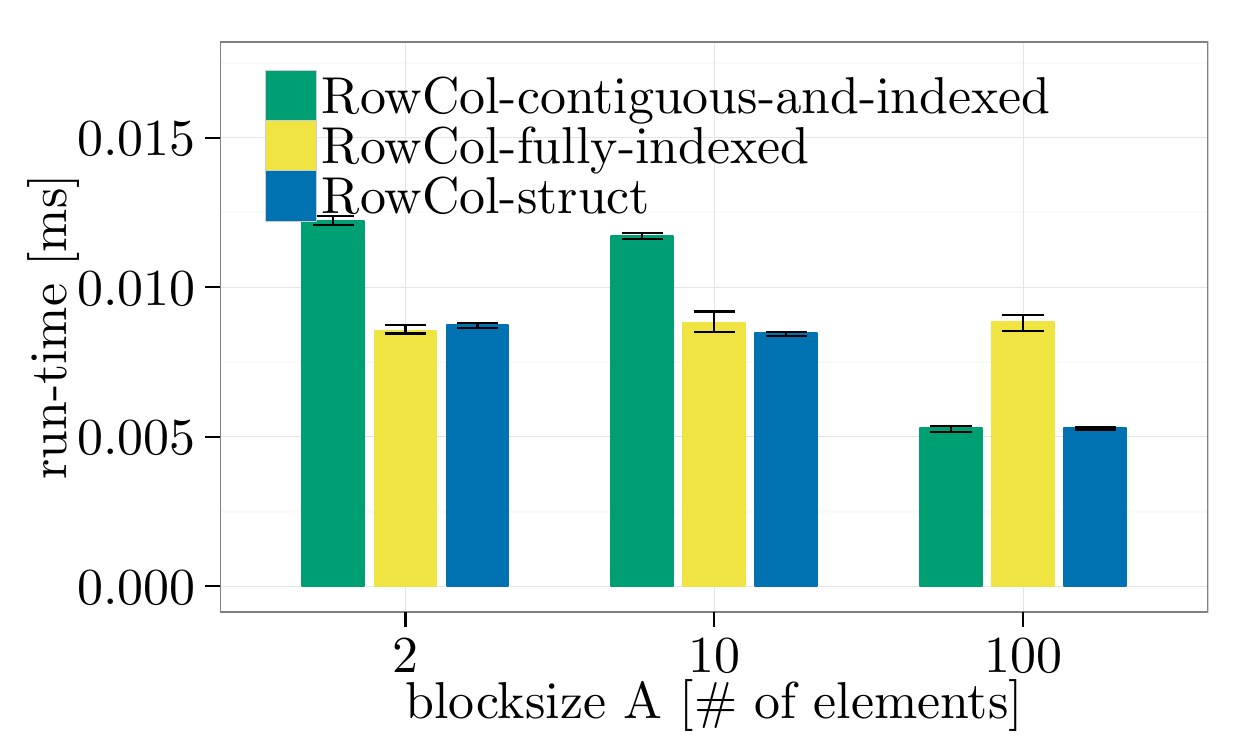}
\caption{%
\label{exp:pingpong-rowcol-small-2x1}%
$n=\num{100}$, \num{2}~nodes%
}%
\end{subfigure}%
\hfill%
\begin{subfigure}{.24\linewidth}
\centering
\includegraphics[width=\linewidth]{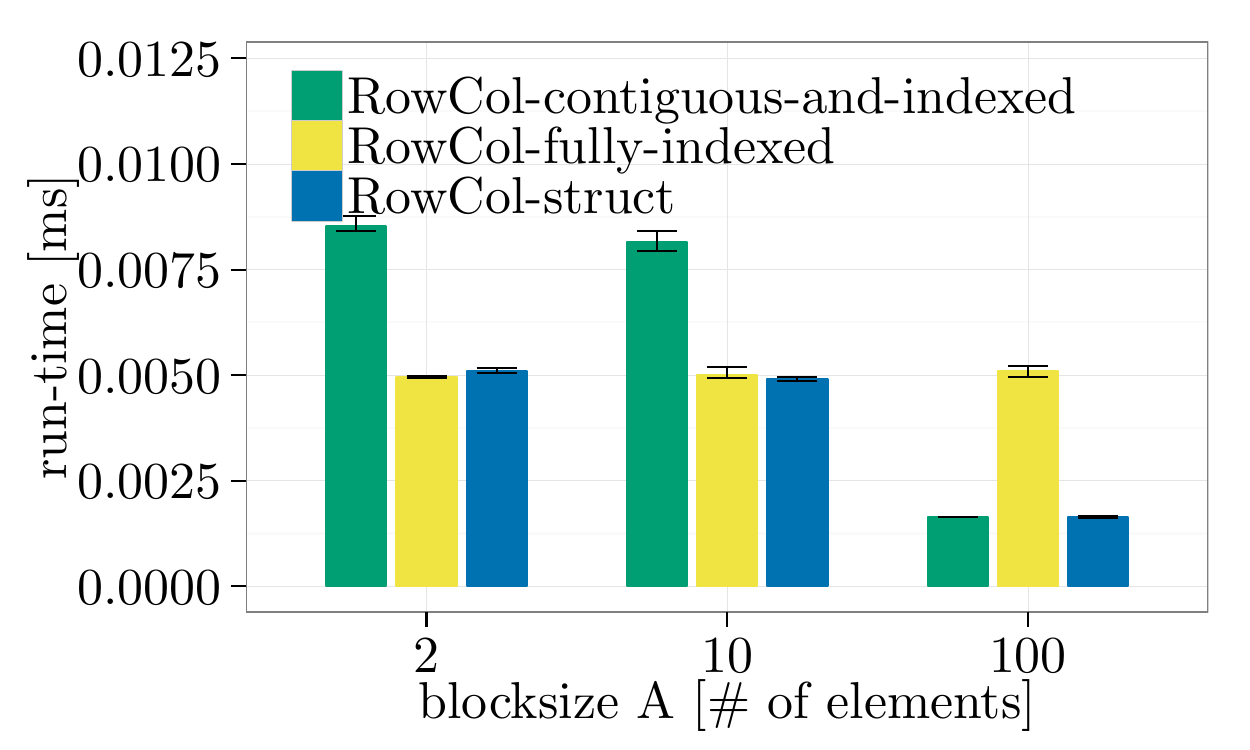}
\caption{%
\label{exp:pingpong-rowcol-small-1x2}%
$n=\num{100}$, same node%
}%
\end{subfigure}%
\hfill%
\begin{subfigure}{.24\linewidth}
\centering
\includegraphics[width=\linewidth]{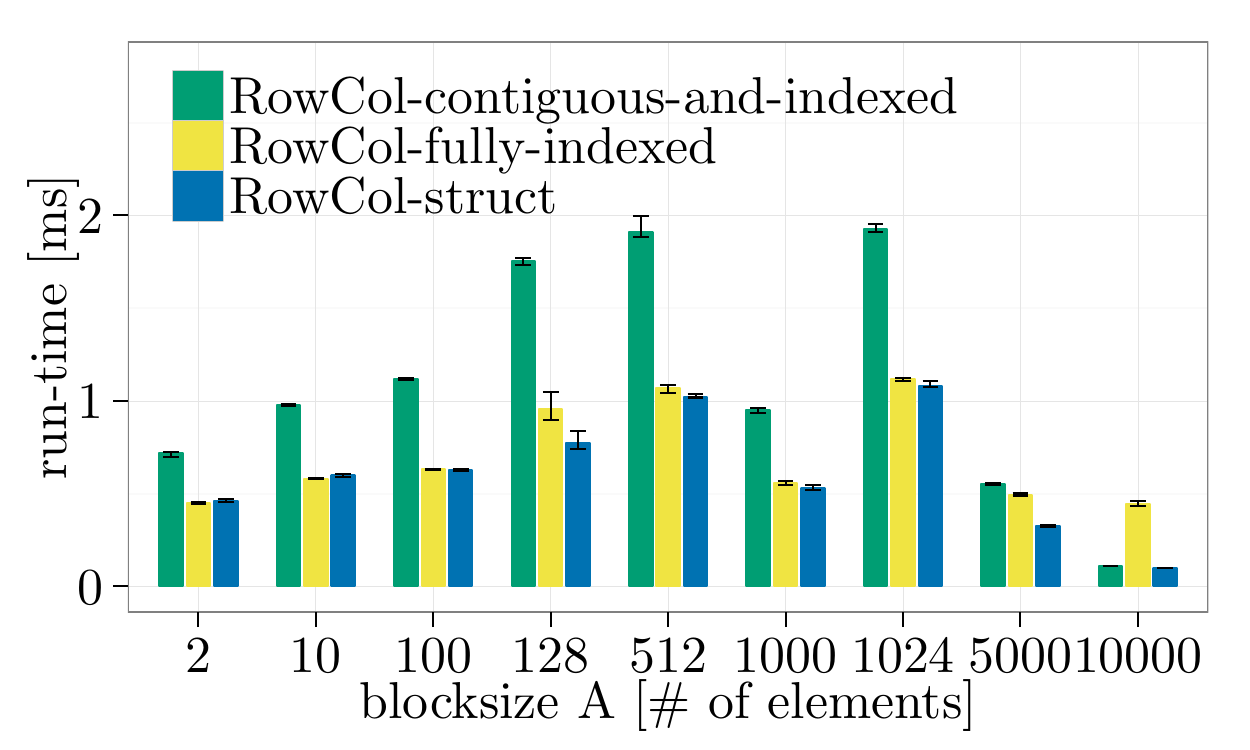}
\caption{%
\label{exp:pingpong-rowcol-large-2x1}%
 $n=\num{10240}$, \num{2}~nodes%
}%
\end{subfigure}%
\hfill%
\begin{subfigure}{.24\linewidth}
\centering
\includegraphics[width=\linewidth]{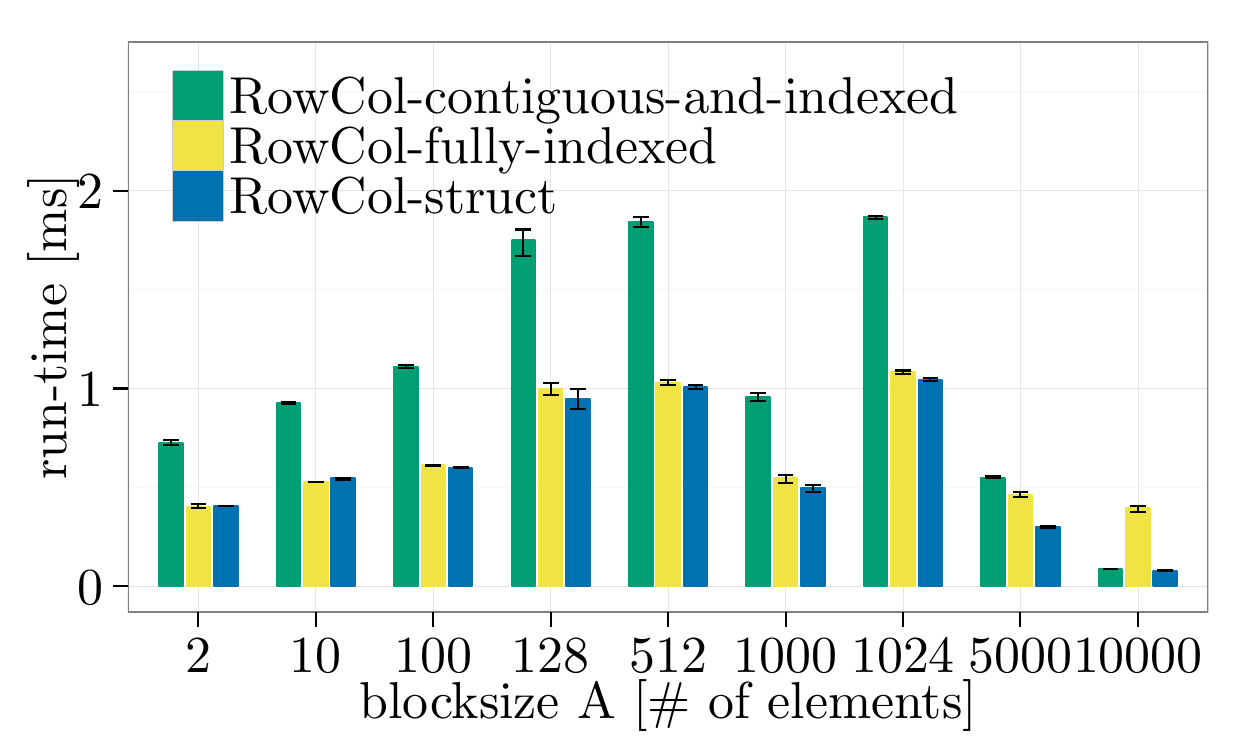}
\caption{%
\label{exp:pingpong-rowcol-large-1x2}%
 $n=\num{10240}$, same node%
}%
\end{subfigure}%
\caption{\label{exp:pingpong-rowcol-nec}  \ddtrowcolfullindexed, \ddtrowcolcontiguousandindexed, \ddtrowcolstruct, element datatype: \mpiint, buffer size (and \extent) increases with \blocksize \VARblocksize, \pingpong, \jupiternecmpi.}
\end{figure*}

\begin{figure*}[htpb]
\centering
\begin{subfigure}{.24\linewidth}
\centering
\includegraphics[width=\linewidth]{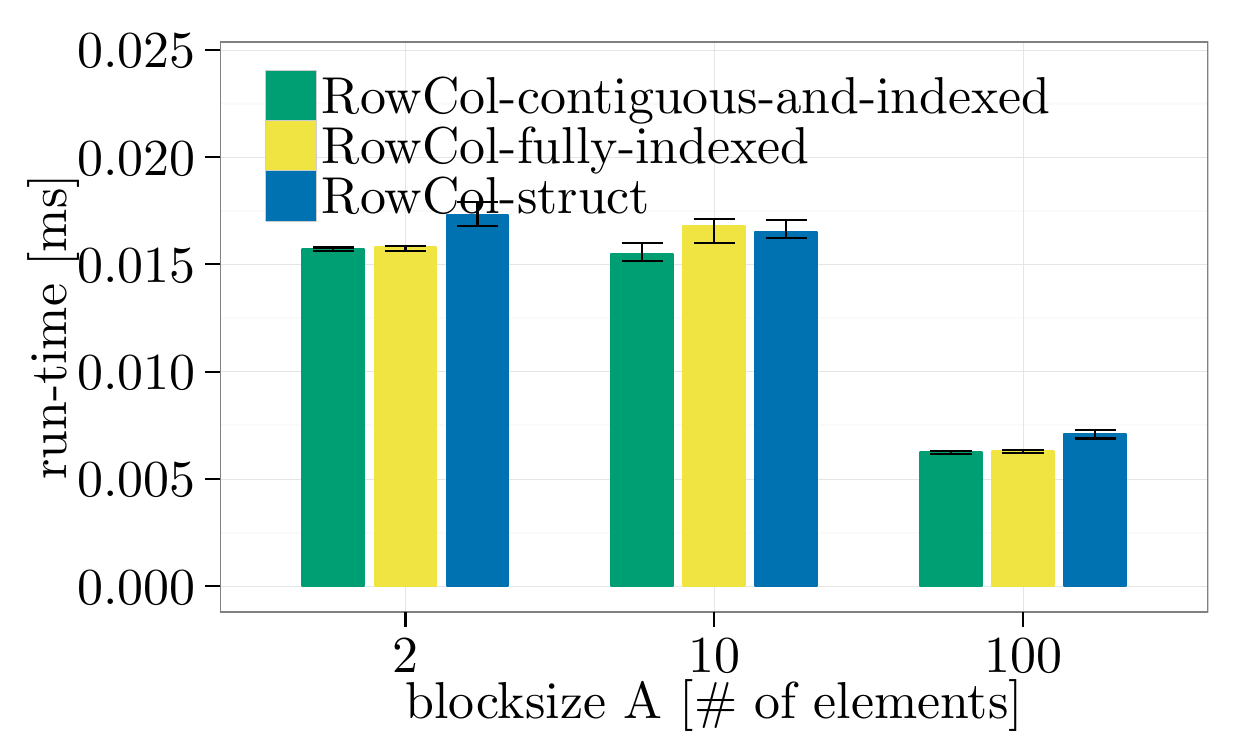}
\caption{%
\label{exp:pingpong-rowcol-small-2x1-mvapich}%
$n=\num{100}$, \num{2}~nodes%
}%
\end{subfigure}%
\hfill%
\begin{subfigure}{.24\linewidth}
\centering
\includegraphics[width=\linewidth]{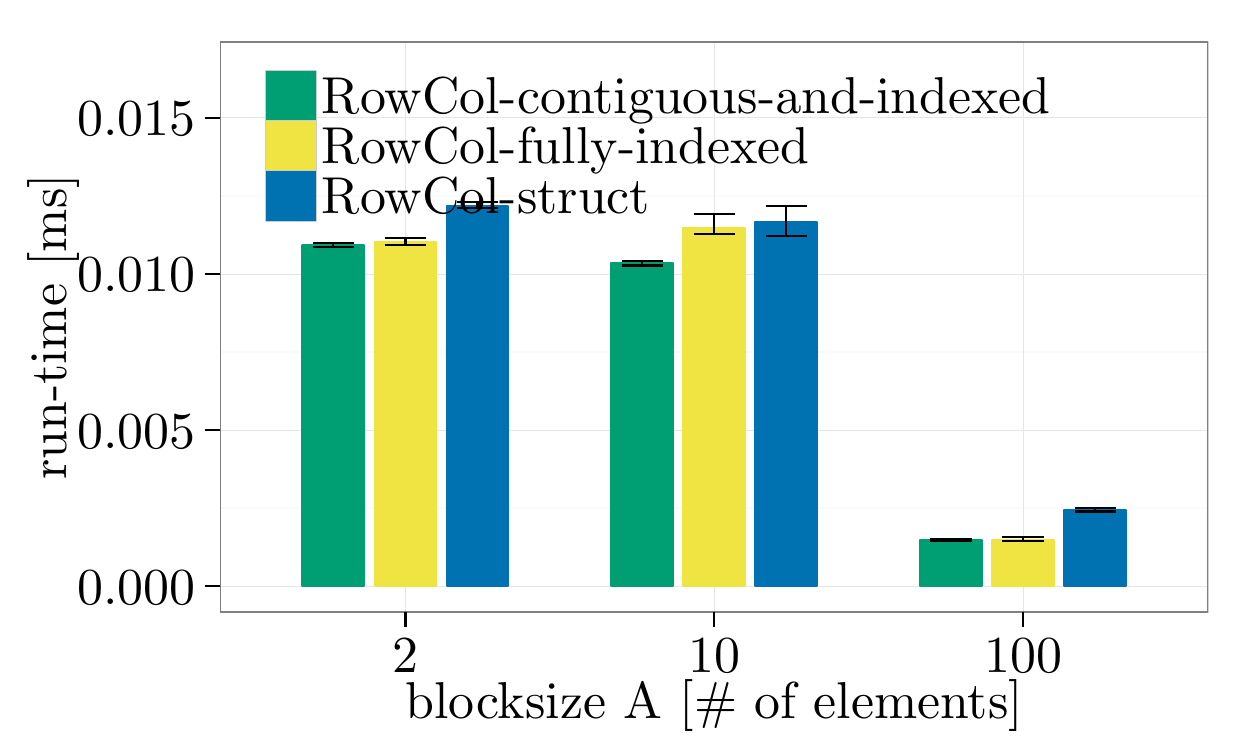}
\caption{%
\label{exp:pingpong-rowcol-small-1x2-mvapich}%
$n=\num{100}$, same node%
}%
\end{subfigure}%
\hfill%
\begin{subfigure}{.24\linewidth}
\centering
\includegraphics[width=\linewidth]{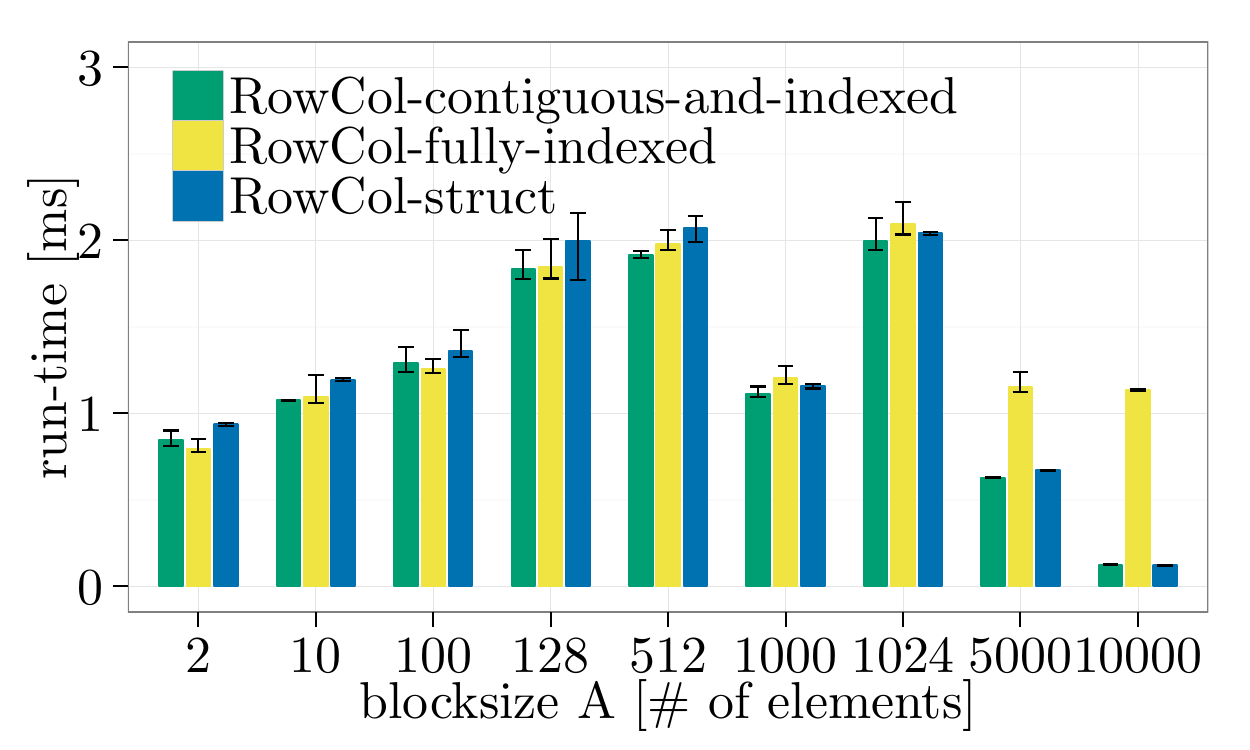}
\caption{%
\label{exp:pingpong-rowcol-large-2x1-mvapich}%
 $n=\num{10240}$, \num{2}~nodes%
}%
\end{subfigure}%
\hfill%
\begin{subfigure}{.24\linewidth}
\centering
\includegraphics[width=\linewidth]{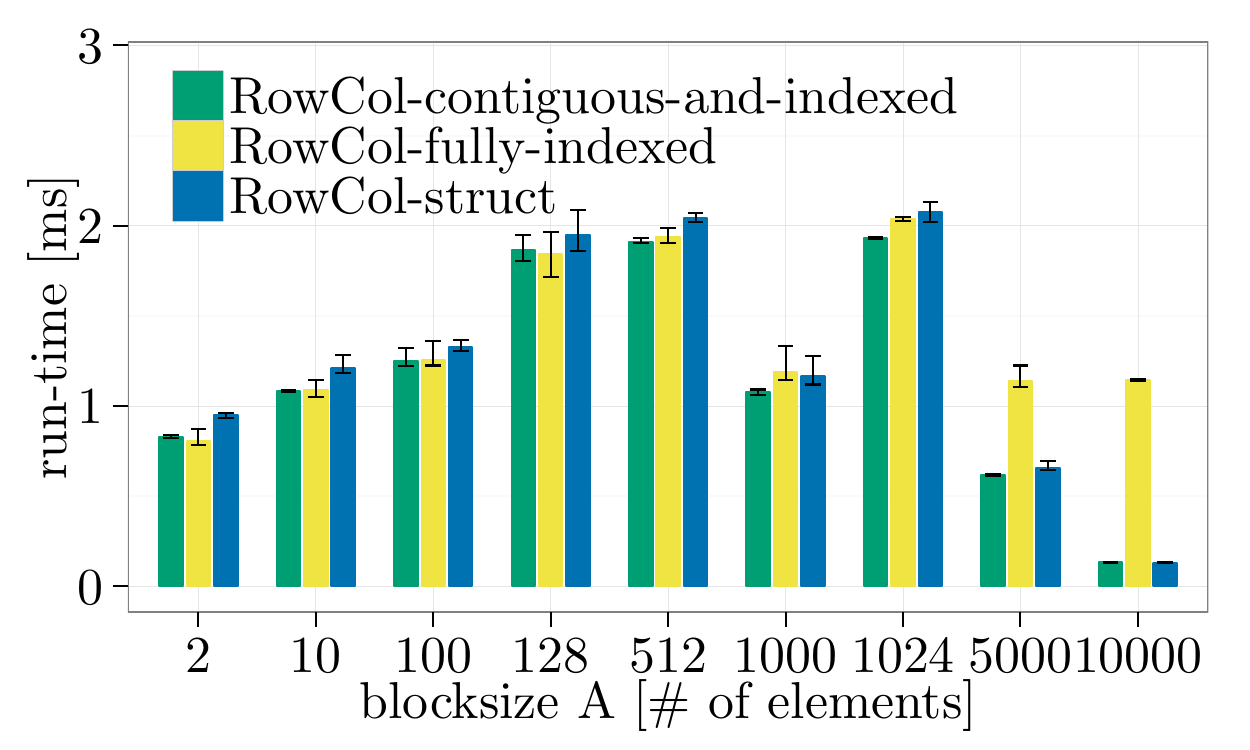}
\caption{%
\label{exp:pingpong-rowcol-large-1x2-mvapich}%
$n=\num{10240}$, same node%
}%
\end{subfigure}%
\caption{\label{exp:pingpong-rowcol-mvapich} \ddtrowcolfullindexed, \ddtrowcolcontiguousandindexed, \ddtrowcolstruct, element datatype: \mpiint, buffer size (and \extent) increases with \blocksize \VARblocksize, \pingpong, \jupitermvapich.}
\end{figure*}

\begin{figure*}[htpb]
\centering
\begin{subfigure}{.24\linewidth}
\centering
\includegraphics[width=\linewidth]{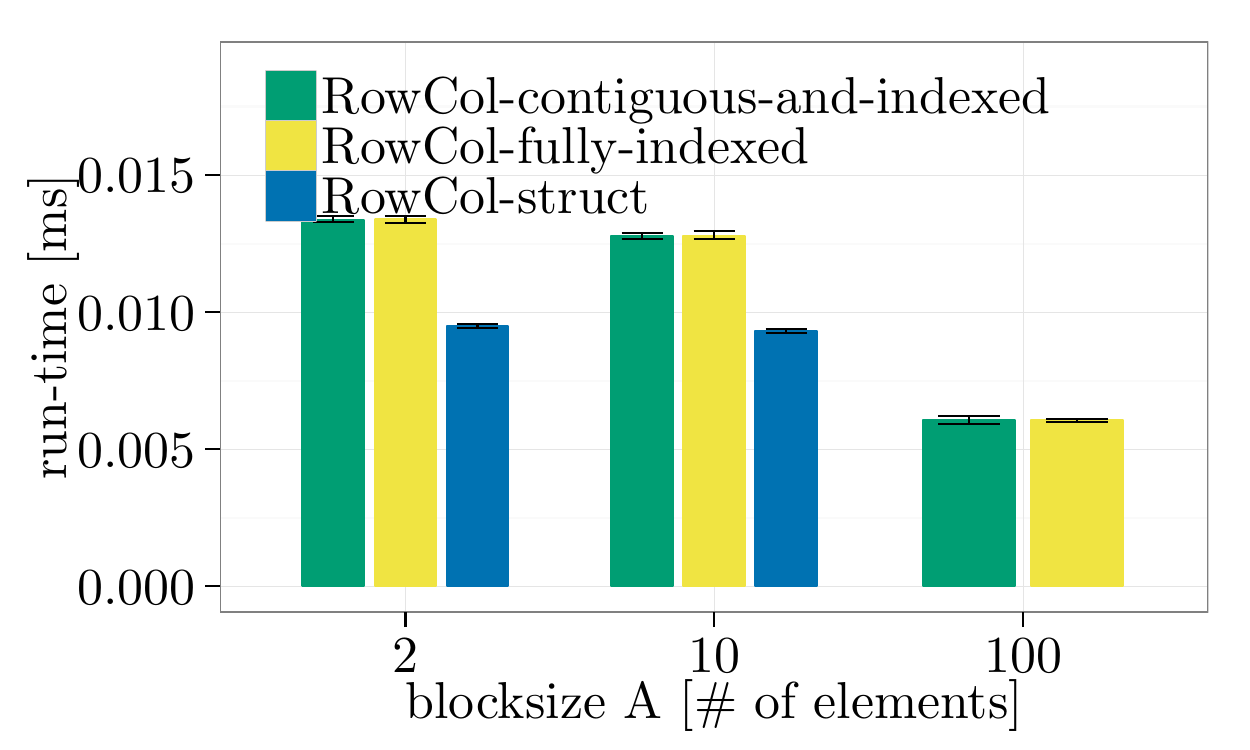}
\caption{%
\label{exp:pingpong-rowcol-small-2x1-openmpi}%
$n=\num{100}$, \num{2}~nodes%
}%
\end{subfigure}%
\hfill%
\begin{subfigure}{.24\linewidth}
\centering
\includegraphics[width=\linewidth]{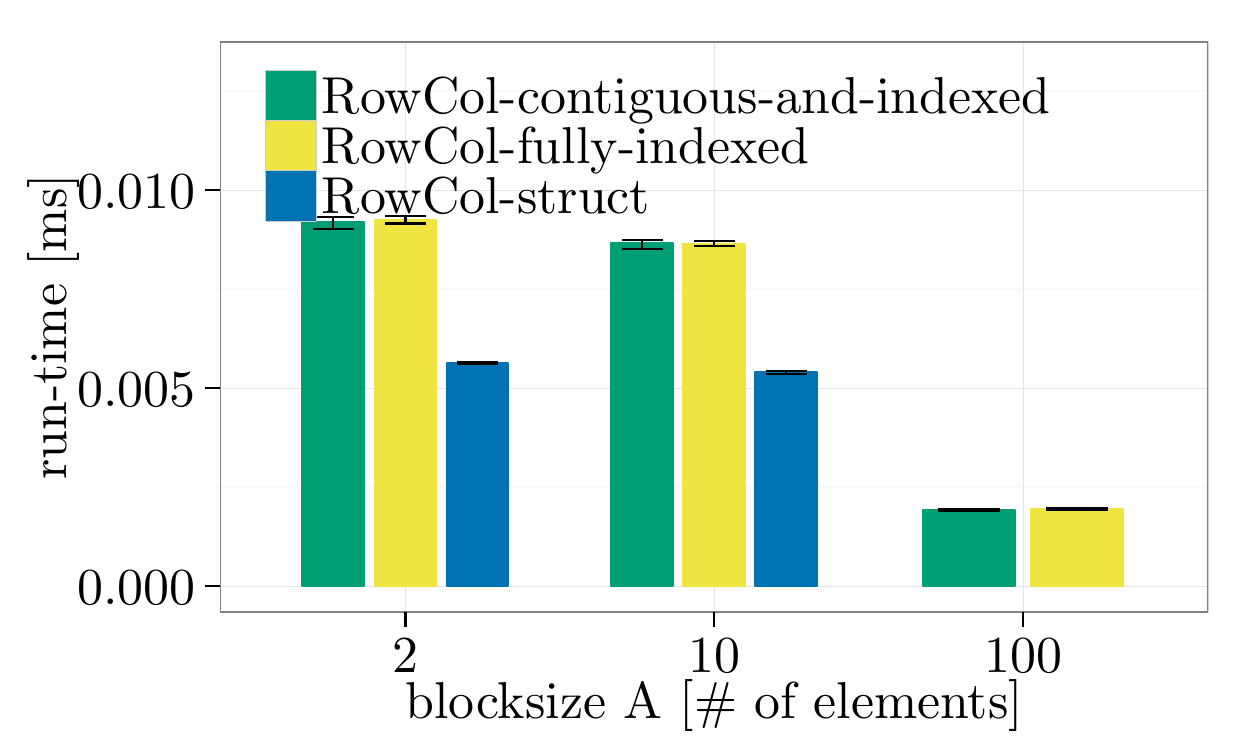}
\caption{%
\label{exp:pingpong-rowcol-small-1x2-openmpi}%
$n=\num{100}$, same node%
}%
\end{subfigure}%
\hfill%
\begin{subfigure}{.24\linewidth}
\centering
\includegraphics[width=\linewidth]{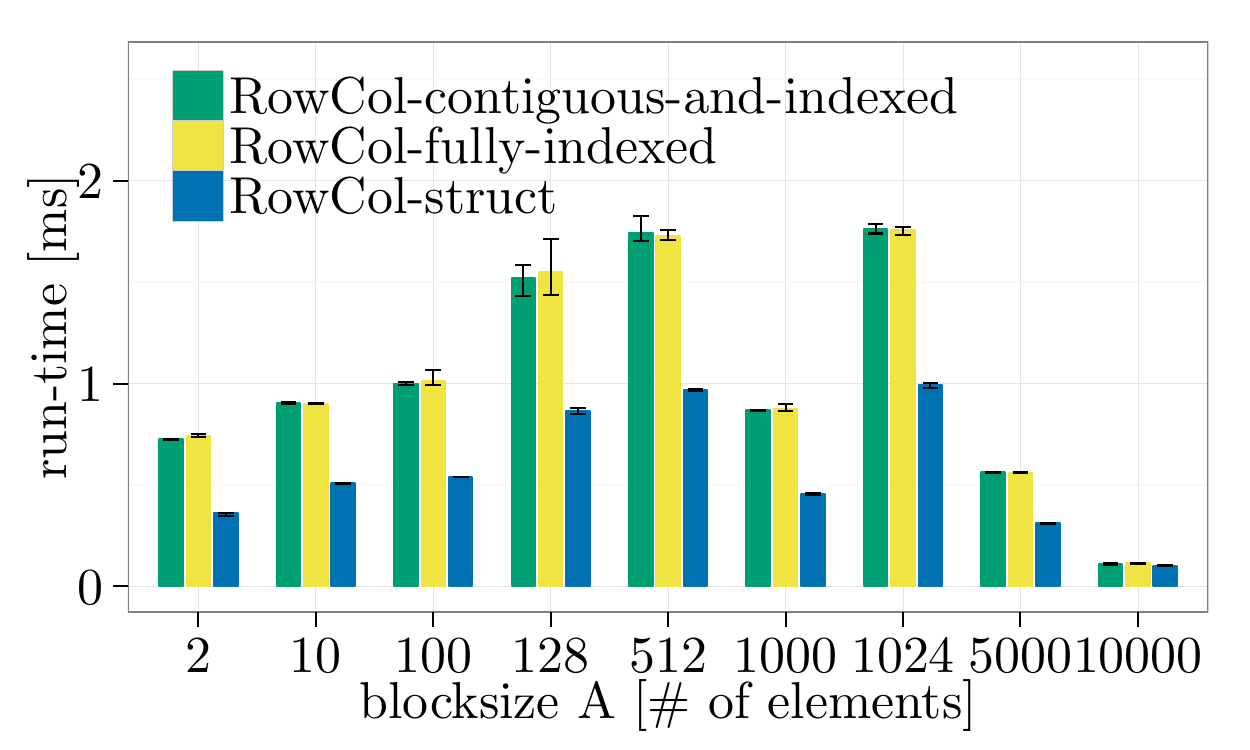}
\caption{%
\label{exp:pingpong-rowcol-large-2x1-openmpi}%
 $n=\num{10240}$, \num{2}~nodes%
}%
\end{subfigure}%
\hfill%
\begin{subfigure}{.24\linewidth}
\centering
\includegraphics[width=\linewidth]{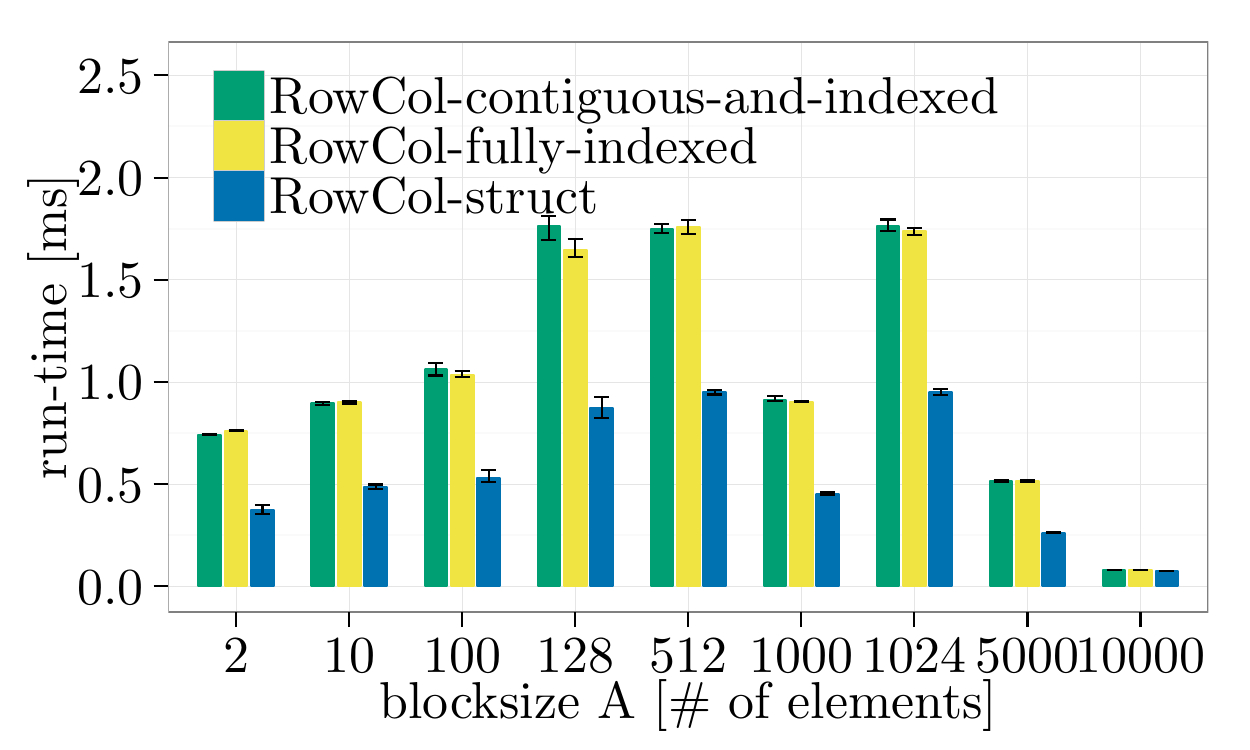}
\caption{%
\label{exp:pingpong-rowcol-large-1x2-openmpi}%
$n=\num{10240}$, same node%
}%
\end{subfigure}%
\caption{\label{exp:pingpong-rowcol-openmpi} \ddtrowcolfullindexed, \ddtrowcolcontiguousandindexed, \ddtrowcolstruct, element datatype: \mpiint, buffer size (and \extent) increases with \blocksize \VARblocksize, \pingpong, \jupiteropenmpi.}
\end{figure*}

\appexp{exptest:rowcol}
\appexpdesc{
  \begin{expitemize}
    \item \ddtrowcolfullindexed, \ddtrowcolcontiguousandindexed, \ddtrowcolstruct
    \item \pingpong
  \end{expitemize}
}{
  \begin{expitemize}
    \item \expparam{\jupiternecmpi}{\fig~\ref{exp:pingpong-rowcol-nec}}
    \item \expparam{\jupitermvapich}{\fig~\ref{exp:pingpong-rowcol-mvapich}}
    \item \expparam{\jupiteropenmpi}{\fig~\ref{exp:pingpong-rowcol-openmpi}}
  \end{expitemize}  
}

\FloatBarrier
\clearpage

\section{Experimental Results on VSC-3}

The experiments in this appendix have been performed using the hardware and software 
setup described below:

\begin{center}
  \begin{scriptsize}
  \begin{tabular}{l@{\hskip .1in}l}
    \toprule
    machine  & \num{2000} $\times$ Dual Intel Xeon E5-2650v2 @ \SI{2.6}{\giga\hertz}  \\
             &  \infiniband QDR-80   \\ 
    machine name & \machtwo \\
    \midrule 
    MPI libraries &  \vscintelmpi \\
    Compiler & \vscintelcomp (flags \texttt{-O3})\\
    \bottomrule
  \end{tabular}
  \end{scriptsize}
\end{center}

\appexp{exptest:basic_layouts} 

\begin{figure*}[htpb]
\centering
\begin{subfigure}{.33\linewidth}
\centering
\includegraphics[width=\linewidth]{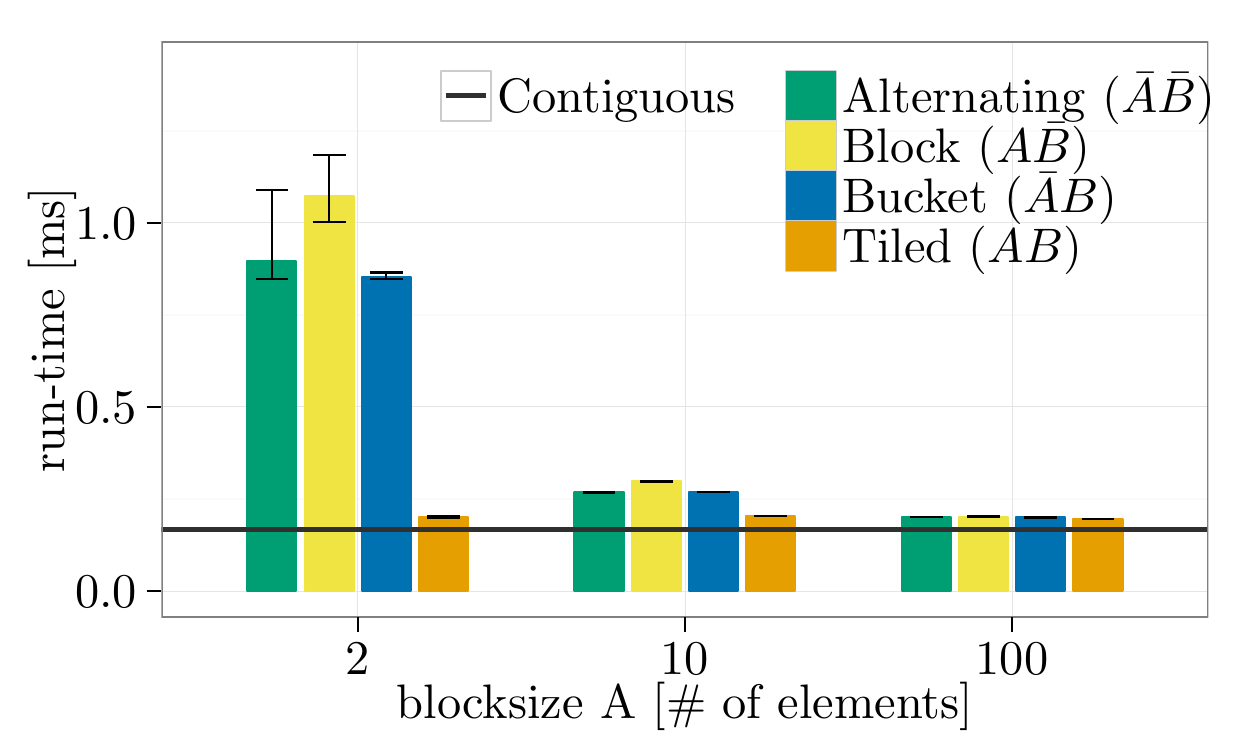}
\caption{%
\label{exp:vsc3-allgather-nsmall-32p}%
\mpiallgather%
}%
\end{subfigure}%
\hfill%
\begin{subfigure}{.33\linewidth}
\centering
\includegraphics[width=\linewidth]{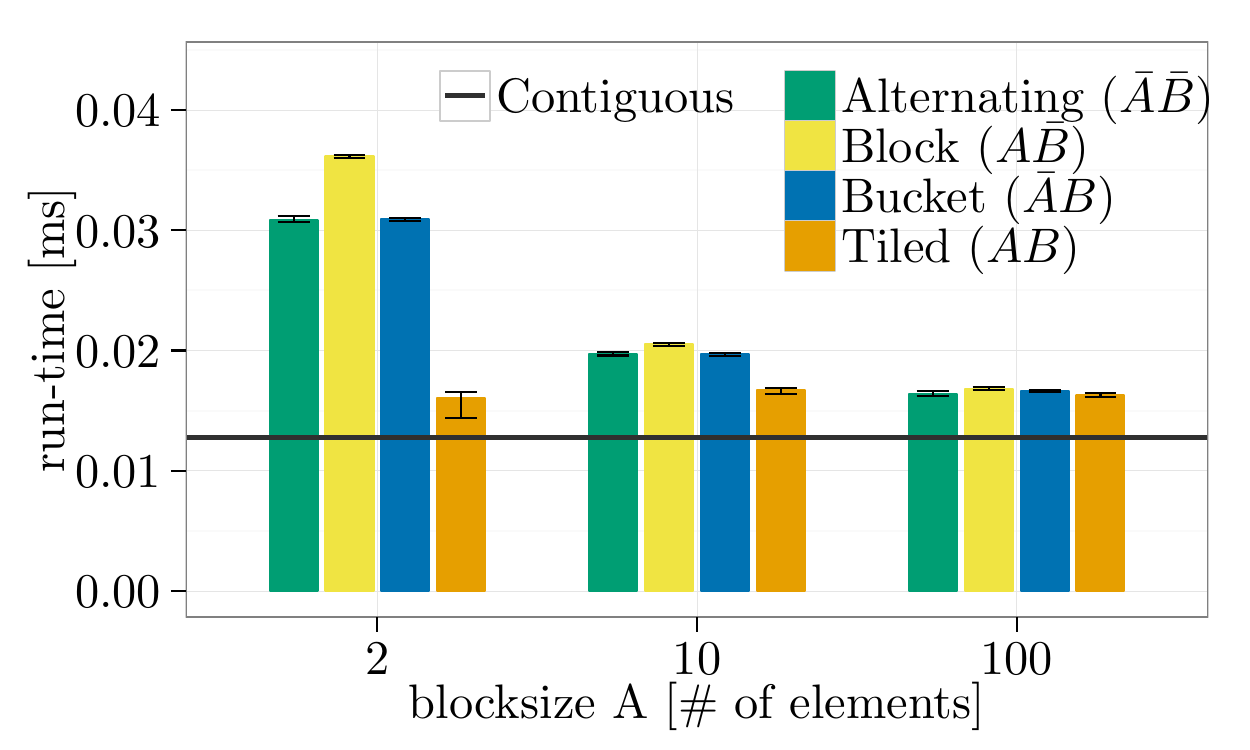}
\caption{%
\label{exp:vsc3-bcast-nsmall-32p}%
\mpibcast%
}%
\end{subfigure}%
\hfill%
\begin{subfigure}{.33\linewidth}
\centering
\includegraphics[width=\linewidth]{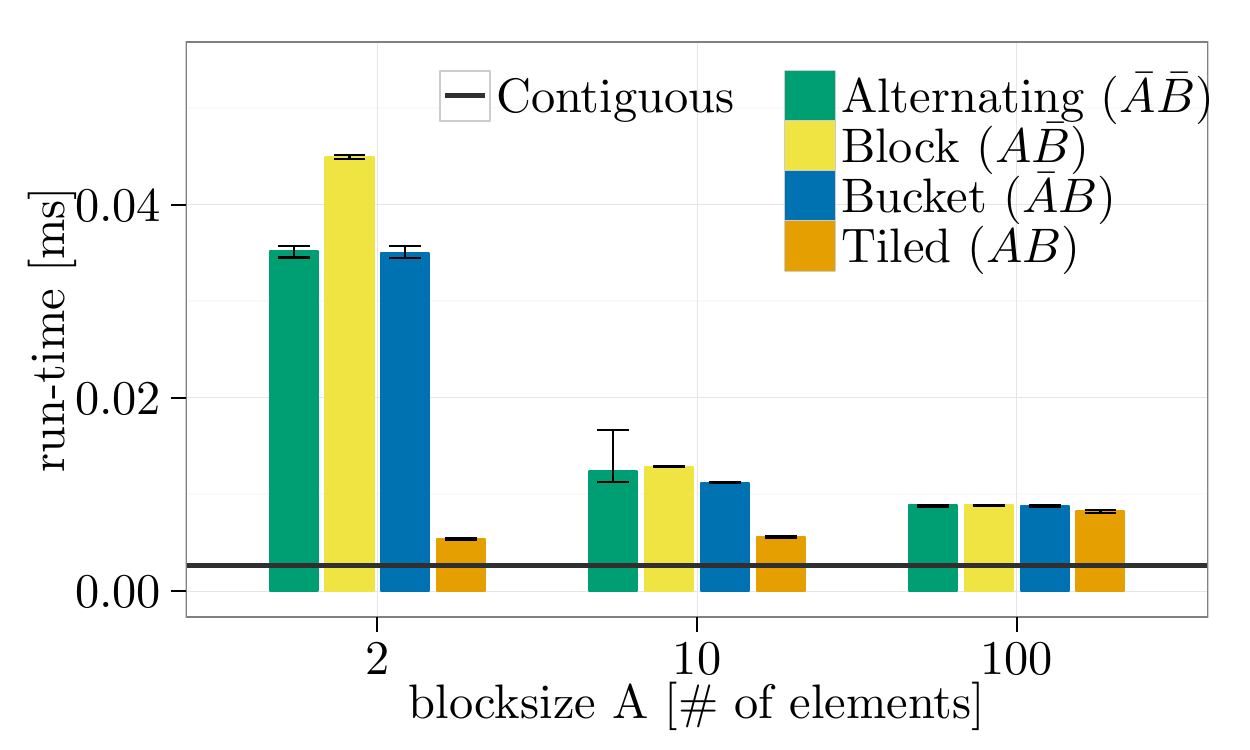}
\caption{%
\label{exp:vsc3-pingpong-nsmall-32p}%
\pingpong%
}%
\end{subfigure}%
\caption{\label{exp:vsc3-layouts-nsmall-32p}  Contiguous \vs typed,  $\VARdatasize=\SI{3.2}{\kilo\byte}$, element datatype: \mpiint, \num{32x1}~processes (\num{2x1} for \pingpong), \vscintelmpi, \variantone.}
\end{figure*}

\begin{figure*}[htpb]
\centering
\begin{subfigure}{.33\linewidth}
\centering
\includegraphics[width=\linewidth]{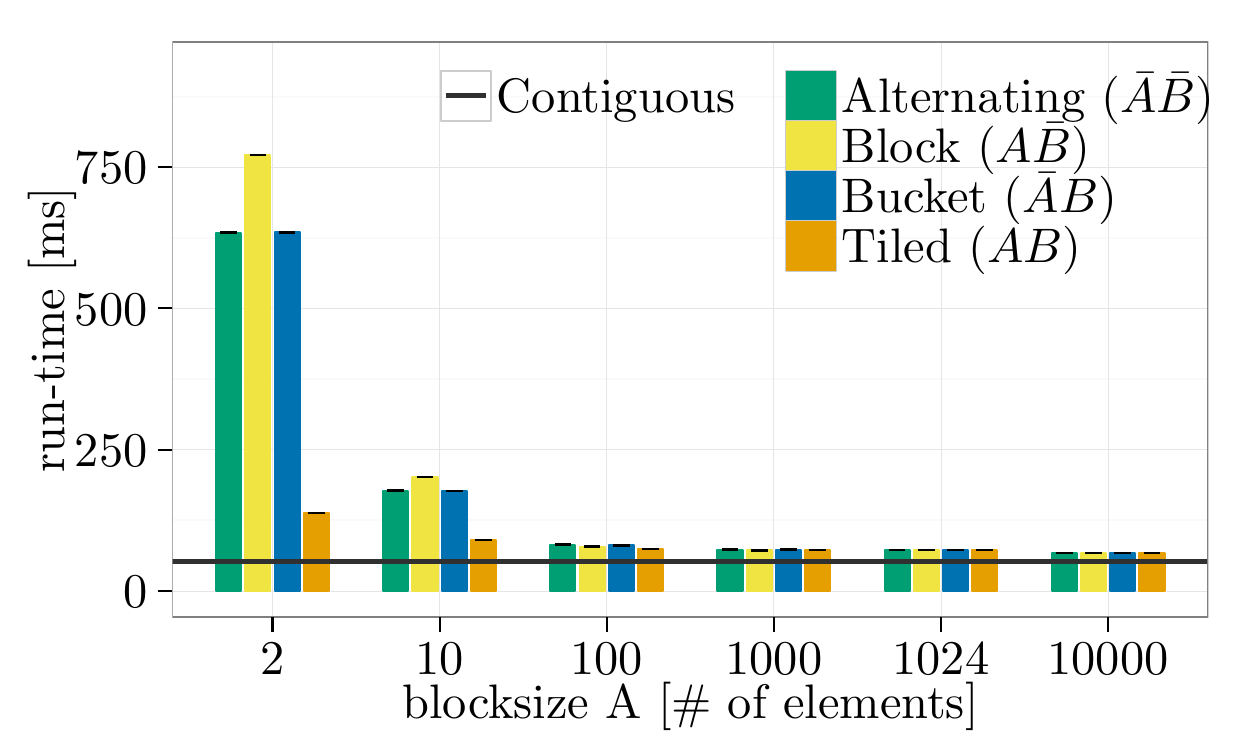}
\caption{%
\label{exp:vsc3-allgather-nlarge-32p}%
\mpiallgather%
}%
\end{subfigure}%
\hfill%
\begin{subfigure}{.33\linewidth}
\centering
\includegraphics[width=\linewidth]{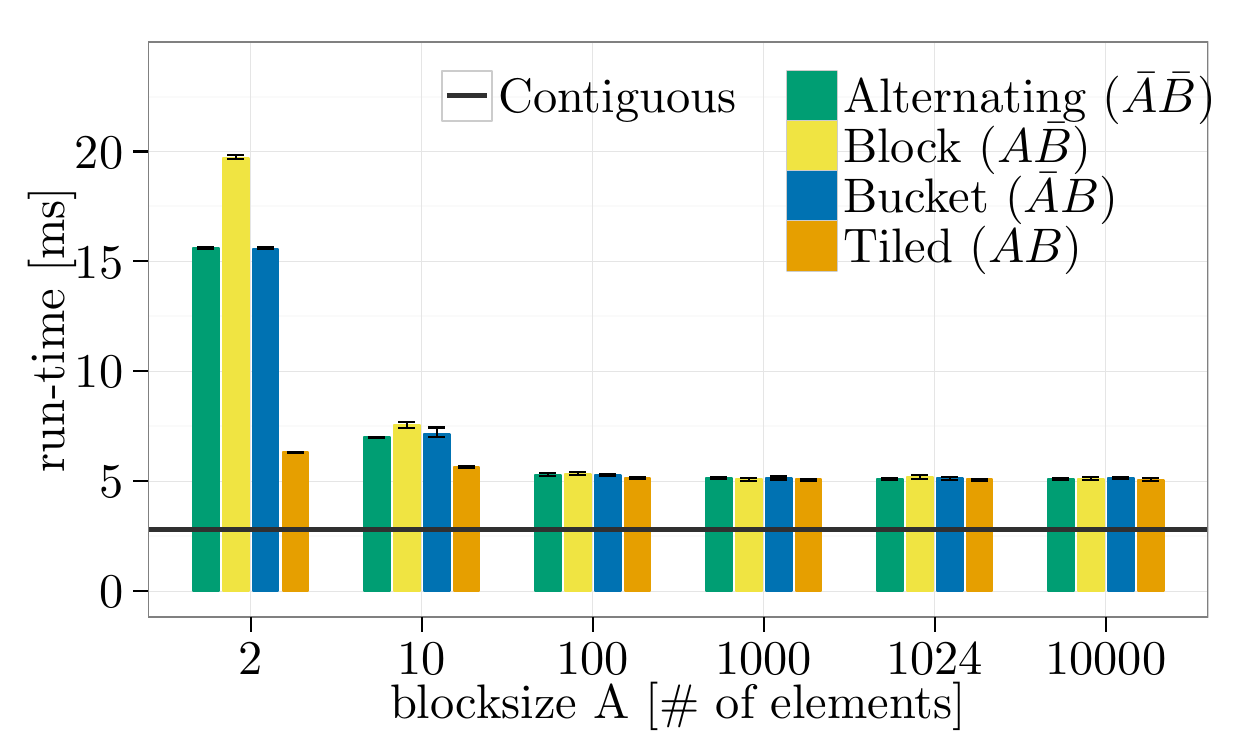}
\caption{%
\label{exp:vsc3-bcast-nlarge-32p}%
\mpibcast%
}%
\end{subfigure}%
\hfill%
\begin{subfigure}{.33\linewidth}
\centering
\includegraphics[width=\linewidth]{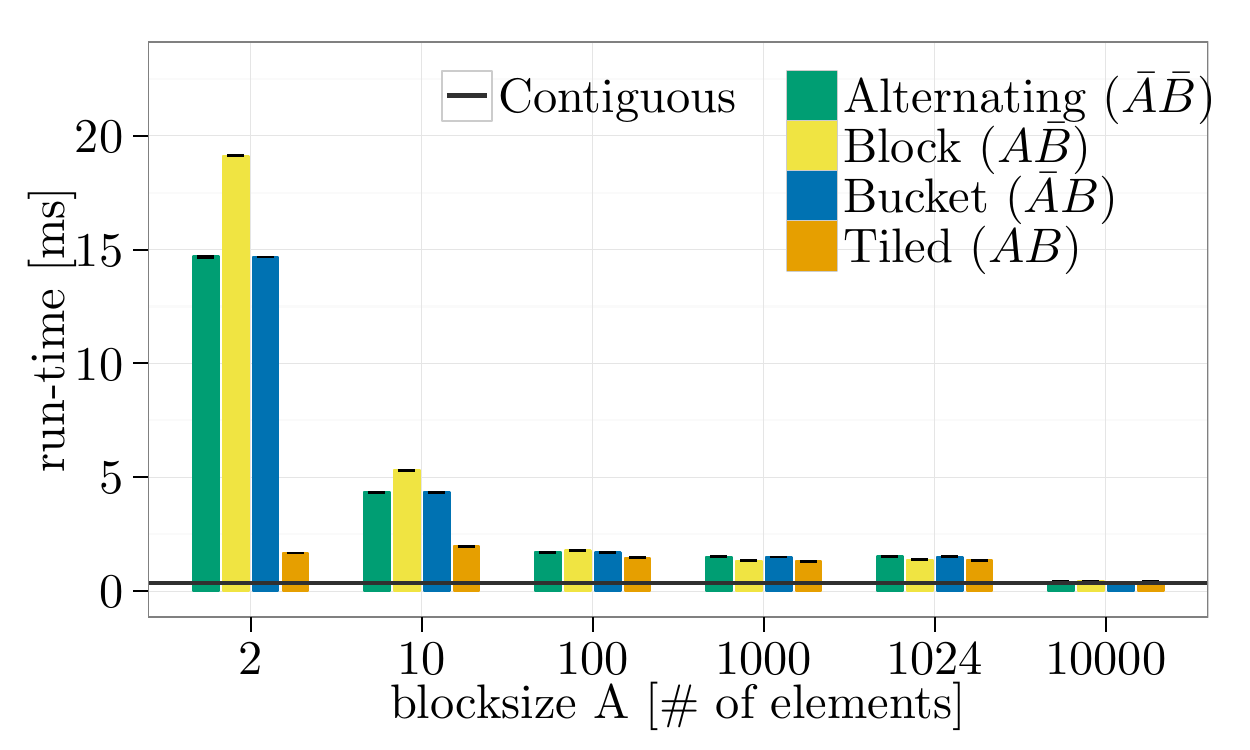}
\caption{%
\label{exp:vsc3-pingpong-nlarge-32p}%
\pingpong%
}%
\end{subfigure}%
\caption{\label{exp:vsc3-layouts-nlarge-32p}  Contiguous \vs typed,  $\VARdatasize=\SI{2.56}{\mega\byte}$, element datatype: \mpiint, \num{32x1}~processes (\num{2x1} for \pingpong), \vscintelmpi, \variantone.}
\end{figure*}

\begin{figure*}[htpb]
\centering
\begin{subfigure}{.33\linewidth}
\centering
\includegraphics[width=\linewidth]{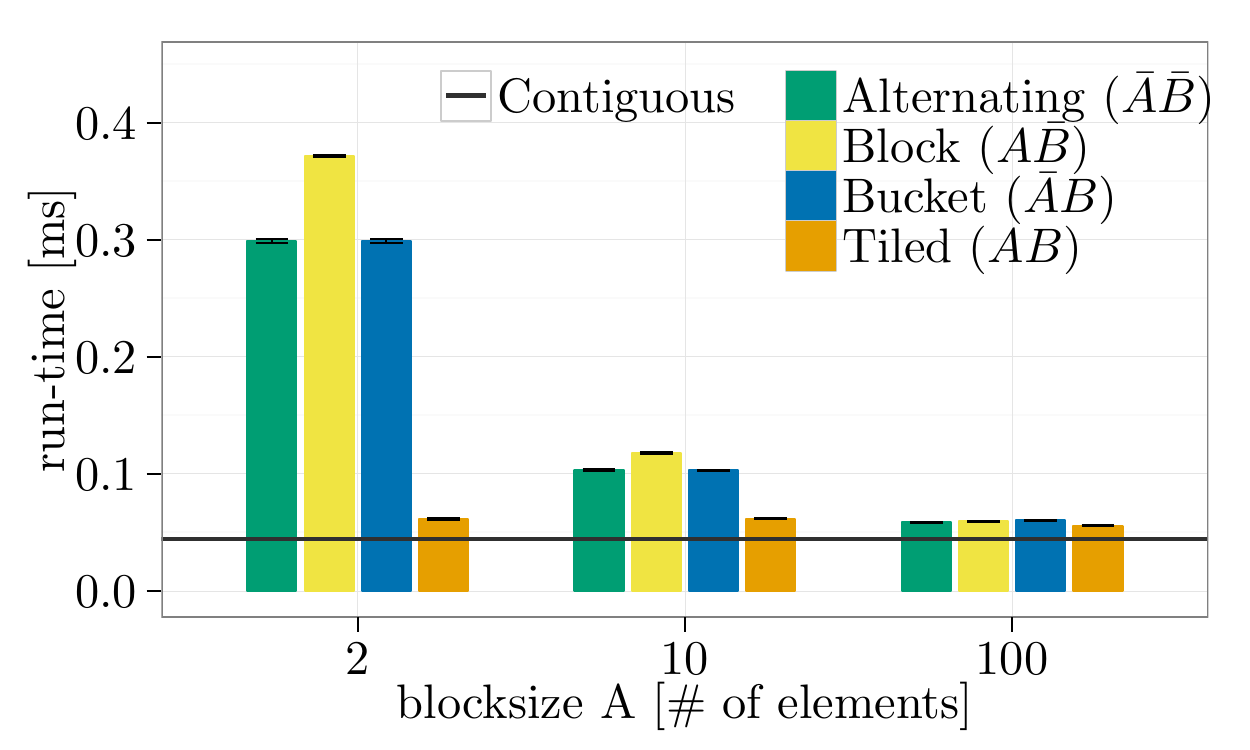}
\caption{%
\label{exp:vsc3-allgather-nsmall-onenode}%
\mpiallgather%
}%
\end{subfigure}%
\hfill%
\begin{subfigure}{.33\linewidth}
\centering
\includegraphics[width=\linewidth]{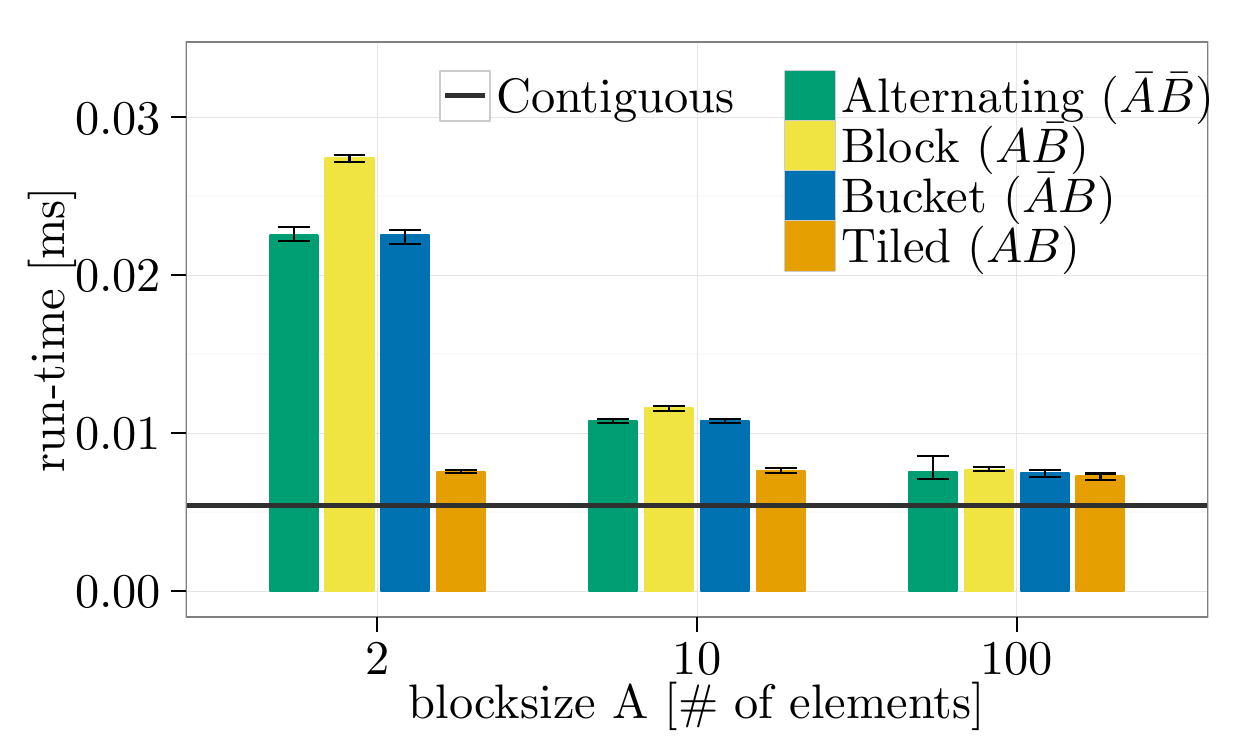}
\caption{%
\label{exp:vsc3-bcast-nsmall-onenode}%
\mpibcast%
}%
\end{subfigure}%
\hfill%
\begin{subfigure}{.33\linewidth}
\centering
\includegraphics[width=\linewidth]{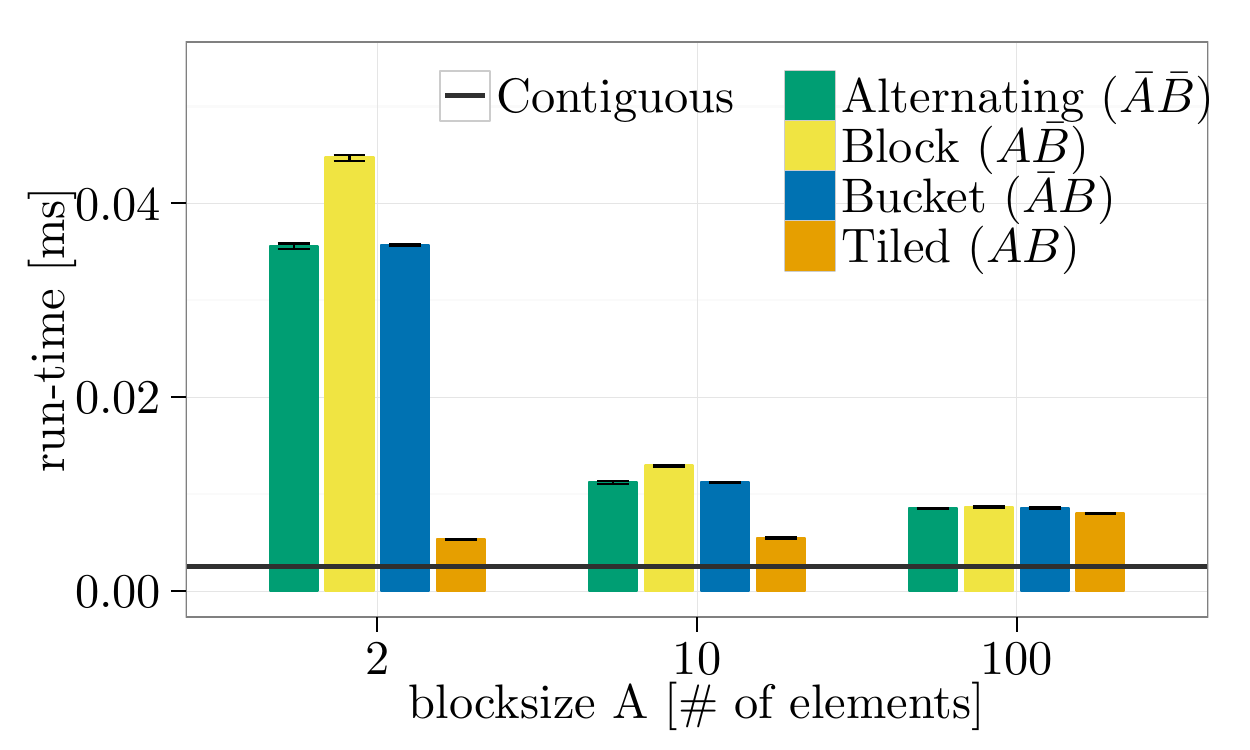}
\caption{%
\label{exp:vsc3-pingpong-nsmall-onenode}%
\pingpong%
}%
\end{subfigure}%
\caption{\label{exp:vsc3-layouts-nsmall-onenode}  Contiguous \vs typed, $\VARdatasize=\SI{3.2}{\kilo\byte}$, element datatype: \mpiint, one node, \num{16}~processes (\num{2} for \pingpong), \vscintelmpi, \variantone.}
\end{figure*}

\begin{figure*}[htpb]
\centering
\begin{subfigure}{.33\linewidth}
\centering
\includegraphics[width=\linewidth]{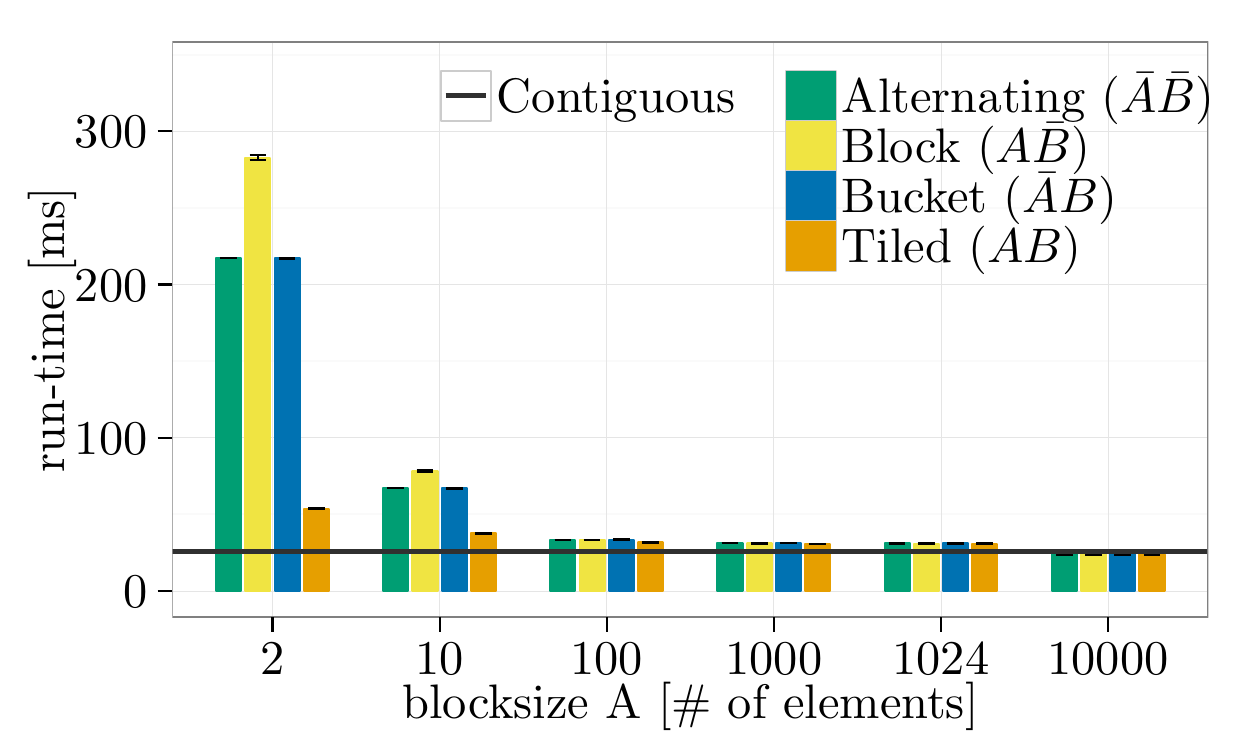}
\caption{%
\label{exp:vsc3-allgather-nlarge-onenode}%
\mpiallgather%
}%
\end{subfigure}%
\hfill%
\begin{subfigure}{.33\linewidth}
\centering
\includegraphics[width=\linewidth]{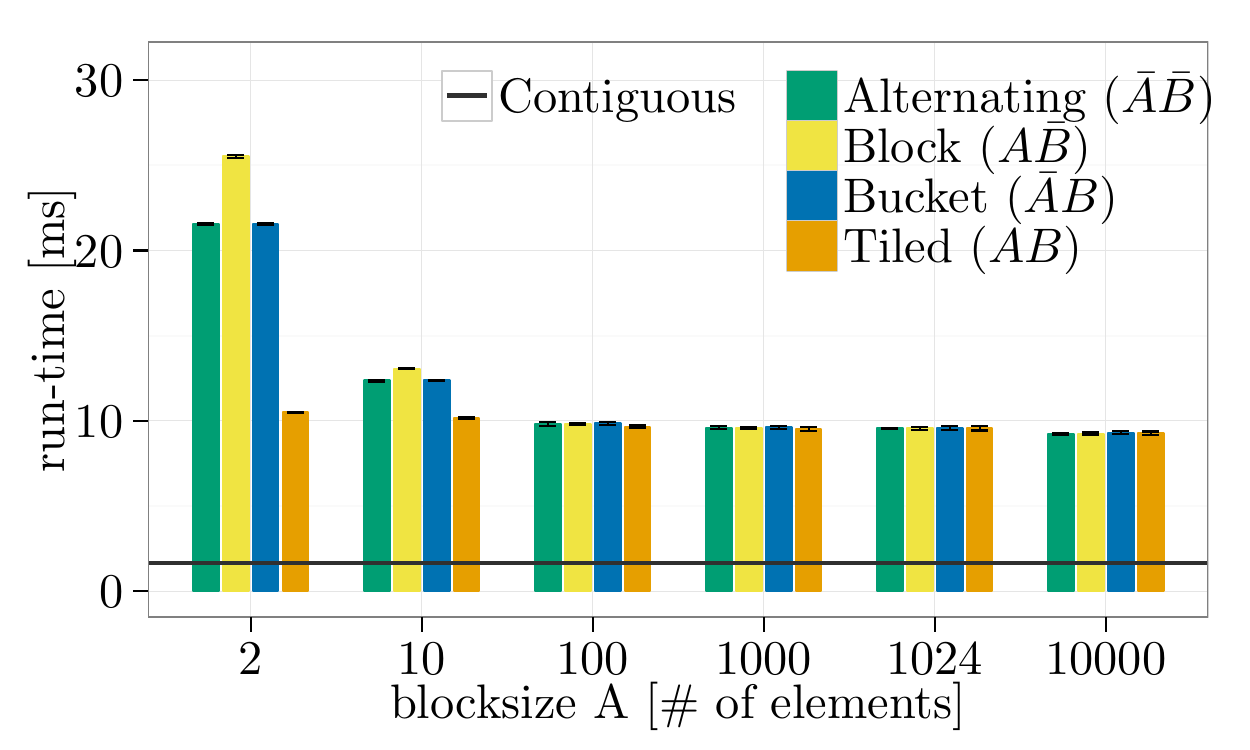}
\caption{%
\label{exp:vsc3-bcast-nlarge-onenode}%
\mpibcast%
}%
\end{subfigure}%
\hfill%
\begin{subfigure}{.33\linewidth}
\centering
\includegraphics[width=\linewidth]{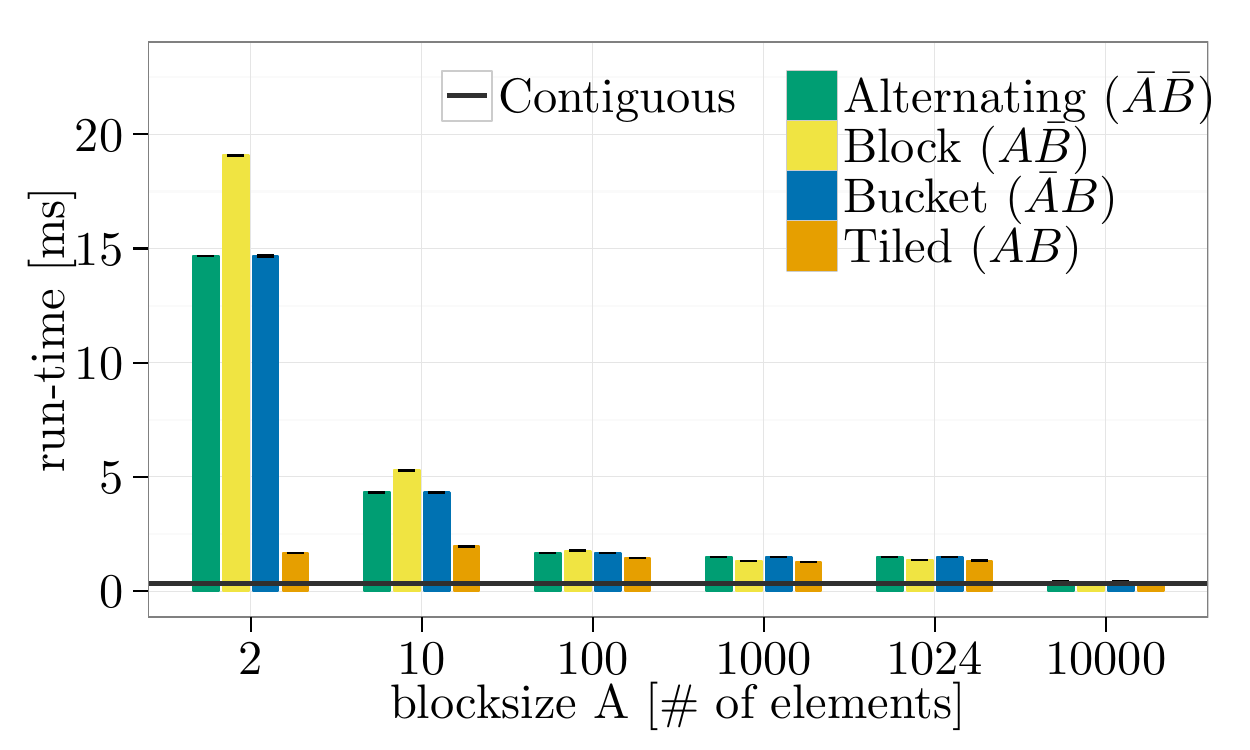}
\caption{%
\label{exp:vsc3-pingpong-nlarge-onenode}%
\pingpong%
}%
\end{subfigure}%
\caption{\label{exp:vsc3-layouts-nlarge-onenode}  Contiguous \vs typed,  $\VARdatasize=\SI{2.56}{\mega\byte}$, element datatype: \mpiint, one~node, \num{16}~processes (\num{2}~processes for \pingpong), \vscintelmpi, \variantone.}
\end{figure*}

\begin{figure*}[htpb]
\centering
\begin{subfigure}{.33\linewidth}
\centering
\includegraphics[width=\linewidth]{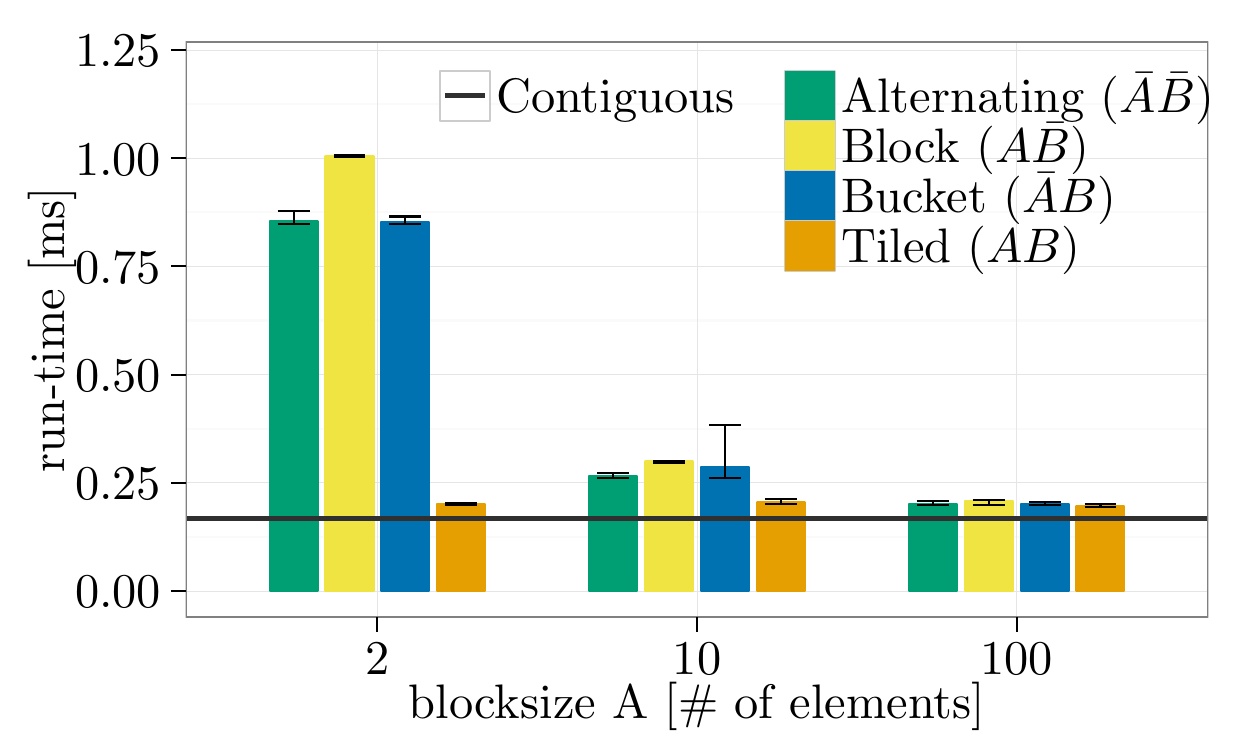}
\caption{%
\label{exp:vsc3-allgather-nsmall-vartwo}%
\mpiallgather%
}%
\end{subfigure}%
\hfill%
\begin{subfigure}{.33\linewidth}
\centering
\includegraphics[width=\linewidth]{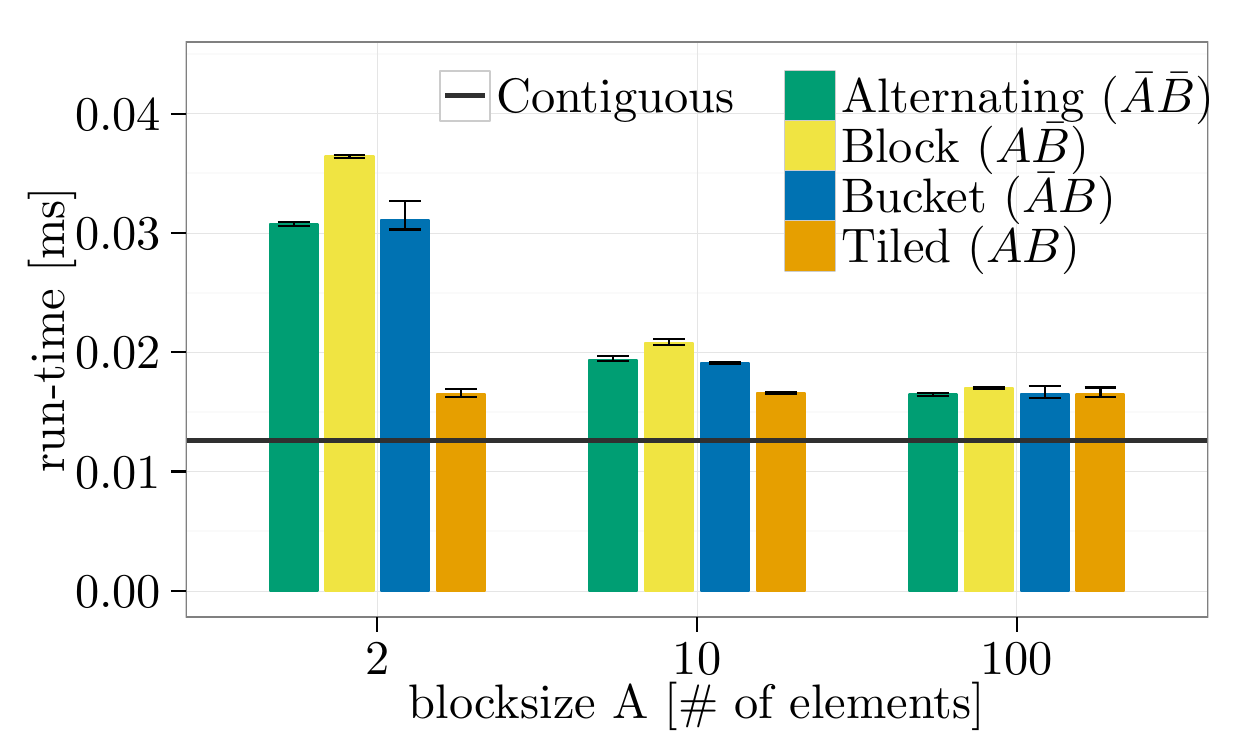}
\caption{%
\label{exp:vsc3-bcast-nsmall-vartwo}%
\mpibcast%
}%
\end{subfigure}%
\hfill%
\begin{subfigure}{.33\linewidth}
\centering
\includegraphics[width=\linewidth]{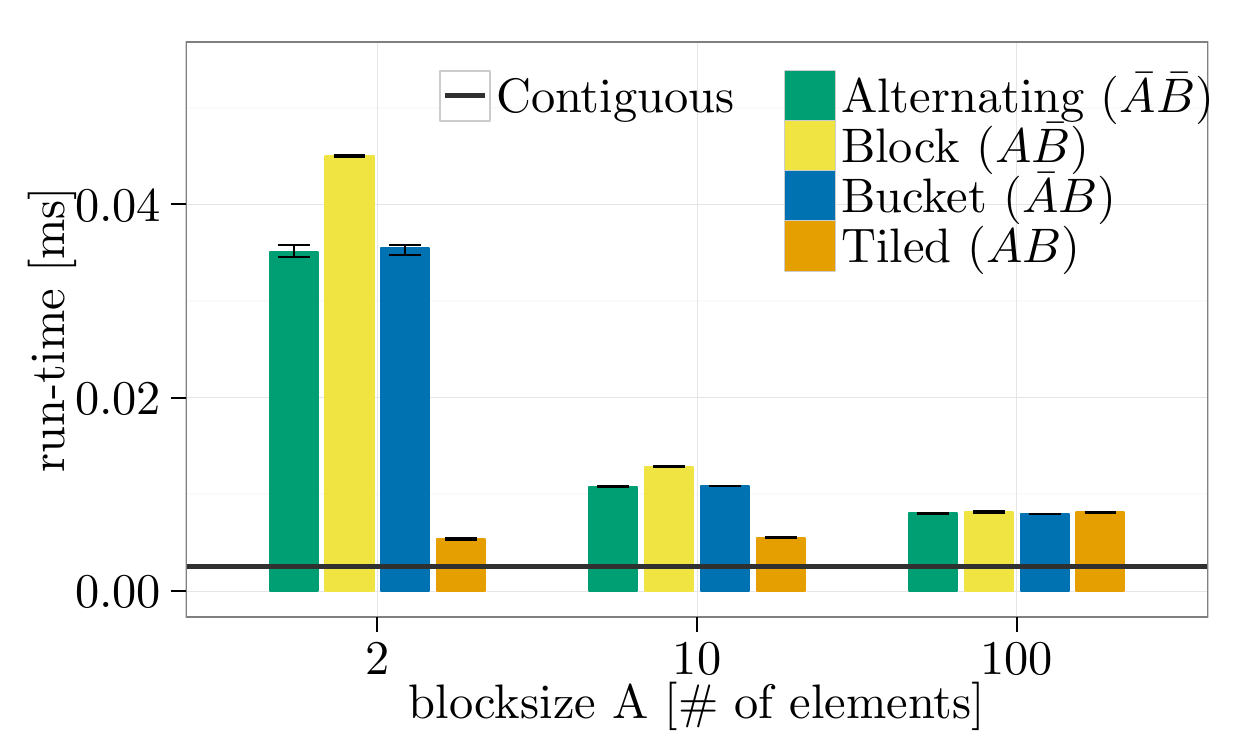}
\caption{%
\label{exp:vsc3-pingpong-nsmall-vartwo}%
\pingpong%
}%
\end{subfigure}%
\caption{\label{exp:vsc3-layouts-small-32p-vartwo}  Contiguous \vs typed, $\VARdatasize=\SI{3.2}{\kilo\byte}$, element datatype: \mpiint, \num{32x1}~processes (\num{2x1} for \pingpong), \vscintelmpi, \varianttwo.}
\end{figure*}

\begin{figure*}[htpb]
\centering
\begin{subfigure}{.33\linewidth}
\centering
\includegraphics[width=\linewidth]{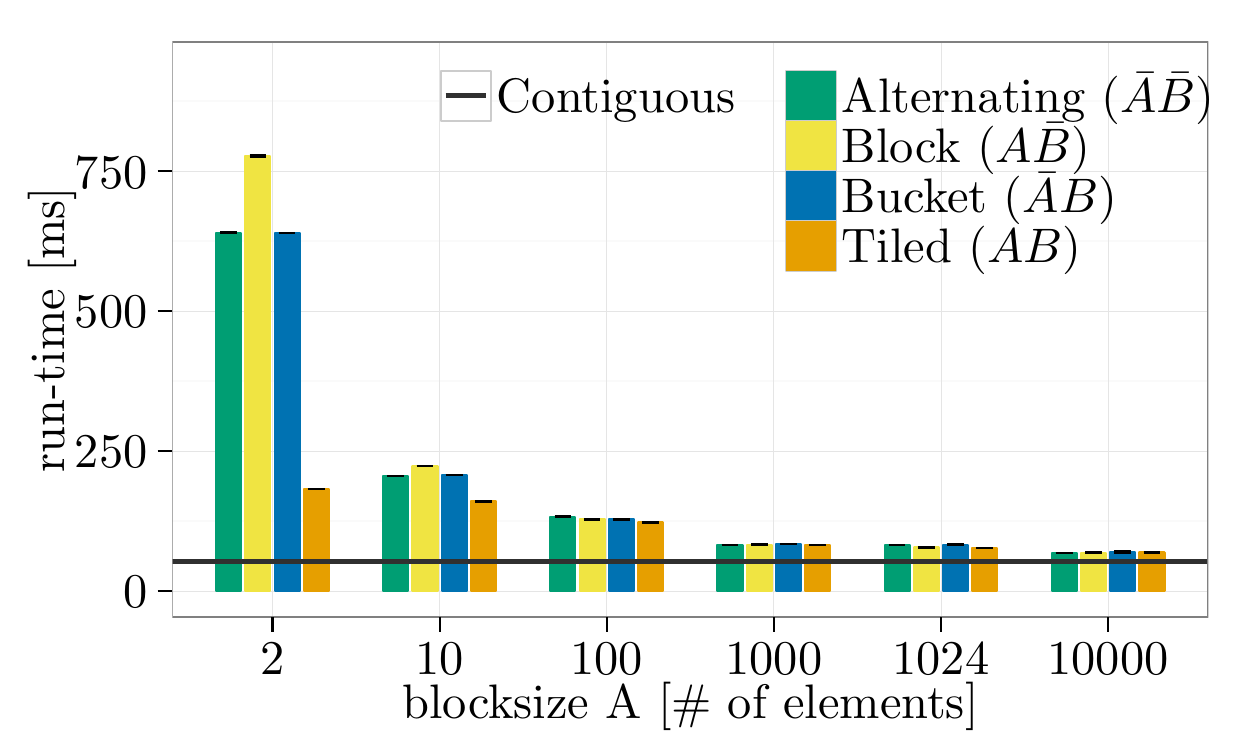}
\caption{%
\label{exp:vsc3-allgather-nlarge-vartwo}%
\mpiallgather%
}%
\end{subfigure}%
\hfill%
\begin{subfigure}{.33\linewidth}
\centering
\includegraphics[width=\linewidth]{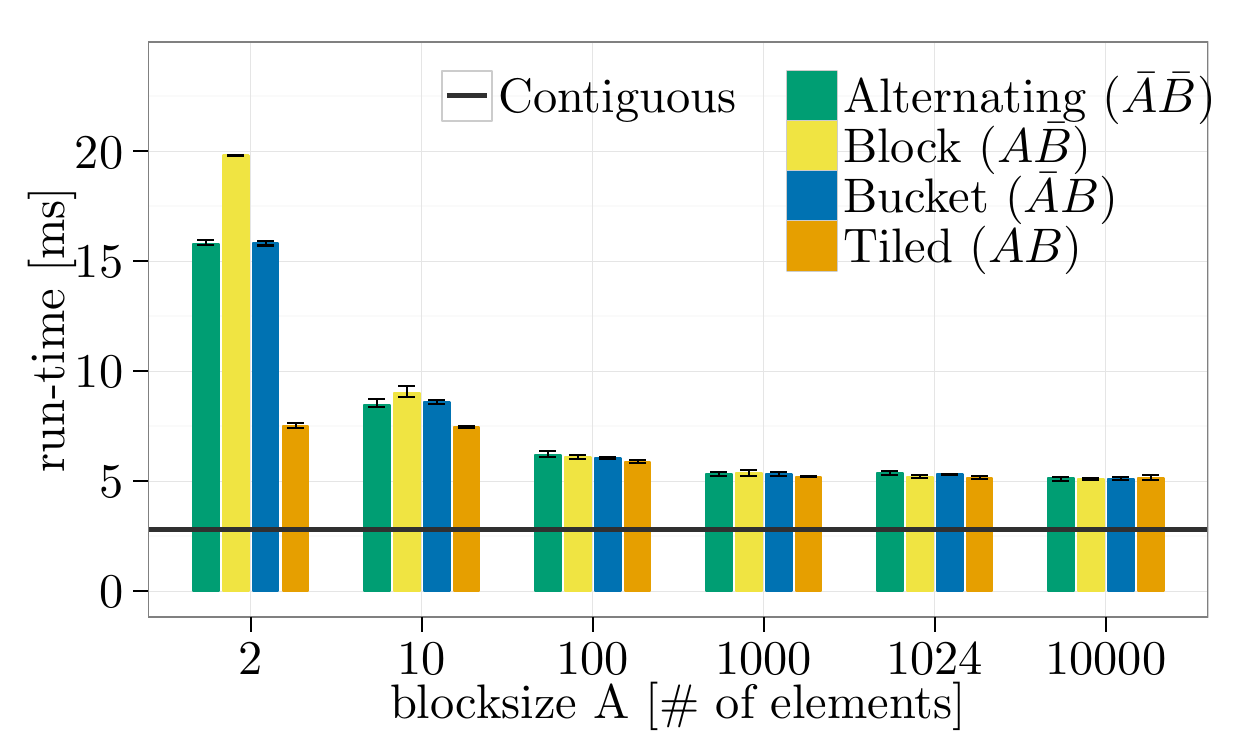}
\caption{%
\label{exp:vsc3-bcast-nlarge-vartwo}%
\mpibcast%
}%
\end{subfigure}%
\hfill%
\begin{subfigure}{.33\linewidth}
\centering
\includegraphics[width=\linewidth]{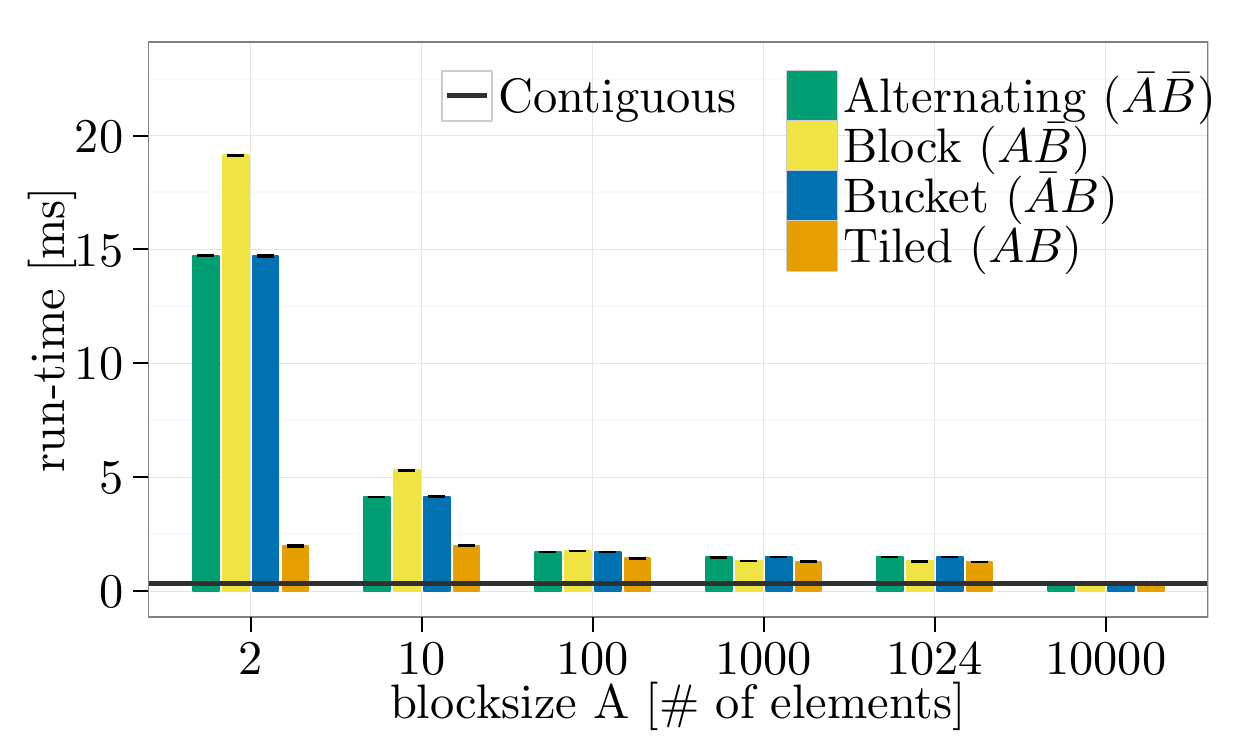}
\caption{%
\label{exp:vsc3-pingpong-nlarge-vartwo}%
\pingpong%
}%
\end{subfigure}%
\caption{\label{exp:vsc3-layouts-large-32p-vartwo} Contiguous \vs typed, $\VARdatasize=\SI{2.56}{\mega\byte}$, element datatype: \mpiint, \num{32x1}~processes (\num{2x1} for \pingpong), \vscintelmpi, \varianttwo.}
\end{figure*}

\begin{figure*}[htpb]
\centering
\begin{subfigure}{.33\linewidth}
\centering
\includegraphics[width=\linewidth]{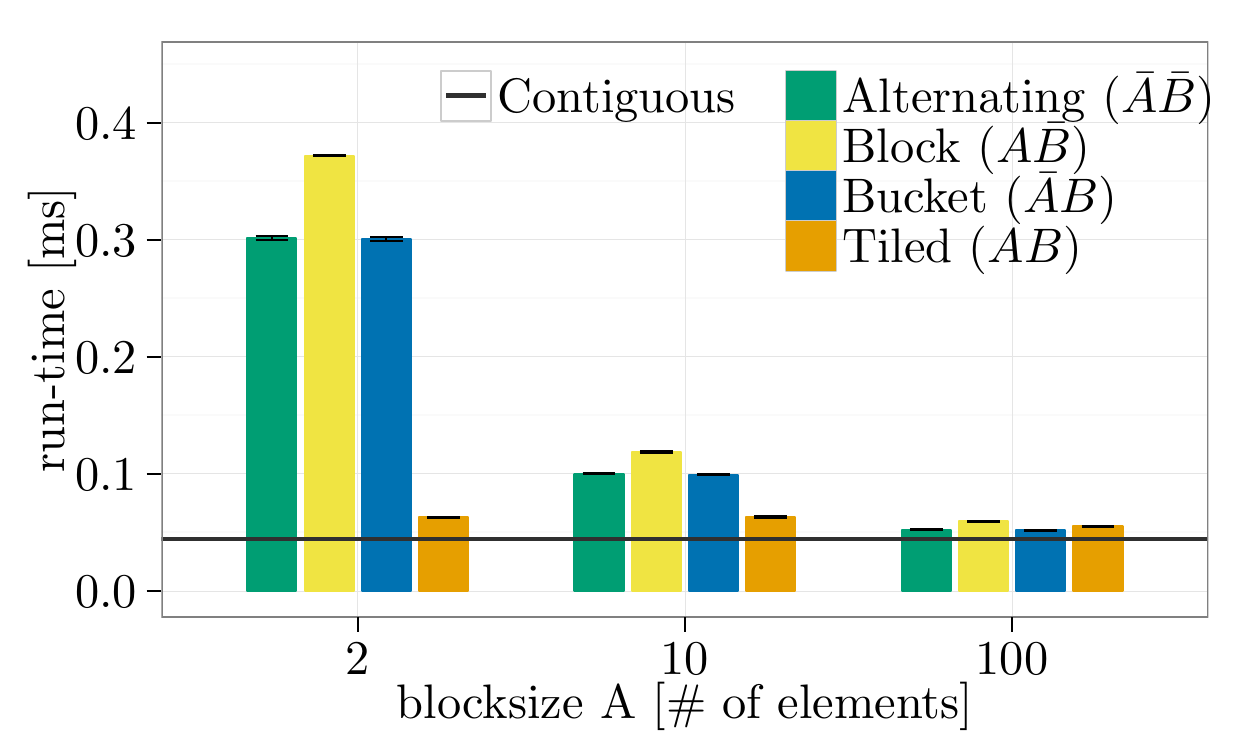}
\caption{%
\label{exp:vsc3-allgather-nsmall-onenode-vartwo}%
\mpiallgather%
}%
\end{subfigure}%
\hfill%
\begin{subfigure}{.33\linewidth}
\centering
\includegraphics[width=\linewidth]{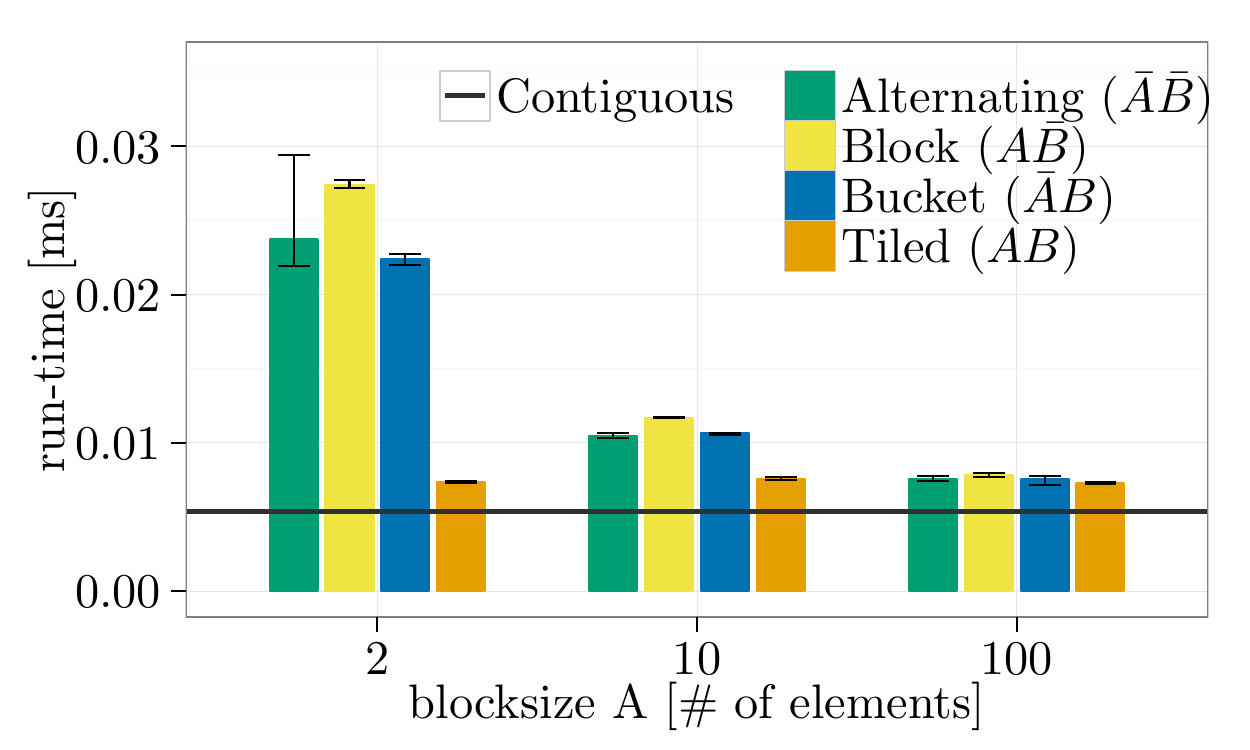}
\caption{%
\label{exp:vsc3-bcast-nsmall-onenode-vartwo}%
\mpibcast%
}%
\end{subfigure}%
\hfill%
\begin{subfigure}{.33\linewidth}
\centering
\includegraphics[width=\linewidth]{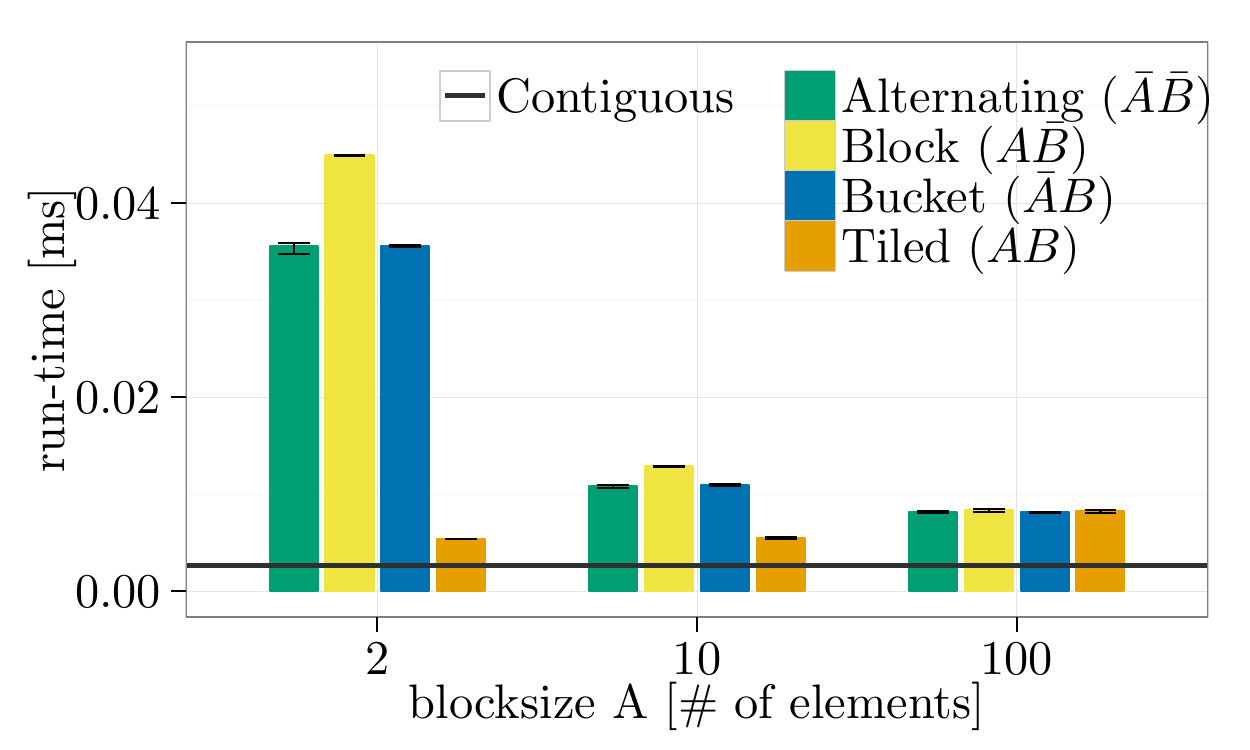}
\caption{%
\label{exp:vsc3-pingpong-nsmall-onenode-vartwo}%
\pingpong%
}%
\end{subfigure}%
\caption{\label{exp:vsc3-layouts-small-onenode-vartwo}  Contiguous \vs typed, $\VARdatasize=\SI{3.2}{\kilo\byte}$, element datatype: \mpiint, one node, \num{16}~processes (\num{2} for \pingpong), \vscintelmpi, \varianttwo.}
\end{figure*}

\begin{figure*}[htpb]
\centering
\begin{subfigure}{.33\linewidth}
\centering
\includegraphics[width=\linewidth]{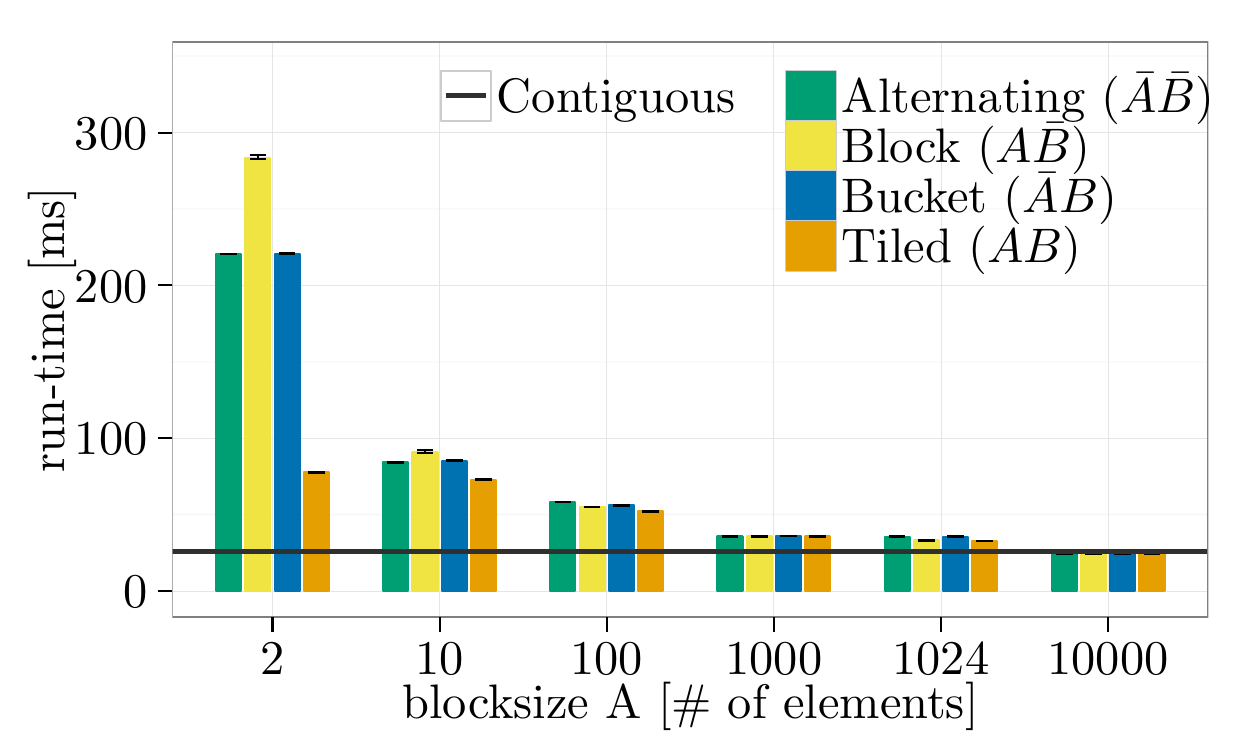}
\caption{%
\label{exp:vsc3-allgather-nlarge-onenode-vartwo}%
\mpiallgather%
}%
\end{subfigure}%
\hfill%
\begin{subfigure}{.33\linewidth}
\centering
\includegraphics[width=\linewidth]{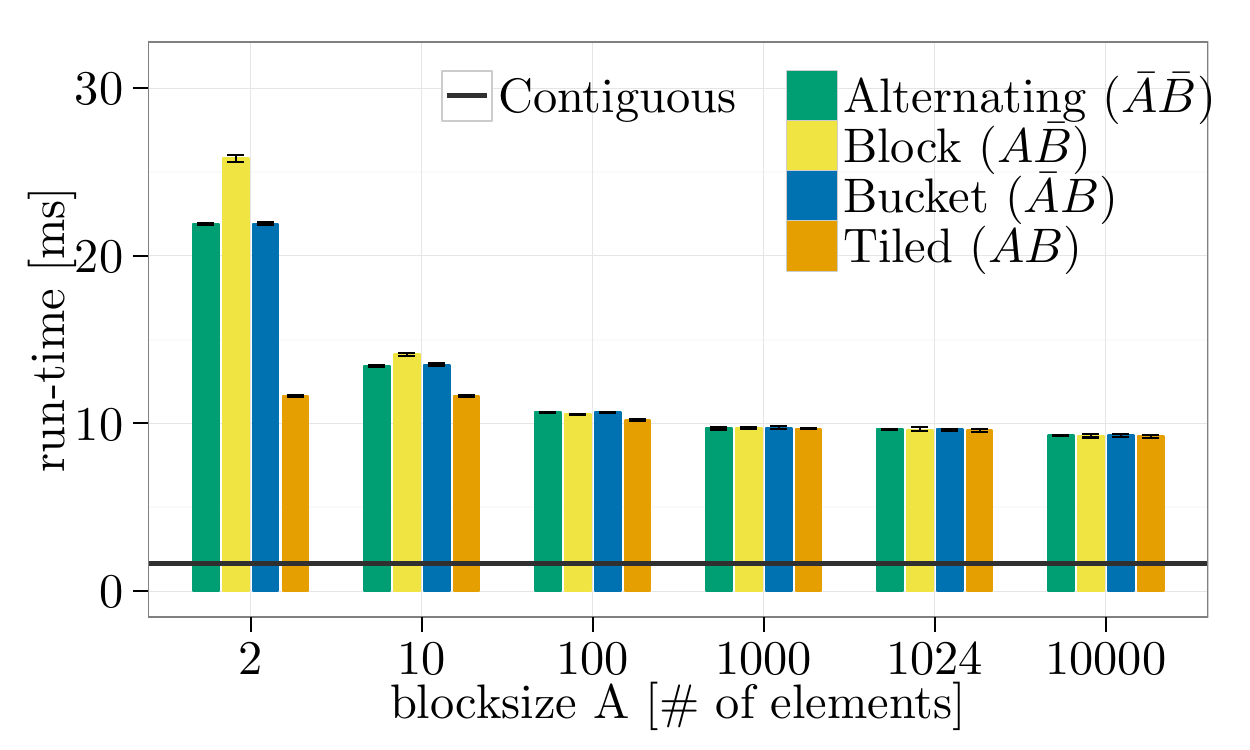}
\caption{%
\label{exp:vsc3-bcast-nlarge-onenode-vartwo}%
\mpibcast%
}%
\end{subfigure}%
\hfill%
\begin{subfigure}{.33\linewidth}
\centering
\includegraphics[width=\linewidth]{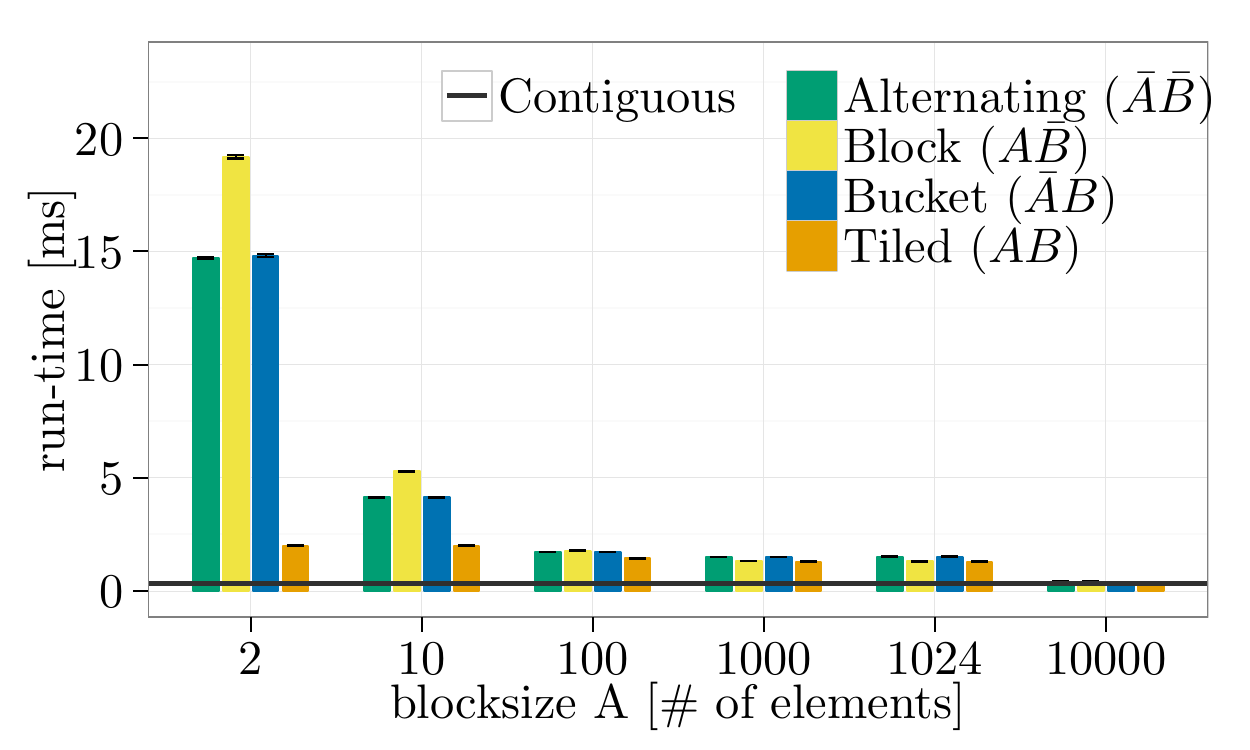}
\caption{%
\label{exp:vsc3-pingpong-nlarge-onenode-vartwo}%
\pingpong%
}%
\end{subfigure}%
\caption{\label{exp:vsc3-layouts-large-onenode-vartwo} Contiguous \vs typed, $\VARdatasize=\SI{2.56}{\mega\byte}$, element datatype: \mpiint, one node, \num{16}~processes (\num{2} for \pingpong), \vscintelmpi, \varianttwo.}
\end{figure*}

\appexpdesc{
  \begin{expitemize}
    \item \dtcontig, \dtdtiled, \dtdblock, \dtdbucket, \dtdalternating
    \item \pingpong, \mpibcast, \mpiallgather
  \end{expitemize}
}{
  \begin{expitemize}
    \item \expparam{\vscintelmpi, small \datasize, \variantone}{\fig~\ref{exp:vsc3-layouts-nsmall-32p}}
    \item \expparam{\vscintelmpi, large \datasize, \variantone}{\fig~\ref{exp:vsc3-layouts-nlarge-32p}}
    \item \expparam{\vscintelmpi, small \datasize, \variantone, one node}{\fig~\ref{exp:vsc3-layouts-nsmall-onenode}}
    \item \expparam{\vscintelmpi, large \datasize, \variantone, one node}{\fig~\ref{exp:vsc3-layouts-nlarge-onenode}}
    \item \expparam{\vscintelmpi, small \datasize, \varianttwo}{\fig~\ref{exp:vsc3-layouts-small-32p-vartwo}}
    \item \expparam{\vscintelmpi, large \datasize, \varianttwo}{\fig~\ref{exp:vsc3-layouts-large-32p-vartwo}}
    \item \expparam{\vscintelmpi, small \datasize, \varianttwo, one node}{\fig~\ref{exp:vsc3-layouts-small-onenode-vartwo}}
    \item \expparam{\vscintelmpi, large \datasize, \varianttwo, one node}{\fig~\ref{exp:vsc3-layouts-large-onenode-vartwo}}
    \end{expitemize}  
}

\FloatBarrier
\clearpage

\appexp{exptest:tiled_het}

\begin{figure*}[htpb]
\centering
\begin{subfigure}{.24\linewidth}
\centering
\includegraphics[width=\linewidth]{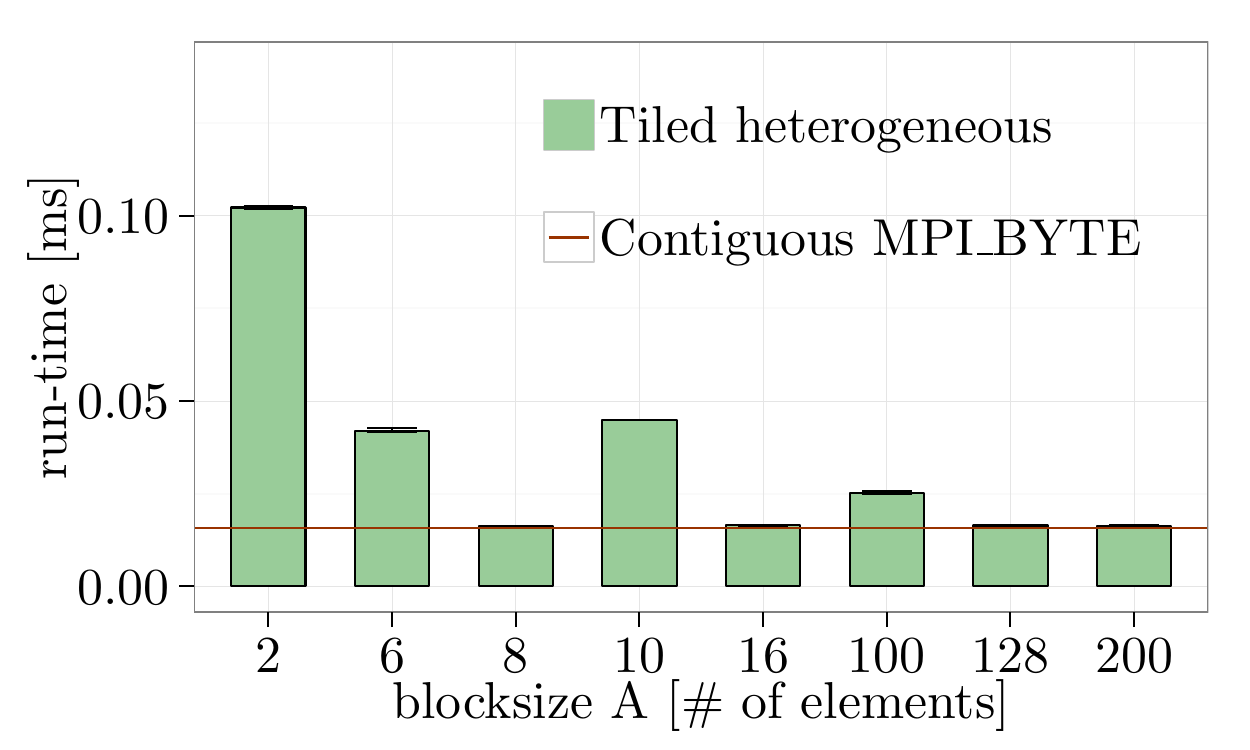}
\caption{%
\label{exp:vsc3-pingpong-heterog-small-2x1}%
$\VARdatasize = \SI{48}{\kilo\byte}$, \num{2}~nodes%
}%
\end{subfigure}%
\hfill%
\begin{subfigure}{.24\linewidth}
\centering
\includegraphics[width=\linewidth]{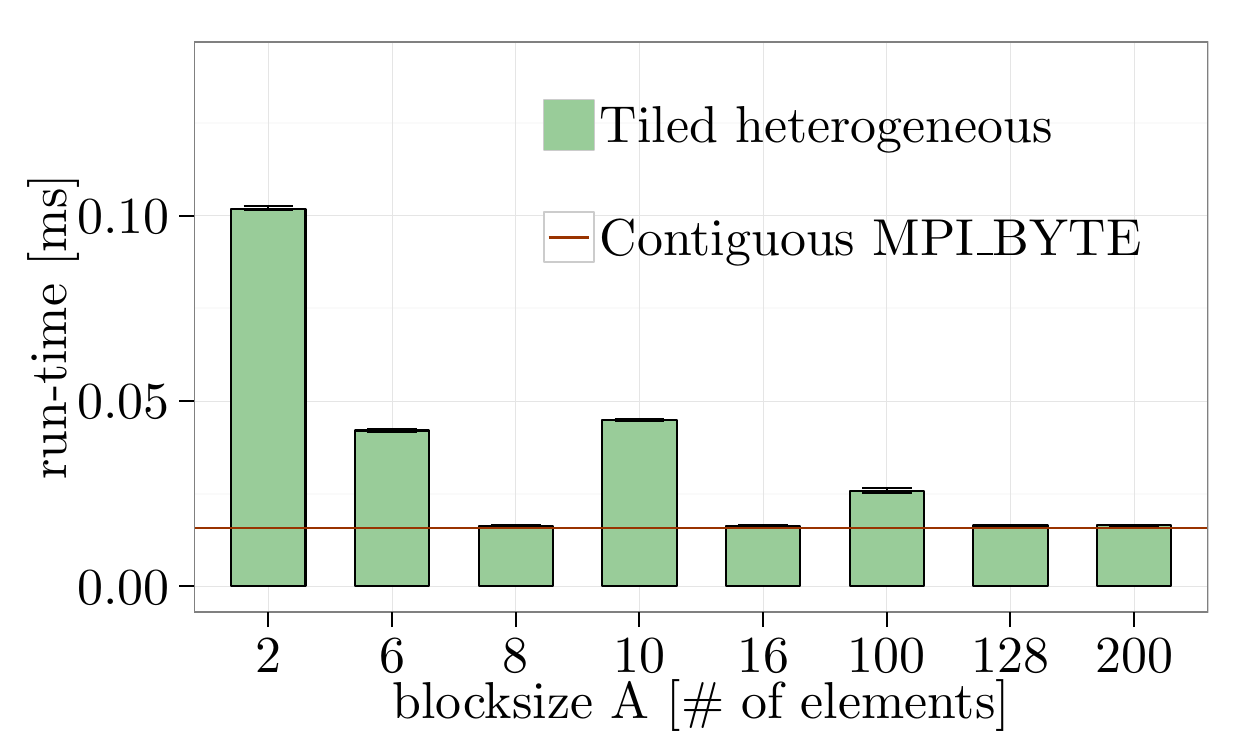}
\caption{%
\label{exp:vsc3-pingpong-heterog-small-1x2}%
$\VARdatasize = \SI{48}{\kilo\byte}$, same node%
}%
\end{subfigure}%
\hfill%
\begin{subfigure}{.24\linewidth}
\centering
\includegraphics[width=\linewidth]{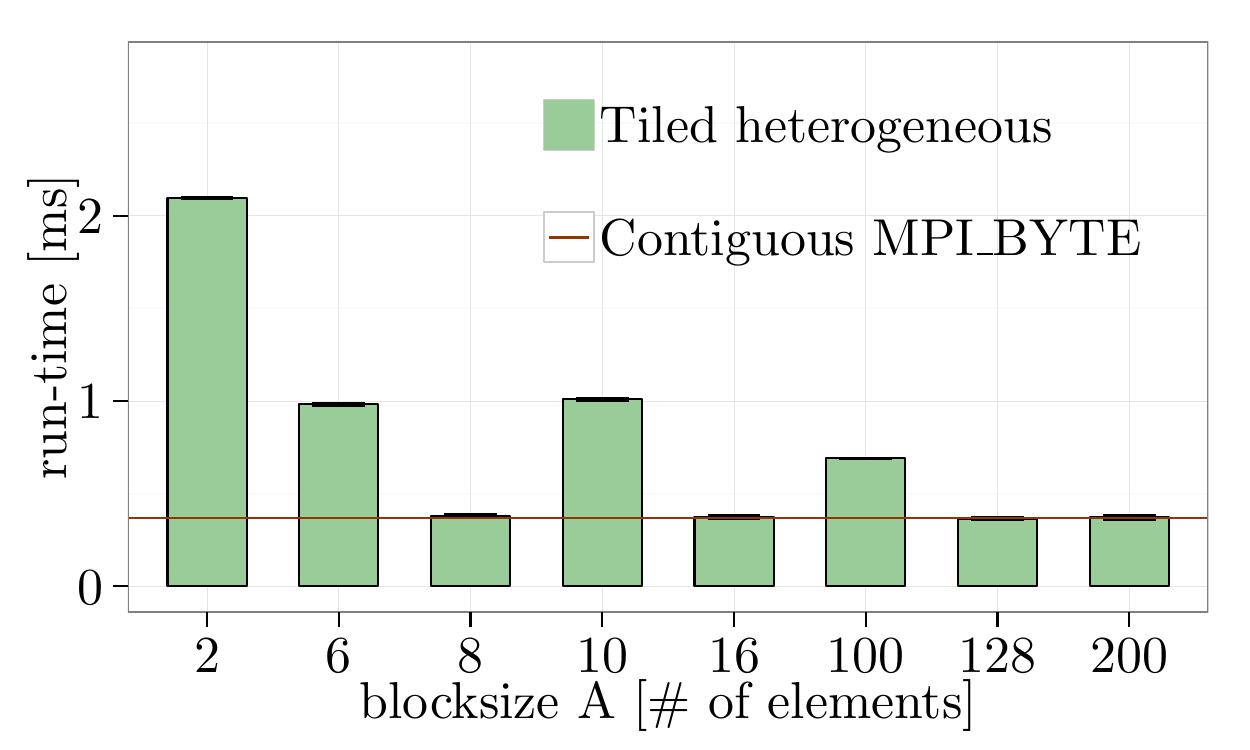}
\caption{%
\label{exp:vsc3-pingpong-heterog-large-2x1}%
$\VARdatasize = \SI{1.5}{\mega\byte}$, \num{2}~nodes%
}%
\end{subfigure}%
\hfill%
\begin{subfigure}{.24\linewidth}
\centering
\includegraphics[width=\linewidth]{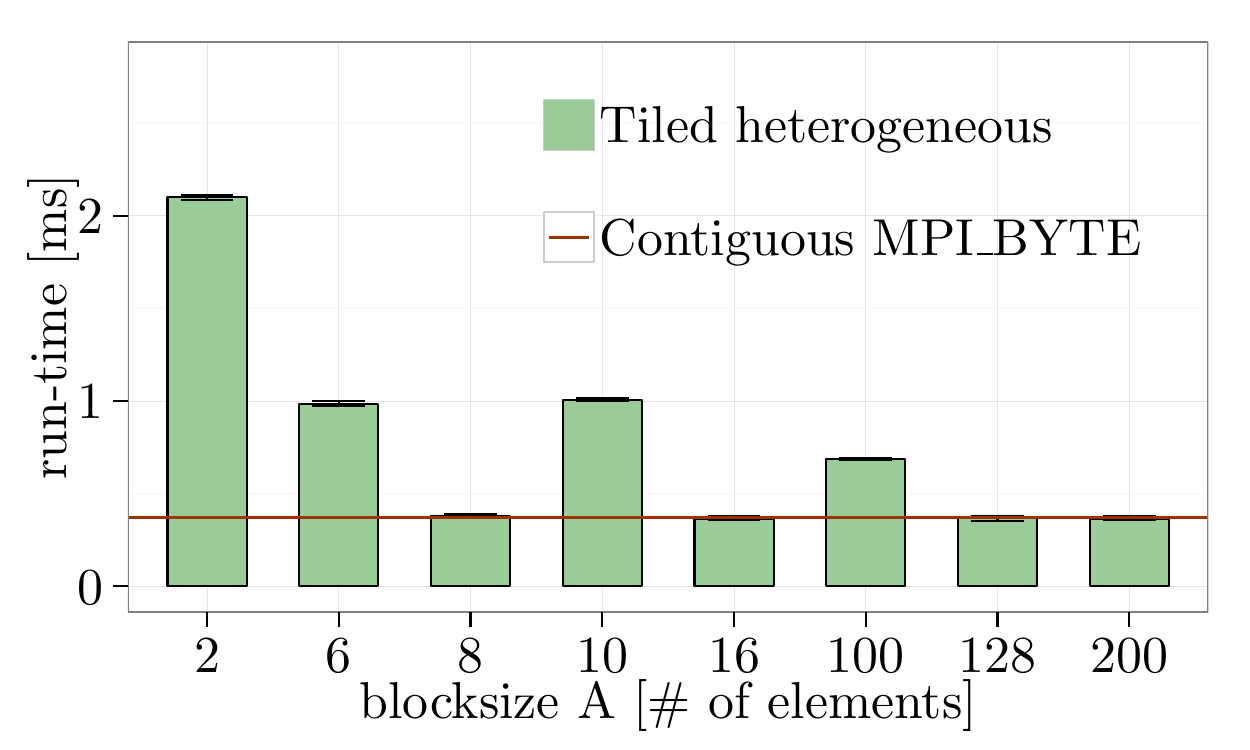}
\caption{%
\label{exp:vsc3-pingpong-heterog-large-1x2}%
$\VARdatasize = \SI{1.5}{\mega\byte}$, same node%
}%
\end{subfigure}%
\caption{\label{exp:vsc3-pingpong-heterog} \dtcontig \vs \dtdtiledhet $A$=$B$, element datatype: \mpiint, \pingpong, \vscintelmpi.}
\end{figure*}

\appexpdesc{
  \begin{expitemize}
    \item \dtcontig, \dtdtiledhet
    \item \pingpong
  \end{expitemize}
}{
  \begin{expitemize}
    \item \expparam{\vscintelmpi}{\fig~\ref{exp:vsc3-pingpong-heterog}}
  \end{expitemize}  
}

\FloatBarrier
\clearpage

\appexp{exptest:pack_unpack}

\begin{figure*}[htpb]
\centering
\begin{subfigure}{.24\linewidth}
\centering
\includegraphics[width=\linewidth]{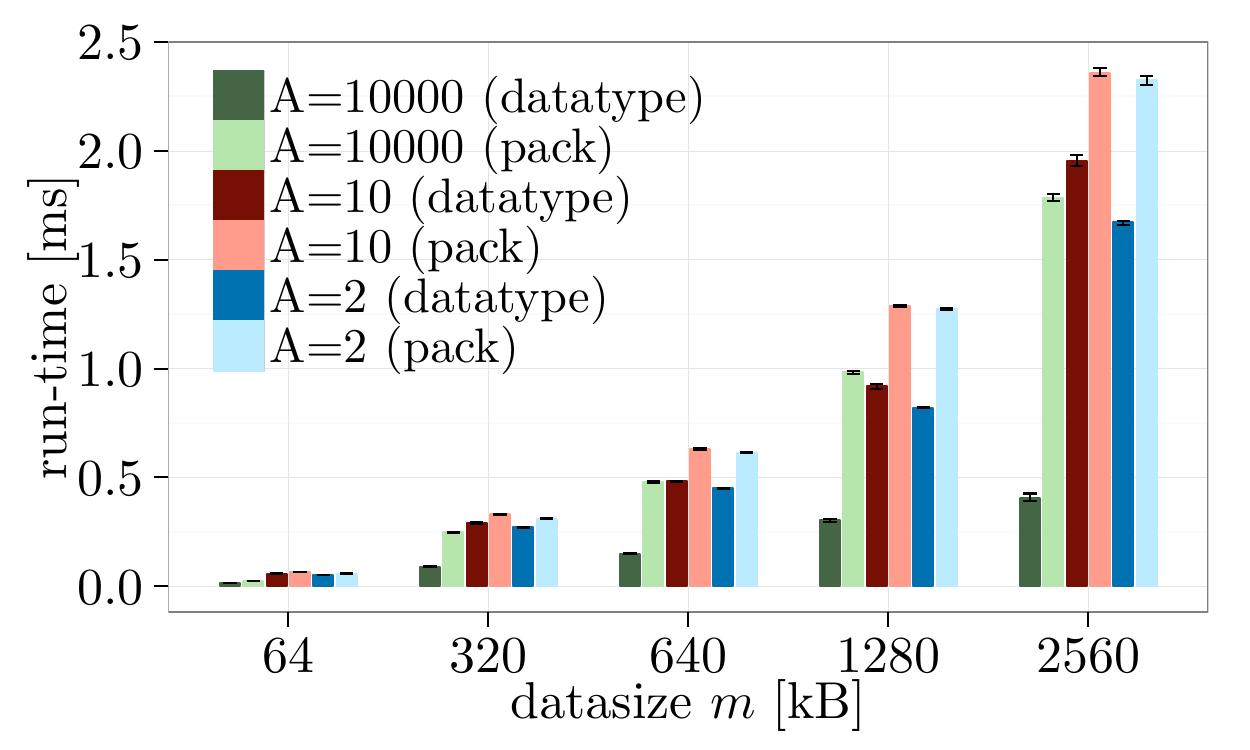}
\caption{%
\label{exp:vsc3-pingpong-pack-tiled-2x1}%
\dttiled%
}%
\end{subfigure}%
\hfill%
\begin{subfigure}{.24\linewidth}
\centering
\includegraphics[width=\linewidth]{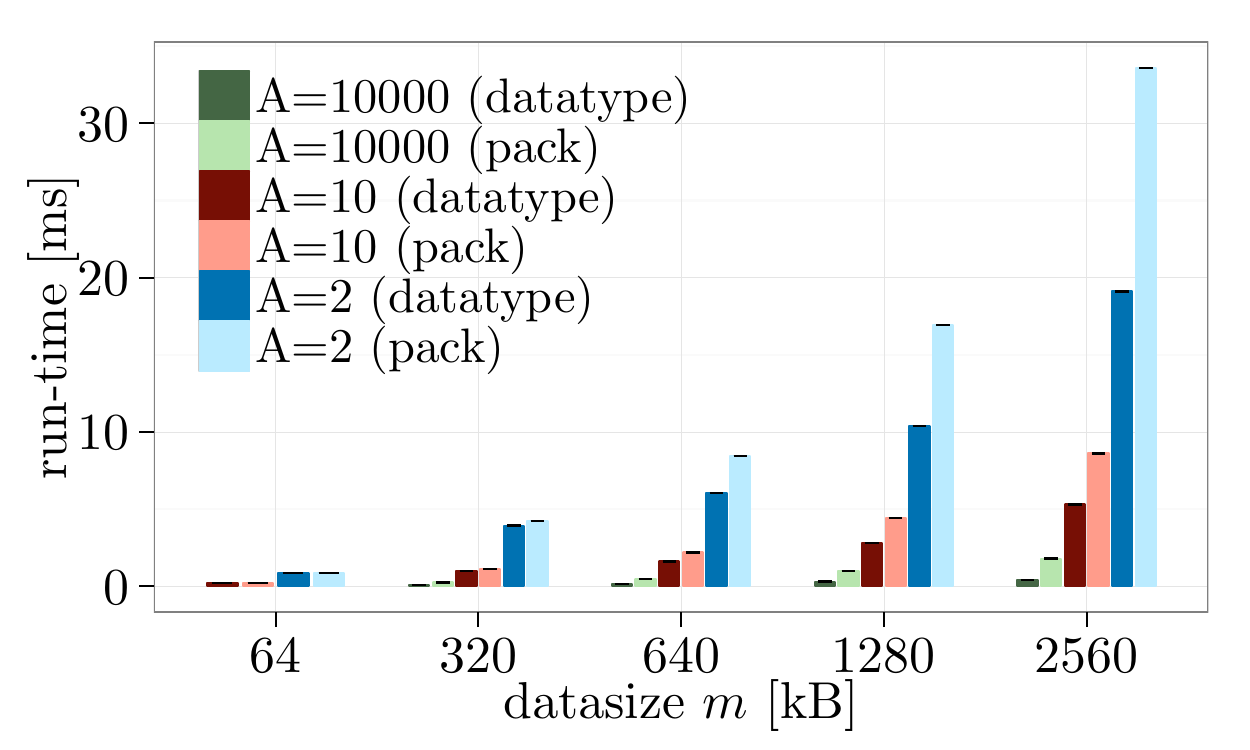}
\caption{%
\label{exp:vsc3-pingpong-pack-block-2x1}%
\dtblock%
}%
\end{subfigure}%
\hfill%
\begin{subfigure}{.24\linewidth}
\centering
\includegraphics[width=\linewidth]{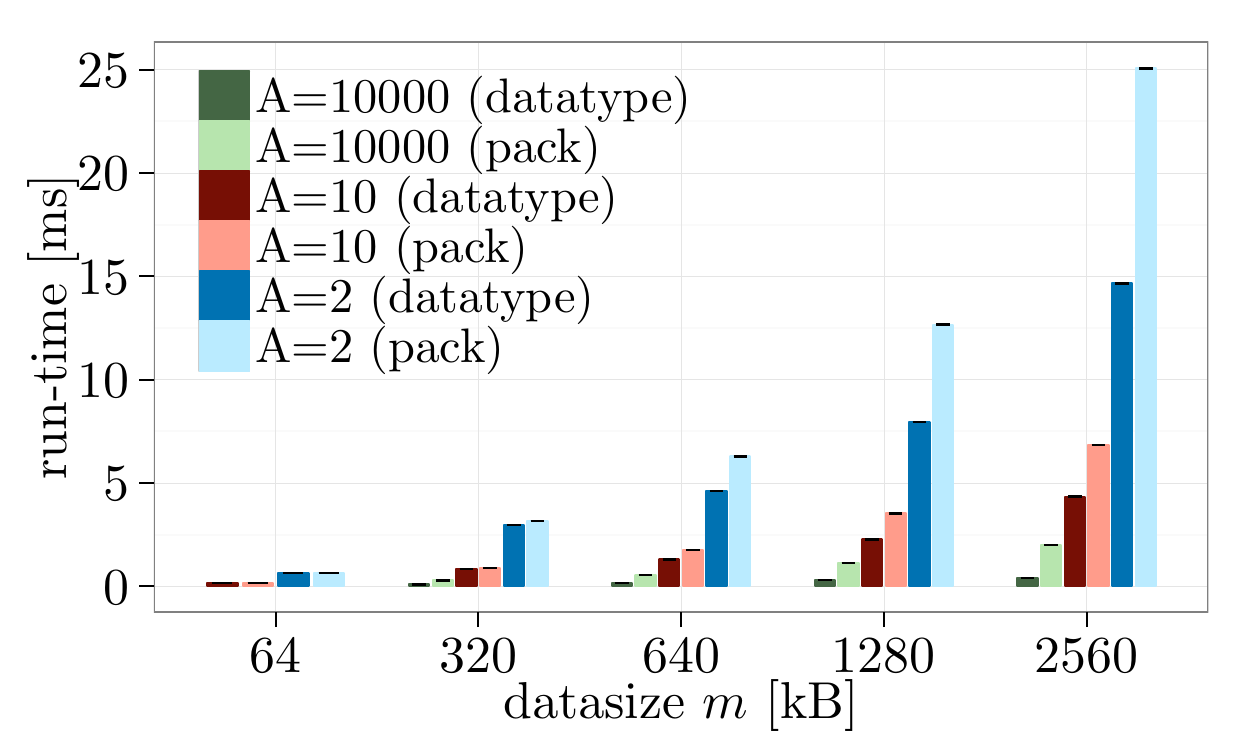}
\caption{%
\label{exp:vsc3-pingpong-pack-bucket-2x1}%
\dtbucket%
}%
\end{subfigure}%
\hfill%
\begin{subfigure}{.24\linewidth}
\centering
\includegraphics[width=\linewidth]{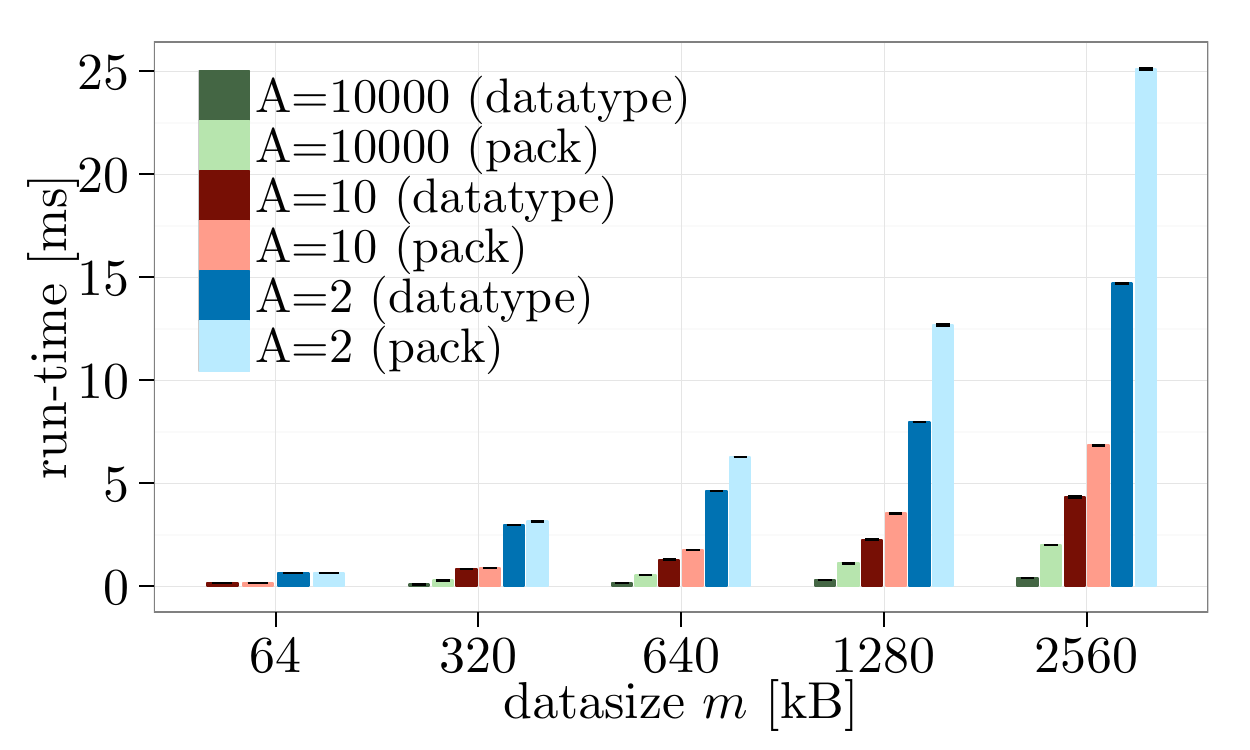}
\caption{%
\label{exp:vsc3-pingpong-pack-alternating-2x1}%
\dtalternating%
}%
\end{subfigure}%
\caption{\label{exp:vsc3-pingpong-pack-2x1}  Basic layouts \vs pack/unpack, element datatype: \mpiint, \num{2x1}~processes, \pingpong, \vscintelmpi.}
\end{figure*}

\begin{figure*}[htpb]
\centering
\begin{subfigure}{.24\linewidth}
\centering
\includegraphics[width=\linewidth]{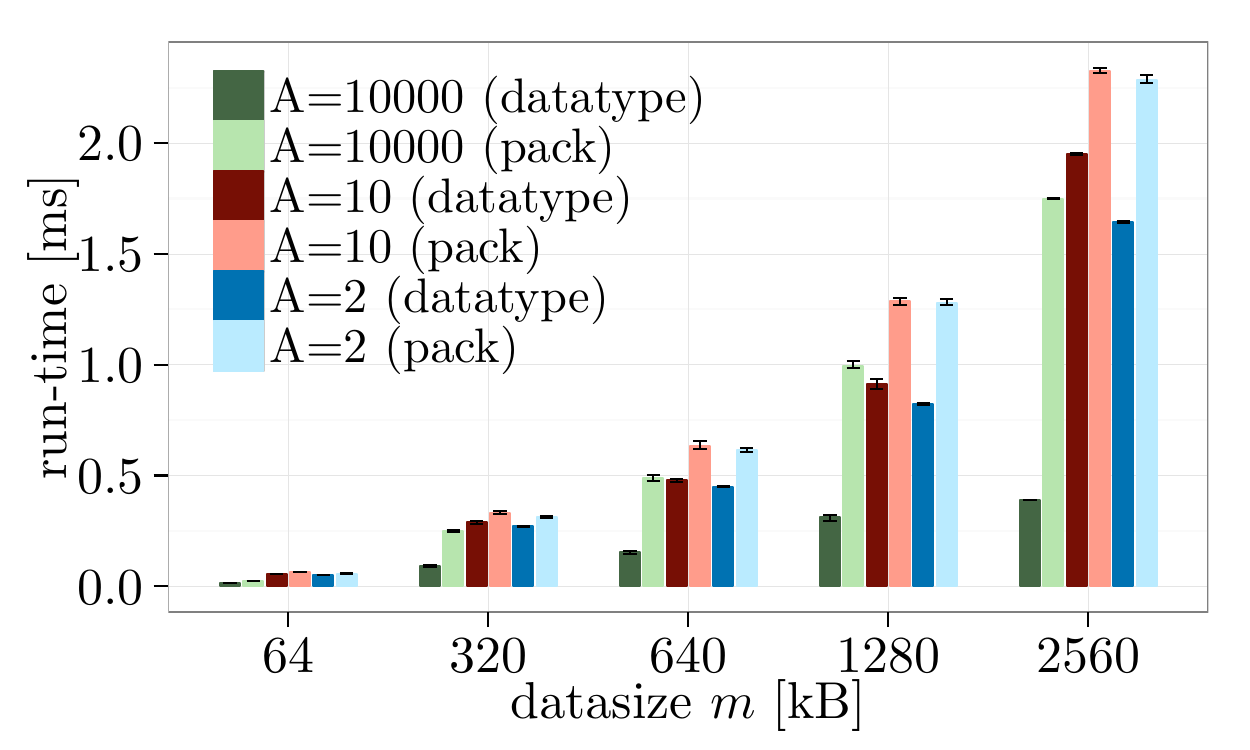}
\caption{%
\label{exp:vsc3-pingpong-pack-tiled-onenode}%
\dttiled%
}%
\end{subfigure}%
\hfill%
\begin{subfigure}{.24\linewidth}
\centering
\includegraphics[width=\linewidth]{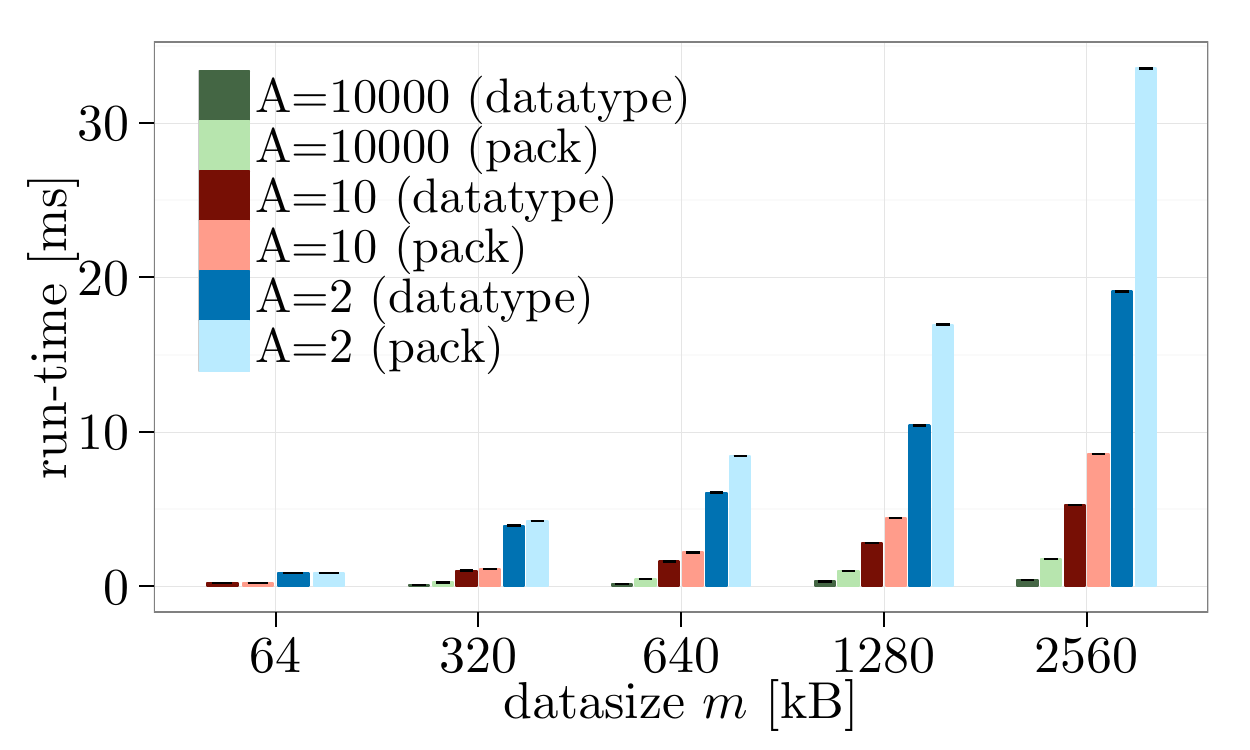}
\caption{%
\label{exp:vsc3-pingpong-pack-block-onenode}%
\dtblock%
}%
\end{subfigure}%
\hfill%
\begin{subfigure}{.24\linewidth}
\centering
\includegraphics[width=\linewidth]{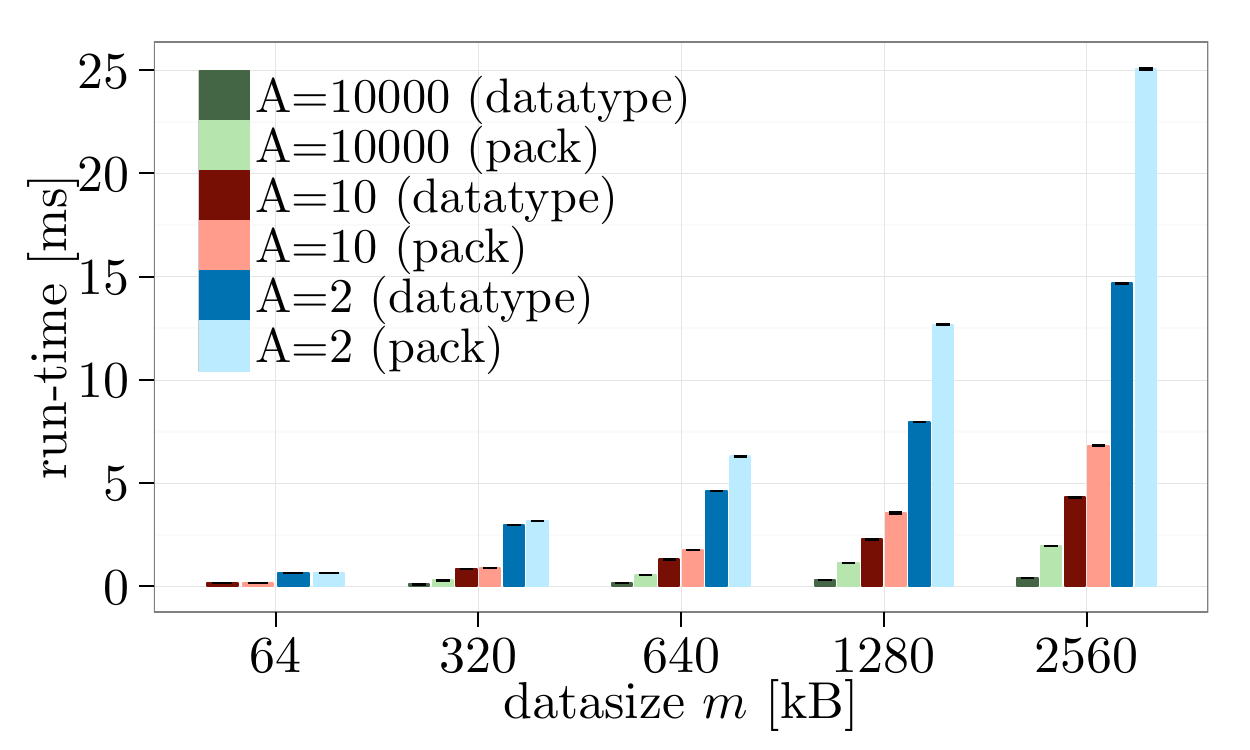}
\caption{%
\label{exp:vsc3-pingpong-pack-bucket-onenode}%
\dtbucket%
}%
\end{subfigure}%
\hfill%
\begin{subfigure}{.24\linewidth}
\centering
\includegraphics[width=\linewidth]{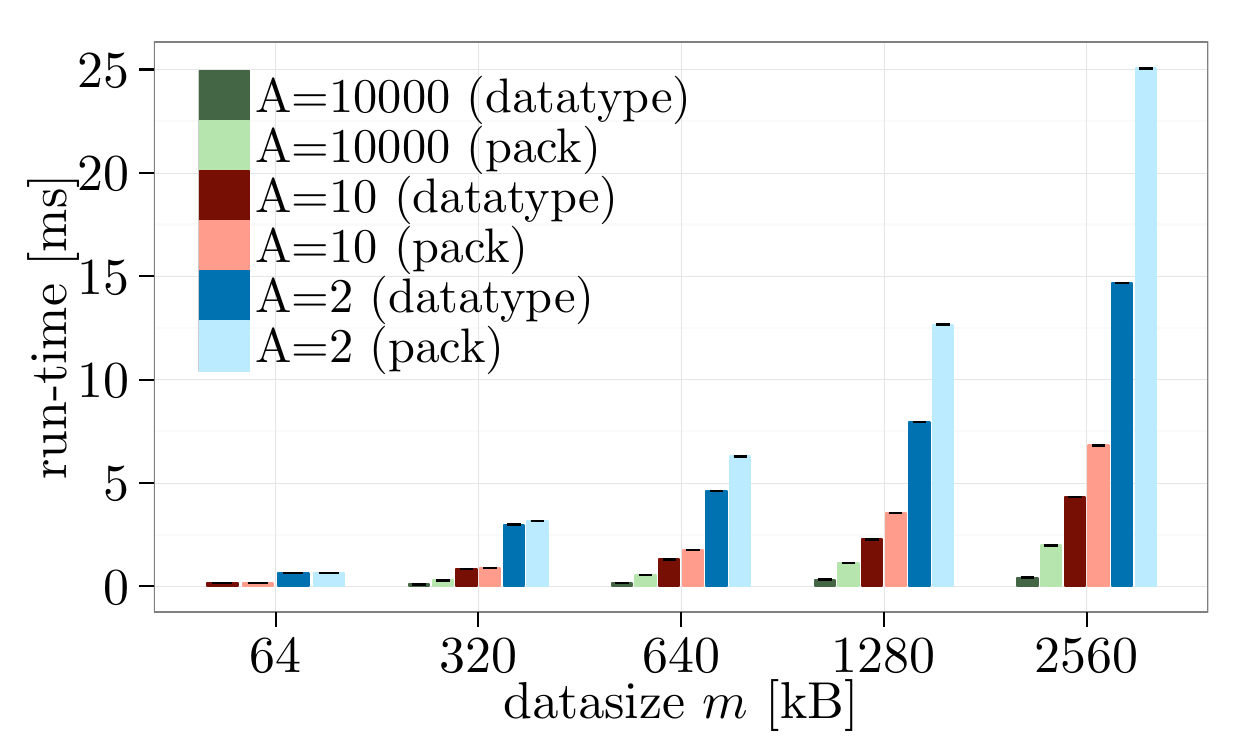}
\caption{%
\label{exp:vsc3-pingpong-pack-alternating-onenode}%
\dtalternating%
}%
\end{subfigure}%
\caption{\label{exp:vsc3-pingpong-pack-onenode}  Basic layouts \vs pack/unpack, element datatype: \mpiint, one~node, \num{2}~processes, \pingpong, \vscintelmpi.}
\end{figure*}

\appexpdesc{
  \begin{expitemize}
    \item pack \vs unpack for basic layouts (\dttiled, \dtblock, \dtbucket, \dtalternating)
    \item \mpiallgather, \mpibcast, \pingpong
  \end{expitemize}
}{
  \begin{expitemize}
    \item \expparam{\vscintelmpi, \pingpong, \num{2x1}~processes}{\fig~\ref{exp:vsc3-pingpong-pack-2x1}}
    \item \expparam{\vscintelmpi, \pingpong, one~node, \num{2}~processes}{\fig~\ref{exp:vsc3-pingpong-pack-onenode}}
    \end{expitemize}  
}

\FloatBarrier
\clearpage

\appexp{exptest:contig}

\begin{figure*}[htpb]
\centering
\begin{subfigure}{.24\linewidth}
\centering
\includegraphics[width=\linewidth]{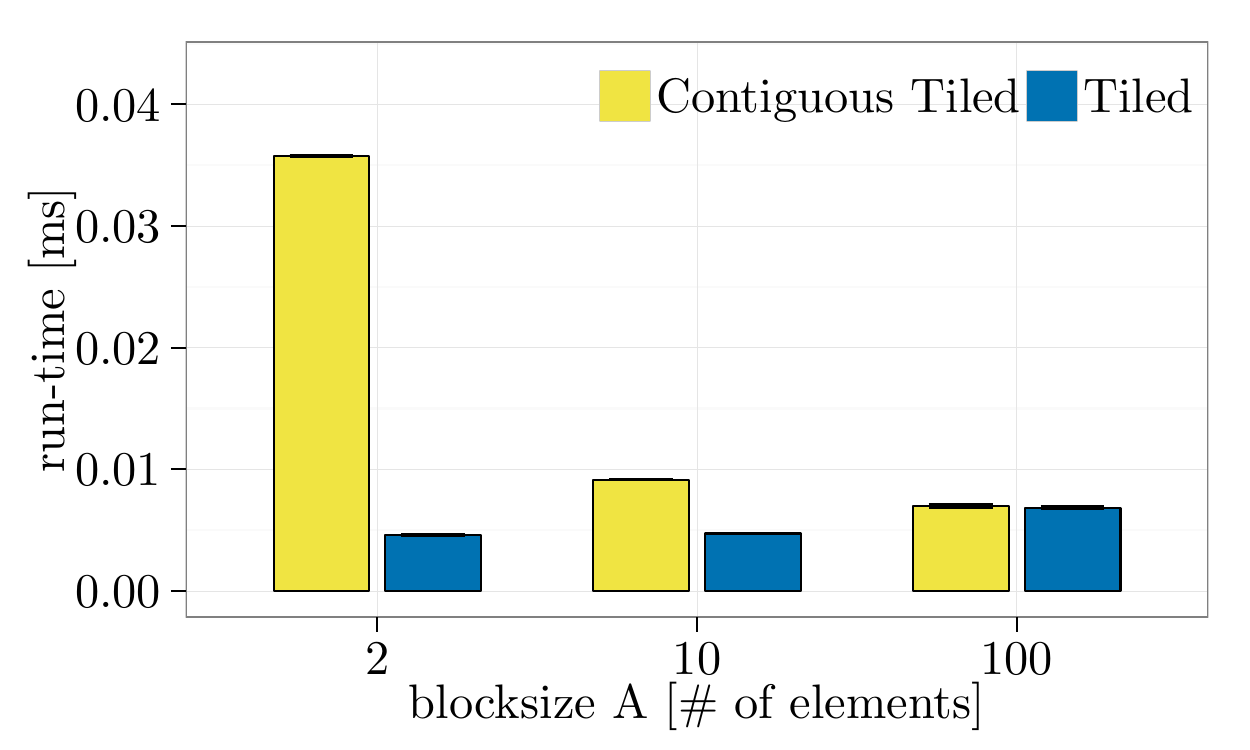}
\caption{%
\label{exp:vsc3-pingpong-contigtiled-2x1}%
\dttiled%
}%
\end{subfigure}%
\hfill%
\begin{subfigure}{.24\linewidth}
\centering
\includegraphics[width=\linewidth]{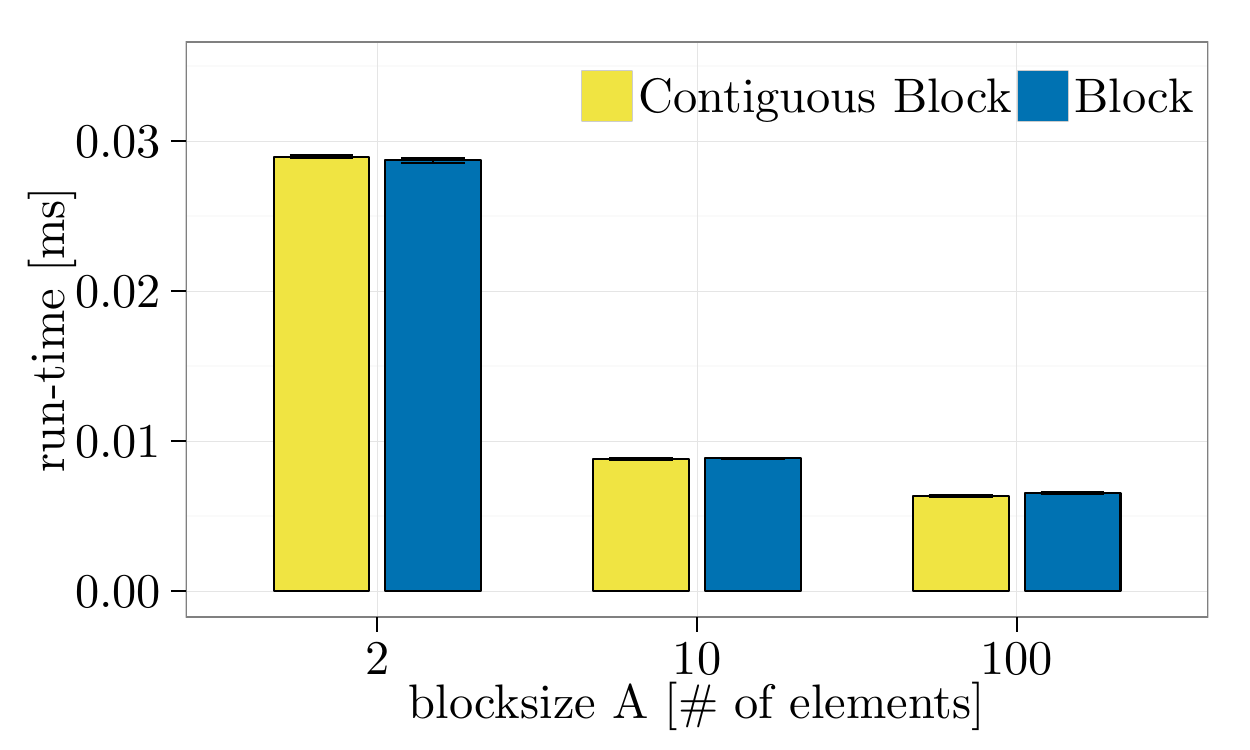}
\caption{%
\label{exp:vsc3-pingpong-contigblock-2x1}%
\dtblock%
}%
\end{subfigure}%
\hfill%
\begin{subfigure}{.24\linewidth}
\centering
\includegraphics[width=\linewidth]{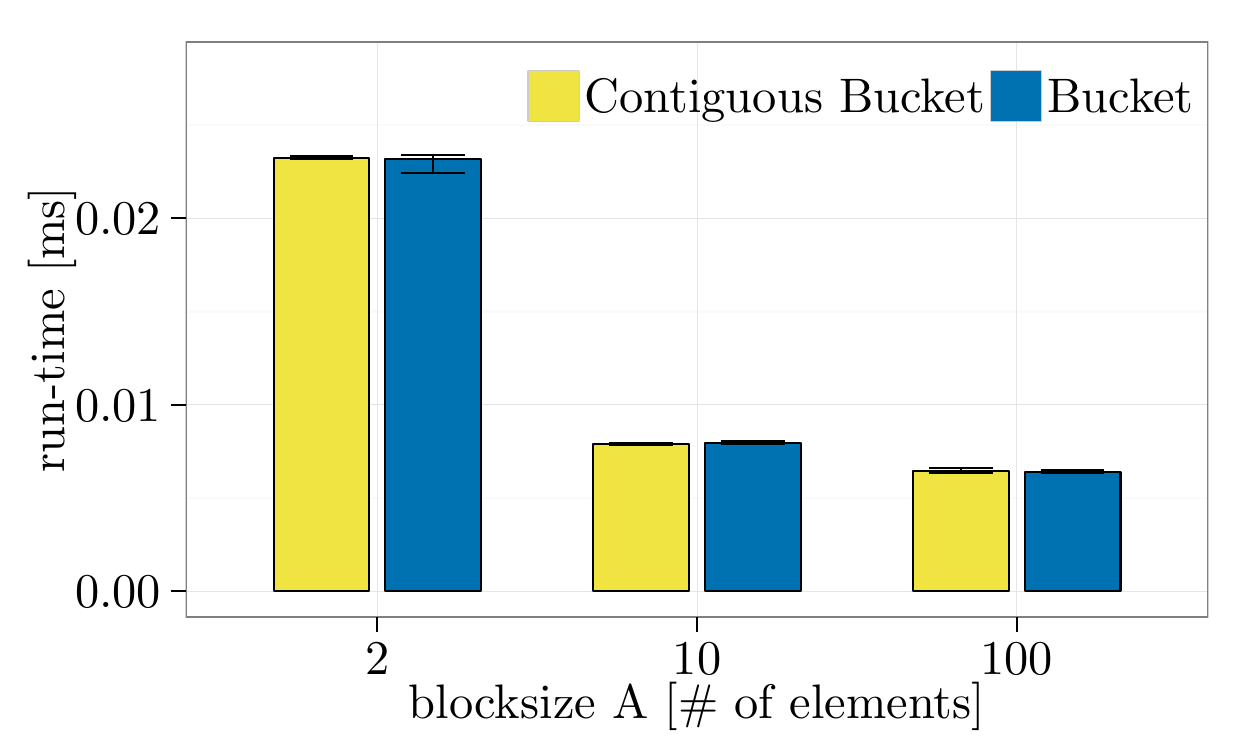}
\caption{%
\label{exp:vsc3-pingpong-contigbucket-2x1}%
\dtbucket%
}%
\end{subfigure}%
\hfill%
\begin{subfigure}{.24\linewidth}
\centering
\includegraphics[width=\linewidth]{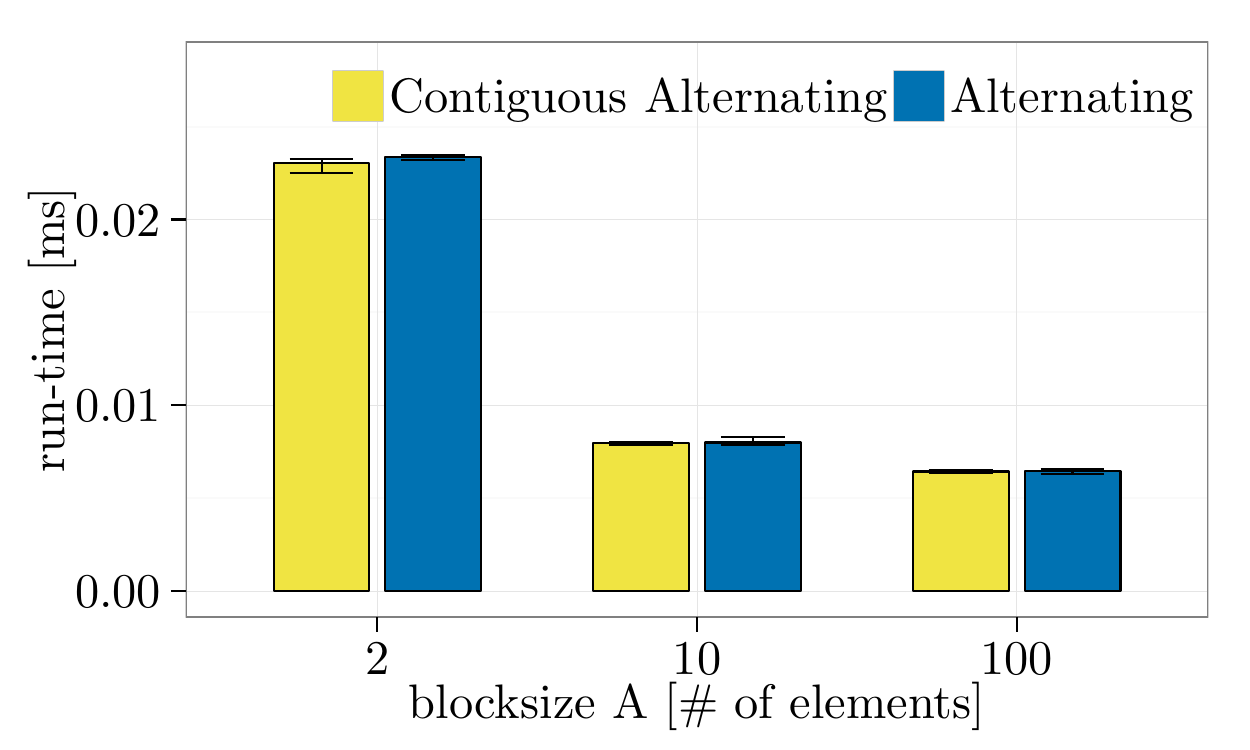}
\caption{%
\label{exp:vsc3-pingpong-contigalternating-2x1}%
\dtalternating%
}%
\end{subfigure}%
\caption{\label{exp:vsc3-pingpong-contig-smallnbytes-2x1}  Basic layouts \vs \ddtcontig, $\VARdatasize=\SI{2}{\kilo\byte}$, element datatype: \mpiint, \num{2x1}~processes, \pingpong, \vscintelmpi (similar results for \num{1x2}~processes).}
\end{figure*}

\begin{figure*}[htpb]
\centering
\begin{subfigure}{.24\linewidth}
\centering
\includegraphics[width=\linewidth]{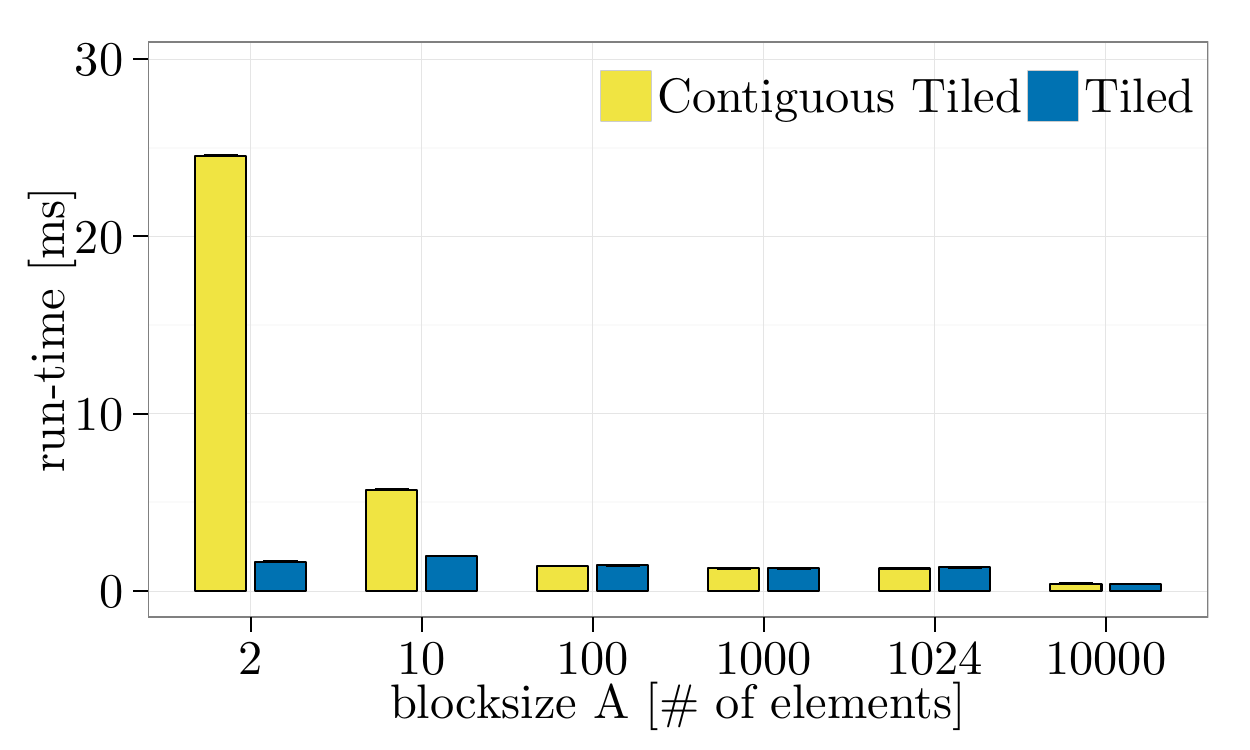}
\caption{%
\label{exp:vsc3-pingpong-contigtiled-2x1-large}%
\dttiled%
}%
\end{subfigure}%
\hfill%
\begin{subfigure}{.24\linewidth}
\centering
\includegraphics[width=\linewidth]{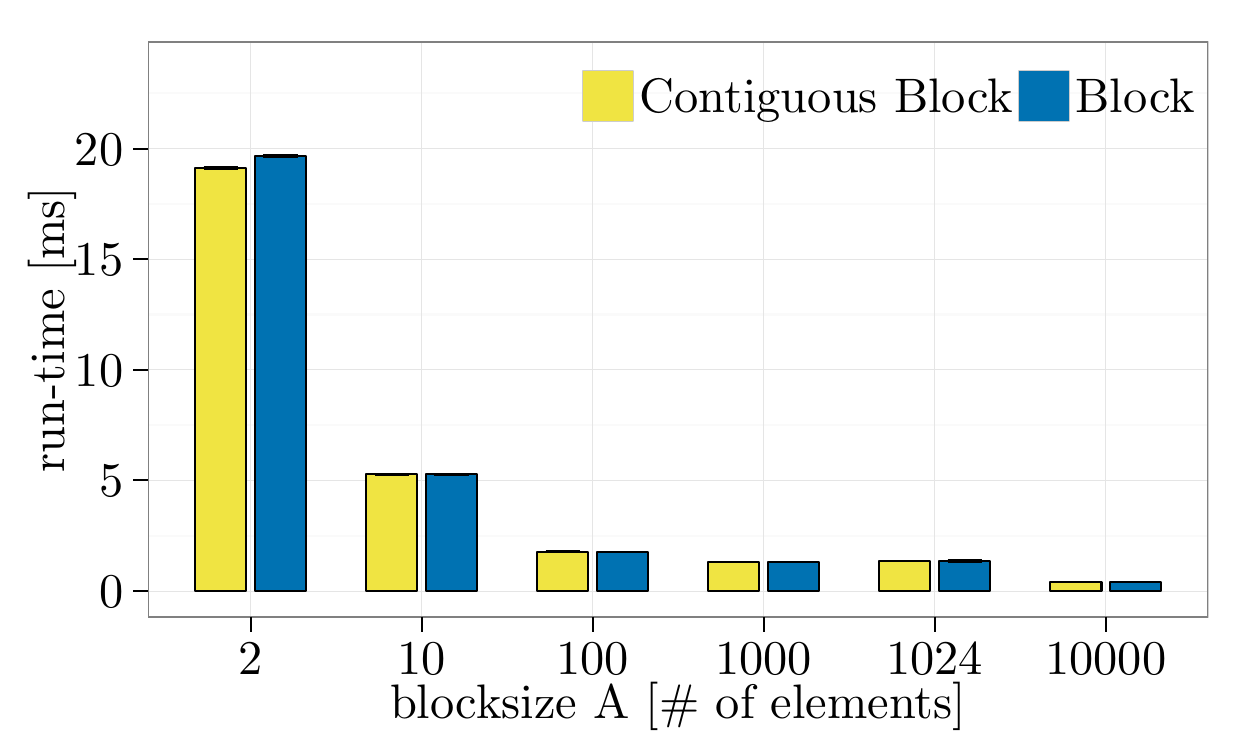}
\caption{%
\label{exp:vsc3-pingpong-contigblock-2x1-large}%
\dtblock%
}%
\end{subfigure}%
\hfill%
\begin{subfigure}{.24\linewidth}
\centering
\includegraphics[width=\linewidth]{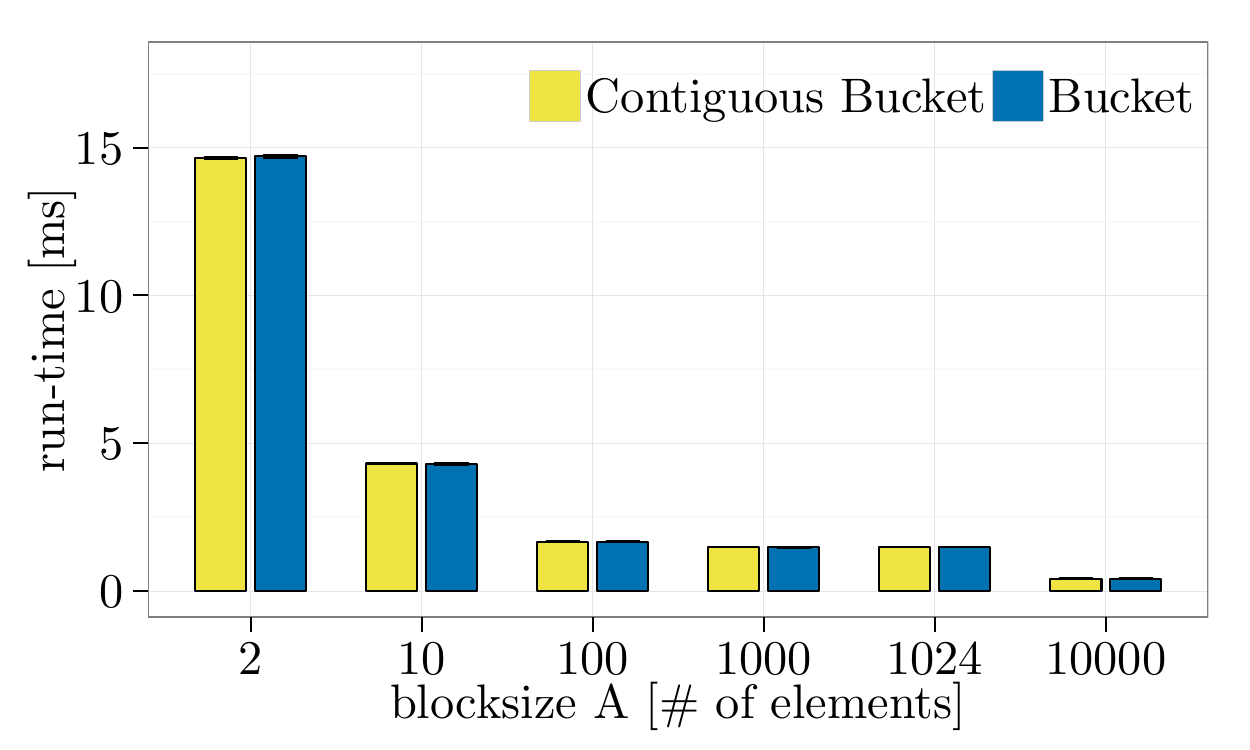}
\caption{%
\label{exp:vsc3-pingpong-contigbucket-2x1-large}%
\dtbucket%
}%
\end{subfigure}%
\hfill%
\begin{subfigure}{.24\linewidth}
\centering
\includegraphics[width=\linewidth]{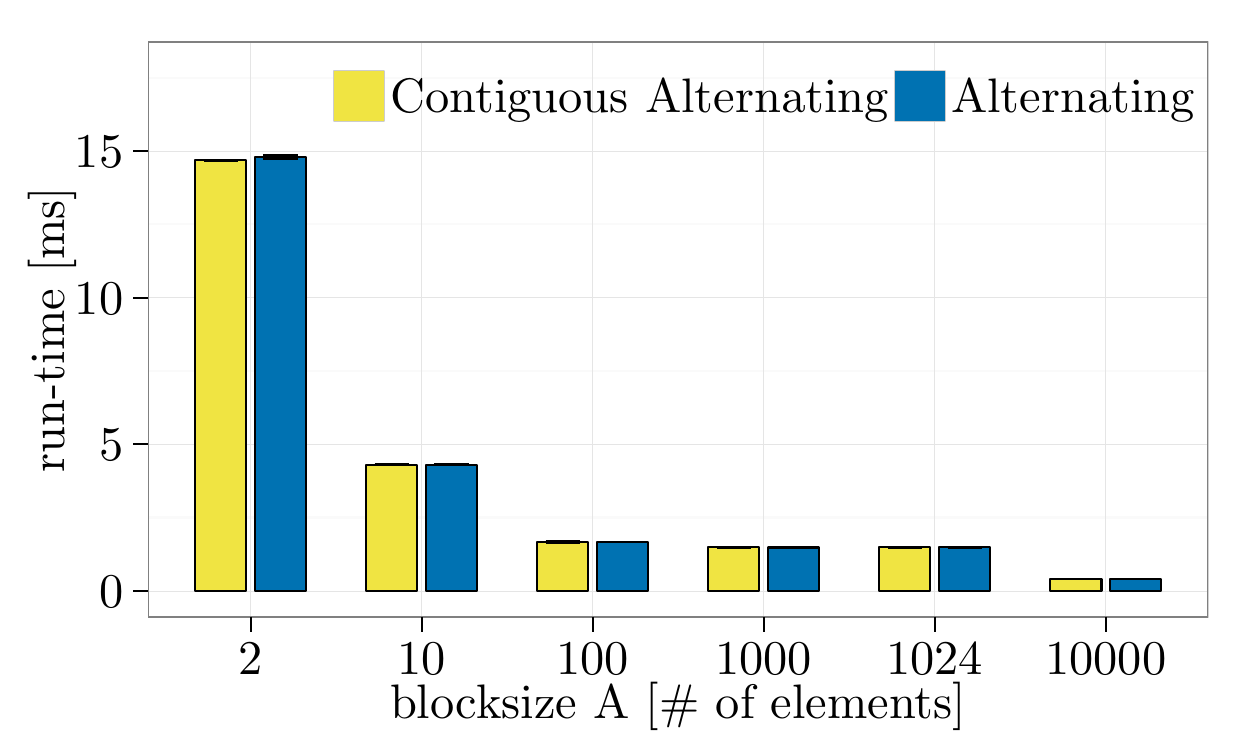}
\caption{%
\label{exp:vsc3-pingpong-contigalternating-2x1-large}%
\dtalternating%
}%
\end{subfigure}%
\caption{\label{exp:vsc3-pingpong-contig-largenbytes-2x1}  Basic layouts \vs \ddtcontig, $\VARdatasize=\SI{2.56}{\mega\byte}$, element datatype: \mpiint, \num{2x1}~processes, \pingpong, \vscintelmpi (similar results for \num{1x2}~processes).}
\end{figure*}

\begin{figure*}[htpb]
\centering
\begin{subfigure}{.24\linewidth}
\centering
\includegraphics[width=\linewidth]{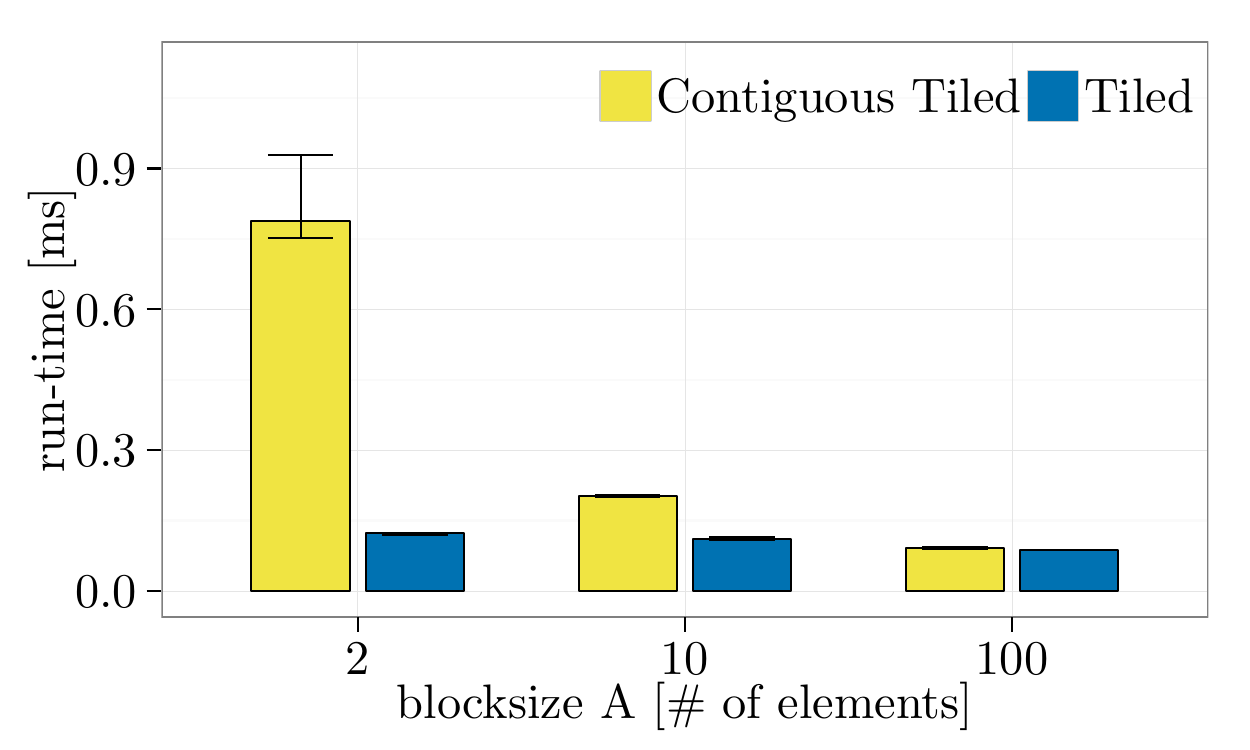}
\caption{%
\label{exp:vsc3-allgather-contigtiled-32x1}%
\dttiled%
}%
\end{subfigure}%
\hfill%
\begin{subfigure}{.24\linewidth}
\centering
\includegraphics[width=\linewidth]{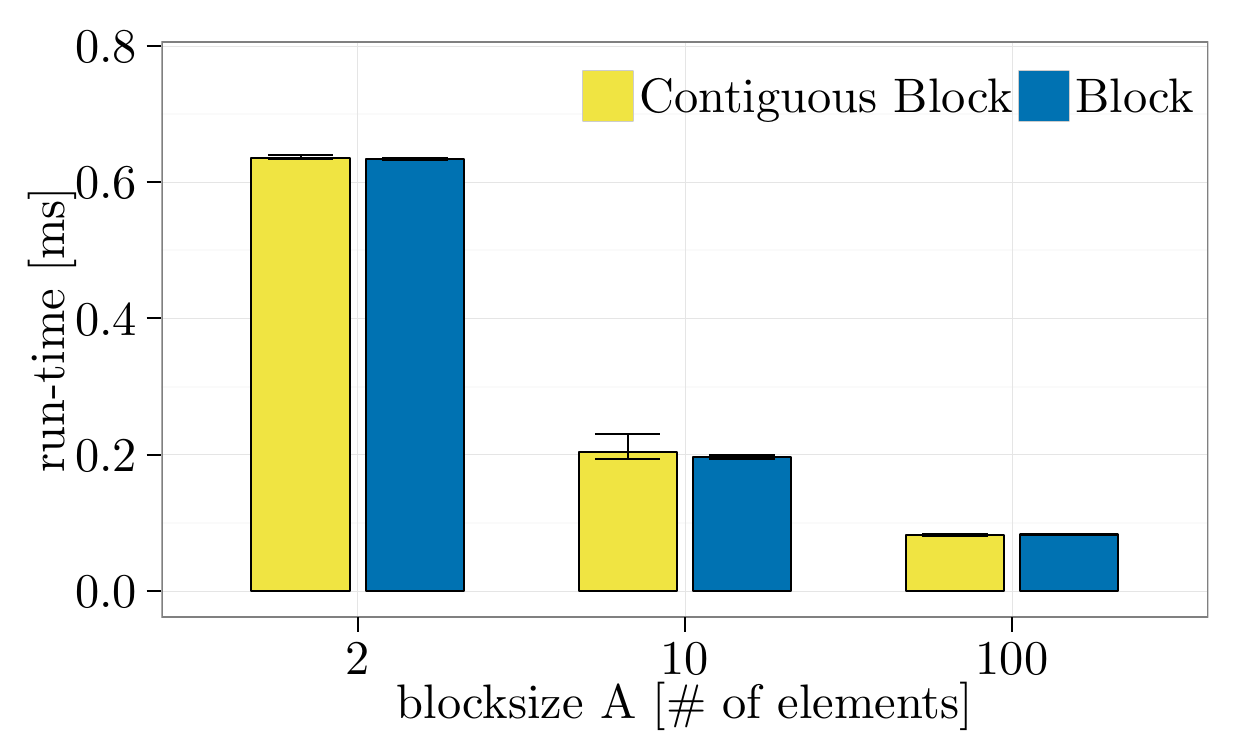}
\caption{%
\label{exp:vsc3-allgather-contigblock-32x1}%
\dtblock%
}%
\end{subfigure}%
\hfill%
\begin{subfigure}{.24\linewidth}
\centering
\includegraphics[width=\linewidth]{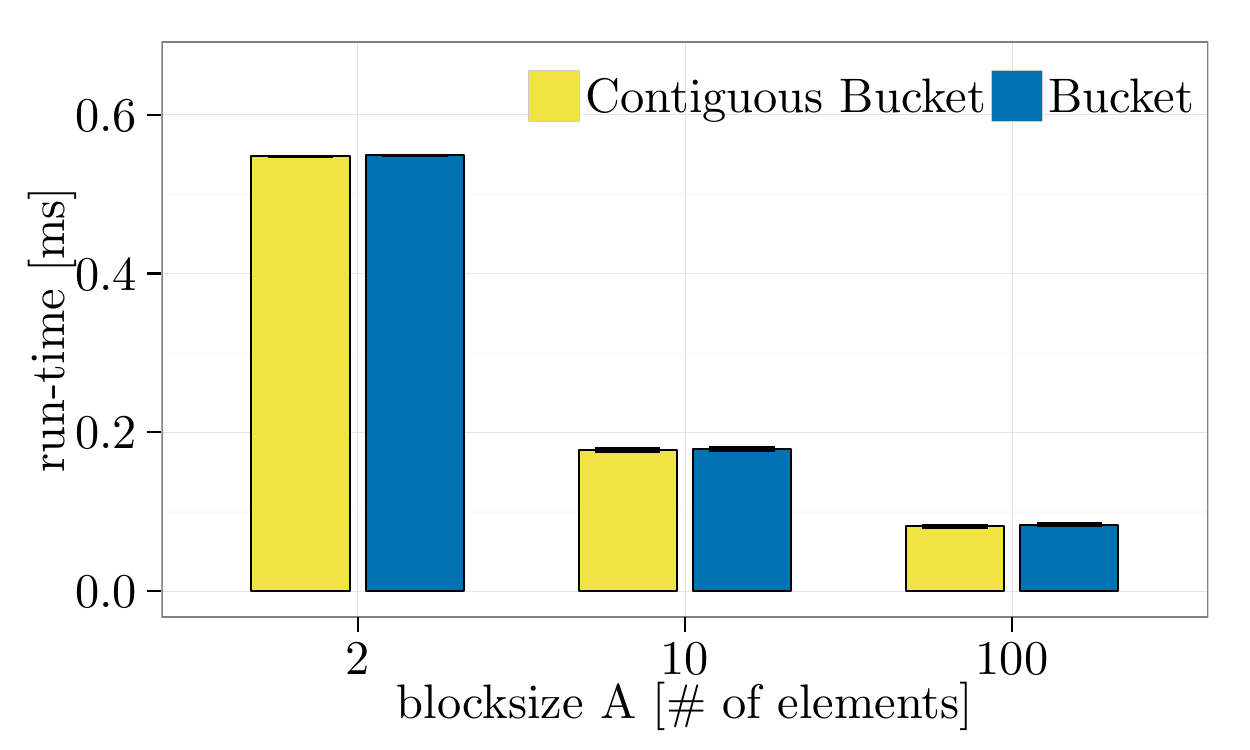}
\caption{%
\label{exp:vsc3-allgather-contigbucket-32x1}%
\dtbucket%
}%
\end{subfigure}%
\hfill%
\begin{subfigure}{.24\linewidth}
\centering
\includegraphics[width=\linewidth]{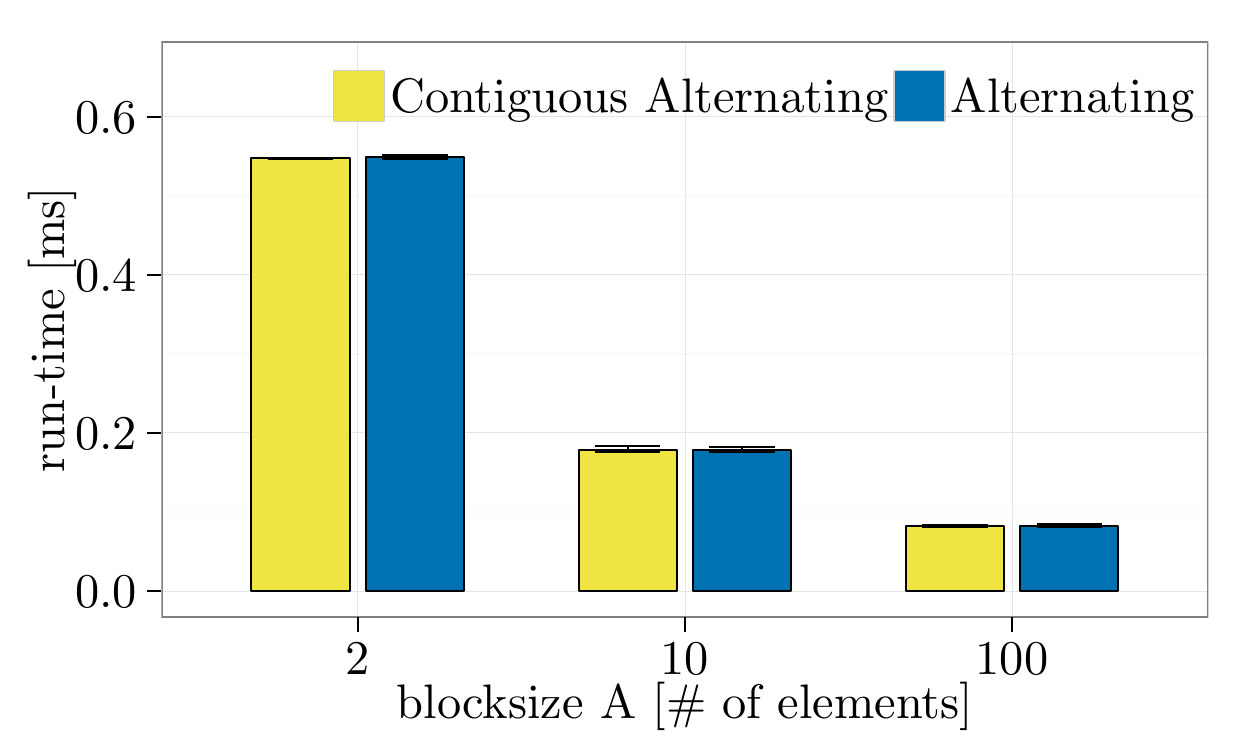}
\caption{%
\label{exp:vsc3-allgather-contigalternating-32x1}%
\dtalternating%
}%
\end{subfigure}%
\caption{\label{exp:vsc3-allgather-contig-smallnbytes-32x1}  Basic layouts \vs \ddtcontig, $\VARdatasize=\SI{2}{\kilo\byte}$, element datatype: \mpiint, \num{32x1}~processes, \mpiallgather, \vscintelmpi (similar results for \num{1x16}~processes).}
\end{figure*}

\begin{figure*}[htpb]
\centering
\begin{subfigure}{.24\linewidth}
\centering
\includegraphics[width=\linewidth]{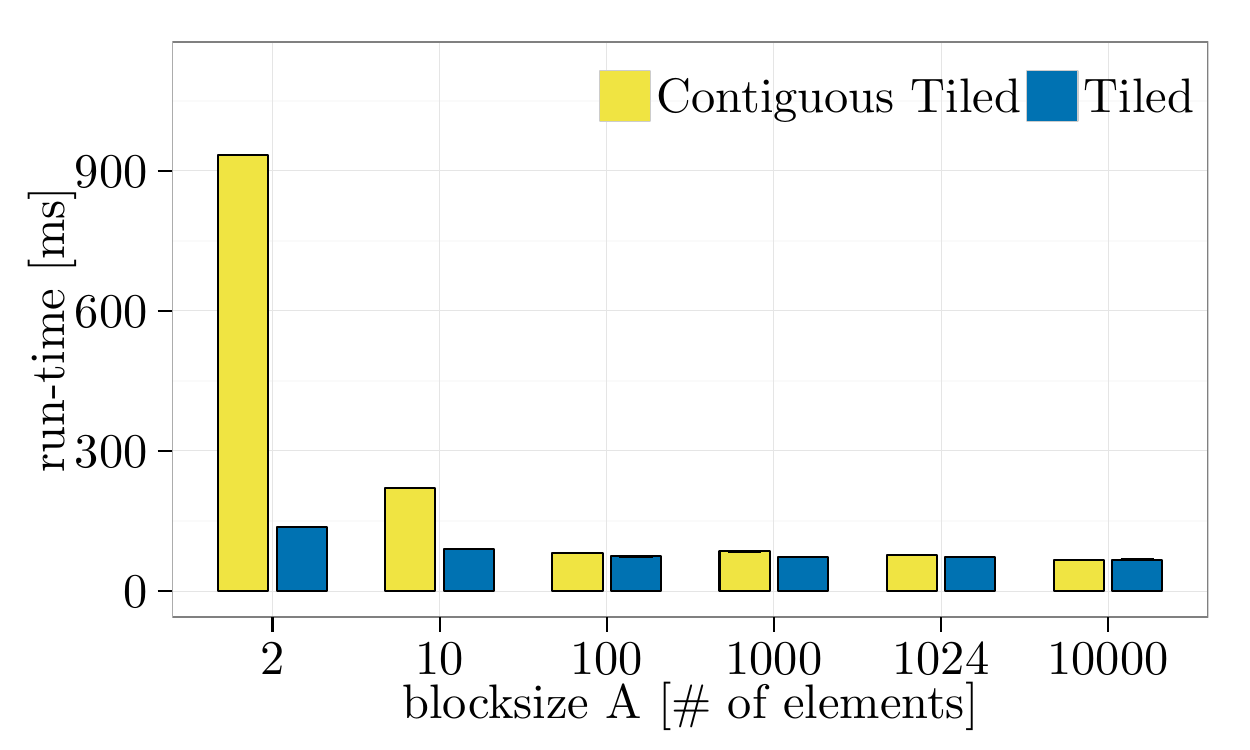}
\caption{%
\label{exp:vsc3-allgather-contigtiled-32x1-large}%
\dttiled%
}%
\end{subfigure}%
\hfill%
\begin{subfigure}{.24\linewidth}
\centering
\includegraphics[width=\linewidth]{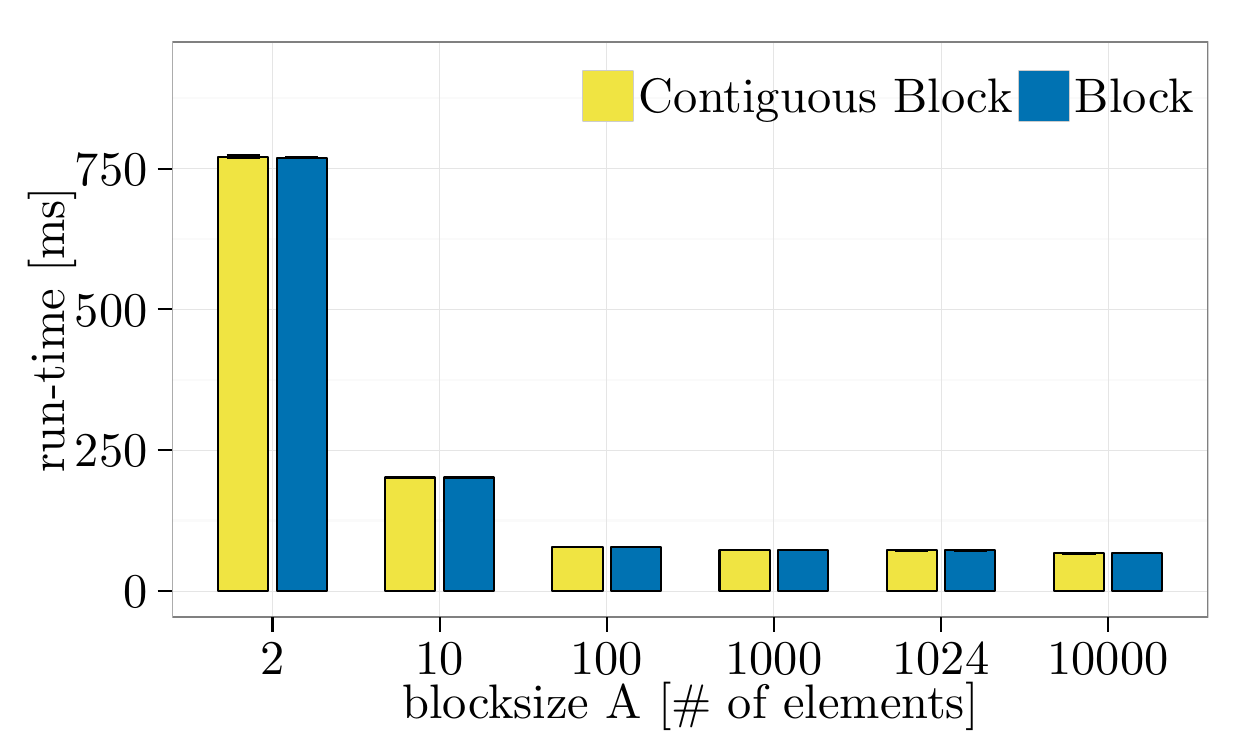}
\caption{%
\label{exp:vsc3-allgather-contigblock-32x1-large}%
\dtblock%
}%
\end{subfigure}%
\hfill%
\begin{subfigure}{.24\linewidth}
\centering
\includegraphics[width=\linewidth]{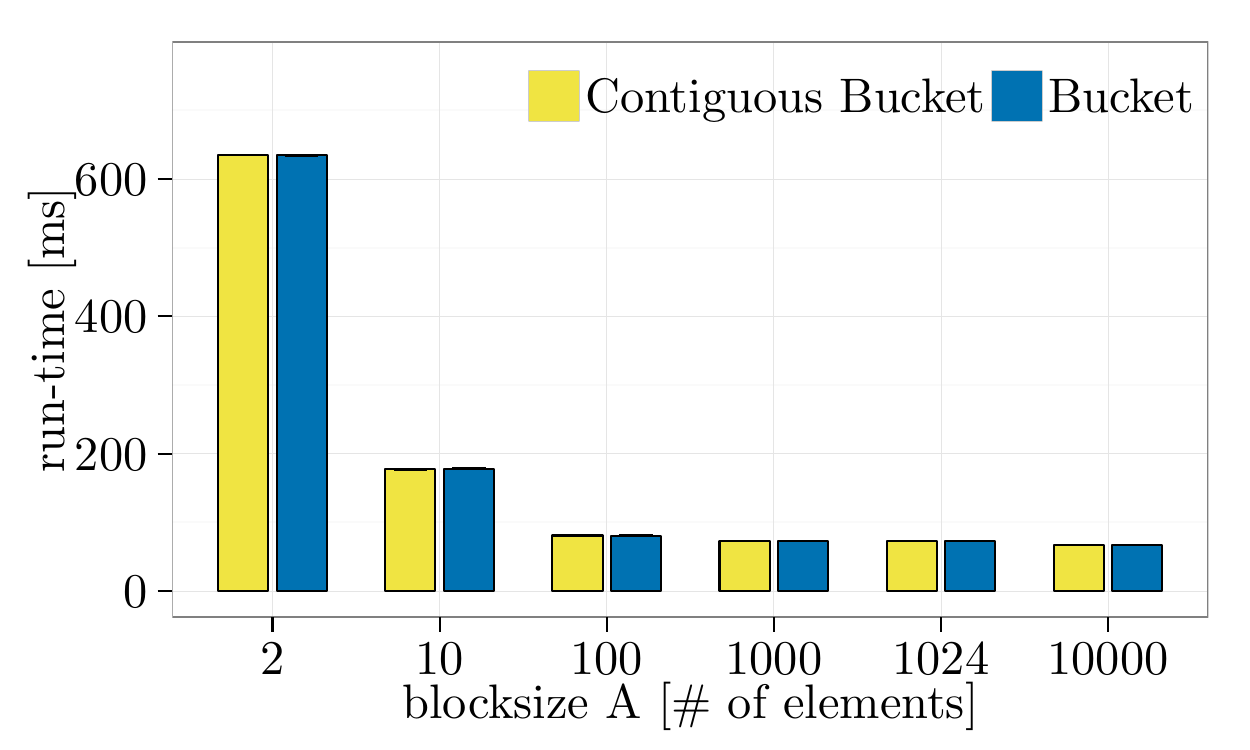}
\caption{%
\label{exp:vsc3-allgather-contigbucket-32x1-large}%
\dtbucket%
}%
\end{subfigure}%
\hfill%
\begin{subfigure}{.24\linewidth}
\centering
\includegraphics[width=\linewidth]{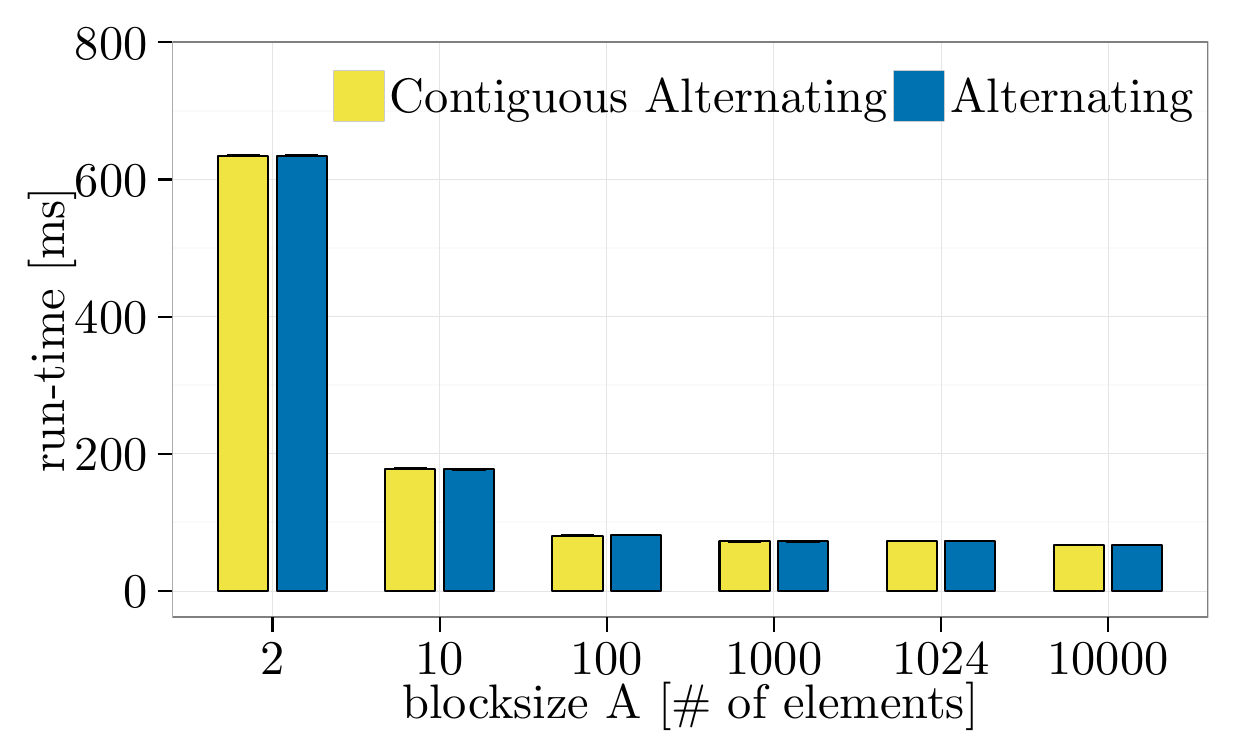}
\caption{%
\label{exp:vsc3-allgather-contigalternating-32x1-large}%
\dtalternating%
}%
\end{subfigure}%
\caption{\label{exp:vsc3-allgather-contig-largenbytes-32x1}  Basic layouts \vs \ddtcontig, $\VARdatasize=\SI{2.56}{\mega\byte}$, element datatype: \mpiint, \num{32x1}~processes, \mpiallgather, \vscintelmpi (similar results for \num{1x16}~processes).}
\end{figure*}

\begin{figure*}[htpb]
\centering
\begin{subfigure}{.24\linewidth}
\centering
\includegraphics[width=\linewidth]{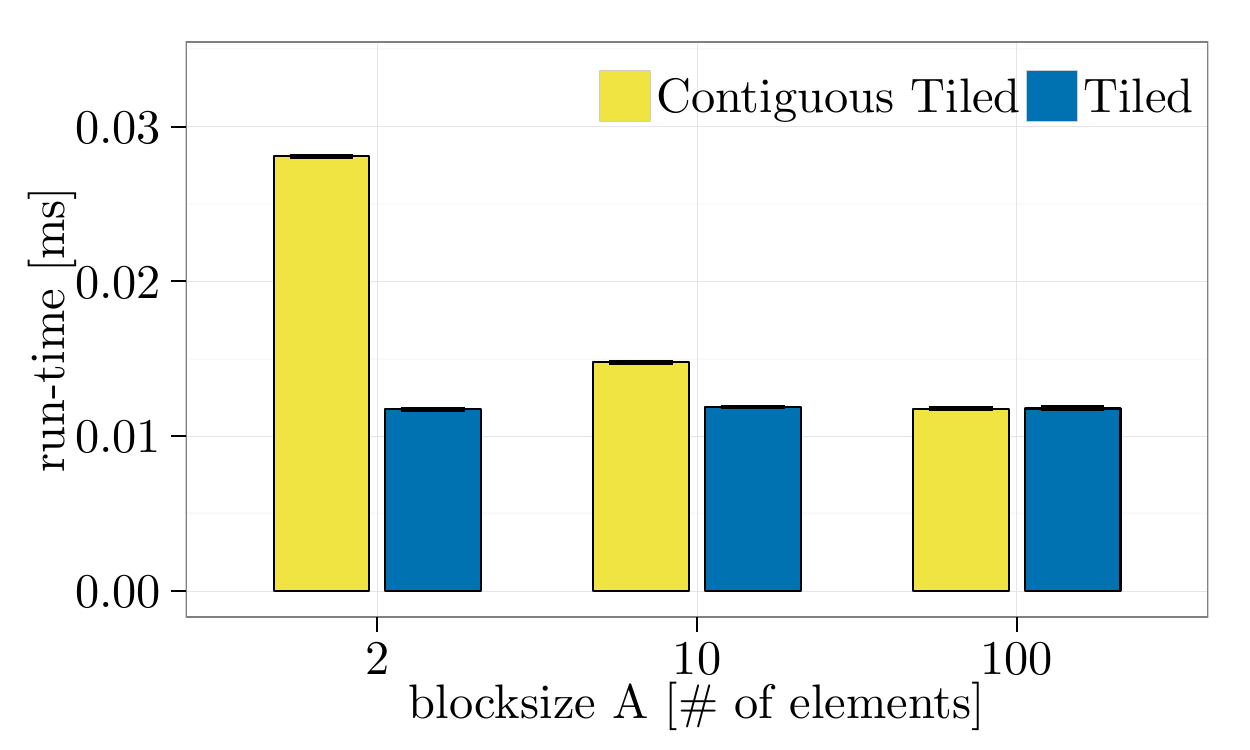}
\caption{%
\label{exp:vsc3-bcast-contigtiled-32x1}%
\dttiled%
}%
\end{subfigure}%
\hfill%
\begin{subfigure}{.24\linewidth}
\centering
\includegraphics[width=\linewidth]{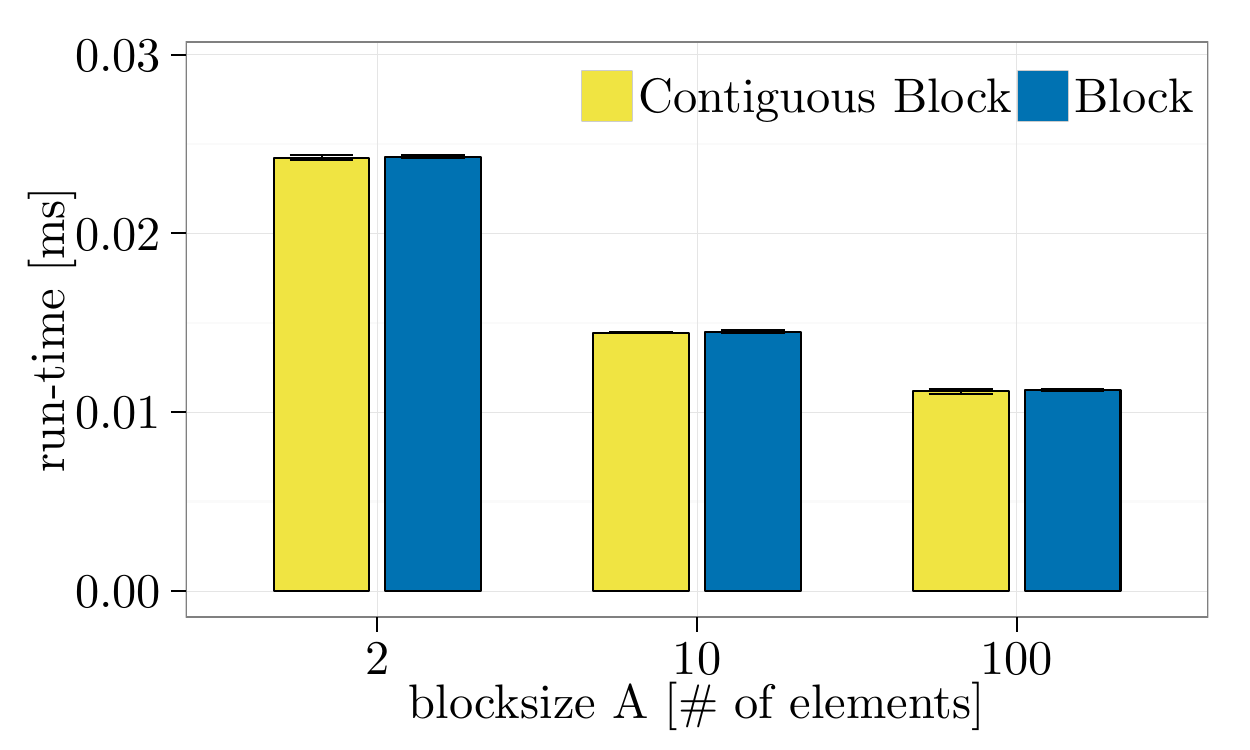}
\caption{%
\label{exp:vsc3-bcast-contigblock-32x1}%
\dtblock%
}%
\end{subfigure}%
\hfill%
\begin{subfigure}{.24\linewidth}
\centering
\includegraphics[width=\linewidth]{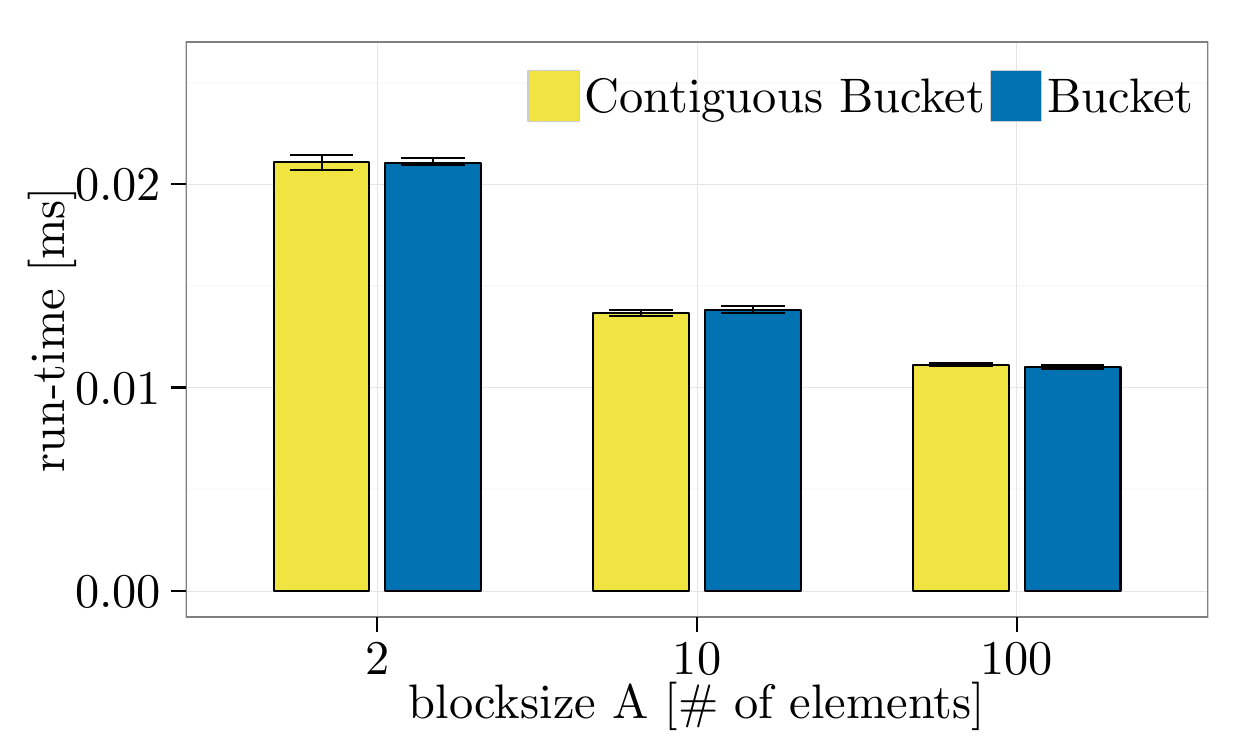}
\caption{%
\label{exp:vsc3-bcast-contigbucket-32x1}%
\dtbucket%
}%
\end{subfigure}%
\hfill%
\begin{subfigure}{.24\linewidth}
\centering
\includegraphics[width=\linewidth]{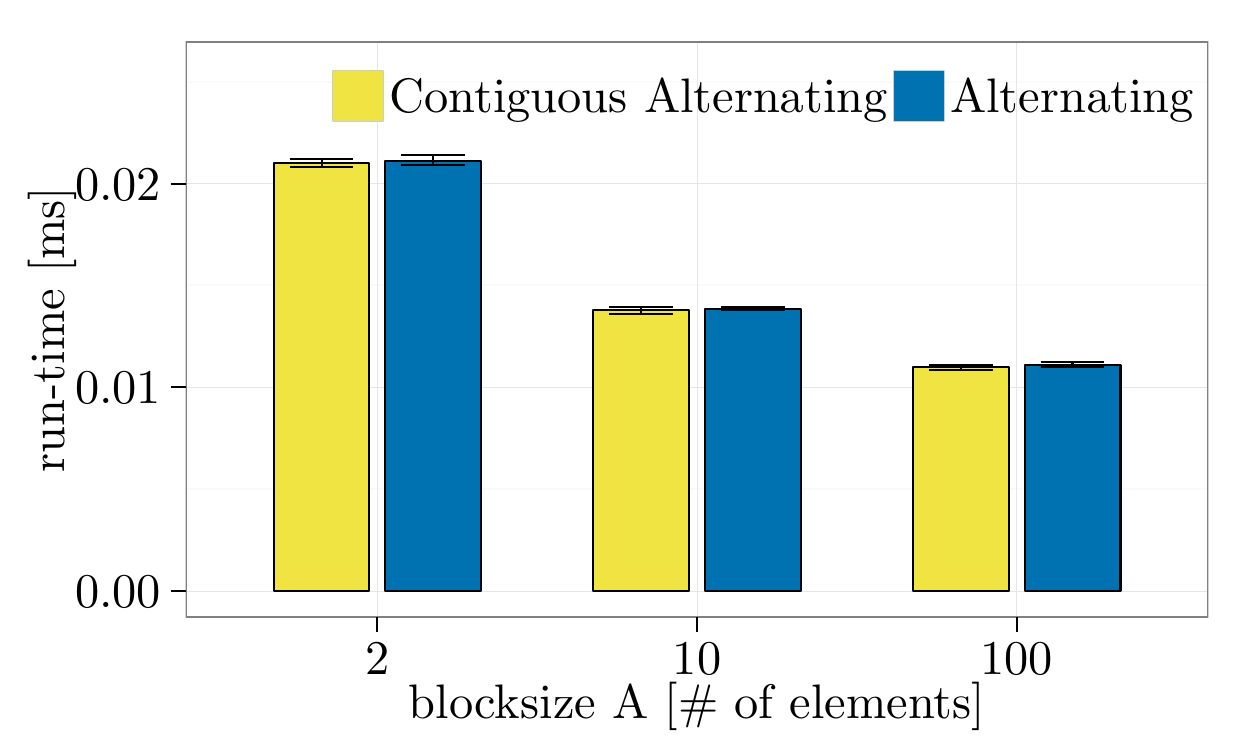}
\caption{%
\label{exp:vsc3-bcast-contigalternating-32x1}%
\dtalternating%
}%
\end{subfigure}%
\caption{\label{exp:vsc3-bcast-contig-smallnbytes-32x1}  Basic layouts \vs \ddtcontig, $\VARdatasize=\SI{2}{\kilo\byte}$, element datatype: \mpiint, \num{32x1}~processes, \mpibcast, \vscintelmpi (similar results for \num{1x16}~processes).}
\end{figure*}

\begin{figure*}[htpb]
\centering
\begin{subfigure}{.24\linewidth}
\centering
\includegraphics[width=\linewidth]{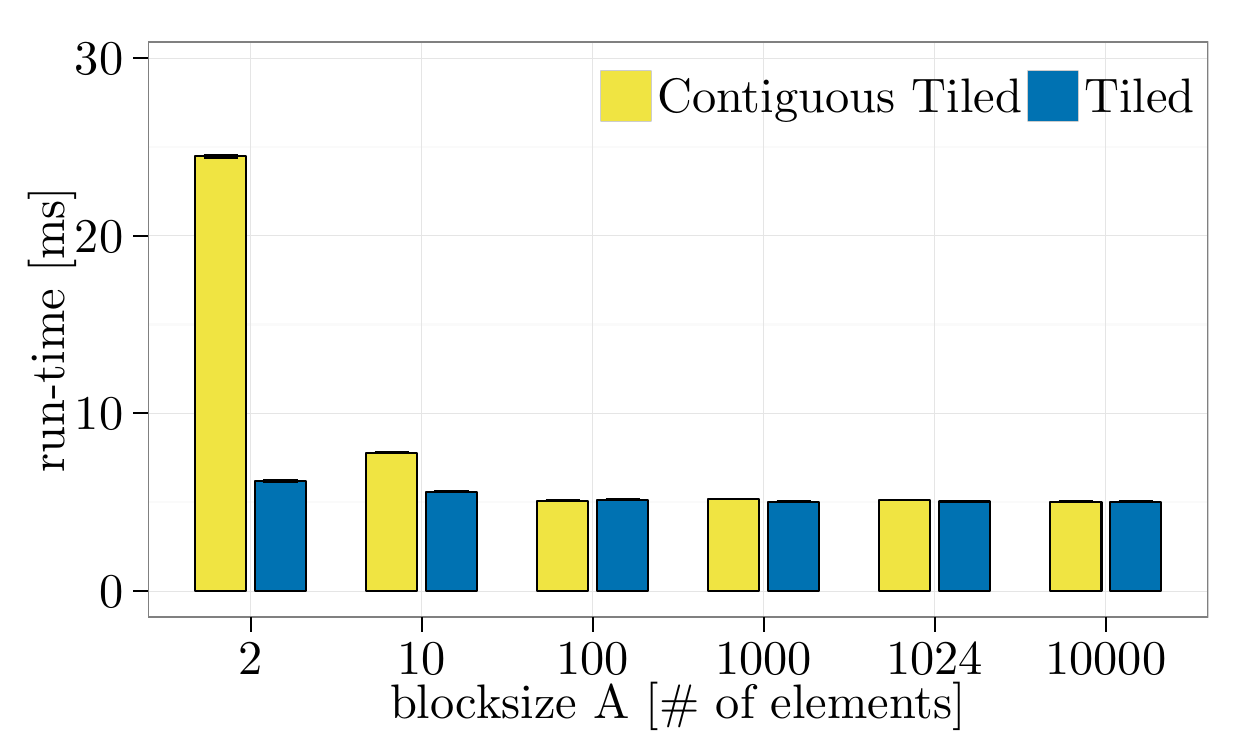}
\caption{%
\label{exp:vsc3-bcast-contigtiled-32x1-large}%
\dttiled%
}%
\end{subfigure}%
\hfill%
\begin{subfigure}{.24\linewidth}
\centering
\includegraphics[width=\linewidth]{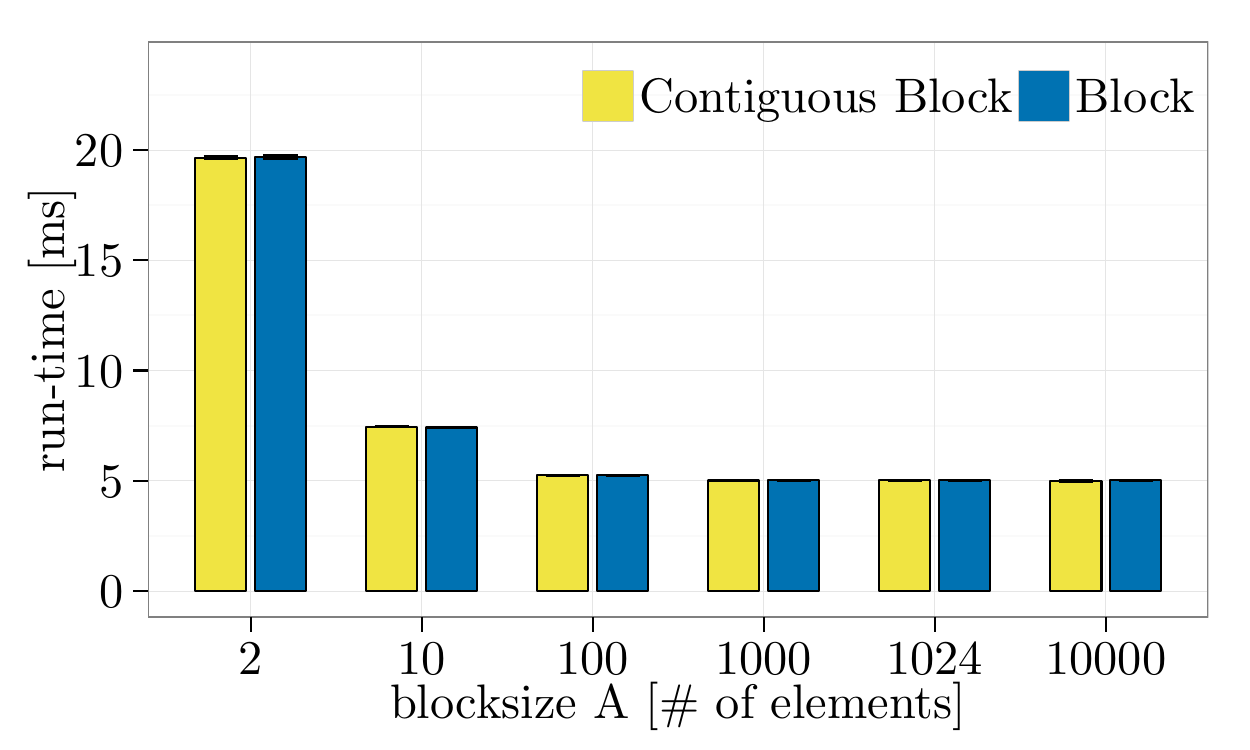}
\caption{%
\label{exp:vsc3-bcast-contigblock-32x1-large}%
\dtblock%
}%
\end{subfigure}%
\hfill%
\begin{subfigure}{.24\linewidth}
\centering
\includegraphics[width=\linewidth]{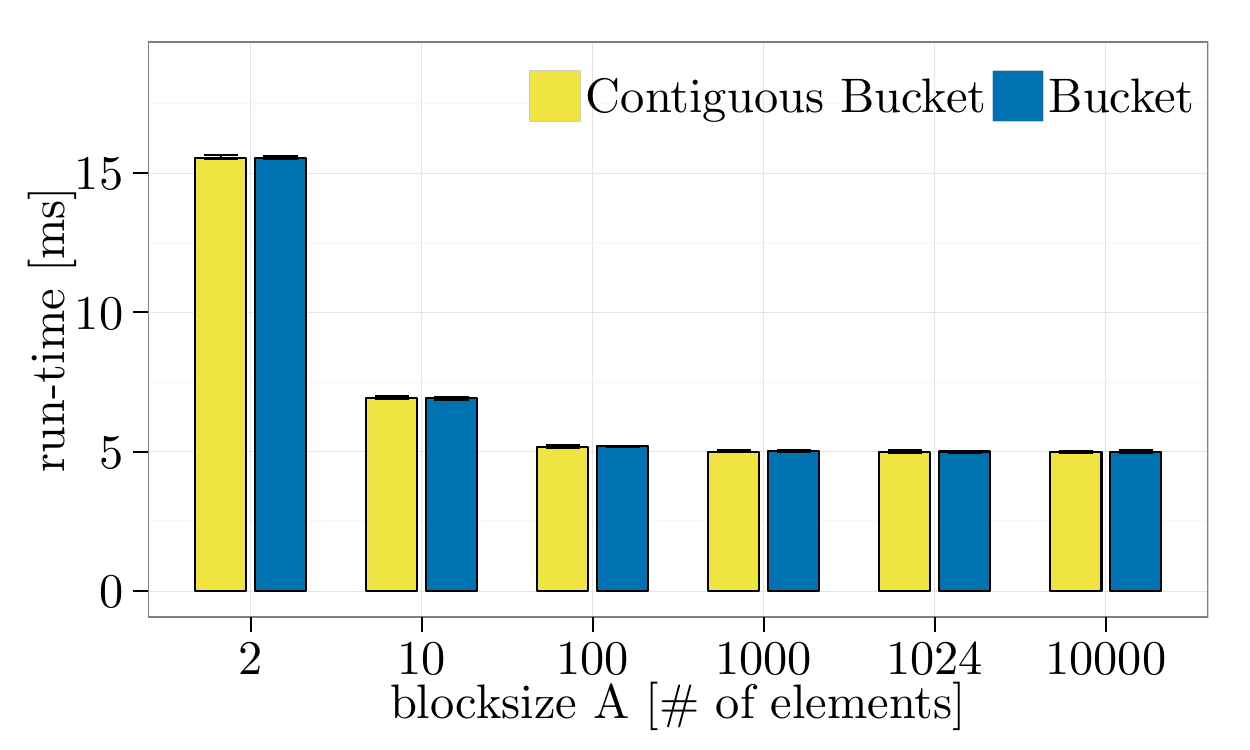}
\caption{%
\label{exp:vsc3-bcast-contigbucket-32x1-large}%
\dtbucket%
}%
\end{subfigure}%
\hfill%
\begin{subfigure}{.24\linewidth}
\centering
\includegraphics[width=\linewidth]{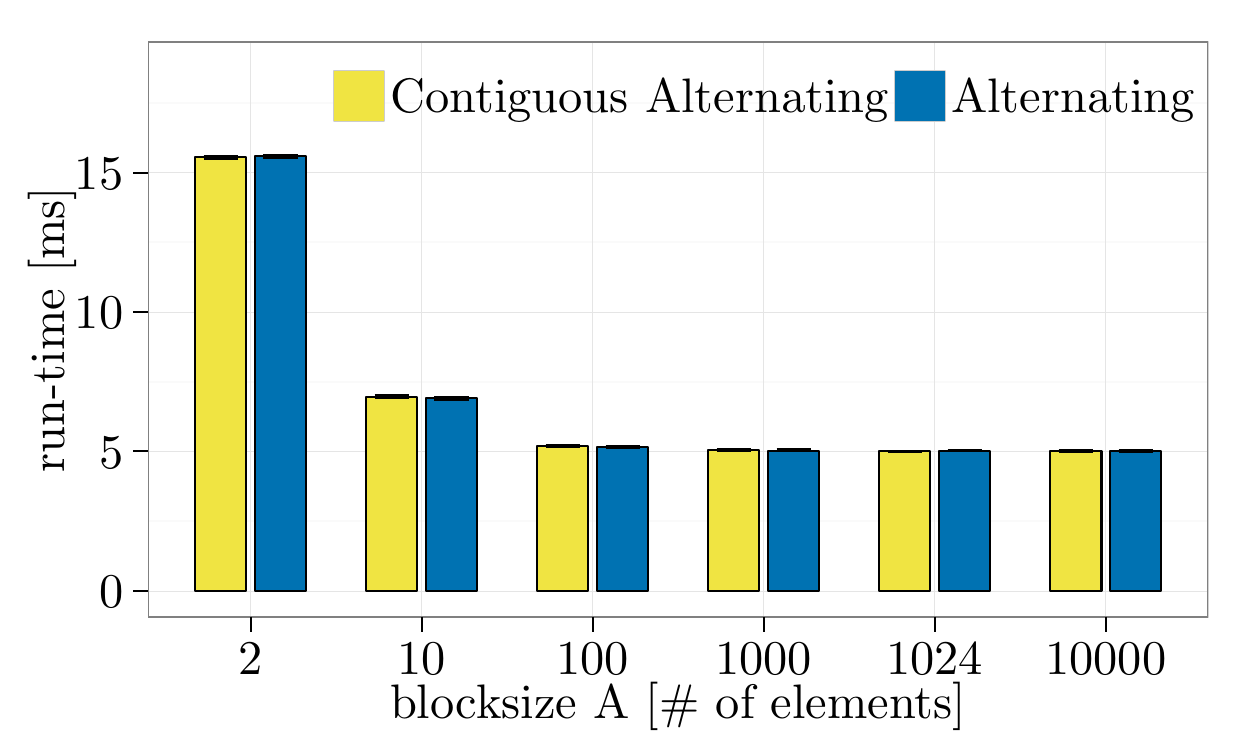}
\caption{%
\label{exp:vsc3-bcast-contigalternating-32x1-large}%
\dtalternating%
}%
\end{subfigure}%
\caption{\label{exp:vsc3-bcast-contig-largenbytes-32x1}  Basic layouts \vs \ddtcontig, $\VARdatasize=\SI{2.56}{\mega\byte}$, element datatype: \mpiint, \num{32x1}~processes, \mpibcast, \vscintelmpi (similar results for \num{1x16}~processes).}
\end{figure*}

\appexpdesc{
  \begin{expitemize}
    \item \ddtcontig \vs basic layouts (\dttiled, \dtblock, \dtbucket, \dtalternating)
    \item \pingpong, \mpiallgather, \mpibcast
  \end{expitemize}
}{
  \begin{expitemize}
    \item \expparam{\vscintelmpi, \pingpong, small \datasize}{\fig~\ref{exp:vsc3-pingpong-contig-smallnbytes-2x1}}
    \item \expparam{\vscintelmpi, \pingpong, large \datasize}{\fig~\ref{exp:vsc3-pingpong-contig-largenbytes-2x1}}
    \item \expparam{\vscintelmpi, \mpiallgather, small \datasize}{\fig~\ref{exp:vsc3-allgather-contig-smallnbytes-32x1}}
    \item \expparam{\vscintelmpi, \mpiallgather, large \datasize}{\fig~\ref{exp:vsc3-allgather-contig-largenbytes-32x1}}
   \item \expparam{\vscintelmpi, \mpibcast, small \datasize}{\fig~\ref{exp:vsc3-bcast-contig-smallnbytes-32x1}}
    \item \expparam{\vscintelmpi, \mpibcast, large \datasize}{\fig~\ref{exp:vsc3-bcast-contig-largenbytes-32x1}}
  \end{expitemize}  
}

\FloatBarrier
\clearpage

\appexp{exptest:vector_tiled}

\begin{figure*}[htpb]
\centering
\begin{subfigure}{.24\linewidth}
\centering
\includegraphics[width=\linewidth]{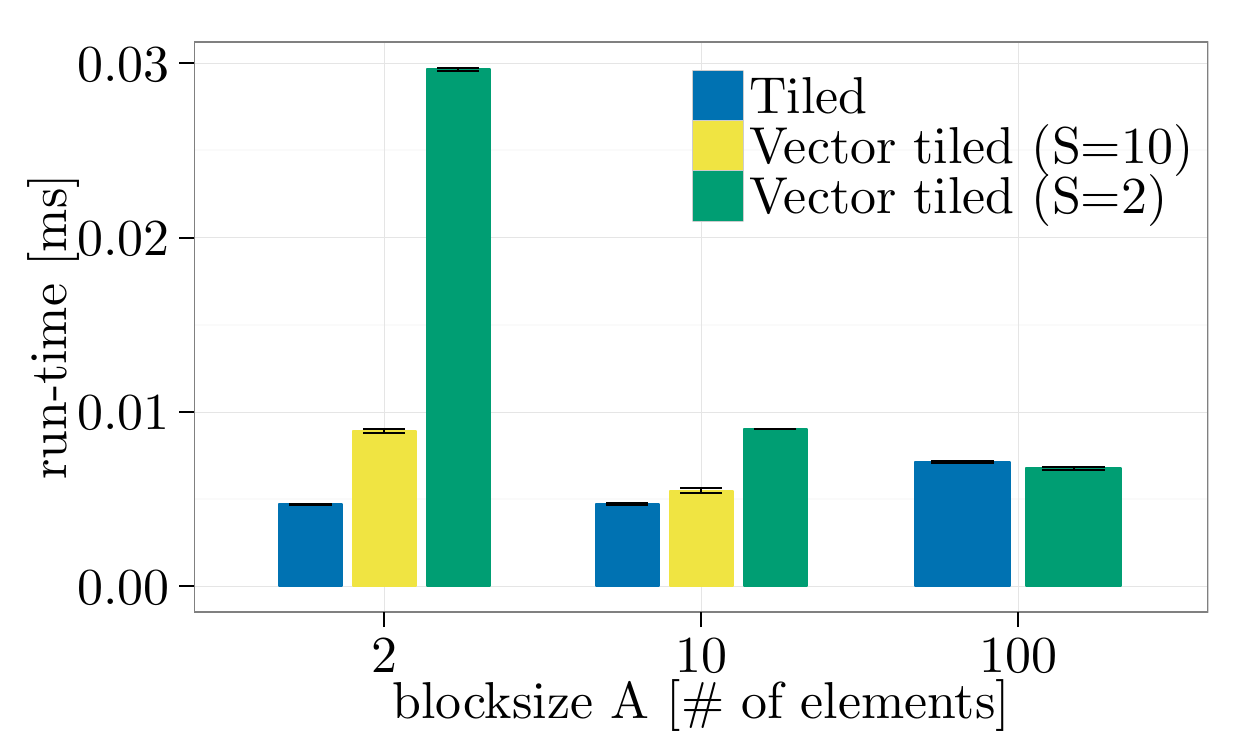}
\caption{%
\label{exp:vsc3-pingpong-vectortiled-small-2x1}%
$\VARdatasize=\SI{2}{\kilo\byte}$, \num{2}~nodes%
}%
\end{subfigure}%
\hfill%
\begin{subfigure}{.24\linewidth}
\centering
\includegraphics[width=\linewidth]{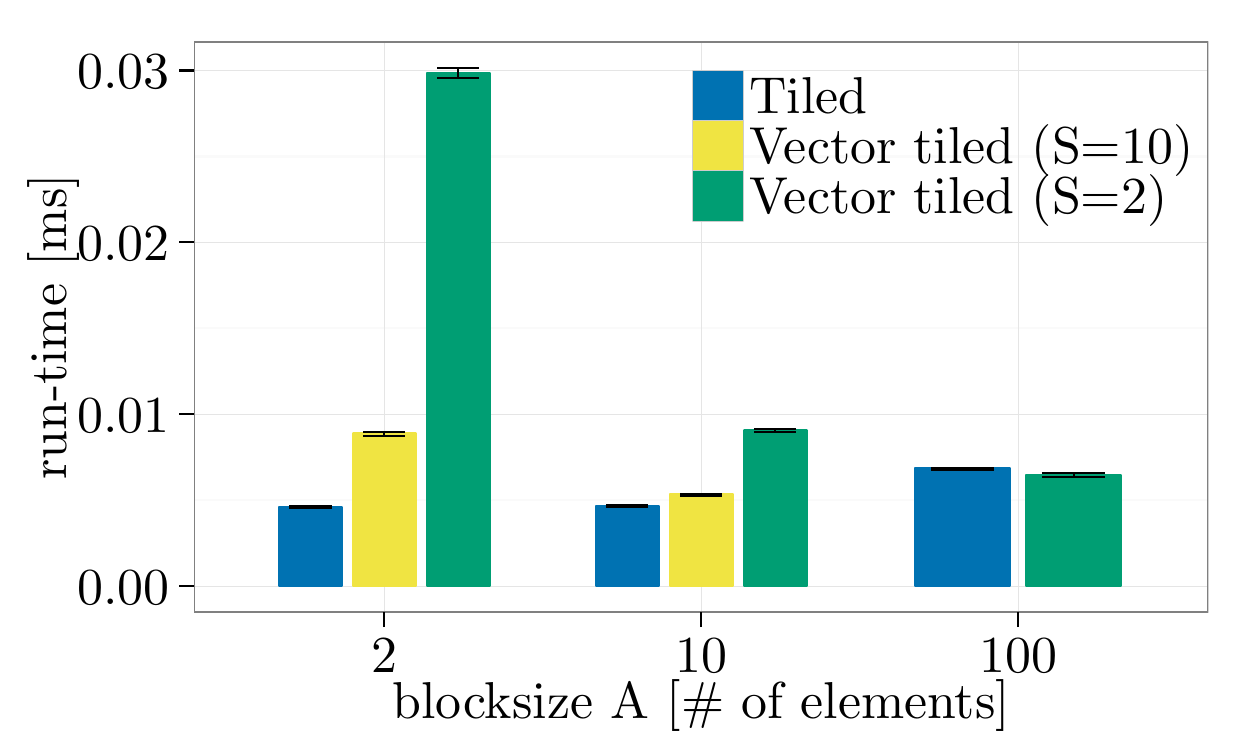}
\caption{%
\label{exp:vsc3-pingpong-vectortiled-small-1x2}%
$\VARdatasize=\SI{2}{\kilo\byte}$, same node%
}%
\end{subfigure}%
\hfill%
\begin{subfigure}{.24\linewidth}
\centering
\includegraphics[width=\linewidth]{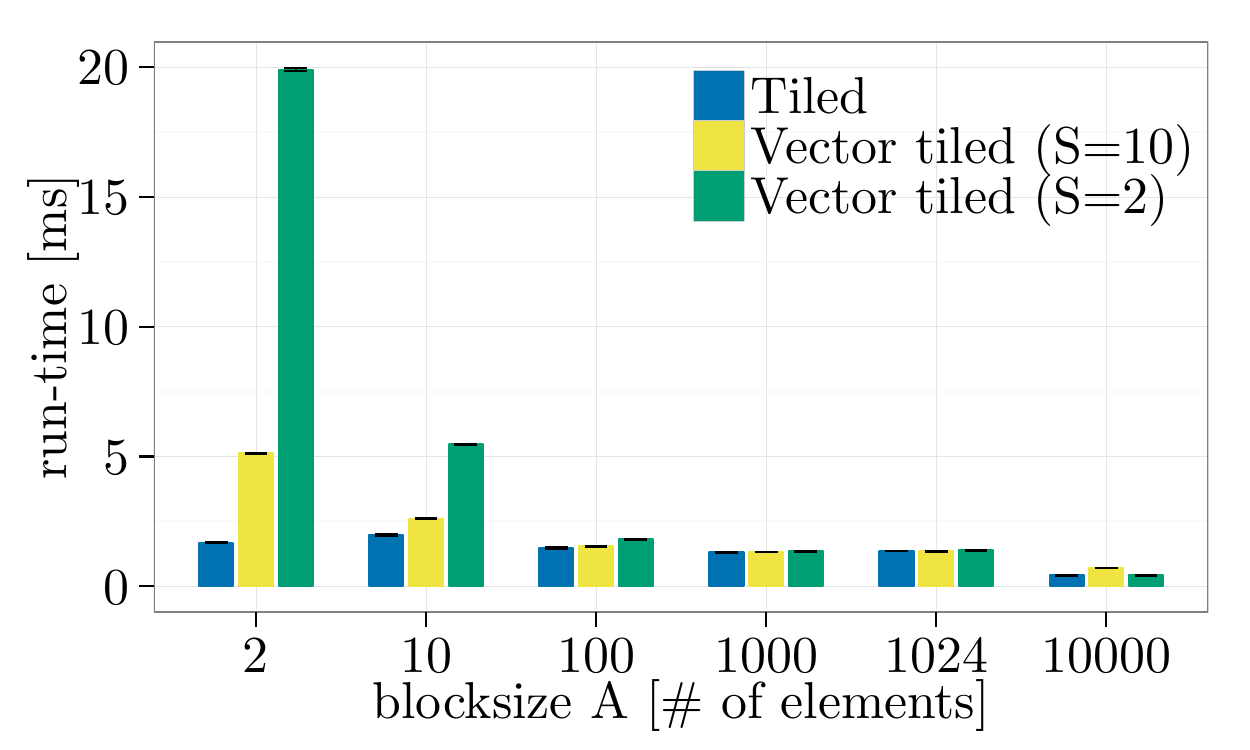}
\caption{%
\label{exp:vsc3-pingpong-vectortiled-large-2x1}%
$\VARdatasize=\SI{2.56}{\mega\byte}$, \num{2}~nodes%
}%
\end{subfigure}%
\hfill%
\begin{subfigure}{.24\linewidth}
\centering
\includegraphics[width=\linewidth]{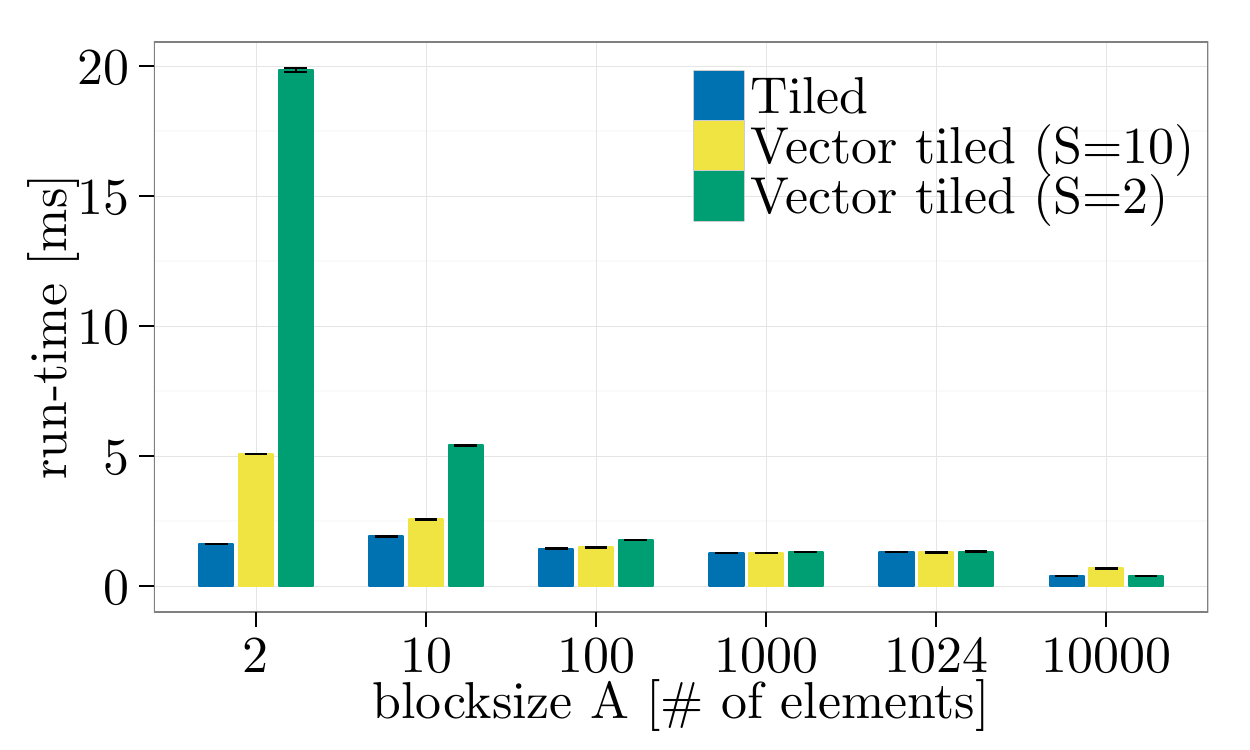}
\caption{%
\label{exp:vsc3-pingpong-vectortiled-large-1x2}%
$\VARdatasize=\SI{2.56}{\mega\byte}$, same node%
}%
\end{subfigure}%
\caption{\label{exp:vsc3-pingpong-vectortiled} \dtdtiled \vs \ddtvectortiled, element datatype: \mpiint, \pingpong, \vscintelmpi.}
\end{figure*}

\appexpdesc{
  \begin{expitemize}
    \item \dtdtiled, \ddtvectortiled
    \item \pingpong
  \end{expitemize}
}{
  \begin{expitemize}
    \item \expparam{\vscintelmpi}{\fig~\ref{exp:vsc3-pingpong-vectortiled}}
  \end{expitemize}  
}

\FloatBarrier
\clearpage

\appexp{exptest:block_indexed}

\begin{figure*}[htpb]
\centering
\begin{subfigure}{.24\linewidth}
\centering
\includegraphics[width=\linewidth]{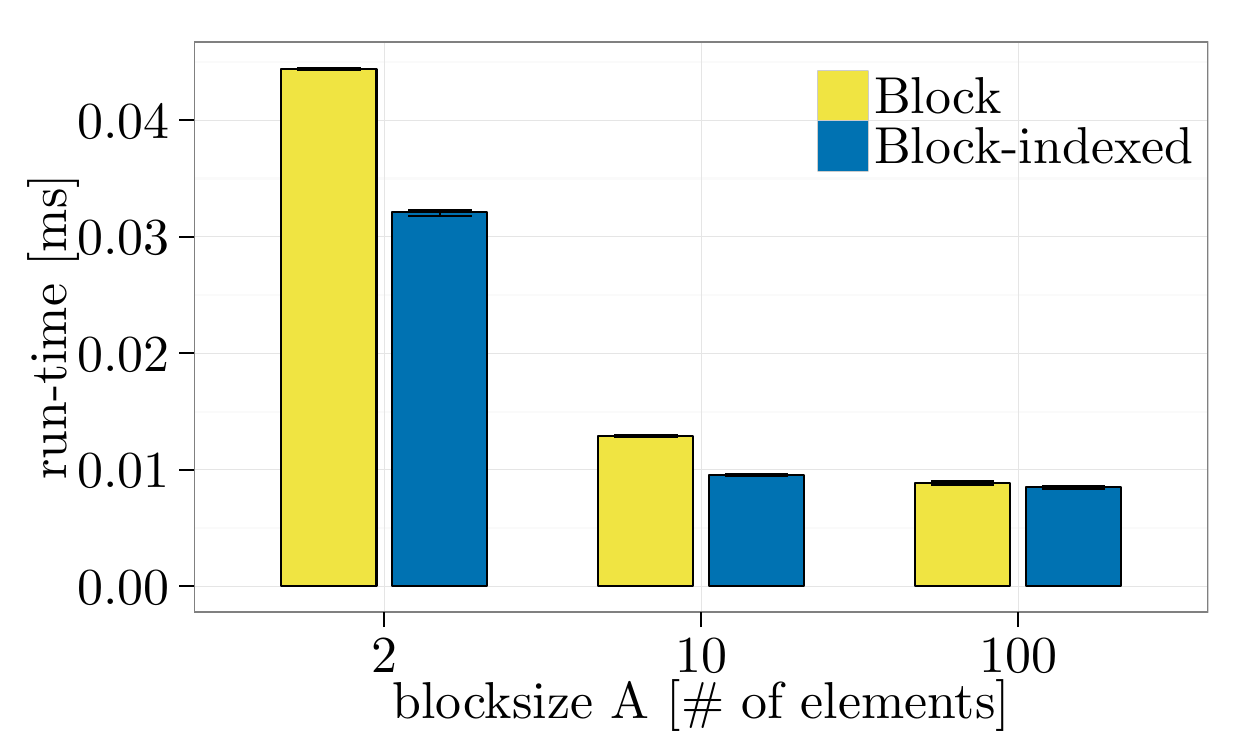}
\caption{%
\label{exp:vsc3-pingpong-blockindexed-small-2x1}%
$\VARdatasize=\SI{3.2}{\kilo\byte}$, \num{2}~nodes%
}%
\end{subfigure}%
\hfill%
\begin{subfigure}{.24\linewidth}
\centering
\includegraphics[width=\linewidth]{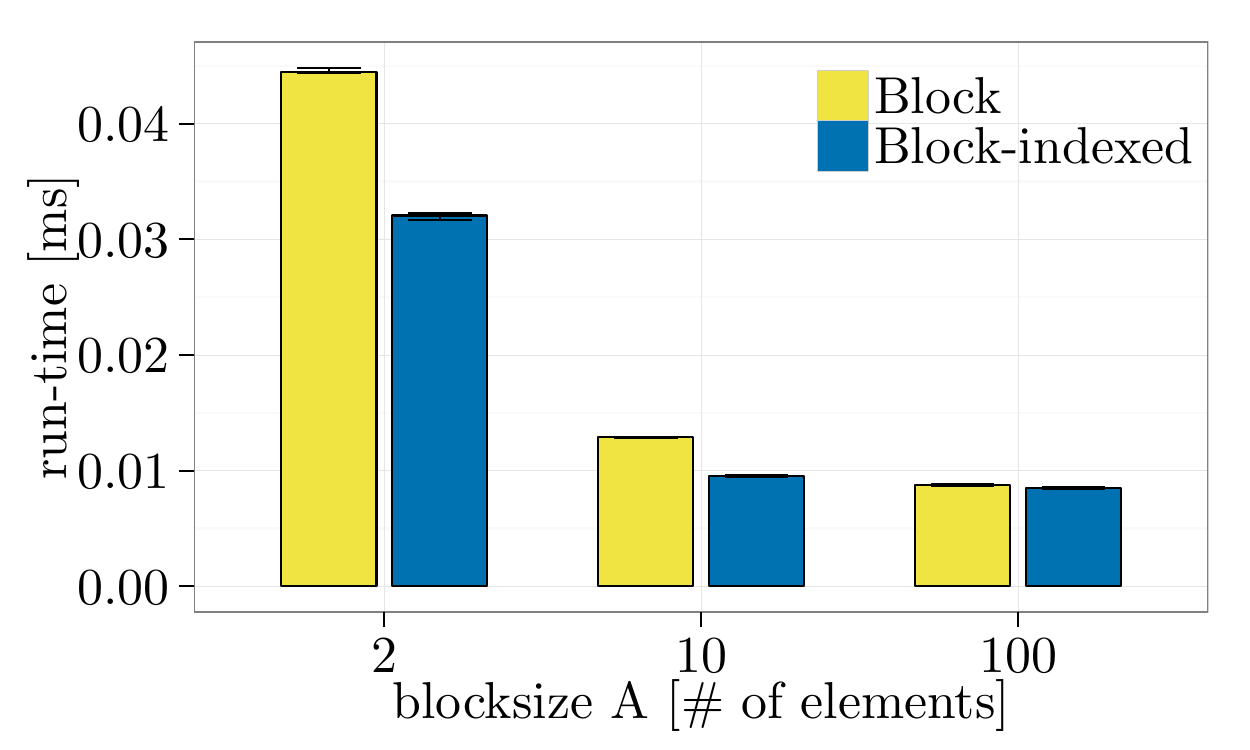}
\caption{%
\label{exp:vsc3-pingpong-blockindexed-small-1x2}%
$\VARdatasize=\SI{3.2}{\kilo\byte}$, same node%
}%
\end{subfigure}%
\hfill%
\begin{subfigure}{.24\linewidth}
\centering
\includegraphics[width=\linewidth]{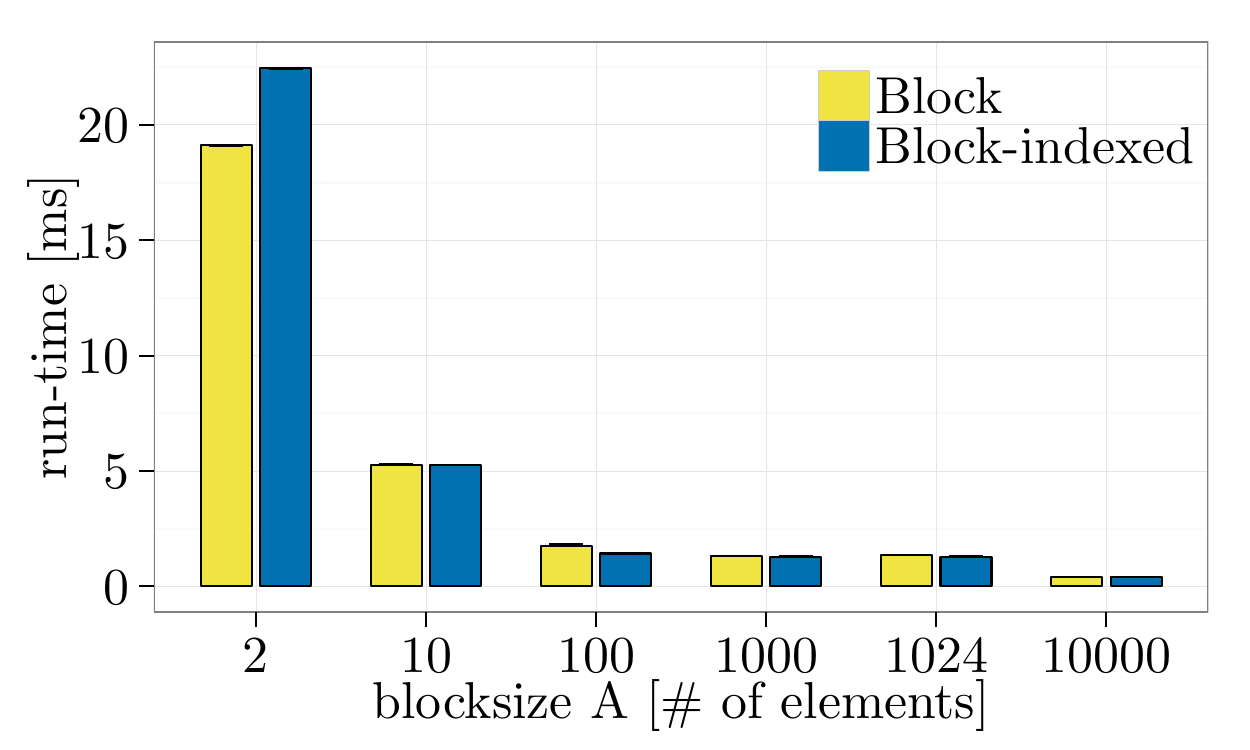}
\caption{%
\label{exp:vsc3-pingpong-blockindexed-large-2x1}%
$\VARdatasize=\SI{2.56}{\mega\byte}$, \num{2}~nodes%
}%
\end{subfigure}%
\hfill%
\begin{subfigure}{.24\linewidth}
\centering
\includegraphics[width=\linewidth]{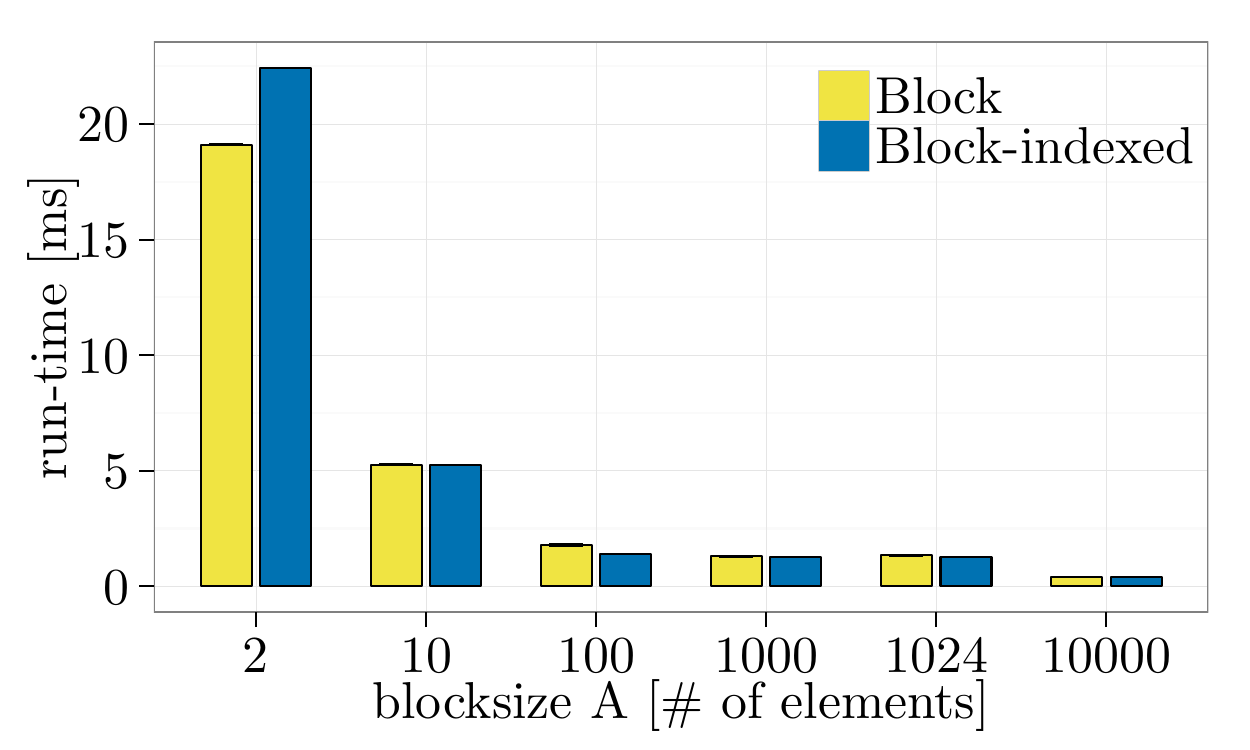}
\caption{%
\label{exp:vsc3-pingpong-blockindexed-large-1x2}%
$\VARdatasize=\SI{2.56}{\mega\byte}$, same node%
}%
\end{subfigure}%
\caption{\label{exp:vsc3-pingpong-blockindexed} \dtblock \vs \ddtblockindexed, element datatype: \mpiint, \pingpong, \vscintelmpi.}
\end{figure*}

\appexpdesc{
  \begin{expitemize}
    \item \dtblock, \ddtblockindexed
    \item \pingpong
  \end{expitemize}
}{
  \begin{expitemize}
    \item \expparam{\vscintelmpi}{\fig~\ref{exp:vsc3-pingpong-blockindexed}}
  \end{expitemize}  
}

\FloatBarrier
\clearpage

\appexp{exptest:alternating_indexed}

\begin{figure*}[htpb]
\centering
\begin{subfigure}{.24\linewidth}
\centering
\includegraphics[width=\linewidth]{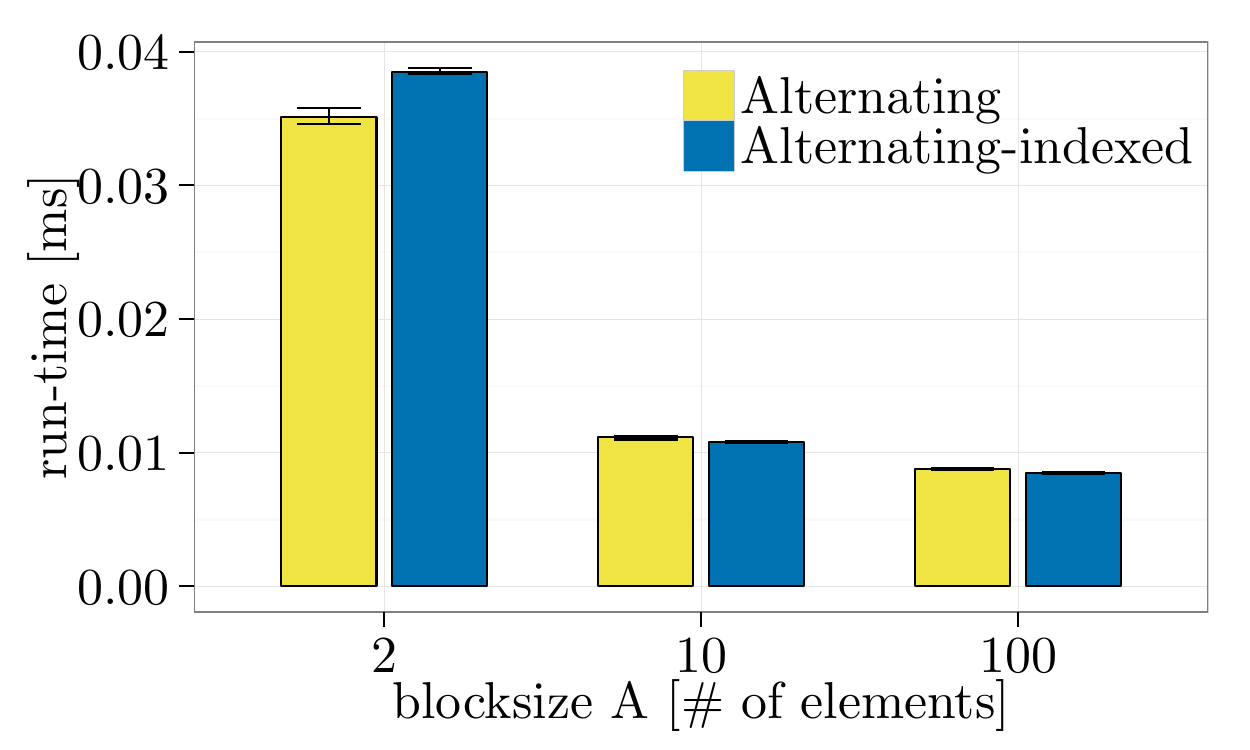}
\caption{%
\label{exp:vsc3-pingpong-alternatingindexed-small-2x1}%
$\VARdatasize=\SI{3.2}{\kilo\byte}$, \num{2}~nodes%
}%
\end{subfigure}%
\hfill%
\begin{subfigure}{.24\linewidth}
\centering
\includegraphics[width=\linewidth]{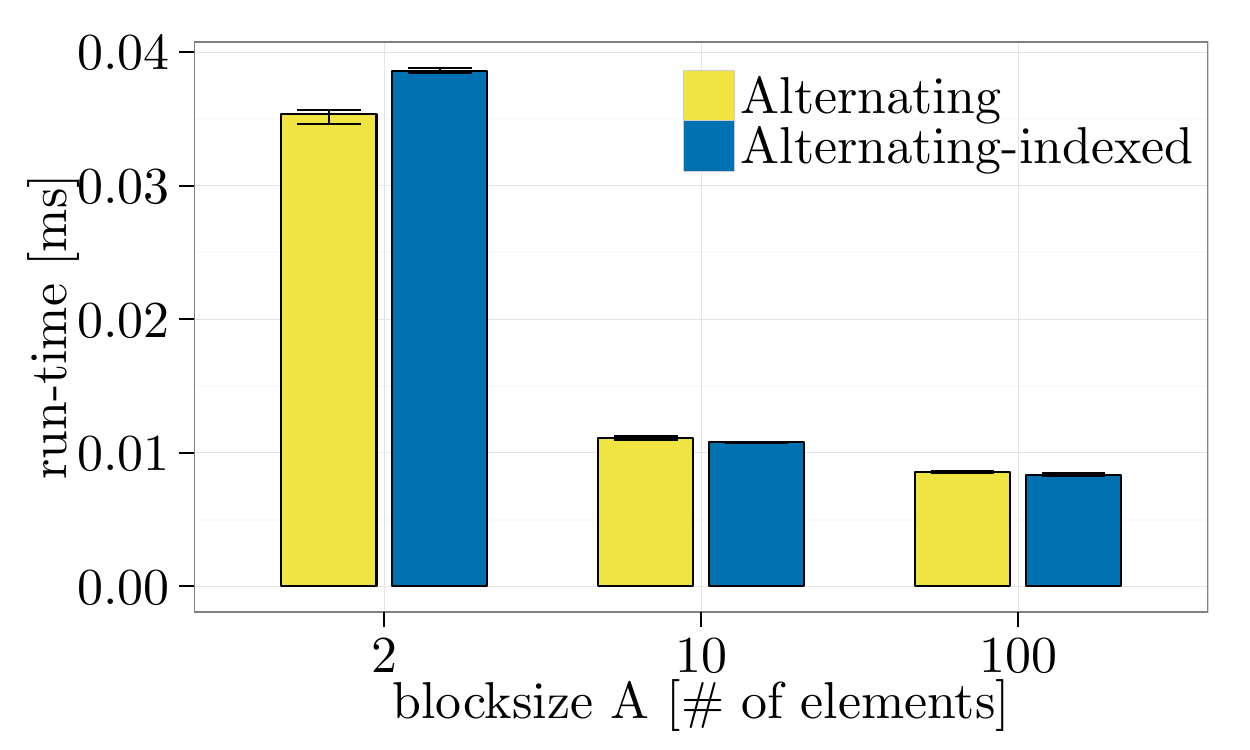}
\caption{%
\label{exp:vsc3-pingpong-alternatingindexed-small-1x2}%
$\VARdatasize=\SI{3.2}{\kilo\byte}$, same node%
}%
\end{subfigure}%
\hfill%
\begin{subfigure}{.24\linewidth}
\centering
\includegraphics[width=\linewidth]{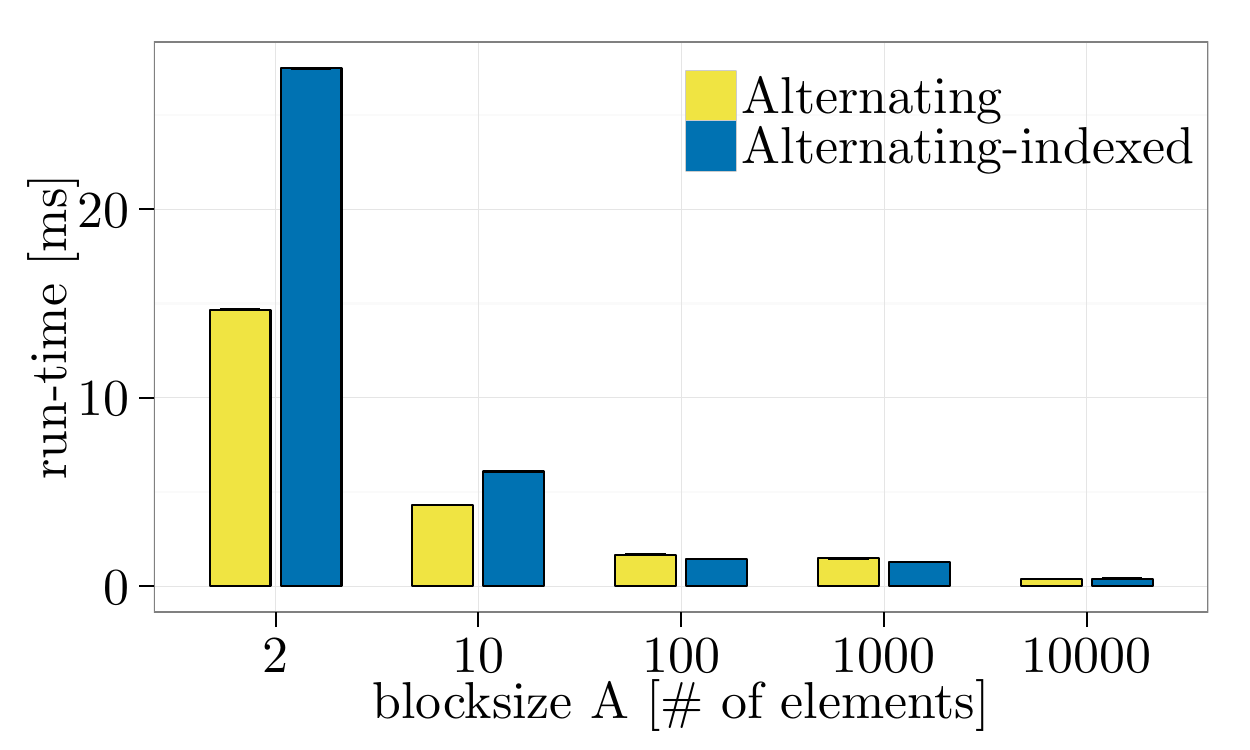}
\caption{%
\label{exp:vsc3-pingpong-alternatingindexed-large-2x1}%
$\VARdatasize=\SI{2.56}{\mega\byte}$, \num{2}~nodes%
}%
\end{subfigure}%
\hfill%
\begin{subfigure}{.24\linewidth}
\centering
\includegraphics[width=\linewidth]{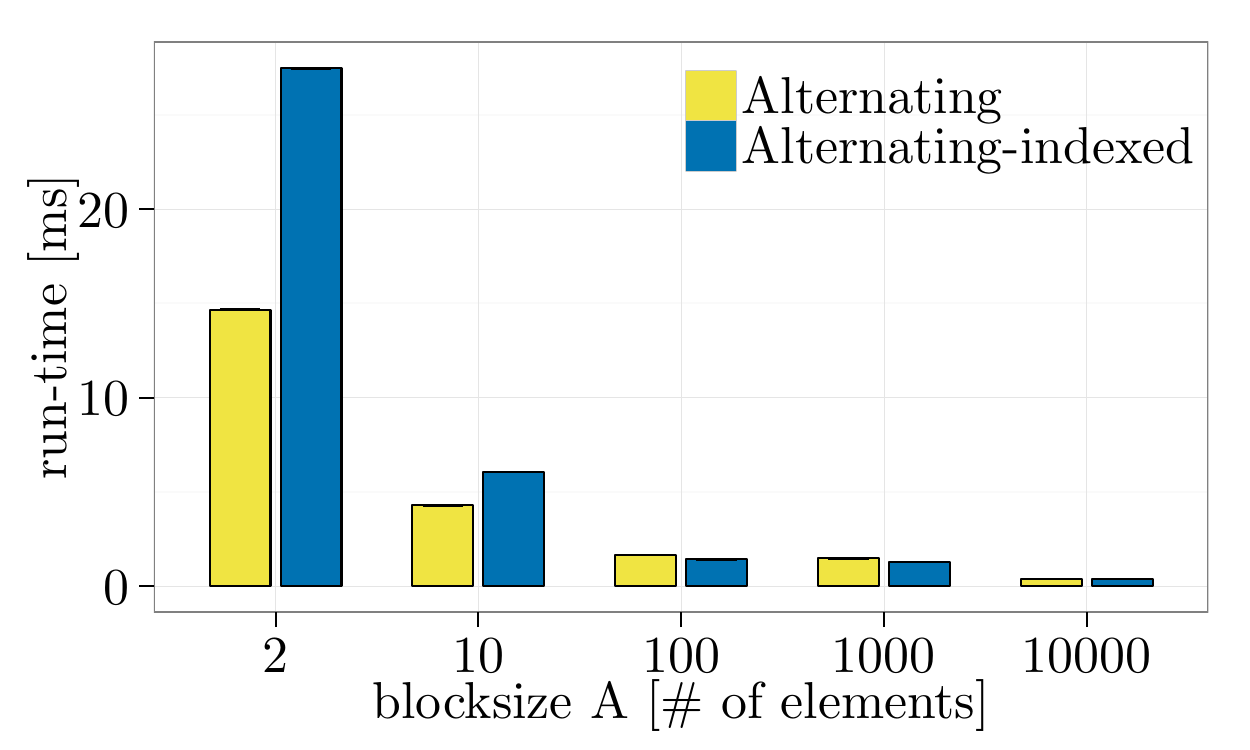}
\caption{%
\label{exp:vsc3-pingpong-alternatingindexed-large-1x2}%
$\VARdatasize=\SI{2.56}{\mega\byte}$, same node%
}%
\end{subfigure}%
\caption{\label{exp:vsc3-pingpong-alternatingindexed} \dtalternating \vs \ddtalternatingindexed, element datatype: \mpiint, \pingpong, \jupiternecmpi.}
\end{figure*}

\appexpdesc{
  \begin{expitemize}
    \item \dtalternating, \ddtalternatingindexed
    \item \pingpong
  \end{expitemize}
}{
  \begin{expitemize}
    \item \expparam{\vscintelmpi}{\fig~\ref{exp:vsc3-pingpong-alternatingindexed}}
  \end{expitemize}  
}

\FloatBarrier
\clearpage

\begin{figure*}[htpb]
\centering
\begin{subfigure}{.24\linewidth}
\centering
\includegraphics[width=\linewidth]{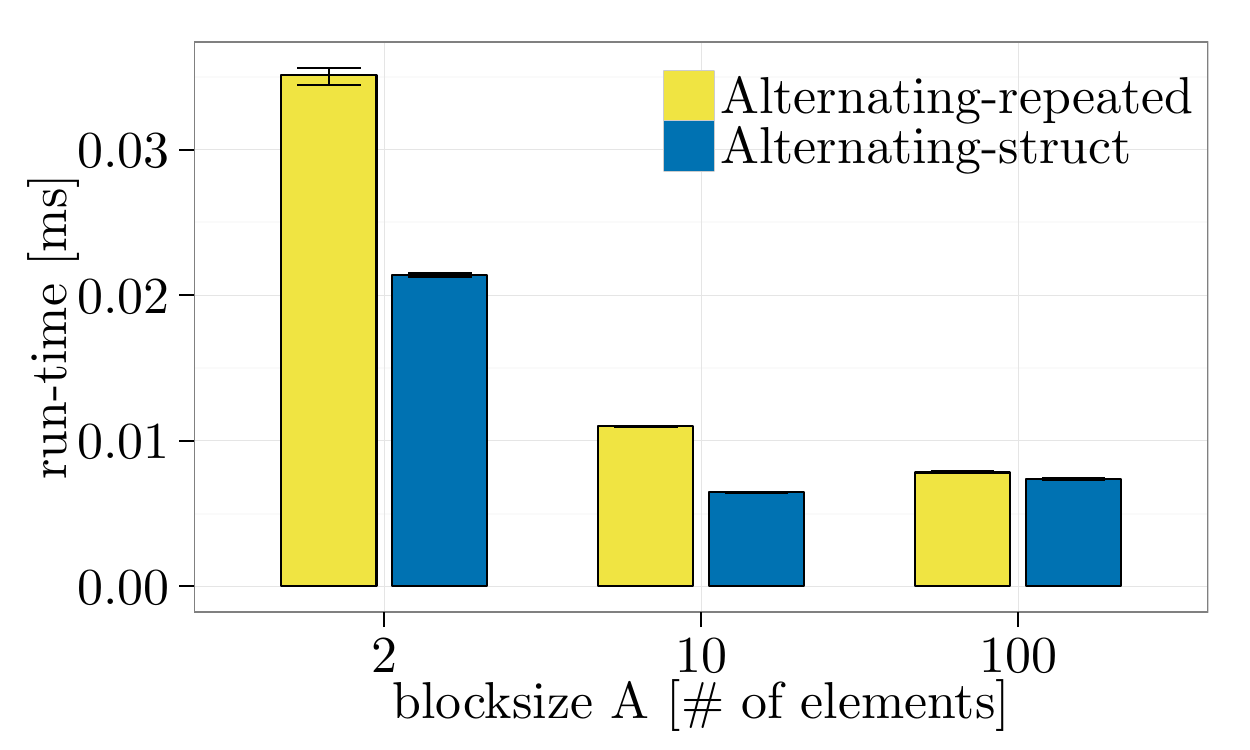}
\caption{%
\label{exp:vsc3-pingpong-alternatingrepeated-small-2x1}%
$\VARdatasize=\SI{3.2}{\kilo\byte}$, \num{2}~nodes%
}%
\end{subfigure}%
\hfill%
\begin{subfigure}{.24\linewidth}
\centering
\includegraphics[width=\linewidth]{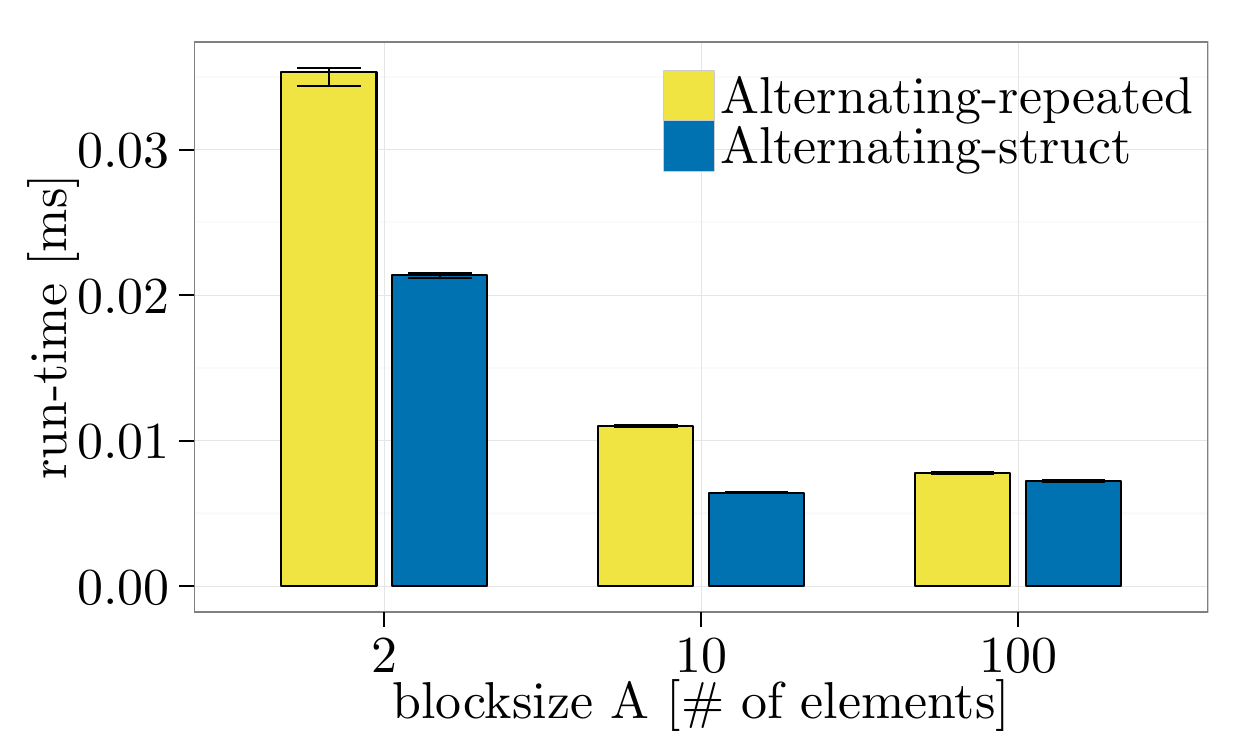}
\caption{%
\label{exp:vsc3-pingpong-alternatingrepeated-small-1x2}%
$\VARdatasize=\SI{3.2}{\kilo\byte}$, same node%
}%
\end{subfigure}%
\hfill%
\begin{subfigure}{.24\linewidth}
\centering
\includegraphics[width=\linewidth]{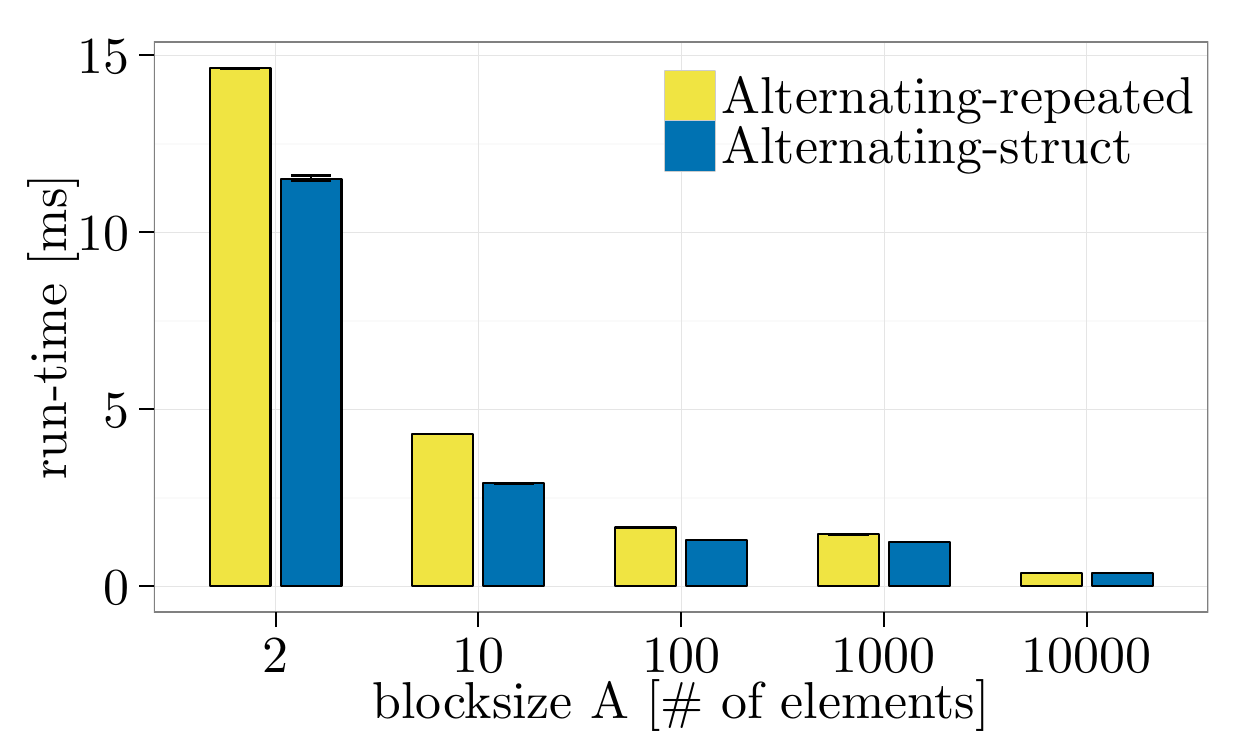}
\caption{%
\label{exp:vsc3-pingpong-alternatingrepeated-large-2x1}%
$\VARdatasize=\SI{2.56}{\mega\byte}$, \num{2}~nodes%
}%
\end{subfigure}%
\hfill%
\begin{subfigure}{.24\linewidth}
\centering
\includegraphics[width=\linewidth]{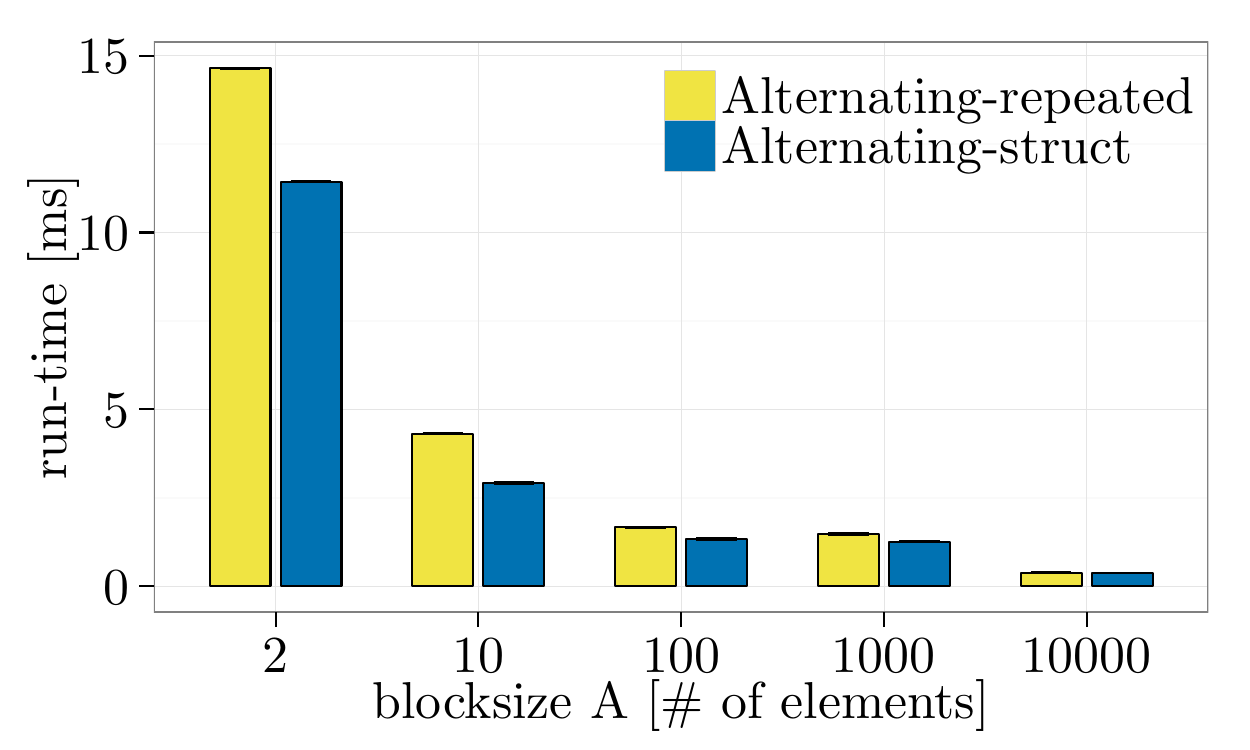}
\caption{%
\label{exp:vsc3-pingpong-alternatingrepeated-large-1x2}%
$\VARdatasize=\SI{2.56}{\mega\byte}$, same node%
}%
\end{subfigure}%
\caption{\label{exp:vsc3-pingpong-alternatingrepeated} \ddtalternatingrepeated \vs \ddtalternatingstruct, element datatype: \mpiint, \pingpong, \vscintelmpi.}
\end{figure*}

\appexp{exptest:alternating_repeated}
\appexpdesc{
  \begin{expitemize}
    \item \ddtalternatingrepeated, \ddtalternatingstruct
    \item \pingpong
  \end{expitemize}
}{
  \begin{expitemize}
    \item \expparam{\vscintelmpi}{\fig~\ref{exp:vsc3-pingpong-alternatingrepeated}}
  \end{expitemize}  
}

\FloatBarrier
\clearpage

\appexp{exptest:tiled_struct}

\begin{figure*}[htpb]
\centering
\begin{subfigure}{.24\linewidth}
\centering
\includegraphics[width=\linewidth]{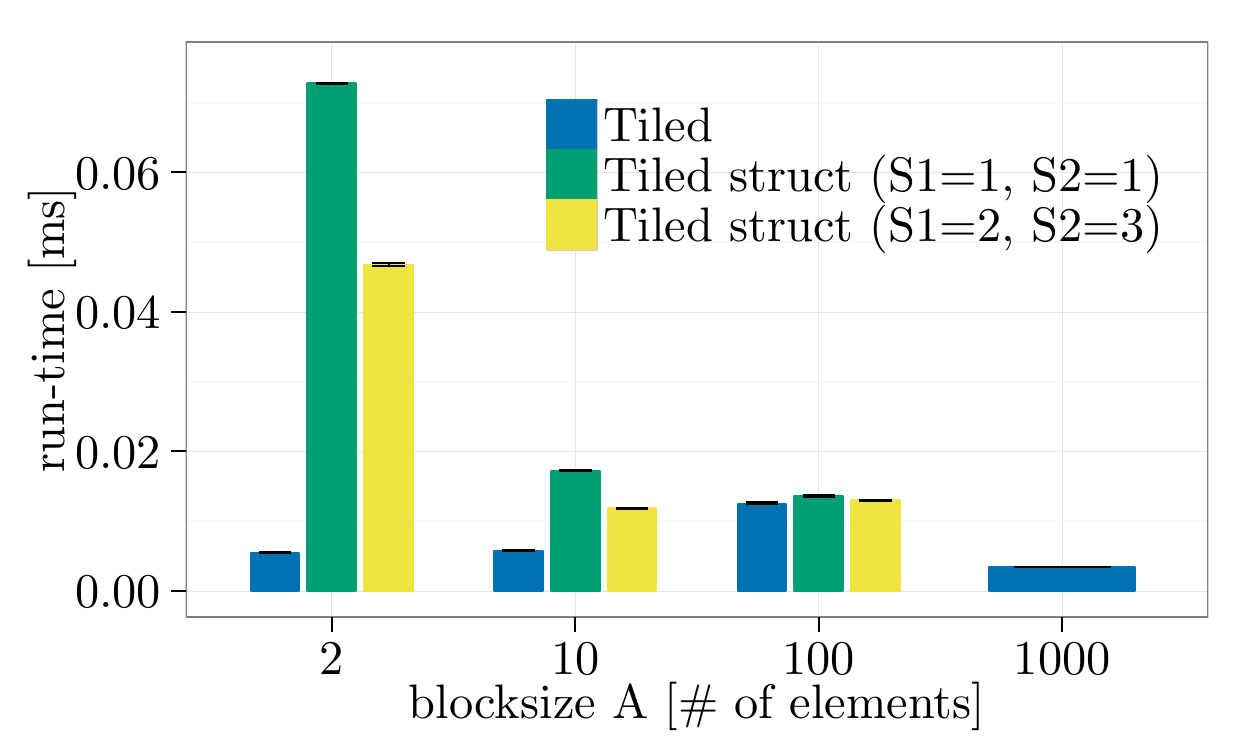}
\caption{%
\label{exp:vsc3-pingpong-tiledstruct-small-2x1}%
$\VARdatasize=\SI{4}{\kilo\byte}$, \num{2}~nodes%
}%
\end{subfigure}%
\hfill%
\begin{subfigure}{.24\linewidth}
\centering
\includegraphics[width=\linewidth]{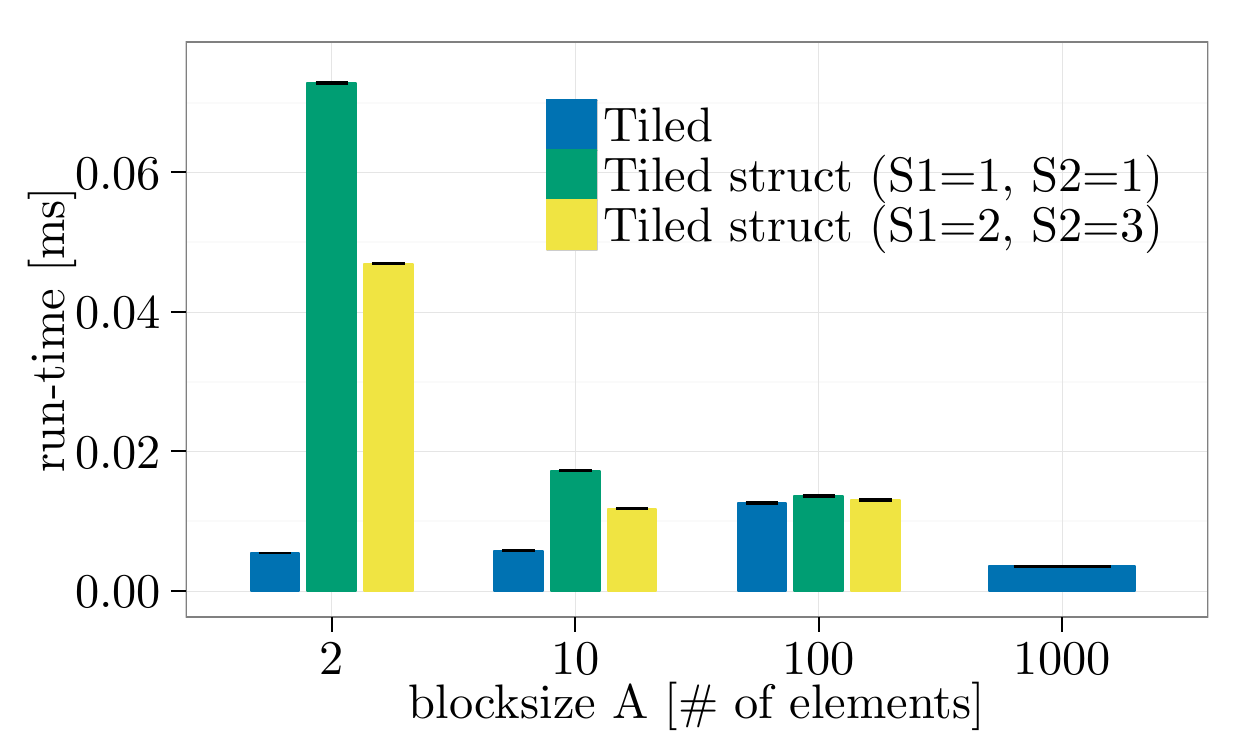}
\caption{%
\label{exp:vsc3-pingpong-tiledstruct-small-1x2}%
$\VARdatasize=\SI{4}{\kilo\byte}$, same node%
}%
\end{subfigure}%
\hfill%
\begin{subfigure}{.24\linewidth}
\centering
\includegraphics[width=\linewidth]{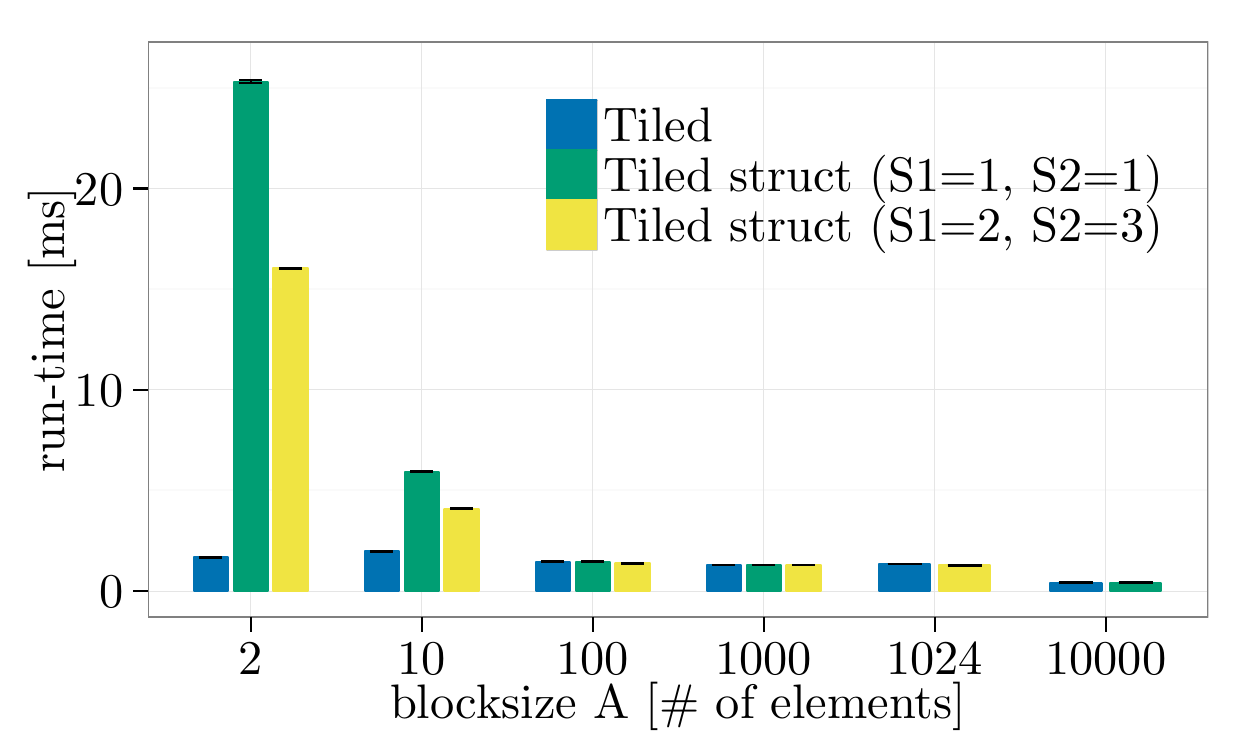}
\caption{%
\label{exp:vsc3-pingpong-tiledstruct-large-2x1}%
$\VARdatasize=\SI{2.56}{\mega\byte}$, \num{2}~nodes%
}%
\end{subfigure}%
\hfill%
\begin{subfigure}{.24\linewidth}
\centering
\includegraphics[width=\linewidth]{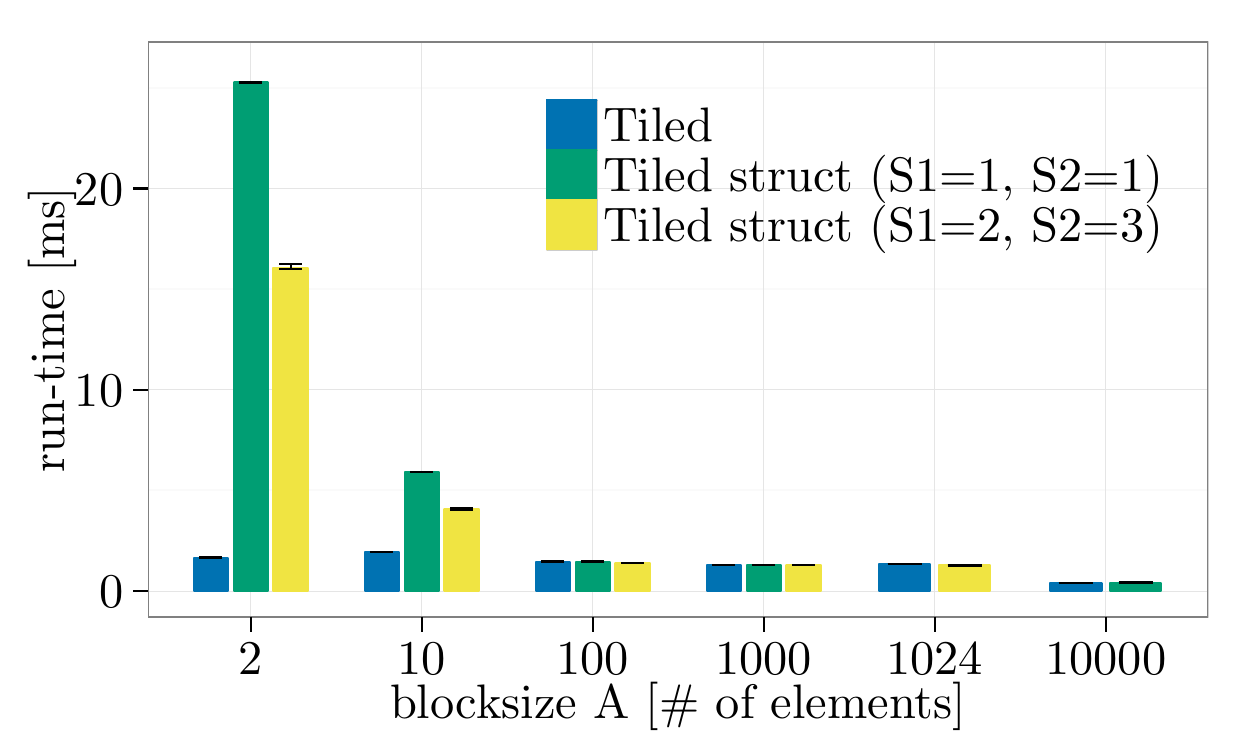}
\caption{%
\label{exp:vsc3-pingpong-tiledstruct-large-1x2}%
$\VARdatasize=\SI{2.56}{\mega\byte}$, same node%
}%
\end{subfigure}%
\caption{\label{exp:vsc3-pingpong-tiledstruct} \dtdtiled \vs \dttiledstruct, element datatype: \mpiint, \pingpong, \vscintelmpi.}
\end{figure*}

\appexpdesc{
  \begin{expitemize}
    \item \dtdtiled, \dttiledstruct
    \item \pingpong
  \end{expitemize}
}{
  \begin{expitemize}
    \item \expparam{\vscintelmpi}{\fig~\ref{exp:vsc3-pingpong-tiledstruct}}
  \end{expitemize}  
}

\FloatBarrier
\clearpage

\appexp{exptest:tiled_vector}

\begin{figure*}[htpb]
\centering
\begin{subfigure}{.24\linewidth}
\centering
\includegraphics[width=\linewidth]{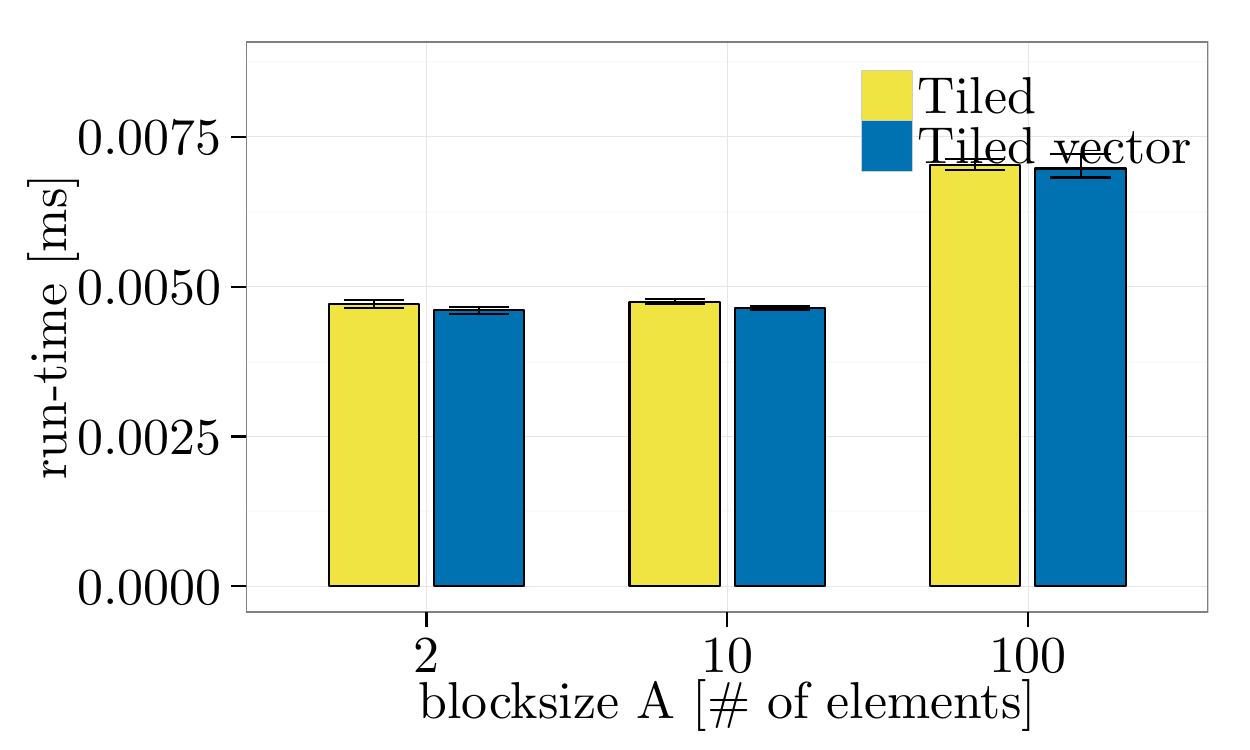}
\caption{%
\label{exp:vsc3-pingpong-tiledvec-small-2x1}%
$\VARdatasize=\SI{2}{\kilo\byte}$, \num{2}~nodes%
}%
\end{subfigure}%
\hfill%
\begin{subfigure}{.24\linewidth}
\centering
\includegraphics[width=\linewidth]{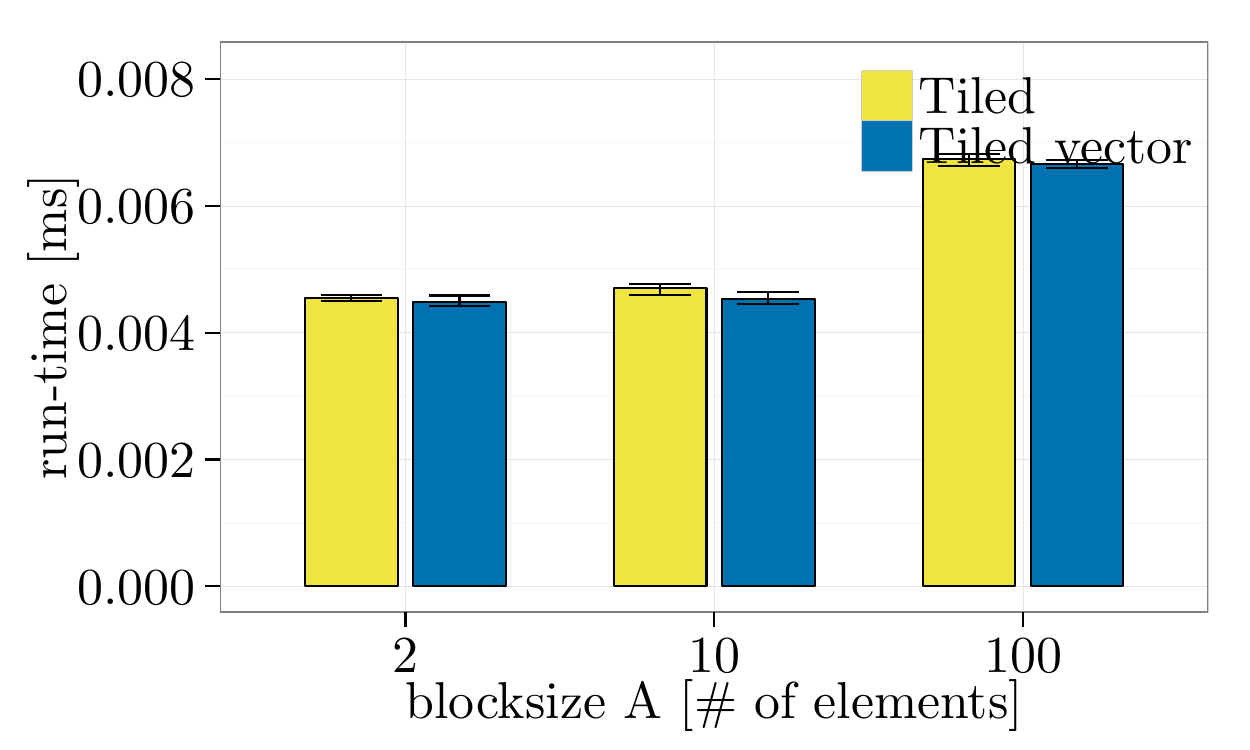}
\caption{%
\label{exp:vsc3-pingpong-tiledvec-small-1x2}%
$\VARdatasize=\SI{2}{\kilo\byte}$, same node%
}%
\end{subfigure}%
\hfill%
\begin{subfigure}{.24\linewidth}
\centering
\includegraphics[width=\linewidth]{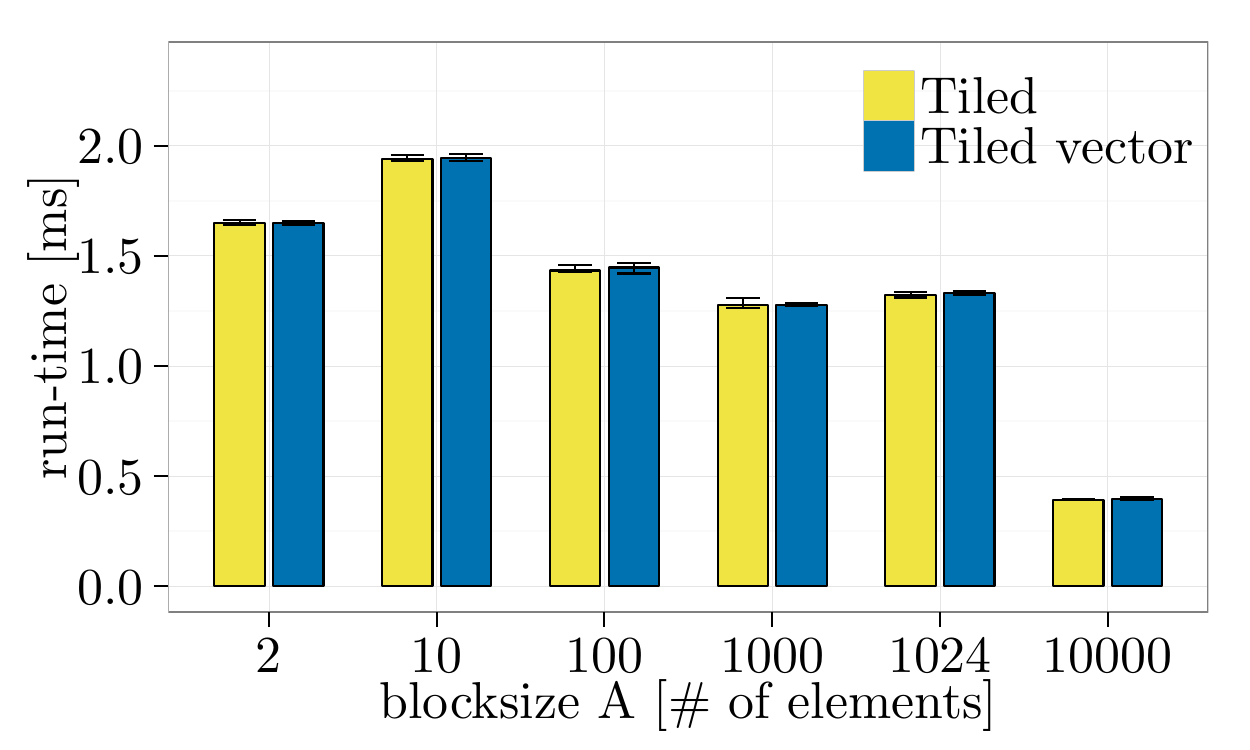}
\caption{%
\label{exp:vsc3-pingpong-tiledvec-large-2x1}%
$\VARdatasize=\SI{2.56}{\mega\byte}$, \num{2}~nodes%
}%
\end{subfigure}%
\hfill%
\begin{subfigure}{.24\linewidth}
\centering
\includegraphics[width=\linewidth]{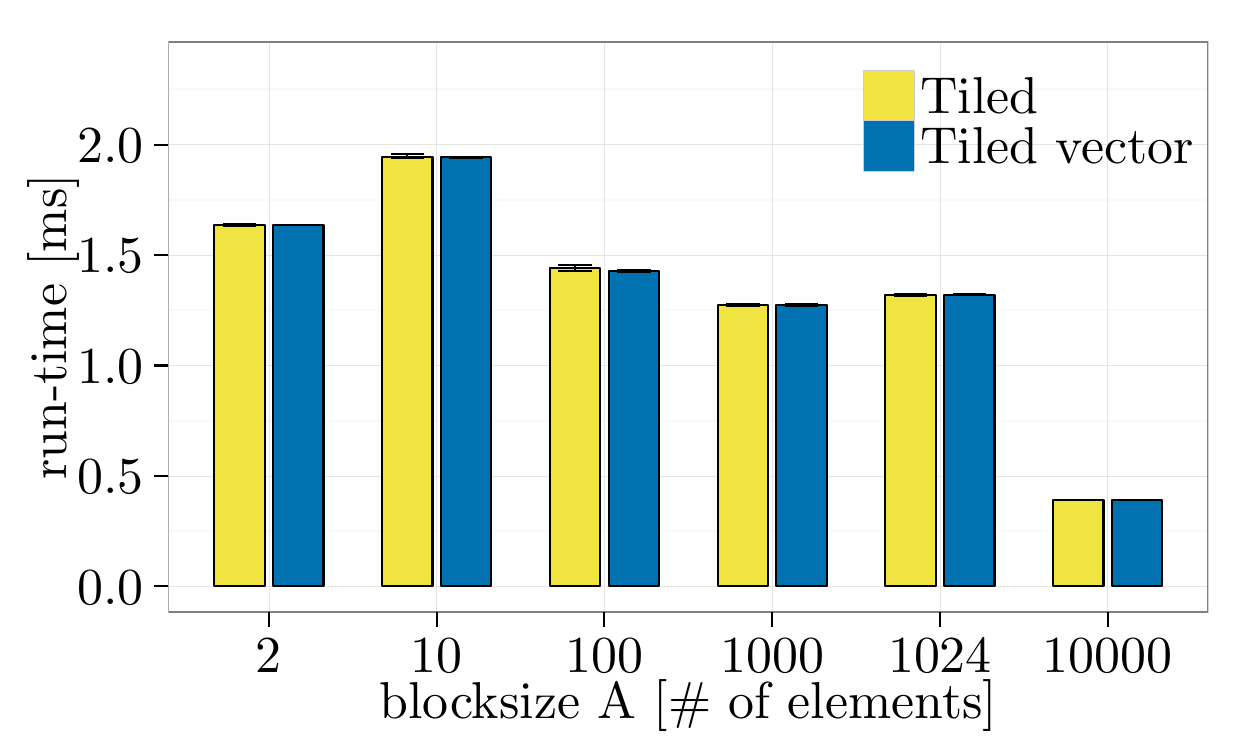}
\caption{%
\label{exp:vsc3-pingpong-tiledvec-large-1x2}%
$\VARdatasize=\SI{2.56}{\mega\byte}$, same node%
}%
\end{subfigure}%
\caption{\label{exp:vsc3-pingpong-tiledvec} \dtdtiled \vs \ddttiledvector, element datatype: \mpiint, \pingpong, \vscintelmpi.}
\end{figure*}

\appexpdesc{
  \begin{expitemize}
    \item \dtdtiled, \ddttiledvector
    \item \pingpong
  \end{expitemize}
}{
  \begin{expitemize}
    \item \expparam{\vscintelmpi}{\fig~\ref{exp:vsc3-pingpong-tiledvec}}
  \end{expitemize}  
}

\FloatBarrier
\clearpage

\begin{figure*}[htpb]
\centering
\begin{subfigure}{.24\linewidth}
\centering
\includegraphics[width=\linewidth]{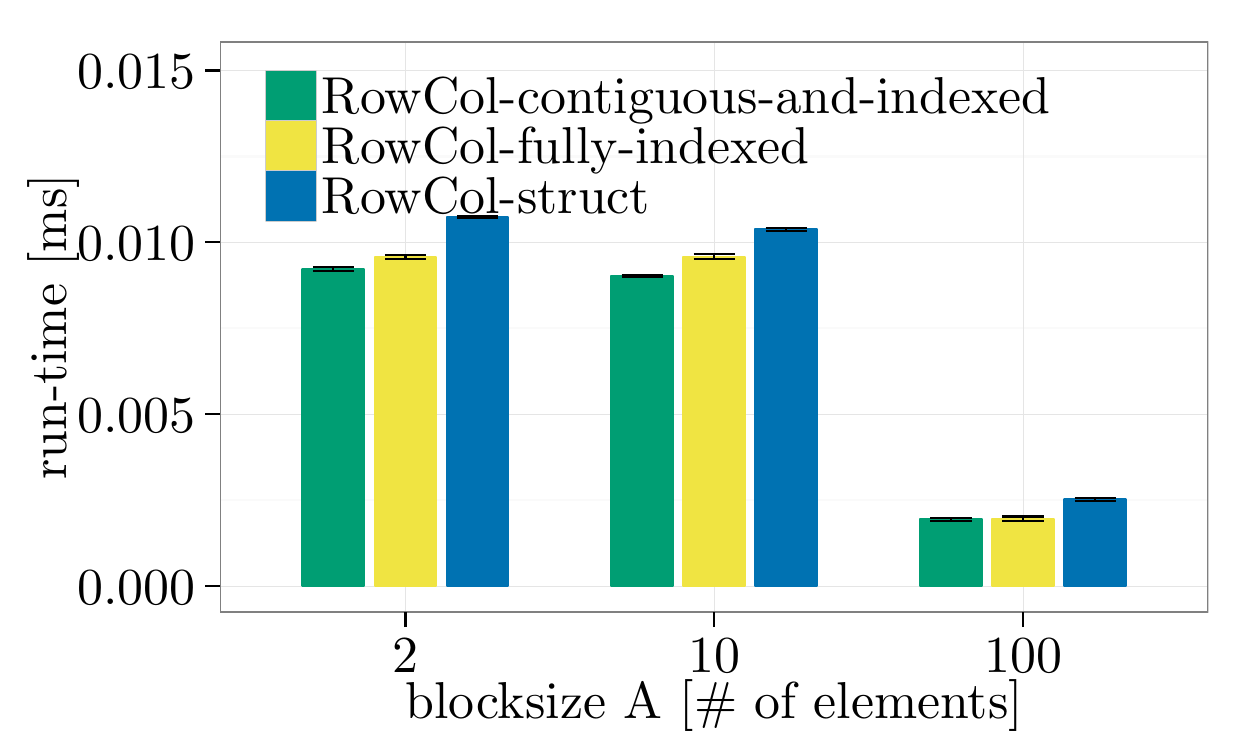}
\caption{%
\label{exp:vsc3-pingpong-rowcol-small-2x1}%
$n=\num{100}$, \num{2}~nodes%
}%
\end{subfigure}%
\hfill%
\begin{subfigure}{.24\linewidth}
\centering
\includegraphics[width=\linewidth]{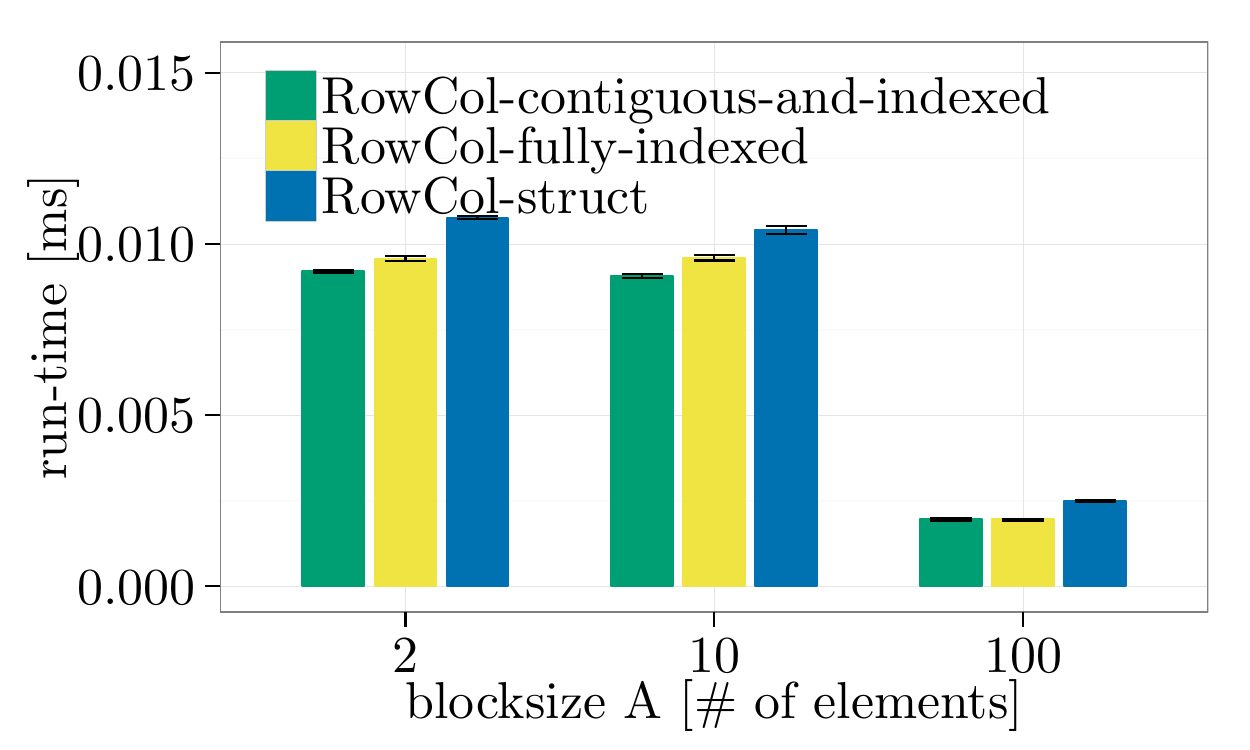}
\caption{%
\label{exp:vsc3-pingpong-rowcol-small-1x2}%
$n=\num{100}$, same node%
}%
\end{subfigure}%
\hfill%
\begin{subfigure}{.24\linewidth}
\centering
\includegraphics[width=\linewidth]{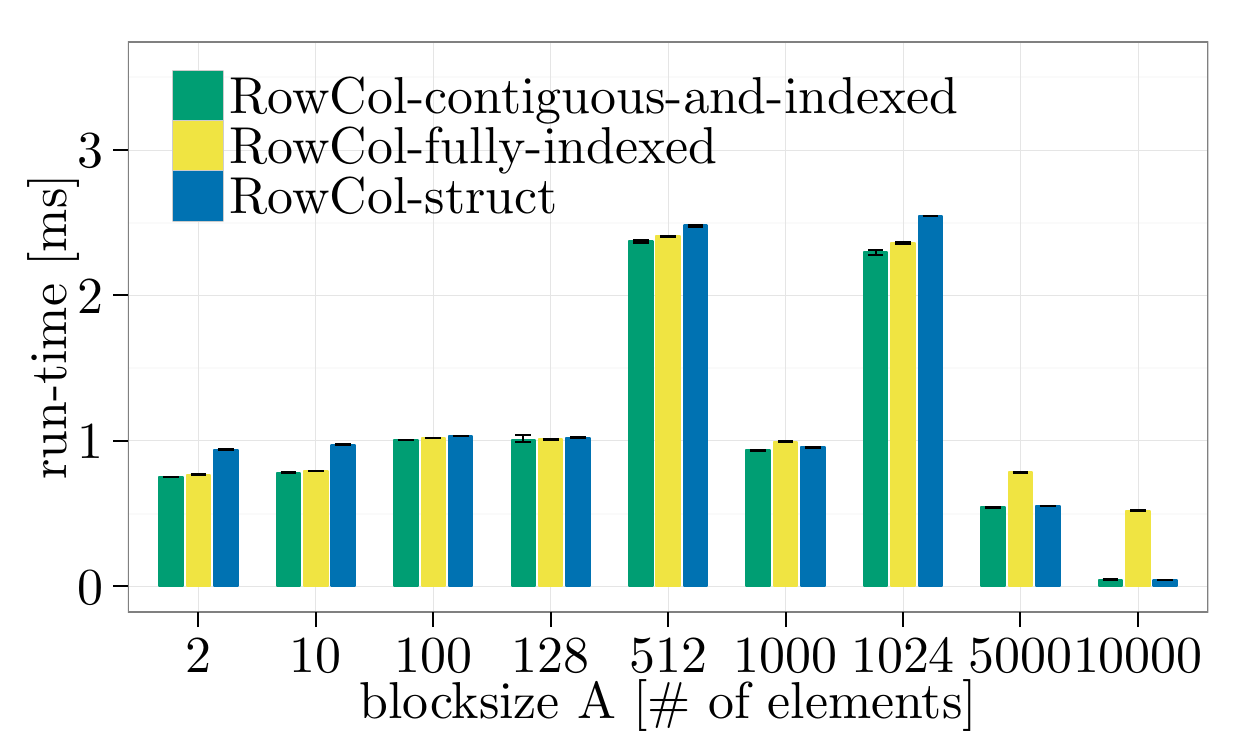}
\caption{%
\label{exp:vsc3-pingpong-rowcol-large-2x1}%
 $n=\num{10240}$, \num{2}~nodes%
}%
\end{subfigure}%
\hfill%
\begin{subfigure}{.24\linewidth}
\centering
\includegraphics[width=\linewidth]{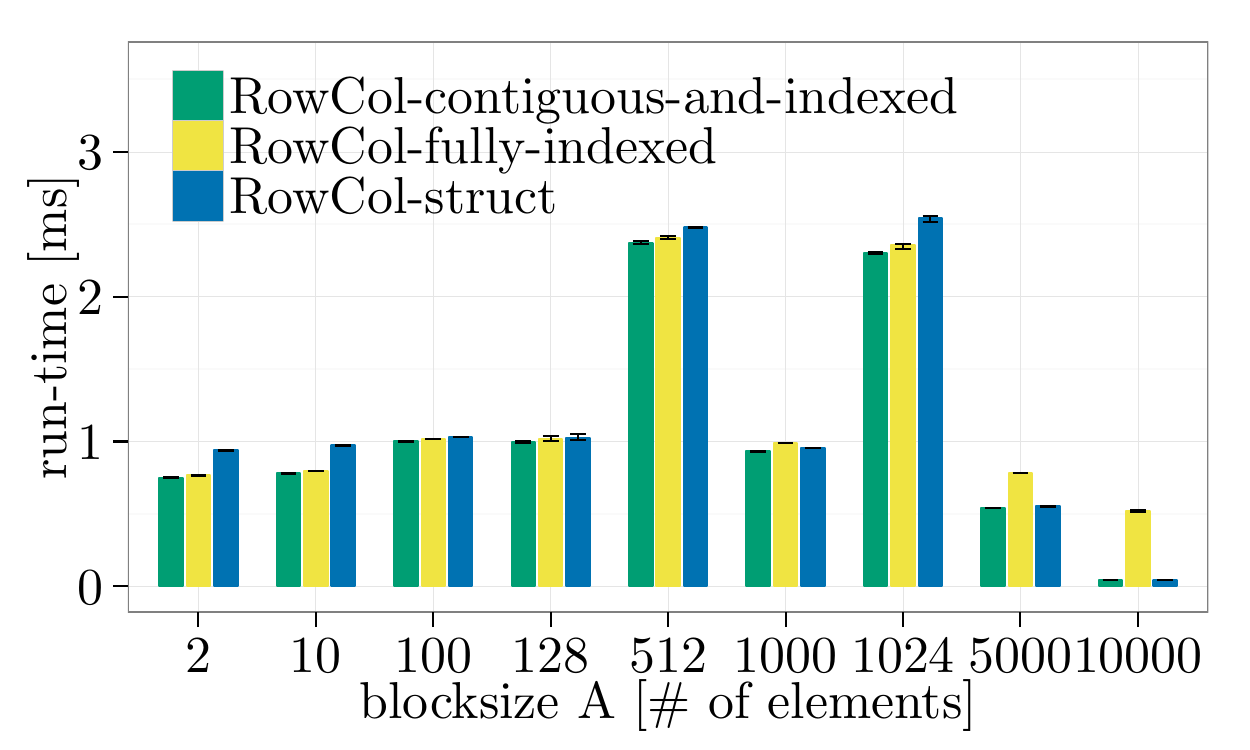}
\caption{%
\label{exp:vsc3-pingpong-rowcol-large-1x2}%
 $n=\num{10240}$, same node%
}%
\end{subfigure}%
\caption{\label{exp:vsc3-pingpong-rowcol}  \ddtrowcolfullindexed, \ddtrowcolcontiguousandindexed, \ddtrowcolstruct, element datatype: \mpiint, \extent increases with \blocksize \VARblocksize, \pingpong, \vscintelmpi.}
\end{figure*}

\appexp{exptest:rowcol}
\appexpdesc{
  \begin{expitemize}
    \item \ddtrowcolfullindexed, \ddtrowcolcontiguousandindexed, \ddtrowcolstruct
    \item \pingpong
  \end{expitemize}
}{
  \begin{expitemize}
    \item \expparam{\vscintelmpi}{\fig~\ref{exp:vsc3-pingpong-rowcol}}
  \end{expitemize}  
}

\end{document}